\documentclass [PhD,nolistoftables,scheader] {uclathes}

\usepackage{amsmath,amssymb,calc, amsthm,bbm, epsfig,psfrag, mathtools,comment}
\usepackage{mathrsfs}
\usepackage{needspace}
\usepackage{graphicx, enumerate}
\usepackage{float}
\usepackage{subcaption}
\usepackage{comment}
\usepackage{hyperref} 
\title          {Holographic Renormalization Group and Stress Tensor Operators}
\author         {Stephen Ebert}
\department     {Physics}
\degreeyear     {2024}

\member         {Zvi Bern}
\chair          {Eric D'Hoker}
\member         {Michael Gutperle}
\chair          {Per J. Kraus}

\dedication     {\textsl{To my parents.}}

\renewcommand   {\acksname}{Acknowledgments}

\acknowledgments{I am grateful to my advisor, Per Kraus, for his support and discussions throughout my graduate school career at UCLA. Additional praise goes towards the high energy theory faculty at UCLA: Zvi Bern, Eric D'Hoker, Thomas Dumitrescu, and Michael Gutperle for facilitating the string theory journal club, organizing the Southern California Strings Seminar at UCLA and Caltech, and making my academic years at UCLA some of the most intellectually stimulating. 

Besides UCLA's high-energy theory faculty, I am also thankful to our theory group's graduate students and postdoctoral fellows from whom I have learned several high-energy theory topics including Sriram Bharadwaj, Kevin Chen, Shouvik Datta, Amey Gaikwad, John Gardiner, Nicholas Geiser, Enrico Herrmann, Justin Kaidi, Seolhwa Kim, Nicholas Klein, Yanyan Li, Lukas Lindwasser, Benjamin Michel, Emily Nardoni, Pierluigi Niro, Julio Parra-Martinez, Konstantinos Roumpedakis, Trevor Scheopner, and Orr Sela. Despite not working from my office frequently, I am happy to have two wonderful office mates, Seolhwa Kim, and Jonah Hyman, to chat with when I was there.

I am incredibly thankful for my excellent collaborators' camaraderie and from whom I have learned a tremendous amount of physics related and unrelated to our projects: Christian Ferko, Eliot Hijano, Per Kraus, Cian Luke Martin, Ruben Monten, Richard Myers, Atul Sharma, Hao-Yu Sun, Zhengdi Sun, Gabriele Tartaglino-Mazzucchelli, Diandian Wang, Ziqi Yan, and Mengyang Zhang. Collaborating with these physicists has enriched my intellectual curiosity and development as a researcher. 

In addition, many other physicists outside of UCLA whom I had the privilege of learning and interacting: Chris Akers, Mykola Dedushenko, Alexander Frenkel, Michael Green, Steven Gubser, Christian Jepsen, Matthew Heydeman, Nathan Haouzi, Fernando Iniguez, Ken Intriligator, Jacob Leedom, John Schwarz, Savdeep Sethi, Alessandro Sfondrini, Dmitri Sorokin, Gabriele Veneziano, and Zhencheng Wang.

I am also thankful to the professors and advisors I had during my time as an undergraduate at UC Berkeley: Ashok Ajoy, James Analytis, Richard Borcherds, Joel Fajans, William Holzapfel, Lawrence Hall, Yury Kolomensky, Daniel Kasen, Alessandra Lanzara, Daniel McKinsey, Joseph Orenstein, Matthias Reinsch, and Peter Sorenson. In particular, I thank Ori Ganor and Petr Ho\v{r}ava for their in-depth courses in string theory, quantum field theory, and general relativity. I especially thank Ori for his mentoring during my numerous visits to the Berkeley Center for Theoretical Physics in the Summer of 2019 when I rode Bay Area Rapid Transit (BART) from Fremont to Berkeley.

I am deeply grateful to my parents, Stephen and Marsha Ebert, siblings Mark and Stephanie Ebert, and grandparents Richard and Maxine Ebert, along with Raffi and Dorthy Krikorian, for their lifelong love and support. Outside of research, I am grateful to the friends I made in San Jose, other cities in the San Francisco Bay Area, and in my academic career: Nijaid Arredondo, Eve Bodnia, Peter Boyle, Evan Bruce, Parker Bruce, Steve Cao, Randall Clark, Andrew Christensen, Tahv Demayo, Jiashu Han, Yifan Hong, Matthew Lee, Ruiyao Liu, and Dhruv Muley.

I acknowledge funding support throughout my Ph.D. from the Bhaumik Institute and the Dissertation Year Fellowship from the UCLA graduate division.

\textbf{Contribution of authors}

Chapter \ref{chFieldTheoryGravitons} of this thesis reports on work with Eliot Hijano, Per Kraus, Ruben Monten, and Richard Myers \cite{Ebert:2022cle}. Chapter \ref{ch:BF} reports on work with Christian Ferko, Hao-Yu Sun and Zhengdi Sun \cite{Ebert:2022ehb}. Chapter \ref{ch:SUSY-QM} reports on work with Christian Ferko, Hao-Yu Sun, and Zhengdi Sun \cite{Ebert:2022xfh}. Chapter \ref{ch:Airy} reports on work with Hao-Yu Sun, and Zhengdi Sun \cite{Ebert:2022gyn}. Chapter \ref{ch:root-TT} reports on work with Christian Ferko and Zhengdi Sun \cite{Ebert:2023tih}. Chapter \ref{ch:ChiralBoson} reports on work with Christian Ferko, Cian Luke Martin, and Gabriele Tartaglino-Mazzucchelli \cite{Ebert:2024zwv}. This thesis omits other published papers \cite{Ebert:2019src,Ebert:2020nqf,Ebert:2020tuy,Ebert:2021mfu,Ebert:2023hba}, in which several authors and I collaborated on.
}

\renewcommand   {\vitaname}{Vita} 

\renewcommand{\vitastretch}{1}

\renewcommand{\vitadatewidth}{1.5in} 

\renewcommand{\vitatextwidth}{4.75in}  
\vitaitem   {2016 -- 2018}
                {B.A. Physics, University of California, Berkeley.}
                \vitaitem{2018 -- 2020}{M.Sc. Physics, University of California, Los Angeles.}
\vitaitem   {2018 -- 2024}
                {Ph.D. Physics, University of California, Los Angeles (Expected).}

\renewcommand{\pubstretch}{1.5}%

\let\oldbibitem\bibitem
\renewcommand{\bibitem}[2][]{\oldbibitem{#2}}

\makeatletter
\providecommand \href@noop [0]{\@secondoftwo}%
\providecommand \href [0]{\begingroup \@sanitize@url \@href}%
\providecommand \@href[1]{\@@startlink{#1}\@@href}%
\providecommand \@@href[1]{\endgroup#1\@@endlink}%
\providecommand \@sanitize@url [0]{\catcode `\\12\catcode `\$12\catcode
  `\&12\catcode `\#12\catcode `\^12\catcode `\_12\catcode `\%12\relax}%
\providecommand \@@startlink[1]{}%
\providecommand \@@endlink[0]{}%
\providecommand \url  [0]{\begingroup\@sanitize@url \@url }%
\providecommand \@url [1]{\endgroup\@href {#1}{\urlprefix }}%
\providecommand \urlprefix  [0]{URL }%
\providecommand \selectlanguage [0]{\@gobble}%
\providecommand \bibinfo  [0]{\@secondoftwo}%
\providecommand \bibfield  [0]{\@secondoftwo}%
\providecommand \BibitemShut  [1]{\csname bibitem#1\endcsname}%
\let\auto@bib@innerbib\@empty

\customCV{
	\vfill
	\begin{mypublications}{9}
\item S. Ebert, C. Ferko, C. L. Martin, and G. Tartaglino-Mazzucchelli, ``Flows in the Space of Interacting Chiral Boson Theories,'' \href{https://arxiv.org/abs/2403.18242}{arXiv: 2403.18242}.
  
\item S. Ebert, Z. Yan, ``Anistropic Compactification of Nonrelativistic M-theory,''\\ \href{https://link.springer.com/article/10.1007/JHEP11(2023)135}{JHEP \textbf{11} (2023) 135}. 

\item S. Ebert, C. Ferko, and Z. Sun, ``Root-$T\overline{T}$ Deformed Boundary Conditions in Holography,'' \href{https://journals.aps.org/prd/abstract/10.1103/PhysRevD.107.126022}{Phys. Rev. D \textbf{107}, 126022}.
 
\item S. Ebert, C. Ferko, H.-Y Sun and Z. Sun, ``$T\overline{T}$ in JT Gravity and BF Gauge Theory,'' \href{https://scipost.org/10.21468/SciPostPhys.13.4.096}{SciPost Phys. \textbf{13} (2022) 096}.

\item S. Ebert, C. Ferko, H.-Y. Sun, and Z. Sun, ``$T\overline{T}$ Deformations of Supersymmetric
Quantum Mechanics,'' \href{https://link.springer.com/article/10.1007/JHEP08(2022)121}{JHEP \textbf{08} (2022) 121}.
 
 \item  S. Ebert, H.-Y. Sun, and Z. Sun, ``$T\overline{T}$-deformed free energy of the Airy model,'' \href{https://link.springer.com/article/10.1007/JHEP08(2022)026}{JHEP \textbf{08} (2022) 026}.
 
 \item  S. Ebert, E. Hijano, P. Kraus, R. Monten, and R. M. Myers, ``Field Theory of
Interacting Boundary Gravitons,'' \href{https://scipost.org/SciPostPhys.13.2.038}{SciPost \textbf{13} (2022) 038}.

\item S. Ebert, H.-Y. Sun, and Z. Yan, ``Dual D-brane actions in nonrelativistic string theory,'' \href{https://link.springer.com/article/10.1007/JHEP04(2022)161}{JHEP \textbf{04} (2022) 161}.

\item S. Ebert, H.-Y. Sun, and Z. Sun, ``$T\overline{T}$ deformation in SCFTs and integrable
supersymmetric theories,'' \href{https://link.springer.com/article/10.1007/JHEP09(2021)082}{JHEP \textbf{09} (2021) 082}.

\item S. Ebert, A. Sharma, and D. Wang, ``Descendants in celestial CFT and emergent
multi-collinear factorization,'' \href{https://link.springer.com/article/10.1007/JHEP03(2021)030}{JHEP \textbf{03} (2021) 030}.

\item S. Ebert, H.-Y. Sun, and M.-Y. Zhang, ``Probing holography in $p$-adic CFT,'' \href{https://journals.aps.org/prd/abstract/10.1103/PhysRevD.107.126011}{Phys. Rev. D \textbf{107}, 126011}.

	\end{mypublications}%
	\vfill
}
\makeatother

\abstract {
The holographic duality conjectures a relation between strongly coupled quantum systems and quantum gravity in higher-dimensional spacetimes. Gravitational theories in two and three dimensions are meaningful examples for classical and quantum exploration due to their unique characteristics, notably the absence of propagating bulk degrees of freedom and the presence of only boundary degrees of freedom, distinguishing them from higher-dimensional counterparts. These gravitational theories exhibit complex interactions when the bulk spacetime has a finite size, regulated by Zamolodchikov's double-trace irrelevant $T\overline{T}$ operator.

This thesis aims to gain a holographic understanding of $\mathrm{AdS}_3$ and JT gravity under the influence of the $T\overline{T}$ deformation. Under a finite radial cutoff, these theories exhibit perturbative behavior that implies the emergence of the Nambu-Goto action for the corresponding boundary graviton action. We also conducted semi-classical calculations of observables related to finite-cutoff gravity and its dual $T\overline{T}$-deformed CFT description, including correlation functions involving stress tensors and gravitational Wilson lines, along with an analysis of their supersymmetric extensions. Additionally, we explored the implications of general stress tensor deformations within field-theoretic and holographic settings.

This thesis integrates previously adapted publications while also pioneering new ground, notably exploring the definition of a quantum $T\overline{T}$ operator beyond two dimensions with $\frac{1}{N}$ corrections, investigating quantum-corrected higher point correlators for a planar boundary, and offering insights into a two-dimensional spherical boundary at a finite cutoff. Furthermore, throughout the thesis, we show more details in calculations at various points.
\phantom{\cite{Ebert:2019src}}
}

\begin {document}

\makeintropages

%
%

\chapter{Introduction}

In contemporary perspectives, the holographic principle provides the most comprehensive framework for understanding quantum gravity, sometimes called gauge/gravity duality or the Maldacena conjecture, which, loosely speaking, states that quantum gravity in $d+1$ dimensions is equivalent to a $d$-dimensional field theory. The AdS/CFT correspondence stands out as a solid and early foundation for this phenomenon, providing what is arguably our most modern advanced understanding of quantum gravity. AdS stands for ``anti-de Sitter,'' which is the maximally symmetric solution of Einstein's equations with a negative cosmological constant. CFT stands for ``conformal field theory,'' which is a type of quantum field theory characterized by its invariance under conformal transformations. The AdS/CFT correspondence is one of the most significant results that string theory produced during the second superstring revolution.

String theory, ambitiously aiming to unify quantum mechanics and general relativity, gained widespread acceptance in the first superstring revolution (1984-1985). Key developments include the Green-Schwarz anomaly mechanism \cite{Green:1984sg}, validating the consistency of supersymmetric gauge theories in ten dimensions when coupled to supergravity with gauge groups $SO(32)$ or $E_8 \times E_8$. The discovery of heterotic string theories with these two gauge groups \cite{Gross:1984dd} and the realization that $E_8 \times E_8$ heterotic string theory allows solutions with a six-dimensional Calabi-Yau space, resulting in a realistic four-dimensional field theory at low energies \cite{Candelas:1985en}, marked significant progress. Eventually, five superstring theories emerged in ten dimensions: type I, type IIA, type IIB, $SO(32)$ heterotic, and $E_8 \times E_8$ heterotic \cite{Green:1981yb, Gross:1984dd}. The understanding of non-perturbative effects awaited the second superstring revolution in the mid-1990s.

The second superstring revolution brought groundbreaking advancements, unveiling perturbative and non-perturbative dualities (i.e. T-duality and S-duality) that connect the five distinct ten-dimensional superstring theories. These theories emerge from a single theory called M-theory found by Edward Witten \cite{Witten:1995ex}, which lacks description in terms of quantized fundamental strings but instead described by non-perturbative solitonic membranes living in eleven dimensions. There were other developments, including the Matrix-theory interpretation of M-theory \cite{Banks:1996vh, Susskind:1997cw,Seiberg:1997ad, Sen:1997we,Banks:1996my,Dijkgraaf:1997vv,Douglas:1997uy,Taylor:1997dy,Taylor:2001vb}, the black hole entropy derived from string theory \cite{Strominger:1996sh}, and the establishment of the AdS/CFT correspondence \cite{Maldacena:1997re,Gubser:1998bc,Susskind:1998dq,Witten:1998qj} were key developments.\footnote{For reviews and textbooks on string/M-theory, see \cite{Green:1987sp, Green:1987mn, Polchinski:1998rq, Polchinski:1998rr, Johnson:2023onr, Becker:2006dvp, Blumenhagen:2013fgp, Kiritsis:2019npv, Schwarz:1996bh, Vafa:1997pm, Taylor:2003gn, Agmon:2022thq}.}

To truly grasp the significance of the AdS/CFT correspondence, it is essential to delve deeper. On one side, we have a theory of quantum gravity, operating in spacetime with $d + 1$ dimensions, encompassing $d$ spatial dimensions and one temporal dimension. On the other side, there is what we call a ``CFT'' also inhabiting $d$ spacetime dimensions. This CFT does not incorporate gravity. In essence, the AdS/CFT correspondence reveals the emergence of an additional spatial dimension as we transition between the gravity-free realm of the CFT and the gravity-filled AdS space.

In 1997, Juan Maldacena proposed groundbreaking dualities that connect specific string and M-theory solutions in AdS$_{d+1}$ times a compact space to a CFT$_d$ \cite{Maldacena:1997re}. The shared $SO(d, 2)$ symmetry between the conformal symmetry group and AdS$_{d+1}$'s isometry group supports these connections. Gubser, Klebanov, Polyakov, and Witten established the GKPW dictionary, precisely relating amplitudes and correlation functions in these theories \cite{Gubser:1998bc,Witten:1998qj}.

Maldacena explored three maximally symmetric backgrounds, with the most studied in the literature being type IIB superstring theory in $\operatorname{AdS}_5 \times S^5$ dual to four-dimensional $\mathcal{N}=4$ super Yang-Mills theory with an $SU(N)$ gauge group, based on $N$ coincident D3-branes. Two other examples linked M-theory in $\operatorname{AdS}_4 \times S^7$ and $\operatorname{AdS}_7 \times S^4$ to three-dimensional and six-dimensional superconformal field theories. A decade later from \cite{Maldacena:1997re}, the understanding of the $\operatorname{AdS}_4 \times S^7$ example was expanded by Aharony, Bergman, Jafferis, and Maldacena (ABJM) \cite{Aharony:2008ug}. They considered a more general setting: M-theory on $\operatorname{AdS}_4 \times S^7/\mathbb{Z}_k$ with $k$ as the level of the Chern-Simons terms, leading to a dual gauge theory, a three-dimensional $\mathcal{N} = 6$ superconformal Chern-Simons theory with gauge group $U(N)_k \times U(N)_{-k}$ and $N$ coincident M2-branes. The $\operatorname{AdS}_7 \times S^4$ example, involving $N$ coincident M5-branes, remains challenging and less understood.

The foundation of the AdS/CFT correspondence comes from a strategy developed by 't Hooft in 1974 for the large-$N$ expansion of quantum chromodynamics (QCD) and similar theories \cite{tHooft:1973alw}. This strategy posits that a gauge theory with large degrees of freedom $N$ can be equated with a string theory. In this context, the primary variables for the gauge theory are $N \times N$ matrices, where $N$ corresponds to the rank of a specific gauge group.

However, while this universal argument provides a starting point for the large-$N$ duality, it doesn't inherently indicate which specific string theory is the dual description of a given gauge theory. Determining the dynamics of the string worldsheet requires separate, independent methods, a task that is feasible only in select instances. One notable example is found in maximally supersymmetric Yang-Mills theories, which possess distinct characteristics that enable the unique identification of their dual string theory, as elucidated by Maldacena \cite{Maldacena:1997re}.

This thesis explores the AdS/CFT correspondence in low dimensions under stress tensor deformations.\footnote{For more in-depth background and applications of the holographic duality, refer to \cite{Schwarz:1998fd,Douglas:1999ww,Aharony:1999ti,Klebanov:2000me,DHoker:2002nbb,Maldacena:2003nj,Kraus:2006wn,Polchinski:2010hw,Harlow:2018fse}.}

\section{AdS/CFT in low dimensions}

Despite the rigorous string-theoretic formulation and motivation of the AdS/CFT correspondence from the previous section, the earliest precursor of the AdS/CFT correspondence comes from the study of pure AdS$_3$ gravity by Brown and Henneaux \cite{Brown:1986nw}. Brown and Henneaux famously discovered the asymptotic spacetime algebra of AdS$_3$ given by two copies of the Virasoro algebra with central charge $c = \frac{3 \ell}{2 G}$. The CFT quantity $c$ is related to two bulk quantities: $\ell$ being the AdS$_3$ length scale and Newton's gravitational coupling $G$. The large central charge limit corresponds to semi-classical behavior, whereas finite central charge leads to fully quantized AdS$_3$ gravity. 

The perturbative expansion around an AdS$_3$ background is well understood: one obtains a theory of boundary gravitons governed by Virasoro symmetry \cite{Brown:1986nw,Maloney:2007ud}.  The quantum theory of these boundary gravitons is perfectly sensible and self-contained, with a well-defined Hilbert space and spectrum of local operators.  Indeed, the boundary graviton theory is simple as the action and stress tensor are rendered quadratic in appropriate field variables \cite{Alekseev:1988ce,Alekseev:1990mp,Cotler:2018zff}.  In CFT parlance, this theory describes the Virasoro vacuum block of some putative CFT with some spectrum of primary operators.  A much-studied problem is how to reconcile the desired modular invariance of such a spectrum with a sum over geometries interpretation in gravity, e.g., \cite{Witten:2007kt,Maloney:2007ud,Hellerman:2009bu}.

There are two salient features of three-dimensional gravity motivating this thesis. 

The first feature is that the three-dimensional Einstein-Hilbert action
\begin{equation}
\label{eq:EH}
    I [g_{\mu \nu}] =- \frac{1}{16 \pi G} \int_{M_3} d^3x \sqrt{g} \left(R - 2\Lambda \right) - \frac{1}{8 \pi G} \int_{\partial M_3} d^2x \sqrt{h} \left( K - 1 \right) \,.
\end{equation}
in the first-order formulation of general relativity; may be written semi-classically as a Chern-Simons gauge theory, as observed by \cite{Achucarro:1986uwr,Witten:1988hc}. One can then calculate fundamental observables in the gravitational Chern-Simons theory, such as Wilson lines and loops obtained from path-ordered exponential integrals of the one-form connection $A_\mu(x)$ along an open interval and a closed contour, respectively.

In the following discussion, we will use the term ``gravitational Wilson lines'' to refer to Wilson lines associated with the Chern-Simons gauge field $A_\mu$. When considering two endpoints, denoted as $Z_1 = (r_1, z_1)$ and $Z_2 = (r_2, z_2)$ located on the AdS$_3$ boundary, the gravitational Wilson line linking these points can be expressed as:
\begin{equation}
 W[Z_2, Z_1] = P \exp \left( \int^{Z_2}_{Z_1} A_\mu(x)  \, dx^\mu \right) \,.
\end{equation}
In the classical (i.e. large-$c$) limit, the object $W[Z_2, Z_1]$ has the peculiar property that it transforms as a bi-local primary operator at its endpoints. In \cite{Besken:2018zro,DHoker:2019clx}, at least in perturbative terms with respect to $\frac{1}{c}$, the \emph{quantum} Wilson line appears to undergo a transformation resembling that of a bi-local primary operator at its endpoints. As argued in \cite{Fitzpatrick:2016mtp,Besken:2016ooo,Hikida:2017ehf,Anand:2017dav,Hikida:2018dxe,Kraus:2018zrn,Besken:2018zro,DHoker:2019clx}, another physical feature of the Wilson line is that it serves as a convenient repackaging of the Virasoro vacuum OPE block
\begin{equation}
\label{eq:defVacuumOPE1}
    \langle T_{zz}(w_1) \cdots T_{zz}(w_n) W[z_{2}, z_{1}] \rangle_0=\langle T_{zz}(w_1)\cdots T_{zz}(w_n) O (z_2) O(z_1)\rangle_0 \,,
\end{equation}
where $\langle W[z_2, z_1] \rangle_0$ will be later defined by \eqref{eq:expandedW} in terms of a path-ordered exponential integral involving only the stress tensor operator's holomorphic component $T_{zz}$.

The Virasoro vacuum OPE block -- whose characteristics are similar to those of the gravitational Wilson line -- precisely captures all operators built out of the stress tensor operator appearing in the OPE of two primary operators $O(z_1)$ and $O(z_2)$. Schematically (suppressing numerical factors, coordinate dependence, and derivatives in the OPE coefficient $C_{OOO_i}$, as well as omitting the $T_{\bar{z} \bar{z}}$ piece), the first term in the following OPE of two primary operators
\begin{equation}
\label{eq:primaryOPE}
    O(z_2) O(z_1) = \left(1 + T_{zz} + T_{zz} T_{zz} + \cdots \right) + \sum_i C_{OOO_i} \left( O_i + O_i T_{zz} + O_i T_{zz} T_{zz} + \cdots \right)\,,
\end{equation}
corresponds to the Virasoro vacuum OPE block.  

From the bulk perspective, the (open) gravitational Wilson line calculates the exponential of the worldline action for a massive point particle, accounting for gravitational self-interaction effects that renormalize its mass \cite{Besken:2017fsj,Besken:2018zro,Iliesiu:2019xuh,DHoker:2019clx}. Conversely, the closed gravitational Wilson line, or the Wilson loop, quantifies the holonomy of the gauge connection. In the context of a BTZ black hole, when the Wilson loop encircles the horizon, its value corresponds to the Bekenstein-Hawking entropy \cite{deBoer:2013gz,Ammon:2013hba}.

The second feature is that upon dimensionally reducing $3d$ gravity \eqref{eq:EH} on a circle with a radius equal to the dilaton $\Phi$, one obtains the JT gravity action
\begin{equation}
\label{eq:JT gravity Action}
I [g^{(2)}_{\mu \nu}, \Phi] = - \frac{1}{16 \pi G} \int_{M_2} d^2x \sqrt{g^{(2)}} \Phi \left( R - 2 \Lambda \right) - \frac{1}{8 \pi G} \int_{\partial M_2} d \tau \sqrt{\gamma} \Phi \left( K - 1 \right)\,,
\end{equation}
where $g^{(2)}_{\mu \nu}$ is the $2d$ bulk metric and $\gamma_{\tau \tau}$ is the $1d$ boundary metric. The holographic dual description of JT gravity is the $1d$ Schwarzian theory \cite{Jensen:2016pah,Maldacena:2016upp,Engelsoy:2016xyb,Stanford:2017thb,Saad:2019lba}.

Just as one can reduce $3d$ gravity on a circle in the metric formalism, one may also perform the dimensional reduction for the Chern-Simons description of $3d$ gravity and obtain the so-called BF gauge theory, which enjoys similar features to $2d$ Yang-Mills gauge theory with a non-compact group. BF theory with gauge group $G$ is holographically dual to the worldline theory of a particle with a free kinetic Lagrangian and moves on target space $G$; we call this the ``particle-on-a-group'' theory. 

Given these complementary perspectives and technical advantages of lower dimensional gravity, it is desirable to understand these corners of the following diagram below in different settings.
\begin{align}\label{gravity_diagram}
\includegraphics[width=.95\linewidth]{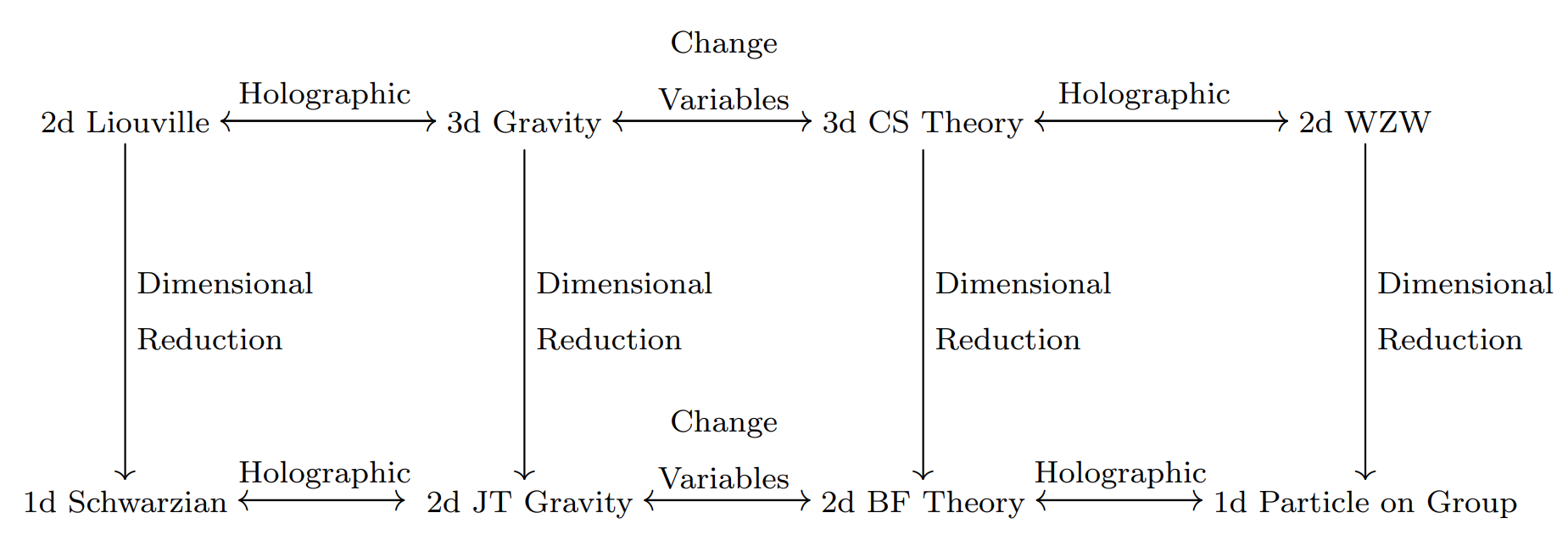}
\end{align}

An area of particular relevance for this thesis involves the examination of diagram \eqref{gravity_diagram} in the presence of Zamolodchikov's double-trace $T\overline{T}$ operator, which is considered an irrelevant operator \cite{Zamolodchikov:2004ce}. Let us now review the $T\overline{T}$ deformation.

\section{Zamolodchikov's $T\overline{T}$ deformation}
Understanding irrelevant deformations are notoriously difficult compared to marginal and relevant deformations. Turning on an irrelevant operator will generically turn on infinitely many additional operators at high energies, which modifies the theory in the UV and leads to a loss of predictive power. However, the $T\overline{T}$ deformation is one irrelevant operator\footnote{The stress tensor has mass dimension $d$ in $d$-dimensions, so the $T\overline{T}$ operator has dimension $2d$ and is irrelevant in any number of spacetime dimensions since $2d > d$. Marginal operators have scaling dimension $\Delta = d$ while relevant operators have scaling dimension $\Delta > d$.} which circumvents these technical difficulties: a $T\overline{T}$-deformed quantum field theory remains under some analytic control and is ``solvable.'' Since this operator is irrelevant, a $T\overline{T}$ flow may seem to be the opposite of a conventional renormalization group flow triggered by the addition of a relevant operator. However, a conventional RG flow connects a family of local QFTs controlled by an RG fixed point in the UV. It is known that a $T\overline{T}$-deformed field theory is not a local QFT and thus not controlled by a CFT in the UV, so this flow is \textit{not} like a conventional RG flow even in reverse.

A remarkable consequence arising from the behavior of $T\overline{T}$-deformed one-point functions is the existence of a differential equation governing the finite-volume spectrum of a $T\overline{T}$-deformed theory. This equation establishes a relationship between energy levels in the $T\overline{T}$-deformed theory and those in the undeformed seed theory \cite{Smirnov:2016lqw,Cavaglia:2016oda}. This relationship between the energy levels exemplifies what we mean by characterizing the deformation as ``solvable.'' The $T\overline{T}$ deformation is solvable because various quantities within the deformed theory, including the torus partition function \cite{Cardy:2018sdv,Datta:2018thy,Aharony:2018bad}, flat space S-matrix \cite{Dubovsky:2013ira,Dubovsky:2017cnj}, and correlation functions \cite{Kraus:2018xrn,Cardy:2019qao,Aharony:2018vux,He:2019vzf,He:2019ahx,He:2020udl,Hirano:2020nwq,Ebert:2020tuy,Ebert:2022gyn,1836175,Ebert:2022cle,Cardy:2022mhn}, can be expressed in terms of corresponding data from the undeformed theory.

It is worth noting that the $T\overline{T}$ operator in two spacetime dimensions retains several symmetries of the original seed theory. These preserved symmetries encompass integrability \cite{Smirnov:2016lqw,Dubovsky:2017cnj} and supersymmetry \cite{Baggio:2018rpv,Chang:2018dge,Jiang:2019hux,Chang:2019kiu,Coleman:2019dvf}. For more information on manifestly supersymmetric $T\overline{T}$-like flows, refer to \cite{Jiang:2019trm,Cribiori:2019xzp,Ferko:2019oyv,Babaei-Aghbolagh:2020kjg,Ferko:2021loo,Ebert:2022xfh}. However, it's important to note that conformal symmetry is not preserved because the flow parameter $\lambda$ introduces a dimensionful scale to the deformed theory.

The $T\overline{T}$ operator is a $2d$ bilinear operator constructed of the stress tensor $T_{\mu \nu}$ which can be expressed as the determinant
\begin{equation}
\label{eq:09iuhy67}
    \det \left( T_{\mu \nu} \right) = \frac{1}{2} \left( \left( T^\mu{}_\mu \right)^2 - T^{\mu \nu} T_{\mu \nu} \right) \,,
\end{equation}
and is unambiguously defined by point-splitting up to total derivatives of local operators that can be neglected \cite{Zamolodchikov:2004ce,Smirnov:2016lqw}. Zamolodchikov assumed the seed QFT has the following properties: 
\begin{enumerate}
    \item \textbf{Local translation and rotation symmetry} 
    
    The stress tensor is conserved $\partial^\mu T_{\mu \nu} = 0$ and symmetric $T_{\mu \nu} = T_{\nu \mu}$.

    \item \textbf{Global translation symmetry} 
    
 One-point functions are independent of the position $     \langle \mathcal{O}_i (z) \rangle = \langle \mathcal{O}_i(0) \rangle$ 
    and any two-point functions $ \langle \mathcal{O}_i (z) \mathcal{O}_j(z') \rangle = F_{ij} (z-z')$.
    
    \item \textbf{Clustering} 
    
    There exists some direction $r \rightarrow \infty$ such that $  \lim_{r \rightarrow \infty} \langle \mathcal{O}_i(r) \mathcal{O}_j(0) \rangle = \langle \mathcal{O}_i \rangle \langle \mathcal{O}_j \rangle$.

    \item \textbf{UV CFT}
    
    The seed QFT is a CFT at short distances.
\end{enumerate}
With these four conditions, we show the coincident point limit defines a local operator:
\begin{equation}
  T\overline{T} = \lim_{z' \rightarrow z} \left( T_{zz} (z') T_{\bar{z} \bar{z}} (z) - T_{z\bar{z}} (z') T_{z\bar{z}}(z') \right)\,.
\end{equation}

Furthermore, the conservation of the stress tensor gives
\begin{equation}
\begin{aligned}
\label{eq:f0e23ersdfs}
&\partial_{\bar{z}}\left(T_{zz}(z) T_{\bar{z} \bar{z}}\left(z^{\prime}\right)-T_{z\bar{z}}(z) T_{z\bar{z}}\left(z^{\prime}\right)\right)=\left(\partial_z+\partial_{z^{\prime}}\right) T_{z\bar{z}}(z) T_{\bar{z} \bar{z}}\left(z^{\prime}\right)-\left(\partial_{\bar{z}}+\partial_{\bar{z}^{\prime}}\right) T_{z\bar{z}}(z) T_{z\bar{z}} \left(z^{\prime}\right) \\
&\partial_z\left(T_{zz}(z) T_{\bar{z} \bar{z}}\left(z^{\prime}\right)-T_{z\bar{z}}(z) T_{z\bar{z}} \left(z^{\prime}\right)\right)=\left(\partial_z+\partial_{z^{\prime}}\right) T_{zz}(z) T_{\bar{z} \bar{z}}\left(z^{\prime}\right)-\left(\partial_{\bar{z}}+\partial_{\bar{z}^{\prime}}\right) T_{zz}(z) T_{z\bar{z}}\left(z^{\prime}\right)\,.
\end{aligned}
\end{equation}
Next, recall the stress tensor's OPEs
\begin{equation}
    \begin{aligned}
    T_{zz} (z) T_{z\bar{z}} (z')    &= \sum_i A_i (z-z') \mathcal{O}_i(z')\,, \quad T_{z\bar{z}} (z) T_{z\bar{z}}(z') = \sum_i C_i (z-z') \mathcal{O}_i(z')\,, \\
    T_{z\bar{z}} (z) T_{\bar{z} \bar{z}} (z') &= \sum_i B_i (z-z') \mathcal{O}_i(z')\,, \quad T_{zz} (z) T_{\bar{z} \bar{z}} (z') = \sum_i D_i (z-z') \mathcal{O}_i(z')\,.
    \end{aligned}
\end{equation}
The stress tensor conservation equations \eqref{eq:f0e23ersdfs} yield 
\begin{equation}
    \begin{aligned}
&\sum_i \partial_{\bar{z}} \left(  D_i (z-z') - C_i (z-z')  \right) \mathcal{O}_i\left(z^{\prime}\right)\\&=\sum_i\left(B_i\left(z-z^{\prime}\right) \partial_{z^{\prime}} \mathcal{O}_i(z')-C_i\left(z-z^{\prime}\right) \partial_{\bar{z}^{\prime}} \mathcal{O}_i(z')\right)
\end{aligned}
\end{equation}
and
\begin{equation}
\begin{aligned}
    &\sum_i \partial_z \left(  D_i (z-z') - C_i (z-z')  \right)  \mathcal{O}_i\left(z^{\prime}\right)\\&=\sum_i\left(D_i\left(z-z^{\prime}\right) \partial_{z^{\prime}} \mathcal{O}_i\left(z^{\prime}\right)-A_i\left(z-z^{\prime}\right) \partial_{\bar{z}^{\prime}} \mathcal{O}_i\left(z^{\prime}\right)\right)\,.
\end{aligned}
\end{equation}

Consequently, this implies that any operator arising in the following OPE
\begin{equation}
    T_{zz} (z) T_{\bar{z} \bar{z}} (z') - T_{z\bar{z}}(z) T_{z\bar{z}}(z') = \sum_i \left(  D_i (z-z') - C_i (z-z')  \right)  \mathcal{O}_i(z')
\end{equation}
must have a coordinate independent coefficient function $D_i (z-z') - C_i (z-z')$ or itself is the derivative of another local operator
\begin{equation}
    T_{zz} (z) T_{\bar{z} \bar{z}} (z') - T_{z\bar{z}}(z) T_{z\bar{z}} (z') = \mathcal{O}_{T\overline{T}} (z') + \sum_\alpha A_\alpha (z-z') \partial_{z'} O_\alpha (z')\,.
\end{equation}
Therefore, we arrive at the composite operator Zamolodchikov \cite{Zamolodchikov:2004ce} found
\begin{equation}
    T\overline{T} = \mathcal{O}_{T\overline{T}}(z)
\end{equation}
up to derivative terms, but these vanish due to global translation symmetry where a one-point function of a total derivative is zero.

At the classical level, given a seed theory's Lagrangian $\mathcal{L}^{(0)}$, the $T\overline{T}$ flow is captured by the differential equation 
\begin{equation}\label{eq:TTb definition}
    \frac{\partial \mathcal{L}^{(\lambda)}}{\partial \lambda} =- \det \left(  T^{(\lambda)}_{\mu \nu}\right)\,,
\end{equation}
where the notation $T^{(\lambda)}_{\mu \nu}$ emphasizes that at each step along the flow, the stress tensor is recomputed from the deformed Lagrangian $\mathcal{L}^{(\lambda)}$. We give an example of $N$ free massless scalars in the subsequent subsection.

\subsection{$T\overline{T}$-deformed free scalars}
In this subsection, we review the solution for the $T\overline{T}$ flow equation for the deformed Lagrangian of multiple free scalar fields\footnote{Appendix \ref{app:no_trace_flow} considers the case for a scalar theory with a potential $V(\phi)$.}
\begin{equation}
    S(0) = \frac{1}{4\pi} \sum^N_{n=1} \int d^2x \sqrt{g} g^{\mu \nu} \partial_\mu \phi_n \partial_\nu \phi_n\,.
\end{equation}

We first consider a single scalar field and write the ansatz for the deformed action
\begin{equation}
\label{eq:pkmngtyw1}
S(\lambda) = \int d^2x \sqrt{g} \frac{F(\lambda \partial^\mu \phi \partial_\mu \phi)}{\lambda}
\end{equation}
and substitute \eqref{eq:pkmngtyw1} into the flow equation
\begin{equation}
    \frac{dS(\lambda)}{d\lambda} = \int d^2x \sqrt{g} T\overline{T}(x)
\end{equation}
to obtain a differential equation for $F(z)$ with solution 
\begin{equation}
    S(\lambda) = \int d^2x \sqrt{g} \left( \frac{1 - \sqrt{1- \pi \lambda \partial^\mu \phi \partial_\mu \phi}}{2\pi^2 \lambda} \right)\,.
\end{equation}

A more sophisticated ansatz is required when dealing with multiple scalars
\begin{equation}
    \mathcal{L}(\lambda) = F(\lambda g^{\mu \nu} \partial_\mu \phi_n \partial_\nu \phi_n, \lambda^2 g^{\mu \sigma} g^{\nu \rho} \partial_\rho \phi_m \partial_\sigma \phi_m \partial_\mu \phi_n \partial_\nu \phi_n  )
\end{equation}
and the solution to the partial differential equation for $F$ is now a determinant 
\begin{equation}
\label{eq:-0iouhygjhefw}
S(\lambda) =  \int d^2x \left( \frac{\sqrt{g}  - \sqrt{\det \left( g_{\mu \nu} - \pi \lambda \partial_\mu \phi_n \partial_\nu \phi_n \right)} }{2\pi^2 \lambda} \right)\,.
\end{equation}
The action \eqref{eq:-0iouhygjhefw} is the Nambu-Goto action written in static gauge 
\begin{equation}
    X^0 = x^0, \quad X^n = \sqrt{-\pi \lambda} \phi_n, \quad X^{N+1} = x^1
\end{equation}
implying 
\begin{equation}
    S(\lambda) = \frac{1}{2\pi^2 \lambda} \int d^2x \sqrt{\det \left( \partial_\mu X^A \partial_\nu X^A \right)} + \operatorname{constant}\,, \quad A =0,1, \cdots, N+1\,.
\end{equation}
The Nambu-Goto action is invariant under an $SO(N+2)$ global symmetry and reparametrization invariance. The nonlinear realization of the $SO(2)$ symmetry in the gauge-fixed form \eqref{eq:-0iouhygjhefw} arises due to the necessity of implementing a compensatory reparametrization to preserve the static gauge.

As we will see in later chapters of this thesis, the same kind of Nambu-Goto action for the deformed scalar theory also appears in the context of low-dimensional $T\overline{T}$ deformed gravitational theories. For example, in Euclidean signature, the resulting boundary graviton action in cutoff AdS$_3$ with planar boundary is
\begin{equation}
\begin{aligned}
\label{eq:90iojkkuhy78d}
    I(r_c) = \frac{1}{32 \pi G} \int d^2x \bigg[& i \left( f' \dot{f} - \bar{f}' \dot{\bar{f}} \right) \\&- \frac{4}{r_c} \left( \sqrt{1-\frac{r_c}{2} \left( f'^2 + \bar{f}'^2 \right) + \frac{r_c^2}{16} \left( f'^2 - \bar{f}'^2 \right)^2  } - 1  \right) \bigg]
\end{aligned}
\end{equation}
and its dimensional reduction yields the deformed Schwarzian theory in JT gravity. To be explained in greater depth, $r_c$ is the radial location of the boundary such that $r_c \rightarrow 0$ (or $\lambda \rightarrow 0$) is the conformal boundary and the fields $(f, \bar{f})$ nonlinearly realize Lorentz symmetry.

Holographically, the $T\overline{T}$ deformation has the interpretation of gravity with Dirichlet boundary conditions for the metric at a finite radial cutoff $r_c$ which is related to $\lambda$ via the $T\overline{T}$ flow equation \eqref{eq:TTb definition}. 

In the forthcoming chapters, we delve into $T\overline{T}$-deformed low-dimensional holography to understand \eqref{gravity_diagram} in greater detail. Nevertheless, before delving into that discussion, we will briefly touch upon some distinctive features of the $T\overline{T}$ deformation. Following this, we will provide the rationale for exploring the AdS/CFT correspondence with a finite radial cutoff in the subsequent two sections.

\section{Factorizaton and flow equation}
\label{sec:factorfloweq}
We derive the $T\overline{T}$ flow equation for the energy levels of a QFT on a cylinder with radius $R$. We demonstrate that in the case of a conformal seed theory, the energy spectrum of the $T\overline{T}$-deformed system exhibits square root behavior. Throughout this thesis, we exclusively examine seed theories characterized by conformal symmetry, and we anticipate observing analogous square root-like behaviors in these cases. 

It is imperative to establish the independence of the expectation value of the $T\overline{T}$ operator with respect to the distance $D(z, w)$ between insertion points
\begin{equation}
    D(z,w) = \langle T_{zz} (z) T_{\bar{z} \bar{z}} (w) \rangle - \langle T_{z\bar{z}} (z) T_{z\bar{z}} (w) \rangle\,.
\end{equation}
Next, use \eqref{eq:f0e23ersdfs} and differentiate with respect to $\bar{z}$
\begin{equation}
    \partial_{\bar{z}} D(z, w) = \partial_z \langle T_{z\bar{z}}(z) T_{\bar{z}\bar{z}}(w) \rangle - \partial_{\bar{z}} \langle T_{z\bar{z}} (z) T_{z\bar{z}}(w) \rangle\,.
\end{equation}
From the assumption of global translation invariance, 
\begin{equation}
    \partial_{\bar{z}} \langle T_{z\bar{z}}(z) T_{z\bar{z}}(w) \rangle = - \partial_{\bar{w}} \langle T_{z\bar{z}}(z) T_{z\bar{z}} (w) \rangle
\end{equation}
and
\begin{equation}
    \partial_z \langle T_{z\bar{z}} (z) T_{\bar{z} \bar{z}} (w) \rangle = - \partial_w \langle T_{z\bar{z}}(w) T_{\bar{z} \bar{z}} (w) \rangle\,.
\end{equation}
Hence, by conservation of the stress tensor,
\begin{equation}
    \partial_{\bar{z}} D(z,w) = - \langle T_{z\bar{z}} (z) \partial_w T_{\bar{z} \bar{z}} (w) \rangle + \langle T_{z\bar{z}} (z) \partial_{\bar{w}} T_{z\bar{z}}(w) \rangle = 0\,.
\end{equation}

Therefore, we have shown that $D(z, w)$ is a constant. When $z \rightarrow w$
\begin{equation}
    \lim_{z \rightarrow w} D(z, w) = \langle T_{zz} T_{\bar{z} \bar{z}} \rangle\,.
\end{equation}

In the context of a QFT on a cylinder, we consider the points $z$ and $w$ infinitely separated along the non-compact dimension of the cylinder. We use the fact that the vacuum two-point functions cluster decompose into products of one-point functions at infinite separation to find 
\begin{equation}
    \lim_{|z-w| \rightarrow \infty} D(z,w) = \langle T_{zz} \rangle \langle T_{\bar{z} \bar{z}} \rangle - \langle T_{z\bar{z}} \rangle^2\,.
\end{equation}
Hence, with the above discussion, we conclude 
\begin{equation}
    \langle T_{zz} T_{zz} \rangle = \langle T_{zz} \rangle \langle T_{\bar{z} \bar{z}} \rangle - \langle T_{z\bar{z}} \rangle^2
\end{equation}
proves the factorization property of the $T\overline{T}$ operator. This derivation assumes we are in the vacuum state $|0\rangle$, but the $T\overline{T}$ operator factorizes for any energy eigenstate $|n\rangle$. 

Proving that factorization holds for any energy eigenstate $|n \rangle$ is straightforward. First, from a spectral expansion for a complete set of states, we have
\begin{equation}
\begin{aligned}
\label{eq:erie}
  &  \langle n | T_{zz} (z) T_{\bar{z} \bar{z}} (w) | n \rangle \\&= \sum_m \langle n| T_{zz} (z) | m \rangle \langle m | T_{\bar{z} \bar{z}} (w) | n \rangle e^{(E_n - E_m) | \operatorname{Im} (z) - \operatorname{Im} (w) | + i (P_n - P_m) | \operatorname{Re} (z) - \operatorname{Re} (w) |}\,.
    \end{aligned}
\end{equation}
The exponential factors contain dependence on coordinates $(z,w)$. However, we know that the function $D(z,w)$ is constant for correlators of any energy eigenstate. Hence, the sum expressed in equation \eqref{eq:erie} is zero for $m \neq n$, rendering the exponential factor equal to one. Consequently, we attain a meaningful limit as the points coincide, i.e. $z \rightarrow w$:
\begin{equation}
    \langle n| T_{zz} T_{zz} | n \rangle = \langle n | T_{zz} | n \rangle \langle n | T_{\bar{z} \bar{z}} | n \rangle - \langle n | T_{z\bar{z}} | n \rangle^2 
\end{equation}
implying that factorization holds for \emph{all} energy eigenstates. 

Now that we have proven factorization, we are prepared to derive the flow equation for a field theory on a cylinder $S^1 \times \mathbb{R}$ with line element
\begin{equation}
\label{cylinder}
    ds^2 = dy^2 + R^2 dx^2\,.
\end{equation}
\begin{figure}[h]
    \centering
    \includegraphics[scale = 0.5]{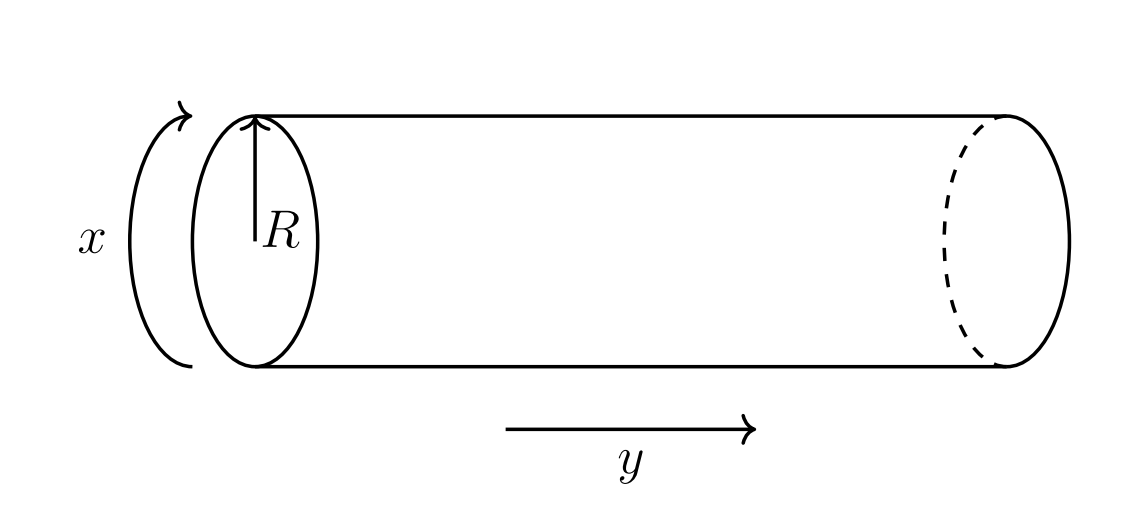}
    \caption{We denote the compact direction by $x \sim x + R$, where $R$ is the radius of the spatial $S^1$, and write $y$ for the non-compact Euclidean time direction.}
\label{fig:cylinder}
\end{figure}

The components of the stress tensor are\footnote{The momentum is quantized in units of $\frac{1}{R}$ and so it does not change with $\lambda$ because $\lambda$ is continuous.}
\begin{equation}
    \begin{aligned}
    \label{eq:r90e90uert90er09uwe}
    \langle n | T_{yy} | n \rangle &= - \frac{E_n(R, \lambda)}{R}\,, \\
    \langle n | T_{xx} | n \rangle &= - \frac{\partial}{\partial R} E_n (R, \lambda)\,, \\
    \langle n | T_{xy} | n \rangle &= \frac{i}{R} P_n (R, 0)\,.
    \end{aligned}
\end{equation}
The governing flow equation \eqref{eq:TTb definition} for the energy levels are
\begin{equation}
    \begin{aligned}
    \label{eq:gr9e13213}
\frac{\partial}{\partial \lambda} E_n (R, \lambda) &=- R \left( \langle n | T_{xx} | n \rangle \langle n | T_{yy} | n \rangle - \langle n | T_{xy} | n \rangle^2 \right)\,.
    \end{aligned}
\end{equation}
Substituting \eqref{eq:r90e90uert90er09uwe} into \eqref{eq:gr9e13213}, we arrive at the same one-dimensional inviscid Burgers’ equation commonly used to describe the dynamics of a diffusionless fluid
\begin{equation}
\label{eq:burgers}
    \frac{\partial}{\partial \lambda} E_n (R, \lambda) = E_n (R, \lambda) \frac{\partial}{\partial R} E_n (R, \lambda) + \frac{P_n(R, 0)^2}{R}\,.
\end{equation}
Before addressing the deformed energy levels, it is essential to establish the initial conditions. While solving the differential equation in a closed form may not be feasible for general seed theories, the situation significantly simplifies when the seed theory conforms to CFT principles. The initial conditions for any CFT on a cylinder of radius $R$ are the following:
\begin{equation}
\begin{aligned}
\label{eq:IC-CFT}
    E_n (R, 0) & = \frac{1}{R} \left( n + \bar{n} - \frac{c}{12} \right)\,, \\
    P_n (R, 0) &= \frac{1}{R} \left( n - \bar{n} \right)\,,
\end{aligned}
\end{equation}
where $n$ and $\bar{n}$ are the eigenvalues of the Virasoro generators $L_0$ and $\bar{L}_0$ respectively. 

With the initial conditions \eqref{eq:IC-CFT}, the solution to \eqref{eq:burgers} is given by 
\begin{equation}
\begin{aligned}
  E_n(R,\lambda) &= \frac{R}{2\lambda} \left( \sqrt{1 + \frac{4\lambda E_n(R,0)}{R}  + \frac{4\lambda^2 P_n^2}{R^2}  } - 1 \right) \\&=  \frac{R}{2\lambda} \left( \sqrt{1 + \frac{4\lambda}{R^2} \left( n + \bar{n} - \frac{c}{12} \right)  + \frac{4\lambda^2}{R^4} \left( n -\bar{n} \right)^2  } - 1 \right)\,.
\end{aligned}
\end{equation}

\section{$T\overline{T}$ and holography}
Besides introducing new states, another route to enriching and extending the theory of boundary gravitons is to move the AdS$_3$ boundary radially inwards, and there are several motivations for doing so. One is a way to access observables that are ``more local'' than the usual asymptotically defined quantities, namely the S-matrix in Minkowski space and boundary correlators in AdS. The need to develop such observables has long been appreciated, particularly in a cosmological context where there may not exist any ``far away spatial region" that an observer at a fixed time can appeal to. In general dimensions, the complications of defining quantum gravity in a finite spatial region are hard to disentangle from the usual UV problems,\footnote{See \cite{Witten:2018lgb} for a review of the boundary value problem in  $D>3$  Euclidean gravity.} but the situation is better in AdS$_3$ since the renormalizability argument of \eqref{eq:EH} applies to the case of a finite boundary.\footnote{At least if the boundary is flat, as will be the main case of interest in this thesis. More generally, we might need boundary counterterms involving boundary curvature. See appendix \ref{sphere} for more discussion.} The problem is also interesting due to its proposed description \cite{McGough:2016lol} as a $T\overline{T}$-deformed CFT \cite{Zamolodchikov:2004ce,Smirnov:2016lqw}. These are theories described in the IR as CFTs perturbed by irrelevant operators; their UV description is not well understood, but they conceivably represent a new type of quantum theory in which locality breaks down in a controlled manner. We take the perspective that these two descriptions -- cutoff AdS$_3$ and $T\overline{T}$-deformed CFT$_2$ --   are mutually illuminating.

Another holographic proposal is for $\lambda < 0$. This sign choice alters the asymptotic boundary conditions for the metric as $r \to \infty$, yet it is crucial to emphasize that it does \textit{not} impose a finite cutoff on the spacetime, as discussed in \cite{Guica:2019nzm}. While this modification enforces the Dirichlet condition at a finite radial cutoff for $\lambda > 0$, it's noteworthy that the same prescription applied in the case of $\lambda < 0$, where the bulk theory extends to $r \to \infty$. In this context, the entire spectrum of the boundary field theory remains real, at least for sufficiently small values of $\lambda$. 

These two holographic interpretations of the $T\overline{T}$ deformation in the metric formalism for $3d$ gravity have been studied in the Chern-Simons formalism by \cite{Llabres:2019jtx,Ouyang:2020rpq,He:2020hhm,He:2021bhj,Ebert:2022cle,Kraus:2022mnu}. In particular, the Chern-Simons analysis of \cite{Llabres:2019jtx} studies $3d$ gravity in Lorentzian signature for $\lambda < 0$ and shows that a $T\overline{T}$ deformation of the dual CFT corresponds to a modification of the boundary conditions for the gauge field $A_\mu$. This interpretation of the modified boundary conditions introduces a novel variational principle wherein specific linear combinations of the undeformed source and expectation value are held constant. Consequently, this combination is a newly defined deformed source.

To complete the corners of low-dimensional gravity in the diagram (\ref{gravity_diagram}), it is desirable to understand the bottom half under the $T\overline{T}$ deformation. The process of dimensional reduction from 3D to JT gravity, along with its dual Schwarzian description and the associated partition functions in the metric formalism, was initially explored by \cite{Gross:2019ach,Gross:2019uxi} in the case of $\lambda < 0$.\footnote{The case for $\lambda > 0$ was studied by \cite{Iliesiu:2020zld}. One immediately encounters a complex-valued energy spectrum and partition function when $E > \frac{1}{8\lambda}$. To cure this problem and obtain a real-valued partition function, \cite{Iliesiu:2020zld} included a non-perturbative contribution.} However, unlike the extensive literature on $T\overline{T}$ deformations in the metric and Chern-Simons descriptions of $3d$ gravity, such $T\overline{T}$-like deformations of the BF theory description of JT gravity (and its dual ``particle-on-a-group'' theory) have received less attention, which is one of the motives for part of this thesis.

\section{Effective field theory interpretation and higher dimensions}
A different perspective to understand the deformation is from the observation that the partition function of the effective field theory (EFT) on a radial slice must be a solution to the radial Wheeler-DeWitt equation to describe gravitational physics. In other words, there is a holographic dictionary for the EFT dual to a sharp radial cutoff in AdS. This interpretation was thoroughly studied in \cite{Hartman:2018tkw}, and a few more intermediate calculations are provided in appendix \ref{app:EFT}. The difference is that the $T^2$ operator is a $d$-dimensional analog of the two-dimensional $T\overline{T}$ operator in the large-$N$ limit so we have factorization. We start with Euclidean Einstein gravity coupled to matter with the appropriate gravitational counterterm
\begin{equation}
\begin{aligned}
\label{eq:0ijhgy6711111}
    I  &= - \frac{1}{16 \pi G}  \int_{M_d} d^dx \sqrt{g} \left( R - 2 \Lambda \right) - \frac{1}{8 \pi G} \int_{\partial M_d} d^{d-1}x \sqrt{g^0} K \\&+ \frac{1}{8 \pi G} \int_{\partial M_d} d^{d-1}x \sqrt{g^0} (d-1 + \mathcal{L}_{\text{Curvature}}) + I_{\text{Matter}}
\end{aligned}
\end{equation}
and the renormalized Brown-York stress tensor of \eqref{eq:0ijhgy6711111} is defined by
\begin{equation}
    \delta I = \frac{1}{2} \int_{\partial M_d} \sqrt{g^0} \widetilde{T}^{ij} \delta g^0_{ij}
\end{equation}
which is determined in \cite{Balasubramanian:1999re}
\begin{equation}
\label{eq:krausT}
    \widetilde{T}_{ij} = \frac{1}{8 \pi G} \left( K_{ij} - K g^{0}_{ij} + (d-1) g^0_{ij} \right) - a_d \widetilde{C}_{ij}\,,
\end{equation}
where $K$ is the trace of the extrinsic curvature $K_{ij} =2 \nabla_{(i} n_{j)}$, $n_j$ is the outward-pointing normal vector, $a_d$ is a constant, $\widetilde{C}_{ij}$ are curvature-dependent counterterms, $g^0_{ij}(x) \equiv g_{ij} (r_c, x)$ is the metric on the cutoff surface, and the tildes are for bulk quantities are related to their undeformed counterparts up to a multiplicative factor of the cutoff. The boundary stress tensor \eqref{eq:krausT} obeys
\begin{equation}
\label{eq:hartmanFlow}
    \widetilde{T}_i^i+a_d \widetilde{C}_i^i=-4 \pi G\left[\left(\widetilde{T}_{i j}+a_d \widetilde{C}_{i j}\right)^2-\frac{1}{d-1}\left(\widetilde{T}_i^i+a_d \widetilde{C}_i^i\right)^2\right]-\frac{\widetilde{R}}{16 \pi G}+\widetilde{t}_r^r\,,
\end{equation}
where $t_{ij}$ represents the stress tensor associated with matter, and one can find the proof provided in appendix \ref{app:EFT1}. Here, \eqref{eq:hartmanFlow} is the $T^2$ flow equation\footnote{We derive this explicit form of $\mathbb{X}$ in appendix \ref{app:EFT} and the explicit form of the holographic cutoff dictionary.}
\begin{equation}
   \frac{\partial S_{\text{EFT}}}{\partial \lambda} = \int d^dx \sqrt{\gamma}  \mathbb{X}\,,
\end{equation}
where
\begin{equation}
\label{eq:X}
 \mathbb{X}=  \left[\left(T_{i j}+a_d \widetilde{C}_{i j}\right)^2-\frac{1}{d-1}\left(T_i^i+a_d \widetilde{C}_i^i\right)^2\right]- \frac{r_c^d}{d\lambda} \left( -\frac{\widetilde{R}}{16 \pi G}+\widetilde{t}_r^r - a_d \widetilde{C}_i^i\right)
\end{equation}
and 
\begin{equation}
\label{eq:TTlambdadef}
    \lambda = \frac{4 \pi G}{d r_c^d}\,.
\end{equation}
\newline
\emph{Torus energy spectrum and quantum corrections}

We can derive the energy spectrum under the $T^2$ deformation in the large-$N$ limit where factorization holds following \cite{Hartman:2018tkw}. An example to consider is a CFT on a manifold $\mathbb{R} \times \mathcal{M}^{d-1}$ with metric
\begin{equation}
    ds^2 = d\tau^2 + h_{ab} dx^a dx^b
\end{equation}
and the flow for the energy levels are
\begin{equation}
    \frac{\partial E}{\partial \lambda} = \int d^{d-1}x \sqrt{\gamma}  \mathbb{X}\,.
\end{equation}
Let's specialize $\mathcal{M}^{d-1}$ to be $(d-1)$-dimensional rectangular torus so the line element is
\begin{equation}
    ds^2  = d\tau^2 + \sum^{d-1}_{i=1} L^2_i dx_i^2\,, 
\end{equation}
where $\tau$ is valued on $\mathbb{R}$, $L_i$ is the radius of the $i^{\text{th}}$ cycle on the torus and $x_i$ are compact coordinates on the torus. For this background, the Einstein tensor vanishes, and there are no trace anomalies on the torus. We furthermore set $\tilde{t}^r_r =0$. Therefore, the operator $ \mathbb{X}$ simplifies to
\begin{equation}
\label{eq:f111a@@}
   \mathbb{X} = T_{i j} T^{i j} -\frac{1}{d-1}\left(T_i^i\right)^2
    \end{equation}
    and the flow equation for the energy becomes\footnote{Here we take the torus to be square $L_1 = L_2 =\cdots = L_{d-1} =L$. See appendix \ref{app:EFT} for more details.}
    \begin{equation}
    \label{eq:pppsa}
        \frac{\partial \varepsilon}{\partial \lambda} = \frac{d-2}{d-1} \varepsilon^2 - \frac{2\varepsilon}{(d-1) L^{d-2} } \frac{\partial}{\partial L} \left( L^{d-1} \varepsilon \right),
    \end{equation}
    where the energy density is $\varepsilon = E/L^{d-1}$. The solution to this differential equation \eqref{eq:pppsa} is 
    \begin{equation}
        E= \frac{(d-1) L^{d-1}}{2d \lambda} \left( 1 -  \sqrt{1 - \frac{4d\lambda}{d-1} E_0} \right),
    \end{equation}
    where the undeformed energy is $E_0 = \frac{M}{L}$.

Related to this discussion, an unanswered question one could ask is if it is possible a higher dimensional $T\overline{T}$-like an operator exists that one can reliably compute $\mathcal{O}(1/N)$ corrections to the energy, such as for a QFT on a torus. For large-$N$ and $d>2$, we have factorization in the operator $ \mathbb{X}$ defined in \eqref{eq:f111a@@}, but can one easily create an operator up to $\mathcal{O}(1/N)$ corrections\footnote{I thank Per Kraus for discussions on this.}
\begin{equation}
    \mathbb{X}(x, y) = T_{i j}(x) T^{i j} (y) -\frac{1}{d-1}T_i^i(x) T_j^j(y)+ \frac{1}{N} \mathcal{O}^{(1)}_{\mathbb{X}}(x, y) + \sum^{\infty}_{p=2} \frac{1}{N^p}\mathcal{O}^{(p)}_{ \mathbb{X}}(x, y) 
\end{equation}
so that $\langle E, P| \mathbb{X}(x, y) | E, P \rangle$ is independent of $(x, y)$? For a straightforward example, consider a theory with $T^i_i = 0$, such as $N$ free massless scalars, then
\begin{equation}
\label{eq:pppppaaaqqweq}
\langle E, P| T_{i j}(x) T^{i j} (y)| E, P  \rangle =  \langle E, P| T_{i j}(x) |E, P \rangle \langle E, P| T^{i j} (y)| E, P \rangle + N  f_{E, P} (x-y)\,. 
\end{equation}
where $f_{E, P}(x-y)$ is a nontrivial energy-momentum state dependent function of $x-y$. The first piece is factorized (i.e. connected) proportional to $N^2$, while the second piece is unfactorized (i.e. disconnected). The unwanted contribution is the connected piece is removed via
\begin{equation}
    \langle E, P| \mathcal{O}^{(1)}_{\mathbb{X}} (x, y) | E, P \rangle = - N^2 f_{E, P}(x-y)\,.
\end{equation}
Now the candidate operator with the normal ordering of stress tensors up to $\mathcal{O}(1/N)$ is
\begin{equation}
\begin{aligned}
 \mathbb{X} (x, y)&=  T_{i j}(x) T^{i j} (y) -\frac{1}{d-1}T_i^i(x) T_j^j(y)\\&- \left[ : T_{ij} (x) : :T^{ij}(y): - \frac{1}{d-1} :T^i_i(x): :T^j_j(y): \right] + \sum^{\infty}_{p=2} \frac{1}{N^p}\mathcal{O}^{(p)}_{ \mathbb{X}}(x, y)\,. 
\end{aligned}
\end{equation}
This operator exhibits the desired properties within the free boson theory's vacuum state as all connected diagrams up to $\mathcal{O}(1/N)$ are removed. However, it appears to extend differently for excited states because the normally ordered operators do not annihilate them.
\newpage 

\section{Outline}
The outline of this thesis is as follows. 

In chapter \ref{chFieldTheoryGravitons}, we determine the boundary graviton action of AdS$_3$ at a finite planar boundary cutoff in the metric and Chern-Simons formalism to calculate the stress tensor correlators at the two-loop order. This analysis extends the previous tree-level discussion. We make new comments on higher point quantum corrected planar correlators. Correlators of the deformed gravitational AdS$_3$ Wilson line are computed semi-classically via conformal perturbation theory. 

In chapter \ref{ch:BF}, we study JT gravity in the BF formalism under the $T\overline{T}$ deformation to determine the dual $T\overline{T}$-deformed Schwarzian theory. Finally, we calculate deformed correlators involving gravitational BF Wilson lines defined up to the Hagedorn temperature. 

In chapter \ref{ch:SUSY-QM}, we define a manifestly supersymmetric $T\overline{T}$ deformation for $(0+1)d$ QFTs with $\mathcal{N} =(0,1)$ and $\mathcal{N} = (1, 1)$ supersymmetry. We determine the $T\overline{T}$-deformed super-Schwarzian theory dual to JT supergravity. These deformations, denoted as $f(\mathcal{Q})$, are explicitly expressed within the Noether currents associated with supertranslations. We show three seemingly different procedures to determine the same deformed supersymmetric quantum mechanics theory.

In chapter \ref{ch:Airy}, we consider the $T\overline{T}$ deformation of JT gravity and its `t Hooft limit, the Airy model. Various aspects of the $T\overline{T}$ deformation are considered, such as complexification of the spectrum at the disk level, negativities in the deformed spectral density, and, in particular, the free energy for $\lambda > 0$ and $\lambda < 0$.

In chapter \ref{ch:root-TT}, we investigate properties of the newly defined root-$T\overline{T}$ operator using holographic properties. In contrast to the better-known $T\overline{T}$ operator, the root-$T\overline{T}$ operator is much harder to make precise in the quantum theory but is also classically marginal, which makes it particularly interesting as a possible tool in CFT. We assume a large-$N$ limit corresponds to a large central charge in the CFT. This limit is needed to perform the path integral to the lowest order in a $1/N$ expansion using large-$N$ factorization. Using this method, a suggestive form for the deformed metric and stress tensor of the CFT in terms of the undeformed quantities is derived. The same results are then derived using the trace of the stress tensor staying zero in the deformed theory. The results require the root-$T\overline{T}$ deformation to commute with the ordinary $T\overline{T}$ deformation. Similar ideas lead one to conjecture the deformed energy levels in the root-$T\overline{T}$ theory. This conjecture is supported by calculating the energy of a class of AdS spacetimes with the root-$T\overline{T}$ boundary conditions derived in the previous sections and showing that it is consistent with the flow of the energy levels. 

In chapter \ref{ch:ChiralBoson}, we conclude this thesis by continuing our previous studies of the root-$T\overline{T}$ deformation from chapter \ref{ch:root-TT} to the context of interacting chiral boson theories and perturbatively study their quantum aspects for particular backgrounds in two dimensions. We extend our interacting chiral boson analysis to the three-dimensional abelian Chern-Simons context.

Details of various calculations, applications, and additional observations are relegated in the appendices.

\chapter{The Field Theory of Interacting Boundary Gravitons}
\label{chFieldTheoryGravitons}

We study pure three-dimensional gravity at a finite radial cutoff in this chapter. Pure three-dimensional gravity is a renormalizable theory with two free parameters labeled by Newton's constant $G$ and the cosmological constant $\Lambda$. As a result, correlation functions of the boundary stress tensor in AdS$_3$ are unambiguously determined solely by the central charge of the Virasoro algebra. The same argument implies that AdS$_3$ gravity at a finite radial cutoff is a renormalizable theory, but now with one additional parameter corresponding to the cutoff location. This theory is conjecturally dual to a $T\overline{T}$-deformed CFT,  assuming these theories exist.   To elucidate this, we study the quantum theory of boundary gravitons living on a cutoff planar boundary and the associated correlation functions of the boundary stress tensor. We compute stress tensor correlation functions to two-loop order (Newton's gravitational coupling $G$ being the loop counting parameter), extending existing tree-level results. This computation is feasible since the boundary graviton action simplifies greatly upon making a judicious field redefinition, turning into the Nambu-Goto action. After imposing Lorentz invariance, the correlators in this order are found to be unambiguous up to a single undetermined renormalization parameter.
\section{Introduction}

We develop the quantum theory of boundary gravitons on a cutoff planar surface, focusing on obtaining the optimal form of the action and using it to compute correlation functions of boundary operators.  

We work in the framework of the covariant phase space formalism \cite{Crnkovic:1986ex,Lee:1990nz}, and in both the metric and Chern-Simons formulation \cite{Achucarro:1986uwr,Witten:1988hc} of $3d$ gravity, since they offer complementary perspectives. This cutoff phase space is established by considering an AdS$_3$ background and applying all coordinate and gauge transformations that maintain a Dirichlet boundary condition. These phase space coordinates consist of two functions defined at some initial time on the boundary; these are the coordinate transformations $(x,t) \rightarrow \big(x+A(x,t), t+B(x,t)\big)$ evaluated at $t=0$. We need a symplectic form and a Hamiltonian on this phase space to construct the canonical formulation, and we develop efficient methods for computing these. In the asymptotically AdS$_3$ case, this procedure is simple to carry out exactly, and we readily arrive at the Alekseev-Shatashvili action \cite{Alekseev:1988ce,Alekseev:1990mp}, as was obtained via the Chern-Simons formulation in \cite{Cotler:2018zff}. At finite cutoff, life is more complicated; we work order-by-order in the $(A, B)$ variables, but the resulting expressions quickly become complicated because the phase space action contains an ever-growing number of higher derivatives acting on these fields.

A pleasant surprise is that a field redefinition $(A, B) \rightarrow (f,\bar{f})$ can be used to remove all higher derivatives from the action, at least to the order we have checked (eighth-order in the fields). The resulting (imaginary time) action is the Nambu-Goto action written in Hamiltonian form \eqref{eq:90iojkkuhy78d}. We obtain further evidence for this action by deriving it to all orders in the special case of linearly varying $(f,\bar{f})$. 

However, the stress tensor is not the canonical stress tensor of the Nambu-Goto theory due to the non-linear action of the Poincar\'e group on the fields. The deformed stress tensor includes a series of higher derivative correction terms reflecting the nonlocal nature of the theory, e.g.\footnote{See \eqref{hi} for all three components.} 
\begin{equation}
\label{zzb}
    4G T_{zz} = \frac{1}{2} f''- \frac{1}{4} f'^2 +\frac{1}{4} r_c f''' \bar{f}'  -\frac{1}{8} r_c \big( f'^2 -2 f'\bar{f}' \big)'~\bar{f}' +\frac{1}{16} r_c^2 \big( f'''' \bar{f}'^2+ (f'^2)''\bar{f}''\big)  + \cdots\,.
\end{equation}

With the expressions for the action and stress tensor, we seek its quantization. The main interest here is in computing two-point functions of the stress tensor order-by-order in the loop counting parameter $G$.\footnote{Previous works \cite{Kraus:2021cwf,Hirano:2020ppu} on this problem in the gravitational formulation stopped at tree level. } There is some tension coming from two perspectives on this problem: on the one hand, the Nambu-Goto action with its square root is problematic to quantize directly without ambiguity; on the other hand, the underlying theory is pure $3d$ gravity, which one expects to be renormalizable.  

The subtlety in reconciling these perspectives has to do with the complicated (nonlinear and nonlocal) manner in which the symmetries of the gravitational description are realized once we pass to the reduced phase space description, and in particular with preserving these symmetries in the quantum theory. In this chapter, we compute the stress tensor correlators to two-loop order using dimensional regularization. At tree level and one-loop, the results are finite and unambiguous. At two loops, we find that a single renormalization of the stress tensor is required, and the divergent part comes as usual with an associated undetermined finite part parametrized here by $\mu$. For example, we find the $\langle T_{zz}T_{zz} \rangle$ correlator at the two-loop order to be\footnote{ The full set of two-point functions is written in  \eqref{jla}.}
\begin{equation}
\begin{aligned}
\label{Ia}
\langle T_{zz}(x)T_{zz}(0)\rangle = \resizebox{.78\hsize}{!}{$\frac{1}{z^4}  \left[ 
\frac{c}{2} 
+ 10(3+4G)\left(\frac{r_c}{z\overline{z}}\right)^2  + 96G\left(8 + 60\ln(\mu^2 z\overline z)\right)\left(\frac{r_c}{z\overline z}\right)^3 
+ 2520G\left(\frac{r_c}{z\overline z}\right)^4
\right]\,.$}
\end{aligned}
\end{equation}
Here $c=c_0+1 = \frac{3 \ell}{2G} +1$ is the one-loop corrected  Brown-Henneaux  central charge \cite{Brown:1986nw,Cotler:2018zff} of the $r_c=0$ theory.\footnote{In this thesis, the symbol 
$c$ will be used, and its meaning -- whether it represents the tree-level central charge ($c_0$) or the loop-corrected central charge -- will be inferred from the context in which it is mentioned.} 
Regarding renormalizability, the main result in this chapter is inconclusive: we suspect that the free parameter reflects that dimensional regularization is not preserving all symmetries, but further work is required to substantiate this by imposing the relevant Ward identities.

While our primary focus centers on a flat planar boundary, it is also valuable to explore the case of a curved boundary metric. In preparation for the curved boundary metric, we meticulously derive the Chern-Simons formulation that accommodates a general boundary metric. As an illustrative example, we demonstrate the procedure for calculating the action for Euclidean AdS$_3$ with a finite $S^2$ boundary, taking into account the substantial radius-related divergence linked to the Weyl anomaly of the boundary theory, as presented in appendix \ref{sphere}. This action is straightforward to obtain in the metric description. However, it is somewhat subtle in the Chern-Simons formulation due to the need to introduce two overlapping patches for the gauge potentials.

Furthermore, we focus on Wilson lines in $2d$ theories deformed by the \textit{double-trace} version of the $T\overline{T}$ operator. A conceptually related analysis involving deformed Wilson loops in the \textit{single-trace} setting was presented in \cite{Chakraborty:2018aji}. However, in that work, the Wilson line was computed for the DBI gauge field living on a D1-brane, rather than the gauge field arising from a gauge theory presentation of a gravity theory. In this chapter, we content ourselves with the double-trace version of the $T\overline{T}$ deformation and Wilson lines for gauge fields associated with gravitational theories. One reason for doing this is that the bulk gravity dual to a single-trace $T\overline{T}$-deformed CFT also involves the dilaton and the three-form flux $H_3$, and writing the kinetic terms for these fields in Chern-Simons variables is somewhat unwieldy. A second reason is that, although the gravity solution relevant for single-trace $T\overline{T}$ is local AdS$_3$ in the deep interior, the solution approaches a linear dilaton spacetime. The linear dilaton region is qualitatively different from AdS$_3$ (in fact, its causal structure is, in some sense, more similar to that of Minkowski space), and the Chern-Simons formulation of $3d$ gravity is not straightforwardly applicable in this regime.

We highlight some prior research in a related vein. In \cite{Kraus:2021cwf}, the authors investigated AdS$_3$ gravity with a finite cylindrical boundary. A result from  \cite{Kraus:2021cwf} was that the asymptotic Virasoro $\times $ Virasoro algebra was deformed by breaking the conformal invariance associated with the finite boundary. Another result was that the free boundary graviton spectrum was deformed in a manner compatible with  $T\overline{T}$ considerations.   In this chapter, the main focus is on a planar boundary; this is simpler, and we make other technical advances that allow us to go further than before.  Stability and causality for gravity with cutoff boundary conditions are discussed in \cite{Marolf:2012dr,Andrade:2015gja,Andrade:2015fna}.  The covariant phase space in the presence of boundaries is reviewed in \cite{Harlow:2019yfa}. Jackiw-Teitelboim gravity \cite{Teitelboim:1983ux,Jackiw:1984je} at a finite radial boundary cutoff was studied in \cite{Stanford:2020qhm,Iliesiu:2020zld,Moitra:2021uiv}, with results relating to the spectrum of $T\overline{T}$-deformed quantum mechanics obtained in \cite{Gross:2019ach,Gross:2019uxi,Ebert:2022ehb}. An important subtlety that arises, discussed in \cite{Stanford:2020qhm}, is the distinction between microscopic versus effective theories of the JT gravity path integral, and the resulting non-trivial relations among the parameters and couplings; presumably these issues are also present in this chapter's context.

Correlation functions in the $2d$ field theory or $3d$ bulk were studied in
\cite{Hirano:2020ppu,Kraus:2018xrn,Cardy:2019qao,Aharony:2018vux,Rosenhaus:2019utc,He:2019vzf,He:2019ahx,He:2020udl,Dey:2020gwm,Hirano:2020nwq,Ebert:2020tuy}. Results in those papers were found either at low order in the $T\overline{T}$ coupling\footnote{In this chapter, we denote the $T\overline{T}$ coupling by $\lambda_{T\overline{T}}$ instead of $\lambda$ to avoid a clash of notation with the AdS$_3$ Gauss parameters $(\lambda, \Psi, F, \bar{\lambda}, \bar{\Psi}, \bar{F})$ defined in \eqref{zm}. In later chapters, we resort to $\lambda$ being the $T\overline{T}$ coupling.} $\lambda_{T\overline{T}} \sim \frac{r_c}{c}$, or lowest order in the $\frac{1}{c}$ expansion (tree level in this chapter's language). An exception is \cite{Cardy:2019qao} that proposed some results of all orders in $\lambda_{T\overline{T}}$. The results in this chapter hold to all orders in $r_c$ but are perturbative in $\frac{1}{c}$ (we go to two-loop order, extending the previous tree-level results). In the context of a massive scalar \cite{Rosenhaus:2019utc} and a Dirac fermion \cite{Dey:2021jyl},  integrability fixed renormalization ambiguities. In \cite{Llabres:2019jtx,Ouyang:2020rpq,He:2020hhm,He:2021bhj}, the $T\overline{T}$-deformed CS formulation of $3d$ gravity was discussed. 

The remaining portion of this chapter unfolds subsequently: In section \ref{gensec}, we lay out some general principles in computing the action for boundary gravitons common to the metric and Chern-Simons formulations. In section \ref{sec:BoundaryGravitonsMetric}, we discuss the metric formulation, developing a streamlined approach to computing the boundary action for a flat radial cutoff boundary. We show how to obtain the Alekseev-Shatashvili action straightforwardly in the $r_c=0$ limit and acquire the all-order action at finite $r_c$ in the special case of constant $(f', \bar{f}')$: the Nambu-Goto action. In section \ref{CSsec}, we turn to the Chern-Simons formulation. Since it is of interest beyond the immediate concerns of this work, we carefully develop the variational principle for CS gravity with a general curved cutoff boundary.    We carry out a perturbative computation of the action for gravitons on a cutoff planar boundary,  obtaining results to eight order in $(f,\bar{f})$; the results turn out to match the expansion of the Nambu-Goto action to this order, leading us to conjecture that this extends to all orders.  In section \ref{corsec}, we turn to correlation functions. We compute correlators of both elementary fields and the stress tensor. The one-loop four-point function of elementary fields is found to require one counterterm in the gravitational action, and the two-loop stress tensor correlators require a single renormalization of the stress tensor. Finally, in section \ref{sec:wil}, we calculate the deformed gravitational AdS$_3$ Wilson line correlators semi-classically. We conclude with a brief discussion in section \ref{dissec}. In appendix \ref{app:ads3}, we provide further details on a general trace flow equation, the Chern-Simons formulation, including the comparison to the metric formulation, relating Chern-Simons theory at finite cutoff and coupling to topological gravity, details regarding the evaluation of Feynman diagrams and a discussion of how to compute the action in the case of a spherical boundary.

\section{Generalities on the phase space formulation of boundary graviton theories} \label{gensec} 

In the upcoming two sections, we obtain the action and stress tensor for boundary gravitons localized on a finite cutoff surface by working in the metric and Chern-Simons formulations. The formalisms provide complementary viewpoints and bring distinct technical advantages to the table, yet it is crucial to note that their outcomes are in complete agreement. To pave the way for the in-depth analysis that ensues, we delve into several fundamental facets of the issue.

The Einstein-Hilbert action \eqref{eq:EH}, or its Chern-Simons equivalent, contains a mixture of ``physical modes" and ``pure gauge modes," and the goal in this chapter is to arrive at a reduced action that omits the latter as much as possible. In general,  one pays a price by reducing the degrees in the form of a loss of manifest symmetry, for example, in the light cone gauge treatment of Yang-Mills theory or string/M-theory.   In Yang-Mills perturbation theory, this price is typically too high, and a Lorentz invariant formulation with unphysical modes is usually adopted.   However, in a topological theory, like pure three-dimensional gravity, the reduction of degrees of freedom is so dramatic (removing all but the boundary modes) that the cost of losing some manifest symmetry is more than repaid.

We will construct a reduced action living on a flat boundary surface with coordinates $(x,t)$. The action is of the phase space variety, built out of a Hamiltonian  $H$ and a ``canonical one-form" $\Upsilon$.\footnote{Note that $\Upsilon $ is a one-form on phase space, not on spacetime.  We use $\delta$ to denote the exterior derivative on phase space, reserving $d$ for the exterior derivative on spacetime. }  The phase space action takes the form
\begin{equation}
    \label{bba}
    I = -\int dt  \big(i_{V_\eta}\Upsilon - H\big)\,,
\end{equation}
where $t$ is Euclidean time and $i_{V_\eta}$ denotes contraction with the phase space vector field $V_\eta$ that implements (Lorentzian) time translation.   For example, for a particle moving in one dimension, we might take $\Upsilon = p\delta q$ and $H(p,q) = \frac{p^2}{2m} +V(q)$. We have $i_{V_\eta}\Upsilon = -i p\dot{q}$ implying that $I= \int\! dt \big(ip\dot{q} +H(p,q)\big)$.

The symplectic form $\Omega$ is given by $\Omega = \delta\Upsilon$. On the true phase space of the theory, $\Omega$ should be nondegenerate, meaning that $i_V\Omega = 0$ if and only if $V = 0$. In gauge theory or gravity, one begins with a larger ``pre-phase space'' with a degenerate, closed two-form $\Omega$. The null directions of $\Omega$ on pre-phase space correspond to small gauge transformations. Part of the task in this chapter will be to remove the pure gauge modes corresponding to these null directions.

In the case of $3d$ gravity, the dynamical variables appearing in the phase space action will be fields $\big(f(x,t),\bar{f}(x,t)\big)$ on the boundary which therefore comprises the physical degrees of freedom.\footnote{More precisely, these fields are subject to residual gauge equivalences associated with isometries of AdS$_3$.}   The route to obtaining the action for these fields differs in the metric versus Chern-Simons descriptions.  

In the metric formulation, the idea is to start with some reference solution and then apply boundary-condition-preserving coordinate transformations to construct a space of solutions.  The symplectic form for gravity on pre-phase space was calculated in \cite{Crnkovic:1986ex} and implies that coordinate transformations that vanish at the boundary correspond to degenerate modes. All that matters is the form of the coordinate transformation near the boundary, and this information is specified by the fields $(f,\bar{f})$. The coordinate transformations preserve the metric on the boundary but change the value of the boundary stress tensor $T_{ij}(x)$.   As we discuss in detail in the next section, the phase space action follows immediately from the expressions for the boundary momentum density  $\mathcal{P}= \frac{i}{2\pi} T_{tx}$ and energy density $ \mathcal{H} = \frac{1}{2\pi} T_{tt}$.

Turning to the Chern-Simons formulation, in this approach, one can pass directly from the Chern-Simons action to the reduced phase space action once one has been sufficiently careful in defining boundary conditions and adding the associated boundary terms in the gravitational action. In the case of an asymptotic AdS$_3$ boundary, previous work on this problem includes \cite{Elitzur:1989nr,Coussaert:1995zp,Cotler:2018zff,Nguyen:2021pdz}.  The general approach in this chapter follows \cite{Cotler:2018zff}. Nonetheless, in our context, we are dealing with a spacetime subject to a finite cutoff. Consequently, our initial task is to establish a rigorously defined variational principle. Additionally, we must incorporate the relevant boundary terms into the action, as they deviate from the conventions typically employed in most of the existing literature. We do this for a general curved boundary geometry, although our primary focus here is the case of a flat boundary. As usual in gauge theory, the time components of the gauge fields $A_t$ act as Lagrange multipliers imposing constraints \cite{Elitzur:1989nr}. Essentially, all one needs to do is to solve these constraint equations in a manner compatible with the boundary conditions, and then substitute back into the action. The Lagrangian density is observed to be a total derivative, and the resulting boundary term is the desired phase space action.  In this approach, the fields $(f,\bar{f})$ appear as free functions that parameterize solutions to the constraint equations and boundary conditions. 

In either approach, obtaining the action (using perturbation theory if necessary) is rather mechanical since the resulting expression may be unwieldy due to a suboptimal choice of coordinates on phase space. Selecting coordinates such that the kinetic term in the gravitational action (the terms involving time derivatives) is purely quadratic in fields is convenient for performing quantum mechanical perturbation theory. This corresponds to choosing ``Darboux coordinates" such that the components of the symplectic form are constant, which is always possible locally.\footnote{In general coordinates, the complication is that the path integral measure is proportional to the nontrivial Pfaffian of the symplectic form.} The $(f,\bar{f})$ are such Darboux coordinates, and part of our analysis will be to identify them.  We also find that it is possible\footnote{To at least eighth order in fields -- we do not yet have a general proof.}   to choose these coordinates such that the Hamiltonian takes a simple form, namely that of the Nambu-Goto action \eqref{eq:90iojkkuhy78d}. The Nambu-Goto action is well-known to be the $T\overline{T}$-deformed action of a free scalar with canonical stress tensor \cite{Cavaglia:2016oda}; the new features here are that the stress tensor derived from the gravity theory is not the canonical stress tensor and the existence of a highly nontrivial field redefinition that relates the natural gravitational variables to those appearing in the Nambu-Goto action.

\section{Metric formulation of boundary graviton action on the plane}
\label{sec:BoundaryGravitonsMetric}

\subsection{Preliminaries}

We start from the Euclidean signature action of $3d$ gravity  with cosmological constant $\Lambda$,
\begin{equation}
\label{eq:ereerer435456}
    I = - \frac{1}{16 \pi G} \int_{M_3} d^3x  \sqrt{g} \left( R-  2 \Lambda \right) + I_{\operatorname{bndy}}\,,
\end{equation}
where the boundary terms are written below. Einstein's equations are 
\begin{equation}
    \label{daa}
    R_{\mu\nu}-{1\over 2}Rg_{\mu\nu} +\Lambda g_{\mu\nu}=0\,.
\end{equation}
For $\Lambda <0$, we define the AdS$_3$ radius as $\Lambda = -\frac{1}{\ell^2}$, and henceforth choose units such that $\ell=1$. We impose Dirichlet boundary conditions on the metric of a two-dimensional boundary surface and pick coordinates $(r,x^i)$ such that this surface lives at $r=r_c$. The interior of the surface is selected to be the region $r>r_c$, so that the vector $\partial_r$ is inward pointing. It is convenient to choose Gaussian normal coordinates in the vicinity of the boundary so that the metric is
\begin{equation}
    \label{db}
    ds^2 =  \frac{dr^2}{4r^2} + g_{ij}(r,x)dx^i dx^j\,. 
\end{equation}
These coordinates may or may not fail away from the boundary, but this is largely immaterial to study the boundary graviton theory. The Dirichlet boundary condition means fixing $g_{ij}(r_c,x^i)$ as well as the form \eqref{db} (we could, in principle, fix only the induced metric on the boundary, but it is convenient to also put ``gauge" conditions on the radial coordinate). The appropriate boundary action appearing in \eqref{eq:ereerer435456} is then
\begin{equation}
    \label{dd}
     I_{\text{bndy}} = - \frac{1}{8 \pi G}  \int_{\partial M_3} \! d^2x \sqrt{ \det g_{ij}} (K-1)  -  \frac{\log r_c}{32 \pi G} \int_{\partial M_3}\! d^2x \sqrt{ \det g_{ij}} R(g_{ij})\,,
\end{equation}
where the extrinsic curvature and its trace are
\begin{equation}
    \label{de}
    K_{ij} = -r \partial_r g_{ij}\,,\quad K = g^{ij}K_{ij}\,.
\end{equation}
The terms in $I_{\text{bndy}}$ not involving $K$ depend only on the Dirichlet boundary data and are unnecessary for a proper variational principle; however, one adds them to ensure the finiteness of the action in the asymptotic AdS$_3$ limit $r_c \rightarrow 0$.

The boundary stress tensor $T_{ij}$ is defined in terms of the on-shell variation of the action \cite{Brown:1992br,Balasubramanian:1999re},
\begin{equation}
    \label{df}
    \delta I = \frac{1}{4\pi}\int_{\partial M_3}\! d^2x \sqrt{\det g_{ij}} T^{ij}\delta g_{ij}\,,
\end{equation}
and works out to be
\begin{equation}
    \label{dg}
    T_{ij} = \frac{1}{4 G} (K_{ij}- K g_{ij} + g_{ij})\,.
\end{equation}
One can easily confirm the boundary stress tensor for three-dimensional gravity \eqref{dg} is conserved
\begin{equation}
    \begin{aligned}
    \nabla^i T_{ij} &= \frac{1}{4G} \left[ \nabla^i \left( K_{ij} - K g_{ij} \right) + \nabla^i g_{ij} \right] = 0\,,
    \end{aligned}
\end{equation}
where the first term $\nabla^i\left(K_{i j}-K g_{i j}\right)$ vanishes from the vacuum Einstein equations \eqref{daa} for metric \eqref{db} 
and the second term always vanishes since $\nabla^{i} g_{ij} = 0$.

To compare to a dual (deformed) CFT, we think of the latter as living on a rescaled metric $\gamma_{ij}$, defined as $\gamma_{ij} = r_c g_{ij}(r_c,x^i)$, which is in particular finite in the asymptotically AdS$_3$ case. The Einstein equations can be used to show that the stress tensor obeys the $T\overline{T}$ trace relation \eqref{eq:TTb definition} which was shown by \cite{Kraus:2018xrn}\footnote{We emphasize \eqref{dh} holds only for AdS$_3$. When $\Lambda > 0$, the trace relation, for example in dS$_3$, differs from AdS$_3$ \eqref{dh} by an additional constant $\propto \frac{1}{\lambda_{T\overline{T}}}$. See \cite{Gorbenko:2018oov,Lewkowycz:2019xse,Shyam:2021ciy,Coleman:2021nor,Coleman:2022lii} and appendix \ref{eq:flowAdS3dS3}.}
\begin{equation}
    \label{dh}
     \gamma^{ij} T_{ij} = \pi \lambda_{T\overline{T}} \det( \gamma^{ik}T_{kj}) - \frac{1}{8G}  R(\gamma)
\end{equation}
with \cite{McGough:2016lol}
\begin{equation}
    \label{di}
    \lambda_{T\overline{T}} = \frac{4G r_c}{\pi}\,.
\end{equation}
On a flat surface, the $T\overline{T}$ deformation of the action is defined as 
\begin{equation}
    \partial_{\lambda_{T\overline{T}}} I_{\lambda_{T\overline{T}}} = -\frac{1}{4} \int \! d^2x \sqrt{\gamma } \det (\gamma^{ik} T_{kj})\,
\end{equation}
We now specialize in the case of a flat boundary metric,
\begin{equation}
    \label{dia}
    g_{ij}(r_c) dx^i dx^j = \frac{dt^2+ dx^2}{r_c}\,,
\end{equation}
where $t$ lives on the real line, and at this stage $x$ is allowed to live on either the line or the circle.

Now, let  $\xi^\mu \partial_\mu $ generate a diffeomorphism vector field that preserves the boundary conditions; namely, $(\nabla_\mu \xi_\nu + \nabla_\nu \xi_\mu)\big|_{r_c} =0$. At this stage, we should emphasize that we do not restrict to vector fields $\xi$ that are tangent to the boundary; we allow for $\xi$ with a nonzero normal component $\xi^r$, which in an active sense corresponds to moving the location of the boundary. To clarify this, note that a more geometrical characterization of our setup consists of finding flat surfaces embedded in an ambient AdS$_3$ background. To translate this picture into the one we use, we note that near each surface, we can construct a Gaussian normal coordinate system, with the surface at $r=r_c$ in these coordinates.    The coordinate transformation needed to relate two such surfaces requires diffeomorphisms that are not tangent to the surfaces. Alternatively, one could take the more algebraic view that the vector field $\xi$ is a computationally efficient way of representing a transformation on field space. Since this canonical formalism only cares that our transformations preserve the boundary conditions, there is no difficulty posed by nonzero $\xi^r$. Associated to each boundary condition preserving vector field is a (not necessarily conserved) boundary charge \cite{Brown:1992br,Kraus:2021cwf}\footnote{These charges are only conserved and only generate symmetries if $\xi$ is tangent to the boundary. It is then also a boundary Killing vector. We should also note that the notation $Q[\xi]$ is not to be confused with Wald's $Q_\xi$ \cite{Iyer:1994ys} as explained in footnote 20 of \cite{Harlow:2019yfa}.}
\begin{equation}
    \label{dj}
    Q[\xi] = \frac{i}{2\pi}  \int\! T_{ti}\xi^i dx\,,
\end{equation}
where the integral evaluates on a constant $t$ slice of the boundary, and the appearance of $i$ is due to our choice of Euclidean signature. Translation invariance of the boundary metric implies the existence of conserved energy and momentum charges,
\begin{equation}
    \begin{aligned}
    \label{dk}
    H&= Q[-i\partial_t] = \frac{1}{2\pi}  \int T_{tt} dx\,, \\
    P &= Q[\partial_x] = \frac{i}{2\pi} \int T_{tx} dx\,.
    \end{aligned}
\end{equation}
\subsection{Phase space}
Adopting the framework of covariant phase space \cite{Crnkovic:1986ex,Lee:1990nz}, we think of phase space as the space of classical solutions that obey the boundary conditions \eqref{dia} where we identify solutions related by ``small gauge transformations,'' made precise momentarily.  By definition, a phase space equips to a symplectic form $\Omega$, which is a non-degenerate, closed two-form.  For pure gravity described by the Einstein-Hilbert action in arbitrary dimension, a closed (and conserved) two-form was found in \cite{Crnkovic:1986ex} as an integral over a Cauchy surface $\Sigma$ of an expression involving the metric and its derivatives
\begin{equation}
    \Omega = \frac{1}{16 \pi G} \int_\Sigma  d\Sigma_\alpha \sqrt{g} \left[ \delta \Gamma^{\alpha}{}_{\mu \nu} \wedge \left( \delta g^{\mu \nu} + \frac{1}{2} g^{\mu \nu} \delta \ln g \right) -\delta \Gamma^\nu{}_{\mu \nu} \wedge \left( \delta g^{\alpha \mu} + \frac{1}{2} g^{\alpha \mu} \delta \ln g \right) \right]\,.
\end{equation}

This object is degenerate on the space of all classical solutions since it gives zero when contracted against an infinitesimal displacement in solution space corresponding to a coordinate transformation that vanishes at the boundary.  To obtain the symplectic form, we must mod out by such coordinate transformations so that the two-form becomes non-degenerate on the quotient space.  In the context of AdS$_3$ gravity, this procedure is straightforward.

We denote $\delta_\xi g_{\mu\nu}$ as the change of the metric under an infinitesimal coordinate transformation, $\delta_\xi g_{\mu\nu} = \nabla_\mu \xi_\mu + \nabla_\nu \xi_\mu$.  We let $V_\xi$ denote the corresponding vector field on the space of solutions; in terms of the Lie derivative, this corresponds to the statement $\delta_\xi g_{\mu\nu} = \mathcal{L}_{V_\xi} g_{\mu\nu}$. A key relation, verified by direct computation in \cite{Kraus:2021cwf}, is\footnote{Note that this result is valid even when $\xi$ has a component normal to the boundary, unlike what is often assumed, e.g., in \cite{Harlow:2019yfa}. This relation holds in pure gravity, where one can choose a gauge so that the quantity $C$ defined in \cite{Harlow:2019yfa} vanishes.} 
\begin{equation}
    \label{dl}
    i_{V_\xi}\Omega  =-\delta Q[\xi]\,,
\end{equation}
where $i$ denotes the contraction operation and $Q[\xi]$ is given by \eqref{dj}. Since $Q[\xi]$ takes the form of a boundary integral, this makes explicit the statement that diffeomorphisms that vanish at the boundary correspond to null directions of $\Omega$.

To provide an explicit expression for the symplectic form, we need a description for the phase space with small diffeomorphisms properly quotient out. Given that \eqref{dl} establishes that diffeomorphisms persisting at the boundary correspond to non-degenerate directions, it becomes a logical choice to employ these diffeomorphisms as our coordinate system on phase space. We then begin with some chosen reference solutions and perform all possible diffeomorphisms that preserve the boundary conditions. This construction gives us the ``boundary graviton phase space" associated with the chosen reference solution. It may not coincide with the complete phase space if the latter encompasses distinct solutions that cannot be connected through finite diffeomorphisms.

Within the realm of pure AdS$_3$ gravity, our point of reference will be the AdS$_3$ solution in global or Poincaré coordinates. As for potential alternative solutions, we consider BTZ black holes and conical defects with varying masses and angular momenta.  However, BTZ black holes obey different boundary conditions than vacuum AdS$_3$ --- i.e. the BTZ black hole has two asymptotic boundaries in Lorentzian signature and a periodic time direction in Euclidean signature --- while conical defects have singular metrics.  We are interested in the phase space of boundary gravitons living on the boundary of vacuum AdS$_3$.

We start by writing down the reference metric, which, in the vicinity of the boundary, takes the form \eqref{db}.   We then change coordinates,  $r=r(r',x'^i)$, $x^i=x^i(r',x'^j)$ and demand that the metric at the boundary is unchanged,
\begin{equation}
    \label{dm}
     ds^2\big|_{r'=r_c} = \frac{dr'^2}{4r_c^2} + g_{ij}(r_c,x'^i)dx'^i dx'^j\,.
\end{equation}
We demand that the metric components at the boundary take the same form in the primed coordinates as in the original coordinates.  Note, in particular, that in the new coordinates, the location of the boundary is at $r'=r_c$; since this, in general, differs from $r=r_c$ and can think of the coordinate transformation as actively changing the location of the boundary.    Imposing the boundary conditions in the new coordinates amounts to solving a system of PDEs for $r(r',x'^i)$  and $x^i(r',x'^j)$.   Given the nature of the problem, it is natural to expand the coordinates transformation near the boundary, and so we write
\begin{equation}
    \begin{aligned}
    \label{dn}
     x & = x' +A(x',t')+ (r'-r_c)U(x',t')  +\cdots\,,  \\
     t & = t' +B(x',t')  + (r'-r_c) V(x',t') +\cdots\,, \\
r & =r' + r'C(x',t')+ r'^2 W(x',t')  +\cdots\,.
    \end{aligned}
\end{equation}
We then further expand around an  initial time surface  (taken to be $t=0$) on the boundary,
\begin{equation}
    \begin{aligned}
    \label{dna}
    x & =x'+A(x') + \sum_{n=1}^\infty A_n(x')t'^n  + (r'-r_c) \sum_{n=0}^\infty U_n(x')t'^n +\cdots\,,\\
    t & = t'+B(x') + \sum_{n=1}^\infty B_n(x')t'^n   +(r'-r_c) \sum_{n=0}^\infty V_n(x')t'^n +\cdots\,,\\
r & =r' + r'\sum_{n=0}^\infty  C_n(x') t'^n+ r'^2 \sum_{n=0}^\infty W_n(r') t'^n +\cdots\,.
    \end{aligned}
\end{equation}
The reason for writing things in this way is that an inspection of the PDEs reveals that one can take as ``initial data" any freely chosen functions $(A(x'), B(x'))$  (which respect periodicity conditions, if any, on $x$) and then determine the remaining functions in terms of these.  At finite $r_c$, it seems rather arduous to solve this problem in closed form, so we will work perturbatively and treat the amplitudes of $(A, B)$ small. This analysis is equivalent to a small $G$ expansion.   Perturbation theory is straightforward to carry out since the functions $(U_n, V_n, C_n, W_n)$ are determined algebraically in terms of $(A, B)$ --- no differential equations need to be solved --- and one only needs a finite number of these functions at any given order.

The functions $(A(x'), B(x'))$ will (modulo the gauge invariances discussed below) serve as coordinates on phase space. These functions determine the location of the boundary in the original reference spacetime. The new boundary at $r'=r_c$ is located at $r= r_c +r_cC(x',t')+r_c^2 W(x',t')+\cdots$ with $(C,W,\cdots)$ are functions of $(A,B)$.

\subsection{Boundary stress tensor}

Proceeding, the output of this analysis that we will need is an expression for the boundary stress tensor $T_{ij}$, evaluated at $t'=0$, in terms of $(A, B)$. For this procedure to work, we need the radial derivatives of the metric components evaluated at the boundary.  One can straightforwardly work this out to any desired order. We focus on the case that the reference solution is Poincar\'{e} AdS$_3$ given by \eqref{dia} with $x$ non-compact. Up to the quadratic order in fields, the stress tensor works out to be
\begin{equation}
    \begin{aligned}
    \label{do}
    4G T_{tt}& = -A''' + \frac{1}{2}   \big( (A'^2)''+A''^2\big) -\frac{1}{2} \big( (B'^2)''+B''^2\big)  - \frac{1}{2} (A''^2)'' r_c + \cdots\,,  \\
4G T_{xt} & = -B'''+(A'B')'' +A''B'' +(A''' B'')' r_c + \cdots\,,  \\
4G T_{xx}& =A''' - \frac{1}{2} \big( (A'^2)''+A''^2\big) + \frac{1}{2} \big( (B'^2)''+B''^2\big)  +(A'' A''''-B''^2)r_c+ \cdots\,.  
    \end{aligned}
\end{equation}
At higher orders in the fields, the stress tensor components are more complicated but, as we discuss below, are greatly simplified after making an appropriate field redefinition.

The general structure of the stress tensor is fixed by gauge invariance (see section \ref{sec:sym} for more detail). Consider the four-parameter subgroup of the entire six-dimensional isometry group corresponding to translations, rotations, and dilatations of $(x,t)$ (with the dilatation accompanied by a rescaling of $r$). Since the stress tensor vanishes for the vacuum solution at $A=B=0$, this means it vanishes for any $(A, B) = (a + bx, c+dx)$ because any such coordinate transformation is expressed as some combination of these isometries.  These choices of $(A, B)$ are ``pure gauge."  From this, we deduce that $(A, B)$ can only appear with at least one derivative and that every term in the stress tensor must have at least one factor with at least two derivatives.

\subsection{Symplectic form}
We now discuss how to compute the symplectic form $\Omega$. We can use the relation $i_{V_\chi}\Omega = -\delta P$, where $\chi$ acts as an $x$-translation on the boundary, to efficiently deduce $\Omega$ given $P$. We first note that the gauge invariance of the preceding paragraph implies that $P$ can always be put in the form, possibly after integrating by parts,
\begin{equation}
    \label{dp}
    P = \int\! dx \Big( P_A(A,B) A' + P_B(A,B)B'\Big)\,,
\end{equation}
where $P_{A, B}(A, B)$ are local functions of $(A, B)$ and their derivatives (regular as $A, B \rightarrow 0$), and are furthermore total derivatives of such local functions.  The latter statement follows from the fact that each term in $P$ contains at least one factor with at least two derivatives.   We now argue that the symplectic form is given by $\Omega = \delta \Upsilon$ with
\begin{equation}
    \label{dq}
     \Upsilon = \int\! dx \Big( P_A(A,B)\delta A + P_B(A,B)\delta B\Big)\,.
\end{equation}
To compute $ i_{V_\chi}\Omega$ and verify equality with $\delta P$, we need expressions for $ i_{V_\chi} \delta A $ and  $ i_{V_\chi} \delta B$.  To this end, we start from \eqref{dn} and perform a subsequent infinitesimal translation $(x', t') = (x''+\chi^x, t'')$ and evaluate the effect at $t''=0$.  For an $x$-translation, take $\chi=\chi^i \partial_i = \partial_x$. This $x$-translation acts as
\begin{equation}
\begin{aligned}
\label{dr}
A(x) &\rightarrow A(x) + \chi^x +A'(x)\chi^x\,,\\ B(x) &\rightarrow B(x) + B'(x) \chi^x\,,
\end{aligned}
\end{equation}
so that
\begin{equation}
\label{ds}
  i_{V_\chi} \delta A = \chi^x +A'(x)\chi^x\,,\quad  i_{V_\chi} \delta B =  B'(x) \chi^x\,.
\end{equation}
Using this to evaluate  $i_{V_\chi}\Omega $, the first term in $ i_{V_\chi} \delta A$ gives zero since $P_A$ is a total derivative.   The remaining terms give
\begin{equation}
\begin{aligned}
\label{dt}
i_{V_\chi}\Omega  &  = -\int\! dx \Big( P_A \delta A' +P_B \delta B' + \delta P_A A' + \delta P_B B'  \Big)  \\
& = -\delta P\,,
\end{aligned}
\end{equation}
as desired.

We also need to establish that $\Omega $ defined above is the unique object obeying all requirements.  Consider replacing $\Omega \rightarrow \Omega + \Delta \Omega$.  We can always write $\Delta \Omega $ in the form
\begin{equation}
    \label{du}
     \Delta \Omega = \int\! dx \Big( \delta X_A(A,B)\wedge \delta A+ \delta X_B(A,B)\wedge \delta B\Big)
\end{equation}
with $X_{A, B}(A, B)$ being local functions with a perturbative expansion
by the following argument.  First, $\Delta \Omega $ is closed.  Second, by gauge invariance each  $A$ or $B$ in $X_{A, B}$ appears with at least one derivative, and $X_{A, B}$ must be total derivatives of local functions.  Using these facts, we contract with the $x$-translation vector field and find
\begin{equation}
\begin{aligned}
    \label{dv}
     i_{V_\chi} \Delta\Omega & = \delta \int\! dx \Big( X_A A' + X_B B' \Big) \\
& = - \delta \int\! dx \Big( X_A' A + X_B' B\Big)\,.
\end{aligned}
\end{equation}
We proceed to assert that this quantity must equal zero to maintain $i_{V_\chi}\Omega = -\delta P$. Using that $X_{A, B}$ admit a perturbative expansion and does not involve undifferentiated $A$ or $B$, it is easy to see that  $  i_{V_\chi} \Delta\Omega =0$ requires $X'_A= X'_B=0$.  Since $X_A$ and $X_B$ are constants, they obey $\delta X_A = \delta X_B=0$, which then implies $\Delta \Omega =0$.

\subsection{Action and equations of motion}
From \eqref{dn}, the canonical equations of motion are
\begin{equation}
    \label{dw}
     \dot{A} = A_1\,,\quad \dot{B} = B_1\,, 
\end{equation}
where we recall that $(A_1, B_1)$ are functions of $(A, B)$ and their derivatives. Related to this, if we consider a diffeomorphism by the time translation vector field $\eta^i \partial_i = \eta^t \partial_t$, we have
\begin{equation}
\label{dx}
 i_{V_\eta}\delta A = A_1 \eta^t  \,,\quad i_{V_\eta}  \delta B =\eta^t+ B_1 \eta^t\,. 
\end{equation}
The equations of motion \eqref{dw} are equivalent to the statement that $V_\eta$ is the Hamiltonian vector field corresponding to $Q[\eta]$, i.e. $i_{V_\eta} \Omega = -\delta Q[\eta]$, as this is a special case of \eqref{dl}. In particular, taking $\eta^t = -i$ gives $Q[\eta]=H$.

We now seek an action whose Euler-Lagrange equations coincide with these equations of motion,  $i_{V_\eta} \Omega = -\delta H$. The answer is
\begin{equation}
\begin{aligned}
\label{dy}
 I  &= -\int\! dt  \big(i_{V_\eta} \Upsilon -H \big) \\
& =  \int\! d^2x \big( i P_A\dot{A} +i P_B \dot{B} +\mathcal{H} \big)\,,
\end{aligned}
\end{equation}
where $\mathcal{H} = {1\over 2\pi}T_{tt} $ so that $H=\int\! dx \mathcal{H}$.  The equality between the two lines makes use of the fact that $P_{A, B}$ are total derivatives so that the $\eta^t$ term in $i_{V_\eta}\delta B$ does not contribute.   To see that this is the correct action, we evaluate $i_{V_\eta} \Omega$  as we did in \eqref{dt} in the case of an $x$-translation.  We now get
\begin{equation}
\begin{aligned}
    \label{dz}
      i_{V_\eta} \Omega& =- i \int\! dx \big( \dot{P}_A \delta A   - \delta P_A \dot{A} +   \dot{P}_B \delta B   - \delta P_B \dot{B}\big)\\
& = i \int\! dx  \delta  \big( P_A \dot{A} + P_B \dot{B} \big)\,,
\end{aligned}
\end{equation}
which implies
\begin{equation}
    \label{daa2}
 i_{V_\eta} \Omega + \delta H = i \int\! dx \delta \big( P_A \dot{A} +P_B\dot{B} -i\mathcal{H} \big)\,.
\end{equation}
The vanishing of $ i_{V_\eta} \Omega + \delta H$ is equivalent to the Euler-Lagrange equations for \eqref{dy}. From \eqref{do}, the action in quadratic order is 
\begin{equation}
\label{dab}
I = {1\over 16\pi G} \int\! d^2x \Big( A'''(\dot{B}-A') + B''' (\dot{A}+B') + \operatorname{cubic} + \cdots  \Big)\,.
\end{equation}
The cubic and higher order terms are complicated when expressed in terms of $(A, B)$, but we will later find a field redefinition that significantly simplifies the action.

\subsection{Symmetries}
\label{sec:sym}

The action \eqref{dy} is invariant under certain gauge and global symmetries. We begin with the former, which originates due to the isometries of the reference solution. AdS$_3$ has a six-parameter group of coordinate transformations that leave the metric invariant, which in Lorentzian signature is $\text{SL}(2,\mathbb{R})\times \text{SL}(2,\mathbb{R})$. Applying one of these followed by a general coordinate transformation has the same effect as applying only the latter, and this statement implies an equivalence relation between distinct $(A, B)$.

We will adopt more compact notation, with $x^i=(x,t)$, writing \eqref{dn} as
\begin{equation}
\begin{aligned}
\label{fa}
 x^i & = x'^i +A^i(x'^j)+ (r'-r_c)U^i(x'^j)  +\cdots\,,  \\
r & =r' + r'C(x'^j)+ r'^2 W(x'^j)  +\cdots\,.
\end{aligned}
\end{equation}
Let $\xi^\mu$ obey $\nabla^0_\mu \xi_\nu + \nabla^0_\nu \xi_\mu =0$, where the derivatives are defined with respect to the reference solution $g^0_{\mu\nu}$. Writing the coordinate transformation $x^\mu = x'^\mu+ \xi^\mu(x') $ in the form \eqref{fa} defines $(A^i_\xi,U^i_\xi,C_\xi,F_\xi)$. We compose this with the subsequent transformation
\begin{equation}
\begin{aligned}
    \label{fa2}
     x'^i & = x''^i +A^i(x''^i)+ (r''-r_c)U^i(x''^i)  +\cdots\,,  \\
r' & =r'' + r''C(x''^i)+ r''^2 W(x''^i)  +\cdots \,,
\end{aligned}
\end{equation}
and evaluate at $(t'', r'') = (0, r_c)$. This defines a transformation between  $x^\mu$ and $x''^\mu$ labeled by some modified $A^i$ functions and is equivalent to the transformation \eqref{fa} in which $(x^i,r)$ appears on the left-hand side. Explicitly, the gauge equivalence is
\begin{equation}
    \label{fb}
     A^i(x) \cong A^i(x) + \left[ A^i_\xi\big(  x^j+ A^j(x^k) \big) + \big( r_c C(x^j)+r_c^2 W(x^j)\big)U^i_\xi \big(  x^j+ A^j(x^k)   \big) \right]\Big|_{t=0}\,. 
\end{equation}
Recalling that $C$ and $W$ are nonlinear functions of $A^i$ and their derivatives, we see that gauge transformations act in a complicated nonlinear way. The stress tensor is invariant under these transformations. The symplectic form is also invariant; this is not entirely obvious from our somewhat indirect method for extracting the symplectic form, but it is manifest when one expresses (as in \cite{Crnkovic:1986ex}) the symplectic form as an integral over a spacelike surface since that expression is expressed in terms of the metric, which is by definition invariant under the isometries.  Furthermore, both the stress tensor and the symplectic form do not depend on time derivatives of $(A, B)$, so the invariance extends to transformations in which the parameters are allowed to have arbitrary dependence on time. This type of gauge symmetry was referred to as ``quasi-local" in \cite{Cotler:2018zff}. The phase space action is determined from the symplectic form and Hamiltonian, which implies that the action is also gauge invariant. As noted, the SL($2,\mathbb{R}) \times $SL$(2,\mathbb{R})$ gauge transformations act on $(A,B)$ in a complicated nonlinear fashion. The formulas of course simplify markedly for $r_c=0$, and we write out the corresponding transformations explicitly in the next section.

Global symmetries correspond to isometries of the metric on the boundary, which are simply translations and rotations (i.e. Poincar\'{e} transformations in real-time). In this case, we first apply \eqref{fa} followed by the infinitesimal transformation
\begin{equation}
    \label{fc}
    x'^i = x''^i+\epsilon^i + \theta^{ij} x''^j\,,\quad r'=r''\,,
\end{equation}
with $\theta^{ij}=-\theta^{ji}$.  

Composing these transformations, we find
\begin{equation}
\begin{aligned}
    \label{fd}
     A(x) &\rightarrow A(x) +(1+A')\epsilon^x +  A_1 \epsilon^t +  A_1 x \theta^{tx}\,, \\
 B(x) &\rightarrow B(x) +(1+B_1)\epsilon^t +  B' \epsilon^x +  B_1 x \theta^{tx}\,, 
\end{aligned}
\end{equation}
where we used $\dot{A}|_{t=0}= A_1$ and $\dot{B}|_{t=0}= B_1$. These transformations are again highly nonlinear due to the appearance of $(A_1, B_1)$. The stress tensor transforms into a symmetric tensor under these translations and rotations.
\subsection{Asymptotic AdS$_3$ case:  Alekseev-Shatashvili action}
\label{ASsec}

For illustrative purposes, we consider the asymptotic case of $r_c=0$ where it is simple to carry out our general procedure in closed form. Starting from
\begin{equation}
\label{ga}
 ds^2 = {dr^2 \over 4r^2} +{1\over r} dzd\bar{z}\,,
\end{equation}
the coordinate transformation  \cite{Roberts:2012aq}
\begin{equation}
\begin{aligned}
    \label{gb}
     z& =  F(z') -{2r' F'^2 \bar{F}'' \over 4F'\bar{F}' +r' F''\bar{F}'' }\,, \\
 \bar{z}& =  \bar{F}(\bar{z}') -{2r' \bar{F}'^2 F'' \over 4F'\bar{F}' +r' F'' \bar{F}'' }\,, \\
 r& = {16r' F'^3 \bar{F}'^3 \over ( 4F'\bar{F}' +r' F''\bar{F}'' )^2}\,,
\end{aligned}
\end{equation}
gives
\begin{equation}
\label{gc}
ds^2 =  {dr'^2 \over 4r'^2} +{1\over r'} dz'd\bar{z}' -{1\over 2} \{F,z'\} dz'^2  -{1\over 2} \{\bar{F},\bar{z}'\} d\bar{z}'^2 +{1\over 4} r'  \{F,z'\}\{\bar{F},\bar{z}'\} dz' d\bar{z}' \,. 
\end{equation}
Here $F=F(z')$, $\bar{F}=\bar{F}(\bar{z}')$, primes on $(F,\bar{F})$ denote derivatives, and the Schwarzian derivative is
\begin{equation}
\label{gd}
 \{F,z'\} = {F''' \over F'} -{3\over 2} {F''^2 \over F'^2}\,.
\end{equation}
Writing $z=x+it$, and comparing \eqref{gb}  to \eqref{dn} we read off
\begin{equation}
\begin{aligned}
\label{ge}
 & A+iB = F -z'\,,\\
 & A-iB = \bar{F} -\bar{z}'\,,\\
 & U+iV =- {1\over 2} {F' \bar{F}'' \over \bar{F}'}\,,   \\
 &U -iV = - {1\over 2} {F'' \bar{F}' \over F'}\,, \\
 & C = F' \bar{F}'\,,  \\
 & W = -{1\over 4} F''\bar{F}''\,. 
\end{aligned}
\end{equation}
The stress tensor is
\begin{equation}
\begin{aligned}
\label{fg}
 T_{zz}& = {c_0\over 12}  \{F,z'\} \,,\quad T_{\bar{z} \bar{z}}=   {c_0\over 12} \{\bar{F},\bar{z}'\}\,,\quad T_{z\bar{z}}=0\,,
\end{aligned}
\end{equation}
where $c_0={3\over 2G}$ is the Brown-Henneaux central charge.

In this case, we, of course, could have written down \eqref{fg} directly from knowledge of the asymptotic Virasoro symmetry of  AdS$_3$ with $r_c=0$, but, here, we are emulating the procedure we carry out for the general cutoff case. Expressing $T_{ij}$ in terms of $(A, B)$ gives the all-order version of the expressions \eqref{do} at $r_c=0$.

From here, it is straightforward to calculate the Alekseev-Shatashvili action for $(F,\bar{F})$.  It is useful to generalize a bit by taking the stress tensor to be 
\begin{equation}
\begin{aligned}
    \label{gg}
     T_{zz}& = {c_0\over 12} \left( {a\over 2} F'^2+ \{F,z'\}\right) \,,\quad T_{\bar{z} \bar{z}}=  {c_0\over 12}\left( {a\over 2} \overline F'^2+  \{\bar{F},\bar{z}'\}\right)\,,\quad T_{z\bar{z}}=0\,,
\end{aligned}
\end{equation}
where $a$ is a parameter that can be thought of as the stress tensor of a more general reference solution with $T_{zz} =  T_{\bar{z} \bar{z}}= {ac_0\over 24}$. For example, $a=1$ corresponds to global AdS$_3$  provided $x\cong x+2\pi$, while values $0<a<1$ correspond to conical defect solutions. The energy and momentum from \eqref{dk} are
\begin{equation}
\begin{aligned}
    \label{gh2}
H& =    -{1\over 2\pi}\int\! dx (T_{zz} +T_{\bar{z} \bar{z}} ) = -{c_0\over 24\pi}\int \!dx \left( {a\over 2} F'^2+ \{F,z'\} +  {a\over 2} \overline F'^2+  \{\bar{F},\bar{z}'\} \right)\,, \\ P& = -{1\over 2\pi}\int\! dx (T_{zz} -T_{\bar{z} \bar{z}} ) = -{c_0\over 24\pi}\int \!dx \left( {a\over 2} F'^2+ \{F,z'\} -  {a\over 2} \overline F'^2-  \{\bar{F},\bar{z}'\} \right)\,. 
\end{aligned}
\end{equation}
We now apply our general procedure to compute the symplectic form.  This procedure was previously stated in terms of $(A, B)$ but applies equally in terms of $(F,\bar{F})$.   We write $P$ in the form
\begin{equation}
\label{gi2}
 P = \int\! dx \left( P_F F' + P_{\bar{F}} \bar{F}' \right)\,,
\end{equation}
where $P_F$ and $P_{\bar{F}}$ are total derivatives. This is easily achieved using
\begin{equation}
\begin{aligned}
    \label{gj}
\{ F,z\} =- {1\over 2} \left({1\over F'}\right)'' F' + \operatorname{total~derivative}
\end{aligned}
\end{equation}
yielding
\begin{equation}
\label{gk}
 P_F  = -{c_0\over 48\pi} \left(a F' - \left({1\over F'}\right)'' \right)\,,\quad
 P_{\bar{F}}  = {c_0\over 48\pi} \left(a \bar{F}' - \left({1\over \bar{F}'}\right)'' \right)\,. 
\end{equation}
The value of $\Upsilon$ \eqref{dq} in this case gives
\begin{equation}
\label{gl}
 \Upsilon =- {c_0\over 48\pi} \int\! dx \left[ \left(a F' - \left({1\over F'}\right)''\right) \delta F -\left( a \bar{F}' - \left({1\over \bar{F}'}\right)'' \right) \delta \bar{F} \right]\,.
\end{equation}
The general formula  \eqref{dy} then  yields the Alekseev-Shatashvili action \cite{Alekseev:1988ce}
\begin{equation}
\label{gm}
 I= -{c_0\over 24\pi}\int\! d^2x \left[ a F' \partial_{\bar{z}} F - \left({1\over F'}\right)'' \partial_{\bar{z}}F +  a \bar{F}' \partial_{z} \bar{F} - \left({1\over \bar{F}'}\right)'' \partial_{z}\bar{F}\right]\,,
\end{equation}
with
\begin{equation}
\label{gn}
 \partial_z = {1\over 2} (\partial_x -i \partial_t)\,,\quad \partial_{\bar{z}} = {1\over 2} (\partial_x +i \partial_t)\,.
\end{equation}
The theory \eqref{gm} describes a single (left and right moving) boson with variable central charge. As such, one expects it to be equivalent to the standard action for a free boson with a linear dilaton (or background charge) contribution to its stress tensor.  Indeed, as noted in \cite{Alekseev:1988ce,Alekseev:1990mp}  the field redefinition
\begin{equation}
\label{go}
 \left( e^{i\sqrt{a}F}\right)' = \sqrt{a}e^f\,,\quad  \left( e^{-i\sqrt{a}\bar{F}}\right)' =\sqrt{a}e^{\bar{f}}\,.
\end{equation}
yields
\begin{equation}
\begin{aligned}
    \label{gp}
     T_{zz}& = -{c_0\over 12}\left({1\over 2} f'^2-f''\right)\,, \\
 T_{\bar{z} \bar{z}}& = -{c_0\over 12}\left({1\over 2} \bar{f}'^2-\bar{f}''\right)\,, \\
 \Upsilon & = {c_0\over 48\pi}\int\! dx \left( f'\delta f - \bar{f}' \delta \bar{f}\right)\,, \cr
 I& ={c_0\over 96\pi} \int\! d^2x \left( f' \partial_{\bar{z}} f + \bar{f}' \partial_z \bar{f} \right)\,.   
\end{aligned}
\end{equation}
As $a\rightarrow 0$, the field redefinition reads
\begin{equation}
\label{gq}
 F'=e^f\,,\quad \bar{F}' = e^{\bar{f}} 
\end{equation}
and relations \eqref{gp} continue to hold.   Each chiral half of the action $I$ in \eqref{gp} is the Floreanini-Jackiw action \cite{Floreanini:1987as}; the two halves combined give a standard free scalar action in Hamiltonian form. On the other hand, the stress tensor in \eqref{gp} coming from gravity includes an improvement term (unlike considered in \cite{Ouyang:2020rpq}). The improvement term is crucial since without it, the central charge would be fixed at $c=1$.

Restricting now to $a=0$, the gauge transformations act as
\begin{equation}
\label{gh}
 \delta_\epsilon F = \sum_{n=0}^2 \epsilon_n(t)F^n\,,\quad  \delta_\epsilon \bar{F} = \sum_{n=0}^2 \overline{\epsilon}_n(t)\bar{F}^n\,.
\end{equation}
To compute the gauge variation of the stress tensor, symplectic form, and the action, we only need the transformations of $(f',\bar{f}')$ which are
\begin{equation}
\label{gi}
 \delta_\epsilon f' = 2\epsilon_2(t) e^f\,,\quad  \delta_\epsilon \bar{f}' = 2\overline{\epsilon}_2(t) e^{\bar{f}}\,.
\end{equation}
Gauge invariance of the stress tensor fixes the relative coefficient of the two terms in $T_{zz}$ and $T_{\bar{z} \bar{z}}$. Using standard formulas from free boson CFT, it follows that correlators of stress tensors are those of a CFT with central charge $c=c_0+1$; we elaborate on this more in the course of our $r_c\neq 0$ discussion below.

It is also instructive to see how the stress tensor in \eqref{gp} arises from Noether's theorem applied to the quadratic action in \eqref{gp}. An infinitesimal rigid translation of the boundary coordinates, $x^i \rightarrow x^i +\epsilon^i$ acts on $(F,\bar{F})$ as
\begin{equation}
\label{gia}
\delta_\epsilon F =  \partial_i F \epsilon^i\,,\quad  \delta_\epsilon \bar{F} =  \partial_i \bar{F} \epsilon^i\,,
\end{equation}
derived using the same reasoning that led to \eqref{fd}. We want a transformation on phase space, and so we use the equations of motion $\partial_{\bar{z}} F = \partial_z \bar{F}=0$ to trade away time derivatives and  obtain
\begin{equation}
\label{gib}
\delta_\epsilon F = F' \epsilon^z\,,\quad \delta_\epsilon = \bar{F}' \epsilon^{\bar{z}}\,.
\end{equation}
As usual in the derivation of Noether's theorem, we now consider the transformation \eqref{gib} that $\epsilon^i$ depends arbitrarily on $x^i$. We then work out the transformation of $(f,\bar{f})$ as
\begin{equation}
\label{gic}
\delta_\epsilon f = f' \epsilon^z+ (\epsilon^z)'\,,\quad \delta_\epsilon \bar{f} = \bar{f}' \epsilon^{\bar{z}}+ (\epsilon^{\bar{z}})'\,.
\end{equation}
We finally compute the variation of the action and write this in the form
\begin{equation}
    \label{gid}
    \delta I =-{1\over 2\pi} \int\! d^2x T_{ij} \partial^i \epsilon^j\,,
\end{equation}
yielding the stress tensor in \eqref{gp}.

\subsection{Exact action for constant  $(f',\bar{f}')$}
Besides the asymptotic $r_c=0$ limit, there is another special case in which we can derive the action to all orders. This is a consequence of the fact that we can find exact solutions to the boundary value problem in the case that second and higher derivatives of $f$ and $\bar{f}$ vanish.  This leads to a result for the action which captures all dependence on $(f',\bar{f}')$, but not on higher derivatives of these fields. 

We start from
\begin{equation}
\label{hia}
ds^2 = {d\rho^2 \over 4\rho^2}+{1\over \rho} dw d\bar{w}
\end{equation}
and first, perform the coordinate change
\begin{equation}
\begin{aligned}
\label{hib}
 w& = {1-r\over 1+r} e^{ 2 \left({ x \over 1+r_c}+{it \over 1-r_c} \right) }\,, \\
\bar{w} & =  {1-r\over 1+r} e^{ 2 \left({ x \over 1+r_c}-{it \over 1-r_c} \right) }\,.  \\
\rho& = {4r\over (1+r)^2} e^{ {4x \over 1+r_c} }\,,  
\end{aligned}
\end{equation}
which gives the line element
\begin{equation}
    \label{hic}
    ds^2 = {dr^2 \over 4r^2} +  {1\over r} \left[  \left(1 -r\over 1-r_c\right)^2 dt^2 +  \left(1 +r\over 1+r_c\right)^2 dx^2\right]\,.
\end{equation}
We now perform a further two-parameter coordinate redefinition that preserves the form of the metric at $r=r_c$. This corresponds to a rescaling by a parameter $a$,
\begin{equation}
\label{hid}
r\rightarrow  ar\,,\quad t \rightarrow  {\sqrt{a}(1-r_c)\over 1-ar_c}t\,,\quad x \rightarrow  {\sqrt{a}(1+r_c)\over 1+ar_c} x\,,
\end{equation}
followed by a rotation by angle $\theta$,
\begin{equation}
\label{hie}
t \rightarrow  t\cos \theta  + x \sin\theta\,,\quad   x \rightarrow  x\cos \theta -t \sin\theta\,.
\end{equation}
Writing the combined transformation in the form $w=A(r)e^{f(x,t)}$ and $\bar{w} = \overline{A}(r)e^{\bar{f}(x,t)}$, we compute
\begin{equation}
\begin{aligned}
    \label{hif}
     f' &= {2\sqrt{a}\over 1-a^2 r_c^2} (e^{i\theta}-ar_c e^{-i\theta})\,, \\
\bar{f}'  &= {2\sqrt{a}\over 1-a^2 r_c^2} (e^{-i\theta}-ar_c e^{i\theta})\,.  
\end{aligned}
\end{equation}
On the other hand, it is straightforward to compute the stress tensor in terms of $(a,\theta)$ and re-express the results in terms of $(f',\bar{f}')$.  Writing $P= \int\! dx p = {i\over 2\pi} \int\! T_{tx} dx$, we find
\begin{equation}
\begin{aligned}
    \label{hig}
    p = {i\over 2\pi} T_{tx} = {1\over 32 \pi G} (f'^2 -\bar{f}'^2) 
\end{aligned}
\end{equation}
as in \eqref{hc}. Further, writing $H= \int\! dx h = {1\over 2\pi} \int\! T_{tt} dx$ we find that $h$ obeys
\begin{equation}
\label{hih}
h-4 \pi G r_c (h^2-p^2)=h_0\,,
\end{equation}
where
\begin{equation}
\begin{aligned}
    \label{hii}
    h_0 =  {1\over 32 \pi G} (f'^2 +\bar{f}'^2)\,.
\end{aligned}
\end{equation}
Solving \eqref{hih} for $h$ gives,
\begin{equation}
\begin{aligned}
    \label{hij}
     h& = -{1\over 8\pi G r_c} \left( \sqrt{ 1- 16\pi G r_c h_0 +(8 \pi G r_c p)^2}-1 \right) \\
& =  -{1\over 8\pi G r_c} \left( \sqrt{1-{1\over 2}r_c(f'^2+\bar{f}'^2)+{1\over 16}r_c^2(f'^2-\bar{f}'^2)^2  }-1 \right)\,,
\end{aligned}
\end{equation}
where we chose the root which obeys $\lim_{r_c\rightarrow 0} h = h_0$.

This result implies that the full action  takes the form
\begin{equation}
\begin{aligned}
    \label{hik}
    I =& {1\over 32 \pi G} \int\! d^2x \left[  if'\dot{f}-i\bar{f}' \dot{\bar{f}} - 4
{ \sqrt{1-{1\over 2}r_c(f^{\prime 2}+\bar{f}^{\prime 2})+{1\over 16}r_c^2(f'^2-\bar{f}'^2)^2  }-1\over r_c}  \right] \\
&+ {\text{higher~derivatives}}\,,  
\end{aligned}
\end{equation}
where the higher derivative terms vanish when $f''=0$ and $\bar{f}''=0$. 

From \cite{Kraus:2022mnu}, we define
\begin{equation}
    \phi = \frac{1}{\sqrt{32 \pi G}} \left(f+ \bar{f} \right), \qquad \Pi =\frac{1}{\sqrt{32 \pi G}} \left( f' - \bar{f}' \right)
\end{equation}
so that the full action \eqref{hik} becomes the Nambu-Goto action in static gauge\footnote{The static gauge is the gauge choice that one identifies two of the target space coordinates with the worldsheet coordinates. More specifically: $  X^0 = t, ~X^1 = x, ~X^2 = \sqrt{\lambda} \phi$ and the direction $X^1$ is taken to be compact $X^1(x+2\pi, t) = X^1(x,t) + 2\pi$ for unit winding sector.
}
\begin{equation}
\begin{aligned}
\label{eq:o0eruioioueriuojeroijeroi}
    I &= \int d^2x \left[ i \Pi \dot{\phi} - \frac{1}{\lambda} \left( 1- \sqrt{1  -  \lambda \left( \phi'^2 + \Pi^2 \right) + \lambda^2 \phi'^2 \Pi^2} \right) \right]\\&= \int d^2x \left[ i \Pi \dot{\phi} - \frac{1}{\lambda} + \mathcal{L}_{\operatorname{Nambu-Goto}}\right]
\end{aligned}
\end{equation}
and
\begin{equation}
    \mathcal{L}_{\operatorname{Nambu-Goto}} = \frac{1}{\lambda} \sqrt{-\det \left(  \eta^{\mu \nu} \partial_i  X_\mu  \partial_j X_\nu \right)}\,, 
\end{equation}
with target space indices being $\mu, \nu = 0, 1, 2$, worldsheet indices being $i,j =0,1$ and Minkowski target space metric $\eta_{\mu \nu} = \operatorname{diag} \left( -1, 1, 1\right)$.

In the $\lambda \rightarrow 0$ limit, or going from the ``string'' to the particle $\alpha' \rightarrow 0$ limit, of \eqref{eq:o0eruioioueriuojeroijeroi}, we obtain\footnote{The quotation marks are to remind us from section \ref{sec:factorfloweq} that the $T\overline{T}$-deformed theory is not a full-fledged string theory due to the absence of dynamical gravity.} 
\begin{equation}
    I = \int d^2x \left( i \dot{\phi} \Pi + \frac{1}{2} \left( \phi'^2 + \Pi^2\right) \right)\,.
\end{equation}

In the next section, we will provide strong evidence that the higher derivative terms are absent in general provided we define $(f,\bar{f})$ appropriately.

In this context, we can address the fact that the square root can become imaginary in some region of the $(f',\bar{f}')$ plane. From \eqref{hif}, we have
\begin{equation}
\label{him}
 1-{1\over 2}r_c(f'^2+\bar{f}'^2)+{1\over 16}r_c^2(f'^2-\bar{f}'^2)^2  =  \left( { 1-2a r_c \cos(2\theta) +a^2 r_c^2 \over 1-a^2 r_c^2 }\right) ^2
\end{equation}
or in terms of $(\phi, \Pi)$
\begin{equation}
1 - 4 \pi G r_c\left( \phi'^2 + \Pi^2 \right) + ( 8 \pi G r_c)^2 \phi'^2 \Pi^2  =  \left( { 1-2a r_c \cos(2\theta) +a^2 r_c^2 \over 1-a^2 r_c^2 }\right) ^2\,.
\end{equation}

Therefore, the square root is real for any real value of $(a,\theta)$. It is natural to expect that in general (i.e. dropping the linearity assumption)  the domain of $(f,\bar{f})$ is bounded such that the integrand in \eqref{hik} is real, but this remains to be shown.  

\subsection{Boundary gravity action on planar cutoff}

We now consider the general case, a planar boundary at $r=r_c$ with arbitrary functions $f$ and $\bar{f}$. As explained, the strategy is to start from the reference solution
\begin{equation}
\label{ha}
 ds^2 ={dr^2 \over 4r^2}+{dx^2+dt^2\over r}
\end{equation}
and look for coordinate transformations, expressed in the form \eqref{dn}-\eqref{dna}, such that
\begin{equation}
    \label{hb}
     g_{r'r'}|_{r'=r_c}  = {1\over 4r_c^2}\,,\quad g_{r' i' }|_{r'=r_c}  = 0\,,\quad  g_{i' j' }|_{r'=r_c}={\delta_{ij}\over r_c}\,,
\end{equation}
where $i$ and $j$ run over $(x,t)$. This problem can be solved order-by-order as an expansion in the freely specifiable functions $(A(x'), B(x'))$.   Only algebraic equations need to be solved in each order, and the procedure is easily automated on the computer. What we need from this procedure are expressions for the components of the boundary stress tensor evaluated at $t'=0$, which in turn depend on  $\partial_{r'}g_{i'j'}|_{r'=r_c}$. The resulting expressions to quadratic order were written in \eqref{do}.

At $r_c=0$, we saw that the expressions for the stress tensor simplify dramatically under the field redefinition \eqref{gq}, and so we seek a version of this at nonzero $r_c$. As a criterion for what constitutes an optimal field redefinition, we note that the symplectic form $\Omega = \delta \Upsilon$ will have a complicated expansion in $(A, B)$. Quantization of the phase space action uses the natural measure  Pf$ (\Omega)$. A nontrivial measure is incorporated by expressing the Pfaffian as a fermionic path integral, but life is much simpler if the Pfaffian is constant, which is indeed the case at $r_c=0$ after making the field redefinition. We therefore try to generalize this feature to $r_c \neq 0$. Recall that the symplectic form is obtained from the momentum $P$ via the formulas \eqref{dp}-\eqref{dq}.\footnote{We argued for this using the $(A, B)$ fields, but we will see below that the argument also holds after the field redefinition to new fields $(f,\bar{f})$.} We look to define new fields $(f,\bar{f})$ such that
\begin{equation}
\label{hc}
 P= {i\over 2\pi} \int\! T_{tx}dx =  {1\over 32\pi G} \int\! dx (f'^2-\bar{f}'^2)\,,
\end{equation}
which implies
\begin{equation}
\label{hd}
\Upsilon =  {1\over 32\pi G} \int\! dx (f' \delta f -\bar{f}' \delta \bar{f} )\,.
\end{equation}
In particular, we want
\begin{equation}
\label{he}
T_{tx} = -{i\over 16G}(f'^2-\bar{f}'^2) + {\text{total~derivative}}\,.
\end{equation}
By explicit computation, we find that this is achieved by taking
\begin{equation}
\begin{aligned}
    \label{hf}
    A'+iB' &= \exp\left[f-{1\over 4} r_c \bar{f}'^2-{1\over 8}r_c^2 \bar{f}'(f'\bar{f}')'-{1\over 16}r_c^2 f'(\bar{f}'^2)'+ \cdots \right] -1\,,  \\
 A'-iB' &= \exp\left[\bar{f}-{1\over 4} r_c f'^2-{1\over 8}r_c^2 f'(f'\bar{f}')'-{1\over 16} r_c^2 \bar{f}'(f'^2)'+\cdots \right]-1\,.  
\end{aligned}
\end{equation}
This redefinition involves the spatial derivatives of $(A, B)$, so $(A, B)$ are non-locally related to $(f,\bar{f})$. However, no undifferentiated $(A, B)$ will ever appear in the stress tensor, and so the stress tensor and action will be local in terms of $(f,\bar{f})$. The terms written in \eqref{hf} are sufficient to work out the stress tensor and action up to quartic order in the new fields, which is sufficient for the two-loop computations we perform in this work. For $T_{xt}$, we have
\begin{equation}
\begin{aligned}
    \label{bia}
     4G T_{xt}& = -{i\over 4}(f'^2 - \bar{f}'^2) + \frac{i }{16}r_c^2 \big(2 f' f'' \bar{f}'' - f'^2 \bar{f}''' - 2 f'' \bar{f}' \bar{f}'' + f''' \bar{f}'^2\big)'\\&+ \text{quartic~total~derivatives} \,.
\end{aligned}
\end{equation}
$T_{tt}$ is found to be
\begin{equation}
\begin{aligned}
    \label{hfa}
 4GT_{tt}& = {1\over 4}(f'^2+\bar{f}'^2)-{1\over 16} r_c^2 \left( f'' \bar{f}'^2 + f'^2 \bar{f}''\right)''+{1\over 8}r_c f'^2 \bar{f}'^2\\&+ \text{quartic~total~ derivatives}\,.
\end{aligned}
\end{equation}
The Hamiltonian to quartic order is 
\begin{equation}
    \label{hg}
     H = \int\! dx \mathcal{H} = {1\over 2\pi} \int\! dx T_{tt} = {1\over 16\pi G} \int\! dx \left( {1\over 2}  f'^2+ {1\over 2}  \bar{f}'^2 +{1\over 4}r_c f'^2 \bar{f}'^2 + \cdots \right)\,, 
\end{equation}
leading to the  simple action
\begin{equation}
\begin{aligned}
    \label{hh}
     I  &= -\int\! dt  \big(i_{V_\eta} \Upsilon -H \big) \\
& = {1\over 16\pi G} \int\! d^2x \Big( f'\partial_{\bar{z}}f+\bar{f}' \partial_z \bar{f}  +{1\over 4} r_c f'^2 \bar{f}'^2 + \cdots  \Big)\,,
\end{aligned}
\end{equation}
where as usual $(z, \bar{z}) =(x+it, x-it)$. This agrees with the expansion of \eqref{hik} to this order, with no higher derivative terms present.\footnote{In the next section, we use the Chern-Simons formulation to verify \eqref{hik} to eighth order.} Coming back to our choice of the $1/16$ in \eqref{hf}, for any other choice of coefficient the Hamiltonian includes a term proportional to  $ r_c^2 f'\bar{f}' f''\bar{f}''$.
The full expressions for the stress tensor components to cubic order are
\begin{equation}
\begin{aligned}
    \label{hi}
     4G T_{zz}& ={1\over 2}f''-{1\over 4} f'^2 +{1\over 4} r_c f''' \bar{f}'  -{1\over 8}r_c \big(f'^2 -2 f'\bar{f}' \big)'\,\bar{f}' \cr
& +{1\over 16} r_c^2 \big( f'''' \bar{f}'^2+ (f'^2)''\bar{f}''\big) +{\text{quartic}}\,, \cr
 4G T_{\bar{z} \bar{z}}& ={1\over 2}\bar{f}''-{1\over 4} \bar{f}'^2 +{1\over 4} r_c \bar{f}''' f'  -{1\over 8}r_c \big(\bar{f}'^2 -2 \bar{f}'f' \big)'\,f' \\
 & +{1\over 16}r_c^2\big( \bar{f}'''' f'^2 +(\bar{f}'^2)'' f''\big)+{\text{quartic}}\,,  \cr
 4G T_{z \bar{z}} & =  -{1\over 4} r_c f''\bar{f}''+ {1\over 8}r_c (f''\bar{f}'^2+f'^2 \bar{f}'')  -{1\over 8} r_c^2 ( f''' \bar{f}' \bar{f}''+ f'f''\bar{f}''') +  {\text{quartic}}\,.
\end{aligned}
\end{equation}
The action and stress tensor are invariant under the gauge transformation discussed in section \ref{sec:sym}. The complicated form of these transformations, along with the need to re-express them in terms of $(f,\bar{f})$, make these symmetries difficult to use in practice. However, we expect that the full expressions for the stress tensor are fixed by gauge invariance in terms of their leading terms. Let us also comment that the stress tensor in principle can be derived via Noether's theorem using the transformations \eqref{fd}, as was done at $r_c=0$ at the end of section \ref{ASsec}.

Finally, we justify why we can pass from the momentum \eqref{hc} to the canonical one-form $\Upsilon$ in \eqref{hd}. As before, the question is whether the equation $i_{V_\chi}\Omega =-\delta P$ fixes the symplectic form $\Omega$ according to our rule, or whether there is an ambiguity of the form $\Delta \Omega = \int\! dx (\delta X_f\wedge \delta f + X_{\bar{f}}\wedge \delta \bar{f})$.  To this end, we note that we can invert \eqref{hf} order-by-order to obtain local expressions for $(f,\bar{f})$ in terms of $(A, B)$. $(A, B)$ will always appear with a least one derivative. Given this, it follows that any candidate $\Delta \Omega$ of the form just noted will, under the field redefinition to $(A, B)$ turn into a $\Delta \Omega$ of the sort that we previously excluded. This justifies the procedure in the $(f,\bar{f})$ frame.

\section{Chern-Simons formulation of boundary graviton action}
\label{CSsec}

Classically or in quantum perturbation theory, $3d$ Einstein gravity can be formulated as a gauge theory, namely a Chern-Simons theory whose connections are constructed from the spin connection and vielbein of the first-order formulation of general relativity. The relation between Chern-Simons theory and Einstein gravity at the non-perturbative level is unclear. One of the main ingredients in a non-perturbative theory of gravity is a sum over topologies, while such a sum is not natural from the perspective of gauge theory. These issues are, however, beyond the scope of this work. In this section, we are interested in understanding the perturbative theory of boundary gravitons from the Chern-Simons perspective.

The general strategy will be similar to section \ref{sec:BoundaryGravitonsMetric}: to identify the boundary phase space, we consider all gauge transformations of a chosen solution that preserve the boundary conditions and then quotient out small gauge transformations. Having identified the phase space, we evaluate the action. This will be done for the case of boundary conditions imposed at the asymptotic boundary of AdS$_3$, as well as the case of a finite cutoff boundary. 

We start with a quick review of Chern-Simons gravity in three dimensions. In this section, we work in  Lorentzian signature, and only Wick rotate to Euclidean signature at the end of the computation to connect to the results of section \ref{sec:BoundaryGravitonsMetric}.\footnote{To be precise, we will relate Lorentzian to Euclidean time by $t_L = i t_E$ and Lorentzian actions $S$ to Euclidean actions $I$ through $I = i S|_{t_L \to i t_E}$.}

\subsection{Action and boundary conditions}
As mentioned above, Einstein gravity in $3d$ is classically equivalent to a gauge theory. For negative cosmological constant, the gauge group is $SO(2,2)=\operatorname{SL}(2,\mathbb{R})\times \operatorname{SL}(2,\mathbb{R})/\mathbb{Z}_2$. 

We denote the generators of $\operatorname{sl}(2,\mathbb{R})$ by $L_{0,\pm1}$ and take them to obey
\begin{equation}
\label{qqa}
 [L_m, L_n] = (m-n) L_{m+n}\,.
\end{equation}
An explicit representation is 
\begin{equation}
\label{qqb}
L_0 = \left(
                 \begin{array}{cc}
                   \frac{1}{2} & 0 \\
                   0 & -\frac{1}{2} \\
                 \end{array}
               \right)\,,\quad  L_1 = \left(
                 \begin{array}{cc}
                   0 & 0 \\
                   -1 & 0 \\
                 \end{array}
               \right)\,,\quad
                L_{-1} = \left(
                 \begin{array}{cc}
                   0 & 1 \\
                   0 & 0 \\
                 \end{array}
               \right) \,,
\end{equation}
which obey
\begin{equation}
    \label{qqba}
    \operatorname{Tr} (L_0^2 )= {1\over 2}\,,\quad \operatorname{Tr} (L_1 L_{-1} ) =-1\,, 
\end{equation}
with other traces vanishing.

The Chern-Simons connections are related to the dreibein and the spin connection of the first-order formulation of gravity as follows
\begin{equation}
\label{ppw}
A=L_a \left(   \omega_\mu^a +  e^a_\mu  \right)dx^{\mu}
\,, \quad
\bar{A} = L_a \left(   \omega_\mu^a -  e^a_\mu  \right)dx^{\mu}\,.
\end{equation}
The base manifold $M_3$ equips with coordinates $x^{\mu}=\{r, t, x\}$, where $r$ is the holographic coordinate for which $r \to 0$ at the conformal boundary. Greek indices are reserved for the boundary $M_3$, which is equipped with coordinates $x^{i}=\{ t, x \}$. The metric extracted from the Chern-Simons connections is
\begin{equation}
g_{\mu\nu} = 2 \operatorname{Tr} \left( e_{\mu}e_{\nu} \right) = {1\over 2} \operatorname{Tr}\left[ (A-\bar{A})_{\mu}(A-\bar{A})_{\nu}  \right]
\end{equation}
The gravitational action is written in terms of the connections as 
\begin{equation}
\label{eq:ConsistentAction}
S_{\text{grav}} = S_{\text{CS}}[A] - S_{\text{CS}}[\bar{A}] + S_{\text{bndy}}\,,
\end{equation}
where the Chern-Simons action at level $k$ reads
\begin{equation}
    S_{\text{CS}}[A] = {k \over 4\pi} \int_{M_3} \operatorname{Tr} \left(  A\wedge dA + {2\over 3}A\wedge A\wedge A  \right)\,.
\end{equation}
Here $k$ is related to Newton's constant $G$ and the AdS$_3$ length scale as%
\footnote{We are again working with $\ell=1$ in this section.} 
\begin{equation}
    k = \frac{\ell}{4G}\,.
\end{equation}
The equations of motion imply the flatness of the Chern-Simons connections, which correspond to the Einstein equations and the vanishing of torsion.

We now turn to the choice of boundary conditions and the associated boundary term in the action. In complete generality, we write the connections as
\begin{equation}
\begin{aligned}
    \label{aama}
     A& = E^+ L_1 + \Omega L_0 + f^- L_{-1} \,,  \\
  \bar{A}& = f^+ L_1 + \bar{\Omega} L_0 + E^- L_{-1} \,,
\end{aligned}
\end{equation}
where at this stage, all functions depend arbitrarily on all three coordinates. The corresponding metric is
\begin{equation}
\label{aamb}
ds^2 =    {1\over 4}(\Omega -\bar{\Omega})^2 +(E^+-f^+)(E^--f^-)  \,.  
\end{equation}
We choose boundary conditions that mimic our construction in the metric formulation, where we choose to fix all metric components at $r=r_c$. We write  
\begin{equation}
\label{aamc}
\left(  \Omega-\bar{\Omega} \right)\vert_{r_c}  = {1\over r_c}dr
\,, \quad
( E-f)^{\pm}\vert_{r_c} = \pm 2 e^{\pm}_{i}  dx^i\,,
\end{equation}
so  that
\begin{equation}
    \label{aamd}
     ds^2\big|_{r_c} = {dr^2 \over 4r_c^2} -4e^+_i e^-_j dx^i dx^j\,.
\end{equation}
Here, $e^{\pm}_{i}$ is the fixed boundary zweibein. A boundary term compatible with these boundary conditions is\footnote{This boundary term appears in a different form in \cite{Llabres:2019jtx}.}
\begin{equation}
    S_{\text{bndy}} = {k \over 4\pi} \int_{\partial M_3} \operatorname{Tr} A\wedge \bar{A} - {k \over 4\pi} \int_{\partial M_3} \operatorname{Tr} \left[ L_0 (A-\bar{A})\wedge (A-\bar{A})     \right]\,.
\end{equation}
In particular, a straightforward computation yields the following on-shell variation of the action 
\begin{equation}
    \label{alg}
     \delta S_{\text{bulk}}+\delta S_{\text{bndy}} = {k\over 2\pi }\int_{\partial M_3}  \Big( f^-\wedge \delta e^+ +f^+\wedge \delta e^- \Big)\,. 
\end{equation}
Since only variations of the fixed quantities $e^\pm$ appear, our variational principle is consistent.  

In practice, it is convenient to impose additional boundary conditions that are compatible with the variational principle and incorporate all solutions of interest. The boundary spin connection $\omega$, given by 
\begin{equation}
    {1\over 2}\left( \Omega + \bar{\Omega} \right)\vert_{r_c} = \omega\,,
\end{equation}
is so far unfixed.  However, the vanishing of the Chern-Simons field strength implies
\begin{equation}
\label{aamda}
 de^+ -\omega \wedge e^+ = de^- +\omega \wedge e^- =0\,,
\end{equation}
which are the usual torsionless conditions that uniquely fix the boundary spin connection $\omega$ in terms of the vielbein $e^\pm$. We therefore impose \eqref{aamda} as a boundary condition. The remaining flatness conditions evaluated at the boundary impose conservation of the stress tensor (defined below) and also fix its trace. 

\subsection{Stress tensor} 

The boundary stress tensor is defined in terms of the on-shell variation of the action as\footnote{In this formula $\sqrt{g}$ and $\det e$ refer to boundary quantities,  related by $\sqrt{g} = 2\det e$ according to our convention \eqref{ppw}.}  
\begin{equation}
    \begin{aligned}
        \label{qqc}
        \delta S &=   {1\over 4\pi} \int\! d^2x \sqrt{g} \, T^{ij}\delta g_{ij} \\
& = {1\over \pi} \int\! d^2x  \det e \, T^i_a \delta e^a_i\,.
    \end{aligned}
\end{equation}
Comparing to \eqref{alg}, we read off
\begin{equation}
    \label{algg}
     T_+^j ={k} \epsilon^{ij} f^-_i\,,\quad   T_-^j ={k} \epsilon^{ij} f^+_i\,,
\end{equation}
where our boundary orientation is defined by 
\begin{equation}
\label{qqd}
\epsilon^{tx} = -\epsilon^{xt}  = {1\over \det e}\,.
\end{equation}

\subsection{Relation to metric formulation}

We write the relation between the bulk and boundary terms in the metric versus Chern-Simons descriptions. For the bulk Einstein-Hilbert action, we have
\begin{equation}
    \label{qqe}
     {1\over 16 \pi G} \int_{M_3} \sqrt{g}(R+2) =   S_{\text{CS}}[A] - S_{\text{CS}}[\bar{A}] - {k\over 4\pi} \int_{\partial M_3} d \operatorname{Tr} (A\wedge \bar{A})\,.
\end{equation}
For the Gibbons-Hawking terms, where $h_{\mu\nu}$ is the boundary metric, we explain in appendix \ref{sec:GHY_CS} how it can be rewritten as 
\begin{equation}
    \label{qqf}
     {1\over 8\pi G} \int_{\partial M_3} \! d^2x \sqrt{h} K = {k\over 2\pi} \int_{\partial M_3} \operatorname{Tr} (A\wedge \bar{A})\,,
\end{equation}
under the condition $\partial_a n^a = 0$, which is satisfied given our choice of gauge \eqref{aamc}.
We finally have the boundary area counterterm, 
\begin{equation}
\label{qqg}
-{1\over 8\pi G} \int_{\partial M_3} \sqrt{h} = -{k\over 4\pi} \int_{\partial M_3} \operatorname{Tr} [L_0(A-\bar{A})\wedge (A-\bar{A})]\,.
\end{equation}
The relationship between the complete actions is therefore\footnote{Note that we are defining the action with an overall sign flip compared \eqref{eq:EH}; this has to do with the fact that \eqref{eq:EH} defined in Euclidean signature.}
\begin{equation}
\begin{aligned}
\label{qqi}
 &{1\over 16 \pi G} \int_{M_3} \sqrt{g}(R+2)   +  {1\over 8\pi G} \int_{\partial M_3} \! d^2x \sqrt{h} K - {1\over 8\pi G} \int_{\partial M_3} \sqrt{h}  \\
& =   S_{\text{CS}}[A] - S_{\text{CS}}[\bar{A}] + {k\over 4\pi} \int_{\partial M_3}  \operatorname{Tr} (A\wedge \bar{A})~  -{k\over 4\pi} \int_{\partial M_3} \operatorname{Tr} [L_0(A-\bar{A})\wedge (A-\bar{A})].
\end{aligned}
\end{equation}
So, our Chern-Simons action agrees with the standard gravity action.

\subsection{Boundary action}\label{sec:CSBoundaryAction}

We now reduce the bulk theory to the boundary by solving the constraints of the theory and substituting it back in. 
The Chern-Simons connections can be written in a space-time split as
\begin{equation}
    \label{eq:CSActionSplit}
    A=A_t dt + \tilde{A}\,, \quad \bar{A}=\bar{A}_t dt+\tilde{\bar{A}}
\end{equation}
and we similarly write the exterior derivative on spacetime as $d = dt \, \partial_t + \tilde{d}$. 
The components $A_t$ and $\bar{A}_t$ appear in the action as  Lagrange multipliers,
\begin{equation}
   S_{\text{CS}}[A]={k\over 4\pi} \int_{M_3} \operatorname{Tr}\left[
2A_t dt\wedge \tilde{\mathcal{F}} + \tilde{A}\wedge dt\wedge \partial_t \tilde{A} - \tilde{d}\left(    \tilde{A}\wedge A_t dt   \right)
    \right]\,, 
\end{equation}
where the spatial field strength is $\tilde{\mathcal{F}} = \tilde{d} \tilde{A} + \tilde{A}\wedge \tilde{A}$. 
The $A_t$ equation imposes the constraint that the spatial components of the field strength (and its barred counterpart)  must vanish,
\begin{equation}
    \tilde{\mathcal{F}} = \tilde{\overline{\mathcal{F}}}=0\,.
\end{equation}
These constraints are solved by writing 
\begin{equation}
    \label{eq:ConstrainedConnections}
    \tilde{A} = g^{-1}\tilde{d}g\,, \quad
\tilde{\bar{A}} = \bar{g}^{-1}\tilde{d}\bar{g}\,.
\end{equation}
We write the group elements in a Gauss parametrization
\begin{equation}
\begin{aligned}
    \label{zm}
&g = \left(\begin{array}{cc}1 & 0 \\-F & 1 \\\end{array}\right)
 \left(\begin{array}{cc}\lambda  & 0 \\ 0 & \lambda^{-1} \\\end{array}\right)
  \left(\begin{array}{cc}1 & \Psi \\0  & 1 \\\end{array}\right)\,, \\
  &
  \bar{g} = \left(\begin{array}{cc}1 & \bar{F} \\ 0 & 1 \\ \end{array}\right)
 \left(\begin{array}{cc}\bar{\lambda}^{-1}   & 0 \\ 0 & \bar{\lambda} \\\end{array}\right)
  \left(\begin{array}{cc}1 & 0 \\  -\bar{\Psi}  & 1 \\\end{array}\right)\,. 
    \end{aligned}
\end{equation}
It is a straightforward exercise to rewrite the boundary conditions \eqref{aamc} in terms of the functions appearing in \eqref{zm}. The $(\Psi,\bar{\Psi})$  are determined as 
\begin{equation}
\begin{aligned}
    \label{zq}
    \Psi  &= -{\lambda' \over \lambda^3 F'} +{\omega_x\over 2\lambda^2 F'}\,, \cr
 \bar{\Psi}  &= -{\bar{\lambda}' \over \bar{\lambda}^3 \bar{F}'} -{\omega_x \over 2\bar{\lambda}^2 \bar{F}'}  \,,
\end{aligned}
\end{equation}
where $\omega_x$ is the space component of the boundary spin connection, fixed in terms of the boundary vielbein. The remaining boundary conditions amount to the following differential equations
\begin{equation}
\begin{aligned}
        \label{zo}
    2e^+_x& = \lambda^2  F' - \bar{\Psi}' - \bar{\lambda}^2 \bar{\Psi}^2 \bar{F}' -\frac{2\bar{\Psi}}{ \bar{\lambda}}\bar{\lambda}'\,,\\
-2e^-_x &=  \bar{\lambda}^2 \bar{F}' -  \Psi' - \lambda^2 \Psi^2  F' -{2\Psi\over \lambda}\lambda'\,.
\end{aligned}
\end{equation}
The equations \eqref{zq} and \eqref{zo} are to be imposed at the boundary surface $r=r_c$.  Having chosen the Gauss parametrization, one finds that the bulk Lagrangian becomes a total derivative and so the complete action takes the form of a boundary term. After performing some algebra (detailed in appendix \ref{app:ActionAsBoundaryTerm}), we obtain:
\begin{equation}
\begin{aligned}
    \label{zs}
    S_{\text{grav}} &= -{k\over 2\pi} \int_{\partial M_3}\! d^2x  \left( {\lambda '\partial_t \lambda \over \lambda^2}-\lambda^2 F'\partial_t\Psi\right) -{k\over \pi}\int_{\partial M_3}\! d^2x \Big(\lambda^2 \Psi^2 F'+\Psi'+{2\Psi\lambda'\over \lambda}\Big)e^+_t \cr
& \quad + {k\over 2\pi} \int_{\partial M_3}\! d^2x  \left( {\bar{\lambda}'\partial_t \bar{\lambda} \over \bar{\lambda}^2}-\bar{\lambda}^2 \bar{F}'\partial_t\bar{\Psi}\right) -{k\over \pi}\int_{\partial M_3}\! d^2x\Big(\bar{\lambda}^2 \bar{\Psi}^2 \bar{F}'+\bar{\Psi}'+{2\bar{\Psi}\bar{\lambda}'\over \bar{\lambda}}\Big)e^-_t \,.
\end{aligned}
\end{equation}

The boundary conditions \eqref{zq}-\eqref{zo} imply four equations for the six Gauss functions,  leaving two free functions, which we can take to be $(F,\bar{F})$.  So, in principle, we should use   \eqref{zq} and \eqref{zo}  to obtain $(\Psi,\bar{\Psi},\lambda,\bar{\lambda})$  in terms of $(F,\bar{F})$ and substitute into \eqref{zs} to obtain the reduced action. However, in practice, it is not possible to carry this out analytically\footnote{
Even though it is not possible to solve the boundary conditions analytically, they do have a beautiful physical interpretation. They correspond to the definition of the stress tensor in a $T\overline{T}$-deformed theory, understood as a theory coupled to topological gravity. See appendix \ref{app:AdSasTopo} for more details.}. To obtain explicit results we either need to consider the asymptotic AdS$_3$ case of $r_c\rightarrow 0$, or use perturbation theory.  We discuss these in turn below. 

One feature to keep in mind is that we only need to solve for the Gauss functions on the cutoff surface.  These functions determine the connections restricted to that surface, which we call $(a,\overline{a})$.   The full connections $(A,\bar{A})$ may then be determined away from the boundary by the construction
\begin{equation}
\label{za}
 A =b^{-1}ab +b^{-1}db\,,\quad \bar{A}= b\bar{a} b^{-1}+bdb^{-1}\,,
\end{equation}
where
\begin{equation}
\label{zb}
b= e^{-{1\over 2}\ln\left( {r\over r_c} \right) L_0}
\end{equation}
and with $a$ and $\bar{a}$ functions of only the boundary coordinates. This is the Chern-Simons equivalent of radial gauge, which we can always choose at least in a neighborhood of the boundary.
Flat boundary connections $(a,\overline{a})$ are thereby promoted to flat bulk connections $(A,\overline{A})$. 

\subsection{Asymptotic boundary} \label{sec:AsymptoticFT}
In this subsection, we consider imposing boundary conditions at the asymptotic boundary of AdS$_3$. The results obtained here are found in \cite{Cotler:2018zff}.

Asymptotically AdS$_3$ boundary conditions correspond to taking $r_c \rightarrow 0$ with boundary vielbein $e^a_\mu \sim  r_c^{-\frac{1}{2}}$. The boundary conditions \eqref{zq} and \eqref{zo} imply $(\Psi,\bar{\Psi}) \sim  r_c^{\frac{1}{2}} $ and $(\lambda,\overline \lambda) \sim r_c^{-\frac{1}{4}}$, while $(F,\bar{F})$ stay finite. The solution for \eqref{zo} reads
 \begin{equation}
    \label{zt}
     \lambda = \sqrt{2 e_x^+ \over F'}\,,\quad \bar{\lambda}=\sqrt{-\frac{2e_x^-}{\bar{F}'}}\,,
 \end{equation}
while the boundary action evaluates to 
\begin{equation}
\begin{aligned}
    \label{zw}
     S_{\text{grav}} & =  -{k\over \pi} \int_{\partial M_3}\! d^2x  \left( {\lambda' D \lambda \over \lambda^2}-\lambda^2 F' D\Psi\right) + {k\over \pi} \int_{\partial M_3}\! d^2x \left( {\bar{\lambda}'\bar{D} \bar{\lambda} \over \bar{\lambda}^2}-\bar{\lambda}^2 \bar{F}'\bar{D}
     \bar{\Psi} \right) \cr
& \quad -{k\over 8\pi} \int_{\partial M_3}\!  d^2x  { e_t^+ e_x^- - e_t^- e_x^+ \over e_x^+ e_x^- } \omega_x^2 
\end{aligned}
\end{equation}
with
\begin{equation}
\label{zx}
 D= {1\over 2} \left(  \partial_t - {e^+_t\over e^+_x} \partial_x \right)  \,, \quad \text{and}\quad 
\bar{D} = {1\over 2}\left( \partial_t - {e^-_t \over e^-_x} \partial_x \right)\,.
\end{equation}
The term in the second line of \eqref{zw} is a constant determined by the boundary conditions. For a flat planar boundary, we arrive at the  Alekseev-Shatashvili action
\begin{equation}
    S_{\text{grav}} = S_{\text{AS}}[F] + S_{\text{AS}}[\bar{F}]\,,
\end{equation}
with
\begin{equation}
\begin{aligned}
    \label{zxa}
     S_{\text{AS}}[F] & = {k\over 4\pi}\int_{\partial M_3}\! d^2x  \left({1\over F'}\right)'' \partial_{\bar{z}}F \\
& = {k\over 4\pi} \int_{\partial M_3} d^2x \left[  { \dot{F}\over F'} \left(  {F'''\over F'} - {F^{\prime\prime 2}\over 2 F^{\prime 2}}   \right) + {F'''\over F'} - {3\over 2 }{F^{\prime\prime 2}\over  F^{\prime 2}}  \right] \,.
\end{aligned}
\end{equation}
As noted previously by \eqref{gq}, the field redefinition 
\begin{equation}
   F' = e^{f}\,, \quad \bar{F}' = e^{\bar{f}}\,, 
\end{equation}
yields the free boson action 
\begin{equation}
\label{eq:ActionFlatInfinityFR}
S_{\text{grav}} [f, \bar{f}] = -{k\over 4\pi} \int_{\partial M_3} d^2 x\, \left[
f'\partial_z f + \bar{f}'\partial_{\bar{z}}\bar{f}
\right]\,.
\end{equation}
\subsection{Perturbation theory for planar cutoff boundary}

We now consider the case of a boundary at a finite cutoff $r=r_c$, with the simplifying assumption of a flat boundary geometry. We will be able to solve the boundary conditions \eqref{zo} by perturbing around a reference solution. Explicitly, we keep $r_c$ finite and fixed and  take the boundary vielbein corresponding to a flat plane
\begin{equation}
    \label{zxba}
     e^+ = {1\over 2\sqrt{r_c} }(dx+dt)\,,\quad e^- = - {1\over 2\sqrt{r_c} }(dx-dt) \,.
\end{equation}
The corresponding solution to \eqref{aamda} is $\omega_x = 0$.
We will perturb around the solution 
\begin{equation}
\label{zxb}
 F^{(0)} = \bar{F}^{(0)} = x\,,
\end{equation}
which implies $\Psi^{(0)} = \bar{\Psi}^{(0)} =0$ and  $\lambda^{(0)}= \bar{\lambda}^{(0)} = r_c^{-\frac{1}{4}} $.  This solution corresponds to the Poincar\'e AdS$_3$ background metric
\begin{equation}
    ds^2 = {dr^2 \over 4r^2}+\frac{dx^2-dt^2}{r}\,.
\end{equation}

Having identified a  background field configuration,  we expand around it order by order.   We adopt the following notation for the perturbations: 
\begin{equation}
\begin{aligned}
    \label{eq:ExpCS}
    \lambda^2 =&  {1\over \sqrt{r_c}} \left(  1 +  f + f^{(2)} + f^{(3)} + \cdots    \right) \,, \\
\bar{\lambda}^2 =&  {1\over \sqrt{r_c}} \left(  1 + \bar{f} + \bar{f}^{(2)} +  \bar{f}^{(3)} + \cdots    \right) \,, \\
F' = & 1 +  g^{(1)}+ g^{(2)}+ g^{(3)} + \cdots \,, \\
\bar{F}' = & 1 +  \bar{g}^{(1)}+  \bar{g}^{(2)}+  \bar{g}^{(3)} + \cdots \,.
\end{aligned}
\end{equation}
We will regard $f$ and $\bar{f}$ as the fundamental fields of our perturbative action, while $f^{(i)}$, $g^{(i)}$ and their barred counter-parts will be chosen so that the boundary conditions \eqref{zo} are satisfied perturbatively. The boundary conditions fully determine $g^{(i)}$ while the functions $f^{(i)}$  can be chosen freely. This freedom amounts to a field redefinition of $(f,\bar{f})$, which will be used to obtain the simplest action possible.

Solving the boundary conditions perturbatively, which means working order-by-order in the amplitudes of $(f,\bar{f})$,  we find the following expressions for the first few functions $g^{(i)}$:
\begin{equation}
\begin{aligned}
    \label{eq:BCCS}
g^{(1)} =& - f - {r_c\over 2} \bar{f}''\,, \\
g^{(2)} = & f^2 - f^{(2)} + {r_c\over 4} \left(  \bar{f}^{\prime 2} +2f\bar{f}'' +2 \bar{f} \bar{f}'' - 2f^{(2)\prime\prime}  \right) - {1\over r_c^2}\left(  f''\bar{f}''+\bar{f}' f'''  \right) \,,
\end{aligned}
\end{equation}
and similarly for $\bar{g}^{(i)}$. The formulas for higher order terms are easily found since the boundary conditions amount to linear equations for $g^{(i)}$. However, their expressions are not illuminating and get messy at higher orders, so we do not write them explicitly.

As mentioned above, we are free to choose the functions $f^{(i)}$, which amounts to a choice of field redefinition. Just as we found in the metric formulation, a judicious choice simplifies the expression of the action greatly. First of all, demand 
\begin{equation}
    4G T_{xt} = {1\over 4} \left(  \bar{f}^{\prime 2} - f^{\prime 2}   \right)   + \text{total derivatives}\,,
\end{equation}
which implies a simple expression for the part of the action involving time derivatives, agreeing with \eqref{eq:ActionFlatInfinityFR}, and essentially corresponds to choosing Darboux coordinates.  
The field redefinition that achieves this reads 
\begin{equation}
\begin{aligned}
\label{eq:FRCS}
f^{(2)} =& {1\over 2}f^2 - {r_c \over 2 }f'\bar{f}' +\cdots \,, \\
f^{(3)} =& {1\over 6}f^3
- {r_c \over 2} f f'\bar{f}'
+ {r_c^2\over 4} \left[
{1\over 2}\bar{f}^{\prime 2}f'' +  f^{\prime 2}\bar{f}'' +\cdots 
\right]\,,
\end{aligned}
\end{equation}
and similarly for the barred functions.
The second condition that can be satisfied is that all higher derivatives of $f$ and $\overline f$ can be canceled in the action, i.e. only powers of the first derivatives appear. This condition first appears in the fourth order, where the appropriate choice of field redefinition reads
\begin{align}
f^{(4)}
&= \frac1{4!} f^4 - \frac{r_c}4 f^2 f' \bar{f}' - \frac{r_c^2}8 ( f'^3 \bar{f}' - f f'' \bar{f}'^2 - 2f f'^2 \bar{f}'' + f' \bar{f}'^3 - 2 f'^2 \bar{f}'^2 )
\nonumber \\
& \mathrel{\phantom{=}} - \frac{r_c^3}{16} ( 4 f' f'' \bar{f}' \bar{f}'' + \tfrac13 f''' \bar{f}'^3 + f'^3 \bar{f}''' ) \,.
\end{align}
Perturbation theory subject to these conditions can be automated using computer algebra software (we used \texttt{Mathematica}) and performed to higher orders. One useful observation is that the terms needed in the choice of $f^{(n)}$ and $\bar{f}^{(n)}$ to satisfy the two aforementioned conditions already appear (with different coefficients) in the Hamiltonian density at order $n-1$. More specifically, the Hamiltonian density has a simple expression up to a total double derivative contribution, and the terms that appear in this double derivative are exactly the ones that make up our choice of $f^{(n)}$ and $\bar{f}^{(n)}$.  

We carried out this perturbation theory to the eighth order (i.e. computing all terms of the schematic form $f^n \bar{f}^m$ with $n+m\leq 8$). The result coincides with the expansion to this order of the Nambu-Goto action \eqref{eq:90iojkkuhy78d}. We naturally conjecture that this result extends to all orders, but we do not have proof.  

This analysis also yields expressions for the boundary stress tensor $T_{ij}$ to eighth order. These agree with the expressions found in the metric formulation (up to quartic order, which is as far as we pushed the computation in the metric formulation). Since our computations below only use the stress tensor up to cubic order, written in \eqref{hi}, we refrain from writing the higher order expressions, which rapidly become complicated.

\section{Correlation functions}
\label{corsec}
In this section, we discuss the computation of correlation functions of the fundamental fields $(f,\bar{f})$ and the stress tensor $T_{ij}$. We will work up to a two-loop order where, as seen from  \eqref{hh}, $G$ acts as a loop counting parameter.

Some subtleties have to do with the realization of symmetries in this theory. For example, the action is not manifestly Lorentz invariant, even though the underlying theory is Lorentz invariant since it was obtained by expanding around a Lorentz invariant background (the flat plane). We expect the stress tensor should behave in correlators like a Lorentz tensor. As was discussed above, Lorentz symmetry is realized nonlinearly on the $(f,\bar{f})$ fields. A general phenomenon that can occur when doing perturbation theory in a QFT with a nonlinearly realized symmetry is that one encounters divergent terms that are not invariant under the symmetry. One then needs to perform a field redefinition to restore the symmetry (or equivalently, to modify the symmetry transformation), e.g., \cite{Appelquist:1980ae}. Another approach is to modify the theory off-shell to preserve the symmetry, e.g., \cite{Honerkamp:1971sh}. Our approach is to modify perturbation theory in a way that maintains Lorentz invariance while only changing contact terms in correlators. In particular, correlation functions of stress tensors at non-coincident points will respect Lorentz invariance.

\subsection{Action}

We found  that the action to quartic order is
\begin{equation}
    \label{ia}
 I = {1\over 16\pi G} \int_{\partial M_3}\! d^2x \Big( f'\partial_{\bar{z}}f+\bar{f}' \partial_z \bar{f}  +{1\over 4} r_c f'^2 \bar{f}'^2 + \cdots  \Big)\,.  
\end{equation}
Recall that  $z=x+it$ and $\bar{z} = x-it$ so that
\begin{equation}
    \label{ib}
     \partial_z = {1\over 2}(\partial_x -i\partial_t)\,,\quad \partial_{\bar{z}}={1\over 2}(\partial_x +i\partial_t)\,.
\end{equation}
Here $G$ is the loop counting parameter.  In particular, since the stress tensor also has a $1/G$ prefactor, it follows that an $L$ loop contribution to a stress tensor correlator has dependence $G^{L-1}$.

\subsection{Propagator}

Let's first discuss the propagators in momentum space using the Fourier transform convention
\begin{equation}
\label{ic}
 \psi(x,t) = \int\! {d^2p \over (2\pi)^2 }\psi(p) e^{ip_t t + i p_x x}
\end{equation}
or in complex coordinates
\begin{equation}
\label{id}
 \psi(z,\bar{z}) = \int \!{d^2p \over (2\pi)^2 }\psi(p) e^{ip_z z + i p_{\bar{z}} \bar{z}} 
\end{equation}
with
\begin{equation}
\label{ie}
p_x = p_z + p_{\bar{z}}\,,\quad p_t = i(p_z - p_{\bar{z}})\,.
\end{equation}
Note also that
\begin{equation}
\label{if}
 p^2 = p_t^2+ p_x^2 = 4p_z p_{\bar{z}}\,.
\end{equation}
The free two-point functions are 
\begin{equation}
\label{ig}
 \langle f'(p) f'(p')\rangle_0 = 32\pi G{p_x p_z \over p^2}(2\pi)^2\delta^2(p+p')\,,\quad \langle \bar{f}'(p) \bar{f}'(p')\rangle_0 = 32\pi G{p_x p_{\bar{z}} \over p^2}(2\pi)^2\delta^2(p+p')\,. 
\end{equation}
We wrote the results for the fields with an $x$-derivative since $(f,\bar{f})$  always appears in the action and stress tensor with at least one $x$-derivative.

 We will be using dimensional regularization to compute loop diagrams. Our convention for going from two to $d$ dimensions is that we introduce $d-2$ new spatial dimensions. We continue to refer to momenta in the original two dimensions by $(p_x, p_t)$ or $(p_z,p_{\bar{z}})$, but $p^2$ is taken to run over all dimensions: $p^2 = p_t^2+ p_x^2 + \sum_{i=2}^d p_i^2$.   In particular, the relation \eqref{if} only holds in $d=2$.

Coming back to the propagators, after stripping off delta functions and using \eqref{ie},  we have
\begin{equation}
\label{ih}
 \langle f'(p) f'(-p)\rangle_0 = 32\pi G\left( {p_z^2 \over p^2}+ {p_z p_{\bar{z}} \over p^2} \right)\,,\quad \langle \bar{f}'(p) \bar{f}'(-p)\rangle_0 = 32\pi G\left( { p_{\bar{z}}^2 \over p^2}+ {p_z p_{\bar{z}} \over p^2}  \right)\,. 
\end{equation}
We now argue that we can drop the ${ p_z p_{\bar{z}} \over p^2}$ terms. First, note that in $d=2$ this term is constant in momentum space and corresponds to a delta function contribution to the propagator in position space. Including such delta functions in propagators is equivalent to a redefinition of couplings and operators since they contract lines down to points, thereby inducing new vertices. The situation in dimensional regularization with $d=2+\varepsilon$  is a bit more subtle. While the violation of $p^2 = 4p_z p_{\bar{z}}$ is morally proportional to $\varepsilon$, this can, of course, be compensated by factors of $\frac{1}{\varepsilon}$ arising from divergent loop integrals. Nonetheless, as shown by explicit computation (see appendix \ref{Integrals:compare}), the effect of including or excluding the ${ p_z p_{\bar{z}} \over p^2}$ terms in the propagator is the same as changing the coupling in front of some local operator. 

In general, this local operator will be non-Lorentz invariant. We will allow ourselves to add local operators to maintain Lorentz invariance, and what we see from the present discussion is that the simplest way to do this is to drop the ${ p_z p_{\bar{z}} \over p^2}$ terms from the propagators. This should be thought of as part of our renormalization scheme.

We take the propagators to be
\begin{align}\label{ii}
\includegraphics[width=.8\linewidth]{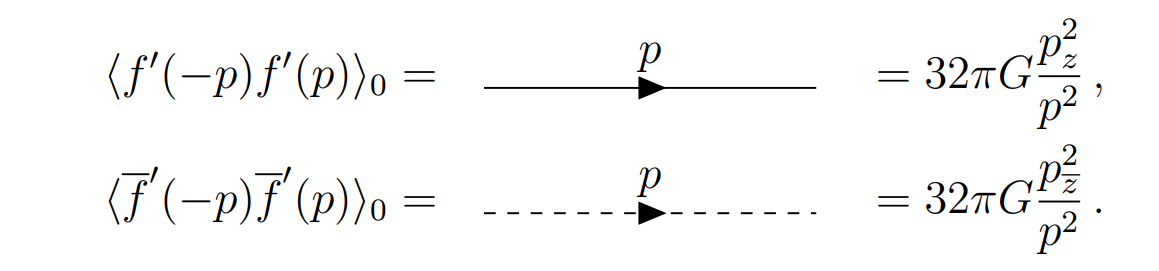}
\end{align}
Arrows indicate momentum flow.   With this propagator rule, $f'$ is effectively the same as $\partial_z f$, and $\bar{f}'$ is effectively the same as $\partial_{\bar{z}} \bar{f}$. We then see from \eqref{hi}  that the stress tensor components have indices that match the $\partial_z$ and $\partial_{\bar{z}}$ derivatives that appear. This implies that stress tensor correlators will be Lorentz covariant.

It will also be useful to Fourier transform back to position space. To perform the $d$-dimensional Fourier transform of the propagators \eqref{ii} is a straightforward application of the integral \eqref{D:a1}, the result of which produces the position space propagators
\begin{equation}
\begin{aligned}
    \label{ij}
  \langle f'(x) f'(0) \rangle_0 &= -8G\pi^{1-{d\over 2}}  \Gamma\left({d\over 2}+1\right)  {\bar{z}^2 \over (x\cdot x)^{{d\over 2}+1} }\,,  \cr
 \langle \bar{f}'(x) \bar{f}'(0) \rangle_0 &= -8G\pi^{1-{d\over 2}}   \Gamma\left({d\over 2}+1\right) {z^2 \over (x\cdot x)^{{d\over 2}+1} }\,,
\end{aligned}
\end{equation}
where  $x\cdot x = z\bar{z} + \sum_{i=2}^d (x^i)^2$. In $d=2$, \eqref{ij} becomes
\begin{equation}
\begin{aligned}
    \label{ija}
     \langle f'(x) f'(0) \rangle_0 &= -{8G\over z^2}\,, \\
 \langle \bar{f}'(x) \bar{f}'(0) \rangle_0 &= -{8G\over \bar{z}^2} \,.
\end{aligned}
\end{equation}
We take $\langle f'f'\rangle_0$ and $\langle \bar{f}' \bar{f}'\rangle_0$ as the propagators. When we refer to an ``amputated" diagram, we mean that we have divided by these propagators.

\subsection{Interaction vertex}

To the order we work, there is a single quartic interaction vertex whose Feynman rule is
\begin{align}\label{ik}
\includegraphics[width=.25\linewidth]{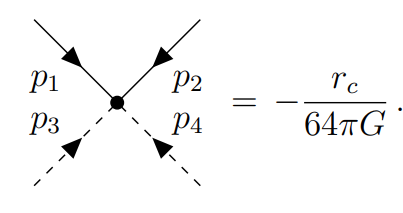}
\end{align}
\subsection{Stress tensor in terms of Feynman diagrams}
From \eqref{hi} and the Feynman rules we have derived, we can express the components of the deformed stress tensor $T_{\mu \nu}$ in terms of the following Feynman diagrams:
\begin{align}
\includegraphics[width=.75\linewidth]{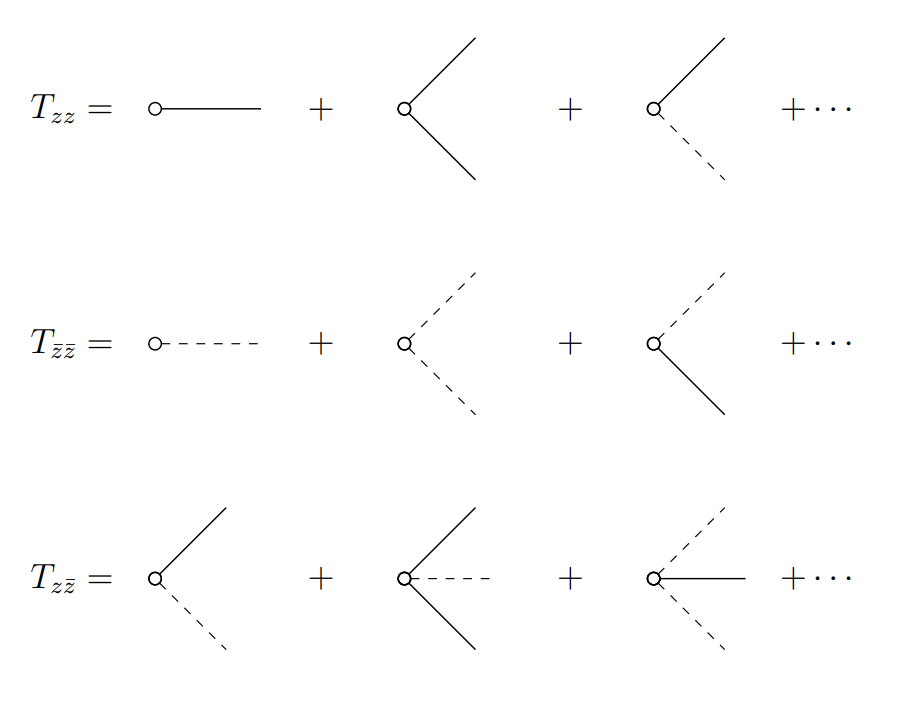}
\end{align}

\subsection{Structure of stress tensor two-point function}

The general stress tensor two-point function can be reconstructed from $\langle T_{z\bar{z}} T_{z\bar{z}}\rangle$, as in \cite{Kraus:2018xrn}. To see this, note that Lorentz invariance and parity implies
\begin{equation}
\begin{aligned}
    \label{il}
     \langle T_{zz}(x)T_{zz}(0)\rangle & = {1\over z^4} f_1(y)\,, \\
\langle T_{zz}(x)T_{z\bar{z}}(0) \rangle& = {1\over z^3 \bar{z}} f_2(y)\,, \\
\langle T_{zz}(x)T_{\bar{z}\bar{z}}(0) \rangle& = {1\over z^2 \bar{z}^2} f_3(y)\,, \\
\langle T_{z\bar{z}}(x)T_{z\bar{z}}(0)\rangle & = {1\over z^2 \bar{z}^2} f_4(y)\,,
\end{aligned}
\end{equation}
where the dimensionless variables $y$ is 
\begin{equation}
\label{im}
 y = {z\bar{z} \over r_c}\,.
\end{equation}
Stress tensor conservation  implies
\begin{equation}
\begin{aligned}
    \label{in}
    f_1'+y^3 \left(f_2 \over y^3\right)'&=0\,,\\ \left(f_2 \over y\right)'+ y \left(f_3 \over y^2\right)'&=0\,, \\
 \left(f_2 \over y\right)'+ y \left(f_4 \over y^2\right)'&=0\,.
\end{aligned}
\end{equation}
As $r_c\rightarrow 0$, we should recover the usual CFT correlators, which implies that we are looking for solutions with $ f_1 \rightarrow {c\over 2}$ as $y \rightarrow \infty$, and with the other functions vanishing in this limit. The central charge $c$ will be computed in terms of $G$ momentarily.
Note that  $f_3=f_4$, which implies that $\langle  T_{zz}(x)T_{\bar{z} \bar{z}}(y) - T_{z\bar{z}}(x)T_{z\bar{z}}(y)\rangle=0$.   This is compatible with the trace relation  $T_{z\bar{z}} = \pi \lambda_{T\overline{T}} \det T$ given that $\langle T_{z\bar{z}}\rangle =0$. 

We find the central charge $c$ by computing correlators at $r_c=0$, where the stress tensor is
\begin{equation}
    \label{ip}
    T_{zz}\vert_{r_c = 0}  = {1\over 8G} \left(f''-{1\over 2}f'^2\right)
\,,\quad
T_{\bar{z} \bar{z}}\vert_{r_c = 0}  = {1\over 8G} \left(\bar{f}''-{1\over 2}\bar{f}'^2\right)
\,, \quad
T_{z\bar{z}}\vert_{r_c = 0}=0\,.
\end{equation}
Using \eqref{ija}, we have
\begin{equation}
\begin{aligned}
    \label{iq}
    \langle T_{zz}(x)T_{zz}(0)\rangle\vert_{r_c = 0} ={c \over 2z^4}
\,,\quad
\langle T_{\bar{z} \bar{z}}(x)T_{\bar{z} \bar{z} }(0)\rangle\vert_{r_c = 0} = {c \over 2 \bar{z}^4}
\,, \\ \langle T_{z\bar{z}}(x)T_{z\bar{z} }(0)\rangle\vert_{r_c = 0}= \langle T_{zz}(x)T_{z \bar{z} }(0)\rangle\vert_{r_c = 0}  = 0
\,,
\end{aligned}
\end{equation}
with
\begin{equation}
\label{ir}
 c =  {3\over 2G}+1 = c_0 +1\,.
\end{equation}
This one-loop correction to the Brown-Henneaux formula is the same as in \cite{Cotler:2018zff}.
We display the contributing diagrams as
\begin{align}
\label{is}
\includegraphics[width=.5\linewidth]{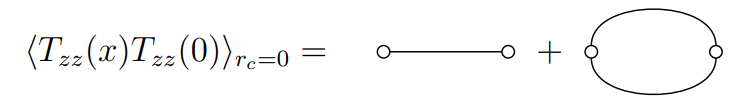}
\end{align}
where the unfilled circles denote stress tensor insertions. In particular, the details of the calculation in \eqref{is} are straightforward
\begin{align}
\includegraphics[width=.75\linewidth]{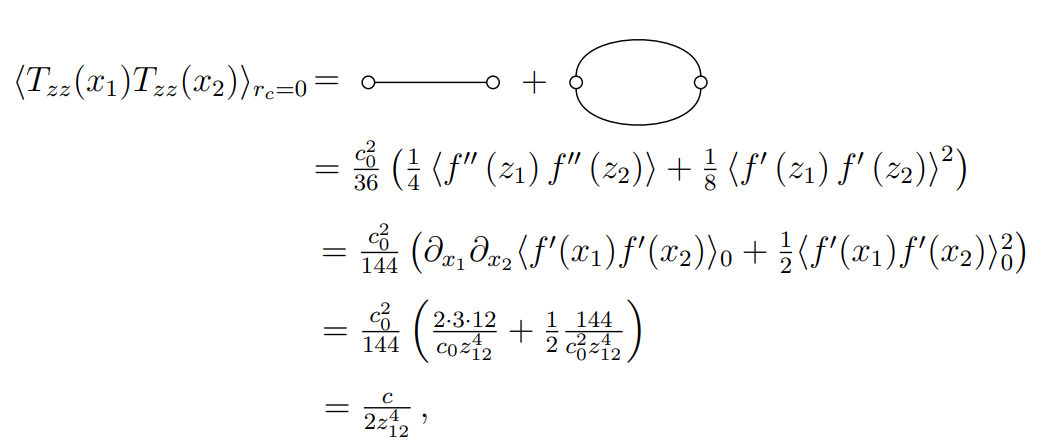}
\end{align}
where the tree-level propagators in terms of the central charge ($G = \frac{3}{2c_0}$) are 
\begin{equation}
    \begin{aligned}
    \langle f'(z_1) f'(z_2) \rangle = - \frac{12}{c_0} \frac{1}{z_{12}^2}, \quad \langle \bar{f}'(\bar{z}_1) \bar{f}'(\bar{z}_2) \rangle = - \frac{12}{c_0} \frac{1}{\bar{z}_{12}^2}.
    \end{aligned}
\end{equation}
For the deformation included at $O(r_c^2)$, then:
\begin{equation}
    \begin{aligned}
\left\langle T_{z z}\left(z_1\right) T_{z z}\left(z_2\right)\right\rangle & =\frac{c_0^2}{36}\bigg(\frac{1}{4}\left\langle f^{\prime \prime}\left(z_1\right) f^{\prime \prime}\left(z_2\right)\right\rangle+\frac{1}{8}\left\langle f^{\prime}\left(z_1\right) f^{\prime}\left(z_2\right)\right\rangle^2\\&+\frac{1}{16} r_c^2\left\langle f^{\prime \prime \prime}\left(z_1\right) f^{\prime \prime \prime}\left(z_2\right)\right\rangle\left\langle\bar{f}^{\prime}\left(\bar{z}_1\right) \bar{f}^{\prime}\left(\bar{z}_2\right)\right\rangle\bigg) \\
& =\frac{c_0+1}{2 z_{12}^4}+\frac{30 r_c^2}{z_{12}^6 \bar{z}_{12}^2}
\end{aligned}
\end{equation}
and using the fact that $\lambda = \frac{6 r_c}{\pi}$, we arrive at 
\begin{equation}
    \left\langle T_{zz} (z_1) T_{zz} (z_2) \right\rangle = \frac{c_0+1}{2z_{12}^4} + \frac{5 \pi^2 \lambda^2}{6 z_{12}^6 \bar{z}_{12}^2}
\end{equation}
which matches \cite{Kraus:2018xrn}.

\subsection{Correlators of elementary fields}

To determine any needed counterterms in the action, we now consider the one-loop four-point and two-loop two-point correlators of $(f,\bar{f})$.

\subsubsection{ $\langle f'(p_1)f'(p_2)f'(p_3)f'(p_4)\rangle $ }

The basic diagram is
\begin{align}
\label{it}
\includegraphics[width=.75\linewidth]{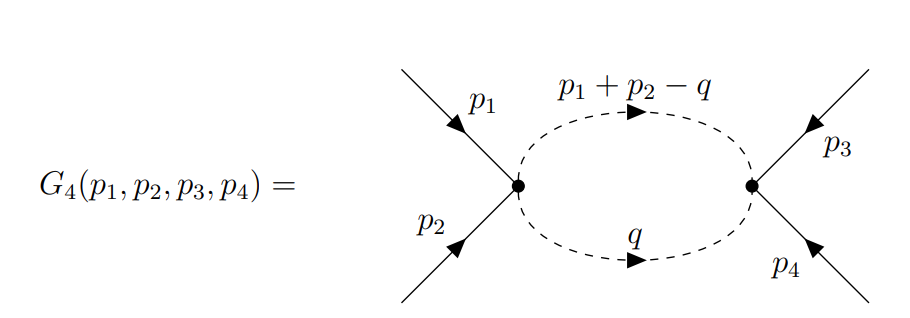}
\end{align}
The full correlator is then
\begin{equation}
    \label{iu}
   \langle f'(p_1)f'(p_2)f'(p_3)f'(p_4)\rangle = G_4(p_1,p_2,p_3,p_4) +G_4(p_1,p_3,p_2,p_4) + G_4(p_1,p_4,p_3,p_2)\,. 
\end{equation}
The amputated  diagram is 
\begin{equation}
\begin{aligned}
    \label{iv}
     G^{{\text{amp}}}_4(p_1,p_2,p_3,p_4) & = 2r_c^2 \int\! {d^dq \over (2\pi)^d} {   (p_{1,\bar{z}}+p_{2,\bar{z}} -q_{\bar{z}} )^2 q_{\bar{z}}^2 \over   (p_{1}+p_{2} -q )^2 q^2 }  \\
& = {r_c^2 \over 12\pi}  {(p_{1,\bar{z}}+p_{2,\bar{z}})^4 \over (p_{1}+p_{2})^2 }\,.
\end{aligned}
\end{equation}
This diagram is in particular finite, hence requires no $(f')^4$ counterterm.

\subsubsection{$\langle f'(p_1)f'(p_2)\bar{f}'(p_3)\bar{f}'(p_4)\rangle $}

The correlator has an (amputated) tree-level contribution
\begin{equation}
     \langle f'(p_1)f'(p_2)\bar{f}'(p_3)\bar{f}'(p_4)\rangle_{\text{tree}}  =- {r_c\over 16\pi G} \,.
\end{equation}
The one-loop diagram is
\begin{align}
\label{iw}
\includegraphics[width=.75\linewidth]{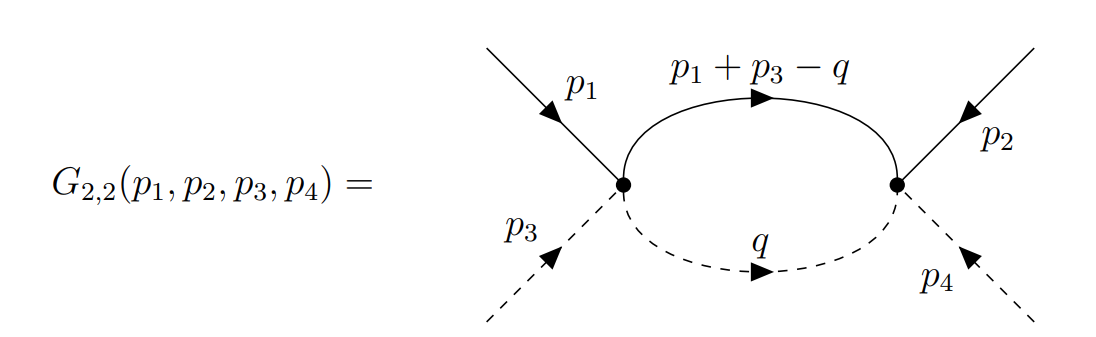}
\end{align}
which we need to evaluate to compute the one-loop contribution to the correlator
\begin{equation}
\label{ix}
f'(p_1)f'(p_2)\bar{f}'(p_3)\bar{f}'(p_4)\rangle_{1-{\text{loop}}} = G_{2,2}(p_1,p_2,p_3,p_4) + G_{2,2}(p_1,p_2,p_4,p_3)\,. 
\end{equation}
Employing the shorthand $p_{ij} = p_i + p_j$, the result computed using dimensional regularization and setting $d=2+\epsilon$ reads
\begin{equation}
\begin{aligned}
    \label{iy}
    G^{\text{amp}}_{2,2}(p_1,p_2,p_3,p_4)&=  4 r_c^2 \int\! {d^d q\over (2\pi)^d} { (p_{1,z}+p_{3,z}-q_z)^2 q_{\bar{z}}^2   \over (p_{1}+p_{3}-q)^2 q^2   }\\
& = 4r_c^2 \left[ \frac{p^2_{13}}{32\pi \epsilon} + \frac{6\gamma - 11 - 6\ln(4\pi)}{384\pi} p^2_{13} + \frac{p^2_{13}}{64\pi}\ln p^2_{13} \right]\,.
\end{aligned}
\end{equation}
The amputated correlator works out to be 
\begin{equation}
\begin{aligned}
    \label{iz}
    \langle f'(p_1)f'(p_2)\bar{f}'(p_3)\bar{f}'(p_4)\rangle_{1-{\text{loop}}}^{\text{amp}}
&= \frac{r_c^2(d + 2)(d + 4)}{4^{d + 5/2}\pi^{\frac{d-3}{2}}}\frac{(p_{13}^2)^{d/2} + (p_{14}^2)^{d/2}}{\sin\left(\frac{\pi d}{2}\right)\Gamma(d/2 + 3/2)}\,.
\end{aligned}
\end{equation}
This amputated correlator \eqref{iz} has a pole at $\varepsilon=0$,  
\begin{equation}
\label{ja}
 \langle f'(p_1)f'(p_2)\bar{f}'(p_3)\bar{f}'(p_4)\rangle_{1-{\text{loop}}}^{\text{amp}} \sim  {1\over 8\pi \varepsilon } r_c^2 \big[  p_{13}^2 +p_{14}^2\big]\,. 
\end{equation}
This divergence \eqref{ja} is canceled by the counterterm:
\begin{equation}
\label{jb}
I_{\text{ct}} = {r_c^2\over 4\pi \varepsilon}  \int_{\partial M_3}\! d^2x \partial_z (f'\bar{f}')\partial_{\bar{z}}(f'\bar{f}')\,.
\end{equation}
The original action has no term of the form \eqref{jb}. One interpretation is that this implies the existence of a new parameter in our theory corresponding to including an undetermined finite term along with \eqref{jb}. On the other hand, as discussed in this chapter's introduction, the $3d$ gravity origin of this theory indicates that no such new parameters should be needed. We thus suspect that the appearance of the undetermined parameter may reflect that our renormalization scheme has not incorporated all symmetries of the $3d$ gravity theory.  

\subsection{\texorpdfstring{$\langle f'(x) f'(0)\rangle$}{<f'(x) f'(0)>} at two-loops}
We will first compute the correlator in momentum space. The relevant Feynman diagram to compute $\langle  f'(p)  f'(-p)\rangle$ is a sunset-type diagram
\begin{align}
\label{jc}
\includegraphics[width=.75\linewidth]{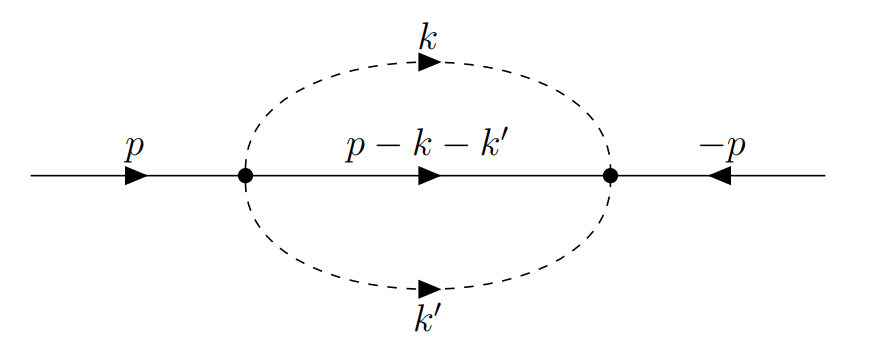}
\end{align}
The two-loop contribution to the amputated correlator is then 
\begin{equation}
    \langle  f'(p)  f'(-p) \rangle_{{\text{two-loop}}}^{\text{amp}}  
=
8 
\left(   {r_c\over 64\pi G}  \right)^2
\left( 32\pi G \right)^3 \int {d^2 k\over (2\pi)^2} \int {d^2 k'\over (2\pi)^2} {k_{\bar{z}}^2 k_{\bar{z}}^{\prime 2}  (p-k-k')_z^2   \over k^2 k^{\prime 2}(p-k-k')^2 }\,,
\end{equation}
where the overall normalization involves two vertex factors as in \eqref{ik}, the normalization of the three internal propagators as in formulas \eqref{ii}, and a symmetry factor of $8$. The integrals over the internal momenta $k$ and $k'$ are computed in dimensional regularization in appendix \ref{app:2Loopff}. The result reads
\begin{equation}
\label{eq:2LoopInt}
\int {d^2 k\over (2\pi)^2} \int {d^2 k'\over (2\pi)^2} {k_{\bar{z}}^2 k_{\bar{z}}^{\prime 2}  (p-k-k')_z^2   \over k^2 k^{\prime 2}(p-k-k')^2 } 
=
{1 \over 2^7 3 \pi^2 }    p_z p_{\bar{z}}^3 \log p^2+ {\text{polynomial}}\,.
\end{equation}
Attaching the external legs and Fourier transforming back to position space using formulas \eqref{D:12}, we conclude
\begin{equation}
    \langle  f'(x)  f'(0) \rangle_{{\text{two-loop}}} 
= -  {64  r_c^2 G^3\over z^4 \bar{z}^2 }\,.
\end{equation}
This diagram is, in particular, finite (up to contact terms at $x=0$), so no wavefunction renormalization is required.\footnote{
As seen in \eqref{app:2Loopff}, the integral \eqref{eq:2LoopInt} does have a divergence in dimensional regularization. However, the divergence is a polynomial in the momentum, which only leads to delta function contact terms in position space.
}

\subsection{\texorpdfstring{$\langle T_{z\bar{z}} f'\bar{f}'\rangle$}{<T{z zbar} f' fbar'>}}

To identify the need for a counterterm for  $T_{z\bar{z}}$,  we consider the correlator of the stress tensor with two elementary fields
\begin{align}
\label{ida}
\includegraphics[width=.65\linewidth]{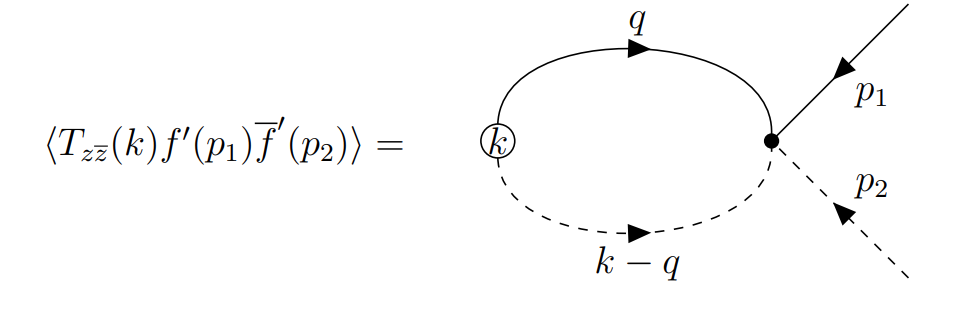}
\end{align}
The amputated diagram is 
\begin{equation}
\begin{aligned}
    \label{idb}
    \langle T_{z\bar{z}}(k) f'(p_1)\bar{f}'(p_2) \rangle_{\operatorname{amp}} &= 4 \pi  r_c^2  \int\! {d^dq \over (2\pi)^d}  { q_z^3 (k_{\bar{z}}-q_{\bar{z}})^3 \over q^2(k-q)^2 }  \\
& = {1\over 32 \varepsilon} r_c^2 (k^2)^2 + {\operatorname{finite}}\,. 
\end{aligned}
\end{equation}
To cancel this divergence we need to redefine this stress tensor component as
\begin{equation}
    \label{idc}
     T_{z\bar{z}} \rightarrow T_{z\bar{z}}  -{1\over 2\varepsilon}r_c^2f''' \bar{f}'''\,.
\end{equation}
Here, we have adopted a minimal subtraction scheme.  Of course, we are free to also add a finite contribution, which will appear below as an undetermined constant in the stress tensor correlator.
\subsection{\texorpdfstring{$\langle T_{z\bar{z}}T_{z\bar{z}}\rangle$}{< T{z zbar} T{z zbar}>}}

To compute $\langle T_{z\bar{z}}T_{z\bar{z}}\rangle$ to two-loop order, we recall
\begin{equation}
\label{je}
4G T_{z\bar{z}}  =  -{1\over 4} r_c f''\bar{f}''+ {1\over 8}r_c (f''\bar{f}'^2+f'^2 \bar{f}'')  -{1\over 8} r_c^2 ( f''' \bar{f}' \bar{f}''+ f'f''\bar{f}''') +  {\operatorname{quartic}}\,.
\end{equation}
The contributing diagrams to the two-loop order are
\begin{align}
\label{jf}
\includegraphics[width=.85\linewidth]{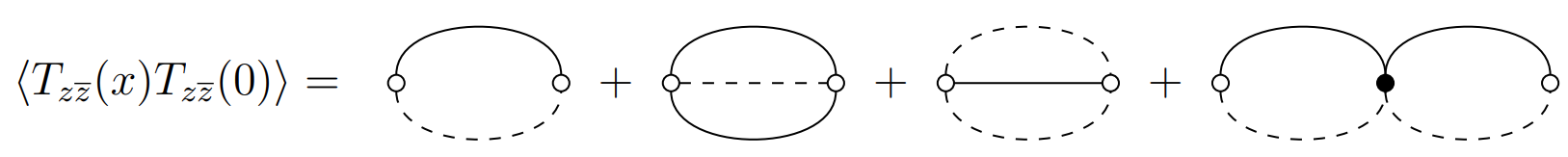}
\end{align}
The first three diagrams are trivially computed by Wick contraction in position space. The one-loop diagram is 
\begin{align}
\label{jg}
\includegraphics[width=.3\linewidth]{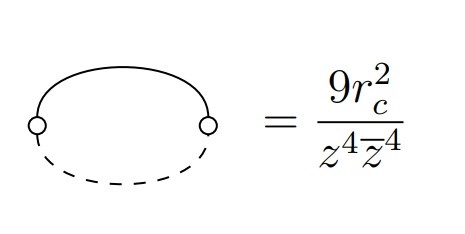}
\end{align}
and the two simple two-loop diagrams sum to
\begin{align}
\label{jh}
\includegraphics[width=.75\linewidth]{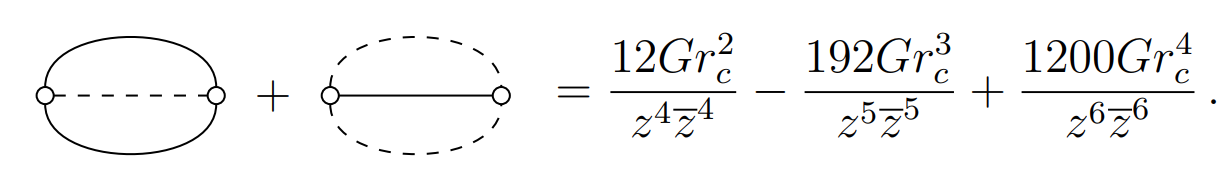}
\end{align}
We next turn to the two-loop diagram in  \eqref{jf}. Working in momentum space, the contribution to $\langle T_{z\bar{z}}(-k) T_{z\bar{z}}(k)\rangle$ is 
\begin{align}
\label{ji}
\includegraphics[width=1\linewidth]{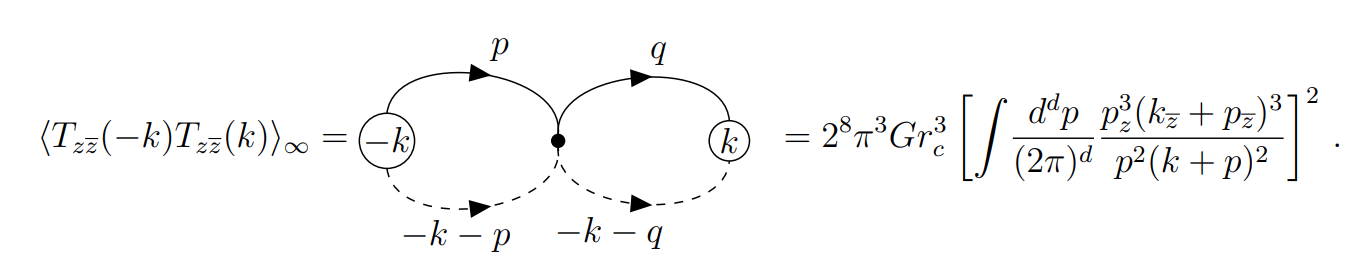}
\end{align}
This diagram has double and single pole divergences in $\varepsilon$. The double pole is polynomial in $k$ and can be ignored as it won't contribute to the two-point function at finite spatial separation. The simple pole is canceled, by design, via the stress tensor counterterm \eqref{idc}; i.e. by the two one-loop diagrams in which one of the stress tensor insertions given by the counterterm in \eqref{idc}. The resulting finite part is 
\begin{equation}
\label{jj}
 \langle T_{z\bar{z}}(-k) T_{z\bar{z}}(k)\rangle_{\infty } =  2^{-7} \pi r_c^3 G \left(  a \ln k^2+ (\ln k^2)^2 \right) (k^2)^4 + {\operatorname{polynomial}}\,.
\end{equation}
The constant $a$ is left unspecified since it can be shifted arbitrarily due to the freedom in including a finite counterterm in \eqref{idc}. Fourier transforming back to position space, we obtain 
\begin{equation}
    \label{jk}
    \langle T_{z\bar{z}}(x) T_{z\bar{z}}(0)\rangle_{\infty } = 2^8 \cdot 3^2  r_c^3 G {\ln (\mu^2 z\overline z) \over (z\overline z)^5}\,,
\end{equation}
where we now traded the arbitrary constant $a$ for a renormalization scale $\mu$.\footnote{Logarithms also appear in the $T \bar{T}$ deformed correlation functions of \cite{Kraus:2018xrn,Cardy:2019qao}.}

\subsection{Summary of two-point deformed correlators at two-loop order}

Combining results to the two-loop order, we have found 
\begin{equation}
\label{jl}
\resizebox{.9\hsize}{!}{$\langle T_{z\bar{z}}(x) T_{z\bar{z}}(0)\rangle  =  \frac{3}{(z\overline z)^2}\left[  (3+4 G)\left( r_c\over z\overline z\right)^2   -64 G \left(1 - 12 \ln(\mu^2 z\overline z)\right)\left(r_c \over z\overline z\right)^3 + 400 G \left(r_c\over z\overline z\right)^4  \right]$}\,. 
\end{equation}

Using the Ward identities \eqref{il} and \eqref{in}, we read off the other two-point functions 
\begin{equation}
\begin{aligned}
    \label{jla}
    \langle T_{zz}(x)T_{zz}(0)\rangle &= \resizebox{.78\hsize}{!}{$\frac{1}{z^4}\left[
\frac{c}{2} + 10(3+4G)\left(\frac{r_c}{z\overline z}\right)^2 + 96G\left(8 + 60\ln(\mu^2 z\overline z)\right)\left(\frac{r_c}{z\overline z}\right)^3 + 2520G\left(\frac{r_c}{z\overline z}\right)^4
\right]\,,$} \\
\langle T_{zz}(x)T_{z\overline z}(0)\rangle &= \frac{4}{z^3\overline z}\left[
 - (3+4G)\left(\frac{r_c}{z\overline z}\right)^2 + 24G\left(1 - 30\ln(\mu^2 z\overline z)\right)\left(\frac{r_c}{z\overline z}\right)^3 - 360G\left(\frac{r_c}{z\overline z}\right)^4
\right]\,, \\
\langle T_{zz}(x)T_{\overline z\overline z}(0)\rangle &= \frac{3}{(z\overline z)^2}\left[
(3+4G)\left(\frac{r_c}{z\overline z}\right)^2 - 64G\left(1 - 12\ln(\mu^2 z\overline z)\right)\left(\frac{r_c}{z\overline z}\right)^3 + 400G\left(\frac{r_c}{z\overline z}\right)^4
\right]\,,
\end{aligned}
\end{equation}

where $c= c_0 +1 =  {3\over 2G}+1$.

\subsection{Higher point correlators}
We have acquired a systematic method to compute any $n$-point stress tensor correlator at any given order in $\lambda$ and $c$. Despite contenting ourselves to two-point correlators in detail so far, studying higher-point correlators is straightforward. 

For example, at the one-loop level at $r_c = 0$, the three-point correlator is
\begin{align}
\includegraphics[width=.75\linewidth]{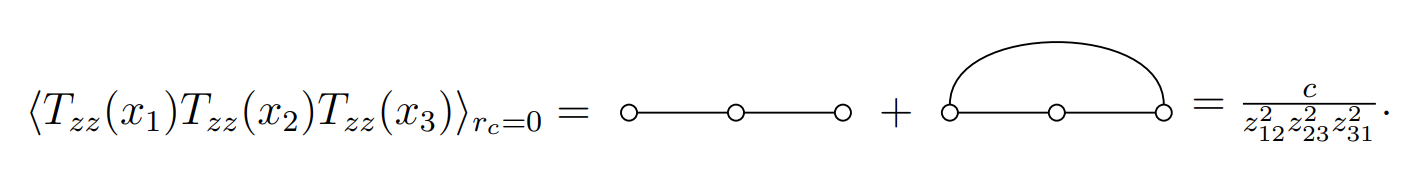}
\end{align}
Another example is the tree-level deformed three-point correlator at $\mathcal{O}(r_c)$:
\begin{align}
\label{eq:3ptzbzzzbzb}
\includegraphics[width=.65\linewidth]{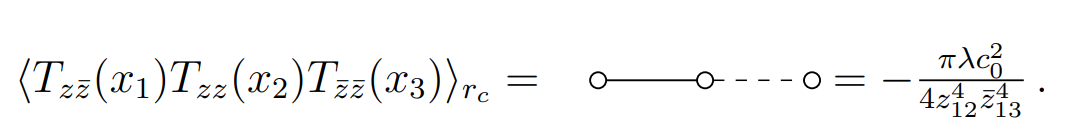}
\end{align}
We can express \eqref{eq:3ptzbzzzbzb} in terms of a product of undeformed tree-level two-point correlators with the trace flow equation 
\begin{equation}
\begin{aligned}
    \langle T_{z\bar{z}} (x_1) T_{zz} (x_2) T_{\bar{z} \bar{z}} (x_3) \rangle_{r_c}&= \langle \left(-\pi \lambda T_{zz} (x_1) T_{\bar{z} \bar{z}} (x_1) \right) T_{zz} (x_2) T_{\bar{z} \bar{z}}(x_3) \rangle_0 + \mathcal{O}(\lambda^2)  
    \\&=- \pi \lambda \langle T_{zz} (x_1) T_{zz} (x_2) \rangle_0 \langle T_{\bar{z} \bar{z}} (x_1) T_{\bar{z} \bar{z}} (x_3) \rangle_0 + \mathcal{O}(\lambda^2)
   \\&=- \frac{\pi \lambda c_0^2}{4} \frac{1}{z^4_{12} \bar{z}^{4}_{13}} + \mathcal{O}(\lambda^2)\,,
    \end{aligned}
\end{equation}
where
\begin{align}
\includegraphics[width=.9\linewidth]{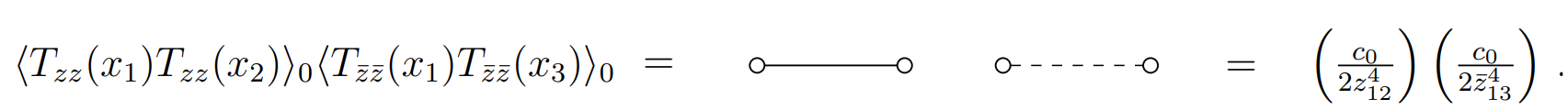}
\end{align}

Moreover, from perturbation theory at the tree-level and $\mathcal{O}(r_c)$, we may compute 
\begin{equation}
 \begin{aligned}
 \langle T_{zz} (x_1) T_{\bar{z} \bar{z}} (x_2) T_{\bar{z} \bar{z}}(x_3) \rangle_{r_c}.
 \end{aligned}
\end{equation}
Using the fact that $\sqrt{g} d^2x = \frac{1}{2} d^2z$, $\partial_{\bar{z}_1} \frac{1}{z_1 - z_2} = 2 \pi \delta^2(z_1 - z_2)$ and integration by parts, we arrive at the following at $\mathcal{O}(\lambda)$
\begin{equation}
    \begin{aligned}
   & \langle T_{zz} (x_1) T_{\bar{z} \bar{z}} (x_2) T_{\bar{z} \bar{z}}(x_3) \rangle_{r_c} \\&= \left \langle T_{zz} (x_1) T_{\bar{z} \bar{z}} (x_2) T_{\bar{z} \bar{z}}(x_3) \left( - \lambda \int d^2x \sqrt{g} T_{zz} (x) T_{\bar{z} \bar{z}} (x) \right) \right \rangle_0
    \\&= -\lambda \int d^2x \sqrt{g} \langle T_{zz} (x_1) T_{\bar{z} \bar{z}}(x_2) T_{\bar{z} \bar{z}}(x_3) T_{zz} (x) T_{\bar{z} \bar{z}}(x) \rangle_0
    \\&= -\lambda \int d^2x \sqrt{g} \langle T_{zz} (x_1) T_{zz} (x) \rangle_0 \langle T_{\bar{z} \bar{z}}(x_2) T_{\bar{z} \bar{z}}(x_3) T_{\bar{z} \bar{z}}(x) \rangle_0
        \\&= -\lambda \int \left( \frac{1}{2} d^2z \right) \left( \frac{c_0}{2} \frac{1}{ (z-z_1)^4} \right) \left( \frac{c_0}{(\bar{z} - \bar{z}_2)^2 (\bar{z} - \bar{z}_3)^2 (\bar{z_2} - \bar{z}_3)^2} \right) 
        \\&= - \frac{\lambda c_0^2}{4} \int \frac{d^2z}{(z-z_1)^4(\bar{z} - \bar{z}_2)^2 (\bar{z} - \bar{z}_3)^2 (\bar{z_2} - \bar{z}_3)^2}
        \\&= \frac{\lambda c_0^2}{12} \int d^2z \left[\partial_z \frac{1}{(z-z_1)^3} \right] \frac{1}{(\bar{z} - \bar{z}_2)^2 (\bar{z}- \bar{z}_3)^2 (\bar{z}_2 - \bar{z}_3)^2}
            \\&= -\frac{\lambda c_0^2}{12} \int d^2z  \frac{1}{(z-z_1)^3} \partial_z\left[\frac{1}{(\bar{z} - \bar{z}_2)^2 (\bar{z}- \bar{z}_3)^2 (\bar{z}_2 - \bar{z}_3)^2} \right]
            \\&=-\frac{\lambda c_0^2}{12} \int d^2z  \frac{1}{(z-z_1)^3} \left( -2 \pi \partial_{\bar{z}} \delta^2 (z-z_2) \right) \frac{1}{(\bar{z} - \bar{z}_3)^2 (\bar{z}_2 - \bar{z}_3)^2  } + (x_2 \leftrightarrow x_3)
         \\&= - \frac{\pi \lambda c_0^2}{3} \frac{1}{(\bar{z}_2 - \bar{z}_3)^5} \left( \frac{1}{(z_1 - z_2)^3 }  - \frac{1}{(z_1 - z_3)^3 } \right).
    \end{aligned}
\end{equation}
These are just a few sample diagrams related to three-point correlators. Combinatorially, there are various ways of constructing three-point diagrams than there are for two-point correlators, as we saw earlier. For illustrative purposes, a few three-point diagrams are:
\begin{align}
\includegraphics[width=.9\linewidth]{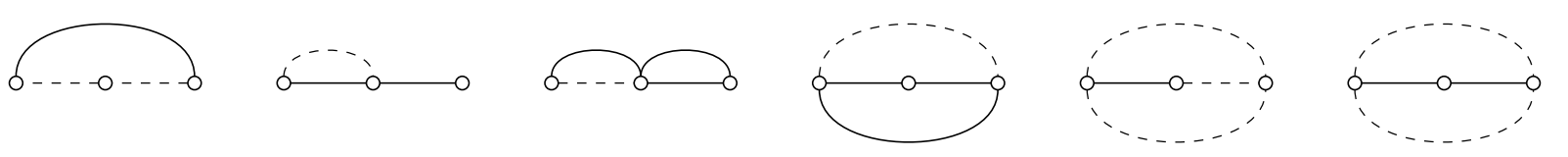}
\end{align}

\section{Gravitational AdS$_3$ Wilson lines}
\label{sec: Gravitational Wilson Lines}

To close this chapter, we perturbatively compute the deformed classical and quantum gravitational Wilson line and its correlators in AdS$_3$. As a consistency check, our classical gravitational Wilson line correlator analysis is consistent with previous results on $T\overline{T}$-deformed scalar correlators \cite{Kraus:2018xrn,He:2019vzf,Cardy:2019qao} for constant stress tensor backgrounds.

\subsection{The classical AdS$_3$ Wilson line}
The gravitational AdS$_3$ Wilson line anchored at the endpoints $z_1$ and $z_2$ is conjectured to be dual to a bi-local primary operator:
\begin{equation}
   \langle W[z_2, z_1] \rangle_0 \longleftrightarrow \langle O(z_2) O(z_1) \rangle_0\,.
\end{equation}
Given two arbitrary AdS$_3$ bulk points $Z_1 = (r_1, z_1)$ and $Z_2 = (r_2, z_2)$, the classical Wilson line is defined as the path-ordered integral
\begin{equation}
   W[(r_2, z_2; r_1, z_1)]_0 = P \exp \left( \int^{(r_2, z_2)}_{(r_1, z_1)} A \right)\,.
\end{equation}
Under a gauge transformation, the Wilson line transforms as
\begin{equation}
   W[(r_2, z_2; r_1, z_1)]_0 \rightarrow g(r_2, z_2)^{-1} W[(r_2, z_2; r_1, z_1)]_0 g(r_1, z_1)\,,
\end{equation}
with $g \in \text{SL}(2, \mathbb{R})$ and $A \rightarrow g^{-1} \left(d + A \right) g$. In particular, the radial dependence of the connection
\begin{equation}
    \begin{aligned}
    \label{eq:GeneralSols}
    A&= b(r) \left( d + a(z) \right) b(r), \quad b(r) = r^{L_0} \,,  \overline{A} &= \overline{b}(r) \left( d + \bar{a} (\bar{z})\right) \overline{b}(r)^{-1}, \quad \overline{b}(r) = r^{\bar{L}_0}\,,
    \end{aligned}
\end{equation}
arises through a gauge transformation:
\begin{equation}
   W [(r_2, z_2; r_1, z_1)]_0 = b(r_2)^{-1} P \exp \left( \int_{z_{1}}^{z_{2}} a \right) b(r_1) \,.
\end{equation}
The matrix elements of $W[z_2, z_1]$ between the lowest and highest weight states are
\begin{equation}
\begin{aligned}
\label{eq:WdefT}
\langle W[z_2, z_1] \rangle_0 &= \langle j, -j \mid  P \exp \left( \int^{z_2}_{z_1} a \right) \mid j, j \rangle_0 \\&= \langle j, -j \mid P \exp \left[\int^{z_2}_{z_1} dz \left( L_1 + \frac{6}{c} T_{zz}(z) L_{-1} \right) \right]  \mid j, j\rangle_0\,,
    \end{aligned}
\end{equation}
where $|j, m \rangle$ is the state of weight $m$ in the spin-$j$ representation of $SL(2, \mathbb{R})$. To see the bi-localness of the classical Wilson line, first consider the vacuum state of $3d$ gravity. In the vacuum state, the path-ordered integral reduces to an ordinary integral 
\begin{equation}
  \langle  W[z_2, z_1]  \rangle_0  \big\vert_{T_{zz} = 0}= \langle j, -j \mid \exp \left( \int^{z_2}_{z_1} dz \, L_1 \right) \mid j, j \rangle_0 = z_{21}^{2j}\,,
\end{equation}
where $z_{ij} = z_i - z_j$ and the bi-local primary field has dimension $h = -j$.

One can recover the case when $T_{zz} \neq 0$ through a local conformal transformation $z \rightarrow f(z)$ which is given by inverting the Schwarzian 
\begin{equation}\label{T_is_schwarzian}
    T_{zz} = \frac{c}{12} \{f(z), z\}\,.
\end{equation}
As a result, the classical Wilson line for a general background is
\begin{equation}
 \langle   W[z_2, z_1] \rangle_0 \big\vert_{T_{zz} \neq 0} = \frac{\left[ f(z_2) - f(z_1) \right]^{2j}}{\left[ f'(z_2) f'(z_1) \right]^{j}} \,, 
\end{equation}
and behaves as a bi-local primary operator at the endpoints. Intuitively, a way to argue for the bi-locality of the Wilson line is because the Chern-Simons equations of motion for the connections are flat. Consequently, this makes the Wilson line path-independent between the two endpoints.

\subsection{The quantum $T\overline{T}$-deformed AdS$_3$ Wilson line}
\label{Sec:Quantum deformed AdS$_3$ Wilson line}
The quantum Wilson line is obtained by beginning with the definition of the classical Wilson line, where the stress tensor $T_{zz}$ is thought of as a commuting number and promoting the stress tensor to an operator of the CFT. The resulting object is conjectured to behave as a bi-local primary operator at its endpoints, $\langle  W[z_2, z_1] \rangle_0 = z_{21}^{-2h(j,c)}$. Because the stress tensor is now an operator, short-distance singularities arise from the stress tensor OPE, so the scaling dimension $h(j,c)$ of the Wilson line experiences quantum corrections of the form\footnote{The specific values of $h_n(j)$ are easily calculable from \cite{Besken:2017fsj,Besken:2018zro,DHoker:2019clx}. For instance, a few values are:
\begin{equation}
   \resizebox{.8\hsize}{!}{$  h_0(j) = -j, \quad h_1(j) = -6j(j+1), \quad h_2(j) = -78j(j+1), \quad h_3(j) = -1230j(j+1), \quad h_4(j) = -21606j(j+1)\,.$}
\end{equation}
 }
\begin{equation}
\label{eq:scalingdimensions}
    h(j,c) = \sum^\infty_{n = 0} \frac{h_n(j)}{c^n} \,.
\end{equation}
Due to these short-distance singularities from the stress tensor OPE, we must regularize the gravitational Wilson line to verify that the quantum Wilson line
\begin{equation}
\begin{aligned}
    \label{eq:expandedW}
\langle W[z_2, z_1] \rangle_0 &= \langle j, -j \mid P \exp \left( \int^{z_2}_{z_1} dz \left( L_1 + \frac{6}{c} T_{zz}(z) L_{-1} \right) \right) \mid j, j \rangle_0 \\
 &= \sum_{n=0}^{\infty} \int_{z_{1}}^{z_{2}} d y_{n} \int_{z_{1}}^{y_{n}} d y_{n-1} \cdots \int_{z_{1}}^{y_{2}} d y_{1}\\&\langle j,-j\mid \left( L_1 + \frac{6}{c} T_{zz}(y_n) L_{-1} \right) \cdots \left( L_1 + \frac{6}{c} T_{zz}(y_1) L_{-1} \right)\mid j, j \rangle_0
\end{aligned}
\end{equation}
captures the correct scaling dimension \eqref{eq:scalingdimensions} as a bi-local primary operator. Further perturbative evidence of the Wilson line's bi-localness was provided in \cite{Besken:2017fsj,Besken:2018zro,DHoker:2019clx}. The authors of \cite{Besken:2017fsj} successfully calculated the undeformed quantum Wilson line $\langle W[z; 0] \rangle_0$ up to $\mathcal{O}(\frac{1}{c})$ and encountered some ambiguities in the coefficients at the two-loop order, $\mathcal{O}(\frac{1}{c^2})$, due to the absence of a systematic renormalization scheme that preserves conformal invariance. The most promising scheme is the dimensional regularization approach used in \cite{Besken:2018zro}, where an overall multiplicative renormalization $N(\varepsilon)$ and a renormalization of the vertex factor $\alpha(\varepsilon)$ were needed in $d = 2- \varepsilon$ dimensions:
\begin{equation}
\label{eq:D'Hoker-Kraus}
   \lim_{\varepsilon \rightarrow 0}  \langle W_{\varepsilon}[z_2, z_1] \rangle_0 = z_{21}^{2j}  \lim_{\varepsilon \rightarrow 0} N(\varepsilon) \langle j,-j \mid  P \exp \left( \frac{6 \alpha(\varepsilon)}{c} \int_{z_1}^{z_2} d y \left( L_1 + \frac{6}{c} T_{zz}(y) L_{-1} \right) \right) \mid j, j \rangle_0\,.
\end{equation}
Here $N(\varepsilon)$ and $\alpha(\varepsilon)$ are chosen order-by-order in $\frac{1}{c}$ to cancel the poles in $\varepsilon$. The authors in \cite{Besken:2018zro} corrected the issue which arose at $\mathcal{O}(\frac{1}{c^2})$ in \cite{Besken:2017fsj}. They also carefully calculated and confirmed the $\mathcal{O}(\frac{1}{c^3})$ corrections to the Wilson line. 

Using the systematic renormalization approach in \cite{Besken:2018zro}, the authors of \cite{DHoker:2019clx} calculated Wilson line correlators with multiple stress tensors insertions $\big\langle \prod^n_{i=1} T_{zz}(w_i) W [z_2, z_1] \big\rangle$ and found results consistent with the expectation that the Wilson line yields the vacuum Virasoro OPE block \eqref{eq:defVacuumOPE1}-\eqref{eq:primaryOPE}. However, whether the quantum Wilson line behaves as a bi-local primary operator non-perturbatively in $\frac{1}{c}$ is still unknown as dimensional regularization may violate conformal invariance. This completes our review of the quantum Wilson line; we now set up the necessary formalism to compute the deformed quantum Wilson line.

We begin with the Wilson line in terms of the boundary stress tensor, which is valid in the undeformed theory because the connections can be brought into Ba\~nados form: 
 \begin{equation}
 \label{defWilson}
     W[z_2, z_1] = P \exp \left( \int^{z_2}_{z_1} dy \left( L_1 + \frac{6}{c} T_{zz}(y) L_{-1} \right) \right).
 \end{equation}
Following \cite{Besken:2017fsj}, we write \eqref{defWilson} in a more convenient form by defining 
\begin{align}
    V[z_1, z_2]&= \exp \left(-L_1 z_{21} \right) W[z_1, z_2]\,,
\end{align}
so that
\begin{equation}
\begin{split}
\label{eq:path-ordered exp}
    \frac{d}{dz_2} V[z_1, z_2] &= \exp \left( -L_1 z_{21} \right) \frac{6}{c} T_{zz}(z_2) L_{-1} \exp \left(L_1 z_{21}\right) V[z_1, z_2] \\&= \frac{6}{c} \left( (1-L_1 z_{21}) L_{-1} (1+L_1 z_{21}) \right) T_{zz}(z_2) V[z_1, z_2]  \\&= \frac{6}{c} \left( L_{-1} + z_{21} [L_{-1}, L_1] - z_{21}^2 L_1 L_{-1} L_1\right) T_{zz}(z_2) V[z_1, z_2] \\&= \frac{6}{c} \left(L_{-1} - 2z_{21} L_0 + z_{21}^2 L_1 \right) T_{zz}(z_2) V[z_1, z_2]\,,
\end{split}
\end{equation}
where we have used the facts $L_{\pm 1}^2 = 0$, $L_1 L_{-1} L_1 = -L_1$, and $[L_{-1}, L_1] = -2L_0$.

Here \eqref{eq:path-ordered exp} is solved by the usual path-ordered exponential and this allows us to write the Wilson line in a more convenient form to systematically implement a $\frac{1}{c}$ expansion:
\begin{equation}
\begin{aligned}
\label{eq:GravWilson}
\hspace{-10pt}&\langle W[z_2, z_1] \rangle\\&= \langle j, -j \mid \exp \left(z_{21} L_1\right) P \exp \left( \frac{6}{c} \int^{z_2}_{z_1} \left( L_{-1} - 2 (y - z_1)L_0 + (y - z_1)^2 L_1 \right) T_{zz}(y) dy \right) \mid j, j \rangle\,.
 \end{aligned}
\end{equation} 
The gravitational Wilson line in this form \eqref{eq:GravWilson} can be understood as a perturbative expansion in $\frac{1}{c}$ of self-energy Feynman diagrams:
\begin{align}
\label{diagrams_oneoverc_expansion}
\includegraphics[width=.75\linewidth]{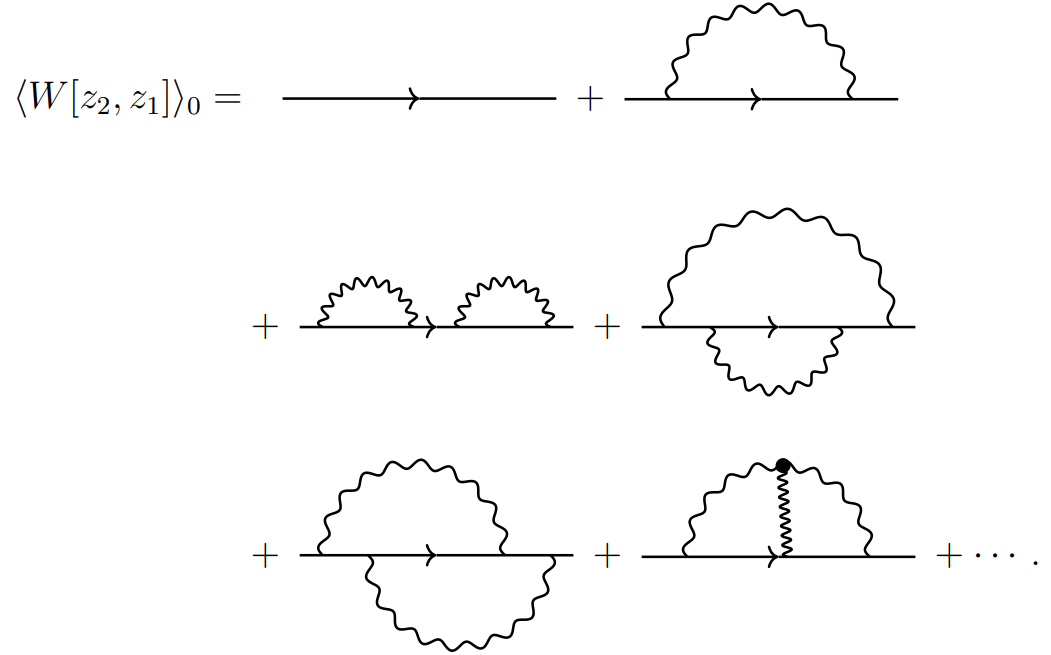}
\end{align}
The first diagram in (\ref{diagrams_oneoverc_expansion}) contributes at $\mathcal{O}(\frac{1}{c^0})$, the second diagram contributes at $\mathcal{O}(\frac{1}{c})$, the final four diagrams contribute at $\mathcal{O}(\frac{1}{c^2})$, and the ellipsis denotes higher order quantum corrections past $\mathcal{O}(\frac{1}{c^2})$.

For every vertex, in the undeformed case, we have holomorphic stress tensor insertions. For $n$ vertices, we have an $n$-point correlator of holomorphic stress tensors to integrate over. Using this setup for the quantum gravitational Wilson line, writing down formal expressions for the deformed corrections from the $n$-point deformed stress tensor correlators is straightforward. 

One can then use the Feynman rules of the fundamental fields derived in this chapter \eqref{ija} and \eqref{ik} to calculate the deformed stress tensor correlators. Intuitively, the quantum corrections to the deformed gravitational Wilson line $\langle W [z_2, z_1] \rangle_\lambda$ involve non-vanishing self-energy interactions between both holomorphic and anti-holomorphic exchanges of the fundamental fields denoted by solid and dashed propagators respectively.

Determining the deformed quantum Wilson line at a given order in $\lambda$ and $\frac{1}{c}$ is computationally complicated for two reasons. The first reason is that the $n$-point stress tensor correlator is subject to both quantum corrections in $\frac{1}{c}$ and $\lambda$ corrections. To be more precise, we notice that the expectation value of the undeformed Wilson line $W_\varepsilon[z, 0]$ has the expansion
\begin{equation}
    \langle W_\varepsilon[z;0] \rangle_0 = z^{2j} N(\varepsilon) \sum_{n=0}^\infty \frac{(6\alpha (\varepsilon))^n}{c^n} \int_0^z dy_n \cdots \int_0^{y_2} dy_1 F_n(z;y_n,\dots, y_1) \langle T_{zz}(y_n) \cdots T_{zz}(y_1) \rangle_0
\end{equation}
from \eqref{eq:D'Hoker-Kraus}. Here the SL$(2, \mathbb{R})$ group theory factor $F_n\left(z ; y_{n}, \dots, y_{1}\right)$ is defined by the following homogeneous polynomial in the variables $z, y_n, \cdots, y_1$ of degree $n$:
\begin{equation}
    z^{2 j} F_{n}\left(z ; y_{n}, \dots, y_{1}\right)=\left\langle j,-j\mid e^{z L_{1}} \left( L_{-1}-2 y_n L_{0}+y_n^{2} L_{1} \right) \cdots \left( L_{-1}-2 y_1 L_{0}+y_1^{2} L_{1} \right) \mid j, j\right\rangle\,.
\end{equation}
Computing $\langle W_\varepsilon[z;0] \rangle_\lambda$ via conformal perturbation theory in $\lambda$ involves an infinite $\lambda$ expansion at each order of the $\mathcal{O} \left(\frac{1}{c}\right)$ expansion of the undeformed Wilson line $\langle W_\varepsilon[z;0]\rangle_0$. For instance, the $\mathcal{O} \left(\frac{1}{c^2}\right)$ term in the $\frac{1}{c}$ expansion\footnote{We start with $O\left(\frac{1}{c^2}\right)$ because at $\mathcal{O}\left(\frac{1}{c}\right)$, the one-point planar correlator $\langle T_{zz}(y_1) \rangle_\lambda$ vanishes identically by Lorentz and translational invariance. In the case of non-planar backgrounds, such as the cylinder or torus \cite{Cotler:2018zff,Nguyen:2021dpa}, the one-point function is nonzero.} of $\langle W_\varepsilon[z;0]\rangle_0$ is
\begin{equation}
\begin{aligned}
    & z^{2j}N(\varepsilon) \frac{(6\alpha (\varepsilon))^2}{c^2}\int_0^z dy_1 \int_0^{y_2} dy_1 F_2(z;y_1,y_2) \langle T_{zz}(y_1) T_{zz}(y_2)\rangle_0  \\
   & \rightarrow  z^{2j}N(\varepsilon) \frac{(6\alpha(\varepsilon))^2}{c^2} \sum_{p=0}^\infty \int_0^z dy_1 \int_0^{y_2}  F_2(z;y_1,y_2) \\& \cdot \left\langle T_{zz}(y_1) T_{zz}(y_2) \frac{\lambda^p}{p!}\left(\int d^2w T_{zz}(w) T_{\bar{z} \bar{z}}(\overline{w})\right)^p\right\rangle\,.
\end{aligned}
\end{equation}

The divergences from the integrals are handled by the dimensional regularization scheme, which we mentioned in the discussion of the renormalized vertex factor and multiplicative renormalization around equation \eqref{eq:D'Hoker-Kraus}. For exposition's sake, we content ourselves with determining the leading order corrections to the quantum Wilson line \eqref{eq:GravWilson}. Expanding the exponential \eqref{eq:GravWilson}, we find
 \begin{equation}
 \begin{aligned}
 \label{eq:QWilson1}
      &\langle W[z_2, z_1]\rangle_\lambda \\
      &\,\,\,= z^{2j} \bigg[ 1 + \left( \frac{6}{c} \right)^2 \int^{z_2}_{z_1} dy_1 \int^{y_2}_{z_1} dy_2   \langle j, -j \mid \exp \left( L_1 z_{21} \right)  \left( L_{-1} - 2 (y_1 - z_1)L_0 + (y_1 - z_1)^2 L_1 \right)\\
      & \hspace{130pt}\times \left( L_{-1} - 2 (y_2 - z_1)L_0 + (y_2 - z_1)^2 L_1 \right) \mid j, j\rangle \langle T_{zz}(y_1) T_{zz}(y_2) \rangle_\lambda \bigg]\,.
      \end{aligned}
 \end{equation}
The tree-level deformed planar stress tensor two-point function at $\mathcal{O}(\lambda^2 c^2)$ was determined first by \cite{Kraus:2018xrn} via translational/rotational invariance and stress tensor conservation. Alternatively, using the approach in this chapter, this is easily understood by the propagators of the fundamental fields as
 \begin{equation}
 \begin{aligned}
 \label{eq:deformed 2pt}
     \langle T_{zz}(y_1) T_{zz}(y_2) \rangle_\lambda &= \frac{1}{(8G)^2} \partial_{y_1} \partial_{y_2} \langle f'(y_1) f'(y_2) \rangle_0 \\&+ \frac{r_c^2}{\left( 16 G \right)^2} \left( \partial^2_{y_1} \partial^2_{y_2} \langle f'(y_1) f'(y_2) \rangle_0 \right) \langle \bar{f}'(\bar{y}_1) \bar{f}'(\bar{y}_2) \rangle_0  \\&= \frac{c}{2 \left( y_1 -y_2 \right)^4} + \frac{5 \pi^2 \lambda^2 c^2}{6 \left( y_1 - y_2 \right)^6 \left( \bar{y}_1 - \bar{y}_2 \right)^2}\,.
     \end{aligned}
 \end{equation}
Thus
 \begin{equation}
 \begin{aligned}
     &\langle j, -j \mid \exp \left( L_1 z \right)  \left( L_{-1} - 2 y_1 L_0 + y_1^2 L_1 \right) \left( L_{-1} - 2 y_2 L_0 + y_2^2 L_1 \right) \mid j, j\rangle \\&= z^{2j-2} \left[ 2j y_2 \left( z - y_1 \right) \left( 2jy_1 (z-y_2) - y_2 (z - y_1) \right) \right]\,.
\end{aligned}
 \end{equation}
 The integral \eqref{eq:QWilson1} reduces to
 \begin{equation}
     \begin{aligned}
     \label{eq:deformedWilatO(lambdasqc0)}
     \langle W[z, 0] \rangle_\lambda &= z^{2j} \bigg[ 1 + \frac{36j}{cz^2} \int^{z}_{0} dy_1 \int^{y_1}_{0} dy_2 \frac{y_2 (z-y_1) \left( 2jy_1 (z-y_2) - y_2 (z-y_1) \right)}{( y_1 - y_2)^4} \\&+ \frac{60 \pi^2 \lambda^2 j}{z^2}  \int^{z}_{0} dy_1 \int^{y_1}_{0} dy_2 \frac{y_2 (z-y_1) \left( 2jy_1 (z-y_2) - y_2 (z-y_1) \right)}{( y_1 - y_2)^6  ( \bar{y}_1 - \bar{y}_2 )^2} \bigg]\,, 
     \end{aligned}
 \end{equation}
which clearly diverges when $y_2 \rightarrow y_1$ or $\bar{y}_2 \rightarrow \bar{y}_1$.

We dimensionally regularize the stress tensor correlators to evaluate these divergent integrals \eqref{eq:deformedWilatO(lambdasqc0)}. The $O(\frac{\lambda^0}{c})$ integral has already been evaluated in \cite{Besken:2018zro} via dimensional regularization, which gives
\begin{equation}
    \frac{36j}{cz^2} \int^{z}_{0} dy_1 \int^{y_1}_{0} dy_2 \, \frac{y_2 (z-y_1) \left( 2jy_1 (z-y_2) - y_2 (z-y_1) \right)}{( y_1 - y_2)^4} = \frac{12j(j+1)}{c} \ln z\,.
\end{equation}
To deal with the $\mathcal{O}(\lambda^2 c^0)$ integral in \eqref{eq:deformedWilatO(lambdasqc0)}, we first specify an integration contour in the complex plane that is a straight line along the direction towards $z$:
\begin{equation}
    \begin{aligned}
&y_{2}=y_{1} t, \quad 0 \leq t \leq 1  \,, \\
&y_{1}=z T, \quad 0 \leq T \leq 1  \,,
\end{aligned}
\end{equation}
and we find
\begin{equation}
    \begin{aligned}
   &  \frac{60 \pi^2 \lambda^2 j}{z^2}  \int^{z}_{0} dy_1 \int^{y_1}_{0} dy_2 \, \frac{y_2 (z-y_1) \left( 2jy_1 (z-y_2) - y_2 (z-y_1) \right)}{( y_1 - y_2)^6  ( \bar{y}_1 - \bar{y}_2 )^2} \\=&  \frac{60 \pi^2 \lambda^2 j}{|z|^4}  \int^{1}_{0} dT \, \frac{1 - T}{T^5} \int^{1}_{0} dt \, \frac{2j (t-t^2 T) - t^2 (1-T)}{(1-t)^8}\,.
    \end{aligned}
\end{equation}
The above integral is evaluated via dimensional regularization:
\begin{equation}
    \begin{aligned}
    \label{eq:dimreglambda}
   & \frac{60 \pi^2 \lambda^2 j}{|z|^4}  \int^{1}_{0} dT \, \frac{1 - T}{T^{5-2\varepsilon}} \int^{1}_{0} dt \, \frac{2j (t-t^2 T) - t^2 (1-T)}{(1-t)^{8-2\varepsilon}} \\=& \frac{\pi^2 j(9j-1)  \lambda^2}{21 |z|^4}\,.
    \end{aligned}
\end{equation}
In summary, the leading order correction to the Wilson line is 
\begin{equation}
    \langle W[z,0] \rangle_\lambda = z^{2j} \left(1 + \frac{12j(j+1)}{c} \ln z + \frac{\pi^2 j (9j-1) \lambda^2}{21 |z|^4} \right)\,.
\end{equation}
An alternative renormalization approach, which yields the same numerical coefficient as \eqref{eq:dimreglambda}, is to introduce cutoffs $\varepsilon_1$ and $\varepsilon_2$ as
\begin{equation}
\begin{aligned}
\label{eq:alternativeregulator}
&\frac{60 \pi^2 \lambda^2 j}{|z|^4}  \int^{1}_{\varepsilon_2} dT \, \frac{1 - T}{T^5} \int^{1-\varepsilon_1}_{0} dt \, \frac{2j (t-t^2 T) - t^2 (1-T)}{(1-t)^8}\,,
\end{aligned}
\end{equation}
and perform ``minimal subtraction'' to remove the divergent terms. Evaluating \eqref{eq:alternativeregulator} gives
\begin{equation}
\frac{\pi^2 j (9j-1) \lambda^2}{21 |z|^4}\,.
\end{equation}

\subsection{AdS$_3$ Wilson line correlators}
\label{sec:wil}

The correlator involving the product of a holomorphic and anti-holomorphic Wilson line is a scalar correlator. A meaningful check is to compute this Wilson line product correlator and see if it is consistent with $T\overline{T}$-deformed scalar correlators \cite{Kraus:2018xrn,He:2019vzf,Cardy:2019qao}. We use conformal perturbation theory at $\mathcal{O}(\lambda)$ to find
\begin{align}
\begin{split}
\label{wfirstthenyintegral}
 \langle W[z, 0] \overline{W} [\bar{z}, 0] \rangle_\lambda &= \left \langle W[z, 0] \overline{W}[\bar{z}, 0] \exp \left( \lambda \int d^2w ~ T_{zz} (w) T_{\bar{z} \bar{z}} (\bar{w}) \right)\right \rangle \\&= \langle W[z, 0] \overline{W}[\bar{z}, 0]  \rangle_0 + \lambda \int d^2w  \left \langle T_{zz}(w) T_{\bar{z} \bar{z}}(\bar{w}) W[z, 0] \overline{W}[\bar{z}, 0]  \right \rangle_0 \\&= |z|^{-4h(j)} + \lambda \int d^2w \left \langle T_{zz}(w) W[z, 0] \right \rangle_0 \left \langle T_{\bar{z} \bar{z}}(\bar{w})  \overline{W}[\bar{z}, 0] \right \rangle_0 \\&= |z|^{-4h(j)} + \lambda h^2 |z|^{4} \int  \frac{d^2w}{|w|^4|w-z|^4} \left \langle W [z, 0] \right \rangle_0 \left \langle \overline{W} [\bar{z}, 0] \right \rangle_0 \\&=  |z|^{-4h} + \lambda h^2 |z|^{4} \int \frac{d^2w}{|w|^4|w-z|^4} |z|^{-4h} \\&= |z|^{-4h(j)}\left( 1 + \lambda h^2 |z|^{4} \mathcal{I}_{2222} (0,z,0,\bar{z}) \right)\,.
 \end{split}
 \end{align}
Here, we have used the Ward identity in \cite{DHoker:2019clx}
 \begin{align}
 \begin{split}
     &\left\langle T_{zz}(w) W\left[z, 0\right]\right\rangle_0 = \frac{h(j) z^{2}}{\left(z-w\right)^{2}w^{2}}\left\langle W\left[z, 0\right]\right\rangle_0 \,, \\
     &\left\langle T_{\bar{z} \bar{z}}(\bar{w}) \overline{W}\left[\bar{z}, 0\right]\right\rangle_0 =\frac{h(j)\bar{z}^{2}}{\left(\bar{z}-\bar{w}\right)^{2}\bar{w}^{2}}\left\langle \overline{W}\left[\bar{z}, 0\right]\right\rangle_0 \,, 
 \end{split}
 \end{align}
 which displays the bi-local structure of the gravitational Wilson line. From appendix A in \cite{He:2019vzf}, the integral \eqref{wfirstthenyintegral} is of the form 
\begin{align}
\begin{aligned}
\label{MasterIntegralEquation}
    &\mathcal{I}_{a_{1}, \cdots, a_{m}, b_{1}, \cdots, b_{n}}\left(z_{i_{1}}, \cdots, z_{i_{m}}, \bar{z}_{j_{1}}, \cdots, \bar{z}_{j_{n}}\right) = \int \frac{d^{2} z}{ \prod^m_{k=1}\left(z-z_{i_{k}}\right)^{a_{k}} \prod^n_{p=1}\left(\bar{z}-\bar{z}_{j_{p}}\right)^{b_{p}}}\,, 
    \end{aligned}
\end{align}
and is evaluated via dimensional regularization. In particular, 
 \begin{align}
 \begin{split}
 \label{I2222}
     \mathcal{I}_{2222} (0, z, 0, \bar{z}) &= \int  \frac{d^2w}{|w|^4|w-z|^4} \\&= \frac{4 \pi}{|z|^6} \left( \frac{4}{\varepsilon} + 2 \ln |z|^2 + 2\ln \pi +2 \gamma - 5 \right) \\&= \frac{1}{|z|^6} \left( C_1 + C_2 \ln |z|^2 \right)\,, 
\end{split}
 \end{align}
where $C_1$ and $C_2$ are constant coefficients. We arrive
\begin{align}
\label{match}
    \langle W[z, 0] \overline{W} [\bar{z}, 0] \rangle_\lambda &=|z|^{-4 h(j)} \left( 1+ \frac{\lambda  h(j)^2\left(C_1 + C_2 \ln |z|^2 \right)}{|z|^2}  \right)\,, 
\end{align}
 which exactly matches what we expect at $\mathcal{O}(\lambda c^0)$ from previous analyses of $T\overline{T}$-deformed scalar correlators \cite{Kraus:2018xrn,He:2019vzf,Cardy:2019qao}. This confirms the claim that the correlator of two Wilson lines behaves as a scalar correlator, at least in this order.
 
Additionally, at leading order in $\lambda$ and in the large-$c$ limit, \eqref{wfirstthenyintegral} agrees with the structure one would expect from the linear mixing of sources and expectation values discussed above. Schematically, in the leading order,
\begin{align}
\label{eq:mixingscalarcorrelators}
    &\left \langle P \exp \left[ \int^z_0 dy \left( e_i(\lambda) L_1 + \frac{6}{c} T_{zz}(y) L_{-1}\right) \right]  P \exp \left[ \int^{\bar{z}}_0 d\bar{y} \left( \bar{e}_i (\lambda) \bar{L}_1 + \frac{6}{c} T_{\bar{z} \bar{z}}(\bar{y}) \bar{L}_{-1} \right) \right] \right \rangle_\lambda \nonumber \\
    &= \Bigg \langle P \exp \left[ \int^z_0 dy \left( (1+\lambda T_{\bar{z} \bar{z}}(y) ) L_1 + \frac{6}{c} T_{zz}(y) L_{-1} \right) \right]  \nonumber \\
    &\qquad \qquad \cdot P \exp \left[ \int^{\bar{z}}_0 d\bar{y} \left( (1+ \lambda T_{zz} (y)) \bar{L}_1 + \frac{6}{c} T_{\bar{z} \bar{z}}(\bar{y}) \bar{L}_{-1} \right) \right] \Bigg \rangle_\lambda \nonumber \\
    &= \langle W[z,0]\rangle_0 \langle \overline{W}[\bar{z}, 0] \rangle_0 + \lambda  \langle \exp \left( zL_1\right) L_1 \rangle \int^z_0 dy  \left \langle T_{\bar{zz}} (\bar{y}) \overline{W} [\bar{z}, 0] \right \rangle_0 +  \mathcal{O}(\lambda^2) \nonumber \\
    &= |z|^{-4h(j)} + \lambda  \partial_z \langle \exp \left( zL_1 \right) \rangle \int^z_0 dy  \left \langle T_{\bar{zz}} (\bar{y}) \overline{W} [\bar{z}, 0] \right \rangle_0 +  \mathcal{O}(\lambda^2) \nonumber  \\
    &= |z|^{-4h(j)} -2 \lambda h(j) z^{-2h(j)-1} \int^z_0 dy \frac{h(j) \bar{z}^2}{(\bar{z} - \bar{y} )^2 \bar{y}^2} \langle \overline{W} [\bar{z}, 0] \rangle_0 +  \mathcal{O}(\lambda^2) \nonumber \\& = |z|^{-4h(j)} -2 \lambda h(j)^2 z^{-2h(j)-1} \bar{z}^{-2h(j)+2} \int^{z}_{0} \frac{ dy }{(\bar{z} - \bar{y} )^2 \bar{y}^2} +  \mathcal{O}(\lambda^2)  \nonumber \\
    & = |z|^{-4h(j)} \left( 1 + \lambda h(j)^2 z^{-1} \bar{z}^2 \left( \frac{c_1 + c_2 \ln |z|^2}{\bar{z}^3} \right) + \mathcal{O}(\lambda^2) \right) \nonumber  \\
    &=  |z|^{-4h(j)} \left( 1 + \lambda h(j)^2 \left( \frac{c_1 + c_2 \ln |z|^2}{|z|^2} \right) +  \mathcal{O}(\lambda^2) \right)\,,
\end{align}
where in the large-$c$ limit, the quantum corrections to the Wilson line's scaling dimension  $h(j) = - j$ are suppressed and $\langle \exp \left( zL_1 \right) \rangle = z^{-2h} = z^{2j}$. The integral in \eqref{eq:mixingscalarcorrelators} may be evaluated via the integration cutoff introduced in \eqref{eq:alternativeregulator} or by dimensional regularization. Using either method, one finds that the result has a similar structure as \eqref{match}, where $c_{1}$ and $c_2$ are constant coefficients. 

We emphasize that if one had not used the linear mixing or conformal perturbation theory, but rather expanded each path-ordered exponential in $\langle W[z_2, z_1] \overline{W}[\bar{z}_2, \bar{z}_1] \rangle_\lambda$, then the leading contribution in $\lambda$ would be at $\mathcal{O}(\lambda^2 c^0)$, which arises from integrating the tree-level deformed stress tensor two-point function. To see this, let us compute the correction
\begin{equation}
\delta \langle W[z,0] \overline{W}[\bar{z}, 0] \rangle_\lambda =  \langle W[z,0] \overline{W}[\bar{z}, 0] \rangle_\lambda - \langle W[z,0] \overline{W}[\bar{z}, 0] \rangle_0 
\end{equation}
to the correlator using this prescription. We expand the path-ordered exponential\footnote{For the single Wilson line \eqref{eq:deformedWilatO(lambdasqc0)}, we expanded up to $\mathcal{O}(\frac{1}{c^2})$ in the path-ordered exponential since the planar one-point function vanishes. At $\mathcal{O}(\frac{1}{c})$, the path-ordered exponential reduces to a regular integral.} for $W[z, 0]$ and $\overline{W}[\bar{z}, 0]$ up to $\mathcal{O}(\frac{1}{c})$ in \eqref{eq:GravWilson}, which gives
\begin{equation}\hspace{-18pt}
    \begin{aligned}
    \label{eq:wrongway}
\left(\frac{6}{c} \right)^2 \int^{z}_{0} dy_1 \int^{\bar{z}}_{0} d\bar{y}_2 &\langle j, -j \mid e^{L_1 z} (L_{-1} - 2y_1 L_0 + y_1 ^2) e^{ \bar{L}_1 \bar{z}} ( \bar{L}_{-1} - 2\bar{y}_2  \bar{L}_0 + \bar{y}_2^2) \mid j, j \rangle \\& \cdot \langle T_{zz} (y_1) T_{\bar{z} \bar{z}} (y_2)  \rangle_\lambda \,.
    \end{aligned}
\end{equation}
Using \eqref{hi} and the Feynman rules for the relevant tree diagrams,
\begin{equation}
\begin{aligned}
    \langle T_{zz} (y_1) T_{\bar{z} \bar{z}} (y_2) \rangle_\lambda &= \frac{r_c^2}{(16G)^2} \left( \partial^2_{y_1} \langle f'(y_1) f'(y_2) \rangle_0 \right) \left( \partial^2_{\bar{y}_2} \langle \bar{f}'(\bar{y}_1) \bar{f}'(\bar{y}_2) \rangle_0 \right) +O(r_c^3)\\ &= \frac{\pi^2 \lambda^2 c^2}{4} \frac{1}{\left( y_1 - y_2 \right)^4 \left( \bar{y}_1 - \bar{y}_2 \right)^4} +\mathcal{O}(\lambda^3)\,.
    \end{aligned}
\end{equation}
Thus the above prescription involving correlators of Wilson lines \eqref{eq:wrongway} is incorrect because the leading correction enters at $\mathcal{O}(\lambda^2 c^0)$, rather than the expected order of $\mathcal{O}(\lambda c^0)$ for scalar two-point correlators.

Furthermore, one may also consider a string of holomorphic and anti-holomorphic stress tensor insertions in correlators involving a Wilson line. For instance, we can calculate this kind of correlator via conformal perturbation theory at $\mathcal{O}(\lambda)$:
\begin{equation}
\label{eq:stringofT}
    \left\langle T_{zz}\left(w_{1}\right) T_{\bar{z}\bar{z}}\left(\bar{w}_{2}\right) W[z, 0] \right \rangle_\lambda = \lambda \int d^{2} y \left \langle T_{zz}(y) T_{zz}\left(w_{1}\right) W [0, z] \right \rangle_{0} \left \langle T_{\bar{z}\bar{z}}(\bar{y}) T_{\bar{z} \bar{z}}\left(\bar{w}_{2}\right)\right\rangle_{0}\,.
\end{equation}
In \cite{DHoker:2019clx}, the following tree-level correlator to $O(1/c^0)$ was derived:
\begin{equation}
\begin{aligned}
    &\left\langle T_{zz}\left(w_{1}\right) T_{zz}\left(w_{2}\right) W[z, 0]\right\rangle_0\\&=\frac{j^{2} z^{2 j+4}}{w_{1}^{2}\left(z-w_{1}\right)^{2} w_{2}^{2}\left(z-w_{2}\right)^{2}}+\frac{j z^{2 j+2}}{w_{1}\left(z-w_{1}\right) w_{2}\left(z-w_{2}\right)\left(w_{1}-w_{2}\right)^{2}}\,,
    \end{aligned}
\end{equation}
in agreement with the predictions from the conformal Ward identities. Therefore, using the fact that
\begin{equation}
    \left \langle T_{\bar{z}\bar{z}}(\bar{y}) T_{\bar{z}\bar{z}}\left(\bar{w}_{2}\right)\right\rangle_{0} = \frac{c}{2 (\bar{y}-\bar{w}_2)^4}\,,
\end{equation}
the integral \eqref{eq:stringofT} is reduced to 
\begin{equation}
\begin{aligned}
   \left\langle T_{zz} \left(w_{1}\right) T_{\bar{z}\bar{z}}\left(\bar{w}_{2}\right) W[z, 0] \right \rangle_{\lambda} &=  \frac{c j \lambda z^{2j}}{2} \int d^{2} y\bigg[\frac{j z^{4}}{y^{2}(y-z)^{2} w_{1}^{2}\left(z-w_{1}\right)^{2}\left(\bar{y}-\bar{w}_{2}\right)^{4}}\\&-\frac{z^{2}}{y(y-z) w_{1}\left(z-w_{1}\right)\left(y-w_{1}\right)^{2}\left(\bar{y}-\bar{w}_{2}\right)^{4}}\bigg]\,,
   \end{aligned}
\end{equation}
and is evaluated in terms of the integrals defined in \eqref{MasterIntegralEquation}:
\begin{equation}
\begin{aligned}
    &\left\langle T_{zz}\left(w_{1}\right) T_{\bar{z}\bar{z}}\left(\bar{w}_{2}\right) W[z, 0] \right \rangle_{\lambda}\\&=\frac{j \lambda c z^{2 j+2}}{2 w_{1}\left(z-w_{1}\right)}\left[\frac{j z^{2}}{w_{1}\left(z-w_{1}\right)} \mathcal{I}_{224}\left(0, z, \bar{w}_{2}\right)- \mathcal{I}_{1124}\left(0, z, w_{1}, \bar{w}_{2}\right)\right]\,.
    \end{aligned}
\end{equation}
Another example is a correlator involving two insertions of anti-holomorphic stress tensors, a holomorphic stress tensor, and a holomorphic Wilson line. The desired correlator 
\begin{equation}
\label{eq:stringantiT's}
     \left \langle  T_{zz}\left(w_{1}\right)  T_{\bar{z}\bar{z}}\left(\bar{w}_{2}\right) T_{\bar{z}\bar{z}}\left(\bar{w}_{3}\right) W [0, z] \exp \left( \lambda \int d^2y T_{zz}(y) T_{\bar{z}\bar{z}}(\bar{y}) \right) \right \rangle
\end{equation}
is easily computable at $\mathcal{O}(\lambda)$ via conformal perturbation theory. 

Noting that the undeformed tree-level planar three-point stress tensor correlator is
\begin{equation}
\langle T_{\bar{z}\bar{z}} (\bar{y}) T_{\bar{z}\bar{z}}\left(\bar{w}_{2}\right) T_{\bar{z}\bar{z}}\left(\bar{w}_{3}\right)  \rangle_0 = \frac{c}{\left( \bar{y} - \bar{w}_2 \right)^2 \left( \bar{w}_2 - \bar{w}_3 \right)^2 \left( \bar{w}_3-\bar{y} \right)^2}\,, 
\end{equation}
then the leading order correction to the integral \eqref{eq:stringantiT's} at $\mathcal{O}(\lambda c)$ is
\begin{align}
    \label{eq:integral312W}
&\left\langle T_{zz}\left(w_{1}\right) T_{\bar{z}\bar{z}}\left(\bar{w}_{2}\right) T_{\bar{z}\bar{z}}\left(\bar{w}_{3}\right) W[z, 0] \right \rangle_\lambda \nonumber \\
  &= \left\langle T_{zz}\left(w_{1}\right)W[z, 0] \right \rangle_0  \left \langle T_{\bar{z}\bar{z}}\left(\bar{w}_{2}\right) T_{\bar{z}\bar{z}}\left(\bar{w}_{3}\right) \right \rangle_0 \nonumber \\
  & +  \lambda \int d^{2} y \, \left \langle  T_{zz} (y) T_{zz}\left(w_{1}\right)  W [0, z] \right \rangle_0 \langle T_{\bar{z}\bar{z}} (\bar{y}) T_{\bar{z}\bar{z}}\left(\bar{w}_{2}\right) T_{\bar{z}\bar{z}}\left(\bar{w}_{3}\right)  \rangle_0 \nonumber \\
  & = \frac{h(j) z^{2}}{\left(z-w_1\right)^{2}w_1^{2}}\left\langle W\left[z, 0\right]\right\rangle_0 \frac{c}{2 (\bar{w}_2 - \bar{w}_3)^4} \nonumber \\
  &+ c j \lambda z^{2j} \int d^{2} y \,  \Bigg[\frac{j z^{4}}{y^{2}(y-z)^{2} w_{1}^{2}\left(z-w_{1}\right)^{2} \left(\bar{y}-\bar{w}_{2}\right)^{2} \left(\bar{w}_2-\bar{w}_{3}\right)^{2} \left(\bar{w}_3-\bar{y}\right)^{2}} \nonumber \\
  & -\frac{z^{2}}{y(y-z) w_{1}\left(z-w_{1}\right)\left(y-w_{1}\right)^{2} \left(\bar{y}-\bar{w}_{2}\right)^{2} \left(\bar{w}_2-\bar{w}_{3}\right)^{2} \left(\bar{w}_3-\bar{y}\right)^{2}}\Bigg]\,.
\end{align}
Evaluating \eqref{eq:integral312W} in terms of the integrals defined in \eqref{MasterIntegralEquation}, we find
\begin{align}
    &\left\langle T_{zz}\left(w_{1}\right) T_{\bar{z}\bar{z}}\left(\bar{w}_{2}\right) T_{\bar{z}\bar{z}}\left(\bar{w}_{3}\right)  W[z, 0] \right\rangle_\lambda \nonumber  = \frac{h(j) c z^{2-2h(j)}}{2(z-w_1)^2 w_1 (\bar{w}_2 - \bar{w}_3)^4} \nonumber \\
    & +\frac{j \lambda c z^{2j+2}}{w_1 \left(z-w_1 \right)  (\bar{w}_2 - \bar{w}_3)^2 } \left[ \frac{jz^2}{w_1 (z-w_1)} \mathcal{I}_{2222} \left(0 , z, \bar{w}_2, \bar{w}_{3} \right) - \mathcal{I}_{11222} \left(0, z, w_1, \bar{w}_2, \bar{w}_{3} \right) \right]\,.
\end{align}
The integrals presented here, which are of the general form given in \eqref{MasterIntegralEquation} but with higher-valued indices, can be expressed in terms of derivatives and linear combinations of known integrals with lower-valued indices. See appendix A in \cite{He:2019vzf} for several detailed examples.

One can automate the above perturbative analysis in $\lambda$ to produce more complicated expressions for correlators involving products of $m$-insertions of holomorphic stress tensors, $n$-insertions of anti-holomorphic stress tensors, and a network of Wilson lines (e.g. $p$-insertions of the holomorphic Wilson line and $q$-insertions of the anti-holomorphic Wilson line) following \cite{DHoker:2019clx}. The leading correction for such a general correlator takes the form
\begin{equation}
    \begin{aligned}
  &\left \langle \prod^{m}_{i=1} T_{zz} (x_i) \prod^{n}_{j=1} T_{\bar{z}\bar{z}} (\bar{w}_j) \prod^p_{k=1} W[z_{k+1}, z_k] \prod^q_{l=1} \bar{W}[\bar{r}_{l+1}, \bar{r}_{l}] \exp \left( \lambda \int d^2y  T_{zz} (y) T_{\bar{z}\bar{z}} (\bar{y}) \right) \right \rangle \\&=\left \langle \prod^{m}_{i=1} T_{zz} (x_i) \prod^p_{k=1} W[z_{k+1}, z_k] \right \rangle_0 \left \langle \prod^{n}_{j=1} T_{\bar{z}\bar{z} } (\bar{w}_j) \prod^q_{l=1} \bar{W}[\bar{r}_{l+1}, \bar{r}_{l}] \right \rangle_0 \\&+ \lambda \int d^2 y \left \langle T_{zz} (y) \prod^{m}_{i=1} T_{zz} (x_i)  \prod^p_{k=1} W[z_{k+1}, z_k] \right \rangle_0  \left \langle T_{\bar{z}\bar{z}} (\bar{y}) \prod^{n}_{j=1} T_{\bar{z}\bar{z} } (\bar{w}_j) \prod^q_{l=1} \bar{W}[\bar{r}_{l+1}, \bar{r}_{l}] \right \rangle_0 \\
  &\quad\,+ \mathcal{O}(\lambda^2)\,.
    \end{aligned}
\end{equation}
\section{Conclusion}
\label{dissec}

We gave evidence for the Nambu-Goto action (in Hamiltonian form) as the all-order action for $3d$ gravity with a cutoff planar boundary. Secondly, we used the action to compute correlators of the stress tensor operator to two-loop order. The proposal for the action was based on finding a suitable field redefinition yielding Nambu-Goto up eighth order in fields. Proving this conjecture and determining the explicit form of the field redefinition to all orders is desirable. Although the action takes the familiar Nambu-Goto form, the stress tensor is not the canonical one, which is due to the way that the original translation symmetries of the AdS$_3$ background act on the redefined fields.   Our computation of stress tensor correlators to two-loop order revealed the need for one stress tensor counterterm with an associated undetermined finite part. As mentioned in the introduction, considering the overarching arguments supporting the renormalizability of pure $3d$ gravity, even when accounting for a finite planar cutoff boundary, it is anticipated that symmetries will play a crucial role in determining all the parameters involved. The implementation of these symmetries is complicated by the non-Lorentz invariant form of the action and by the nonlocal field redefinition that puts the action in Nambu-Goto form. A task for the future is to systematically implement the Ward identities corresponding to these symmetries and check if these yield unique results for stress tensor correlators. The ultimate goal here is to get sufficient control over the stress tensor correlators to say something about their short-distance structure, since this gets to the heart of the nature of this theory, including its anticipated nonlocal character; e.g. \cite{Cardy:2020olv,Jiang:2020nnb}. Third is that in the $3d$ Chern-Simons setting, we studied modifications to correlators involving boundary-anchored Wilson lines, which were induced by a $T\overline{T}$ deformation on the $2d$ boundary; results were presented at both the classical level (using modified boundary conditions) and the quantum-mechanical level (using conformal perturbation theory).

Developing cases with curved cutoff boundaries would also be worthwhile. The Chern-Simons computation of the action for a finite $S^2$ boundary is considered in appendix \ref{sphere}, and it should be possible to extend this to one-loop and compare with results in \cite{Mazenc:2019cfg}; see also \cite{Caputa:2019pam,Jiang:2019tcq} for related results. The technical complication here is the two patches needed to define the gauge connections on the sphere. 

We close this chapter by commenting on the appearance of the Nambu-Goto action in our analysis. By construction, solutions of our Nambu-Goto equations of motion yield flat two-dimensional surfaces embedded in AdS$_3$. On the other hand, the precise Nambu-Goto action that arises is that of a string worldsheet embedded in flat $\mathbb{R}^3$, with $\alpha'$ controlled by the cutoff $r_c$. One usually thinks of the solutions as describing extremal area surfaces embedded in this flat spacetime. There is a correspondence between flat surfaces embedded in AdS$_3$ and extremal area surfaces embedded in $\mathbb{R}^3$. 

\chapter{$T\overline{T}$ in JT Gravity and BF Gauge Theory}
 \label{ch:BF}
JT gravity can be represented using a first-order formulation akin to a two-dimensional BF theory. This formulation can be perceived as the dimensional reduction of the Chern-Simons description of $3d$ gravity. We consider $T\overline{T}$-type deformations of the $(0+1)$-dimensional dual to this $2d$ BF theory and interpret the deformation as a modification of the BF theory boundary conditions. The fundamental observables in this deformed BF theory and its $3d$ Chern-Simons lift are Wilson lines and loops. In the last chapter, we studied the $3d$ Chern-Simons setting and the modifications to correlators involving boundary-anchored Wilson lines induced by a $T\overline{T}$ deformation on the $2d$ boundary. In this chapter, we determine the $T\overline{T}$-deformed boundary conditions in the BF description of JT gravity. We discuss Wilson lines in the BF theory and calculate the analogous deformed Wilson line correlators in $2d$ BF theory below the Hagedorn temperature, where the principal series dominates over the discrete series. 

\section{Introduction} \label{intro33333}
In this chapter, we consider the $T\overline{T}$-deformation of two-dimensional gauge or gravity theories, which are constructed in the following way. We begin with a three-dimensional bulk gravity theory dual to a $2d$ CFT. Then, we deform the boundary CFT by the $T\overline{T}$ operator and interpret this deformation as a modification of the bulk gravity theory. Finally, we dimensionally reduce this scenario on the circle to obtain a correspondence between a deformed $2d$ gravity theory and a dual one-dimensional theory. One can also rewrite the gravity theory in gauge theory variables and study the deformation of the $2d$ gauge theory. In the diagram (\ref{gravity_diagram}), this corresponds to deforming the $2d$ WZW model in the top-right corner and then studying the image of this deformation under the sequence of maps relating this theory to $2d$ JT gravity and $2d$ BF theory.

We emphasize that this deformation is \textit{not} the same as directly applying the $T\overline{T}$ deformation in the JT gravity or BF theory itself. Indeed, in the JT case, it is unclear how to define a local stress tensor in a theory of gravity, and in the BF case, the theory is topological, so the stress tensor vanishes. We also note that, although we consider $T\overline{T}$-like deformations of two-dimensional $\mathrm{AdS}$ gravity theories, our procedure is quite different from defining the $T\overline{T}$ deformation for a $2d$ field theory on a fixed $\mathrm{AdS}_2$ geometry. The latter problem has been considered in \cite{Jiang:2019tcq,Brennan:2020dkw}. Likewise, although the deformation of BF gauge theory treated in this manuscript is not the same as performing a $T\overline{T}$ deformation of a $2d$ gauge theory directly, such direct deformations of gauge theories have been considered for $2d$ Yang-Mills both with and without matter \cite{Conti:2018jho,Ireland:2019vvj,Brennan:2019azg,Griguolo:2022xcj}. Instead, we study a deformation holographically dual to a $T\overline{T}$-like deformation of the boundary $(0+1)$-dimensional theory rather than a $T\overline{T}$-deformation of the $2d$ gauge theory itself.

The layout of this chapter is as follows. Sections \ref{sec:Various Presentations} and \ref{sec:JT grav Review} are relevant reviews for this chapter of standard results about $3d$ gravitational Chern-Simons theory and $2d$ JT gravity, respectively, including the interpretation of a boundary $T\overline{T}$ deformation in both theories. In section \ref{sec: Deformed BF BCs}, we find the change in BF theory boundary conditions corresponding to a $T\overline{T}$-like deformation of the dual $1d$ theory for two different choices of boundary conditions in the seed theory. In section \ref{sec:JTgravWil}, we first study the deformed BF theory's boundary spectrum to find that the contribution from the principal series dominates the discrete series only below the Hagedorn temperature. The $T\overline{T}$-deformed Schwarzian theory description of the boundary spectrum is only valid below the Hagedorn temperature. We conclude the section by computing deformed Wilson lines and their correlators in the BF theory below the Hagedorn temperature. In section \ref{sec: Discussion11111111111111111111111}, we conclude with a summary and discussions on possible extensions of the results presented in this chapter.

\section{$T\overline{T}$ deformations in $3d$ Chern-Simons theory} \label{BF}
\label{sec:Various Presentations}

In this section, we review the presentation of the Chern-Simons formulation of $\mathrm{AdS}_3$ gravity, which will be relevant for later sections. In particular, we recall that the bulk interpretation of a $T\overline{T}$ deformation in the boundary CFT is a change in the boundary conditions for the Chern-Simons gauge field \cite{Llabres:2019jtx}. 

\subsection{Revisiting $T\overline{T}$-deformed $3d$ SL$(2,\mathbb{R})$ gravitational Chern-Simons}
\label{subsec:general_cs_review}

The most general asymptotically $\mathrm{AdS}_3$ metric is described by a Fefferman-Graham expansion of the metric. It was shown in  \cite{Llabres:2019jtx} that, in Chern-Simons variables, such an expansion corresponds to a solution where the connections $a$ and $\bar{a}$ take the more general form
\begin{equation}
 \begin{aligned}
 \label{eq:genLlabres}
 a_i &= 2 e^+_i L_+ - f^-_i L_- + \omega_i L_0 \,, \\
 \bar{a}_i &= f^+_i L_+ - 2 e^-_i L_- + \omega_i L_0 \,.
 \end{aligned}   
\end{equation}
The connections \eqref{eq:genLlabres} are solutions to the equations of motion when 
\begin{equation}
\begin{aligned}
\label{eq:boundaryEOMfora}
 &da+a\wedge a = 0\,, \\
 &d\bar{a} + \bar{a} \wedge \bar{a} = 0\,.
\end{aligned}
\end{equation}
Substituting \eqref{eq:genLlabres} into \eqref{eq:boundaryEOMfora},  we find
\begin{equation}
\begin{aligned}
d \omega-2 \varepsilon_{a b} e^{a} \wedge f^{b}&=0\,,\\
d e^{a} - \varepsilon^{a}{}_{b} e^{b} \wedge \omega&=0\,, \\ d f^{a} - \varepsilon^{a}{}_{b} f^{b} \wedge \omega &=0\,, \\ e^{a} \wedge f_{a} &= 0\,,
\end{aligned}
\end{equation}
which are the zero torsion conditions for the frame $e^a$ with spin connection $\omega$. Since by our conventions early Latin indices $a,b$ are flat while middle Latin indices $i,j$ are curved, $\varepsilon^{ab}$ is the Levi-Civita symbol with constant entries $\varepsilon_{+-} = - \varepsilon_{-+} = 1$, while $\varepsilon^{ij}$ is the Levi-Civita tensor with curved indices $\varepsilon^{x^+ x^-} = - \varepsilon^{x^- x^+} = \frac{1}{2e}$.


In the presence of a boundary, additional boundary terms are needed in the action to have a well-defined variational principle. Varying the Einstein-Hilbert action written in terms of the Chern-Simons connections gives
\begin{equation}
\begin{aligned}
\label{eq:variationofEHCS}
    \delta  S_{\operatorname{EH}} &= \delta S_{\operatorname{CS}}[A] - \delta S_{\operatorname{CS}}[\bar{A}] \\&= \frac{1}{8 \pi G} \int_{M_3} \operatorname{Tr} \left( F \wedge \delta A - \bar{F} \wedge \delta \bar{A} \right) - \frac{1}{16\pi G} \int_{\partial M_3} \operatorname{Tr} \left( A \wedge \delta A-  \bar{A} \wedge \delta \bar{A} \right)\,.
    \end{aligned}
\end{equation}
We desire a variational principle with Dirichlet boundary conditions for the metric, which corresponds to holding $e^a$ fixed at the boundary but letting $f^a$ vary. However, going on-shell by using the connections in \eqref{eq:genLlabres}, we find that \eqref{eq:variationofEHCS} reduces to 
\begin{equation}\label{on_shell_delta_S_EH}
\delta S_{\operatorname{CS}}[A] - \delta S_{\operatorname{CS}}[\bar{A}] = - \frac{1}{8\pi G} \int_{\partial M_3} \varepsilon_{ab} \left( e^a \wedge \delta f^b - f^a \wedge \delta e^b \right)\,,
\end{equation}
which does not vanish and is inconsistent with the specified boundary conditions. We must add the following boundary term to  \eqref{eq:variationofEHCS}:
\begin{equation}
    \label{boundary_action_As}
    S_{\text{bdry}} = - \frac{1}{8 \pi G} \int_{\partial M_3} \varepsilon_{ab} \left( A^a \wedge A^b + \bar{A}^a \wedge \bar{A}^b \right)\,.
\end{equation}
The result now is consistent with Dirichlet boundary conditions since
\begin{equation}
\label{eq:undeformedLlabres}
    \delta S_{\text{EH}} + \delta S_{\text{bdry}} = \frac{1}{4 \pi G} \int_{\partial M_3} \varepsilon_{ab} f^a \wedge \delta e^b\,.
\end{equation}
From the GKPW dictionary \cite{Gubser:1998bc,Witten:1998qj}, it is understood that $e^a$ is the source and $f^a$ is the expectation value of the dual operator. In particular, the operator dual to the boundary vielbein is the stress tensor. By identifying
\begin{align}
    \delta  S=4 \int_{\partial M_3} \, d^{2} x \, \left( \det e \right) \, T^i{}_a \, \delta e_{i}^{a}\,,
\end{align}
we find that
\begin{align}
\label{eq:Llabres Stress Tensor}
   T^i{}_a = \frac{1}{4\pi G} \varepsilon_{ab} \varepsilon^{ij} f^b_j\,,
\end{align}
with $ \nabla_{[i} f^a_{j]} =0$. 

When we turn to the two-dimensional BF theory in section \ref{sec: Deformed BF BCs}, it will be convenient to refer to the dimensional reduction of the $3d$ Chern-Simons action on a circle. The resulting dimensionally reduced theory is equivalent to BF theory with a particular choice of boundary term. To perform this reduction, we first write
\begin{equation}
\begin{aligned}
    8 \pi G S_{\text{CS}} &= \frac{1}{2}\int_{M_3} \operatorname{Tr} \left( A \wedge dA + \frac{2}{3} A \wedge A \wedge A  \right) \\&= \frac{1}{2} \int_{M_3} \, d^3x \, \varepsilon^{\mu \nu \rho} \operatorname{Tr} \left( A_\mu \partial_\nu A_\rho + \frac{2}{3} A_\mu A_\nu  A_\rho  \right) \\&= \int_{M_3} \operatorname{Tr} \left( A_\varphi F_{tr} + A_r \partial_\varphi A_t \right) + \frac{1}{2} \oint_{\partial M_3} \operatorname{Tr} A_t^2\,.
\end{aligned}
\end{equation}
Next we impose the boundary condition $A_t = A_\varphi \vert_{\partial M_3}$ so that $ \phi \equiv  A_\varphi$ and $\partial_\varphi = 0$ (see \cite{Fan:2021bwt}). Doing this yields
\begin{equation}
\label{eq:dimredCStoBF}
    S_{\text{BF}} =  \int_{M_2} \operatorname{Tr} \left( \phi F \right) + \frac{1}{2} \oint_{\partial M_2} \operatorname{Tr} \left( \phi^2 \right)\,.
\end{equation}
The first term is the usual action for $2d$ BF theory, which in this case has gauge group $G = \operatorname{SL} ( 2, \mathbb{R} )$. The degrees of freedom in this theory are a gauge field $A_\mu$ with field strength $F_{\mu \nu}$ along with an $\operatorname{SL} ( 2, \mathbb{R} )$-valued scalar field $\phi$. We will again consider this action in section \ref{sec:jt_as_bf}, where we will recall that the theory is equivalent to JT gravity. The second term of (\ref{eq:dimredCStoBF}) controls the dynamics of a boundary degree of freedom, which can be described via the Schwarzian theory or the particle-on-a-group theory. We refer to this as a ``Schwarzian-type'' boundary term, which will be revisited in section \ref{sec:schwarzian_type}.

\subsection{Interpretation of the $T\overline{T}$ deformation}\label{subsec:chern_simons_linear_mixing}
\label{sec:3dTTbar}

The $3d$ gravitational Chern-Simons theory is dual to a conformal field theory on the $2d$ boundary of the spacetime via the usual $\mathrm{AdS}/\mathrm{CFT}$ correspondence. On the other hand, in any two-dimensional field theory enjoying translation invariance, one can define a deformation by the double-trace $T\overline{T}$ operator. Our goal in the present section is to apply this deformation to the boundary CFT and interpret the resulting flow in terms of bulk Chern-Simons variables. We follow the discussion of \cite{Llabres:2019jtx} where this analysis first appeared.

We must first express the $T\overline{T}$ deformation in terms of the asymptotic expansion coefficients for the Chern-Simons gauge fields. We have already seen, for instance, in (\ref{eq:undeformedLlabres}) and (\ref{eq:Llabres Stress Tensor}), that the functions $e_i^a$ correspond to the boundary vielbein (or equivalently the metric) and that the $f_i^a$ are the dual expectation values which encode the stress tensor as
\begin{align}
  T^i{}_a = \frac{1}{4 \pi G} \varepsilon_{ab} \varepsilon^{ij} f_j^b\,.
\end{align}
On the other hand, using the definition of the determinant in terms of the Levi-Civita symbol, the $T\overline{T}$ operator can be written as
\begin{equation}
\label{eq: 3D TT}
    T\overline{T}= -2\varepsilon^{ab} \varepsilon_{ij} T^i{}_a T^j{}_b\,. 
\end{equation}
In terms of the one-forms $f^-$ and $f^+$, one therefore has
\begin{align}
    T\overline{T} = \frac{1}{(4 \pi G)^2} f^- \wedge f^+\,.
\end{align}
The flow equation for the boundary action can be written as
\begin{align}\label{boundary_llabres_flow}
    \frac{\partial S}{\partial \lambda} = \frac{1}{(4 \pi G)^2} \int_{\partial M_3} f^- \wedge f^+\,.
\end{align}
We note that this is a flow equation for the \emph{combined} boundary action, which in the undeformed case is a sum of three terms:
\begin{align}\label{sum_three_terms}
    S ( \lambda = 0 ) = S_{\text{CS}} [ A ] - S_{\text{CS}} [ \bar{A} ] + S_{\text{bdry}}\,.
\end{align}
In section \ref{subsec:general_cs_review}, we saw that variation of the first two terms $S_{\text{CS}} [ A ] - S_{\text{CS}} [ \bar{A} ]$ generated a boundary variation of the form $\varepsilon_{ab} ( e^a \wedge \delta f^b - f^a \wedge \delta e^b )$. The first term involving $\delta f^b$ was unsuitable for our desired variational principle, so we added $S_{\text{bdry}}$ to cancel this variation.

We will make the ansatz that the finite-$\lambda$ deformed boundary action has the same structure as a sum of three terms involving sources $e_i^a ( \lambda )$ and dual expectation values $f_i^a$. In this ansatz we allow the sources to acquire $\lambda$ dependence under the flow, but not the expectation values. As a result, the total boundary variation (\ref{eq:undeformedLlabres}) of our $\lambda$-dependent ansatz takes the form
\begin{align}\label{deformed_varied_ansatz}
    \delta S = \frac{1}{4 \pi G} \int_{\partial M_3} \varepsilon_{ab} f^a \wedge \delta e^b ( \lambda )\,.
\end{align}
We now substitute this ansatz into the flow equation (\ref{boundary_llabres_flow}). More precisely, if the boundary action $S$ satisfies (\ref{boundary_llabres_flow}), then its variation satisfies
\begin{align}\label{boundary_llabres_flow_varied}
    \frac{\partial ( \delta S )}{\partial \lambda} = \frac{1}{(4 \pi G)^2} \int_{\partial M_3} \delta ( f^- \wedge f^+ )\,.
\end{align}
This then implies
\begin{align}\label{boundary_llabres_flow_subbed}
    \int_{\partial M_3} \varepsilon_{ab} f^a \wedge \delta \left( \frac{ \partial e^b ( \lambda ) }{\partial \lambda} \right) = \frac{1}{4 \pi G} \int_{\partial M_3} \varepsilon_{ab} f^a \wedge \delta f^b\,.
\end{align}
We see that (\ref{boundary_llabres_flow_subbed}) will be satisfied if
\begin{align}
    \frac{\partial e^b ( \lambda ) }{\partial \lambda} = \frac{1}{4 \pi G} f^b\,.
\end{align}
Since $f^b$ is independent of $\lambda$ by assumption, this equation can be trivially integrated to find
\begin{align}\label{eia_soln_first_way}
    e_i^a ( \lambda ) = e_i^a ( 0 ) + \frac{\lambda}{4 \pi G} f_i^a\,,
\end{align}
and $f_i^a ( \lambda ) = f_i^a ( 0 )$. One can show that if the spin connection $\omega$ vanishes in the seed theory (as we will typically assume), then $\omega ( \lambda ) = 0$ along the flow. We have characterized the full solution to the flow equation.

\emph{Remarks on deformed boundary conditions}

We now pause to make several comments on this interpretation. We see that the effect of a boundary $T\overline{T}$ deformation is to rotate our undeformed source $e_i^a$ into a new source $e_i^a ( \lambda )$, which depends linearly on the corresponding undeformed expectation value. Since $e_i^a$ determines the boundary metric, this means that the deformed theory sees an effective stress-tensor-dependent metric. This is reminiscent of the result of $T\overline{T}$-deforming a two-dimensional field theory defined on a cylinder of radius $R$. As we will review around (\ref{2d_tt_burgers_implicit}), in the zero-momentum sector, this deformation has the interpretation of placing the theory on a cylinder with an effective energy-dependent radius $\widetilde{R} (R, E_n)$.

Next, we note that although the sources $e_i^a$ have been modified. The variational principle defining our theory has not changed when expressed in terms of the new sources. The deformed boundary variation solving our flow is written as (\ref{deformed_varied_ansatz}), which vanishes if the sources $e_i^a ( \lambda )$ are held fixed. Therefore, the theory described by these $T\overline{T}$-deformed boundary conditions still corresponds to a variational principle where the metric is held fixed at the boundary, but the dual expectation value is free to fluctuate. All that has changed is the expression for this fixed metric in terms of the undeformed metric and stress tensor.

A third remark concerns the trace flow equation for the $T\overline{T}$ deformation. Because there is no dimensionful scale in a CFT, if one solves a $T\overline{T}$ flow beginning from a CFT seed then the resulting theory has a single effective energy scale $ \Lambda = \frac{1}{\sqrt{\lambda}}$ set by the length dimension $2$ parameter $\lambda$. By noting that the derivative of the action with respect to this single scale $\lambda$ is controlled by the trace of the stress tensor,
\begin{align}
    \Lambda \frac{d}{d \Lambda} S = \int d^2 x \, T^\mu{}_\mu \,.
\end{align}
while on the other hand, the derivative of the action is related to the $T\overline{T}$ operator by the definition of the flow \eqref{eq:TTb definition}
\begin{align}
    \Lambda \frac{d}{d \Lambda} S &= \frac{1}{\sqrt{\lambda}} \frac{d}{d \left( \frac{1}{\sqrt{\lambda}} \right)} S \nonumber \\
    &= - 2 \lambda \int d^2 x \, \left( T^{\mu \nu} T_{\mu \nu} - \left( T^\mu{}_\mu \right)^2 \right) \,,
\end{align}
one finds the relation
\begin{align}\label{trace_flow_eqn}
T^\mu{}_\mu ( \lambda ) = - 2 \lambda T\overline{T} ( \lambda ). 
\end{align}

Since the modified boundary conditions (\ref{eia_soln_first_way}) correspond to a $T\overline{T}$-deformation of a CFT, it is an instructive check to verify explicitly that the trace flow equation (\ref{trace_flow_eqn}) holds. Indeed, the trace of the deformed stress tensor with respect to the deformed metric is
\begin{align}\label{T_trace_flow}
   T^i{}_i &= \eta^{ij} e_i^a T_{j a} \nonumber \\
    &= \frac{1}{4 \pi G} \left( e_i^a ( 0 ) + \frac{\lambda}{4 \pi G} f_i^a \right) \left( \varepsilon_{ab} \varepsilon^{ij} f_j^b \right) \nonumber \\
    &= \frac{\lambda}{(4 \pi G)^2} \varepsilon_{ab} \varepsilon^{jk} f_i^a f_j^b \,, 
\end{align}
where in the last step, we have used that the undeformed stress tensor is traceless by assumption. On the other hand, at finite $\lambda$ the combination $T\overline{T}$ is given by (\ref{eq: 3D TT}):
\begin{align}\label{TT_trace_flow}
    T\overline{T} &= - 2 \varepsilon^{ab} \varepsilon_{ij} T^i{}_a T^j{}_b \nonumber \\
    &=- \frac{2}{(4 \pi G)^2} \varepsilon^{ab} \varepsilon_{ij} \left( \varepsilon_{ac} \varepsilon^{ik} f_k^c \right) \left( \varepsilon_{bd} \varepsilon^{jn} f_n^d \right) \nonumber \\
    &= -\frac{2}{(4 \pi G)^2} \varepsilon_{ab} \varepsilon^{jk} f_i^a f^b_j \,,
\end{align}
where we have repeatedly used the $2d$ contracted epsilon identity $\varepsilon^{in}  \varepsilon_{ij} = \delta^n{}_j$. Comparing (\ref{T_trace_flow}) to (\ref{TT_trace_flow}), we see that the trace flow equation $T^\mu{}_\mu ( \lambda ) = - 2 \lambda T\overline{T} ( \lambda ) $ holds as expected.

We make a fourth and final comment, which is a trivial observation in this case but could conceivably be relevant for generalizations of the procedure described here. We emphasized around (\ref{sum_three_terms}) that the undeformed action $S ( \lambda = 0 ) = S_{\text{EH}} + S_{\text{bdry}}$ includes a boundary term which was added by hand to give a particular variational principle. Since the process of deforming the action by $T\overline{T}$ and the process of adding the boundary term $S_{\text{bdry}}$ are two distinct steps, there are na\"ively two ways to proceed:

\begin{enumerate}
    \item \label{first_way} First add the boundary term $S_{\text{bdry}}$ to get the total boundary action $S$. Then solve the flow equation (\ref{boundary_llabres_flow}) for this combined action.
    
    \item \label{second_way} First solve the flow equation $\frac{\partial S_{\text{EH}}}{\partial \lambda} \Big\vert_{\text{bdry}} = \frac{1}{(4 \pi G)^2} \int_{\partial M_3} f^- \wedge f^+$ which only deforms the first contribution to the action. Solve this by identifying new sources $e_i^a ( \lambda )$. After doing this, add a new boundary term $S_{\text{bdry}} ( \lambda )$ by hand to restore the desired variational principle.
\end{enumerate}
In the discussion above, we performed the deformation described by \ref{first_way}. However, it is straightforward to see that procedure \ref{second_way} gives the same result precisely because the dual expectation values $f_i^a$ do not flow according to our ansatz. To show this, we recall from (\ref{on_shell_delta_S_EH}) that
\begin{align}\label{illustrating_two_ways}
    \delta S_{\text{EH}} \Big\vert_{\text{on-shell}} = - \frac{1}{8\pi G} \int_{\partial M_3} \varepsilon_{ab} \left( e^a \wedge \delta f^b - f^a \wedge \delta e^b \right) \,.
\end{align}
Suppose that we had allowed both $e_i^a$ and $f_i^a$ to acquire $\lambda$ dependence along our flow. Then the derivative of this boundary variation would be
\begin{equation}
\begin{aligned}
    \frac{\partial ( \delta S_{\text{EH}} ) }{\partial \lambda}  = - \frac{1}{8\pi G} \int_{\partial M_3} \varepsilon_{ab} &\Bigg( \frac{\partial e^a ( \lambda )}{\partial \lambda} \wedge \delta f^b ( \lambda ) + e^a ( \lambda ) \wedge \frac{\partial ( \delta f^b ( \lambda ) ) }{\partial \lambda} \\
    & - \frac{\partial  f^a ( \lambda )}{\partial \lambda} \wedge \delta e^b ( \lambda )  - f^a ( \lambda ) \wedge \frac{\partial ( \delta e^b ( \lambda )  ) }{\partial \lambda}  \Bigg) \,.
\end{aligned}
\end{equation}
In order to satisfy the flow equation $\frac{\partial ( \delta S_{\text{EH}} )}{\partial \lambda} \Big\vert_{\text{on-shell}} = \frac{1}{(4 \pi G)^2} \int_{\partial M_3} \delta ( f^- \wedge f^+ )$, whose right side is again
\begin{align}
    \frac{1}{(4 \pi G)^2} \int_{\partial M_3} \delta ( f^- \wedge f^+ ) = \frac{1}{(4 \pi G)^2} \int_{\partial M_3} \varepsilon_{ab} f^a \wedge \delta f^b \,,
\end{align}
we must have
\begin{equation}
\begin{aligned}
\label{both_lambda_dep_equation}
    \hspace{-7pt} &\frac{\partial e^a ( \lambda )}{\partial \lambda} \wedge \delta f^b ( \lambda ) + e^a ( \lambda ) \wedge \frac{\partial ( \delta f^b ( \lambda ) ) }{\partial \lambda}  - \frac{\partial  f^a ( \lambda )}{\partial \lambda} \wedge \delta e^b ( \lambda ) - f^a ( \lambda ) \wedge \frac{\partial ( \delta e^b ( \lambda )  ) }{\partial \lambda} \\&= - \frac{1}{2 \pi G} f^a \wedge \delta f^b \,.
\end{aligned}
\end{equation}
The left side involves both $\delta e^a$ and $\delta f^a$, whereas the right side only involves $\delta f^a$. If these two variations are both independent, nonzero, and $\lambda$-dependent, it seems that we cannot have a solution. However, if we assume that $f^a$ and therefore $\delta f^a$ are independent of $\lambda$ as we did before, in addition to imposing that $\delta e^a ( \lambda ) = 0$ according to our choice of deformed variational principle, the equation \eqref{both_lambda_dep_equation} reduces to
\begin{align}
    \frac{\partial e^a ( \lambda )}{\partial \lambda} \wedge \delta f^b ( \lambda ) = - \frac{1}{2 \pi G} f^a \wedge \delta f^b \,.
\end{align}
The solution to this simple flow is
\begin{align}
    e^a_i ( \lambda ) = e_i^a ( 0 ) - \frac{\lambda}{2 \pi G} f_i^a \,.
\end{align}
Up to an overall rescaling of $\lambda$ by a factor of $- \frac{1}{2}$, this is the same solution as (\ref{eia_soln_first_way}). This completes the first step of the alternate deformation procedure described in \ref{second_way}, but we must still add a new boundary term so that the combined boundary action is consistent with the variational principle $\delta e^a = 0$ that we have assumed. Our $\lambda$-dependent deformed boundary variation before adding this boundary term is
\begin{align}
    \delta S_{\text{EH}} ( \lambda ) \Big\vert_{\text{on-shell}} = - \frac{1}{8\pi G} \int_{\partial M_3} \varepsilon_{ab} \left( e^a ( \lambda ) \wedge \delta f^b - f^a \wedge \delta e^b ( \lambda )  \right) \,.
\end{align}
But because this has the same form as the variation (\ref{on_shell_delta_S_EH}) which we saw in the undeformed case, we may repeat the same procedure and add the term $S_{\text{bdry}}$ as defined in (\ref{boundary_action_As}), except replacing $e_i^a$ with $e_i^a ( \lambda )$ everywhere that it appears in the expansions of $A^a$ and $A^b$. The result is, again,
\begin{equation}
\label{eq:undeformedLlabresLater}
    \delta S_{\text{EH}} ( \lambda )  \Big\vert_{\text{on-shell}} + \delta S_{\text{bdry}} ( \lambda ) = \frac{1}{4 \pi G} \int_{\partial M_3} \varepsilon_{ab} f^a \wedge \delta e^b ( \lambda ) \,, 
\end{equation}
exactly as we found before.

The upshot of this simple calculation is that the two processes described above -- first adding a boundary term and then deforming, or first deforming and then adding a boundary term -- commute in the calculation we consider here. However, in another setting where both the sources and expectation values become $\lambda$-dependent, performing the second deformation procedure \ref{second_way} would produce a flow equation analogous to (\ref{both_lambda_dep_equation}), which is not equivalent to the flow of procedure \ref{first_way}. In such cases, one must choose a prescription to define the deformation.

\section{$T\overline{T}$ deformations in $2d$ JT gravity}
\label{sec:JT grav Review}

We now review features of JT gravity and its BF gauge theory description, which are relevant to later sections when we study the $T\overline{T}$ deformation in BF theory. As in section \ref{sec:Various Presentations}, none of the material in this discussion is new. For instance, the interpretation of a $T\overline{T}$-like deformation in the boundary dual to $2d$ JT gravity was considered in \cite{Gross:2019ach,Gross:2019uxi}, and we follow their discussion closely in section \ref{subsec:2d_jt_TT_interpretation_review}. We include a reminder of these results here to facilitate comparison with the new results of section \ref{sec: Deformed BF BCs}, where we present an analogous interpretation of the $T\overline{T}$ deformation in BF variables.

\subsection{JT gravity as a BF gauge theory}\label{sec:jt_as_bf}

In the introduction of this chapter, we mentioned that one salient feature of $3d$ gravity motivating the present work is that it can be dimensionally reduced on a circle to yield JT gravity as described by the action (\ref{eq:JT gravity Action}). This subsection's goal is to recall the standard statement that this $2d$ dilaton gravity theory can be, equivalently, written in gauge theory variables as a BF theory. Our treatment will follow \cite{Iliesiu:2019xuh}.

One way of motivating this reformulation is to note that $3d$ gravity is equivalent to a Chern-Simons theory, as we reviewed in section \ref{sec:Various Presentations}, and that the dimensional reduction of this $3d$ Chern-Simons theory is a BF gauge theory. Indeed, we saw this reduction explicitly around (\ref{eq:dimredCStoBF}). These observations are summarized by the sub-diagram formed by the second and third columns of (\ref{gravity_diagram}):

\begin{align}
\label{gravity_diagram_later}
\includegraphics[width=.5\linewidth]{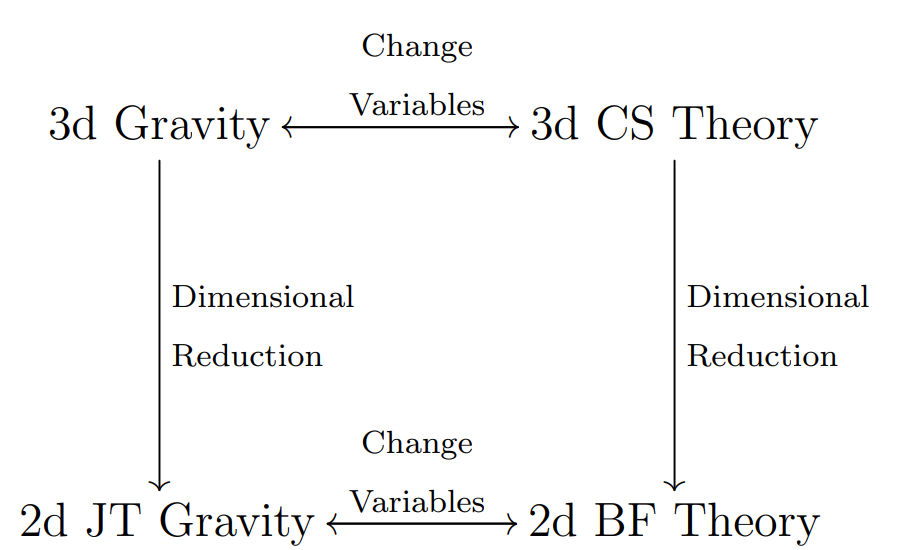}
\end{align}

We have reviewed all of the arrows in (\ref{gravity_diagram_later}) except for the change of variables linking the two theories in the bottom row. Although such a change of variables must exist by the consistency of the diagram, it is instructive to spell out the map explicitly.

Recall that the BF theory in Euclidean signature is described by the action
\begin{equation}
\label{eq:BF Action}
    I_{\text{BF}} = -i \int_{M_2} \operatorname{Tr} \left( \phi F \right)\,,
\end{equation}
where $\phi$ is a scalar field and $F$ is the field strength of the gauge field $A_\mu$. At the moment, we will only be concerned with the bulk equations of motion and will not include any additional boundary term like the one that appeared in (\ref{eq:dimredCStoBF}). The equations of motion arising from \eqref{eq:BF Action} are
\begin{equation}
    \begin{aligned}
    \label{eq: BF eom}
    \phi&: \quad F = 0\,, \\
    A_\mu &: \quad D_\mu \phi = \partial_\mu \phi - [A_\mu, \phi] = 0 \,.
    \end{aligned}
\end{equation}
On the other hand, beginning from the action \eqref{eq:JT gravity Action} of JT gravity and setting $\Lambda = -1$, one finds the equations of motion
\begin{equation}\label{eq:JT eom}
    \begin{aligned}
  \Phi&: \quad  R = - 2\,, \\
    g_{\mu \nu}  &: \quad   \nabla_\mu \nabla_\nu \Phi = g_{\mu \nu} \Phi\,.
    \end{aligned}
\end{equation}
Next, we will argue that the JT equations of motion in (\ref{eq:JT eom}) are equivalent to the BF equations of motion in (\ref{eq: BF eom}). To accomplish this, we first expand the BF fields in terms of generators:
\begin{equation}
\begin{aligned}
\label{eq:ExpansionsForFields}
    A(x)= e^2(x) P_2 + e^1(x) P_1 + \omega (x) P_0\,, \quad \phi(x)=\phi^{1}(x) P_{1} + \phi^{2}(x) P_{2} +\phi^{0}(x) P_{0}\,,
    \end{aligned}
\end{equation}
where
\begin{equation}
\begin{aligned}
P_0 =\left(\begin{array}{cc}
0 & \frac{1}{2} \\
-\frac{1}{2} & 0
\end{array}\right)\,, \quad P_1  =\left(\begin{array}{cc}
0 & \frac{1}{2} \\ 
\frac{1}{2} & 0
\end{array}\right)\,, \quad 
P_2 =\left(\begin{array}{cc}
\frac{1}{2} & 0 \\
0 & -\frac{1}{2}
\end{array}\right)\,.
\end{aligned}
\end{equation}

Written in differential form notation, the equation of motion for $A_\mu$ in (\ref{eq: BF eom}) becomes $d\phi - A \wedge \phi =0$.  The exterior derivative of the scalar $\phi$ is 
\begin{equation}
    d\phi =d\phi^0(x) P_0 + d\phi^1(x) P_1 +d\phi^2(x) P_2\,.
\end{equation}
Meanwhile, a short calculation gives
\begin{equation}
\begin{aligned}
    A \wedge \phi =  \left( e^2 \wedge \phi^1 - e^1 \wedge \phi^2 \right) P_0  + \left( e^2 \wedge \phi^0 - \omega \wedge \phi^2 \right) P_1 + \left( \omega \wedge \phi^1 -  e^1 \wedge \phi_0 \right) P_2\,.
\end{aligned}
\end{equation}
Putting everything together, we find
\begin{equation}
    \begin{aligned}
    \label{eq:sfdo}
    d\phi^0(x) &= e^2 (x) \wedge \phi^1(x) -e^1(x) \wedge \phi^2(x)\,, \\
    d\phi^1(x) &= e^2(x) \wedge \phi^0(x) - \omega(x) \wedge \phi^2(x)\,, \\
    d\phi^2(x) &= \omega(x) \wedge \phi^1(x) - e^1(x) \wedge \phi^0(x)\,.
    \end{aligned}
\end{equation}
We now act with the covariant derivative $\nabla_\mu$ on the equation for $d \phi^0$ in (\ref{eq:sfdo}). At the risk of being pedantic, we pause to clarify one point of possible confusion. When acting on a generalized tensor with both curved (spacetime) indices and flat (tangent space) indices, the action of the covariant derivative $\nabla_\mu$ involves Christoffel symbol terms associated with the curved indices and spin connection terms associated with the flat indices. For instance, on the vielbein $e_\nu^a$ with one curved and one flat index, one has
\begin{align}\label{cov_deriv_vielbein}
    \nabla_\mu e_\nu^a = \partial_\mu e_\nu^a + \omega_\mu{}^a{}_b  e_\nu^b - \Gamma^\sigma{}_{\nu \mu} e_\sigma^a \,.
\end{align}
Since the covariant derivative annihilates the vielbein by the zero-torsion constraint $\tau^a = de^a + \omega^a_b \wedge e^b = 0$, the combination (\ref{cov_deriv_vielbein}) vanishes.

However, the equations (\ref{eq:sfdo}) are covariant with respect to their single curved index but not with respect to the implicit flat index on the vielbeins. It is easiest to see this by writing the equations in components. For instance, the $\phi^0$ equation is
\begin{align}
    \partial_\mu \phi^0 = \phi^1 e^2_\mu - \phi^2 e^1_\mu \,.
\end{align}
Although this equation has a free $\mu$ index, there is no free $a$ index in the $e_\mu^a$ factors. Indeed, this equation could never have been covariant with respect to such a tangent space index since the quantity $\partial_\mu \phi^0$ on the left has no flat indices. Therefore, when we act with the covariant derivative, there will be no spin connection terms introduced in the derivatives of vielbein factors. One has
\begin{align}
    \nabla_\mu e^2_\nu = \partial_\mu e_\nu^2  - \Gamma^{\sigma}{}_{\nu \mu} e_\sigma^2 \,,
\end{align}
and likewise for $\nabla_\nu e^1_\mu$. However $\nabla_\mu e_\nu^a = 0$, (\ref{cov_deriv_vielbein}) implies
\begin{align}
    \partial_\mu e_\nu^2  - \Gamma^\sigma{}_{\nu \mu} e_\sigma^2 = - \omega_\mu{}^2{}_b e_\nu^b \,, 
\end{align}
and again a similar equation for $\nabla_\nu e^1_\mu$. Using $\omega^1{}_2= - \omega^2{}_1= \omega$, we find
\begin{align}
    \nabla_\mu e_\nu^1 &= - \omega_\mu{}^1{}_b e_\nu^b = - \omega_\mu e_\nu^2 \,, \nonumber \\
    \nabla_\mu e_\nu^2 &= - \omega_\mu{}^2{}_b  e_\nu^b = \omega_\mu e^1_\nu \,.
\end{align}
Now we are prepared to act with the covariant derivative on the $\phi^0$ equation of motion. On the left, the result is a two-tensor with components $\nabla_\mu \nabla_\nu \phi^0$. One finds
\begin{align}\label{phi_eom_intermediate}
    \nabla_\mu \nabla_\nu \phi^0 &= \left( \partial_\mu \phi^1 \right) e^2_\nu - \left( \partial_\mu \phi^2 \right) e^1_\nu + \phi^1 \left( \nabla_\mu e^2_\nu \right) - \phi^2 \left( \nabla_\mu e^1_\nu \right) \, \nonumber \\
    &= \left( \partial_\mu \phi^1 \right) e^2_\nu - \left( \partial_\mu \phi^2 \right) e^1_\nu + \phi^1 \omega_\mu e^1_\nu + \phi^2 \omega_\mu e_\nu^2
\end{align}
On the other hand, writing the second and third equations of (\ref{eq:sfdo}) in components gives $\partial_\mu \phi^1 = \phi^0 e^2_\mu - \phi^2 \omega_\mu$ and $\partial_\mu \phi^2 = \phi^1 \omega_\mu - \phi^0 e^1_\mu$. Substituting these into (\ref{phi_eom_intermediate}) gives
\begin{align}\label{phi_eom_intermediate_two}
    \nabla_\mu \nabla_\nu \phi^0 &= \left( \phi^0 e^2_\mu - \phi^2 \omega_\mu \right) e^2_\nu - \left( \phi^1 \omega_\mu - \phi^0 e^1_\mu \right) e^1_\nu + \phi^1 \omega_\mu e^1_\nu + \phi^2 \omega_\mu e_\nu^2 \nonumber \\
    &= \left( e^1_\mu e^1_\nu + e^2_\mu e^2_\nu \right) \phi^0 \,.
\end{align}
If we identify the metric as $g_{\mu \nu} = e^1_\mu e^1_\nu + e^2_\mu e^2_\nu$ and assume that the JT dilaton $\Phi$ is proportional to the BF field $\phi^0$, then this is the metric equation of motion in (\ref{eq:JT eom}):
\begin{align}
    \nabla_\mu \nabla_\nu \Phi = g_{\mu \nu} \Phi \,.
\end{align}
We, therefore, have demonstrated that the JT gravity equations of motion (\ref{eq:JT eom}) are recovered from the BF equations of motion in (\ref{eq: BF eom}) after making the change of variables
\begin{align}
    \phi^0 = \frac{i}{4\pi G} \Phi \,, \qquad g_{\mu \nu} = e^1_\mu e^1_\nu + e^2_\mu e^2_\nu \,.
\end{align}
Here, the choice of the proportionality factor $\frac{i}{4 \pi G}$ between $\phi^0$ and $\Phi$ is required by our normalizations for the BF and JT actions in (\ref{eq:BF Action}) and (\ref{eq:JT gravity Action}), respectively. Under this identification, we see that the expansion coefficients $e^a$ appearing in the BF gauge field $A_\mu$ are interpreted as the frame fields in the JT gravity theory, whereas the field $\omega$ defines the spin connection, which satisfies $d \omega = \frac{R}{2} e^1 \wedge e^2$ for a $2d$ manifold. In this correspondence, the $\phi^1, \phi^2$ equations of motion are mapped onto the torsionless conditions $\tau^a = de^a + \omega^a_b \wedge e^b = 0$. This completes our review of the final arrow on the bottom row of (\ref{gravity_diagram_later}) linking JT gravity with BF gauge theory.

Next, we explain the boundary conditions and the choice of boundary term for the BF gauge theory, which recovers the Schwarzian action. Variation of the BF action on-shell yields the boundary action 
\begin{equation}
\label{eq:on-shell boundary}
    \delta I_{\text{BF}} = -i \int_{\partial M_2} \, d \tau \, \operatorname{Tr} \left( \phi \, \delta A_\tau \right)\,,
\end{equation}
with $\tau$ parametrizing the one-dimensional boundary $\partial M_2$. 

Thus, the variation (\ref{eq:on-shell boundary}) of the BF action vanishes if $A_\tau$ is held fixed on $\partial M_2$. In fact, from JT gravity's first-order formulation, the spin connection and frame are already fixed, so no boundary term is required to have a well-defined variational principle. Unfortunately, this means the BF theory cannot be holographically dual to the Schwarzian because the theory is topologically trivial. In particular, the observables of the theory would depend on the holonomy around the boundary rather than depending on the local value of $A_\tau$. To recover the Schwarzian dynamics, one includes a string defect $I_{\text{string}}$ around a loop $L \subset M_2$, which yields the modified action
\begin{equation}
\label{eq:string defect}
    I = -i \int_{M_2} \operatorname{Tr} \left( \phi F \right) - \oint^\beta_0 du \, V(\phi)\,, \quad V(\phi) = \frac{\nu}{4} \operatorname{Tr} \phi^2 \,.
\end{equation}
The second term in \eqref{eq:string defect} is the string defect with coupling $\nu$, and $u$ is the proper length parametrization of the loop with circumference $\beta$. This form of $V(\phi)$ is consistent with the boundary term in (\ref{eq:dimredCStoBF}), which we expect from the dimensional reduction of Chern-Simons and, as we will see, correctly recovers the Schwarzian action. 

The overall action \eqref{eq:string defect} preserves the defect diffeomorphisms and the degrees of freedom from the string defect are realized by the Schwarzian theory as \cite{Iliesiu:2019xuh} showed by evaluating the action \eqref{eq:string defect} using the solution to the equation of motion \eqref{eq: BF eom} for $\phi(u)$ along $L$. To see the derivation more explicitly, we parametrize the boundary fields $\phi$ and $A_\tau$ by\footnote{Note that a different representation of the generators when solving the equations of motion for the field $\phi$ at the loop.} 
  \begin{equation}\label{BF_sl2_expansions}
      A_\tau = \omega \ell_0 + e_+ \ell_+ + e_- \ell_-\,, \quad \phi = \phi_+ \ell_+ + \phi_- \ell_- + \phi_0 \ell_0\,,
  \end{equation}
  where
  \begin{equation}\label{e_P_ell_conventions}
  \begin{aligned}
      &\ell_0 =iP_0, \quad \ell_+ = -iP_1 - P_2\,, \quad \ell_- = -iP_1 + P_2\,, \\
      & \omega = -i\omega_\tau \bigg|_{\partial M_2}\,, \quad e_+ = \frac{ie^1_\tau - e^2_\tau}{2} \bigg|_{\partial M_2}\,, \quad e_- = \frac{ie^1_\tau +e^2_\tau}{2} \bigg|_{\partial M_2}\,.
\end{aligned}
  \end{equation}
  We compute the commutator 
  \begin{equation}
   \left[A_\tau, \phi \right] = \left(e_+ \phi_0 - \omega \phi_+ \right) \ell_+ + \left( \omega \phi_- -  e_- \phi_0 \right) \ell_- + 2 \left( e_+ \phi_- - e_- \phi_+  \right) \ell_0 
  \end{equation}
  to write the complete set of equations of motion $D_\tau \phi = 0$ at the loop
  \begin{equation}
      \begin{aligned}
      \label{eq:Loop eom}
      \ell_0 &: \quad \partial_\tau \phi_0 = 2 \left( e_+ \phi_- - e_- \phi_+ \right)\,, \\
      \ell_{-}&: \quad \partial_\tau \phi_- = \omega \phi_- -  e_- \phi_0\,, \\
      \ell_+&: \quad \partial_\tau \phi_+ = e_+ \phi_0 - \omega \phi_+\,.
      \end{aligned}
  \end{equation}
  To solve the equations at the loop \eqref{eq:Loop eom}, we perform the same change of variables as \cite{Iliesiu:2019xuh}
  \begin{equation}
      \phi_- (\tau) =  \frac{2e_-}{\partial_\tau u(\tau)} \implies \phi_- (u) = 2 e_- \tau'\,, 
  \end{equation}
  where
  \begin{equation}
    \partial_\tau  \phi_i = \left( \partial_\tau u \right) \left( \partial_u \phi_i \right) = \frac{\partial_u \phi_i}{\tau'}\,.
  \end{equation}
Substituting the above into the equation of motion for the $\ell_-$ component, we find
  \begin{equation}
      \phi_0 = -\frac{\partial_\tau \phi_- - \omega \phi_-}{e_-} = - \frac{2\tau''}{\tau'} + 2 \omega \tau'\,. 
  \end{equation}
  Then, solving for the $\ell_0$ component, one uses $\phi_-$ and $\phi_0$ to find 
  \begin{equation}
  \begin{aligned}
      \phi_+ &= \frac{1}{e_-} \left( - \frac{1}{2} \partial_\tau \phi_0 + e_+ \phi_- \right) \\&= 2 \left( e_+ \tau' + \frac{\tau'''}{2 e_- \tau^{\prime 2}} - \frac{\omega \tau''}{2 e_- \tau'} - \frac{\tau^{\prime \prime 2}}{2 e_- \tau^{\prime 3}} \right)\,.
\end{aligned}
  \end{equation}
We have found all the components for the field $\phi(u)$:
\begin{equation}
    \nu \phi(u) = 2 e_- \tau' \ell_- + 2 \left( \omega \tau' - \frac{\tau''}{\tau'} \right) \ell_0 + 2 \left( e_+ \tau' + \frac{\tau'''}{2 e_- \tau^{\prime 2}} - \frac{\omega \tau''}{2e_- \tau'} - \frac{\tau^{\prime \prime 2}}{2e_- \tau^{\prime 3}} \right) \ell_+\,.
\end{equation}
Here $\tau(u)$ is further constrained by the $\ell_+$ component of the equation $D_u \phi = 0$, which gives
\begin{equation}
    4 \left( \operatorname{det} A_{\tau} \right) \tau^{\prime 4} \tau^{\prime \prime}+3\tau^{\prime \prime 3}-4 \tau^{\prime} \tau^{\prime \prime} \tau^{\prime \prime \prime}+ \tau^{\prime 2} \tau^{\prime \prime \prime \prime} = 0\,,
\end{equation}
where $\tau (u)$ is monotonic so $\tau'(u) \neq 0$ and $\det A_\tau =  e_- e_+ - \frac{\omega^2}{4}$.

Now we are ready to evaluate the string defect action by computing $\operatorname{Tr} \phi^2$. This computation is straightforward as
\begin{equation}
    \phi^2 = \frac{1}{\nu^2} \left(\begin{array}{cc}
-4 e_{-} e_{+} \tau^{\prime 2}+\omega^{2} \tau^{\prime 2}+\frac{3 \tau^{\prime \prime 2}}{\tau^{\prime 2}}-\frac{2 \tau^{\prime \prime  \prime}}{\tau^{\prime}} & 0 \\
0 & -4 e_{-} e_{+} \tau^{\prime 2}+\omega^{2} \tau^{\prime 2}+\frac{3 \tau^{\prime \prime 2}}{\tau^{\prime 2}}-\frac{2 \tau^{\prime \prime \prime}}{\tau^{\prime}}
\end{array}\right)
\end{equation}
and
\begin{equation}
    \begin{aligned}
    V(\phi) &= \frac{\nu}{4} \operatorname{Tr} \phi^2 \\&= \frac{1}{\nu} \left( \{\tau(u), u\} + 2 \tau'(u)^2 \det A_\tau \right) \\&= \frac{1}{\nu} \left\{\tan \left( \sqrt{\left( \det A_\tau \right) \tau(u) } \right), u\right\}\,.
    \end{aligned}
\end{equation}
As expected, we have recovered the Schwarzian action\footnote{One can show that this derivation of the Schwarzian theory holds for any $\Lambda$ by using the more general parameterization $A_\tau = \omega \ell_0 + \sqrt{\Lambda} e_+ \ell_+ + \sqrt{\Lambda} e_- \ell_-$. Equivalently, this corresponds to replacing the determinant in \eqref{eq:derivedSchwarzianfromBF} by $\det A_\tau =\Lambda e_- e_+ - \frac{\omega^2}{4} $.} from including the string defect in the BF action \eqref{eq:string defect}, which gives
\begin{equation}
\label{eq:derivedSchwarzianfromBF}
    I = - \frac{1}{\nu} \int^\beta_0 du \left\{\tan \left( \sqrt{\left( \det A_\tau \right) \tau(u) } \right)\,, u\right\}\,.
\end{equation}

\subsection{Interpretation of $T\overline{T}$ deformation}\label{subsec:2d_jt_TT_interpretation_review}

Before we begin with the $T\overline{T}$ deformation in JT gravity, we first recall how a general class of related deformations is defined and their physical meaning in the $\mathrm{AdS}/\mathrm{CFT}$ correspondence. Following \cite{Gross:2019uxi}, we deform a seed action $I(0)$ via a generic operator $M_\lambda$ as
\begin{equation}
\label{eq:generalclass0}
    I(\lambda) = I(0) + \int d\tau \sqrt{\gamma} \,  M_\lambda (T_{\tau \tau},\gamma^{\tau \tau})\,,
\end{equation}
where the variational principle in the undeformed theory (where $M_0 = 0$)  is defined by
\begin{equation}
    \delta I(0) = \frac{1}{2} \int d\tau \, \sqrt{\gamma} \, T_{\tau \tau} \delta \gamma^{\tau \tau}\,.
\end{equation}
With the deformation \eqref{eq:generalclass0}, one finds the following variation:
\begin{equation}
\label{eq:generalclassBC1}
    \delta I(\lambda) = \delta I(0) + \int \, d\tau \, \Big[ \delta \left( \sqrt{\gamma} \right) M_\lambda + \sqrt{\gamma} \delta M_\lambda \Big]\,.
\end{equation}
Using the facts that
\begin{equation}
    \delta M_\lambda = \frac{\partial M_\lambda}{\partial T_{\tau \tau}} \delta T_{\tau \tau} + \frac{\partial M_\lambda}{\partial \gamma^{\tau \tau}} \delta \gamma^{\tau \tau}
\end{equation}
and 
\begin{equation}
    \delta \left( \sqrt{\gamma} \right) = - \frac{\sqrt{\gamma}}{2\gamma^{\tau \tau}} \delta \gamma^{\tau \tau}
\end{equation}
we find \eqref{eq:generalclassBC1} is written as
\begin{equation}
\label{eq:generalclass}
    \delta I(\lambda) = \frac{1}{2} \int d\tau \sqrt{\gamma} \left[ \left( T_{\tau \tau} - \frac{M_\lambda}{\gamma^{\tau \tau}} + 2 \frac{\partial M_\lambda}{\partial \gamma^{\tau \tau}} \right) \delta \gamma^{\tau \tau} + 2 \frac{\partial M_\lambda}{\partial T_{\tau \tau}} \delta T_{\tau \tau} \right]\,.
\end{equation}
We wish to identify sources and expectation values by demanding that we can rewrite \eqref{eq:generalclass} in terms of $\lambda$-dependent quantities $\widetilde{T}_{\tau \tau}$, $\widetilde{\gamma}_{\tau \tau}$ as
\begin{equation}
\begin{aligned}
\label{eq:generalclass1}
    \delta I(\lambda) &= \frac{1}{2} \int d\tau \frac{\widetilde{T}_{\tau \tau}}{\sqrt{\widetilde{\gamma}^{\tau \tau}}} \delta \widetilde{\gamma}^{\tau \tau} \\&= \frac{1}{2} \int d\tau \frac{\widetilde{T}_{\tau \tau}}{\sqrt{\widetilde{\gamma}^{\tau \tau} }} \left(  \frac{\partial \widetilde{\gamma}^{\tau \tau}}{\partial T_{\tau \tau}} \delta T_{\tau \tau} + \frac{\partial \widetilde{\gamma}^{\tau \tau}}{\partial \gamma^{\tau \tau}} \delta \gamma^{\tau \tau} \right) \,.
\end{aligned}
\end{equation}
Here the operator $\widetilde{T}_{\tau \tau}$ is sourced by $\widetilde{\gamma}^{\tau \tau}$. In other words, the deformation changes the variational principle from one where $\gamma^{\tau \tau}$ is held fixed to one where $\widetilde{\gamma}^{\tau \tau}$ is fixed.

Comparing \eqref{eq:generalclass} and \eqref{eq:generalclass1}, we find the following coupled PDEs for the deformed boundary stress tensor and metric:
\begin{equation}
    \begin{aligned}
    \label{eq:coupledsystemofpdes}
     \frac{\widetilde{T}_{\tau \tau}}{\sqrt{\widetilde{\gamma}^{\tau \tau}}} \frac{\partial \widetilde{\gamma}^{\tau \tau}}{\partial T_{\tau \tau}} &= 2\sqrt{\gamma} \frac{\partial M_\lambda}{\partial T_{\tau \tau}} \,, \\\frac{\widetilde{T}_{\tau \tau}}{\sqrt{\widetilde{\gamma}^{\tau \tau}}} \frac{\partial \widetilde{\gamma}^{\tau \tau}}{\partial \gamma^{\tau \tau}} &= \sqrt{\gamma} \left( T_{\tau \tau} - \frac{M_\lambda}{\gamma^{\tau \tau}} + 2 \frac{\partial M_\lambda}{\partial \gamma^{\tau \tau}} \right) \,, 
    \end{aligned}
\end{equation}
with the initial conditions $\widetilde{T}_{\tau \tau}(\lambda = 0) = T_{\tau \tau}$ and $\widetilde{\gamma}^{\tau \tau}(\lambda = 0) = \gamma^{\tau \tau}$.

To further illustrate, we focus on a specific class of deformations that only depend on the trace of the stress tensor $T_{\tau \tau} \gamma^{\tau \tau}$. It is convenient to express our ansatz in terms of the dimensionless combination
\begin{align}
    X = \lambda T_{\tau \tau} \gamma^{\tau \tau} \,.
\end{align}
We assume
\begin{equation}
\label{eq:ansatzTrace}
  \widetilde{T}_{\tau \tau} = T_{\tau \tau} \xi \left( X \right)\,, \quad \widetilde{\gamma}^{\tau \tau} = \gamma^{\tau \tau} \chi \left( X \right)\,,
\end{equation}
where $\xi(0) = \chi(0) = 1$ so that we recover the undeformed stress tensor and metric as $\lambda \to 0$. On the other hand, by dimensional analysis, we can write the function $M_\lambda ( \lambda, T_{\tau \tau}, \gamma^{\tau \tau} )$ in the form
\begin{align}\label{Mlambda_dimensional}
    M_\lambda = \frac{1}{\lambda} m_\lambda ( X ) \,.
\end{align}
By substituting (\ref{eq:ansatzTrace})-(\ref{Mlambda_dimensional}) into the system of coupled PDEs \eqref{eq:coupledsystemofpdes}, we find the pair of equations
\begin{align}\begin{split}
\label{eq:ODE and algebraic}
   \chi' ( X ) &= \frac{2 \sqrt{\chi(X)} m_\lambda' ( X )}{X \xi ( X ) } \,, \\
   \sqrt{\chi(X)} \left( X - m_\lambda'(X) + 2 X m_\lambda'(X) \right) &= X \xi(X) \left( \chi(X) + X \chi'(X) \right) \,.
\end{split}\end{align}
The usual double-trace $T\overline{T}$ deformation is quadratic in stress tensors, so one might be interested in studying a deformation that is proportional to the combination $X^2 = \left( \lambda T_{\tau \tau} \gamma^{\tau \tau} \right)^2$ since this is the only dimensionless and reparameterization-invariant stress tensor bilinear in $(0+1)$-dimensions. This corresponds to a deformation of the form
\begin{align}
\label{eq:TT def}
    M_\lambda &= \frac{1}{\lambda} X^2 \, \nonumber \\
    &=  \lambda \left( T_{\tau \tau} \gamma^{\tau \tau} \right)^2\,.
\end{align}
Using the form \eqref{eq:TT def} of the deformation, the equations (\ref{eq:ODE and algebraic}) become
\begin{align}
    \xi ( X ) = \frac{(3X + 1) \sqrt{\chi(X)}}{\chi(X) + X \chi'(X)} \,, \qquad \chi'(X) = \frac{4 \sqrt{\chi(X)}}{\xi(X)} \,,
\end{align}
which have the solutions
\begin{align}
    \chi ( X ) = \frac{1}{(1 - X)^4} \,, \qquad \xi ( X ) = ( 1 - X )^3 \,.
\end{align}
We have therefore found that, for the form of the deformation $M_\lambda = \lambda ( T_{\tau \tau} \gamma^{\tau \tau} )^2$ motivated by the usual $T\overline{T}$ deformation,\footnote{For a multi-trace deformation $  M^{(n)}_{\lambda} = \lambda_n  \left( T_{\tau \tau} \gamma^{\tau \tau} \right)^{2n}$ with coupling $\lambda_n$, one finds via solving \eqref{eq:ODE and algebraic}
\begin{equation}
  \widetilde{T}_{\tau \tau}(\lambda_n) = T_{\tau \tau} \left( 1- \lambda_n \left( T_{\tau \tau} \gamma^{\tau \tau}   \right)^{2n-1} \right)^{\frac{4n-1}{2n-1}}\,, \quad  \widetilde{\gamma}^{\tau \tau}(\lambda_n) =\gamma^{\tau \tau}\left( 1 - \lambda_n \left( T_{\tau \tau} \gamma^{\tau \tau}  \right)^{2n-1} \right)^{\frac{4n}{1-2n}}\,.
\end{equation}
}
the solution is
\begin{equation}
    \widetilde{T}_{\tau \tau}(\lambda) = T_{\tau \tau} \left( 1 - \lambda T_{\tau \tau} \gamma^{\tau \tau} \right)^3\,, \quad \widetilde{\gamma}^{\tau \tau}(\lambda) = \frac{\gamma^{\tau \tau}}{(1-\lambda T_{\tau \tau} \gamma^{\tau \tau})^4}\,.
\end{equation}
However, as mentioned in section \ref{intro33333}, and derived in appendix A of \cite{Gross:2019ach}, despite \eqref{eq:TT def} being proportional to a double-trace operator $T^2$, it is not suitable as a $T\overline{T}$ deformation for JT gravity with a Dirichlet cutoff. The following choice of operator is suitable for the $T\overline{T}$ deformation found by \cite{Gross:2019ach}. In \cite{Gross:2019ach}, the operator which yields the correct deformed energy spectrum is:
\begin{equation}
    M_\lambda = -2\lambda O T_{\tau \tau} \gamma^{\tau \tau}\,,
\end{equation}
where the operator $O$ (i.e. the dilaton momentum) is sourced by the boundary dilaton $\Phi_b$ as
\begin{equation}\label{O_dilaton_operator_def}
    O = \frac{1}{\sqrt{\gamma}} \frac{\delta I}{\delta \Phi_b}\,.
\end{equation}
 The seed theory action is now deformed as
\begin{equation}
   I(\lambda) = I(0) + \int d\tau \sqrt{\gamma} M_\lambda \left(T_{\tau \tau}, \gamma^{\tau \tau}, O, \Phi_b \right)\,,
\end{equation}
where the variation of the undeformed theory is
\begin{equation}\label{JT_boundary_term}
    \delta I(0) = \int d\tau \sqrt{\gamma} \left( \frac{1}{2} T_{\tau \tau} \delta \gamma^{\tau \tau} + O \delta \Phi_b \right) \,.
\end{equation}
To identify the variational principle of the deformed theory, we demand that $\delta I ( \lambda )$ can be written in terms of $\lambda$-dependent sources and expectation values as
\begin{equation}
    \delta I(\lambda) =\int d\tau \sqrt{\gamma} \left( \frac{1}{2} T_{\tau \tau} (\lambda) \delta \gamma^{\tau \tau}(\lambda) + \mathcal{O}(\lambda) \delta \Phi_b(\lambda) \right)\,.
\end{equation}
Following the same procedure as in the previous example with $M_\lambda (T_{\tau \tau}, \gamma^{\tau \tau})$, we find the sources and expectation values transform as
\begin{equation}\label{JT_TT_deformed_solution}
    \begin{aligned}
    \gamma_{\tau \tau}(\lambda) &= \gamma_{\tau \tau}(0) \left( 1 + 2 \lambda O(0) \right)^2\,,\quad
    T_{\tau \tau} (\lambda) = T_{\tau \tau} (0) \left(1+2\lambda O(0) \right)^2\,,\\
   \Phi_b (\lambda) &= \Phi_b(0) -2\lambda T(0)\,, \quad
   \mathcal{O}(\lambda) = \frac{O(0)}{1+2\lambda O(0)} \,,
    \end{aligned}
\end{equation}
which satisfy 
\begin{equation}
    \delta I(\lambda) = \delta I(0) -2 \lambda \delta \left( \int d\tau \sqrt{\gamma} O(0)T_{\tau \tau}(0) \gamma^{\tau \tau}(0)  \right)\,.
\end{equation}
The $\lambda$-dependent sources and expectation values (\ref{JT_TT_deformed_solution}) describe the full solution for the bulk JT gravity fields, which corresponds to performing a $T\overline{T}$-like deformation of the $1d$ boundary theory. As in the analogous deformation of $3d$ gravitational Chern-Simons reviewed in section \ref{subsec:chern_simons_linear_mixing}, we note that the result can be interpreted as a linear mixing of sources and expectation values, although in this case each source becomes a function of the dual expectation value for a \emph{different} operator -- for instance, the metric becomes dependent on the field $O$ which is dual to the dilaton $\Phi_b$.

\section{$T\overline{T}$-deformed boundary conditions in BF theory}\label{sec: Deformed BF BCs}

In the previous section, we have seen that the interpretation of a boundary $T\overline{T}$ deformation in JT gravity is a particular $\lambda$-dependent mixing (\ref{JT_TT_deformed_solution}) of the metric $\gamma_{\tau \tau}$, dilaton $\Phi_b$, and their dual operators. Since JT gravity can also be written in BF variables, there must be an analogous interpretation of the boundary $T\overline{T}$ deformation. The goal of the current section is to make this BF interpretation explicit.

Because BF gauge theory is topological, all of the dynamics of the theory occur at the boundary. As a result, the choice of boundary term -- and the variational principle -- is an important input for defining the theory. We consider $T\overline{T}$-type deformations for two choices of boundary terms: one which gives a variational principle analogous to that of the JT gravity theory and one whose boundary theory is the Schwarzian.

\subsection{Deformation with JT-type boundary term}\label{subsec:deformed_jt_like_bc}

First, we will determine the choice of boundary term in BF theory, which gives a variational principle most analogous to that of the JT gravity theory. We saw in (\ref{JT_boundary_term}) that the on-shell variation of the JT gravity action is
\begin{align}
    \delta I \Big\vert_{\text{on-shell}} = \int_{\partial M_2} \, d \tau \, \sqrt{\gamma} \, \left( \frac{1}{2} T_{\tau \tau} \, \delta \gamma^{\tau \tau} + O \, \delta \Phi_b  \right) \,.
\end{align}
This boundary term vanishes if we fix the value of the (inverse) metric $\gamma^{\tau \tau}$ and the dilaton $\Phi_b$ at the boundary. The operators dual to the metric and dilaton are the boundary stress tensor $T_{\tau \tau}$ and the operator $O$, respectively.

On the other hand, the variation of the BF action $I_{\text{BF}} = - i \int_{M_2}  \operatorname{Tr} ( \phi F )$ was given in (\ref{eq:on-shell boundary}) as
\begin{equation}\label{BF_boundary_later}
    \delta I_{\text{BF}} \Big\vert_{\text{on-shell}} = -i \int_{\partial M_2} d\tau\operatorname{Tr} \left( \phi \delta A_\tau \right)\,.
\end{equation}
We parameterize the BF theory fields in terms of $SL(2, \mathbb{R})$ generators as
\begin{align}
    A_\mu(x) = e_\mu^+ ( x ) L_+ + e_\mu^- ( x ) L_- + \omega_\mu ( x ) L_0 \,, \qquad \phi ( x ) = \phi^+ ( x ) L_+ + \phi^- ( x ) L_- + \phi^0 ( x ) L_0 \,,
\end{align}
Note that we use the notation $L_+$ for $L_{1}$ and $L_-$ for $L_{-1}$. In terms of the functions appearing in this expansion, the boundary term (\ref{BF_boundary_later}) is
\begin{equation}\label{BF_boundary_later_two}
    \delta I_{\text{BF}} \Big\vert_{\text{on-shell}} = - i \int_{\partial M_2} \, d \tau \, \left( \frac{1}{2} \phi^0 \delta \omega_\tau - \phi^+ \delta e^-_\tau - \phi^- \delta e^+_\tau \right) \,.
\end{equation}
The asymptotic values of the expansion coefficients $e_\tau^{\pm}$ in the BF fields are interpreted as the einbein for the one-dimensional boundary theory. These fields are the BF analog of the boundary metric $\gamma_{\tau \tau}$. Likewise, the boundary value of the BF variable $\phi^0$ is proportional to the boundary dilaton $\Phi_b$ in JT variables.

Thus we see that the na\"ive BF action, without any added boundary term, corresponds to a different variational principle than that of JT gravity. For the variation (\ref{BF_boundary_later_two}) to vanish, we must fix the boundary values of $e^{\pm}_\tau$ (which corresponds to fixing the boundary metric) but \textit{not} the boundary value of $\phi^0$; rather the asymptotic value of $\omega_\tau$ is held fixed. In JT gravity language, this corresponds to a variational principle where the value of the dual operator $O$ is held fixed, but the boundary dilaton $\Phi_b$ is free to vary.

We can, of course, modify the BF variational principle by adding an appropriate boundary term. Suppose that we choose the BF action to be
\begin{align}\label{schwarzian_appropriate_boundary}
    I = I_{\text{BF}} + I_{\text{bdry}} \,, \qquad I_{\text{bdry}} = \frac{i}{2} \int_{\partial M_2}  d^2 x \, \sqrt{g} \, n_\mu \partial^\mu \left( \phi^0 \omega_\tau \right) \,,
\end{align}
where $n_\mu$ is a unit normal vector in the radial direction. The corresponding contribution to the boundary variation is
\begin{align}\label{JT_type_boundary_term}
    \delta I_{\text{bdry}} = \frac{i}{2} \int_{\partial M_2}  \, d \tau \, \left( \omega_\tau \delta \phi^0 + \phi^0 \delta \omega_\tau \right) \,.
\end{align}
This cancels the $\phi^0 \delta \omega_\tau$ term appearing in (\ref{BF_boundary_later}). The total boundary variation is now
\begin{align}\label{final_jt_like_bc}
    \delta I \Big\vert_{\text{on-shell}} = i \int_{\partial M_2}  \,  d \tau \, \left( \frac{1}{2} \omega_\tau \, \delta \phi^0 + \phi^+ \delta e_\tau^- + \phi^- \delta e_\tau^+ \right) \,.
\end{align}
Demanding that this boundary term vanish leads us to a variational principle where $e_{\tau}^{\pm}$ and $\phi^0$ are held fixed at the boundary. This is the direct BF theory analog of the variational principle in JT gravity, where the boundary metric and dilaton are held fixed, so we will refer to this choice as ``JT-type boundary conditions.''

We now wish to identify the modification of these JT-type boundary conditions, which corresponds to a $T\overline{T}$-like deformation of the dual $(0+1)$-dimensional theory. There are two ways one might identify the appropriate form of the deforming operator. One way is to dimensionally reduce the $T\overline{T}$ operator written in $3d$ Chern-Simons variables, which takes the form $f^- \wedge f^+$ as reviewed in section \ref{subsec:chern_simons_linear_mixing}. Recall that, in Chern-Simons language, the operators $f^{a}$ are dual to the boundary vielbeins $e^a$, and, therefore, the $f^a$ contain the boundary stress tensor. Upon such a reduction, one component of $f$ reduces to the one-dimensional stress tensor $T_{\tau \tau}$, which is dual to the boundary einbein $e_\tau$. Since the component of the metric in the direction along which we reduce is identified with the field $\phi^0$, the other component of $f$ reduces to the operator dual to $\phi^0$, which is $\omega_\tau$. Therefore, the dimensional reduction instructs us to deform the boundary action by an operator constructed from the combination $T_{\tau \tau} \omega_\tau$ (contracted with the appropriate einbein factors to yield a quantity which is a scalar under diffeomorphisms).

The other way to identify the deforming operator is by using the combination $OT$, which defines the $T\overline{T}$ deformation in JT variables and converts all expressions to BF variables. We now carry out this procedure and demonstrate that it produces an operator of the schematic form $T_{\tau \tau} \omega_\tau$ suggested by dimensional reduction. The (Hilbert) definition of the boundary stress tensor is
\begin{align}\label{hilbert_bdry_stress}
    T_{\tau \tau} &= - \frac{2}{\sqrt{\gamma_{\tau \tau}}} \frac{\delta I}{\delta \gamma^{\tau \tau}} \nonumber \\
    &= - \frac{2}{\sqrt{\gamma_{\tau \tau}}} \left( \frac{\delta I}{\delta e^+_\tau} \frac{\delta e^+_\tau}{\delta \gamma^{\tau \tau}} \Big\vert_{e_\tau^-} + \frac{\delta I}{\delta e^-_\tau} \frac{\delta e^-_\tau}{\delta \gamma^{\tau \tau}} \Big\vert_{e_\tau^+} \right) \,.
\end{align}
The map from the metric $\gamma_{\tau \tau}$ to the boundary BF fields $e^{\pm}_\tau$ is simply
\begin{align}\label{gamma_e_def}
    \gamma_{\tau \tau} = - 4 e^+_\tau e^-_\tau \,, \qquad \gamma^{\tau \tau} = - \frac{1}{4 e^+_\tau e^-_\tau} \,.
\end{align}
Note that, according to our conventions (\ref{e_P_ell_conventions}), the relative minus sign in the definition (\ref{gamma_e_def}) of $\gamma_{\tau \tau}$ is required to have a positive-definite worldline metric since
\begin{align}
    e_\tau^+ e_\tau^- = \frac{1}{4} \left( i e_\tau^1 - e_\tau^2 \right) \left( i e_\tau^1 + e_\tau^2 \right) = - \frac{1}{4} \left( \left( e_\tau^1 \right)^2 + \left( e_\tau^2 \right)^2 \right) \,.
\end{align}
Thus, the derivatives appearing in the stress tensor can be written as
\begin{align}
    \frac{\delta e^+_\tau}{\delta \gamma^{\tau \tau}} \Big\vert_{e_\tau^-} &= \frac{1}{\left( \gamma^{\tau \tau} \right)^2} \cdot \frac{1}{4 e_\tau^-} = - \frac{e_\tau^+}{\gamma^{\tau \tau}} \,, \nonumber \\
    \frac{\delta e^-_\tau}{\delta \gamma^{\tau \tau}} \Big\vert_{e_\tau^+} &= \frac{1}{\left( \gamma^{\tau \tau} \right)^2} \cdot \frac{1}{4 e_\tau^+} = - \frac{e_\tau^-}{\gamma^{\tau \tau}} \,.
\end{align}
Meanwhile, from (\ref{final_jt_like_bc}) we see that $\frac{\delta I}{\delta e^+_\tau} = i \phi^-$ and $\frac{\delta I}{\delta e^-_\tau} = i \phi^+$. So, the stress tensor is
\begin{align}
    T_{\tau \tau} = \frac{2 i}{\sqrt{\gamma^{\tau \tau}}} \left(  \phi^- \cdot \frac{e_\tau^+}{\gamma^{\tau \tau}}  + \phi^+ \cdot \frac{e_\tau^-}{\gamma^{\tau \tau}} \right) \,,
\end{align}
and its trace is
\begin{align}
    T = T_{\tau \tau} \gamma^{\tau \tau} = \frac{i}{\sqrt{- e^+_\tau e^-_\tau}} \left( e^+_\tau \phi^- + \phi^+ e^-_\tau \right) \,.
\end{align}
Next, we express the operator $O$ dual to the dilaton in BF variables. Using the map
\begin{align}
    \Phi = - \frac{i}{4} \phi^0 \,, 
\end{align}
one has from (\ref{O_dilaton_operator_def}) that 
\begin{align}\label{O_op_bf_variables_def}
    O &= \frac{1}{\sqrt{\gamma}} \frac{\delta I}{\delta \Phi_b} \nonumber \\
    &= \frac{2 i}{\sqrt{- e_\tau^+ e_\tau^-}} \frac{\delta I}{\delta \phi^0} \nonumber \\
    &= - \frac{1}{\sqrt{- e_\tau^+ e_\tau^-}} \omega_\tau \,.
\end{align}
We conclude that, in BF variables, the combination which corresponds to a boundary $T\overline{T}$ deformation is
\begin{align}
    OT = \frac{i}{e_\tau^+ e_\tau^-} \left( e^+_\tau \phi^- + \phi^+ e^-_\tau \right) \omega_\tau \,.
\end{align}
As claimed, this matches the expectation described above from dimensionally reducing the boundary deformation of $3d$ Chern-Simons.

Now that we have identified the appropriate $T\overline{T}$ operator for a boundary deformation of our BF theory, we will apply the deformation and study the resulting modified boundary conditions. To do this, we promote the sources $\phi^0$ and $e^{\pm}_\tau$ to $\lambda$-dependent quantities and attempt to solve for the $\lambda$ dependence. The boundary variation now takes the form
\begin{align}\label{total_delta_I}
    \delta I ( \lambda ) \Big\vert_{\text{on-shell}} = i \int_{\partial M_2}  \, d \tau \, \left( \frac{1}{2} \omega_\tau \, \delta \phi^0 ( \lambda ) + \phi^+ \delta e_\tau^- ( \lambda ) + \phi^- \delta e_\tau^+ ( \lambda ) \right) \,.
\end{align}
As we discussed around (\ref{illustrating_two_ways}), there are na\"ively two ways to deform the boundary action: we can either deform the combined action (\ref{total_delta_I}) which already includes a boundary term, or we can deform the action without the boundary term which includes variations of both the sources and expectation values. Here, we will take the latter strategy. Without the boundary term, the full variation of the action is
\begin{align}\label{full_off_shell_lambda_bdry}
    \delta I_{\text{BF}} ( \lambda ) &= i \int_{\partial M_2}  \, d \tau \, \bigg( \frac{1}{2} ( \delta \omega_\tau ) \phi^0 ( \lambda ) + \frac{1}{2} \omega_\tau ( \delta \phi^0 ( \lambda ) ) + ( \delta \phi^+ ) e_\tau^- ( \lambda ) + \phi^+ ( \delta e_\tau^- ( \lambda ) ) \\
    &\qquad \qquad \qquad + ( \delta \phi^- ) e_\tau^+ ( \lambda ) + \phi^- ( \delta e_\tau^+ ( \lambda ) ) \bigg) \,.
\end{align}
We will re-scale our flow equation by a factor of $- \frac{1}{4}$ for convenience, writing
\begin{align}
    \frac{\partial I_{\text{BF}}}{\partial \lambda} = - \frac{1}{4} \int_{\partial M_2}  \, \sqrt{\gamma_{\tau \tau}} \, d \tau \, O T = i \int_{\partial M_2}  \, \sqrt{\gamma_{\tau \tau}} \, d \tau \, \frac{1}{ ( - 4 e_\tau^+ e_\tau^- ) } \left( e^+_\tau \phi^- + \phi^+ e^-_\tau \right) \omega_\tau \,.
\end{align}
This implies that the variation of the action also satisfies the flow
\begin{equation}
\begin{aligned}\label{variation_flow}
    \frac{\partial ( \delta I_{\text{BF}} ( \lambda ) ) }{\partial \lambda}  &= - \frac{1}{4} \int_{\partial M_2}   \, d \tau \ \delta ( O T \, \sqrt{\gamma_{\tau \tau}} ) \\&=  i \int_{\partial M_2}  \, d \tau \, \delta \left[ \, \sqrt{\gamma_{\tau \tau}}  \, \frac{1}{ ( - 4 e_\tau^+ e_\tau^- ) } \left( e^+_\tau \phi^- + \phi^+ e^-_\tau \right) \omega_\tau \right] \,.
\end{aligned}
\end{equation}
We impose the variational principle that, at any point along the $T\overline{T}$ flow, the $\lambda$-dependent expressions for the sources $e_\tau^{\pm} ( \lambda )$ and $\phi^0 ( \lambda )$ are held fixed. In particular, this means that $\delta e_\tau^{\pm} ( \lambda ) = 0$ and thus no terms are generated on the right side of (\ref{variation_flow}) from $\delta$ acting on $e_\tau^{\pm}$ or on $\sqrt{\gamma_{\tau \tau}}$. Likewise, the terms involving the variations of these sources in (\ref{full_off_shell_lambda_bdry}) also vanish. We then take the $\lambda$ derivative of the surviving terms in (\ref{full_off_shell_lambda_bdry}) and set the result equal to the right side of (\ref{variation_flow}) to obtain

\begin{equation}
\begin{aligned}\label{jt_like_bc_flow_equation_read_off}
    & i \int_{\partial M_2} \, ( e ( \lambda ) \, d \tau ) \, \frac{1}{e ( \lambda ) } \, \left( \frac{1}{2} \delta  \omega_\tau \, \frac{\partial (\phi^0 ( \lambda ))}{\partial \lambda} + \delta  \phi^+ \frac{\partial (e_\tau^- ( \lambda ) )}{\partial \lambda} + \delta  \phi^- \frac{\partial ( e_\tau^+ ( \lambda ) )}{\partial \lambda} \right)  \\
    & = i \int_{\partial M_2}  \, ( e ( \lambda ) \, d \tau )  \, \frac{1}{e ( \lambda ) ^2} \left( \left( e^+_\tau ( \lambda ) \, \delta  \phi^- +  e^-_\tau ( \lambda ) \,  \delta \phi^+ \right) \omega_\tau + \left( e^+_\tau ( \lambda ) \phi^- + e^-_\tau  ( \lambda ) \phi^+ \right) \delta \omega_\tau \right)\,.
\end{aligned}
\end{equation}
To ease notation we have defined $e ( \lambda ) = \sqrt{ - 4 e_\tau^+ ( \lambda ) e_\tau^- ( \lambda )}$, so that $\gamma_{\tau \tau} (\lambda ) = e ( \lambda )^2$, and we have multiplied and divided by $e ( \lambda )$ on the left side. We can now read off the differential equations for the $\lambda$-dependent sources from (\ref{jt_like_bc_flow_equation_read_off}), finding
\begin{align}
    \frac{\partial ( \phi^0 ( \lambda ))}{\partial \lambda} &= \frac{2}{e(\lambda)} \left( e^+_\tau (\lambda) \phi^- + e^-_\tau  (\lambda) \phi^+  \right) \,, \nonumber \\
    \frac{\partial ( e_\tau^- ( \lambda ) )}{\partial \lambda} &= \frac{1}{e(\lambda)} e_\tau^- (\lambda)  \omega_\tau \,, \nonumber \\
    \frac{\partial (  e_\tau^+ ( \lambda ) )}{\partial \lambda} &= \frac{1}{e(\lambda)} e_\tau^+ (\lambda) \omega_\tau \,.
\end{align}
We note that these differential equations imply that the ratios 
\begin{align}
    \hat{e} = \sqrt{- \frac{e_\tau^+}{e_\tau^-}} \,, \qquad \hat{e}^{-1} = \sqrt{- \frac{e_\tau^-}{e_\tau^+}}
\end{align}
do not flow with $\lambda$:
\begin{align}
    \frac{\partial (\hat{e}^2)}{\partial \lambda} &= - \frac{( \partial_\lambda e_\tau^+ ) e_\tau^- - ( \partial_\lambda e_\tau^- ) e_\tau^+ }{\left( e_\tau^- \right)^2} \nonumber \\
    &= - \frac{ e_\tau^+  e_\tau^- \omega_\tau -  e_\tau^- e_\tau^+ \omega_\tau }{e \left( e_\tau^- \right)^2} \nonumber \\
    &= 0 \,.
\end{align}
Since $\frac{e_\tau^+}{e} = \frac{1}{2} \hat{e}$ and $\frac{e_\tau^-}{e} = \frac{1}{2} \hat{e}^{-1}$, we can write the flow equations as
\begin{align}\begin{split}\label{jt_like_flow_simplified}
    \frac{\partial ( \phi^0 ( \lambda ))}{\partial \lambda} &= \left( \hat{e} \phi^- + \hat{e}^{-1} \phi^+  \right) \,,  \\
    \frac{\partial ( e_\tau^- ( \lambda ) )}{\partial \lambda} &= \frac{1}{2} \hat{e}^{-1}  \omega_\tau \,,  \\
    \frac{\partial (  e_\tau^+ ( \lambda ) )}{\partial \lambda} &= \frac{1}{2} \hat{e}  \omega_\tau \,.
\end{split}\end{align}
Because the right sides of the three differential equations in (\ref{jt_like_flow_simplified}) are independent of $\lambda$, they can be trivially solved to find
\begin{align}\label{schwarzian_lambda_dependent_solutions}
    \phi^0 ( \lambda ) &= \phi^0 ( 0 ) + \lambda \left( \hat{e} \phi^- + \hat{e}^{-1} \phi^+  \right) \,, \nonumber \\
    e_\tau^{+} ( \lambda ) &= e_\tau^{\pm} ( 0 ) + \frac{\lambda}{2} \hat{e}^{-1} \omega_\tau \,, \nonumber \\
    e_\tau^{-} ( \lambda ) &= e_\tau^{\pm} ( 0 ) + \frac{\lambda}{2} \hat{e} \omega_\tau \,. 
\end{align}
Replacing hatted quantities with the original variables, we conclude
\begin{align}\label{jt_like_bc_final_soln}
    \phi^0 ( \lambda ) &= \phi^0 ( 0 ) + \frac{2 \lambda}{e(0)} \left( e_\tau^+ (0) \phi^- + e_\tau^- (0) \phi^+  \right) \,, \nonumber \\
    e_\tau^{\pm} ( \lambda ) &= e_\tau^{\pm} ( 0 ) + \frac{\lambda}{e(0)} e_{\tau}^{\pm} ( 0 ) \omega_\tau \,.
\end{align}
The rotated sources (\ref{jt_like_bc_final_soln}) define the full solution to the $T\overline{T}$ flow at finite $\lambda$. Similarly to the case of $3d$ Chern-Simons variables, the $T\overline{T}$ deformation of $2d$ BF theory corresponds to a linear mixing of the expectation values $\phi^{\pm}$ and $\omega_\tau$ into the sources $e_{\tau}^{\pm}$ and $\phi^0$. We reiterate that the variational principle remains unchanged along the $T\overline{T}$ flow; the new $\lambda$-dependent source fields $e_{\tau}^{\pm} ( \lambda )$ and $\phi^0 ( \lambda )$ remain fixed at the boundary at any $\lambda$. The expressions for these sources in terms of the sources in the undeformed theory are modified according to (\ref{jt_like_bc_final_soln}). Note that we have chosen to solve the flow equation for the action $I_{\text{BF}}$ without the boundary term. To complete the solution and ensure the correct variational principle, we must go back and add a $\lambda$-dependent boundary term $I_{\text{bdry}} ( \lambda )$ which takes the same form as that given in (\ref{schwarzian_appropriate_boundary}) except with $\phi^0$ replaced by $\phi^0 ( \lambda )$ as in (\ref{schwarzian_lambda_dependent_solutions}).

Although this result is morally analogous to the mixing of sources and expectation values in the $3d$ Chern-Simons context, we briefly comment on two superficial differences. The first is that, in the Chern-Simons context reviewed in section \ref{subsec:chern_simons_linear_mixing}, both the variation of the action $\delta S = \frac{1}{4\pi G} \int_{\partial M_3} \varepsilon_{ab} f^a \wedge \delta e^b$ and the $T\overline{T}$ deformation $S_{f^+ f^-} = \frac{1}{32 \pi^2 G^2}  \int_{\partial M_3} f^- \wedge f^+$ are written as integrals of differential two-forms, and therefore the integrals do not contain any measure factors. However, in our deformation, we have chosen to write the deforming operator $OT$ as a scalar rather than as a one-form. Since the variation of the on-shell action (\ref{final_jt_like_bc}) is written as the integral of a boundary one-form rather than a scalar, our solution (\ref{jt_like_bc_final_soln}) to the flow equation must introduce extra factors of $e$ to compensate for the difference in measure between the two integrals. No such measure factors appeared in the solution to the Chern-Simons flow. In principle, the presence of such $\lambda$-dependent factors could have spoiled the conclusion that the $T\overline{T}$ flow generates only a linear mixing of sources and expectation values, rather than some more complicated behavior. However, as we saw in (\ref{jt_like_flow_simplified}), only certain $\lambda$-independent combinations enter the flow equation, so the simple linear form of the solution is preserved.

The second difference is that, in the Chern-Simons context, each source $e_i^a$ became a function of its dual expectation value $f_i^a$. In the BF theory solution, however, each source field became a function of the dual expectation value for a \textit{different} field. From the perspective of dimensional reduction, this difference simply arises because we have split the two-dimensional metric into a one-dimensional metric and a scalar field $\phi^0$. This splitting causes the dual operators $T$ and $O$ to appear asymmetrically in the action, even though both expectation values descend from the two-dimensional stress tensor on the boundary of the Chern-Simons theory. Therefore the apparent difference that each source rotates into an expectation value dual to a different operator is merely an artifact of the splitting performed as part of the reduction.

To summarize this section, we have found that the addition of the appropriate $OT$-like operator in BF variables can be interpreted as a linear mixing of sources and expectation values, in the same way as the $f^- \wedge f^+$ deformation implemented such a linear mixing in $3d$ Chern-Simons variables. It is worth pointing out that this feature is fairly generic: the addition of a double-trace operator constructed out of the expectation values dual to certain sources should generally correspond to precisely this type of change in boundary conditions, in which the sources become dependent upon the expectation values. Indeed, part of the standard lore from $\mathrm{AdS}/\mathrm{CFT}$ is that the addition of a double-trace operator in the field theory corresponds to a modification of the boundary conditions for the bulk fields \cite{Klebanov:1999tb,Witten:2001ua}, although this behavior has more often been studied in the case of relevant or marginal operators rather than irrelevant operators.

By way of analogy, we mention that there is another example where the addition of a double-trace operator leads to such a linear mixing. Let us return to the context of JT gravity and consider the operator $O$ dual to the dilaton as defined in (\ref{O_dilaton_operator_def}). One could consider adding a boundary term to the JT gravity Lagrangian of the form
\begin{align}
    I_{O^2} = \mu \int_{\partial M_2} \, d \tau \, \sqrt{\gamma_{\tau \tau}} \, O^2 \,.
\end{align}
As discussed in \cite{Goel:2020yxl}, the combined on-shell variation of the JT gravity action plus $I_{O^2}$ is
\begin{align}
    \delta ( I_{\text{BF}} + I_{O^2} ) \Big\vert_{\text{on-shell}} = \int_{\partial M_2}  \, d \tau \, \sqrt{\gamma_{\tau \tau}} \left( \frac{1}{2} \left( T_{\tau \tau} - \mu g_{\tau \tau} O^2 \right) \delta g^{\tau \tau} + O \delta \left( \Phi + 2\mu O \right) \right) \,.
\end{align}
This corresponds to a variational principle where both the metric $g_{\tau \tau}$ and the combination
\begin{align}
    \Phi ( \mu ) = \Phi ( 0 ) + 2 \mu O
\end{align}
are held fixed on the boundary. Of course, this has an identical interpretation as a linear mixing of the operator $\Phi$ into its dual expectation value $O$. We, therefore, see that the $f^- \wedge f^+$ deformation of $3d$ Chern-Simons, the $OT$ operator written in $2d$ BF variables, and the $O^2$ deformation of JT gravity are all of this qualitatively similar form.

\subsection{Deformation with Schwarzian-type boundary term}\label{sec:schwarzian_type}

Although the JT-like boundary conditions considered in the preceding subsection led to an especially simple modification under a $T\overline{T}$-like deformation, these are in some sense not the most natural boundary conditions to consider in BF theory. For instance, we saw in (\ref{eq:dimredCStoBF}) that the dimensional reduction of $3d$ Chern-Simons theory produces a BF theory action of the form
\begin{align}\label{schwarzian_type_bdry_term_def}
    I &=  - i \int_{M_2} \operatorname{Tr} \left( \phi F \right) - \frac{i}{2} \oint_{\partial M_2} \, d \tau \, \operatorname{Tr} \left( \phi^2 \right) \nonumber \\
    &\equiv I_{\text{BF}} + I_{\text{bdry}} \,,
\end{align}
which contains an additional boundary term compared to the bare BF action (\ref{eq:BF Action}) and where we have introduced a factor of $\frac{1}{2}$ by convention. In addition to its emergence from dimensional reduction, this boundary term is also of interest since it gives rise to a dual Schwarzian theory on the $1d$ boundary, as we reviewed in section \ref{sec:jt_as_bf}. We would, therefore, like to study the $T\overline{T}$ deformation of the theory with this choice of boundary term as well.

First, it is instructive to see what variational principle this additional boundary term yields in the undeformed case. We have already seen that the on-shell variation of the first term $I_{\text{BF}}$ gives
\begin{equation}
\label{eq:on-shell boundary later}
    \delta I_{\text{BF}} \Big\vert_{\text{on-shell}} = i \int_{\partial M_2} \, d \tau \, \operatorname{Tr} \left( \phi \delta A_\tau \right) \,, 
\end{equation}
which means that the combined boundary variation including contributions from both the bulk and boundary terms is
\begin{align}
    \delta I \Big\vert_{\text{on-shell}} = \frac{i}{2} \int_{\partial M_2} \, d \tau \,  \operatorname{Tr} \left( \phi \, \delta A_\tau - \phi \, \delta \phi \right) \,.
\end{align}
This boundary variation will vanish if we demand that the value of the one-form $A_\mu$ on the boundary is equal to the combination $\phi \, d \tau$ where $d \tau$ is the one-form appearing in the boundary length element. We write this condition as
\begin{align}\label{A_is_phi_bdry}
    A_\tau \big\vert_{\text{bdry}} = \phi \big\vert_{\text{bdry}} \,.
\end{align}
After imposing this condition, the boundary term can be written as 
\begin{align}\label{group_bdry_intermediate}
    I_{\text{bdry}} = - \frac{i}{2} \int_{\partial M_2} \, d \tau \,  \operatorname{Tr} ( A_\tau^2 ) \,.
\end{align}
On the other hand, the equation of motion $F = 0$ requires that $A_\mu$ is pure gauge and can be written as
\begin{align}\label{Amu_is_pure_gauge}
    A_\mu = g^{-1} \partial_\mu g
\end{align}
for some group element $g \in \operatorname{SL} ( 2, \mathbb{R} )$. This means that the boundary term (\ref{group_bdry_intermediate}) becomes
\begin{align}\label{group_bdry_intermediate_two}
    I_{\text{bdry}} &= - \frac{i}{2} \int_{\partial M_2} \, d \tau \, \mathrm{Tr} \left( ( g^{-1} \partial_\mu g ) ( g^{-1} \partial^\mu g ) \right) \, \nonumber \\
    &= - \frac{i}{2} \int_{\partial M_2} \, d \tau \, g_{ij} ( x ) \, \dot{x}^i \dot{x}^j \,,
\end{align}
where in the last step we have introduced coordinates $x^i (t)$ on the three-dimensional group manifold $M = \operatorname{SL}(2, \mathbb{R})$ and the canonical metric $g_{ij}$ on $M$. The equivalence of the two lines of (\ref{group_bdry_intermediate_two}) is a standard result concerning the Cartan metric tensor on a Lie group $G$ which is induced by the Killing form on the Lie algebra $\mathfrak{g}$ of $G$.

We, therefore, see that, with the choice of boundary term appearing in (\ref{schwarzian_type_bdry_term_def}), the BF theory has a boundary degree of freedom whose dynamics are described by the theory of a free particle moving on the $SL(2, \mathbb{R})$ group manifold. This theory is equivalent to the Schwarzian theory in its usual presentation \cite{Mertens:2017mtv,Mertens:2018fds}, so this derivation provides a complementary way to see that BF gauge theory is dual to the Schwarzian, although in a somewhat different language than that used in section \ref{sec:jt_as_bf}.

We now turn to the issue of $T\overline{T}$ deforming these boundary conditions and, by extension, the dual $1d$ theory. We first note that the procedure we followed in the case of JT-type boundary conditions will not work here because we have modified the variational principle of the theory. In section \ref{subsec:deformed_jt_like_bc}, since the value of the boundary field $\phi^0$ was held fixed at the boundary, we were free to define the operator $O$ dual to $\phi^0$ as in (\ref{O_op_bf_variables_def}) and then construct the deformation $OT$. However, with our Schwarzian-type boundary conditions, the value of $\phi^0$ is not held fixed to some constant value at the boundary but is rather related to the value of $A_\tau$ via (\ref{A_is_phi_bdry}), and $A_\tau$ is viewed as a boundary degree of freedom. We cannot define the operator $O$ in the same way and must identify a suitable replacement for $O$ in some other way.

We propose that the correct scalar which replaces $O$ for this choice of boundary conditions is the Lagrangian itself. To motivate this choice, we must take a brief detour and explain a rewriting of the $T\overline{T}$ deformation for $(0+1)$-dimensional theories, which will appear in the next chapter.

We now recall certain facts about the dimensional reduction of the $T\overline{T}$ operator from $(1+1)$ dimensions to $(0+1)$ dimensions. We follow the discussion in \cite{Gross:2019ach}, where these results first appeared. The flow equation \eqref{eq:TTb definition} determines a one-parameter family of Lagrangians $\mathcal{L}^{(\lambda)}$ with the initial condition that $\mathcal{L}^{(0)}$ matches the CFT Lagrangian $\mathcal{L}_0$. We note that $\lambda$ has length dimension two, which means that there is an effective energy scale $\Lambda$ set by
\begin{align}
    \Lambda = \frac{1}{\sqrt{\lambda}} \,.
\end{align}
Because the seed theory was conformal and hence had no dimensionful parameters, in the deformed theories, the quantity $\Lambda$ is the only scale in the problem. This means that an infinitesimal change in $\Lambda$ is equivalent to an infinitesimal scale transformation of the theory, and on general grounds, we know that the response of the action $I^{(\lambda)} = \int d^2 x \, \mathcal{L}^{(\lambda)}$ to such a scale transformation is controlled by the trace of the stress tensor. We therefore have
\begin{align}
    \Lambda \frac{d}{d \Lambda} I = \int d^2 x \, T^\mu{}_\mu \,.
\end{align}
On the other hand, by comparing to equation \eqref{eq:TTb definition}, we see that by definition the response of the action $S^{(\lambda)}$ to a change in $\lambda$ is given by the integral of the $T\overline{T}$ operator. That is,
\begin{align}
    \Lambda \frac{d}{d \Lambda} I &= \frac{1}{\sqrt{\lambda}} \frac{d}{d \left( \frac{1}{\sqrt{\lambda}} \right)} I \nonumber \\
    &= - 2 \lambda \int d^2 x \, \left( T^{\mu \nu} T_{\mu \nu} - \left( T^\mu{}_\mu \right)^2 \right) \,.
\end{align}
We therefore conclude
\begin{align}
\label{trace_flow}
  T^\mu{}_\mu = 2 \lambda \, \left( \left( T^\mu{}_\mu \right)^2 - T^{\mu \nu} T_{\mu \nu} \right) \,,
\end{align}
according to our normalization of $T\overline{T}$. We emphasize that this relationship, namely the trace flow equation, between the components of the stress tensor, holds at any point along the trajectory of $T\overline{T}$-flow without imposing any equations of motion or conservation equations. We may solve (\ref{trace_flow}) for a component of the stress tensor and use the resulting equation to eliminate this component in the definition of the deformation.

Suppose our $2d$ theory has coordinates $x$ and $t$, where we take $x$ to parameterize a circular direction. We wish to dimensionally reduce along this circle to determine an effective deformation in the resulting $(0+1)$-dimensional theory. To do this, it is natural to solve for the spatial component $T_{xx}$. Writing out both sides of (\ref{trace_flow}) in components yields
\begin{align}
    T^x{}_x + T^t{}_t = 2 \lambda \left( \left( T^x{}_x + T^t{}_t \right)^2 - \left( T^{xx} T_{xx} + 2 T^{xt} T_{xt} + T_{tt} T_{tt} \right) \right) \,,
\end{align}
which can be solved to find
\begin{align}\label{Txx_soln}
T^x{}_x = \frac{T^t{}_t + 4 \lambda T^{xt} T_{xt}}{-1 + 4 \lambda T^t{}_t} \,.
\end{align}
To dimensionally reduce, we assume that the mixed component $T_{xt}$ of the stress tensor vanishes. We would then like to replace the spatial component $T^x{}_x$ and write a flow equation that depends only on the ``stress scalar'' $T \equiv T^t{}_t$ for the $(0+1)$-dimensional theory. After replacing $T^x{}_x$ and setting $T_{xt} = 0$ in this way, the flow equation for the Lagrangian becomes
\begin{align}\label{gross_reduction_intermediate}
    \frac{\partial \mathcal{L}^{(\lambda)}}{\partial \lambda} &= \left(  \left( \frac{T^t{}_t}{-1 + 4 \lambda T^t{}_t} \right)^2 + \left( T^t{}_t \right)^2 - \left( T^t{}_t + \frac{T^t{}_t}{-1 + 4 \lambda T^t{}_t} \right)^2 \right) \nonumber \\
    &= \frac{\left( T^t{}_t \right)^2}{\frac{1}{2} - 2 \lambda T^t{}_t} \,.
\end{align}
We now assume that $T \equiv T^t{}_t$ is independent of the spatial coordinate $x$ and perform the integral over the $x$ circle. This will introduce an irrelevant length factor, which can be scaled away. The result is a deformation for the action of the $(0+1)$-dimensional quantum mechanics theory:
\begin{align}\label{reduced_qm_flow_no_susy}
    \frac{\partial I}{\partial \lambda} = \int \, dt \, \frac{T^2}{\frac{1}{2} - 2 \lambda T} \,.
\end{align}
Interpreting the Euclidean Lagrangian as the Hamiltonian\footnote{In our conventions, $I = \int \, d \tau \, H$, the Hamiltonian is equivalent to the Euclidean Lagrangian, although the Hamiltonian is written in terms of canonical momenta $p_\mu$ rather than $\dot{X}^\mu$. This agrees with the conventions of \cite{Gross:2019ach}, but differs from the common convention $H ( X^\mu , p^\mu ) = - L_E ( X^\mu, \dot{X}^\mu)$.}, we can evaluate (\ref{reduced_qm_flow_no_susy}) in an energy eigenstate $| n \rangle$ to find
\begin{align}
    \frac{\partial \langle n \, \vert \, H \, \vert \, n \rangle}{\partial \lambda} = \frac{ \left( \langle n \, \vert \, H \, \vert \, n \rangle \right)^2 }{\frac{1}{2} - 2 \lambda \langle n \, \vert \, H \, \vert \, n \rangle} \,,
\end{align}
or more simply writing $E_n$ for $\langle n \, \vert \, H \, \vert \, n \rangle$,
\begin{align}
    \frac{\partial E_n}{\partial \lambda} = \frac{E_n^2}{\frac{1}{2} - 2 \lambda E_n} \,.
\end{align}
This differential equation has the familiar solution from the introduction in section \ref{sec:factorfloweq}
\begin{align}
\label{eq:erergw3q1}
    E_n ( \lambda ) = \frac{1 - \sqrt{1 - 8 \lambda E_n^{(0)}}}{4 \lambda} \,.
\end{align}

We note in passing that, for free kinetic seed theories of the form we are interested in here, both the deformed Hamiltonian and the deformed Lagrangian have a similar square root form \eqref{eq:erergw3q1}. Beginning from a generic non-linear sigma model with the Lagrangian
\begin{equation}
    L = \frac{1}{2}G_{\mu\nu}(X)\dot{X}^\mu \dot{X}^\nu \,,
\end{equation} 
the Hamiltonian is given by
\begin{equation}
    H = \frac{1}{2}G^{\mu\nu}(X) p_\mu p_\nu\,. 
\end{equation} 
Under the $T\overline{T}$-deformation, the Hamiltonian becomes
\begin{equation}
    H_\lambda = \frac{1-\sqrt{1-4\lambda G^{\mu\nu}(X) p_\mu p_\nu}}{4\lambda}\,.
\end{equation}
The deformed (Lorentzian) Lagrangian is recovered by Legendre transformation from the deformed Hamiltonian:
\begin{equation} 
\label{eq:particle-on-a-group-X}
L_\lambda = p_\mu \dot{X}^\mu - H_\lambda = \frac{\sqrt{1+4\lambda G_{\mu\nu}(X)\dot{X}^\mu \dot{X}^{\nu}} - 1}{4\lambda} \,.
\end{equation}
Thus, both $H_\lambda$ and $L_\lambda$ are determined by a square root-type function of their undeformed values, up to sending $\lambda \to - \lambda$, which changes the operator driving the flow by a sign.

Next, one can define the (Euclidean) Hilbert stress tensor associated with $I$ via
\begin{align}\label{one_dim_hilbert}
    T^{(\text{Hilb})} = - \frac{2}{\sqrt{g^{\tau \tau}}} \frac{\delta I}{\delta g^{\tau \tau}} = H - 2 \frac{\partial H ( g^{\tau \tau} )}{\partial g^{\tau \tau}} \Big\vert_{g^{\tau \tau} = 1} \,.
\end{align}
We note that this is a variation with respect to the worldline metric $g_{\tau \tau}$, not to be confused with the target-space metric $g_{ij} ( x )$ appearing in (\ref{group_bdry_intermediate_two}). The expression \eqref{one_dim_hilbert} for $T^{(\text{Hilb})}$ simplifies in the case of $T\overline{T}$ deformations of seed theories, which are ``purely kinetic'' in the following sense. Suppose that we begin with an undeformed Hamiltonian $H_0$ with the property that, when $H_0$ couples to worldline gravity, it takes the form
\begin{align}
    H_0 ( g^{\tau \tau} ) = g^{\tau \tau} \mathcal{H} \,, 
\end{align}
where $\mathcal{H} \equiv H_0 ( g^{\tau \tau} = 1 )$ is some expression which is independent of the worldline metric. For instance, $\mathcal{H} = g_{ij} ( x ) \dot{x}^i \dot{x}^j$ in (\ref{group_bdry_intermediate_two}). Under the $T\overline{T}$ flow (\ref{one_dim_hilbert}), the deformed theory $H ( \lambda )$ will also only depend on the worldline metric through the combination $g^{\tau \tau} \mathcal{H}$. For such theories, the Hilbert stress tensor is
\begin{align}\label{explicit_onedim_hilbert}
       T^{(\text{Hilb})} &= H ( \lambda ) - 2 \frac{\partial H}{\partial H_0} \frac{\partial H_0 ( g^{\tau \tau} ) }{\partial g^{\tau \tau} } \Big\vert_{g^{\tau \tau}=1} \nonumber \\
    &= H ( \lambda ) - 2 \mathcal{H} \frac{\partial H}{\partial \mathcal{H}} \,.
\end{align}
In particular, substituting the explicit solution \eqref{eq:erergw3q1} for a $T\overline{T}$-like flow into (\ref{explicit_onedim_hilbert}) and multiplying by $H$ itself, one can compute the combination
\begin{align}
    H    T^{(\text{Hilb})} &= H \left( H - 2 H_0 \frac{\partial H}{\partial H_0} \right) \nonumber \\
    &= - \frac{ \left( \sqrt{1 - 8 \lambda H_0} - 1 \right)^2}{16 \lambda^2 \sqrt{1 - 8 \lambda H_0} } \,.
\end{align}
On the other hand, substituting the same solution \eqref{eq:erergw3q1} into the definition of the deforming operator in \eqref{reduced_qm_flow_no_susy} gives
\begin{align}
    \frac{H^2}{\frac{1}{2} - 2 \lambda H} = \frac{\left( \sqrt{ 1 - 8 \lambda H_0} - 1 \right)^2}{8 \lambda^2 \sqrt{1 - 8 \lambda H_0}} \,.
\end{align}
Thus, we find that for purely kinetic seed theories one has the equivalence
\begin{align}\label{HTrelation}
    H    T^{(\text{Hilb})}= - \frac{1}{2} \left( \frac{H^2}{\frac{1}{2} - 2 \lambda H} \right) \,.
\end{align}
Up to a constant rescaling, we can, therefore, view the $T\overline{T}$ flow as being driven by the combination $H    T^{(\text{Hilb})}$. We will refer to this as the $HT$ deformation for simplicity. This therefore suggests that the replacement for the operator $O$ in the $OT$ deformation is the Hamiltonian (or Euclidean Lagrangian) itself, since the object $T \equiv T_{\tau \tau} \gamma^{\tau \tau}$ defined in (\ref{hilbert_bdry_stress}) corresponds to the Hilbert stress tensor $T^{(\text{Hilb})}$. 

We now demonstrate explicitly that the proposed deformation yields the expected square root form of the solution for the boundary action. After coupling the boundary action (\ref{group_bdry_intermediate}) to worldline gravity, one has
\begin{align}
    I_{\text{bdry}} = - \frac{i}{2} \int_{\partial M_2}  \sqrt{g_{\tau \tau}} \, d \tau \, g^{\tau \tau}  \operatorname{Tr} ( A_\tau^2 ) \,.
\end{align}
This is a seed theory of purely kinetic form since $H_0 = g^{\tau \tau}  \operatorname{Tr} ( A_\tau^2 ) = g^{\tau \tau} \mathcal{H}$. We make an ansatz for the finite-$\lambda$ deformed (Euclidean) Lagrangian as
\begin{align}
    I_{\text{bdry}} ( \lambda ) = - \frac{i}{2} \int_{\partial M_2}   \, d \tau \, \frac{1}{\lambda} f ( \lambda \mathcal{H} )  \,.
\end{align}
The Hilbert stress tensor at finite $\lambda$ is therefore given by
\begin{align}
    T^{(\text{Hilb})}= \frac{1}{\lambda} \left( f ( \lambda \mathcal{H} ) - 2 \lambda \mathcal{H} f' ( \lambda \mathcal{H} ) \right) \,.
\end{align}
Substituting this into the flow equation $\partial_\lambda H ( \lambda) = H ( \lambda )    T^{(\text{Hilb})}$, we then find
\begin{align}
    \lambda \mathcal{H} f' ( \lambda \mathcal{H} ) - f ( \lambda \mathcal{H} ) = f ( \lambda \mathcal{H} ) \left( f ( \lambda \mathcal{H} ) - 2 \lambda \mathcal{H} f' ( \lambda \mathcal{H} ) \right) \,, 
\end{align}
which has the solution
\begin{align}
    f ( \lambda \mathcal{H} ) = \frac{1}{2} \left( \sqrt{1 + 4 \lambda \mathcal{H} } - 1 \right) \,.
\end{align}
The deformed boundary action is therefore
\begin{align}\label{Atau_deformed_boundary_action}
    I_{\text{bdry}} ( \lambda ) = - \frac{i}{2} \int_{\partial M_2}   \, d \tau \, \frac{1}{2 \lambda} \left( \sqrt{1 + 4 \lambda  \operatorname{Tr}( A_\tau^2 ) } - 1 \right)  \,,
\end{align}
as expected.

We would also like an interpretation of this $HT$ deformation in terms of a rotation between sources and expectation values, as we had in the case of the $OT$ deformation of section \ref{subsec:deformed_jt_like_bc}. However, with this choice of boundary conditions, such an interpretation is somewhat obscured because the two objects in the product $HT$ are not both expectation values dual to some sources, as the objects $O$ and $T$ were in the case of JT-type boundary conditions. One partial interpretation is the following. The flow equation induced by $HT$ can be written as
\begin{align}\label{HT_flow_burgers}
    \frac{\partial H}{\partial \lambda} = H^2 - 2 H \frac{\partial H ( g^{\tau \tau} )}{\partial g^{\tau \tau}} \Big\vert_{g^{\tau \tau} = 1} \,.
\end{align}
This equation is functionally similar to the inviscid Burgers' equation determining the cylinder energy levels of a $T\overline{T}$-deformed QFT in two dimensions from this thesis' introduction \eqref{eq:burgers}. As mentioned in the introduction of this thesis, this has the interpretation that the energy eigenstates of the deformed theory see a cylinder with an effective energy-dependent radius.

The analogous manipulation of (\ref{HT_flow_burgers}) is to ignore the $H^2$ term on the right side. This is harder to justify, but we will return to this assumption in a moment. The resulting equation is
\begin{align}
    \frac{\partial H}{\partial \lambda} = - 2 H \frac{\partial H ( g^{\tau \tau} )}{\partial g^{\tau \tau}} \Big\vert_{g^{\tau \tau} = 1} \,,
\end{align}
which again has the implicit solution
\begin{align}\label{qm_implicit_burgers_solnxxx}
    H ( g^{\tau \tau} , \lambda ) = H ( g^{\tau \tau} - 2 \lambda H ( g^{\tau \tau} , \lambda ) , 0 ) \,.
\end{align}
The interpretation of (\ref{qm_implicit_burgers_solnxxx}) is very similar to that of the solution to the $T\overline{T}$ flow for $3d$ Chern-Simons theory reviewed in section \ref{subsec:chern_simons_linear_mixing}. In that case, the solution for the deformed boundary vielbein is
\begin{align}\label{llabres_mix_comp}
    e_i^a ( \lambda ) = e_i^a (0) + \frac{\lambda }{8 \pi G} f_i^a \,.
\end{align}
Here $e_i^a$ is the source that determines the metric, and $f_i^a$ is the operator dual to $e_i^a$, which is related to the boundary stress tensor. Note (\ref{llabres_mix_comp}) represents a rotation from the undeformed metric (controlled by $e_i^a ( 0 )$) to a deformed metric (determined by $e_i^a ( \lambda )$) which is a function of the stress tensor. But a stress-tensor-dependent (or Hamiltonian-dependent) metric is exactly what we see in the implicit solution (\ref{qm_implicit_burgers_solnxxx}).

Although this is an appealing interpretation, it is obstructed by the presence of the $H^2$ source term in (\ref{HT_flow_burgers}). The full interpretation of this flow equation may involve simultaneously introducing a Hamiltonian-dependent metric and allowing another source-like quantity to depend on its dual expectation value. The identity of this other source-like quantity is not obvious since it seems that the metric and its dual stress tensor are the only pair of such fields that remain.

In our case, the stress tensor has only a single component, so this proposal reduces to the statement that the (Euclidean) Lagrangian $L_E = H$ is dual to $T$. If so, our deforming operator $HT$ might admit an interpretation as a product of two expectation values dual to sources, with the first dual to the stress tensor and the second dual to the metric. In analogy with the $OT$ deformation of section \ref{subsec:deformed_jt_like_bc}, one might expect this deformation to simultaneously introduce linear $\lambda$-dependence of the stress tensor on the Hamiltonian and dependence of the metric on the stress tensor. 

We conclude this section by offering two interpretations of the deformation in the case of Schwarzian-type boundary conditions.

The first interpretation is motivated by the fact that a $T\overline{T}$ deformation of the boundary theory of $3d$ Chern-Simons is seen as replacing the field $e_i^a$ appearing in the expansion of $A_\mu$ with a $\lambda$-dependent version $e_i^a ( \lambda )$. That is, the deformation corresponds to replacing the Chern-Simons gauge field $A_\mu$ with some $A_\mu ( \lambda )$. This can be viewed as a field redefinition from an undeformed gauge field to a deformed gauge field.

Similarly, in the case of BF gauge theory with Schwarzian-type boundary conditions, we see that the boundary action for $A_\tau$ has been modified by making the replacement
\begin{align}
   \operatorname{Tr} ( A_\tau^2 ) \longrightarrow \frac{1}{2 \lambda} \left( \sqrt{1 + 4 \lambda  \operatorname{Tr} ( A_\tau^2 ) } - 1 \right) \,.
\end{align}
We may, of course, interpret this replacement by defining an effective $\lambda$-dependent gauge field $A_\tau ( \lambda )$, which satisfies $ \operatorname{Tr} ( A_\tau ( \lambda )^2 ) = \frac{1}{2 \lambda} \left( \sqrt{1 + 4 \lambda  \operatorname{Tr}( A_\tau ( 0 ) ^2 ) } - 1 \right)$. From this perspective, the functional form of the boundary action
\begin{align}
    I_{\text{bdry}} ( \lambda ) = - \frac{i}{2} \int_{\partial M_2} \, d \tau \,  \operatorname{Tr}( A_\tau ( \lambda ) ^2 ) \, 
\end{align}
remains unchanged, but it appears different when we express the Lagrangian in terms of the undeformed gauge field $A_\tau ( 0 )$, which yields
\begin{align}\label{Atau_deformed_boundary_action_later}
    I_{\text{bdry}} (   \lambda ) = - \frac{i}{2} \int_{\partial M_2}  \, d \tau \, \frac{1}{2 \lambda} \left( \sqrt{1 + 4 \lambda  \operatorname{Tr} ( A_\tau ( 0 ) ^2 ) } - 1 \right)  \,,
\end{align}
as before. 

In the above discussion, we assumed that the boundary condition (\ref{A_is_phi_bdry}) relating $A_\tau$ to $\phi$ on the boundary did not flow with $\lambda$. That is, the gauge field $A_\tau$ underwent a field redefinition, but its relationship with $\phi$ was unmodified. However, a second interpretation is that the $T\overline{T}$ flow modifies the relationship between these fields.

To see this, we again consider the deformed Euclidean Lagrangian (\ref{Atau_deformed_boundary_action_later}). Rather than thinking of this as an action defining a one-dimensional theory in isolation, suppose we consider the bulk theory with this choice of boundary term as
\begin{align}
    I &= I_{\text{BF}} + I_{\text{bdry}} \nonumber \\
    &=- i \int_{M_2} \operatorname{Tr} \left( \phi F \right) - \frac{i}{4 \lambda} \int_{\partial M_2}  \, d \tau \,  \left( \sqrt{1 + 4 \lambda  \operatorname{Tr}( A_\tau^2 ) } - 1 \right) \,.
\end{align}
The combined on-shell variation is
\begin{align}
    \delta I \Big\vert_{\text{on-shell}} = i \int_{\partial M_2}  \operatorname{Tr} \left(  \phi \, \delta A_\tau - \frac{A_\tau \, \delta A_\tau }{\sqrt{1 + 4 \lambda  \operatorname{Tr} ( A_\tau^2 )}}  \right) \,.
\end{align}
This boundary term vanishes if we impose
\begin{align}
    \phi \Big\vert_{\text{bdry}} = \frac{A_\tau }{\sqrt{1 + 4 \lambda  \operatorname{Tr} ( A_\tau^2 )}} \Big\vert_{\text{bdry}} \,,
\end{align}
which is a $\lambda$-dependent modification of (\ref{A_is_phi_bdry}). We, therefore, see that we can either interpret the boundary deformation as (1) redefining $A_\tau$ to $A_\tau ( \lambda )$ while leaving the boundary condition unchanged, or (2) leaving the field $A_\tau$ unchanged but modifying the boundary condition which relates $A_\tau$ to $\phi$.

\subsection{Reducing the boundary graviton to $T\overline{T}$-deformed Schwarzian}
We can also make contact with the $T\overline{T}$-deformed Schwarzian quantum mechanics from the dimensional reduction of the boundary graviton action we found in chapter \ref{chFieldTheoryGravitons}. Beginning from \eqref{hik}, we force $(f(x,t), \bar{f}(x,t)) = (f(x), \bar{f}(x))$ and set the left-movers equal to the right-movers $f(x) = \bar{f}(x)$ to obtain
\begin{equation}\label{assdadddddadad}\begin{split}
I &= {1\over 32 \pi G} \int_{\partial M_3}\! d^2x \bigg[  if'\dot{f}-i\bar{f}' \dot{\bar{f}} \\&- 4
{ \sqrt{1-{1\over 2}r_c(f^{\prime 2}+\bar{f}^{\prime 2})+{1\over 16}r_c^2(f'^2-\bar{f}'^2)^2  }-1\over r_c}  \bigg] + \text{higher derivs.} 
 \bigg|_{d^2x = dx, f= \bar{f}}   
 \\&= \frac{1}{32 \pi G} \int_{\partial M_2} dx \left[- 4 { \sqrt{1-r_c f^{\prime 2} }-1\over r_c} \right]  + \text{higher derivs.}  
 \\&= \frac{1}{8 \pi G r_c} \int_{\partial M_2} dx \left(1 - \sqrt{1 - r_c f'^2} \right) + \text{higher derivs.} 
\end{split}
\end{equation}
To relate this to the Schwarzian theory, define the following field redefinition from \cite{Kraus:2022mnu}
\begin{equation}
    F'(x)=e^{f(x)}
\end{equation}
which yields
\begin{equation}
    \{F(x), x\} = f''(x) - \frac{1}{2} f'(x)^2.
\end{equation}
For constant $f'$, then $f'' = 0$ so
\begin{equation}
    f'(x)^2 = -2 \{F(x), x\}\,.
\end{equation}
Therefore \eqref{assdadddddadad}, up to higher Schwarzian derivatives, becomes
\begin{equation}
    I =\frac{1}{8 \pi G r_c} \int dx \left(1 - \sqrt{1 +2 r_c \{F(x), x\}  } \right) 
\end{equation}
which is identical to what \cite{Iliesiu:2020zld} found provided that we identify $r_c = \varepsilon^2$ and $\frac{1}{8\pi G} = \phi_r$.
\section{Gravitational BF Wilson lines}
\label{sec:JTgravWil}

To complete our study of Wilson lines in low dimensional gravity, we conclude by investigating Wilson lines and their correlators in BF theory under the $T\overline{T}$ deformation. 

We first study the boundary spectrum of the BF theory under the $T \overline{T}$ deformation. In the undeformed theory, between the two classes of irreducible representations of the gauge group $\operatorname{SL}(2,\mathbb{R})$ with normalizable characters \cite{Iliesiu:2019xuh}, the principal series dominates over the discrete series. This domination leads to the spectrum of the Schwarzian theory and matches with the result from the metric formalism of JT gravity.

Following the same treatment in the deformed BF theory, we find the principal series remains dominant compared to the discrete series \emph{below} the Hagedorn transition temperature, defined after \eqref{eq:Hagedorn}. The deformed theory's dynamics are captured by the $T\overline{T}$-deformed Schwarzian theory. We find that above the Hagedorn transition temperature, the discrete series dominates over the principal series,  implying the boundary theory's dynamics are no longer captured by the $T\overline{T}$-deformed Schwarzian theory.\footnote{See \cite{Barbon:2020amo,Chakraborty:2020xwo} for detailed discussions on the thermodynamics of $T\overline{T}$-deformed $1d$ quantum mechanical systems.} The correct description above the transition temperature should be some effective field theory that captures the discrete series contribution. This is consistent with the fact that the deformed partition function of the Schwarzian theory diverges at the Hagedorn temperature indicating the breakdown of the deformed Schwarzian description. Our analysis provides a glimpse into what happens across the Hagedorn transition, and understanding the entire phase diagram is an interesting and important future direction.

We then move on to study Wilson lines in BF theory. Due to the subtleties mentioned above, we will only study the correlators of Wilson lines \textit{below} the Hagedorn transition temperature where the boundary theory is captured by the deformed Schwarzian theory.

Just as the Wilson line in $3d$ Chern-Simons is conjectured to transform as a bi-local primary operator at its endpoints, a Wilson line in $2d$ BF theory in representation $\eta$ is also believed to transform as a bi-local primary operator. We indicate this schematically by writing
\begin{equation}\label{bf_theory_bilocal}
   \langle W_\eta [\mathcal{C}_{\tau_1, \tau_2}] \rangle \longleftrightarrow   \langle O_\eta (\tau_2) O_\eta(\tau_1) \rangle
\end{equation}
for a boundary-anchored path $\mathcal{C}_{\tau_1, \tau_2}$ on the disk $D$ which intersects the string defect $L$ mentioned in section \ref{sec:jt_as_bf} at points $\tau_1$ and $\tau_2$.

In the context of $3d$ gravitational Chern-Simons theory, we saw that Wilson line operators admitted interpretations from both the bulk perspective and the boundary perspective. Similar interpretations exist in the case of BF theory. In the $1d$ boundary theory, we can view the bi-local operator (\ref{bf_theory_bilocal}) as a two-point function for an operator $O_\eta$ in some matter CFT which has been coupled to the Schwarzian theory. From the viewpoint of the $2d$ bulk, the Wilson line computes a certain path integral
\begin{equation}
\label{eq:WilsonJTpointparticle}
W_\eta [\mathcal{C}_{\tau_1, \tau_2}] \cong  \int_{\operatorname{paths} \sim \mathcal{C}_{\tau_1, \tau_2}} \mathcal{D} x\, \exp \left( - m \int_{\mathcal{C}_{\tau_1, \tau_2}} ds \, \sqrt{ g_{\mu \nu} \frac{dx^\mu}{ds} \frac{dx^\nu}{ds}  } \right)
\end{equation}
involving the action for a probe particle coupled to gravity with mass $m^2 = \eta \left( \eta - 1 \right) = - C_2(\eta) $. The right-hand-side of \eqref{eq:WilsonJTpointparticle} is a functional integral weighted by the point particle action over all paths $x(s)$ diffeomorphic to $\mathcal{C}_{\tau_1, \tau_2}$.

\subsection{Single BF Wilson line}

Motivated by \eqref{eq:WilsonJTpointparticle}, one notices that the basic building block for constructing the gravitational Wilson line in BF theory is the disk partition function \cite{Mertens:2017mtv,Iliesiu:2019xuh}. We, therefore, would like to develop a formulation of the $T\overline{T}$ deformation, which is convenient for computing these disk partition functions. We first illustrate this formalism for compact groups and then generalize to the non-compact group $\operatorname{SL}(2,\mathbb{R})$. 

To produce the deformed theory whose seed is given by \eqref{eq:string defect}, we consider deforming the boundary term $\oint_L du \, V(\phi(u))$, where $V(\phi(u)) = \nu \operatorname{Tr}\phi^2$ with a constant $\nu$, 
\begin{equation} \label{eq:defVlambda} \oint_L du  \, V (\phi(u)) \longrightarrow \oint_L du \, V_\lambda(\phi(u))\,, \quad V_\lambda(\phi(u)) = \frac{1 - \sqrt{1 - 8\lambda V(\phi)}}{4\lambda}\,.
\end{equation}
To compute the disk partition function with a fixed holonomy $g = P \exp \left( \oint A \right)$, we fix the boundary value of $A_\tau$ accordingly such that \eqref{eq:on-shell boundary} vanishes to guarantee a well-defined variational principle. It is important to note that the string defect $L$ supporting the potential $\oint du \, V(\phi)$ is arbitrarily close to the boundary, but not actually \emph{on} the boundary, so that no new boundary terms appear and spoil the variational principle emphasized in \cite{Iliesiu:2019xuh}.

For a compact group with a generic potential $V(\phi)$, the disk partition function has been computed in \cite{Iliesiu:2019xuh}. Specializing to the potential $V_\lambda(\phi)$ in \eqref{eq:defVlambda}, the partition function is 
\begin{align}
    Z_\lambda(g, \nu, \beta)=\sum_{R} \left( \operatorname{dim} R \right) \chi_{R}(g) \exp \left( - \beta f^-_{\lambda}\left(\frac{\nu C_{2}(R)}{N}\right) \right) \,,
    \label{eq:BFPF}
\end{align}
where $f^-_\lambda \left( \frac{\nu C_{2}(R)}{N} \right)$  is the negative branch of the deformed JT gravity's spectrum \cite{Gross:2019ach,Gross:2019uxi}
\begin{equation}
 f_{\lambda}^{-}\left( \frac{\nu C_{2}(R)}{N} \right)=\frac{1-\sqrt{1-8 \lambda \nu C_{2}(R)/N }}{4 \lambda}\,,
\end{equation}
$C_2(R)$ is the quadratic Casimir in the representation $R$, and $\chi_R(g)$ is a character serving as a wavefunction in canonical quantization. Compared to the formula for a $2d$ Yang-Mills partition function \cite{Witten:1991we,Witten:1992xu}, the absence of surface area in the exponent in \eqref{eq:BFPF} shows that the BF theory is truly topological.

Taking the boundary holonomy $g \rightarrow \mathbb{I}$, we recover the partition function of the $T\overline{T}$-deformed quantum mechanics describing a particle-on-a-group\footnote{See \cite{Blommaert:2018oue} for more detailed discussions of Wilson lines in theories with compact gauge groups.}
\begin{equation}
    Z_\lambda(\mathbb{I},\nu,\beta) = \sum_{R}(\dim R)^2 \exp \left( -\beta f^-_\lambda\left(\frac{\nu  C_{2}(R)}{N}\right) \right)\,.
\end{equation}
We next work out the generalization of the above to non-compact groups following \cite{Iliesiu:2019xuh}. We first choose the gauge group to be 
\begin{equation} \mathcal{G}_B = \frac{\widetilde{\operatorname{SL}}(2,\mathbb{R})\times \mathbb{R}}{\mathbb{Z}}\,,
\end{equation}
as in \cite{Iliesiu:2019xuh}. The identification associated to the quotient $\mathbb{Z}$ is given by
\begin{equation}
(\tilde{g}, \theta) \sim (h_n \tilde{g}, \theta + Bn) \,,
\end{equation}
where $\tilde{g} \in \widetilde{\operatorname{SL}}(2,\mathbb{R})$, the universal cover of $SL(2,\mathbb{R})$, $\theta \in \mathbb{R}$, $h_n$ is the $n$-th element of $\mathbb{Z} \subset \widetilde{SL}(2,\mathbb{R})$, and $B\in \mathbb{R}$ defines the extension.

The irreducible representations of $\mathcal{G}_B$ are given by the irreducible representations of $\widetilde{\operatorname{SL}}(2,\mathbb{R})\times \mathbb{R}$ which are invariant under the action of elements $(h_n, B n)$, for $n \in \mathbb{Z}$, in the $\mathbb{Z}$ subgroup of $\widetilde{\operatorname{SL}}(2,\mathbb{R})\times \mathbb{R}$. The irreducible representations of $\widetilde{\operatorname{SL}}(2,\mathbb{R})$ are labeled by two quantum numbers $\eta$ and $\mu$, and the irreps of $\mathbb{R}$ are labelled by $k \in \mathbb{R}$. The irreducible representations of $\mathcal{G}_B$ are given by the irreps of $\widetilde{\operatorname{SL}}(2,\mathbb{R})\times \mathbb{R}$ satisfying
\begin{equation} \mu = - \frac{Bk}{2\pi} + p, \quad p \in \mathbb{Z}\,.
\end{equation}
The boundary BF action is modified accordingly to
\begin{equation}
I = - i \int_{M_2} \operatorname{Tr}(\phi F) - \oint^\beta_0 du V_\lambda(\phi)\,,
\end{equation}
where 
\begin{equation} A = e^1 P_1 +  e^2 P_2 + \omega P_0 + \frac{B^2}{\pi^2} A^{\mathbb{R}} \mathbb{I}, \quad \phi = \phi^1 P_1 + \phi^2 P_2 + \phi^0 P_0 + \phi^{\mathbb{R}} \mathbb{I}\,.
\end{equation}
Motivated by the results of section \ref{sec:schwarzian_type}, we choose the deformed potential $V_\lambda (\phi)$ to be
\begin{equation}
\begin{aligned}
    \label{eq:deformedpotentialsection6} V_\lambda(\phi) &= \frac{1-\sqrt{1-8\lambda \widetilde{V}(\phi^0,\phi^\pm,\phi^{\mathbb{R}})}}{4\lambda} \\&= \frac{1 - \sqrt{1 - 8\nu \lambda\left(\frac{1}{2} + \frac{1}{4} \operatorname{Tr}_{\left(2,-\frac{\pi}{B}\right)}\phi^2\right)}}{4\lambda} + O\left(\frac{1}{B}\right)\,,
\end{aligned}
\end{equation}
where we used that, in the large-$B$ limit, the potential $\widetilde{V}(\phi)$ in \eqref{eq:deformedpotentialsection6} is
\begin{equation} \widetilde{V}(\phi^0,\phi^\pm,\phi^{\mathbb{R}}) = \frac{\nu}{2} + \frac{\nu}{4} \operatorname{Tr}_{\left(2,-\frac{\pi}{B}\right)} \phi^2 + O\left(\frac{1}{B}\right)\,.
\end{equation}
By $\operatorname{Tr}_{\left(2,-\frac{\pi}{B}\right)}$, we mean the trace is taken over the 2-dimensional representation with $k = - \frac{\pi}{B}$. In the large-$B$ limit, the trace is only over $\mathfrak{sl}(2,\mathbb{R})\subset \mathfrak{sl}(2,\mathbb{R})\oplus \mathbb{R}$. For the boundary conditions, we add a boundary term 
\begin{equation} i \oint_{\partial \Sigma} \phi^{\mathbb{R}}A^{\mathbb{R}}\,, 
\end{equation}
and fix the boundary value of $\phi^{\mathbb{R}}$ to be $k_0$. The partition function $Z_{k_0}(\tilde{g},\nu, \beta)$ that we find is related to the partition function $Z((\tilde{g},\theta),\nu \beta)$ with a fixed holonomy $\tilde{g} \in \widetilde{\operatorname{SL}}(2,\mathbb{R})$ and $\theta \in \mathbb{R}$ via the Fourier transform
\begin{equation} Z_{k_0}(\tilde{g},\nu, \beta) = \int_{-\infty}^\infty d\theta Z((\tilde{g},\theta),\nu, \beta) \exp \left( -i k_0 \theta \right)\,.
\end{equation}
We are ready to compute the disk partition function $Z_{k_0}(\tilde{g},\nu, \beta)$ with $k_0 = - i$ and  $B\rightarrow \infty$ in the deformed theory.

The non-trivial irreducible unitary representations of $\widetilde{\operatorname{SL}}(2,\mathbb{R})$ consist of three types, and among the three, only the principal unitary series $C^\mu_{\eta = \frac{1}{2}+is}$ with $\mu\in[-1/2,1/2]$ and the positive/negative discrete series $D_\eta^\pm$ with $\eta=\pm\mu>0$ admit a well-defined Hermitian inner product allowing one to define a density of states given by the Plancherel measure. Taking the $\mathbb{Z}$ quotient fixes $\exp \left( 2\pi i\mu \right) = \exp \left(-iBk \right)$ (see (3.19) in \cite{Iliesiu:2019xuh}) and the disk partition function $Z(g,\nu \beta)$ receive contributions only from the above two types of representations:
\begin{equation}
\begin{aligned}
     Z(g,\nu, \beta) \propto &\int_{-\infty}^\infty dk \int_0^\infty ds \frac{s\sinh(2\pi s)}{\cosh(2\pi s) + \cos(Bk)} \chi_{(s,\mu = -\frac{Bk}{2\pi},k)}(g) e^{ -\beta f_\lambda^{-}\left(\frac{\nu s^2}{2} \right)} \nonumber \\
    &+ \sum_{n=1}^{n_{\max}} \frac{1}{2\pi^2}\left(-\frac{Bk}{2\pi} + n - \frac{1}{2}\right) \chi(g)  e^{ -\beta f_\lambda^-\left(\frac{\nu}{2}\left(\left(-\frac{Bk}{2\pi}+n\right)\left(1+\frac{Bk}{2\pi}-n\right)-\frac{1}{4}\right)\right)}\,,
\end{aligned}
\end{equation}
where the first term is the contribution from the principal series representations, the second term is the contribution from the discrete series representations, and $n_{\max}$ is a cutoff on the discrete series representations.

We consider the boundary condition $k_0 = - i$ and the limit $B\rightarrow \infty$ to compute $Z_{k_0}(\tilde{g},\nu,\beta)$. We arrive at important subtleties. In the undeformed theory, the leading order contribution in this limit comes from the principal series and scales as
\begin{equation} \frac{1}{\cosh(2\pi s) + \cos(Bk_0)} \sim e^{-B} \,.
\end{equation}
The contribution from the discrete series scales as 
\begin{equation} e^{-\frac{\nu \beta}{2} \left(\left(-\frac{Bk_0}{2\pi}\right)\left(1+\frac{Bk_0}{2\pi} - n\right)-\frac{1}{4}\right) } \sim e^{ -\frac{\nu \beta}{8\pi^2}B^2 }\,,
\end{equation}
which is subleading and can be dropped.
In the deformed theory, the scaling of the contribution from the principal series remains the same while the scaling of the contribution from the discrete series can change depending on the sign of the deformation. For $\lambda < 0$, we have
\begin{equation} e^{-\beta f_\lambda^-\left(\frac{\nu}{2}\left(\left(-\frac{Bk}{2\pi}+n\right)\left(1+\frac{Bk}{2\pi}-n\right)-\frac{1}{4} \right)\right) } \sim e^{ -\frac{\beta B}{4\pi }\sqrt{\frac{\nu}{-\lambda}} }\,.
\end{equation}
Comparing with the suppression $\exp \left( -B \right)$ from the principal series, we find the principal series remains dominant as long as 
\begin{equation} 
\label{eq:Hagedorn}
\frac{\beta}{4\pi}\sqrt{\frac{\nu}{-\lambda}} > 1\,.
\end{equation}
Consequently, we identify the critical temperature $T_c = \frac{1}{4\pi} \sqrt{\frac{\nu}{-\lambda}}$ as the temperature for the Hagedorn transition. 

For $\lambda >0$, the function
\begin{align}
    f_\lambda^-\left(\frac{\nu}{2}\left(\left(-\frac{Bk_0}{2\pi}\right)\left(1+\frac{Bk_0}{2\pi} - n\right)-\frac{1}{4}\right)\right)
\end{align}
becomes complex when $B\rightarrow \infty$, so we will not find the desired suppression. We suspect that the resolution to this issue for $\lambda >0$ is to follow an analysis similar to that of \cite{Iliesiu:2020zld}, where including the non-perturbative contribution $f^+_\lambda (E)$ makes the partition function real.

For a non-compact group, which is relevant for JT gravity, the corresponding expression for the partition function is
\begin{equation}
\begin{aligned}
    Z_\lambda (g, \nu, \beta)&=\int d R \, \rho(R) \, \chi_{R}(g) e^{- \beta f^-_{\lambda} \left( \frac{\nu C_{2}(R)}{N} \right) }\,.
    \end{aligned}
\end{equation}
Here $g$ is the holonomy, $R$ is the representation, $\chi_R$ is the character, and $\rho$ is the density of states. We note that only the energy flows via $f^-_\lambda (E)$, but the other factors in the integrand are $\lambda$-independent. This result is reminiscent of the expression for the deformed partition function in terms of an integral transformation involving the undeformed partition function and kernel discussed in \cite{Gross:2019ach,Gross:2019uxi} (see also \cite{Hashimoto:2019wct} for analogous integral kernel expressions in $2d$ theories). We write the principal series portion of the deformed Wilson line in terms of the un-normalized Wilson line anchored at $\tau_1$ and $\tau_2$ on the boundary as (schematically $E = \frac{\nu s^2}{2}$)
\begin{equation}
 \begin{aligned}
 \label{eq:singleWilson}
 \hspace{-15pt}&\langle W \left( \tau_1, \tau_2 \right) \rangle_\lambda (g) \\&= \int dh Z_\lambda (h, \nu, \tau_{21}) \chi (h) Z_\lambda \left( gh^{-1}, \nu, \tau_{12} \right)  \\
 &= \int_0^\infty ds^2_1 ds^2_2 \sinh (2\pi s_1) \sinh (2\pi s_2)  N^{s_2}_{s_1,\eta^\pm} e^{ -  \left[ \tau_{21} f^-_{\lambda} \left(\frac{\nu s_1^2}{2} \right) + \tau_{12} f^-_{\lambda}\left(\frac{\nu s_2^2}{2}\right) \right]} \\
 &= \prod_{i=1}^2 \mathscr{D}_{y_i;\lambda}|_{y_i = \tau_{i+1,i}} \int ds^2_1 ds^2_2 \sinh (2\pi s_1) \sinh (2\pi s_2) N^{s_2}_{s_1,\Lambda^\pm} e^{ - \frac{\nu}{2} \left[  y_1 s_1^2 + y_2 s_2^2 \right] } \,, 
\end{aligned}
\end{equation} 
where\footnote{$\tau_{i+1,i} \equiv \tau_{i+1} - \tau_{i}$. Here, for two-point function, $\tau_{32}  = \beta - \tau_1 + \tau_2$. In general, for $n$-point function, we would have $\tau_{n+1,n} = \beta - \sum_{i=1}^{n-1} \tau_{i+1,i}$.} the differential operator $\mathscr{D}_{y ; \lambda}$, also defined in \cite{Gross:2019uxi,Ebert:2022gyn} and later used in chapter \ref{ch:Airy} to compute the quenched free energy, is given by the infinite series of $y$-derivatives as follows\footnote{We slightly abuse the notation here. Strictly speaking, we should add an another subscript $\tau_{i+1,i}$ such that $\mathscr{D}_{y;\tau_{i+1,i};\lambda} = \exp(-\tau_{i+1,i} \sum_{m=1}^\infty c_m \lambda^m (-\partial_y)^{m+1})$, but since we will then later fix $y = \tau_{i+1,i}$, we will drop the additional subscript $\tau_{i+1,i}$.}:
\begin{equation}
    \begin{aligned}
    \label{eq:derivative}
e^{- \tau_{i+1,i} f_{\lambda}^{-}\left( \frac{\nu s_i^2}{2} \right)} &=
e^{-\tau_{i+1,i} \sum_{m=1}^{\infty} c_{m} \lambda^{m} (\nu s_i^2)^{m+1}} e^{ -\frac{\tau_{i+1,i}\nu s_i^2}{2}} \\
&= e^{ -\tau_{i+1,i} \sum_{m=1}^{\infty} c_{m} \lambda^{m}\left(-2 \partial_{y}\right)^{m+1} }  \bigg|_{y=\tau_{i+1,i}} e^{ -\frac{\nu s_i^2 y}{2}} \\
&= \mathscr{D}_{y ; \lambda} |_{y=\tau_{i+1,i}} e^{- \frac{\nu y s_i^2}{2}} \,.
\end{aligned}
\end{equation}
Here $\tau_{21} \equiv \tau_2 - \tau_1$, $\tau_{12} \equiv \beta - \tau_{21}$ and $  N^{s_2}_{s_1,\eta^\pm}$ are fusion coefficients between two continuous series representations and a discrete series representation provided in appendix D of \cite{Iliesiu:2019xuh}:
\begin{equation}
 N^{s_2}_{s_1,\eta^\pm}=\frac{\Gamma(\eta \pm is_1\pm is_2)}{\Gamma(2\eta)}\,.
\end{equation}

\subsection{Non-intersecting BF Wilson lines and local operators}

Additionally, one may also consider other examples. Again with the boundary holonomy $g \rightarrow \mathbb{I}$ and $k_0=-i$ for $n$ non-intersecting Wilson lines, we write the unrenormalized expression
\begin{equation}
    \begin{aligned}
    \label{eq:nonintersectingWilson}
    \left \langle \prod^n_{i=1} W(\tau_{2i-1}, \tau_{2i} ) \right \rangle &= \int  \left( \prod^n_{i=1} dh_i \right) \left( \prod_{i=1}^{n} Z_\lambda\left(h_{i}, \nu, \tau_{2 i, 2 i-1}\right) \bar{\chi} \left(h_{i}\right) \right) Z_\lambda \left(g\left(h_{1} \ldots h_{n}\right)^{-1}, \nu \tau_{1,2 n}\right) \\
    &= \int ds_0 \rho(s_0) \left( \prod^n_{i=1} ds_i \rho(s_i)   N^{s_0}_{s_i,\eta^\pm} \right) \\
    & \hspace{-5pt}\times e^{ - \left[\left(\sum_{i=1}^{n} f^-_\lambda \left(\frac{\nu s_{i}^{2}}{2} \right)\tau_{2 i,2i-1}\right)+f^-_{\lambda} \left(\frac{\nu s_{0}^{2}}{2} \right) \left(\beta-\sum_{i=1}^{n}\tau_{2 i,2i-1}\right)\right]}\,,
    \end{aligned}
\end{equation}
with
\begin{equation}
    \tau_{2i, 2i-1} = \tau_{2i} - \tau_{2i-1}, \quad i=1,\cdots,n
\end{equation}
as defined below \eqref{eq:derivative}, and
\begin{equation}
    \tau_{2 n, 1}\equiv\beta-\tau_{21}-\cdots-\tau_{2 n,2 n-1}
\end{equation}
is the total boundary length not enclosed by $n$ Wilson lines. Equivalent to the single Wilson line case \eqref{eq:singleWilson}, one may also express \eqref{eq:nonintersectingWilson} in terms of a product of the derivative operator defined in \eqref{eq:derivative}.

Moreover, it is interesting to consider correlators involving the topological term\footnote{The reason why it is topological can be seen from the Schwinger-Dyson equation \cite{Iliesiu:2019xuh}.} $\operatorname{Tr} \phi^2(x)$, because they are the zero-length limit of various loop or line operators. This correlator is equivalent to insertions of the Hamiltonian operator $H(x)$ at different points in the path integral:
\begin{equation}
    \begin{aligned}
    \label{eq:correlators of phi-sq}
\left\langle\operatorname{Tr} \phi^{2}\left(x_{1}\right) \cdots \operatorname{Tr} \phi^{2}\left(x_{n}\right)\right\rangle_{k_{0}} &=\left(\frac{\nu}{4}\right)^{-n}\left\langle H\left(x_{1}\right) \cdots H\left(x_{n}\right)\right\rangle_{k_{0}} \\
& = \Xi  \int^\infty_0 d s \rho(s) s^{2 n}e^{ -\frac{\nu \beta s^{2} }{2} }\\&= \left(-2 \right)^n \partial^n_{\nu \beta} Z_{k_0} (\nu \beta)\,,
\end{aligned}
\end{equation}
where the partition function is
\begin{equation}
    Z_{k_0} (\nu \beta) = \Xi \int^\infty_0 ds \rho(s) e^{ - \frac{\nu \beta s^2 }{2} }\,.
    \label{eq:insertion}
\end{equation}
The divergent factor $\Xi = \underset{x \rightarrow 1^+, \, n = 0}{\lim} \chi_{s,\mu} (g)$ is a limit of the character $\chi_{s, \mu}(\widetilde{g})$, related to  $\widetilde{\operatorname{SL}}(2,\mathbb{R})$ principal series representations, which arises from summing over all states in each continuous series irrep $\eta = \frac{1}{2} + is$.\footnote{See \cite{Iliesiu:2019xuh} for more comments on the divergent factor $\Xi$.} The independence of $x_1,\dots,x_n$ in the last line of \eqref{eq:insertion} simply reflects the topological nature of the BF theory.

In the $B \rightarrow \infty$ limit and with $k_0 = -i$, the integral \eqref{eq:correlators of phi-sq} is easily evaluated as
\begin{equation}
\left\langle\operatorname{Tr} \phi^{2}\left(x_{1}\right) \cdots \operatorname{Tr} \phi^{2}\left(x_{n}\right)\right\rangle_{k_{0}} \propto \frac{ 2^{n+\frac{3}{2}} \pi \Xi \Gamma\left(n+\frac{3}{2}\right)}{ (\nu \beta)^{ n + \frac{3}{2}}} ~{}_{1}F_{1} \left( n + \frac{3}{2} ; \frac{3}{2}; \frac{2\pi^2}{\nu \beta} \right) \,,
\end{equation}
where ${}_{1}F_{1}(a;b;z)$ is the Kummer confluent hypergeometric function defined for $n > - 1$ and $\nu \beta > 0$. The disk's density of states and partition function in JT gravity are the usual
\begin{equation}
\rho(s) = s \sinh (2\pi s), \quad    Z_{k_0} (\nu \beta) =  \Xi \left( \frac{2\pi}{\nu \beta} \right)^{\frac{3}{2}} e^{\frac{2\pi^2}{\nu \beta} }\,.
\end{equation}
In the deformed setting, the integral of concern is
\begin{equation}
    \begin{aligned}
    \label{eq:deformed correlators of phi-sq}
\Xi\int^\infty_0 d s \rho(s) s^{2n} e^{ - \beta f^-_\lambda \left(\frac{\nu s^{2}}{2} \right) }\,.
    \end{aligned}
\end{equation}
A similar integral was also evaluated in \cite{Gross:2019ach}, but now the deformed correlator
\begin{equation}
\left\langle\operatorname{Tr} \phi^{2}\left(x_{1}\right) \cdots \operatorname{Tr} \phi^{2}\left(x_{n}\right)\right\rangle_{k_{0},\lambda}    
\end{equation}
involves $\nu \beta'$-derivatives of their deformed partition function. We first express the deformed Boltzmann weight using a kernel $K(\beta, \beta')$ as
\begin{equation}
\begin{aligned}
   e^{ -\beta f^-_\lambda \left( \frac{\nu s^2}{2} \right)  } &= \int^\infty_0 d\beta' K(\beta, \beta') e^{ - \frac{\nu \beta' s^2 }{2}} \,,
\end{aligned}
\end{equation}
so that we can re-express the deformed partition function as \cite{Gross:2019ach,Gross:2019uxi}
\begin{equation}
    Z_{k_0}(\beta)_\lambda = \int^\infty_0 d\beta' K(\beta,\beta') Z_{k_0}(\beta')\,.
\end{equation}
The kernel is the inverse Laplace transform of the Boltzmann weight of the deformed theory:
\begin{equation}
\begin{aligned}
\label{eq:IntegrationKernel}
   K(\beta, \beta') &=  \frac{1}{2\pi i} \int^{i \infty}_{-i \infty} dE e^{ - \beta f^-_\lambda \left( E \right) + \beta' E } \\&=\frac{\beta}{\sqrt{-8\pi \lambda} \left( \beta'\right)^{\frac{3}{2}}} e^{\frac{\left(\beta -  \beta'\right)^2}{8\beta' \lambda} }.
\end{aligned}
\end{equation}
Then, for our integral \eqref{eq:deformed correlators of phi-sq}, we have
\begin{equation}
\begin{aligned}
\label{eq:finalphi-sq}
&\Xi\int^\infty_0 d s \rho(s) s^{2 n} e^{ -\beta f^-_\lambda \left(\frac{\nu s^{2}}{2} \right) } \\
&= \Xi\int^\infty_0 d\beta'  K(\beta, \beta') \int^\infty_0 d s \rho(s) s^{2 n} e^{ -\frac{\nu \beta' }{2} s^2} \\
&= (-2)^n \int^\infty_0 d\beta' K(\beta,\beta') \partial^n_{\nu \beta'} Z_{k_0} (\nu \beta')\,. 
\end{aligned}
\end{equation}
In other words, one can perform an integral transformation for the undeformed correlators $\left\langle\operatorname{Tr} \phi^{2}\left(x_{1}\right) \cdots \operatorname{Tr} \phi^{2}\left(x_{n}\right)\right\rangle_{k_{0}}$ against a kernel \eqref{eq:IntegrationKernel} to obtain the deformed correlators for any $n$ in principle. Equivalent to the above method \eqref{eq:finalphi-sq}, we also derive a recursion relation. We denote 
\begin{equation} \langle X \rangle \equiv \Xi \int_0^\infty ds \rho(s) X e^{-\beta f_\lambda^-\left(\frac{\nu s^2}{2}\right)}
\end{equation}
and define $F_n = \langle s^{2n} \rangle$. Then by the linearity of $\langle X\rangle$, we arrive at 
\begin{equation}
\begin{aligned}
\partial_{\beta} F_n &= - \left\langle s^{2n} \frac{1 - \sqrt{1 - 4\nu\lambda s^2}}{4\lambda} \right\rangle = - \frac{F_n}{4\lambda} + \left\langle \frac{\sqrt{1-4\nu\lambda s^2}}{4\lambda}\right\rangle\,, \nonumber \\  \partial_{\beta}^2 F_n &= \left\langle s^{2n} \left(\frac{1 - \sqrt{1 - 4\nu\lambda s^2}}{4\lambda}\right)^2 \right\rangle = \frac{1}{8\lambda^2} F_n - \frac{\nu}{4\lambda} F_{n+1} - \frac{1}{2\lambda}\left\langle \frac{\sqrt{1-4\nu\lambda s^2}}{4\lambda}\right\rangle \,.
\end{aligned}
\end{equation}
We find
\begin{equation} 
\label{eq:recursionBF}
F_{n+1} = \frac{- 4\lambda \partial_{\beta}^2 F_n - 2\partial_{\beta} F_n}{\nu} \,.
\end{equation}
From this recursion relation \eqref{eq:recursionBF}, one obtains
\begin{equation} F_{n} = \nu^{-n} (-4\lambda \partial_{\beta}^2 - 2\partial_{\beta})^n F_0 \,, 
\end{equation}
where the deformed disk partition function from \cite{Gross:2019ach,Gross:2019uxi} is 
\begin{equation}
F_0 = \Xi \frac{2\pi \beta}{\sqrt{-\nu \lambda}} \frac{e^{ - \frac{ \beta}{4\lambda}}}{\nu \beta^2 + 16 \pi^2 \lambda}  K_{2} \left( - \frac{\sqrt{ \beta^2 + 16 \nu^{-1}\pi^2 \lambda}}{4\lambda} \right) \,. 
\end{equation}
Here $K_2(x)$ is the modified Bessel function of the second kind and is defined up to the inverse Hagedorn temperature
\begin{equation}
    \beta_{H} = 4\pi \sqrt{\frac{-\lambda}{\nu}}\,,
\end{equation}
agreeing with \eqref{eq:Hagedorn}. 

Our analysis leads to a new understanding of the Hagedorn transition in the Schwarzian quantum mechanics. In the $B\rightarrow \infty$ limit, when turning on the $T\overline{T}$ deformation, the contribution from the principal series competes with the discrete series. Below the critical temperature $T_c = \frac{1}{4\pi} \sqrt{\frac{\nu}{-\lambda}}$, the principal series remains dominant over the discrete series, just as in the undeformed theory. Therefore, the effective boundary theory is described by $T\overline{T}$-deformed Schwarzian quantum mechanics. This critical temperature $T_c$ coincides with the critical temperature $T_{\operatorname{H}} = \frac{1}{\beta_{\operatorname{H}}}$ of the Hagedorn transition of the Schwarzian quantum mechanics. In other words, the BF description of JT gravity provides a UV completion which allows us to understand what happens when crossing the transition temperature $T_{\operatorname{H}} = T_{\operatorname{c}}$: the discrete series becomes dominant over the principal series, and therefore the boundary theory is no longer described by the $T\overline{T}$ deformation of the Schwarzian theory but rather some other $T\overline{T}$-deformed theory associated with the discrete series.

\section{Conclusion}
\label{sec: Discussion11111111111111111111111}

In this chapter, we have interpreted the dimensionally reduced $T\overline{T}$ deformation in a $1d$ theory from the perspective of its $2d$ holographic dual, which can be presented as either a JT gravity theory or a BF gauge theory.

In BF variables, we saw that the effect of this deformation depends on the boundary term (and thus the variational principle), which defines the undeformed seed theory. For one choice of boundary term, we find that a $T\overline{T}$-like deformation in the $1d$ dual causes a rotation of the sources and expectation values of the $2d$ BF theory. This matches the expectation from the analogous deformation of the $3d$ gravitational Chern-Simons theory, which is dual to the ordinary $2d$ $T\overline{T}$ deformation of a CFT. For the choice of boundary term which yields the Schwarzian theory as the holographic dual, we find that the $T\overline{T}$-like deformation of the boundary can be expressed in the so-called $HT$ form, where the flow is driven by a product of the Hamiltonian (or Euclidean Lagrangian) and the corresponding Hilbert stress tensor. In the bulk, such a deformation can be interpreted either as an asymptotic field redefinition of the gauge field $A_\tau$, or as a modification of the boundary conditions relating $A_\tau$ to the BF scalar field $\phi$.

As we have stressed throughout this thesis, Wilson lines and loops are natural observables in gauge theories, including the $3d$ Chern-Simons theory from the last chapter, which is classically equivalent to $\mathrm{AdS}_3$ gravity and the analogous $2d$ BF gauge theory which repackages the fields of JT gravity. We computed corrections to the Wilson line and related correlators induced by a $T\overline{T}$ deformation on the boundary. In the context of $2d$ BF theory, the deformed Wilson lines can be expressed in terms of deformed disk partition functions, and an analysis of the contributions from the principal and discrete series allows us to identify a critical temperature which is interpreted as the point of the Hagedorn transition.

We now describe a few interesting directions for future research. 

\emph{Higher order corrections in $\lambda$ and $c$ to the quantum AdS$_3$ Wilson line}

As alluded to in Section \ref{Sec:Quantum deformed AdS$_3$ Wilson line}, the deformed AdS$_3$ quantum Wilson line is computationally difficult due to the double expansion in $\lambda$ and $\frac{1}{c}$ and the regularization of the path-ordered exponential integrals. While we only have considered the leading correction to the quantum AdS$_3$ Wilson line, at $\mathcal{O}(\lambda^2 c^0)$, it is desirable to systematically study higher-order contributions at different orders in $\lambda$ and $\frac{1}{c}$ as automated for $\lambda = 0$ in Section 5 of \cite{Besken:2018zro}. 
 
 For instance, when we expand the path-ordered exponential \eqref{eq:D'Hoker-Kraus} to $\mathcal{O}(\frac{1}{c^2})$ in dimension $2-\varepsilon$, one uses the deformed two-loop two-point planar stress tensor correlator \eqref{Ia} to calculate the Wilson line's loop contributions.\footnote{Here $\frac{3}{2G} + 1$  is the one-loop corrected Brown-Henneaux central charge of the $r_c = 0$ theory following the renormalization conventions in \cite{Ebert:2022cle} and $\mu$ is an unspecified renormalization parameter.}
 
In general, expanding the path-ordered exponential \eqref{eq:D'Hoker-Kraus} in powers of $\frac{\alpha(\varepsilon)}{c}$,
\begin{equation}
\begin{aligned}
 &\langle W_{\varepsilon}[0, z] \rangle_\lambda \\&=z^{2j}  N(\varepsilon) \sum_{n=0}^{\infty} \frac{(6 \alpha(\varepsilon))^{n}}{c^{n}}\int_{0}^{z} d y_{n} \cdots \int_{0}^{y_{2}} d y_{1} F_{n}\left(z ; y_{n}, \dots, y_{1}\right)\left\langle T_{zz}\left(y_{n}\right) \cdots T_{zz}\left(y_{1}\right)\right\rangle_\lambda\,,
\end{aligned}
\end{equation}
one can systematically calculate the quantum gravitational Wilson line to any order in $\lambda$ or $\frac{1}{c}$ by using (loop-corrected) deformed $n$-point planar stress tensor correlators. The tree-level higher point planar stress tensor correlators were found perturbatively in $\lambda$ by \cite{Kraus:2018xrn,Hirano:2020nwq}.

\emph{Charting the phase diagram of deformed Schwarzian theory}

As we found in Section \ref{sec:JTgravWil}, below the Hagedorn transition temperature the principal series dominates over the discrete series in the deformed BF theory, which captures the $T\overline{T}$-deformed Schwarzian theory. However, above the transition temperature, the discrete series dominates the principal series, which is consistent with the breakdown of the deformed Schwarzian theory description at and across the Hagedorn transition. One would naturally expect that, above the Hagedorn transition temperature, the correct effective theory should correspond to the spectrum of the discrete series from the BF theory. At the critical temperature, the boundary theory should capture the contributions from both the principal and discrete series since the contributions from the two are comparable at the Hagedorn temperature. Furthermore, correlation functions in these theories should have the bulk interpretation as correlation functions of Wilson lines. One could then ponder how to find the correct quantum mechanics that describe these boundary theories at and above the Hagedorn temperature.

\emph{Connecting the $2d$ Wilson lines with the 3d Wilson lines in the deformed theory}

Given the intimate and yet subtle relationship between $2d$ JT gravity and $3d$ gravity \cite{Achucarro:1993fd,Das:2017pif,Gaikwad:2018dfc,Maxfield:2020ale}, we expect it is possible to compute the correlation functions involving the Wilson line in $2d$ BF theory from the correlators of Wilson lines in the Chern-Simons description of $3d$ gravity under the $T\overline{T}$ deformation. This has been explored in the undeformed theory by \cite{Mertens:2018fds,Blommaert:2018oro,Ghosh:2019rcj}. To study this relation in the deformed theory, one possible direction is to use the result that the Wilson line in $3d$ gravity corresponds to a bi-local operator in the boundary CFT. Then one can turn on the $T\overline{T}$ deformation in the boundary CFT to study correlation functions of these bi-local operators on the torus in the same limit studied in \cite{Ghosh:2019rcj}, which leads to a Schwarzian sector for any CFT with large central charge $c$. The $T\overline{T}$-deformed CFT correlation functions on the torus were computed via conformal perturbation theory in \cite{He:2020udl}. Determining the deformed correlation functions of $2d$ Wilson lines from the correlation functions of bi-local operators \cite{Ghosh:2019rcj,He:2020udl} is desirable.

\emph{The (graded) Poisson sigma model and generalized dilaton (super)gravity}

Our work focused on a special case of the most general $2d$ dilaton gravity theory, namely JT gravity and its BF theory description. However, one could consider other kinds of models, such as those listed in the bestiary in Appendix A of \cite{Grumiller:2007ju}. One could then study $T\overline{T}$-deformations of these more general models, as \cite{Grumiller:2020fbb} did for a broad class of Maxwell-dilaton-gravity theories and showed that these theories exhibit the typical square root behavior for the deformed energy spectrum. Limiting ourselves to an action functional containing at most two derivatives, the most general bulk (Euclidean) action supplemented with the Gibbons-Hawking-York boundary term is \cite{Grumiller:2002nm,Grumiller:2007ju}
\begin{equation}
\label{eq:gen}
    I[g_{\mu \nu}, \Phi] = - \frac{1}{16 \pi G} \int d^2x \, \sqrt{g} \, \left[ \Phi R - U(\Phi) g^{\mu \nu} \nabla_\mu \Phi \nabla_\nu \Phi -2 V(\Phi) \right] - \frac{1}{8 \pi G} \int \, d \tau \, \sqrt{\gamma} \, \Phi K \,,
\end{equation}
where different $2d$ dilaton gravity models are distinguished by kinetic and potential functions $U(\Phi)$ and $V(\Phi)$ respectively.

Analogously to the Chern-Simons description of $3d$ gravity, one has a gauge-theoretic formulation of \eqref{eq:gen} as the topological Poisson sigma model \cite{Ikeda:1993fh,Schaller:1994es} with $3d$ target space. The gravitational Poisson sigma model action is\footnote{For recent works on general dilaton (super)gravity and its relation to the (graded) Poisson sigma model, see \cite{Fan:2021bwt,Grumiller:2021cwg}.}
\begin{equation}
\label{eq:PSM}
    I_{\mathrm{PSM}}\left[A_{i}, X_i \right]=\frac{1}{8 \pi G} \int\left(A_i \wedge d X^i+\frac{1}{2} P^{ij}\left(X \right) A_{i} \wedge A_{j}\right)\,.
\end{equation}
Here $X_i$ are the set of target space coordinates spanning a Poisson manifold with Poisson tensor $P^{ij} (X) = - P^{ji}(X)$ and $A_i$ are the one-form gauge fields which, in general, transform non-linearly under gauge transformations to preserve the action \eqref{eq:PSM}. Generalizing our $T\overline{T}$-deformed analysis of the BF theory and our previous studies on supersymmetric $\mathcal{N}=1,2$ quantum mechanics \cite{Ebert:2022xfh} under the framework of general dilaton supergravity theories described by a graded Poisson sigma model \eqref{eq:PSM} is an interesting direction.

\emph{Irrelevant Deformations as Recoupling Throat Regions}

The single-trace $T\overline{T}$ deformation has a well-known bulk gravity interpretation in the context of type IIB supergravity on $\mathrm{AdS}_3 \times S^3 \times T^4$ \cite{Giveon:1998ns,Kutasov:1999xu}. More specifically, consider the IIB solution for a bound state of fundamental strings and NS5-branes. The F-strings wrap a circular direction $x_5$ of the $\mathrm{AdS}_3$, whereas the NS5-branes wrap both $x_5$ and all cycles of the $T^4$.

This gravity solution is characterized by two length scales, $r_1$ and $r_5$, associated with the horizons of the F-strings and NS5-branes respectively. If we restrict to the deep bulk, where the radial $\mathrm{AdS}_3$ coordinate $r$ is small compared to both $r_1$ and $r_5$, then we are in the near-horizon region of both the strings and the five-branes. This region looks like a conventional $\mathrm{AdS}_3$ spacetime, which is dual to an ordinary CFT. The other supergravity fields are essentially spectators, since the dilaton is constant in the deep interior while the $H_3$ flux has two terms that thread the $S^3$ and $\mathrm{AdS}_3$ but is otherwise non-dynamical. Thus, this limit is effectively a solution of a three-dimensional pure gravity theory.

On the other hand, suppose that we assume $r \ll r_5$ but not necessarily $r \ll r_1$. In this limit, we are in the near-horizon region of the five-branes but not of the strings. The resulting gravity solution interpolates between a pure $\mathrm{AdS}_3$ solution at small $r$ to a linear dilaton spacetime at large $r$, with an additional parameter $\lambda$ in the solution which characterizes the slope of the linear dilaton. The holographic interpretation of such an interpolating solution is that we have deformed the dual CFT by the single-trace $T\overline{T}$ operator and flowed by the deformation parameter $\lambda$. In other words, the single-trace $T\overline{T}$ deformation has recoupled the linear dilaton throat region of the bulk spacetime.

It would be interesting to explore whether some version of the $T\overline{T}$ deformation has a similar interpretation as recoupling an intermediate region in other gravitational settings. One such setting is the near-horizon region of a near-extremal black hole in four dimensions. It was pointed out in \cite{Maldacena:2016upp} that this region is described by JT gravity on $\mathrm{AdS}_2$, which is, of course, dual to a one-dimensional Schwarzian or particle-on-a-group theory. Is there an irrelevant deformation of this one-dimensional theory, which has the interpretation of recoupling more of the throat between the near-horizon and asymptotically flat regions of the $4d$ black hole? That is, does some irrelevant operator in the $1d$ theory capture the leading corrections as we move away from the limit $\frac{r}{r_H} = 1$ in the gravity solution, where $r_H$ is the horizon radius? If so, this would suggest that irrelevant current-type deformations have a more general holographic interpretation as capturing corrections to near-horizon limits.

\chapter{$T\overline{T}$ Deformations of Supersymmetric Quantum Mechanics}
\label{ch:SUSY-QM}
 In this chapter, we define a manifestly supersymmetric version of the $T\overline{T}$ deformation appropriate for a class of $(0+1)$-dimensional theories with $\mathcal{N} = 1$ or $\mathcal{N} = 2$ supersymmetry, including one presentation of the super-Schwarzian theory which is dual to JT supergravity. These deformations are written in terms of Noether currents associated with translations in superspace, so we refer to them collectively as $f(\mathcal{Q})$ deformations. We provide evidence that the $f(\mathcal{Q})$ deformations of $\mathcal{N} = 1$ and $\mathcal{N} = 2$ theories are on-shell equivalent to the dimensionally reduced supercurrent-squared deformations of $2d$ theories with $\mathcal{N} = (0,1)$ and $\mathcal{N} = (1,1)$ supersymmetry, respectively.
In the $\mathcal{N} = 1$ case, we present two forms of the $f(\mathcal{Q})$ deformation, which drive the same flow, and clarify their equivalence by studying the analogous equivalent deformations in the non-supersymmetric setting.

\section{Introduction} \label{intro}
Supersymmetric quantum mechanics is a fruitful playground for exploring the consequences of supersymmetry in a setting that is simpler than quantum field theory. In particular, since quantum mechanics is a theory with one time dimension and no space dimensions, almost all complications involving Lorentz structure disappear.\footnote{In what follows, we will use the phrases ``supersymmetric quantum mechanics,'' ``SUSY-QM,'' and ``supersymmetric $(0+1)$-dimensional theory'' interchangeably.}

Despite this apparent simplicity, SUSY-QM exhibits great mathematical depth, including rich connections to geometry and topology. Perhaps the most famous example is the relationship between the index of a SUSY-QM theory, which encodes information about the spectrum of bosonic and fermionic ground states, and the Euler characteristic of the target space on which the quantum-mechanical particle moves \cite{WITTEN1982253}. A related, well-known example is the connection between supersymmetric quantum mechanics and Morse theory \cite{Witten:1982im}. For surveys of supersymmetric quantum mechanics, see \cite{Cooper:1994eh,Frohlich:1995mr,Bellucci:2006ts,Gaiotto:2015aoa,Castellani:2017ycm,clay_notes}.

In addition to its surprisingly deep mathematical structure, there are at least two senses in which supersymmetric quantum mechanics is somehow ``generic'' or ``universal'':

\begin{enumerate}

    \item  Such theories encode the worldline dynamics of a supersymmetric particle, like the Brink-Schwarz superparticle and related models, which are pointlike analogs of the superstring \cite{Brink:1976sz,BRINK1981310}. But in fact, the worldline theory of \emph{any} spinning particle is (locally) supersymmetric, even if the target spacetime does not possess any supersymmetries  \cite{Brink:1976uf,CASALBUONI197649,Barducci:1976qu,BEREZIN1977336,vanHolten:1995qt,Gagne:1996gq}. In some sense, SUSY-QM is relevant for any pointlike particle with spin.
    
    \item Supersymmetric quantum mechanics generically arises as the zero-energy sector of supersymmetric QFTs. Thus, although SUSY-QM is a simple $(0+1)$-dimensional theory, it carries information about the vacuum structure of more complicated $(d+1)$-dimensional theories for $d > 0$.
\end{enumerate}

It is desirable to learn more about phenomena in field theory, such as $T\overline{T}$, by studying their analogs in (SUSY) quantum mechanics. In particular, we are interested in a SUSY-QM presentation of the $T\overline{T}$ operator. Such an endeavor requires first understanding how the usual $2d$ $T\overline{T}$ interacts with supersymmetry and second understanding how to dimensionally reduce from $(1+1)$-dimensions to $(0+1)$-dimensions.

Manifestly supersymmetric versions of the $T\overline{T}$ operator has been presented for $2d$ field theories with $(1,1)$, $(0,1)$, $(2,2)$, or $(0,2)$ supersymmetry \cite{Chang:2018dge,Baggio:2018rpv,
Coleman:2019dvf,Chang:2019kiu,Jiang:2019hux}. The second question, about dimensional reduction, has been addressed in the non-manifestly supersymmetric context. In Gross et al. \cite{Gross:2019ach,Gross:2019uxi}, it was shown that one can solve for the spatial component $T_{xx}$ of the two-dimensional stress tensor using the $T\overline{T}$ trace flow equation, which holds for deformations of conformally invariant seed theories. Doing this allows one to dimensionally reduce along the spatial direction and obtain a flow equation for the Euclidean action $I$ of the reduced theory, which takes the familiar form we saw from \eqref{reduced_qm_flow_no_susy}

\begin{align}\label{gross_flow_eqn}
    \frac{\partial I}{\partial \lambda} = \int \, d t \, \frac{T^2}{\frac{1}{2} - 2 \lambda T} \,.
\end{align}
The solution for deformed worldline actions of this form with canonical kinetic terms, including an arbitrary number of fermionic fields $\psi^i$, was also presented in \cite{Gross:2019ach}. This result can be used to understand the deformed versions of a class of supersymmetric quantum mechanical theories, at least in component form. 

However, the additional control provided by supersymmetry is most powerful when the symmetry is made manifest, for example, by a superspace construction that geometrizes the supersymmetry transformations. Thus, it is desirable to have a superfield analog of this deformation. The goal of this chapter is to find such an analog: that is, we wish to combine the two ingredients described above to find a manifestly supersymmetric version of the dimensionally reduced $T\overline{T}$ operator for SUSY-QM theories.

In particular, we obtain versions of the flow equation (\ref{gross_flow_eqn}), which are presented directly in superspace. These deformations will be written in terms of superspace Noether currents, which contain the Hamiltonian, and we typically represent them with variables like $\mathcal{Q}$. For this reason, we refer to this class of operators as $f(\mathcal{Q})$ deformations.

For $\mathcal{N} = 2$ theories, the corresponding Noether currents $\mathcal{Q}, \bar{\mathcal{Q}}$ are complex. Thus, we also refer to the $\mathcal{N} = 2$ version of the $f(\mathcal{Q})$ operator as the $f(\mathcal{Q}, \bar{\mathcal{Q}})$ deformation. We eventually see that it takes the form:
\begin{align}\label{susy_qq_deformation_definition}
    \frac{\partial I}{\partial \lambda} = \int \, d t \, d^2 \theta \, \frac{\mathcal{Q} \bar{\mathcal{Q}}}{\frac{1}{2} - 2 \lambda \bar{D} \mathcal{Q}} \,,
\end{align}
where the precise definition of the superfield $\mathcal{Q}$ will be given later.
 We will refer to the integrand appearing in (\ref{susy_qq_deformation_definition}) as $f(\mathcal{Q}, \bar{\mathcal{Q}})$:
 \begin{align}
     f(\mathcal{Q}, \bar{\mathcal{Q}}) \equiv \frac{\mathcal{Q} \bar{\mathcal{Q}}}{\frac{1}{2} - 2 \lambda \bar{D} \mathcal{Q}} \,.
 \end{align}
For $\mathcal{N} = 1$ theories, we will present two equivalent forms of the appropriate $f(\mathcal{Q})$ flow,
\begin{align}\label{n_equals_one_intro}
    \frac{\partial I}{\partial \lambda}  = \int \, dt \, d \theta \, \frac{\widetilde{\mathcal{Q}_\theta} \mathcal{Q}_t}{1 + 2 \lambda \mathcal{Q}_t} \quad \text{ and } \quad \frac{\partial I}{\partial \lambda} = \frac{1}{2} \int \, dt \, d \theta \, \mathcal{Q}_\theta \mathcal{Q}_t \,,
\end{align}
which will likewise be defined later. Due to the second expression in (\ref{n_equals_one_intro}), the $\mathcal{N} = 1$ version of the $f(\mathcal{Q})$ operator will also be referred to as the $\mathcal{Q}_\theta \mathcal{Q}_t$ deformation. Although the two forms of the deformation in (\ref{n_equals_one_intro}) look very different, we will see that they lead to the same superspace flow equation for a free scalar. This surprising equivalence is held by a rewriting of the non-supersymmetric deformation (\ref{gross_flow_eqn}). In particular, recall from the last chapter \eqref{HTrelation} for a certain class of quantum mechanical theories, it turns out that:
\begin{align}\label{HT_equivalence_intro}
    \frac{H^2}{\frac{1}{2} - 2 \lambda H} = - \frac{1}{2} H T^{(\text{Hilb})} \,,
\end{align}
where $T^{(\text{Hilb})}$ is the (Euclidean) Hilbert stress tensor computed from the Euclidean Lagrangian $H$. Therefore, it is equivalent to deform by either the rational function of $H$ appearing on the left side of (\ref{HT_equivalence_intro}) (whose $\mathcal{N} = 1$ superspace version is $\frac{\widetilde{\mathcal{Q}_\theta} \mathcal{Q}_t}{1 + 2 \lambda \mathcal{Q}_t}$) or to the simple product on the right side of (\ref{HT_equivalence_intro}) (whose $\mathcal{N} = 1$ superspace version is $\mathcal{Q}_\theta \mathcal{Q}_t$), as we explain later.

Besides making the supersymmetry of the deformed theory manifest, this procedure has the additional advantage that the supercharges of the deformed theory continue to act in a canonical way on superfields, whereas, in a component presentation of the deformed quantum mechanics, the supercharges $Q_i$ must be corrected order-by-order in $\lambda$. 

A final piece of motivation for performing this analysis is the relationship between certain $(0+1)$-dimensional theories and higher-dimensional gauge and gravity theories. For instance, the JT gravity theory (which descends via dimensional reduction from $3d$ gravity on $\mathrm{AdS}_3$ \cite{Maxfield:2020ale,Gross:2019ach,Achucarro:1993fd} and shown in the last chapter) is related to the Schwarzian theory as suggested in \cite{Jensen:2016pah,Maldacena:2016upp,Engelsoy:2016xyb}; the Schwarzian itself can be written as the theory of a particle moving on an SL$(2, \mathbb{R})$ group manifold \cite{Mertens:2017mtv,Mertens:2018fds}. JT gravity can also be written in BF variables as a two-dimensional gauge theory \cite{PhysRevLett.63.834,Chamseddine:1989yz}, and the interpretation of the $T\overline{T}$ deformation in this setting was explored in the last chapter. One would like to understand the action of $T\overline{T}$ deformations in all of these related theories, both with and without supersymmetry. The present chapter represents one step towards such an understanding, where we study the manifestly supersymmetric version of the deformation in the simplest member of this family of related theories, i.e. $(0+1)$-dimensional quantum mechanics.

The layout of this chapter is as follows. Section \ref{sec:met1} pursues one method of obtaining deformed SUSY-QM theories, namely first solving the superspace flow equation for a simple class of models in $2d$ and then dimensionally reducing the result to quantum mechanics. In section \ref{method2}, we instead dimensionally reduce the supercurrent-squared operator itself to produce a candidate superspace deformation for theories in $(0+1)$-dimensions. The main result of this chapter is section \ref{method3}, where we use a Noether procedure to construct a superspace deformation directly in the superspace of an $\mathcal{N}=2$ quantum mechanics theory, and check that this deformation is consistent with the dimensional reductions of the preceding sections. In section \ref{sec:n_equals_one}, we present an abridged version of this analysis for theories with $\mathcal{N} = 1$ supersymmetry, including defining two equivalent forms of the appropriate deformation and comparing the deformed theory of a single scalar to the dimensional reduction of the corresponding deformed $2d$ $\mathcal{N} = (0, 1)$ theory. Section \ref{sec: Discussion} concludes with a summary of this chapter's results. We have relegated certain details to appendix \ref{app:susyqms}, including conventions in appendix \ref{sec:conventions},  a change of conventions from real to complex supercurrents in appendix \ref{app:change_to_complex}, and an example of a non-supersymmetric dimensional reduction of a theory with a potential in appendix \ref{app:no_trace_flow}.

\section{Description of models and deformation methods}\label{subsec:description}

In previous chapters, we have motivated the study of purely kinetic Lagrangians of the form:
\begin{align}\label{later_purely_kinetic}
    I = \frac{1}{2} \int \, d t \, g_{ij} ( x ) \dot{x}^i \dot{x}^j \,, 
\end{align}
and noted that they can be deformed by the dimensionally-reduced $T\overline{T}$ operator via
\begin{align}
    \frac{\partial I}{\partial \lambda} = \int \, d t \, \frac{T^2}{\frac{1}{2} - 2 \lambda T} \,.
\end{align}
Next, we will recall how to embed theories whose bosonic parts take the form (\ref{later_purely_kinetic}) into superfields. For concreteness, we will focus on $\mathcal{N} = 2$ supersymmetric quantum mechanics (i.e. $2$ real supercharges or $1$ complex supercharge). Consider a collection of $\mathcal{N} = 2$ superfields with expansions
\begin{align}
    X^i ( t, \theta, \bar{\theta} ) = x^i ( t ) + \theta \, \psi^i ( t ) - \bar{\theta} \bar{\psi}^i ( t ) + \theta \bar{\theta} F^i ( t ) \,.
\end{align}
The superspace action whose bosonic part reduces to (\ref{later_purely_kinetic}) is
\begin{align}\label{susy_particle_on_group}
    I = \frac{1}{2} \int \, dt \, d \bar{\theta} \, d \theta \, g_{ij} ( X ) \, \left( D X^i ( t, \theta , \bar{\theta} ) \right)^{\ast} \, D X^j ( t , \theta , \bar{\theta} ) \,.
\end{align}
By performing the integration over the anticommuting coordinates $\theta, \bar{\theta}$, one can show that the superspace action (\ref{susy_particle_on_group}) reduces to the component form
\begin{align}\label{susy_kinetic_components}
    I = \frac{1}{2} \int \, dt \, &\Big( g_{ij} \dot{x}^i \dot{x}^j + g_{ij} F^i F^j + 2 i g_{ij} \bar{\psi}^i \dot{\psi}^j + \bar{\psi}^i \psi^j \dot{x}^k \left( \partial_k g_{ij} + \partial_j g_{ik} - \partial_i g_{jk} \right) \nonumber \\
    &\quad + \bar{\psi}^i \psi^j F^k \left( \partial_k g_{ij} - \partial_j g_{ik} - \partial_i g_{jk} \right) - \bar{\psi}^i \psi^j \bar{\psi}^k \psi^l \partial_k \partial_l g_{ij} \Big) \,.
\end{align}
From (\ref{susy_kinetic_components}), we see that the equations of motion for the auxiliary fields $F^i$ are purely algebraic. On-shell, they can be eliminated in terms of fermions via the equation of motion:
\begin{align}
    F^i = \Gamma^i{}_{jk} \bar{psi}^j \psi^k \,,
\end{align}
where $\Gamma^i{}_{jk}$ are the Christoffel symbols associated with the metric $g_{ij}$. Similarly, it is convenient to define a covariant derivative $\nabla_t$ with the property that $\psi^i$ transform as vectors:
\begin{align}
    \nabla_t \psi^j = \dot{\psi}^j + \Gamma^j{}_{lm}  \psi^l \dot{x}^m \,.
\end{align}
The terms involving four fermions can be written in terms of the Riemann curvature tensor $R_{ijkl}$ which is computed from the metric $g_{ij}$ in the usual way. In terms of these new quantities, the action can be written more compactly as
\begin{align}\label{susy_kinetic_components_final}
    I = \frac{1}{2} \int \, dt \, \left( g_{ij} \dot{x}^i \dot{x}^j + 2 i g_{ij} \bar{\psi}^i \nabla_t \psi^j + \frac{1}{2} R_{ikjl} \bar{\psi}^i \psi^j \bar{\psi}^k \psi^l \right) \,.
\end{align}
This theory, therefore, reduces to the theory of a collection of bosonic degrees of freedom $x^i$ and their fermionic superpartners $\psi^i$. The $x^i$ are subject to the purely kinetic Lagrangian (\ref{later_purely_kinetic}) as desired whereas the fermions have both kinetic terms and four-fermion couplings determined by the Riemann curvature of the target space.

In the remainder of this work, we will restrict our attention to supersymmetric $T\overline{T}$-type deformations of seed theories which take the form (\ref{susy_particle_on_group}). There are three, na\"ively different ways in which one could study supersymmetric current-squared deformations of this $(0+1)$-dimensional theory.
\begin{enumerate}
    \item Write a flow equation for a $(1+1)$-dimensional field theory which reduces to (\ref{susy_particle_on_group}) using the supercurrent-squared operator. Solve this flow equation in the parent $(1+1)$-dimensional theory, and only after finding the full deformed solution, dimensionally reduce the result to quantum mechanics. This will be explored in section \ref{sec:met1}.

    \item Begin with the definition of the supercurrent-squared operator in a $(1+1)$-dimensional theory. Apply dimensional reduction to this operator itself, thus defining a deformation of the $(0+1)$-dimensional theory. We perform this procedure in section \ref{method2}.

    \item Work directly in the superspace of the quantum mechanics theory. Construct a conserved superfield that contains the Hamiltonian and then define an appropriate superspace deformation using bilinears in this superfield with the property that this flow equation reduces to (\ref{gross_flow_eqn}) after integrating out the anticommuting directions (and imposing on-shell conditions). This is done in section \ref{method3}.
\end{enumerate}
A priori, it is not clear that these three procedures are equivalent since one might imagine that the process of performing dimensional reduction does not commute with the process of deforming by a supercurrent-squared operator. However, in the following sections, we will provide evidence that the three approaches yield the same deformation on-shell.

\section{Dimensional reduction of solution to $2d$ flow}
\label{sec:met1}

In this section, we will directly solve the supercurrent-squared flow equation in the $2d, \, \mathcal{N} = (1, 1)$ field theory and then dimensionally reduce the result. This is a slight generalization of the analysis for a single $\mathcal{N} = (1,1)$ superfield whose flow equation was studied in \cite{Baggio:2018rpv,Chang:2018dge,Ferko:2021loo}. Although much of this analysis has appeared before, we review it here to make the present work self-contained and to provide a check on our results in section \ref{method3}.
\subsection{Definition of supercurrents}

Consider a general superspace Lagrangian $\mathcal{A}$ which depends on a collection of superfields $\Phi^i$ and their derivatives as
\begin{align}
    \mathcal{A} = \mathcal{A}\left( \Phi^i , D_+ \Phi^i, D_- \Phi^i, \partial_{++} \Phi^i, \partial_{--} \Phi^i, D_+ D_- \Phi^i \right) \,.
\end{align}
A general variation $\delta \mathcal{A}$ of this superspace Lagrangian can be written as
\begin{equation}\label{general_superspace_variation} 
\begin{split}
	\delta \mathcal{A} &= D_+ \left( \delta \Phi^i \frac{\delta \mathcal{A}}{\delta (D_+ \Phi^i)} \right) + D_- \left( \delta \Phi^i \frac{\delta \mathcal{A}}{\delta (D_- \Phi^i)} \right) + \partial_{++} \left( \delta \Phi^i \frac{\delta \mathcal{A}}{\delta (\partial_{++} \Phi^i)} \right) \cr
    &+ \partial_{--} \left( \delta \Phi^i \frac{\delta \mathcal{A}}{\delta (\partial_{--} \Phi^i)} \right) \cr & + \frac{1}{2} \left( D_+ \left( \frac{\delta \mathcal{A}}{\delta (D_+ D_- \Phi^i)} D_- \delta \Phi^i \right) + D_- \left( \delta \Phi^i D_+ \frac{\delta \mathcal{A}}{\delta (D_+ D_- \Phi^i)} \right) \right) \cr
    &- \frac{1}{2} \left( D_- \left( \frac{\delta \mathcal{A}}{\delta (D_+ D_- \Phi^i)} D_+ \delta \Phi^i \right) + D_+ \left( \delta \Phi^i D_- \frac{\delta \mathcal{A}}{\delta (D_+ D_- \Phi^i)} \right)  \right) \cr
    &- \delta \Phi^i \left( - \frac{\delta \mathcal{A}}{\delta \Phi^i} + D_+ \frac{\delta \mathcal{A}}{\delta (D_+ \Phi^i)} + D_- \frac{\delta \mathcal{A}}{\delta (D_- \Phi^i)} + \partial_{++} \frac{\delta \mathcal{A}}{\delta (\partial_{++} \Phi^i)} + \partial_{--} \frac{\delta \mathcal{A}}{\delta (\partial_{--} \Phi^i)}  \right. \cr & \left. - D_+ D_- \frac{\delta \mathcal{A}}{\delta (D_+ D_- \Phi^i)} \right) . 
\end{split} 
\end{equation}
First, this general variation (\ref{general_superspace_variation}) can be used to derive the superspace equations of motion for each of the $\Phi^i$. After performing some superspace integrations by parts and collecting the terms proportional to each $\delta \Phi^i$, we find that the overall variation $\delta \mathcal{A}$ will vanish for a general variation of the superfield $\Phi^i$ if
\begin{align}\label{susy_eom}
	\frac{\delta \mathcal{A}}{\delta \Phi^i} & =  D_+ \left( \frac{\delta \mathcal{A}}{\delta (D_+ \Phi^i)} \right) + D_- \left( \frac{\delta \mathcal{A}}{\delta (D_- \Phi^i)} \right) + \partial_{++} \left( \frac{\delta \mathcal{A}}{\delta (\partial_{++} \Phi^i)} \right) \cr & + \partial_{--} \left( \frac{\delta \mathcal{A}}{\delta (\partial_{--} \Phi^i)} \right) - D_+ D_- \left( \frac{\delta \mathcal{A}}{\delta (D_+ D_- \Phi^i)} \right) ,
\end{align}
which is exactly the $\Phi^i$ equation of motion. A related calculation can be used to find the superspace Noether current for spatial translations. Consider a spacetime translation $\delta x^{\pm \pm} = a^{\pm \pm}$ where the parameters $a^{\pm \pm}$ are constants. For such a translation, the variations appearing in (\ref{general_superspace_variation}) can be replaced as $\delta \mathcal{A} = a^{++} \partial_{++} \mathcal{A} + a^{--} \partial_{--} \mathcal{A}$ and likewise for $\delta \Phi^i$, $D_{\pm} \delta \Phi^i$, and so on. Restricting to the case of on-shell variations so that we can discard the term proportional to the superspace equations of motion, one finds that the resulting equation can be written as
\begin{align}
    0 = a^{++} \left( D_+ \mathcal{T}_{++-} + D_- \mathcal{T}_{+++} \right) + a^{--} \left( D_+ \mathcal{T}_{---} + D_- \mathcal{T}_{--+} \right) \,, 
\end{align}
where the components of $\mathcal{T}$ are given by
\begin{align}
	\mathcal{T}_{++-} &= \partial_{++} \Phi^i \frac{\delta \mathcal{A}}{\delta (D_+ \Phi^i)} + D_+ \left(  \partial_{++} \Phi^i \frac{\delta \mathcal{A}}{\delta (\partial_{++} \Phi^i)} \right) + \frac{1}{2} \frac{\delta \mathcal{A}}{\delta (D_+ D_- \Phi^i)} D_- \left( \partial_{++} \Phi^i  \right) \nonumber \\
	&\quad - \frac{1}{2} \partial_{++} \Phi^i D_- \left( \frac{\delta \mathcal{A}}{\delta (D_+ D_- \Phi^i)} \right) - D_+ \mathcal{A} \,, \nonumber \\
    \mathcal{T}_{+++} &= \partial_{++} \Phi^i \frac{\delta \mathcal{A}}{\delta (D_- \Phi^i)} + D_- \left(  \partial_{++} \Phi^i  \frac{\delta \mathcal{A}}{\delta (\partial_{--} \Phi^i)} \right)  - \frac{1}{2} \frac{\delta \mathcal{A}}{\delta (D_+ D_- \Phi^i)} D_+ \left( \partial_{++} \Phi^i \right) \nonumber \\
    &\quad + \frac{1}{2} \partial_{++} \Phi^i D_+ \left( \frac{\delta \mathcal{A}}{\delta (D_+ D_- \Phi^i)} \right)\,, \nonumber \\
    \mathcal{T}_{---} &= \partial_{--} \Phi^i \frac{\delta \mathcal{A}}{\delta (D_+ \Phi^i)} + D_+ \left( \partial_{--} \Phi^i \frac{\delta \mathcal{A}}{\delta (\partial_{++} \Phi^i)} \right) + \frac{1}{2} \frac{\delta \mathcal{A}}{\delta (D_+ D_- \Phi^i)} D_- \left( \partial_{--} \Phi^i \right) \nonumber \label{final_ttbar_general} \\
    &\quad - \frac{1}{2} \partial_{--} \Phi^i D_- \left( \frac{\delta \mathcal{A}}{\delta (D_+ D_- \Phi^i)} \right) \,,\nonumber \\
    \mathcal{T}_{--+} &=  \partial_{--} \Phi^i \frac{\delta \mathcal{A}}{\delta (D_- \Phi^i)} + D_- \left( \partial_{--} \Phi^i \frac{\delta \mathcal{A}}{\delta (\partial_{--} \Phi^i)} \right) - \frac{1}{2} \frac{\delta \mathcal{A}}{\delta (D_+ D_- \Phi^i)} D_+ \left( \partial_{--} \Phi^i \right) \nonumber  \\
    &\quad + \frac{1}{2} \partial_{--} \Phi^i D_+ \left( \frac{\delta \mathcal{A}}{\delta (D_+ D_- \Phi^i)} \right) - D_- \mathcal{A} \,. 
\end{align}
We interpret the superfield $\mathcal{T}$ as a conserved superspace supercurrent since it satisfies the conservation equations:
\begin{align}\label{susy_conservation}
    D_+ \mathcal{T}_{++-} + D_- \mathcal{T}_{+++} = 0 \,, \qquad D_+ \mathcal{T}_{---} + D_- \mathcal{T}_{--+} = 0 \,.
\end{align}
Writing the superfield equation (\ref{susy_conservation}) in components reduces to the usual conservation equation for the stress tensor, $\partial^\mu T_{\mu \nu} = 0$, along with other equations related to this one by supersymmetry.

\subsection{Supercurrent-squared flow for $n$ scalars}

Next, we define a superspace deformation, which is built from bilinears in $\mathcal{T}$. If we write the superspace Lagrangian as $\mathcal{A}$, so that
\begin{align}
    I = \int d^2 x \, d^2 \theta \, \mathcal{A} \,, 
\end{align}
then the flow equation generated by the supercurrent-squared operator is
\begin{align}\label{supercurrent_squared_our_notation}
    \frac{\partial \mathcal{A} ( \lambda )}{\partial \lambda} = \mathcal{T}_{+++}^{(\lambda)} \mathcal{T}_{---}^{(\lambda)} - \mathcal{T}_{--+}^{(\lambda)} \mathcal{T}_{++-}^{(\lambda)} \,, 
\end{align}
where the superscript $(\lambda)$ is meant to emphasize that the supercurrent $\mathcal{T}^{(\lambda)}$ must be re-computed from $\mathcal{A} ( \lambda )$ at each point along the flow, rather than using the supercurrent $\mathcal{T}^{(0)}$ of the undeformed theory, as with the ordinary $T\overline{T}$ flow.

To get intuition for the structures in the superspace Lagrangian which will be generated by this deformation, it is helpful to write out the deforming operator to leading order in $\lambda$ in a particular example. We will focus on the $2d$ field theory whose dimensional reduction produces an undeformed superspace action of the form (\ref{susy_particle_on_group}), which is the theory of a collection of superfields $\Phi^i$ with the superspace Lagrangian:
\begin{align}\label{undeformed_2d_susy_theory}
    I = \int d^2 x \, d^2 \theta \, g_{ij} ( \Phi ) \, D_+ \Phi^i D_- \Phi^j \,.
\end{align}
Computing the supercurrent components for this theory, one finds
\begin{align}
	\mathcal{T}_{++-}^{(0)} &=  \left( \partial_{++} \Phi^i \right) g_{ij} D_- \Phi^j - D_+ \left( g_{ij} D_+ \Phi^i D_- \Phi^j \right)\,, \nonumber \\
    \mathcal{T}_{+++}^{(0)} &= - \left( \partial_{++} \Phi^i \right) g_{ij} D_+ \Phi^j\,, \nonumber \\
    \mathcal{T}_{---}^{(0)} &= \left( \partial_{--} \Phi^i \right) g_{ij} D_- \Phi^j\,, \nonumber \\
    \mathcal{T}_{--+}^{(0)} &=  - \left( \partial_{--} \Phi^i \right) g_{ij} D_+ \Phi^j - D_- \left( g_{ij} D_+ \Phi^i D_- \Phi^j \right) \,.
\end{align}
Therefore, we see that the leading correction to $\mathcal{A}$ from the supercurrent-squared flow equation is $\mathcal{A}^{(0)} \longrightarrow \mathcal{A}^{(0)} + \lambda \mathcal{A}^{(1)}$ where
\begin{align}\label{leading_sc2_2d}
    \mathcal{A}^{(1)} &= \mathcal{T}_{+++}^{(0)} \mathcal{T}_{---}^{(0)} - \mathcal{T}_{--+}^{(0)} \mathcal{T}_{++-}^{(0)} \nonumber \\
    &= - g_{ij} g_{kl} \left( \partial_{++} \Phi^i \right) \left( \partial_{--} \Phi^k \right) D_+ \Phi^j D_- \Phi^l + g_{ij} g_{kl} ( \partial_{--} \Phi^i ) (\partial_{++} \Phi^k ) D_+ \Phi^j D_- \Phi^l \nonumber \\
    & - ( \partial_{--} \Phi^i ) g_{ij} D_+ \Phi^j D_+ \left( g_{kl} D_+ \Phi^k D_- \Phi^l \right) \nonumber \\&- ( \partial_{++} \Phi^i ) g_{ij} D_- \Phi^j D_- \left( g_{kl} D_+ \Phi^k D_- \Phi^l \right) \nonumber \\
    & + D_+ \left(  g_{ij} D_+ \Phi^i D_- \Phi^j \right) D_- \left( g_{kl} D_+ \Phi^k D_- \Phi^l \right)  \,.
\end{align}
The leading deformation (\ref{leading_sc2_2d}) contains terms proportional to the undeformed Lagrangian (\ref{undeformed_2d_susy_theory}) in addition to new terms that have more than two fermions. For instance, terms involving $D_+ \Phi^i D_- \Phi^j D_+ \Phi^k D_- \Phi^l$ will be generated. The full solution for the finite-$\lambda$ deformed superspace Lagrangian will, therefore, take the schematic form
\begin{align}\label{finite_lambda_2d_schematic}
    \mathcal{A}^{(\lambda)} = F_1 ( D \Phi )^2 + F_2 ( D \Phi )^4 + \cdots + F_n ( D \Phi )^{2n} \,,
\end{align}
where each of the functions $F_i$ depends on a collection of Lorentz scalars built from the $\Phi^j$ and their derivatives, and the expression $(D \Phi)^{2k}$ is shorthand for a product of the form $D_+ \Phi^{i_1} \cdots D_- \Phi^{i_{2k}}$. This expansion is only schematic; for instance, there can be multiple inequivalent ways of constructing a term $( D \Phi )^{2k}$ by changing which fields in the product are acted on by $D_+$ and which are acted on by $D_-$, and, all such inequivalent combinations can appear in principle. Three examples of Lorentz scalars on which the functions $F_i$ can depend are
\begin{align}\label{xyz_def}
    x &= \lambda g_{ij} ( \Phi ) \partial_{++} \Phi^i \partial_{--} \Phi^j \,, \nonumber \\
    y &= \lambda g_{ij} ( \Phi ) \left( D_+ D_- \Phi^i \right) \left( D_+ D_- \Phi^j \right) \,, \nonumber \\
    z &= \lambda^2 \left( g_{ij} \partial_{++} \Phi^i \partial_{++} \Phi^j \right) \left( g_{mn} \partial_{--} \Phi^m \partial_{--} \Phi^n \right) \,.
\end{align}
The number $n$ appearing in the highest term of (\ref{finite_lambda_2d_schematic}) is the same as the number of scalars $\Phi^i$ since the $2n$ possible derivatives of the form $D_{\pm} \Phi^i$ are all fermionic quantities and thus any term with a product of more than $2n$ such factors must vanish by nilpotency. 

The general flow equation for $\mathcal{A}^{(\lambda)}$ induced by supercurrent-squared will therefore yield a complicated set of partial differential equations relating the various $F_i$ and their derivatives with respect to the several independent scalars. We will not undertake an analysis of this general case here. However, we can make some comments about the most fermionic term in the action. First, note that there is only one independent term that one can write down involving $2n$ copies of $\Phi$, which is simply
\begin{align}
    D_+ \Phi^1 D_- \Phi^1 \, \cdots \, D_+ \Phi^n D_- \Phi^n \,.
\end{align}
This is in contrast to other terms like $( D \Phi )^4$ for which \emph{a priori} it appears that multiple inequivalent expressions can be written down, like
\begin{align}
    D_+ \Phi^i D_- \Phi^j D_+ \Phi^k D_- \Phi^l \text{ and } D_+ \Phi^i D_+ \Phi^j D_- \Phi^k D_- \Phi^l \,, 
\end{align}
which need not yield the same contribution when contracted against a general $f_{ijkl}$ without any special symmetry properties. 

Next, we claim that -- if we are willing to go partially on-shell by imposing one implication of the equations of motion in the Lagrangian -- the function $F_n$ multiplying the unique term $(D \Phi)^{2n}$ can be taken to be independent of the scalar $y$ in (\ref{xyz_def}). To see this, we will begin with the superspace equation of motion (\ref{susy_eom}) and then multiply both sides by the most fermionic term $(D \Phi)^{2n}$. The left side of the equation of motion is $\frac{\delta \mathcal{A}}{\delta \Phi^i}$, which is a sum of terms of the form
\begin{align}\label{dA_dPhi_eqn}
    \frac{\delta \mathcal{A}}{\delta \Phi^i} = \sum_k \left[ \sum_j  \frac{\partial F_k}{\partial x_j} \frac{\partial x_j}{\partial \Phi^i} ( D \Phi )^{2k} + F_k \cdot \frac{\partial ( D \Phi )^{2k}}{\partial g_{nm}} \frac{\partial g_{nm}}{\partial \Phi^i} \right] \,.
\end{align}
Here $x_j$ is the collection of Lorentz scalars that the coefficient functions $F_k$ can depend on. This equation is again only schematic, and the details of these scalars $x_j$ are not important. The only important point is that every term in (\ref{dA_dPhi_eqn}) contains at least two fermions since taking the derivative of any term in $\mathcal{A}$ with respect to some $\Phi^i$ will not change the number of fermions in that term. Therefore, when we multiply by the maximally fermionic term $(D \Phi)^{2n}$, all terms in (\ref{dA_dPhi_eqn}) vanish by nilpotency. Similarly, the two terms
\begin{align}\label{spacetime_deriv_terms}
    \partial_{++} \left( \frac{\delta \mathcal{A}^{(\lambda)}}{\delta (\partial_{++} \Phi^i)} \right) \,, \qquad  \partial_{--} \left( \frac{\delta \mathcal{A}^{(\lambda)}}{\delta (\partial_{--} \Phi^i)} \right)
\end{align}
appearing on the right side of the equation of motion will also vanish when multiplied by $(D \Phi)^{2n}$. This is because every term in either of (\ref{spacetime_deriv_terms}) will be proportional either to some product $D_+ \Phi^i D_- \Phi^j$, or to a factor of the form $ \left( \partial_{++} D_+ \Phi^i \right) D_- \Phi^j$, and in either case such a term is proportional to at least one of the $2n$ fermions $D_{\pm} \Phi^i$.

Dropping these terms that do not contribute, we can write:
\begin{align}\label{susy_eom_multiplied_intermediate}
	0 & =  ( D \Phi )^{2n} \, \left[ D_+ \left( \frac{\delta \mathcal{A}^{(\lambda)}}{\delta (D_+ \Phi^i)} \right) + D_- \left( \frac{\delta \mathcal{A}^{(\lambda)}}{\delta (D_- \Phi^i)} \right) - D_+ D_- \left( \frac{\delta \mathcal{A}^{(\lambda)}}{\delta ( D_+ D_- \Phi^i)} \right) \right] \,.
\end{align}
Furthermore, we claim that the only term in the superspace Lagrangian that affects the right side of (\ref{susy_eom_multiplied_intermediate}) is the lowest term involving $F_1$. For any term involving four or more fermions, the three combinations inside the brackets of (\ref{susy_eom_multiplied_intermediate}) will all contain at least two fermions and therefore vanish when multiplying $(D \Phi)^{2n}$. The only term which survives is the one arising from $F_1$, which gives
\begin{align}\label{susy_eom_multiplied_intermediate_two}
	0 & = 2 ( D \Phi )^{2n} \, ( g_{il} ( D_+ D_- 
\Phi^l ) ) \,  \left[ F_1  + \lambda \frac{\partial F_1}{\partial y}  g_{km} D_+ D_- \Phi^k D_+ D_- \Phi^m \right] \,.
\end{align}
Thus, when multiplying $(D \Phi)^{2n}$ and on-shell, either the combination $F_1  + y \frac{\partial F_1}{\partial y}$ vanishes or the object $g_{il} ( D_+ D_- 
\Phi^l )$ vanishes. The former cannot hold identically since it fails near the free theory where $F_1 = 1$. Therefore we conclude that the combination $g_{il} ( D_+ D_- 
\Phi^l )$ can be set to zero when multiplying $(D \Phi)^{2n}$ as a consequence of the equations of motion, and as a result, the scalar $y$ (which is proportional to this combination) can be set to zero in this context as well. In particular, since we may view the most fermionic term in the Lagrangian as a Taylor series in $y$ via
\begin{align}\label{most_fermionic_taylor}
    F_n ( D \Phi )^{2n} = \left( F_n  \Big\vert_{y=0}  + y \cdot \frac{\partial F_n}{\partial y} \Big\vert_{y=0} + \cdots \right) ( D \Phi )^{2n} \,,
\end{align}
we see that all terms but the first can be set to zero on-shell. Thus, we are free to impose that the function $F_n$ be independent of $y$ when the equations of motion are satisfied. We note that this trick of simplifying $T\overline{T}$-like flows by going partially on-shell using the superspace equations of motion was first used in the series of works \cite{Baggio:2018rpv,Chang:2019kiu,Ferko:2019oyv}. In terms of component fields, imposing this implication of the superspace equations of motion is equivalent to integrating out the auxiliary fields using their (algebraic) equations of motion. 

\subsection{Solution for one scalar}

Finally, we will specialize in a case where we can explicitly solve the flow equation and dimensionally reduce the result, which will provide a check for the $(0+1)$-dimensional deformation that we will introduce in section \ref{method3}. This is the case of a single scalar field $\Phi$. The undeformed Lagrangian is
\begin{align}
    \mathcal{A}^{(0)} = g ( \Phi ) \, D_+ \Phi \, D_- \Phi \,.
\end{align}
Following the definitions (\ref{xyz_def}) in the general case, we define
\begin{align}\label{xyz_def_later}
    x &= \lambda g ( \Phi ) \partial_{++} \Phi \partial_{--} \Phi \,, \nonumber \\
    y &= \lambda g ( \Phi ) \left( D_+ D_- \Phi \right) \left( D_+ D_- \Phi \right) \,, 
\end{align}
and make an ansatz for the finite-$\lambda$ solution of the form
\begin{align}
    \mathcal{A}^{(\lambda)} = F ( x, y ) \, g ( \Phi ) \, D_+ \Phi \, D_- \Phi \,.
\end{align}
In the case of a single scalar, the two-fermion term $D_+ \Phi D_- \Phi$ is also the most fermionic term that one can construct. Therefore, given the general result discussed around (\ref{most_fermionic_taylor}), we can assume that the function $F(x, y)$ is independent of $y$ up to terms which vanish on-shell.

Next we compute the components of the supercurrents $\mathcal{T}_{\pm \pm \pm}$, $\mathcal{T}_{\pm \pm \mp}$. Since we will drop dependence on the variable $y$, we will omit any terms proportional to $y$ in the supercurrents.\footnote{The full flow equation, including dependence on $y$, can be found in \cite{Chang:2018dge}.} We will also drop terms that are proportional to $D_+ \Phi D_- \Phi$ since every term in the supercurrents is at least proportional to either $D_+ \Phi$ or $D_- \Phi$, and therefore terms that contain both fermionic quantities will not contribute to bilinears because they vanish by nilpotency when multiplied against another component of $\mathcal{T}$. For instance, $\mathcal{T}_{++-}$ contains a term
\begin{align}
    D_+ \left(  \partial_{++} \Phi \frac{\delta \mathcal{A}}{\delta \partial_{++} \Phi} \right) = D_+ \left( \left( \partial_{++} \Phi \right) \lambda \frac{\partial F}{\partial x} \partial_{--} \Phi g(\Phi)^2 D_+ \Phi D_- \Phi \right) \,.
\end{align}
Although, in principle, this generates several terms when the $D_+$ acts on each factor, the only relevant one for bilinears is the term where it acts on $D_+ \Phi$ to give $D_+^2 \Phi = \partial_{++} \Phi$. Every other term will either be proportional to $D_+ \Phi D_- \Phi$ or to $y$ and, therefore, can be ignored.

Using the symbol ``$\sim$'' to mean ``equal up to terms which are either proportional to $y$ or do not contribute to bilinears,'' we find
\begin{align}\label{supercurrents_for_bilinears}
    \mathcal{T}_{++-} &\sim ( \partial_{++} \Phi ) F g(\Phi) D_- \Phi + x \frac{\partial F}{\partial x} ( \partial_{++} \Phi ) g ( \Phi ) D_- \Phi - F g ( \Phi ) \partial_{++} \Phi D_- \Phi \,, \nonumber \\
    \mathcal{T}_{+++} &\sim - ( \partial_{++} \Phi ) F g ( \Phi )  D_+ \Phi - x \frac{\partial F}{\partial x} ( \partial_{++} \Phi ) g ( \Phi ) D_+ \Phi \,, \nonumber \\
    \mathcal{T}_{---} &\sim ( \partial_{--} \Phi ) F g ( \Phi ) D_- \Phi + x \frac{\partial F}{\partial x} ( \partial_{--} \Phi ) g ( \Phi ) D_- \Phi \,, \nonumber \\
    \mathcal{T}_{--+} &\sim - ( \partial_{--} \Phi ) F g ( \Phi ) D_+ \Phi - x \frac{\partial F}{\partial x} ( \partial_{--} \Phi ) g ( \Phi ) D_+ \Phi + F g ( \Phi ) ( \partial_{--} \Phi ) D_+ \Phi \,.
\end{align}
On the other hand, when we ignore dependence on $y$ the $\lambda$-derivative of $\mathcal{A}$ is simply
\begin{align}
    \frac{\partial \mathcal{A}^{(\lambda)}}{\partial \lambda} = g ( \Phi )^2  ( \partial_{++} \Phi ) ( \partial_{--} \Phi ) \frac{\partial F}{\partial x} \, D_+ \Phi \, D_- \Phi \,.
\end{align}
Equating this with the combination of supercurrents appearing on the right side of (\ref{supercurrent_squared_our_notation}), using their expressions (\ref{supercurrents_for_bilinears}), we arrive at the simple differential equation:
\begin{align}
    0 = \frac{\partial F}{\partial x} + F^2 + 2 x F \frac{\partial F}{\partial x} \,.
\end{align}
The solution is
\begin{align}
    F(x) = \frac{1}{2x} \left( -1 + \sqrt{ 1 + 4 x } \right) \,.
\end{align}
Therefore, the finite-$\lambda$ solution to the flow equation is on-shell equivalent to
\begin{align}
    \mathcal{A}^{(\lambda)} = \frac{1}{2 x} \left( -1 + \sqrt{ 1 + 4 x } \right) g ( \Phi ) D_+ \Phi D_- \Phi \,.
\end{align}
For $g(\Phi) = 1$, this solution was first obtained in \cite{Baggio:2018rpv}. It is easy to dimensionally reduce this result to quantum mechanics. We assume that the superfield $\Phi ( x, t )$ is independent of the spatial coordinate $x$. It is convenient to express the spacetime derivatives $\partial_{\pm \pm}$ acting on $\Phi$ in terms of supercovariant derivatives using the algebra $D_{\pm} D_{\pm} = \partial_{\pm \pm}$, so that
\begin{align}
    x = \lambda g ( \Phi ) ( D_+ D_+ \Phi ) ( D_- D_- \Phi ) \,.
\end{align}
Following our conventions for dimensional reduction in appendix \ref{sec:conventions}, we will rotate to complex supercovariant derivatives defined by
\begin{align}
    D = \frac{1}{\sqrt{2}} \left( D_+ + i D_- \right) \,, \quad \bar{D} = \frac{1}{\sqrt{2}} \left( D_+ - i D_- \right) \,, 
\end{align}
Then, one finds that:
\begin{align}
    D_+ \Phi D_- \Phi &= i \, D \Phi \, \bar{D} \Phi \,, \nonumber \\
    D_+ D_+ \Phi = D_-D_- \Phi &= \frac{1}{2} ( D \bar{D} + \bar{D} D ) \Phi \,, \nonumber \\
    D_+ D_- \Phi &= \frac{i}{2} ( \bar{D} D - D \bar{D} ) \Phi \,.
\end{align}
In this notation, the on-shell condition which allows us to set $D_+ D_- \Phi = 0$ in terms that multiply $D_+ \Phi D_- \Phi$ means that we can replace $D \bar{D} \Phi$ with $\bar{D} D \Phi$ (and vice-versa) in terms which multiply $D \Phi \bar{D} \Phi$. Therefore, we can write $x$ in several on-shell equivalent ways as
\begin{align}
    x = \lambda g ( \Phi ) \left( D \bar{D} \Phi \right)^2 = \lambda g ( \Phi ) \left( \bar{D} D \Phi \right)^2 = \lambda g ( \Phi ) \left( D \bar{D} \Phi \right) \left( \bar{D} D \Phi \right) \,.
\end{align}
We will choose the last of these rewritings because it is more symmetrical. After dimensionally reducing and absorbing some irrelevant constant factors, we arrive at a deformed $(0+1)$-dimensional theory with the action:
\begin{align}\label{dimensionally_reduced_answer}
    I = \int \, dt \, d \theta \, d \bar{\theta} \, \frac{\left( -1 + \sqrt{ 1 + 4 \lambda g ( \Phi ) \left( D \bar{D} \Phi \right) \left( \bar{D} D \Phi \right) } \right) g ( \Phi ) \, D \Phi \, \bar{D} \Phi}{2 \lambda g ( \Phi ) \left( D \bar{D} \Phi \right) \left( \bar{D} D \Phi \right)}  \,.
\end{align}

\section{Reduction of $2d$ supercurrent-squared operator}\label{method2}

In this subsection, we will follow a slightly different strategy. Rather than solving the flow driven by supercurrent-squared in two dimensions and then dimensionally reducing the solution, we will aim to dimensionally reduce the supercurrent-squared operator itself. This will suggest a supersymmetric version of the $f(H)$ operator, which can be applied directly in the superspace of a $(0+1)$-dimensional theory. Later in section \ref{method3}, we will see how to identify this dimensionally reduced operator as a function of certain conserved superfields that can be obtained from a Noether procedure.

\subsection{Trace flow equation}

To dimensionally reduce supercurrent-squared, we would like to eliminate some of the components of the superfield $\mathcal{T}$. This process is analogous to that reviewed in section \ref{sec:schwarzian_type} for the dimensional reduction of the non-supersymmetric $T\overline{T}$. In that context, it was convenient to use the trace flow equation:
\begin{align}\label{trace_flow_later}
T^\mu{}_\mu = - 2 \lambda \left( T^{\mu \nu} T_{\mu \nu} - \left( T^\mu{}_\mu \right)^2 \right) \,, 
\end{align}
in order to solve for the spatial component $T_{xx}$ of the stress tensor as
\begin{align}\label{solve_trace}
    T^x{}_x = \frac{ T^t{}_t + 4 \lambda T^{tx} T_{tx} }{4 \lambda T^t{}_t - 1 } \,, 
\end{align}
where the coordinates of the $2d$ spacetime $(t, x)$ are related to the light-cone coordinates by $x^{\pm \pm} = \frac{1}{\sqrt{2}} \left( t \pm x \right)$. We note that the trace flow equation only holds if the seed theory is conformal.

Next, we will motivate a superspace analog of this trace flow relation. First, we recall the interpretation of the components in the expansion of the supercurrents $\mathcal{T}_{\pm \pm \pm}$, $\mathcal{T}_{\mp \mp \pm}$. It was argued in \cite{Chang:2018dge} that, on-shell, these superfields can be written as
\begin{align}
\begin{split}\label{s-multiplet}
    \mathcal{T}_{+++} &= - S_{+++} - \theta^+ T_{++++} - \theta^- Z_{++} + \theta^+ \theta^- \partial_{++} S_{-++} \,,  \\
    \mathcal{T}_{---} &= S_{---} + \theta^+ Z_{--} + \theta^- T_{----} + \theta^+ \theta^- \partial_{--} S_{+--} \,, \\
    \mathcal{T}_{++-} &= S_{-++} + \theta^+ Z_{++} + \theta^- T_{++--} - \theta^+ \theta^- \partial_{++} S_{+--} \,, \\
    \mathcal{T}_{--+} &= - S_{+--} - \theta^+ T_{++--} - \theta^- Z_{--} - \theta^+ \theta^- \partial_{--} S_{-++}\,. 
\end{split}
\end{align}
Here $T_{\mu \nu}$ is the stress tensor, $S_{\mu \alpha}$ is the conserved current associated with supersymmetry transformations,\footnote{$S_{\mu \alpha}$ is often called the supercurrent, although we reserve that term for superfields like $\mathcal{T}_{\pm \pm \pm}$.} and $Z_\mu$ is a vector which is associated with a scalar central charge.

Because we will reduce along the spatial coordinate $x$, it will be convenient to change from $x^{\pm \pm}$ to $x, t$ coordinates. First, we want to act with various $D$ operators to construct superfields whose lowest components are stress tensors. If we define
\begin{align}
    \widetilde{\mathcal{T}}_{++++} = - D_+ \mathcal{T}_{+++} \,, \qquad \widetilde{\mathcal{T}}_{----} = D_- \mathcal{T}_{---} \,, \nonumber \\
    \widetilde{\mathcal{T}}_{++--} = D_- \mathcal{T}_{++-} \,, \qquad \widetilde{\mathcal{T}}_{--++} = - D_+ \mathcal{T}_{--+} \,,
\end{align}
then the lowest components of these superfields are simply
\begin{align}\label{calt_4_index_defn}
    \widetilde{\mathcal{T}}_{++++} \Big\vert_{\theta =0 } = T_{++++} \,, \qquad \widetilde{\mathcal{T}}_{----} \Big\vert_{\theta =0 } = T_{----} \,, \nonumber \\
    \widetilde{\mathcal{T}}_{++--} \Big\vert_{\theta = 0 } = T_{++--} \,, \qquad \widetilde{\mathcal{T}}_{--++} \Big\vert_{\theta =0 } = T_{++--} \,.
\end{align}
Note that symmetry of the stress tensor implies $\tilde{\mathcal{T}}_{--++} = \tilde{\mathcal{T}}_{++--}$. Another way to see this is to note that the supercurrents are related to fields of the $S$ multiplet by $\mathcal{T}_{++-} = \chi_+$, $\mathcal{T}_{--+} = - \chi_-$, and the $S$ multiplet fields satisfy the constraint $D_- \chi_+ = D_+ \chi_-$. One can then change coordinates to $(t, x)$ to find
\begin{gather}
   \tilde{\mathcal{T}}_{tt} = \frac{1}{2} \left( \tilde{\mathcal{T}}_{++++} + 2 \tilde{\mathcal{T}}_{++--}  + \tilde{\mathcal{T}}_{----} \right) \,, \qquad \tilde{\mathcal{T}}_{tx} = \frac{1}{2} \left( \tilde{\mathcal{T}}_{++++} - \tilde{\mathcal{T}}_{----} \right) \,, \nonumber \\
    \tilde{\mathcal{T}}_{xx} = \frac{1}{2} \left( \tilde{\mathcal{T}}_{++++} - 2 \tilde{\mathcal{T}}_{++--}  + \tilde{\mathcal{T}}_{----} \right) \,.
\end{gather}
The superfield equation whose lowest component is (\ref{solve_trace}) is
\begin{align}\label{elim_cal_Txx}
    \tilde{\mathcal{T}}_{xx} = \frac{\tilde{\mathcal{T}}_{tt} + 4 \lambda \tilde{\mathcal{T}}_{tx}^2}{4 \lambda \tilde{\mathcal{T}}_{tt} - 1 } \,.
\end{align}
This is the desired superspace analog of the trace flow equation. We will assume that it holds as an exact superfield expression, at least on-shell. Furthermore, as in the dimensional reduction of the non-supersymmetric $T\overline{T}$ operator, we will assume that $T_{tx} = 0$ and therefore $\tilde{\mathcal{T}}_{tx} = 0$, which implies that
\begin{align}\label{no_Ttx_condition}
    \tilde{\mathcal{T}}_{++++} = \tilde{\mathcal{T}}_{----} \,.
\end{align}
We note that from this point on, the number of $+$ and $-$ indices in our equations no longer match because we are explicitly breaking Lorentz invariance by forcing $\tilde{\mathcal{T}}_{tx} = 0$. However, this is unproblematic since our goal is to single out the $x$ direction for dimensional reduction, which is also not a Lorentz-invariant procedure.

As a consequence of the condition (\ref{no_Ttx_condition}), we may write the other components of $\tilde{\mathcal{T}}_{\mu \nu}$ as
\begin{align}\label{calTt_simplified_consequence}
    \tilde{\mathcal{T}}_{tt} &= \tilde{\mathcal{T}}_{++++} + \tilde{\mathcal{T}}_{++--} = - D_+ \left( \mathcal{T}_{+++} + \mathcal{T}_{--+} \right) = D_- \left( \mathcal{T}_{---} + \mathcal{T}_{++-} \right)  \,, \nonumber \\
    \tilde{\mathcal{T}}_{xx} &= \tilde{\mathcal{T}}_{++++} - \tilde{\mathcal{T}}_{++--} = D_+ \left( - \mathcal{T}_{+++} + \mathcal{T}_{--+} \right) = D_- \left( \mathcal{T}_{---} - \mathcal{T}_{++-} \right) \,.
\end{align}

\subsection{Rewriting of supercurrent-squared}

Next, we would like to explain a relationship between the product $\tilde{\mathcal{T}}_{tt} \tilde{\mathcal{T}}_{xx}$ and the supercurrent-squared operator, which holds under our assumptions thus far. To organize this calculation, it is helpful to list the on-shell constraints relating to the various objects created from one supercovariant derivative acting on one component of $\mathcal{T}$. There are na\"ively $8$ such objects, but there are four constraints:
\begin{align}
    \label{relation1} &\text{By conservation, } D_- \mathcal{T}_{+++} = - D_+ \mathcal{T}_{++-} \text{ and } D_+ \mathcal{T}_{---} = - D_- \mathcal{T}_{--+} \,. \\
    \label{relation2} &\text{By the assumption that } \tilde{\mathcal{T}}_{tx} = 0 \text{, we have } D_+ \mathcal{T}_{+++} = - D_- \mathcal{T}_{---} \,, \\
    \label{relation3} &\text{By symmetry of the stress tensor, } D_- \mathcal{T}_{++-} = - D_+ \mathcal{T}_{--+}\,.
\end{align}
Therefore, there are, in fact, only four independent objects of the form $D \mathcal{T}$ after imposing these conditions. Further, by acting with a second supercovariant derivative and using the algebra $D_\pm D_{\pm} = \partial_{\pm \pm}$, $\{ D_+, D_- \} = 0$,  we obtain the added constraints that
\begin{align}\begin{split}\label{spacetime_deriv_constraints}
    \partial_{++} \mathcal{T}_{+++} + \partial_{--} \mathcal{T}_{--+} = 0 \,, &\qquad \partial_{++} \mathcal{T}_{++-} + \partial_{--} \mathcal{T}_{---} = 0 \,,  \\
    \partial_{--} \mathcal{T}_{+++} + \partial_{++} \mathcal{T}_{--+} = 0 \,, &\qquad \partial_{++} \mathcal{T}_{---} + \partial_{--} \mathcal{T}_{++-} = 0 \,.
\end{split}\end{align}
Next, recall that, since $D_{\pm} = \frac{\partial}{\partial \theta^{\pm}} + \theta^{\pm} \partial_{\pm \pm}$, we can exchange a superspace integral for a supercovariant derivative, up to a total spacetime derivative. This allows us to write:
\begin{align}\label{scsquare_reduction_intermediate}
    &\int \, d^2 x \, d^2 \theta \, \left( \mathcal{T}_{+++} \mathcal{T}_{---} - \mathcal{T}_{--+} \mathcal{T}_{++-} \right) \nonumber \\
    &\quad \sim \int \, d^2 x \, D_+ D_- \, \left( \mathcal{T}_{+++} \mathcal{T}_{---} - \mathcal{T}_{--+} \mathcal{T}_{++-} \right)  \,.
\end{align}
where $\sim$ means equivalence, assuming we can ignore boundary terms. When we write expressions like those in the last line of (\ref{scsquare_reduction_intermediate}), which involve a superfield expression that is integrated over the spacetime coordinates $d^2 x$ but not over the superspace coordinates $d^2 \theta$, it is always implied that we mean to take the lowest component of the superfield expression.

When the combination $D_+ D_-$ acts on the combination in parentheses, we generate two types of terms: (I) terms with both supercovariant derivatives acting on a single superfield $\mathcal{T}$, and (II) terms involving one supercovariant derivative acting on each factor $\mathcal{T}$. We claim that all terms of type (I) can be ignored using our constraint equations (\ref{relation1}) - (\ref{relation2}) and (\ref{spacetime_deriv_constraints}) above, up to integration by parts. We will show this by considering each term explicitly. The type I terms are
\begin{equation}
\begin{aligned}
    \label{type_I_terms}
    \hspace{-5pt}\int d^2 x \, \Big[ &( D_+ D_- \mathcal{T}_{+++} ) \mathcal{T}_{---} + \mathcal{T}_{+++} ( D_+ D_- \mathcal{T}_{---} ) \\& - ( D_+ D_- \mathcal{T}_{--+} ) \mathcal{T}_{++-} - \mathcal{T}_{--+} ( D_+ D_- \mathcal{T}_{++-} ) \Big] \,.
\end{aligned}
\end{equation}
The first term in (\ref{type_I_terms}) can be rewritten as
\begin{align}
    \int d^2 x \, ( D_+ D_- \mathcal{T}_{+++} ) \mathcal{T}_{---} &= - \int d^2 x \, ( D_+ D_+ \mathcal{T}_{++-} ) \mathcal{T}_{---} \nonumber \\
    &= - \int d^2 x \, ( \partial_{++} \mathcal{T}_{++-} ) ( \mathcal{T}_{---} ) \,.
\end{align}
In the first step, we have used the conservation equation $D_- \mathcal{T}_{+++} = - D_+ \mathcal{T}_{++-}$; in the second step, we use the algebra $D_+ D_+ = \partial_{++}$.

On the other hand, by a similar manipulation, the third term can be written as
\begin{align}
    - \int d^2 x \left( D_+ D_- \mathcal{T}_{--+} \right) \mathcal{T}_{++-} &= \int d^2 x \, \left( D_+ D_+ \mathcal{T}_{---} \right) \mathcal{T}_{++-} \nonumber \\
    &= \int d^2 x \, \left( \partial_{++} \mathcal{T}_{---} \right) \mathcal{T}_{++-} \,.
\end{align}
In the first step, we have used $D_- \mathcal{T}_{--+} = - D_+ \mathcal{T}_{---}$ and in the second step, we have again used the algebra $D_+ D_+ = \partial_{++}$. Therefore, the sum of the first and third terms is
\begin{align}
    &\int d^2 x \Big[ ( \partial_{++} \mathcal{T}_{---} ) \mathcal{T}_{++-} + \mathcal{T}_{---} ( \partial_{++} \mathcal{T}_{++-} )  \Big] \nonumber \\
    &\qquad= \int d^2 x \, \partial_{++} \left( \mathcal{T}_{---} \mathcal{T}_{++-}  \right) \,, 
\end{align}
which is a total spacetime derivative and can be ignored.

We repeat this procedure for the remaining terms. Using similar arguments, the second term can be written as
\begin{align}
    \int d^2 x \, \mathcal{T}_{+++} ( D_+ D_- \mathcal{T}_{---} ) &= - \int d^2 x \mathcal{T}_{+++} ( D_+ D_+ \mathcal{T}_{+++} ) \nonumber \\
    &= - \int d^2 x \mathcal{T}_{+++} ( \partial_{++} \mathcal{T}_{+++} )  \nonumber \\
    &= \int d^2 x \mathcal{T}_{+++} ( \partial_{--} \mathcal{T}_{--+} )\,.
\end{align}
Likewise, the fourth term is on-shell equivalent to
\begin{align}
    - \int d^2 x \, \mathcal{T}_{--+} ( D_+ D_- \mathcal{T}_{++-} ) &= \int d^2 x \, \mathcal{T}_{--+} ( D_- D_+ \mathcal{T}_{++-} ) \nonumber \\
    &= - \int d^2 x \, \mathcal{T}_{--+} ( D_- D_- \mathcal{T}_{+++} ) \nonumber \\
    &= - \int d^2 x \, \mathcal{T}_{--+} \partial_{--} \mathcal{T}_{+++} \,.
\end{align}
Adding the second and fourth terms gives
\begin{align}
    &\int d^2 x \Big[ \mathcal{T}_{+++} ( \partial_{--} \mathcal{T}_{--+} ) + (\partial_{--} \mathcal{T}_{+++} ) \mathcal{T}_{--+} \Big] \nonumber \\
    \qquad &= \int d^2 x \, \partial_{--} \left(  \mathcal{T}_{+++} \mathcal{T}_{--+} \right) \,,
\end{align}
which is again a total spacetime derivative that we will drop.

The upshot is that all terms which involve two supercovariant derivatives acting on a single $\mathcal{T}$ will drop out, up to equations of motion and total derivatives. Therefore, we are left with
\begin{align}\label{scsquare_reduction_intermediate_two}
    &\int \, d^2 x \, d^2 \theta \, \left( \mathcal{T}_{+++} \mathcal{T}_{---} - \mathcal{T}_{--+} \mathcal{T}_{++-} \right) \nonumber \\
    & \sim \int d^2 x  \resizebox{.8\hsize}{!}{$\left( D_- \mathcal{T}_{+++} D_+ \mathcal{T}_{---} - D_+ \mathcal{T}_{+++} D_- \mathcal{T}_{---} - D_- \mathcal{T}_{--+} D_+ \mathcal{T}_{++-} + D_+ \mathcal{T}_{--+} D_- \mathcal{T}_{++-} \right)$}\,.
\end{align}
The first and third terms of (\ref{scsquare_reduction_intermediate_two}) cancel after after using the conservation equation (\ref{relation1}):
\begin{align}
    &\int d^2 x \, \Big( D_- \mathcal{T}_{+++} D_+ \mathcal{T}_{---} - D_- \mathcal{T}_{--+} D_+ \mathcal{T}_{++-} \Big) \nonumber \\
    &\quad \sim \int d^2 x \, \Big( D_+ \mathcal{T}_{++-} D_- \mathcal{T}_{--+} - D_- \mathcal{T}_{--+} D_+ \mathcal{T}_{++-} \Big) = 0 \,.
\end{align}
We then arrive at the conclusion:
\begin{equation}
    \begin{aligned}
        \label{scsquare_reduction_intermediate_three}
    &\int \, d^2 x \, d^2 \theta \, \left( \mathcal{T}_{+++} \mathcal{T}_{---} - \mathcal{T}_{--+} \mathcal{T}_{++-} \right) \\& \sim \int d^2 x \, \left( D_+ \mathcal{T}_{--+} D_- \mathcal{T}_{++-}  - D_+ \mathcal{T}_{+++} D_- \mathcal{T}_{---} \right)\,.
    \end{aligned}
\end{equation}
We now compare this to the combination:
\begin{align}\label{caltsquared_on_shell}
    \tilde{\mathcal{T}}_{tt} \tilde{\mathcal{T}}_{xx} &= \left( - D_+ \mathcal{T}_{+++} - D_+ \mathcal{T}_{--+} \right) \left( D_- \mathcal{T}_{---} - D_- \mathcal{T}_{++-} \right) \, \nonumber \\
    &= D_+ \mathcal{T}_{--+} D_- \mathcal{T}_{++-}  - D_+ \mathcal{T}_{+++} D_- \mathcal{T}_{---} \,, 
\end{align}
where in the last step, we have used the second and third terms in the second line again cancel after using (\ref{relation1}) - (\ref{relation2}):
\begin{align}
    D_+ \mathcal{T}_{+++} D_- \mathcal{T}_{++-} - D_+ \mathcal{T}_{--+} D_- \mathcal{T}_{---} &= - D_- \mathcal{T}_{---} D_- \mathcal{T}_{++-} - D_+ \mathcal{T}_{--+} D_- \mathcal{T}_{---} \nonumber \\
    &= - D_- \mathcal{T}_{---} \left( D_- \mathcal{T}_{++-} + D_+ \mathcal{T}_{--+} \right) \nonumber \\
    &= 0 \,.
\end{align}
In the last step, we have used the symmetry of the stress tensor (\ref{relation3}). Therefore, comparing (\ref{scsquare_reduction_intermediate_three}) to (\ref{caltsquared_on_shell}), we see that on-shell one has
\begin{align}
    \int d^2 x \, d^2 \theta \, \left( \mathcal{T}_{+++} \mathcal{T}_{---} - \mathcal{T}_{--+} \mathcal{T}_{++-} \right) \sim \int d^2 x \, \tilde{\mathcal{T}}_{tt} \tilde{\mathcal{T}}_{xx} \,.
\end{align}
If we now impose the superfield analog of the trace flow relation (\ref{elim_cal_Txx}), setting $\tilde{\mathcal{T}}_{tx} = 0$ as we have already assumed, then on-shell we can write supercurrent-squared as
\begin{equation}
    \begin{aligned}
        \label{scsquare_reduction_intermediate_four}
    &\int d^2 x \, d^2 \theta \, \left( \mathcal{T}_{+++} \mathcal{T}_{---} - \mathcal{T}_{--+} \mathcal{T}_{++-} \right)\\
    &\quad = \int d^2 x \, \frac{\tilde{\mathcal{T}}_{tt}^2}{ 4 \lambda \tilde{\mathcal{T}}_{tt} - 1 } \\
    &\quad = \int d^2 x \, \frac{D_+ ( - \mathcal{T}_{+++} - \mathcal{T_{--+}} ) D_- ( \mathcal{T}_{---} + \mathcal{T}_{++-} ) }{ 4 \lambda D_+ ( - \mathcal{T}_{+++} - \mathcal{T_{--+}} ) - 1 }  \\
    &\quad = \int d^2 x \, D_+ D_- \Bigg[ \frac{(\mathcal{T}_{+++} + \mathcal{T_{--+}} ) ( \mathcal{T}_{---} + \mathcal{T}_{++-} ) }{ 4 \lambda D_+ ( \mathcal{T}_{+++} + \mathcal{T_{--+}} ) + 1 } \Bigg]  \\
    &\quad = \int d^2 x \, d^2 \theta \, \frac{ ( \mathcal{T}_{+++} + \mathcal{T_{--+}} ) ( \mathcal{T}_{---} + \mathcal{T}_{++-} ) }{ 4 \lambda D_+ (  \mathcal{T}_{+++} + \mathcal{T_{--+}} ) + 1 } \,.
    \end{aligned}
\end{equation}
In the second step, we have used the expressions (\ref{calTt_simplified_consequence}) for $\tilde{\mathcal{T}}_{tt}$. In the middle step, we have pulled out an overall pair of supercovariant derivatives; this manipulation relies on the fact that additional type I terms may have been generated when two supercovariant derivatives hit a single factor of $\mathcal{T}$ in the numerator all drop out by a similar calculation as the one presented above, where we saw that such type I terms in (\ref{type_I_terms}) did not contribute. There are also no additional terms generated when the supercovariant derivatives act on the denominator. We will see a simple way to understand why the denominator does not generate additional terms when we present the interpretation of this combination in terms of (complex) conserved charges in the dimensionally reduced theory.

\subsection{Dimensional reduction and interpretation}

We have now written the supercurrent-squared operator in a form that is suitable for dimensional reduction since the combination appearing in (\ref{scsquare_reduction_intermediate_three}) is a function only of $\tilde{\mathcal{T}}_{tt}$ and not of any $x$-components of the supercurrents. We may, therefore, assume that all superfields are independent of $x$ and perform the $dx$ integral, which yields a constant. We then arrive at an expression for a supercurrent-squared operator in the $(0+1)$-dimensional quantum mechanics theory:
\begin{align}\label{scsquare_reduction_intermediate_five}
    \int d t \, d^2 \theta \, \frac{ ( \mathcal{T}_{+++} + \mathcal{T_{--+}} ) ( \mathcal{T}_{---} + \mathcal{T}_{++-} ) }{ 4 \lambda D_+ ( \mathcal{T}_{+++} + \mathcal{T_{--+}} ) + 1 } \,.
\end{align}
To aid interpretation, we will define the auxiliary quantities:
\begin{align}
    \mathcal{Q}_+ = \mathcal{T}_{+++} + \mathcal{T}_{--+} \,, \qquad \mathcal{Q}_- = \mathcal{T}_{---} + \mathcal{T}_{++-} \,.
\end{align}
These satisfy the conservation equation:
\begin{align}
    D_- \mathcal{Q}_+ + D_+ \mathcal{Q}_- = D_- \mathcal{T}_{+++} + D_- \mathcal{T}_{--+} + D_+ \mathcal{T}_{---} + D_+ \mathcal{T}_{++-} = 0 \,,
\end{align}
as a consequence of the conservation equations for $\mathcal{T}$. They also obey
\begin{align}
    D_+ \mathcal{Q}_+ = D_+ \mathcal{T}_{+++} + D_+ \mathcal{T}_{--+} = - D_- \mathcal{T}_{---} - D_- \mathcal{T}_{++-} = - D_- \mathcal{Q}_- \,,
\end{align}
due to the conditions (\ref{relation2}) and (\ref{relation3}). In terms of the $\mathcal{Q}_{\pm}$, the deformation (\ref{scsquare_reduction_intermediate_five}) is
\begin{align}\label{scsquare_reduction_final}
    \int d t \, d^2 \theta \, \frac{ \mathcal{Q}_+ \mathcal{Q}_- }{ 4 \lambda D_+ \mathcal{Q}_+ + 1 } \,.
\end{align}
We define the combination under the integral in (\ref{scsquare_reduction_final}) as $f ( \mathcal{Q}_+ , \mathcal{Q}_- ) = \frac{ \mathcal{Q}_+ \mathcal{Q}_- }{ 4 \lambda D_+ \mathcal{Q}_+ + 1 }$. This is a manifestly supersymmetric deformation of the $(0+1)$-dimensional supersymmetric quantum mechanics theory constructed from objects $\mathcal{Q}_{\pm}$, which satisfy certain conservation equations and therefore resemble conserved currents.

To compare with the common conventions for supersymmetric quantum mechanics, which use a complex Grassmann coordinate $\theta, \bar{\theta}$ rather than $\theta^{\pm}$, we will now translate (\ref{scsquare_reduction_final}) into this new notation. The details of this change of variables are described in appendix \ref{app:change_to_complex}; we summarize the results. In complex coordinates, the operator (\ref{scsquare_reduction_intermediate_five}) is on-shell proportional to
\begin{align}\label{scsquare_reduction_final_complex}
    \int \, dt \, d^2 \theta \, \frac{\mathcal{Q} \bar{\mathcal{Q}}}{\frac{1}{2} - 2 \lambda \bar{D} \mathcal{Q}} \,.
\end{align}
Similarly, we call this combination $f(\mathcal{Q}, \bar{\mathcal{Q}}) = \frac{\mathcal{Q} \bar{\mathcal{Q}}}{\frac{1}{2} - 2 \lambda \bar{D} \mathcal{Q}}$. The new supercurrents $(\mathcal{Q}, \bar{\mathcal{Q}})$ satisfy the conservation equation $\bar{D} \mathcal{Q} + D \bar{\mathcal{Q}} = 0$ as shown in appendix \ref{app:change_to_complex}. Since the complex supercovariant derivatives obey the algebra $D^2 = \bar{D}^2 = 0$, one also has $D \bar{D} \mathcal{Q} = 0$ and $\bar{D} D \bar{\mathcal{Q}} = 0$. This presentation makes it more transparent that no additional terms are generated when the overall $D_+$ and $D_-$ derivatives act on the denominator of (\ref{scsquare_reduction_intermediate_four}). In complex notation, this is simply the statement that:
\begin{align}
    D \bar{D} \left( \frac{1}{4 \lambda \bar{D} \mathcal{Q} - 1} \right) = \bar{D}  D \left( \frac{1}{4 \lambda \bar{D} \mathcal{Q} - 1} \right) = 0 \,,
\end{align}
which is clear since $\bar{D}^2 \mathcal{Q} = D \bar{D} \mathcal{Q} = 0$ as we have pointed out.

Because $\mathcal{Q}, \bar{\mathcal{Q}}$ contain the single component $T$ of the stress tensor in their component expansion, the combination (\ref{scsquare_reduction_final}) can be viewed as a manifestly supersymmetric extension of the deforming operator (\ref{gross_flow_eqn}), which is now written directly in superspace. This gives a prescription for deforming any theory of supersymmetric quantum mechanics that descends via dimensional reduction from a $2d$ superconformal field theory (we have assumed that the seed theory in $2d$ is conformal to use the trace flow equation).

One shortcoming of this presentation is that we have not provided any operational method for computing the objects $\mathcal{Q}, \bar{\mathcal{Q}}$ within a given theory of supersymmetric quantum mechanics. To construct these objects using the procedure described in this section, one would need to lift such a $(0+1)$-dimensional theory to theory in $(1+1)$-dimensions, construct the supercurrents of this parent theory, and then assemble the appropriate combination of the supercurrents which appear upon reducing back down to quantum mechanics. It is, of course, desirable to have a complementary view of $\mathcal{Q}, \bar{\mathcal{Q}}$, which facilitates direct computation of these conserved superfields in a $(0+1)$-dimensional theory, ideally via some Noether procedure that provides an interpretation of these objects as conserved charges associated with time translations. We turn to this issue next.

\section{Direct definition of deformation in $1d$}\label{method3}

In this section, we will define a manifestly supersymmetric deformation directly in the superspace of a $(0+1)$-dimensional quantum mechanics theory. We first develop some formalism for defining a conserved Noether ``current,'' denoted $\mathcal{Q}$, associated with time translation invariance. Because the theory has no spatial dimensions, this conserved quantity is a charge rather than a current; however, we will still use the term ``supercurrent'' rather than ``supercharge'' for this object to avoid confusion with the supercharges $Q_i$, which represents the action of the SUSY algebra on superfields.

\subsection{Noether currents in $\mathcal{N} = 2$ theories}

We will follow the strategy of using a superspace Noether procedure, which closely parallels the discussion of section \ref{sec:met1}; we note that a similar analysis for supersymmetric quantum mechanics appeared in \cite{Clark:2001zv} for a different class of superspace Lagrangians.

Begin by considering a theory for a collection of real scalars $X^i$ described by the action:
\begin{align}\label{general_lag_susy_qm}
    I = \int \, dt \, d \theta \, d \bar{\theta} \, \mathcal{A} \left( X^i, D X^i , \bar{D} X^i , D \bar{D} X^i , \bar{D} D X^i \right) \,,
\end{align}
Although we have not allowed the superspace Lagrangian $\mathcal{A}$ to explicitly depend on the time derivatives $\dot{X}^i = \partial_t X^i$, such dependence is implicitly allowed since
\begin{align}
    D \bar{D} X^i + \bar{D} D X^i = - 2 i \dot{X}^i
\end{align}
according to our conventions for the supersymmetry algebra which are described in appendix \ref{sec:conventions}. Therefore, since $\mathcal{A}$ depends on both $D \bar{D} X^i$ and $\bar{D} D X^i$, arbitrary dependence on $\dot{X}^i$ can also be accommodated.

For the moment, we will make no additional assumptions about the superfields $X^i$ besides the reality condition $\left( X^i \right)^{\ast} = X^i$. We first consider an arbitrary variation of the superspace Lagrangian under a field fluctuation $\delta X^i$:
\begin{equation}
    \begin{aligned}
        \label{general_susy_qm_variation}
    \hspace{-15pt} \delta \mathcal{A} &= \delta X^i \, \frac{\delta \mathcal{A}}{\delta X^i}  + \delta ( D X^i ) \, \frac{\delta \mathcal{A}}{\delta ( D X^i ) }  + \delta ( \bar{D} X^i ) \, \frac{\delta \mathcal{A}}{\delta ( \bar{D} X^i ) } \\&+ \delta ( D \bar{D} X^i ) \, \frac{\delta \mathcal{A}}{\delta ( D \bar{D} X^i ) } + \delta ( \bar{D} D  X^i ) \, \frac{\delta \mathcal{A}}{\delta ( \bar{D} D  X^i ) }  \,.
    \end{aligned}
\end{equation}
It will be convenient to re-express (\ref{general_susy_qm_variation}) by writing each term as the derivative of a product minus an appropriate correction. For instance,
\begin{equation}
    \begin{aligned}
        \label{derivative_rewriting}
    \delta ( D X^i ) \, \frac{\delta \mathcal{A}}{\delta ( D X^i ) } &= D \left( \delta X^i  \, \frac{\delta \mathcal{A}}{\delta ( D X^i ) } \right) - \left( \delta X^i \right) D \left( \frac{\delta \mathcal{A}}{\delta ( D X^i ) } \right) \,,  \\
	\delta \left( D \bar{D} X^i \right) \frac{\delta \mathcal{A}}{\delta ( D \bar{D} X^i ) } &= D \left( \frac{\delta \mathcal{A}}{\delta (D \bar{D} X^i)} \bar{D} \delta X^i \right) + \bar{D} \left( \delta X^i D \frac{\delta \mathcal{A}}{\delta (D \bar{D} X^i)} \right)\\& - ( \delta X^i ) \bar{D} D  \left( \frac{\delta \mathcal{A}}{\delta (D \bar{D} X^i)} \right)\,. 
    \end{aligned}
\end{equation}
This gives
\begin{equation}
    \begin{aligned}
        \label{susy_variation_ibp}
    \hspace{-10pt} &\delta \mathcal{A} = D \left( \delta X^i  \, \frac{\delta \mathcal{A}}{\delta ( D X^i ) } \right) + \bar{D} \left( \delta X^i  \, \frac{\delta \mathcal{A}}{\delta ( \bar{D} X^i ) } \right)  \\
    &+ D \left( \frac{\delta \mathcal{A}}{\delta (D \bar{D} X^i)} \bar{D} \delta X^i \right) + \bar{D} \left( \delta X^i D \frac{\delta \mathcal{A}}{\delta (D \bar{D} X^i)} \right) \\&+ \bar{D} \left( \frac{\delta \mathcal{A}}{\delta  (\bar{D} D  X^i)} D \delta X^i \right) + D \left( \delta X^i \bar{D} \frac{\delta \mathcal{A}}{\delta (\bar{D} D X^i)} \right)  \\
    &- \delta X^i \bigg( - \frac{\delta \mathcal{A}}{\delta X^i} + D \left( \frac{\delta \mathcal{A}}{\delta ( D X^i ) } \right) + \bar{D} \left( \frac{\delta \mathcal{A}}{\delta ( \bar{D} X^i ) } \right) \\&+ D \bar{D} \left( \frac{\delta \mathcal{A}}{\delta ( D \bar{D} X^i ) } \right) + \bar{D} D  \left( \frac{\delta \mathcal{A}}{\delta (  \bar{D} D  X^i ) } \right) \bigg) \,.
    \end{aligned}
\end{equation}
One advantage of the form (\ref{susy_variation_ibp}) is that we can immediately read off the superspace equations of motion. Suppose we consider a linearized fluctuation $\delta X^i$ around a configuration $X^i$ which satisfies the equations of motion, and demand that $\delta S = \int \, dt \, d \theta \, d \bar{\theta} \, \delta \mathcal{A} = 0$. Since the terms in the first two lines are total superspace derivatives, and the final line must vanish for any $\delta X^i$, we obtain the on-shell condition:
\begin{align}\label{susy_qm_eom}
    \frac{\delta \mathcal{A}}{\delta X^i} = D \left( \frac{\delta \mathcal{A}}{\delta ( D X^i ) } \right) + \bar{D} \left( \frac{\delta \mathcal{A}}{\delta ( \bar{D} X^i ) } \right) + D \bar{D} \left( \frac{\delta \mathcal{A}}{\delta ( D \bar{D} X^i ) } \right) + \bar{D} D  \left( \frac{\delta \mathcal{A}}{\delta (  \bar{D} D  X^i ) } \right) \,.
\end{align}
Next, we study the conserved charge associated with time translations $t \to t' = t + \delta t$. Under such a transformation, the superspace Lagrangian varies as
\begin{align}
    \delta \mathcal{A} = \left( \delta t \right) \partial_t \mathcal{A} = \frac{i}{2} \left( \delta t \right) \left( D \bar{D} \mathcal{A} + \bar{D} D \mathcal{A} \right) \,,
\end{align}
where we have again used the algebra $\{ D , \bar{D} \} = - 2 i \partial_t$. Meanwhile, each superfield $X^i$ also transforms as
\begin{align}
    \delta X^i = ( \delta t ) \dot{X}^i = \frac{i}{2} \left( \delta t \right) \left( D \bar{D} X^i + \bar{D} D X^i \right) \,.
\end{align}
We use these expressions in (\ref{susy_variation_ibp}) and also restrict to the case of on-shell variations, which means that the equations of motion are satisfied, and we can discard the term proportional to $\delta X^i$ in the final line. This gives
\begin{equation}
\begin{aligned}
     0 &= - \frac{i}{2} \left( \delta t \right) \left( D \bar{D} \mathcal{A} + \bar{D} D \mathcal{A} \right) + \left( \delta t \right) D \left( \frac{i}{2}  \left( D \bar{D} X^i + \bar{D} D X^i \right)  \, \frac{\delta \mathcal{A}}{\delta ( D X^i ) } \right)  \\
    & + \frac{i}{2} \left( \delta t \right) \bar{D} \left(  \left( D \bar{D} X^i + \bar{D} D X^i \right)  \, \frac{\delta \mathcal{A}}{\delta ( \bar{D} X^i ) } \right)  + \frac{i}{2} \left( \delta t \right) D \left( \frac{\delta \mathcal{A}}{\delta (D \bar{D} X^i )} \bar{D} \left( D \bar{D} X^i  \right) \right) \\
    & + \frac{i}{2} ( \delta t ) \bar{D} \left( \frac{\delta \mathcal{A}}{\delta (\bar{D} D  X^i)} D ( \bar{D} D X^i ) \right) + \frac{i}{2} ( \delta t ) D \left( \left( D \bar{D} X^i + \bar{D} D X^i \right) \bar{D} \frac{\delta \mathcal{A}}{\delta (\bar{D} D X^i)} \right) \\
    &+ \frac{i}{2} ( \delta t ) \bar{D} \left( \left( D \bar{D} X^i + \bar{D} D X^i \right) D \frac{\delta \mathcal{A}}{\delta (D \bar{D} X^i)} \right) \,.
\end{aligned}
\end{equation}
To ease notation, we define $\eta^i = D \bar{D} X^i$, $\tilde{\eta}^i = \bar{D} D X^i$. After simplifying and collecting terms, the resulting equation can be written as
\begin{align}\label{direct_1d_Q_cons_result}
    0 &= \frac{i}{2} ( \delta t ) \, D \left[ ( \eta^i + \tilde{\eta}^i ) \left( \frac{\delta \mathcal{A}}{\delta ( D X^i ) } + \bar{D} \left( \frac{\delta \mathcal{A}}{\delta \tilde{\eta}^i} \right) \right) + \frac{\delta \mathcal{A}}{\delta \eta^i} \bar{D} \eta^i  - \bar{D} \mathcal{A} \right] \nonumber \\
    &\qquad + \frac{i}{2} ( \delta t ) \, \bar{D} \left[ ( \eta^i + \tilde{\eta}^i ) \left( \frac{\delta \mathcal{A}}{\delta ( \bar{D} X^i ) } + D \left( \frac{\delta \mathcal{A}}{\delta \eta^i} \right) \right) + \frac{\delta \mathcal{A}}{\delta \tilde{\eta}^i} D \tilde{\eta}^i - D \mathcal{A} \right] \,.
\end{align}
This can be interpreted as a superspace conservation equation of the form:
\begin{align}\label{susy_conservation_Q}
    D \bar{\mathcal{Q}} + \bar{D} \mathcal{Q} = 0 \,,
\end{align}
where
\begin{align}\label{susy_Q_def}
    \mathcal{Q} &= ( \eta^i + \tilde{\eta}^i ) \left( \frac{\delta \mathcal{A}}{\delta ( \bar{D} X^i ) } + D \left( \frac{\delta \mathcal{A}}{\delta \eta^i} \right) \right) + \frac{\delta \mathcal{A}}{\delta \tilde{\eta}^i} D \tilde{\eta}^i - D \mathcal{A} \,  , \nonumber \\
    \bar{\mathcal{Q}} &=  ( \eta^i + \tilde{\eta}^i ) \left( \frac{\delta \mathcal{A}}{\delta ( D X^i ) } + \bar{D} \left( \frac{\delta \mathcal{A}}{\delta \tilde{\eta}^i} \right) \right) + \frac{\delta \mathcal{A}}{\delta \eta^i} \bar{D} \eta^i  - \bar{D} \mathcal{A} \,.
\end{align}
We note that there is an overall factor of $i$ multiplying each term in (\ref{direct_1d_Q_cons_result}), but we have chosen to strip off this factor in defining the charges (\ref{susy_Q_def}). From the perspective of conservation properties, this is, of course, irrelevant because any scalar multiple of the combination $D \bar{\mathcal{Q}} + \bar{D} \mathcal{Q}$ still vanishes. Thus, we are free to rescale $\mathcal{Q}$ and $\bar{\mathcal{Q}}$ by any constant. However, this means that there will be a relative factor of $i$ when comparing $\mathcal{Q}$, $\bar{\mathcal{Q}}$ to $\mathcal{Q}_+$, $\mathcal{Q}_-$, in which case there was no factor of $i$ naturally appearing in the conservation equation. We will account for this rescaling when converting between conventions in appendix \ref{app:change_to_complex}.

In summary, the Noether procedure leading to (\ref{susy_Q_def}) provides a direct definition of the objects $\mathcal{Q}, \bar{\mathcal{Q}}$ obtained in (\ref{scsquare_reduction_final}) without relying upon dimensional reduction. Since $D^2 = \bar{D}^2 = 0$, the superspace conservation (\ref{susy_conservation_Q}) also implies that $D \bar{D} \mathcal{Q} = \bar{D} D \bar{\mathcal{Q}} = 0$. As mentioned above, we note that the supercurrents $\mathcal{Q}, \bar{\mathcal{Q}}$ are not to be confused with the supercharges $Q, \bar{Q}$ defined by
\begin{align}
    Q = \frac{\partial}{\partial \theta} + i \bar{\theta} \frac{\partial}{\partial t} \,, \qquad \bar{Q} = \frac{\partial}{\partial \bar{\theta}} + i \theta \frac{\partial}{\partial t} \,, 
\end{align}
which represents the action of the supersymmetry algebra on superfields.

\subsection{Definition of $f(\mathcal{Q}, \bar{\mathcal{Q}})$ deformation}

To acquire some intuition for the objects $\mathcal{Q}, \bar{\mathcal{Q}}$, it is useful to consider a simple example. The theory of a single real scalar is described by
\begin{align}
    L = \frac{m}{2} \int \, d \theta \, d \bar{\theta} \, D X \, \bar{D} X \,.
\end{align}
Setting $m=2$ for simplicity, the supercurrent $\mathcal{Q}$ and its conjugate are given by
\begin{align}\label{first_order_free_Qs}
    \mathcal{Q} &= - \left( \eta + \tilde{\eta} \right) D X - D ( D X \, \bar{D} X )  \nonumber \\
    &= - \tilde{\eta} D X \,, \nonumber \\
    \bar{\mathcal{Q}} &= \left( \eta + \tilde{\eta} \right) \bar{D} X - \bar{D} ( D X \, \bar{D} X ) \nonumber \\
    &= \eta \bar{D} X \,,
\end{align}
where we used $\bar{D} D X = - 2 i \dot{X} - D \bar{D} X$. The component expressions for these charges are
\begin{align}
    \mathcal{Q} &= \psi ( i \dot{x} - F ) - 2 i \theta \psi \dot{\psi} + \bar{\theta} \left( \dot{x}^2 + 2 i F \dot{x} - F^2 \right) + \theta \bar{\theta} \left( i \psi \dot{F} + \psi \ddot{x} - 3 \dot{x} \dot{\psi} - 3 i F \dot{\psi} \right) \,, \nonumber \\
    \bar{\mathcal{Q}} &= \bar{\psi} ( F + i \dot{x} ) + 2 i \bar{\theta} \bar{\psi} \dot{\bar{\psi}} + \theta \left( F^2 + 2 i F \dot{x} - \dot{x}^2 \right)  + \theta \bar{\theta} \left( i \bar{\psi} \dot{F} - \bar{\psi} \ddot{x} + 3 \dot{x} \dot{\bar{\psi}} - 3 i F \dot{\bar{\psi}}  \right) \,.
\end{align}
We can also compute the highest component of the product $\mathcal{Q} \bar{\mathcal{Q}}$. For simplicity, we will set the auxiliary field to zero using its equation of motion. Then
\begin{align}
    \mathcal{Q} \bar{\mathcal{Q}} \Big\vert_{\theta^2, F=0} = - 4 \psi \bar{\psi} \dot{\psi} \dot{\bar{\psi}} + 3 i \left( \psi \dot{\bar{\psi}} + \bar{\psi} \dot{\psi} \right) \dot{x}^2 + \dot{x}^4 \,.
\end{align}
We compare this with the Lagrangian of the theory written in components,
\begin{align}
    L = \dot{x}^2 + i \left( \bar{\psi} \dot{\psi} - \dot{\bar{\psi}} \psi \right) + F^2 \,.
\end{align}
To further develop our intuition, we focus on the bosonic sector. Setting $F = 0$ and $\dot{\psi} = \dot{\bar{\psi}} = 0$, we have the relation
\begin{align}
    L^2 = \mathcal{Q} \bar{\mathcal{Q}} \,.
\end{align}
Interpreting the Euclidean Lagrangian as the Hamiltonian, we see that deforming the bosonic sector by the product $\mathcal{Q} \bar{\mathcal{Q}}$ is equivalent to a deformation by $H^2$. Next, the lowest component of $\bar{D} \mathcal{Q}$ is given by
\begin{align}
    \bar{D} \mathcal{Q} \Big\vert_{\theta = \bar{\theta} = 0} = - F^2 + 2 i F \dot{x} + \dot{x}^2 \,.
\end{align}
When the auxiliary field equation of motion is satisfied, the lowest component of $\bar{D} \mathcal{Q}$ is, therefore, $\dot{x}^2$, which is the Hamiltonian for the bosonic degree of freedom.

Using the intuition that $\mathcal{Q} \bar{\mathcal{Q}}$ has $H^2$ as its top component and $\bar{D} \mathcal{Q}$ has $H$ as its bottom component, a natural guess for a combination of superfields which has the deforming operator (\ref{gross_flow_eqn}) as its top component is $\frac{\mathcal{Q} \bar{\mathcal{Q}}}{\frac{1}{2} - 2 \lambda \bar{D} \mathcal{Q}}$, which suggests the flow equation
\begin{align}\label{susy_qm_Qsquare_flow}
    \frac{\partial \mathcal{A}}{\partial \lambda} = \frac{\mathcal{Q} \bar{\mathcal{Q}}}{\frac{1}{2} - 2 \lambda \bar{D} \mathcal{Q}} \,.
\end{align}
This is exactly the form of the deformation (\ref{scsquare_reduction_final}), which we obtained by dimensionally reducing the supercurrent-squared operator in $(1+1)$-dimensions.

\subsection{Solution for one scalar}

We now provide evidence that this is on-shell equivalent to the deformation (\ref{gross_flow_eqn}). In particular, we will check that the flow (\ref{susy_qm_Qsquare_flow}) generates the expected superspace Lagrangian on-shell for the case of a single real scalar field $X$. From the expressions (\ref{first_order_free_Qs}) for the conserved charges in the free theory, we see that the leading deformation is $- \lambda \eta \tilde{\eta} D X \bar{D} X$. Motivated by this, we will make an ansatz for the finite-$\lambda$ solution of the form:
\begin{align}\label{susy_qm_ansatz}
    \mathcal{A} = f ( \lambda \eta \tilde{\eta} ) \, DX \, \bar{D} X \,,
\end{align}
with $f(y) \to 1 - y + \mathcal{O} ( y^2 )$ as $y \to 0$. Using the definition (\ref{susy_Q_def}) we can compute the conserved superspace charges associated with a Lagrangian of this form, which gives
\begin{align}
    \mathcal{Q} &= - ( \eta + \tilde{\eta} ) \left( f D X + \lambda \eta \tilde{\eta} f' DX + \lambda^2 \tilde{\eta} f'' D ( \eta \tilde{\eta} ) DX \bar{D} X \right) + \lambda \eta f' DX \, \bar{D} X \, D \tilde{\eta} \nonumber \\
    &\qquad + f \eta D X - \lambda f' D ( \eta \tilde{\eta} ) DX \, \bar{D} X \,, \nonumber \\
    \bar{\mathcal{Q}} &= ( \eta + \tilde{\eta} ) \left( f \bar{D} X + \lambda \eta \tilde{\eta} f' \, \bar{D} X + \lambda^2 \eta f'' D ( \eta \tilde{\eta} ) DX \bar{D} X \right) + \lambda \tilde{\eta} f' D X \, \bar{D} X \, \bar{D} \eta \nonumber \\
    &\qquad - f \tilde{\eta} \bar{D} X - \lambda f' \bar{D} ( \eta \tilde{\eta} ) DX \, \bar{D} X \,.
\end{align}
The product of these is therefore
\begin{align}\label{qqbar_for_ansatz}
    \mathcal{Q} \bar{\mathcal{Q}} = - \left( \eta \tilde{\eta} \left( f + f' \eta \lambda ( \eta + \tilde{\eta} ) \right) \left( f + f' \lambda \tilde{\eta} ( \eta + \tilde{\eta} \right) \right) DX \, \bar{D} X \,.
\end{align}
We now pause to investigate an implication of the superspace equations of motion, which will allow us to simplify this expression for $\mathcal{Q} \bar{\mathcal{Q}}$ and, therefore, the flow equation. This will be the analog of (\ref{susy_eom_multiplied_intermediate}), which we used to make a similar simplification in the field theory setting. The equation of motion (\ref{susy_qm_eom}) can be written as
\begin{align}
    0 = D \left( \frac{\delta \mathcal{A}}{\delta ( D X ) } \right) + \bar{D} \left( \frac{\delta \mathcal{A}}{\delta ( \bar{D} X ) } \right) + D \bar{D} \left( \frac{\delta \mathcal{A}}{\delta \eta } \right) + \bar{D} D  \left( \frac{\delta \mathcal{A}}{\delta \tilde{\eta} } \right) \,,
\end{align}
which for our ansatz (\ref{susy_qm_ansatz}) is
\begin{align}\label{on_shell_intermediate_step}
    0 = D \left( f \bar{D} X \right) - \bar{D} \left( f D X \right) + D \bar{D} \left( \tilde{\eta} \lambda f' D X \bar{D} X \right) + \bar{D} D \left( \eta \lambda f' DX \, \bar{D} X \right) \,.
\end{align}
Suppose we multiply both sides of (\ref{on_shell_intermediate_step}) by $DX \bar{D} X$. Since $(DX)^2 = ( \bar{D} X )^2 = 0$, several terms vanish by nilpotency, and the surviving contributions are
\begin{align}
    f \eta D X \bar{D} X - f \tilde{\eta} D X \bar{D} X + \lambda \tilde{\eta}^2 \eta f' D X \bar{D} X - \lambda \eta^2 \tilde{\eta} f' DX \bar{D} X = 0 \,.
\end{align}
It follows that
\begin{align}
    \left( f - \lambda f' \eta \tilde{\eta} \right) \left( \eta - \tilde{\eta} \right) DX \bar{D} X = 0 \,.
\end{align}
Therefore, either $\eta - \tilde{\eta}$ or $f - \lambda f' \eta \tilde{\eta}$ vanishes when multiplying $DX \bar{D} X$ . The latter cannot hold identically unless $f(y) = \frac{c}{y}$, which is not consistent with the boundary condition $f(0) = 1$. We conclude that 
\begin{align}
    \left( \eta - \tilde{\eta} \right) D X \, \bar{D} X = 0 \,.
\end{align}
In particular, this means that, on-shell, we can replace $\eta$ with $\tilde{\eta}$ or vice-versa when either is multiplying $DX \, \bar{D} X$. Making this replacement in the expression (\ref{qqbar_for_ansatz}) for the bilinear $\mathcal{Q} \bar{\mathcal{Q}}$ gives
\begin{align}\label{QQbar_eom_replacement}
    \mathcal{Q} \bar{\mathcal{Q}} = - \eta \tilde{\eta} \left( f + 2 f' \lambda \eta \tilde{\eta} \right)^2 DX \, \bar{D} X \,.
\end{align}
Next, to construct our deforming operator (\ref{susy_qm_Qsquare_flow}), we consider the combinations $D \bar{\mathcal{Q}}$ and $\bar{D} \mathcal{Q}$ (which are of course related by the conservation equation). Any term appearing in these combinations which is proportional to $DX$ or $\bar{D} X$ will not contribute to the deformation, since the function of $\bar{D} \mathcal{Q}$ appearing in (\ref{susy_qm_Qsquare_flow}) comes multiplying $\mathcal{Q} \bar{\mathcal{Q}}$, which is already proportional to $DX \, \bar{D} X$. The only terms which we need to retain are therefore
\begin{align}
    \bar{D} \mathcal{Q} &\sim - \eta \tilde{\eta} f - 2 \lambda \eta^2 \tilde{\eta}^2 f' \,, \nonumber \\
    D \bar{\mathcal{Q}} &\sim \eta \tilde{\eta} f + 2 \lambda \eta^2 \tilde{\eta}^2 f' \,, 
\end{align}
where by ``$\sim$'' we mean equivalence up to terms proportional to $DX$ or $\bar{D} X$, which vanish when multiplying $DX \, \bar{D} X$ by nilpotency. We have thus found that, on-shell, the combination of superfields which drives our flow equation can be written as
\begin{align}
    \frac{\mathcal{Q} \bar{\mathcal{Q}}}{\frac{1}{2} - 2 \lambda \bar{D} Q} = - \frac{ \eta \tilde{\eta} ( f + 2 \lambda \eta \tilde{\eta} f' )^2 }{\frac{1}{2} + 2 \lambda \eta \tilde{\eta} ( f + 2 \lambda \eta \eta f' )} \, DX \, \bar{D} X \,.
\end{align}
In terms of the dimensionless variable $y = \lambda \eta \tilde{\eta}$, the flow equation then reduces to an ordinary differential equation for $f(y)$,
\begin{align}
    f'(y) = \frac{2 \left( f(y) + 2 y f'(y) \right)^2}{1 + 4 y \left( f(y) + 2 y f'(y) \right)} \,,
\end{align}
whose solution is
\begin{align}
    f ( y ) = \frac{1}{4 y} \left( \sqrt{ 1 + 8 y } - 1 \right) \,.
\end{align}
We, therefore, conclude that the all-orders solution to the flow equation (\ref{susy_qm_Qsquare_flow}) is on-shell equivalent to the expression:
\begin{align}
    \mathcal{A} ( \lambda ) = \frac{1}{4 \lambda \eta \tilde{\eta}} \left( \sqrt{ 1 + 8 \lambda \eta \tilde{\eta} } - 1 \right) \, DX \, \bar{D} X \,.
\end{align}
To facilitate comparison with our earlier analysis, we re-scale $\lambda \to \frac{\lambda}{2}$ and replace $\eta, \tilde{\eta}$ with their explicit expressions. The resulting deformed quantum mechanics theory is
\begin{align}\label{susy_qm_1_scalar_soln}
    I = \int \, dt \, d \theta \, d \bar{\theta} \frac{1}{2 \lambda (D \bar{D} X) ( \bar{D} D X )} \left(  -1 + \sqrt{ 1 + 4 \lambda ( D \bar{D} X ) ( \bar{D} D X ) } \right) \, D X \, \bar{D} X \,.
\end{align}
We see that this matches (\ref{dimensionally_reduced_answer}) on the nose after identifying $X$ with $\Phi$ and setting the metric to $g(\Phi) = 1$. The case with a non-trivial metric for the $(0+1)$-dimensional theory can be treated similarly.

One could also consider flows driven by other operators constructed from $\mathcal{Q}$ and $\bar{\mathcal{Q}}$. These are supersymmetric versions of the $f(H)$ deformations considered in \cite{Gross:2019ach,Gross:2019uxi}. From the perspective of the quantum mechanics theory, there is no distinguished choice of $f(H)$ since any such function drives a qualitatively similar flow where all energy eigenstates remain eigenstates, and their energy eigenvalues change in a prescribed way. The only reason for treating the particular $f(H)$ corresponding to $T\overline{T}$ is due to the connections to interesting deformations of higher dimensional theories.

As an example of a different supersymmetric $f(H)$ deformation, one could instead study the flow:
\begin{align}
    \frac{\partial \mathcal{A}}{\partial \lambda} = \mathcal{Q} \bar{\mathcal{Q}} \,,
\end{align}
which is analogous to the deformation $\frac{\partial L}{\partial \lambda} = H^2$. If we again restrict to the case of a single real scalar considered above and use the result (\ref{QQbar_eom_replacement}), which is equivalent to $\mathcal{Q} \bar{\mathcal{Q}}$ on-shell, this leads to a differential equation
\begin{align}
    f'(y) = \left( f(y) + 2 y f'(y) \right)^2 \,.
\end{align}
for the function $f(y)$ appearing in the ansatz (\ref{susy_qm_ansatz}). This is a quadratic equation that can be solved for $f'(y)$ as
\begin{align}
    f'(y) = \frac{1 - 4 y f(y) - \sqrt{1 - 8 y f(y)}}{8 y^2} \,.
\end{align}
However, we note that this deformation, and flows driven by other functions of $\mathcal{Q}$ and $\bar{\mathcal{Q}}$, will not be related to the usual two-dimensional $T\overline{T}$ deformation by dimensional reduction. Only the operator appearing in (\ref{susy_qm_Qsquare_flow}) has this property, and even in that case, the relationship only holds for deformations of conformal $2d$ seed theories since the derivation relies on the trace flow equation.

\section{Theories with $\mathcal{N} = 1$ supersymmetry}\label{sec:n_equals_one}

Thus far, we have focused on theories with two real supercharges, such as $2d$ field theories with $\mathcal{N} = (1, 1)$ supersymmetry or quantum mechanical theories with $\mathcal{N} = 2$ supersymmetry. However, one could carry out an analogous study of theories with only a single real supercharge. This would be relevant for theories with either $\mathcal{N} = (0, 1)$ or $\mathcal{N} = (1, 0)$ theories in two dimensions, which then reduce to theories with $\mathcal{N} = 1$ SUSY in $(0+1)$-dimensions.

We will not carry out an extensive analysis of the three different methods for constructing a supersymmetric $T\overline{T}$ deformation in the $\mathcal{N} = 1$ case, as we did in sections \ref{sec:met1} - \ref{method3} for $\mathcal{N} = 2$. However, in this section, we will briefly outline some of the ingredients that would go into such an analysis and argue that similar results hold.

\subsection{Noether currents in $\mathcal{N} = 1$ theories}

Consider a theory of a collection of $\mathcal{N} = 1$ superfields $X^i$ in $(0+1)$-dimensions. The $\mathcal{N} = 1$ superspace has a single anticommuting coordinate $\theta$, so the superfields $X^i$ can be expanded in components as
\begin{align}
    X^i = x^i + i \theta \psi^i \,.
\end{align}
The supercovariant derivative associated with $\theta$ is
\begin{align}
    D = \frac{\partial}{\partial \theta} - i \theta \frac{\partial}{\partial t} \,,
\end{align}
which satisfies the algebra
\begin{align}
    \{ D, D \} = - 2 i \partial_t \,.
\end{align}
As a simple example, the free superspace Lagrangian for such a collection of $\mathcal{N} = 1$ superfields is written:
\begin{align}\label{example_n_equals_one_free}
    \mathcal{A} = \frac{i}{2} \dot{X}^i D X^i \,.
\end{align}
We now carry out a version of the Noether procedure, which was used to obtain expressions for the supercurrents $\mathcal{Q}, \bar{\mathcal{Q}}$ in the $\mathcal{N} = 2$ case. Consider a superspace Lagrangian that depends on the $X^i$, their superspace derivatives $D X^i$, and their time derivatives $\dot{X}^i$:
\begin{align}\label{general_nequals1_action}
    I = \int \, dt \, d \theta \, \mathcal{A} ( X^i, D X^i, \dot{X}^i ) \,.
\end{align}
We note that, unlike in the $\mathcal{N} = 2$ case, the superspace Lagrangian $\mathcal{A}$ in (\ref{general_nequals1_action}) must be fermionic so that the action $S$ itself is bosonic. The variation of the superspace Lagrangian is given by
\begin{align}\label{n_equals_one_first_variation_step}
    \delta \mathcal{A} = \delta X^i \frac{\delta \mathcal{A}}{\delta X^i} + \delta ( D X^i ) \frac{\delta \mathcal{A}}{\delta ( D X^i )} + \delta \dot{X}^i \frac{\delta \mathcal{A}}{\delta \dot{X}^i} \,. 
\end{align}
The only difference in our Noether procedure is that rather than specializing to the case of time translations $\delta X^i = ( \delta t ) \dot{X}^i$ as we did for $\mathcal{N} = 2$, we will now consider both translations along the Grassmann coordinates $\theta$ and along time. The reason for this is that we would like to construct current bilinears, which require the presence of two current-like objects, such as $\mathcal{Q}$ and $\bar{\mathcal{Q}}$. However, for the $\mathcal{N} = 1$ case, there is only a single current associated with time translations, and (as we will see shortly) its square does not have $T\overline{T}$ as its top component. Similarly, there is a single current associated with superspace translations, but because this current is fermionic, we cannot square it to construct bilinears since the result would vanish by nilpotency.\footnote{Another way to see that we need two separate currents is that the superspace Lagrangian for $\mathcal{N} = 1$ is itself fermionic. Thus, we could not have constructed a fermionic current bilinear out of a single conserved current since the square of such a current is necessarily bosonic.}

With this motivation, we will again re-express (\ref{n_equals_one_first_variation_step}) using the product rule as before. Now, we must be careful because $\delta$ does not commute with $D$ since we are allowing translations along the $\theta$ direction as well. One finds
\begin{align}\label{n_equals_one_general_variation}
    \delta \mathcal{A} &= \delta X^i \frac{\delta \mathcal{A}}{\delta X^i} + D \left( \delta X^i \right) \frac{\delta \mathcal{A}}{\delta ( D X^i )} + \left( \big[ \delta, D \big] X^i \right) \left( \frac{\delta \mathcal{A}}{\delta ( D X^i ) } \right) + \delta \dot{X}^i \frac{\delta \mathcal{A}}{\delta \dot{X}^i} \nonumber \\
    &= D \left( \delta X^i \frac{\delta \mathcal{A}}{\delta ( DX^i )} \right) + \partial_t \left( \delta X^i \frac{\delta \mathcal{A}}{\delta \dot{X}^i} \right)  \nonumber \\
    &\qquad- \delta X^i \left( - \frac{\delta \mathcal{A}}{\delta X^i} + D \left( \frac{\delta \mathcal{A}}{\delta ( D X^i ) } \right) + \partial_t \left( \frac{\delta \mathcal{A}}{\delta \dot{X}^i} \right) \right) +  \left( \big[ \delta, D \big] X^i \right) \left( \frac{\delta \mathcal{A}}{\delta ( D X^i ) } \right) \,.
\end{align}
Exactly, as before, in (\ref{susy_qm_eom}), we can read off the superspace equation of motion:
\begin{align}
    \frac{\delta \mathcal{A}}{\delta X^i} = D \left( \frac{\delta \mathcal{A}}{\delta ( D X^i ) } \right) + \partial_t \left( \frac{\delta \mathcal{A}}{\delta \dot{X}^i} \right) \,.
\end{align}
Now consider a combined superspace translation of the form $t \to t + \delta t$, $\theta \to \theta + \delta \theta$ for a commuting constant $\delta t$ and a Grassmann constant $\delta \theta$. The resulting change in the fields is
\begin{align}\label{X_variation_nequalsone}
    \delta X^i = (\delta \theta) \frac{\partial}{\partial \theta} X^i + ( \delta t ) \dot{X}^i \,.
\end{align}
Using the definition $D X^i = \frac{\partial}{\partial \theta} X^i - i \theta \dot{X}^i$, we can rewrite
\begin{align}
    \frac{\partial}{\partial \theta} X^i = D X^i + i \theta \dot{X}^i \,, 
\end{align}
and therefore repackage the variation (\ref{X_variation_nequalsone}) as
\begin{align}
    \delta X^i &= (\delta \theta) D X^i + ( \delta t + i (\delta \theta) \theta ) \dot{X}^i \nonumber \\
    &= (\delta \theta) D X^i + ( \delta \tilde{t} ) \dot{X}^i \,,
\end{align}
where in the last step we have defined $\delta \tilde{t} \equiv \delta t + i (\delta \theta) \theta$. Likewise, the variation $\delta \mathcal{A}$ of the superspace Lagrangian can be written in the same way:
\begin{align}
    \delta \mathcal{A} = ( \delta \theta ) D \mathcal{A} + ( \delta \tilde{t} ) \partial_t \mathcal{A} \,.
\end{align}
We can also compute the commutator:
\begin{align}
    \big[ \delta, D \big] X^i &= \delta ( DX^i ) - D ( \delta X^i ) \nonumber \\
    &= \left( (\delta \theta) D D X^i + ( \delta \tilde{t} ) D \dot{X}^i \right) - D \left( (\delta \theta) D X^i + ( \delta \tilde{t} ) \dot{X}^i \right) \nonumber \\
    &= - i ( \delta \theta ) \dot{X}^i \,.
\end{align}

Substituting these variations into (\ref{n_equals_one_general_variation}) and going on-shell so that we can discard the equation of motion term gives
\begin{align}
    &(\delta \theta) D \mathcal{A} + i ( \delta \tilde{t} ) D D \mathcal{A} \nonumber \\&= D \left( (\delta \theta) ( D X^i ) \frac{\delta \mathcal{A}}{\delta ( D X^i ) } \right) + D D \left( (\delta \theta) ( D X^i )  \left( \frac{\delta \mathcal{A}}{\delta D D X^i} \right) \right) \nonumber \\
    &\quad + D \left( i ( \delta \tilde{t} ) ( D D X^i ) \frac{\delta \mathcal{A}}{\delta ( D X^i ) } \right) + D D \left( i ( \delta \tilde{t} ) (D D X^i ) \left( \frac{\delta \mathcal{A}}{\delta D D X^i} \right) \right) \nonumber \\
    &\quad - i ( \delta \theta ) \dot{X}^i \left( \frac{\delta \mathcal{A}}{\delta ( D X^i ) } \right) \,.
\end{align}
where we have rewritten time derivatives in terms of $D$ using $\partial_t = i D^2$. Collecting terms then gives
\begin{align}\label{nequals_one_noether_intermediate}
    0 &= - (\delta \theta) \Bigg[ D \left( ( D X^i ) \frac{\delta \mathcal{A}}{\delta ( D X^i )} - D \left( ( D X^i ) \frac{\delta \mathcal{A}}{\delta ( D D X^i )} \right) + \mathcal{A} \right) + i  \dot{X}^i \left( \frac{\delta \mathcal{A}}{\delta ( D X^i ) } \right) \Bigg] \, \nonumber \\
    &\quad + i D \left[  ( \delta \tilde{t} ) ( D D X^i ) \frac{\delta \mathcal{A}}{\delta ( D X^i ) } + D \left( ( \delta \tilde{t} ) ( D D X^i ) \frac{\delta \mathcal{A}}{\delta ( D D X^i ) } \right) \right] - i ( \delta \tilde{t} ) D D \mathcal{A}\,.
\end{align}
It is now tempting to commute the $\delta \tilde{t}$ past various instances of $D$ in the second line of (\ref{nequals_one_noether_intermediate}) and define two charge-like objects corresponding to the quantities in brackets, namely
\begin{align}\label{n_equals_one_supercurrents_defn}
    \mathcal{Q}_\theta &= ( D X^i ) \frac{\delta \mathcal{A}}{\delta ( D X^i )} - i D \left( ( D X^i ) \frac{\delta \mathcal{A}}{\delta \dot{X}^i} \right) + \mathcal{A} \,, \nonumber \\
    \mathcal{Q}_t &= - i \dot{X}^i \frac{\delta \mathcal{A}}{\delta ( D X^i ) } + D \left( \dot{X}^i \frac{\delta \mathcal{A}}{\delta \dot{X}^i } \right) - D \mathcal{A} \,.
\end{align}
One might then conclude that $D \mathcal{Q}_t$ must vanish and $D \mathcal{Q}_\theta$ must be related to the remaining term $i  \dot{X}^i \left( \frac{\delta \mathcal{A}}{\delta ( D X^i ) } \right)$, giving us one conserved charge and one object, which is not conserved, but which has a known source. However, this manipulation is not valid because $\delta \tilde{t}$ itself depends on $\theta$ and therefore does not commute with $D$. If we first set $\delta \theta = 0$ and consider only a finite $\delta t$, then $\delta \tilde{t} = \delta t$ does commute with the $D$ operator so we can write
\begin{align}
    0 &= ( \delta t ) D \left[\left( ( D D X^i ) \frac{\delta \mathcal{A}}{\delta ( D X^i ) } + D \left( ( D D X^i ) \frac{\delta \mathcal{A}}{\delta ( D D X^i ) } \right) - D \mathcal{A} \right) \right] \,,
\end{align}
which is interpreted as a conservation equation of the form
\begin{align}
    D \mathcal{Q}_t = 0 \,,
\end{align}
with $\mathcal{Q}_t$ defined in (\ref{n_equals_one_supercurrents_defn}). Next, let us set $\delta t = 0$ and look at a fermionic translation $\delta \theta$. First, we must account for the additional terms introduced when commuting $\delta \tilde{t} = i ( \delta \theta ) \theta$ past the $D$ operators. Note that
\begin{align}
    - i ( \delta \tilde{t} ) D D \mathcal{A} &= ( \delta \theta ) \theta D D \mathcal{A} \nonumber \\
    &= ( \delta \theta ) \left( - D \left( \theta D \mathcal{A} \right) + D \mathcal{A} \right) \nonumber \\
    &= ( \delta \theta ) D \left( \mathcal{A} - \theta D \mathcal{A} \right) \,.
\end{align}
Then one finds
\begin{align}
    (\delta \theta) D \mathcal{Q}_\theta &= - D \left[  ( \delta \theta ) \theta ( D D X^i ) \frac{\delta \mathcal{A}}{\delta ( D X^i ) } + D \left( ( \delta \theta ) \theta \dot{X}^i \frac{\delta \mathcal{A}}{\delta \dot{X}^i } \right) \right] + ( \delta \theta ) D \left( \mathcal{A} - \theta D \mathcal{A} \right) \, \nonumber \\
    &\qquad - i (\delta \theta) \dot{X}^i \left( \frac{\delta \mathcal{A}}{\delta ( D X^i ) } \right) \nonumber \\
    &= ( \delta \theta ) D \left[ \theta ( D D X^i ) \frac{\delta \mathcal{A}}{\delta ( D X^i ) } - D \left( \theta \dot{X}^i \frac{\delta \mathcal{A}}{\delta \dot{X}^i } \right) + \mathcal{A} - \theta D \mathcal{A} \right] \nonumber \\&- i (\delta \theta) \dot{X}^i \left( \frac{\delta \mathcal{A}}{\delta ( D X^i ) } \right) \nonumber \\
    &= ( \delta \theta ) D \left[  \theta \mathcal{Q}_t - \dot{X}^i \frac{\delta \mathcal{A}}{\delta \dot{X}^i} + \mathcal{A} \right] - i (\delta \theta) \dot{X}^i \left( \frac{\delta \mathcal{A}}{\delta ( D X^i ) } \right) \nonumber \\
    &= ( \delta \theta ) \left( \mathcal{Q}_t - \theta D \mathcal{Q}_t - D \left( \dot{X}^i \frac{\delta \mathcal{A}}{\delta \dot{X}^i} \right) + D \mathcal{A} - i \dot{X}^i \left( \frac{\delta \mathcal{A}}{\delta ( D X^i ) } \right) \right) \,.
\end{align}
Using the conservation equation $D \mathcal{Q}_t = 0$, we then have
\begin{align}\label{qtheta_non_conservation_intermediate}
    D \mathcal{Q}_\theta = \mathcal{Q}_t - D \left( \dot{X}^i \frac{\delta \mathcal{A}}{\delta \dot{X}^i} \right) + D \mathcal{A} - i \dot{X}^i \left( \frac{\delta \mathcal{A}}{\delta ( D X^i ) } \right) \,.
\end{align}
Thus, we see that the ``charge'' $\mathcal{Q}_\theta$ is not an independent quantity but is related to the time translation charge $\mathcal{Q}_t$, as one might expect from the intuition that the supersymmetry algebra relates successive superspace translations to time translations through $D^2 = - i \partial_t$. In particular, $\mathcal{Q}_\theta$ itself is not conserved in general. We can quantify this non-conservation by acting again on (\ref{qtheta_non_conservation_intermediate}) with $D$ and using $D \mathcal{Q}_t = 0$ to write
\begin{align}
    \partial_t \mathcal{Q}_\theta = \partial_t \left( \mathcal{A} - \dot{X}^i \frac{\delta \mathcal{A}}{\delta \dot{X}^i}  \right) - i D \left( \dot{X}^i \left( \frac{\delta \mathcal{A}}{\delta ( D X^i ) } \right) \right) \,.
\end{align}
Therefore, one could define a modified charge $\widetilde{Q}_\theta$ and a correction term $\mathcal{Q}_c$ by
\begin{align}
    \widetilde{\mathcal{Q}}_\theta = \mathcal{Q}_\theta + \dot{X}^i \frac{\delta \mathcal{A}}{\delta \dot{X}^i} - \mathcal{A} \,, \qquad \mathcal{Q}_c = i  \dot{X}^i \left( \frac{\delta \mathcal{A}}{\delta ( D X^i ) } \right) \,,
\end{align}
with the property that
\begin{align}
    \partial_t \widetilde{\mathcal{Q}}_\theta + D \mathcal{Q}_c = 0 \,.
\end{align}

\subsection{Definition of $\mathcal{Q}_\theta \mathcal{Q}_t$ deformation and solution for one scalar}

To get some intuition for these objects constructed in the preceding subsection, we compute them for the free theory (\ref{example_n_equals_one_free}):
\begin{align}
    \mathcal{Q}_\theta &= i D X^i \dot{X}^i \,, \nonumber \\
    \widetilde{\mathcal{Q}}_\theta &= i D X^i \dot{X}^i \,, \nonumber \\
    \mathcal{Q}_c &= - \frac{1}{2} \dot{X}^i \dot{X}^i \,, \nonumber \\
    \mathcal{Q}_t &= \frac{1}{2} \dot{X}^i \dot{X}^i  \,.
\end{align}
Note that $\mathcal{Q}_\theta$, $\widetilde{\mathcal{Q}}_\theta$ are fermionic and $\mathcal{Q}_t$ is bosonic, as expected for Noether currents associated with Grassmann translations and time translations respectively. Therefore, the product $\mathcal{Q}_\theta \mathcal{Q}_t $ is a fermion and thus an appropriate quantity to add to the Lagrangian as a deformation. In particular, for the free theory, we note that the top component of $\mathcal{Q}_\theta \mathcal{Q}_t $ is proportional to $(\dot{x}^i \dot{x}^i)^2$, which is the square of the Hamiltonian. 
Using this intuition, we propose an $\mathcal{N} = 1$ version of the SUSY-QM deformation as
\begin{align}\label{n_equals_one_susy_flow}
    \frac{\partial \mathcal{A}}{\partial \lambda} = \frac{1}{2} \mathcal{Q}_\theta \mathcal{Q}_t \,.
\end{align}
In this proposal we do \emph{not} divide by the combination $\frac{1}{2} - 2 \lambda \mathcal{Q}_t$, as one might expect from the analogous $f(\mathcal{Q}, \bar{\mathcal{Q}})$ expression in the $\mathcal{N} = 2$ case. This may seem strange because the form of this deformation is very different than in the preceding cases that we have considered. However, we will later see that there is an equivalent rewriting of this flow equation as
\begin{align}\label{alternate_tilde_nequalsone_flow}
    \frac{\partial \mathcal{A}}{\partial \lambda} = \frac{\tilde{\mathcal{Q}}_\theta \mathcal{Q}_t}{1 + 2 \lambda \mathcal{Q}_t} \,.
\end{align}
For the class of Lagrangians that we focus on in this work, the solution to the flow equation (\ref{alternate_tilde_nequalsone_flow}) is identical to the solution of (\ref{n_equals_one_susy_flow}).  We will explore the reason for this in section \ref{subsec:HT_form}, where we see that there is a simpler way to understand this equivalence by studying an analogous pair of deformations in the non-supersymmetric setting. For the moment, however, we will work with the first deformation (\ref{n_equals_one_susy_flow}).

First, we argue that this is on-shell equivalent to the dimensional reduction of the supercurrent-squared flow for theories with $\mathcal{N} = ( 0, 1)$ supersymmetry. In particular, we will solve the flow equation (\ref{n_equals_one_susy_flow}) for the seed theory of a single free boson and verify that it matches the dimensional reduction of the corresponding $2d$ flow. We make an ansatz for the finite-$\lambda$ deformed superspace Lagrangian of the form:
\begin{align}
    \mathcal{A}^{(\lambda)} = \frac{i}{2} f ( \lambda \dot{X}^2 ) \, \dot{X} D X \,.
\end{align}
Next, we compute the supercurrents. To ease notation, we define the dimensionless combination $\xi = \lambda \dot{X}^2$. Then using (\ref{n_equals_one_supercurrents_defn}) one finds
\begin{align}
    \mathcal{Q}_\theta &= i  f ( \xi ) \dot{X} D X \,, \nonumber \\
    \mathcal{Q}_t &= \frac{1}{2} \dot{X}^2 f ( \xi ) + \frac{i}{2} D \Big( \dot{X} \left( f ( \xi ) + 2 \xi f'(\xi) \right) D X \Big) - \frac{i}{2} D \left( f ( \xi ) \dot{X} D X \right) \,.
\end{align}
Here we have used that $DX$ is fermionic so $(DX)^2 = 0$. Furthermore, since $\mathcal{Q}_\theta$ is proportional to $DX$, when we construct the combination $\mathcal{Q}_\theta \mathcal{Q}_t$, any terms proportional to $DX$ in $\mathcal{Q}_t$ will drop out by nilpotency. Therefore, we can write
\begin{align}\label{calQ_t}
    \mathcal{Q}_t \sim \frac{1}{2} \left( f ( \xi ) +  2 \xi f'(\xi) \right) \dot{X}^2 \,,
\end{align}
where ``$\sim$'' means equality up to terms which will not contribute to the product $\mathcal{Q}_\theta \mathcal{Q}_t$. The flow equation (\ref{n_equals_one_susy_flow}) becomes
\begin{align}
    \frac{i}{2} f'(\xi) \dot{X}^3 D X = \frac{i}{2} \cdot \left( f(\xi) \left( f ( \xi ) + 2 \xi f'(\xi) \right) \right) \, \dot{X}^3 \, D X \,,
\end{align}
whose solution is
\begin{align}\label{calQ_theta_calQ_t_solution}
    f ( \xi ) = \frac{1}{2 \xi} \left( 1 - \sqrt{1 - 4 \xi} \right) \,.
\end{align}
Thus, the full solution for the deformed superspace Lagrangian is
\begin{align}\label{nequalsone_susyqm_full_solution}
    \mathcal{A}^{(\lambda)} = \frac{i}{4 \lambda \dot{X}^2} \left( 1 - \sqrt{1 - 4 \lambda \dot{X}^2 } \right) \, \dot{X} D X \,.
\end{align}
As we mentioned around (\ref{alternate_tilde_nequalsone_flow}), the same flow can be acquired by deforming with another operator
\begin{equation}
\label{eq:newN=1Operator}
    \frac{\widetilde{\mathcal{Q}}_\theta \mathcal{Q}_t}{1 + 2\lambda \mathcal{Q}_t}\,,
\end{equation}
similar to the irrelevant operators used in the previous sections. To see that (\ref{eq:newN=1Operator}) yields the same flow, we first compute
\begin{equation}
\begin{aligned}
    \widetilde{\mathcal{Q}}_\theta &= \mathcal{Q}_\theta + \dot{X} \frac{\delta \mathcal{A}}{\delta \dot{X}} - \mathcal{A} \\ 
    &= if(\xi)\dot{X} DX + \frac{i}{2}\left(f(\xi) + 2f'(\xi)\xi \right)\dot{X} DX - \frac{i}{2}f(\xi)\dot{X} DX \\
    &= i(f(\xi) + \xi f'(\xi)) \dot{X} DX\,.
\end{aligned}
\end{equation}
Then using the expression for $\mathcal{Q}_t$ from (\ref{calQ_t}), we have
\begin{equation}
    \frac{\widetilde{\mathcal{Q}}_\theta \mathcal{Q}_t}{1 + 2 \lambda \mathcal{Q}_t}  = \frac{\left(f(\xi) + \xi f'(\xi)\right)\left(f(\xi) + 2\xi f'(\xi)\right)}{1 + \xi \left(f(\xi) + 2\xi f'(\xi)\right)} \frac{i\dot{X}^3 DX}{2}\,.
\end{equation}
The flow equation $\frac{\partial \mathcal{A}}{\partial \lambda } = \frac{\widetilde{\mathcal{Q}}_\theta \mathcal{Q}_t}{1 + 2 \lambda \mathcal{Q}_t}$ leads to the following differential equation:
\begin{equation}
    f'(\xi) = \frac{(f(\xi) + 2\xi f'(\xi))(f(\xi) + \xi f'(\xi))}{1 + \xi (f(\xi) + 2\xi f'(\xi))}\,,
\end{equation}
which has the same solution as in (\ref{calQ_theta_calQ_t_solution}) from the flow triggered by the operator $\mathcal{Q}_\theta \mathcal{Q}_t$,
\begin{equation}
    f(\xi) = \frac{1 - \sqrt{1 - 4\xi}}{2\xi}\,.
\end{equation}
Therefore, the operator $\frac{\tilde{\mathcal{Q}}_\theta \mathcal{Q}_t}{1 + 2\lambda \mathcal{Q}_t}$ also triggers the same $T\overline{T}$-like flow. Notice that from $\mathcal{Q}_t = D\widetilde{ \mathcal{Q}}_\theta + \mathcal{Q}_c$, we can express the operator $\frac{\widetilde{\mathcal{Q}}_\theta \mathcal{Q}_t}{1 + 2\lambda \mathcal{Q}_t}$ in terms of the conserved currents $\mathcal{Q}_c$ and $\widetilde{\mathcal{Q}}_\theta$, satisfying the conservation equation
\begin{equation}
    \partial_t \widetilde{\mathcal{Q}}_\theta + D \mathcal{Q}_c = 0\,.
\end{equation}
We now argue that this result is on-shell equivalent to the dimensional reduction of the solution to the supercurrent-squared flow for the corresponding $\mathcal{N} = (0, 1)$ theory in $2d$. The dimensional lift of this theory can be written as
\begin{align}\label{2d_nequalszeroone_seed}
I = \int \, d^2 x \, d \theta \, D_+ \Phi \partial_{++} \Phi \,.
\end{align}
A superspace Noether procedure analogous to the one that we have used in the $\mathcal{N} = (1, 1)$ analysis of section \ref{sec:met1} can also be applied here. The input of this process is a superspace Lagrangian $\mathcal{A} ( D_+ \Phi, \partial_{\pm \pm} \Phi )$. The output is a conservation equation
\begin{align}\label{nequalszerooneconservation}
\begin{split}
	& \partial_{--} \mathcal{S}_{+++} + D_+ \mathcal{T}_{++--} = 0\,, \\
	& \partial_{--} \mathcal{S}_{--+} + D_+ \mathcal{T}_{----} = 0\,,
\end{split}
\end{align}
where $\mathcal{S}_{\pm\pm+}$ and $\mathcal{T}_{\pm \pm - -}$ are superfields given by:
\begin{align}
\begin{split}
    & \mathcal{S}_{+++} = \frac{\delta\mathcal{A}}{\delta (\partial_{--} \Phi)}\partial_{++} \Phi\,, \\
    & \mathcal{S}_{--+} = \frac{\delta\mathcal{A}}{\delta (\partial_{--} \Phi)}\partial_{--} \Phi - \mathcal{A}\,, \\
    & \mathcal{T}_{++--} = \frac{\delta\mathcal{A}}{\delta (D_+ \Phi)} \partial_{++} \Phi + D_+ \left( \frac{\delta\mathcal{A}}{\delta (\partial_{++} \Phi)} \partial_{++} \Phi \right) -D_+ \mathcal{A}\,, \\
    & \mathcal{T}_{----} = \frac{\delta\mathcal{A}}{\delta (D_+ \Phi)} \partial_{--} \Phi + D_+ \left( \frac{\delta\mathcal{A}}{\delta (\partial_{++} \Phi)} \partial_{--}  \Phi \right)\,.
\end{split}
\end{align}
The $\mathcal{N} = (0, 1)$ supercurrent-squared flow is defined by
\begin{align}\label{nequalszeroonesupercurrentsquareddefn}
    \frac{\partial}{\partial \lambda} \mathcal{A}^{(\lambda)} = \mathcal{S}_{+++} \mathcal{T}_{----} - \mathcal{S}_{--+} \mathcal{T}_{++--} \,.
\end{align}
Beginning from the seed superspace Lagrangian (\ref{2d_nequalszeroone_seed}), we make an ansatz for the finite-$\lambda$ solution:
\begin{align}
    \mathcal{A}^{(\lambda)} = f(\lambda \partial_{++}\Phi\partial_{--}\Phi)D_+\Phi\partial_{--}\Phi\,.
\end{align}
After evaluating the supercurrents, computing the combination of bilinears (\ref{nequalszeroonesupercurrentsquareddefn}), and simplifying the differential equation, one finds
\begin{align}
    x \frac{\partial f}{\partial x} = - x f^2 - 2 x^2 f \frac{\partial f}{\partial x} \,,
\end{align}
where $x = \lambda \partial_{++}\Phi\partial_{--}\Phi$. The solution is
\begin{align}
    f(x) = \frac{\sqrt{1+4x}-1}{2x}.
\end{align}
Thus, the fully deformed superspace Lagrangian is
\begin{align}\label{full_2d_nequalsone_soln}
    \mathcal{A}^{(\lambda)} = \frac{1}{2 \lambda \partial_{++}\Phi\partial_{--}\Phi} \left( \sqrt{ 1 + 4 \lambda \partial_{++}\Phi\partial_{--}\Phi } - 1 \right) D_+ \Phi \partial_{++} \Phi
\end{align}
giving the $\mathcal{N} = (1, 0)$ superspace action
\begin{align}
    S = \int dt d\theta \frac{\left( \sqrt{ 1 + 4 \lambda \partial_{++}\Phi\partial_{--}\Phi } - 1 \right) D_+ \Phi \partial_{++} \Phi}{2 \lambda \partial_{++}\Phi\partial_{--}\Phi} \,.
\end{align}

Upon dimensional reduction, we identify the superfield $\Phi$ with $X$, and all partial derivatives $\partial_{\pm \pm} \Phi$ are proportional to $\dot{X}$. Doing this, we see that -- up to various constant factors that can be absorbed
into rescalings -- the solution (\ref{full_2d_nequalsone_soln}) exactly matches (\ref{nequalsone_susyqm_full_solution}). 

We will not perform the analog of the analysis in section \ref{method2}, where we dimensionally reduced the supercurrent-squared operator itself using the trace flow equation, in this $\mathcal{N} = 1$ case. However, such a procedure should certainly be possible. One would identify a superfield analog of the trace flow equation which relates $\mathcal{S}_{\pm \pm +}$ and $\mathcal{T}_{\pm \pm - -}$, and then use this to eliminate the appropriate linear combinations of these superfields that correspond to the $x$ directions. One might even expect the process to be simpler, in this case, since the dimensionally reduced deformation should be bilinear of the form $\mathcal{Q}_\theta \mathcal{Q}_t$ rather than a rational function of supercurrents. In particular, it appears $\mathcal{T}_{++--}$ is structurally similar to $\mathcal{Q}_t$, so one might believe that the correct dimensionally reduced deformation would be some product of $\mathcal{T}_{++--}$ with another superfield that plays the role of $\mathcal{Q}_\theta$.


We conclude this subsection with a few comments about the relationship between $\mathcal{N} = 1$ and $\mathcal{N} = 2$ theories.

\begin{enumerate}
    \item Every SUSY-QM theory with $\mathcal{N} = 2$ SUSY can be viewed as a special case of a theory with $\mathcal{N} = 1$ supersymmetry. Therefore, one can always write the $f(\mathcal{Q}, \bar{\mathcal{Q}})$ deformation for a theory with $\mathcal{N} = 2$ supersymmetry and integrate out one of the fermionic directions to obtain a deformation in $\mathcal{N} = 1$ superspace. The resulting $\mathcal{N} = 1$ deformation should be on-shell equivalent to the combination $\mathcal{Q}_\theta \mathcal{Q}_t$ which we described in this section since this generates the appropriate supercurrent-squared flows for $\mathcal{N} = 1$ theories. Evidence for the on-shell equivalence of these two flows in the case of $2d$ field theory was given in \cite{Chang:2018dge}; the SUSY-QM case should be similar.
    
    \item As pointed out in \cite{Combescure:2004ey}, quantum mechanical theories with $\mathcal{N} = 1$ supersymmetry are often equivalent to $\mathcal{N} = 2$ theories because they have a hidden second supersymmetry. In particular, this will be true for any $\mathcal{N} = 1$ theory with a fermion number symmetry. A second hidden supersymmetry of this form was not present in the case of a single $\mathcal{N} = 1$ superfield, which we considered in this section, but it would be present in other cases (such as those with an even number of $\mathcal{N} = 1$ superfields). For those theories, one should be able to present the supercurrent-squared deformation of the theory in either $\mathcal{N} = 1$ or $\mathcal{N} = 2$ language, and we expect the results to be equivalent on-shell.
\end{enumerate}

\subsection{The $HT$ form of the deformation}\label{subsec:HT_form}

We now turn to the question of why our deforming operator in the case of $\mathcal{N} = 1$ supersymmetry could be written either as a bilinear $\mathcal{Q}_\theta \mathcal{Q}_t$ or a rational function of the form
\begin{align}
    \frac{\widetilde{\mathcal{Q}}_\theta \mathcal{Q}_t}{1 + 2 \lambda \mathcal{Q}_t} \,,
\end{align}
although these two expressions appear quite different. This equivalence is related to an exact correspondence between two expressions involving the Hamiltonian, which holds for $T\overline{T}$-like deformations of any kinetic seed theory. It will be simplest to discuss this correspondence in the purely bosonic context first, without any supersymmetry.

To begin, we first point out that there are two natural notions of energy in the theory of quantum mechanics. The first is the Hamiltonian $H$ of the system. Since we work in Euclidean signature and interpret the Euclidean Lagrangian as the Hamiltonian, $H$ is the object that sits under the integral sign in the action:
\begin{align}
I = \int \, dt \, H \,.
\end{align}
The second notion of energy is the (Euclidean) Hilbert stress tensor $T^{(\text{Hilb})}$. In $(0+1)$ dimensions, there is only a single component of the stress tensor. It is defined by coupling the theory to a worldline metric $g_{tt}$, or equivalently an einbein $e_t$, and computing:
\begin{align}
    \label{one_dim_hilbert_stress}
    T^{(\text{Hilb})} = - \frac{2}{\sqrt{g^{tt}}} \frac{\delta I}{\delta g^{tt}} = H - 2 \frac{\partial H ( g^{tt} )}{\partial g^{tt}} \Big\vert_{g^{tt} = 1} \,.
\end{align}
Here, by $H ( g^{tt} )$, we mean the expression obtained by minimally coupling $H$ to a worldline metric. Since generically $\frac{\partial H ( g^{tt} )}{\partial g^{tt}} \neq 0$, the two notions of energy differ. Thus far, we have been somewhat sloppy and used the symbols $H$ and $T$ interchangeably, for instance, in the deformation (\ref{gross_flow_eqn}). Although this deformation is written in terms of $T$, it is more properly a flow equation for the object $H$ appearing under the integral in the Euclidean action. Therefore, we will be more careful and write:
\begin{align}\label{H_T_flow_one}
    \frac{\partial H}{\partial \lambda} = \frac{H^2}{\frac{1}{2} - 2 \lambda H} \,,
\end{align}
whose solution is
\begin{align}\label{ham_sqrt_soln}
    H ( \lambda ) = \frac{1}{4 \lambda} \left( 1 - \sqrt{ 1 - 8 \lambda H_0 } \right) \,.
\end{align}
We recall that (\ref{H_T_flow_one}) was derived using the trace flow equation, which means that it is valid only for theories that descend from CFTs. In particular, it does not hold for theories with potential. We now restrict to a particular class of theories for which (\ref{H_T_flow_one}) is valid, which we will refer to as ``kinetic seed theories.'' Explicitly, we assume that the undeformed Hamiltonian does not depend on any dimensionful scale but depends linearly on the inverse metric $g^{tt}$ when coupled to worldline gravity. For instance, the free scalar Hamiltonian
\begin{align}
    H_0 ( g^{tt} ) = \dot{x}^2 = g^{tt} \partial_t x \partial_t x
\end{align}
belongs to this class of theories. Since $H_0 ( g^{tt} )$ depends linearly on the metric, which is a scalar, the Hilbert stress tensor (\ref{one_dim_hilbert_stress}) associated with $H(\lambda)$ is
\begin{align}\label{explicit_onedim_hilbert_stress}
    T^{(\text{Hilb})} &= H ( \lambda ) - 2 \frac{\partial H}{\partial H_0} \frac{\partial H_0 ( g^{tt} ) }{\partial g^{tt} } \Big\vert_{g^{tt}=1} \nonumber \\
    &= H ( \lambda ) - 2 H_0 \frac{\partial H}{\partial H_0} \,.
\end{align}
We now ask whether we can express the right side of the flow equation (\ref{H_T_flow_one}) more simply in terms of $H$ and $ T^{(\text{Hilb})}$, rather than $H$. One can verify by explicit calculation that, for a Hamiltonian of the form (\ref{ham_sqrt_soln}), the operator appearing in the flow is
\begin{align}
    \frac{H^2}{\frac{1}{2} - 2 \lambda H} = \frac{\left( \sqrt{ 1 - 8 \lambda H_0} - 1 \right)^2}{8 \lambda^2 \sqrt{1 - 8 \lambda H_0}} \,.
\end{align}
On the other hand, using the expression (\ref{explicit_onedim_hilbert_stress}) for the Hilbert stress tensor, one can also compute the combination
\begin{align}
    H  T^{(\text{Hilb})} &= H \left( H - 2 H_0 \frac{\partial H}{\partial H_0} \right) \nonumber \\
    &= - \frac{ \left( \sqrt{1 - 8 \lambda H_0} - 1 \right)^2}{16 \lambda^2 \sqrt{1 - 8 \lambda H_0} } \,.
\end{align}
Therefore, for this class of theories, we conclude
\begin{align}\label{HT_relation}
    H  T^{(\text{Hilb})} = - \frac{1}{2} \left( \frac{H^2}{\frac{1}{2} - 2 \lambda H} \right) \,.
\end{align}
The upshot of this discussion is that, up to an overall constant that can be absorbed into the scaling of $\lambda$, we are free to deform \emph{either} by the combination $\frac{H^2}{\frac{1}{2} - 2 \lambda H}$ or by the combination $H T^{(\text{Hilb})}$. We refer to this latter expression as the $HT$ form of the flow equation (dropping the superscript (Hilb) for simplicity).

This $HT$ deformation has a simple interpretation.\footnote{In the holographic context, we interpret the addition of the double-trace $HT$ operator as a change in boundary conditions for the dual BF gauge theory fields. This is discussed in section \ref{sec:schwarzian_type}.} From the form (\ref{one_dim_hilbert_stress}) of the Hilbert stress tensor, one has
\begin{align}\label{one_dim_burgers}
    \frac{\partial H}{\partial \lambda} = H^2 - 2 H \frac{\partial H}{\partial g^{tt}} \,.
\end{align}
This is reminiscent of the form of the inviscid Burgers' equation \eqref{eq:burgers} for the cylinder energy levels of a $T\overline{T}$-deformed CFT in two dimensions, which we repeat:
\begin{align}
    \frac{\partial E_n}{\partial \lambda} = E_n \frac{\partial E_n}{\partial R} + \frac{1}{R} P_n^2 \,.
\end{align}
In the zero-momentum sector, the inviscid Burgers' equation \eqref{eq:burgers} admits an implicit solution:
\begin{align}\label{2d_tt_burgers_implicit}
    E_n ( R, \lambda ) = E_n ( R + \lambda E_n ( R, \lambda ) , 0 ) \,.
\end{align}
This has the interpretation that the theory has effectively been put on a cylinder with an ``energy-dependent radius.'' That is, energy eigenstates with different energy eigenvalues see different effective geometries.

There is no straightforward analog of the limit $P_n = 0$ in the quantum mechanical case, but if we restrict to the case of small energies so that $H^2$ is negligible compared to $H$, the (\ref{one_dim_burgers}) becomes
\begin{align}\label{one_dim_burgers_QM}
    \frac{\partial H}{\partial \lambda} \approx - 2 H \frac{\partial H}{\partial g^{tt}} \,,
\end{align}
which likewise has the implicit solution
\begin{align}\label{qm_implicit_burgers_soln}
    H ( g^{tt} , \lambda ) = H ( g^{tt} - 2 \lambda H ( g^{tt} , \lambda ) , 0 ) \,.
\end{align}
This has a similar interpretation that different states in the deformed quantum mechanics theory see different effective energy-dependent metrics. Note that the relative factor of $-2$ between the rescalings in (\ref{2d_tt_burgers_implicit}) and (\ref{qm_implicit_burgers_soln}) is because the relation (\ref{HT_relation}) between $HT$ and the dimensionally reduced $T\overline{T}$ operator required us to re-scale $\lambda$ by a factor of $-\frac{1}{2}$.

We now see why there were also two equivalent ways of writing the deformation in the $\mathcal{N} = 1$ case. On-shell, the quantity $\mathcal{Q}_\theta$ is always proportional to the superspace Lagrangian $\mathcal{A}$ for deformations of a free seed theory, whereas the time translation current $\mathcal{Q}_t$ contains the Hilbert stress tensor. Therefore, the top component of their product is proportional to
\begin{align}
    \mathcal{Q}_\theta \mathcal{Q}_t \Big\vert_{\theta} \sim H  T^{(\text{Hilb})}\,.
\end{align}
That is, the bilinear $\mathcal{Q}_\theta \mathcal{Q}_t$ is the superspace analog of the $HT$ deformation. On the other hand, the second form of the deformation
\begin{align}
    \frac{\widetilde{Q}_\theta \mathcal{Q}_t}{1 + 2 \lambda \mathcal{Q}_t} \,,
\end{align}
is the $\mathcal{N} = 1$ superspace analog of the $\frac{H^2}{\frac{1}{2} - 2 \lambda H}$ form of the deformation. The fact that we obtained square root solutions to the two flows driven by $\mathcal{Q}_\theta \mathcal{Q}_t$ and $\frac{\widetilde{Q}_\theta \mathcal{Q}_t}{1 + 2 \lambda \mathcal{Q}_t}$ is therefore expected since this is related to the statement that we likewise obtain square roots solutions in the bosonic sector using either $H T$ or $\frac{H^2}{\frac{1}{2} - 2 \lambda H}$.

\section{Conclusion}
\label{sec: Discussion}

In this chapter, we have proposed a manifestly supersymmetric deformation of the superspace Lagrangian for a theory of $\mathcal{N} = 2$ quantum mechanics:
\begin{align}
    \frac{\partial \mathcal{A}}{\partial \lambda} = f(\mathcal{Q}, \bar{\mathcal{Q}}) \equiv \frac{\mathcal{Q} \bar{\mathcal{Q}}}{\frac{1}{2} - 2 \lambda \bar{D} \mathcal{Q}} \,.
\end{align}
The conserved superfields $\mathcal{Q}, \bar{\mathcal{Q}}$ are computed using a Noether prescription, for which we have given explicit formulas that apply to a class of theories involving scalar superfields $X^i$. We have also performed several non-trivial checks that this superspace deformation is on-shell equivalent to the dimensional reduction of the $\mathcal{N} = (1, 1)$ supercurrent-squared deformation of two-dimensional field theories, at least for conformally invariant seed theories. For such conformal seeds, this deformation is, therefore, a natural candidate for the appropriate supersymmetric version of $T\overline{T}$ for $(0+1)$-dimensional theories.

Additionally, we proposed two manifestly supersymmetric deformations for an $\mathcal{N} =1$ quantum mechanics theory
\begin{align}
    \frac{\partial \mathcal{A}}{\partial \lambda} = \frac{1}{2} \mathcal{Q}_\theta \mathcal{Q}_t \quad  \text{ and } \quad \frac{\partial \mathcal{A}}{\partial \lambda} = \frac{\widetilde{\mathcal{Q}_\theta} \mathcal{Q}_t}{1 + 2 \lambda \mathcal{Q}_t} \,.
\end{align}
Although the forms of this deformation appear different, we showed that they produce the same flow equation when applied to the seed theory of a single free scalar. This flow equation matches the dimensional reduction of the $\mathcal{N} = (0, 1)$ supercurrent-squared operator. We also interpreted the equivalence of these deformations by pointing out an analogous rewriting that holds for deformations of the bosonic sector of kinetic seed theories, namely $-\frac{1}{2} H T$ and $\frac{H^2}{\frac{1}{2} - 2\lambda H}$.

We close this chapter with future research.

\emph{More Supersymmetry}

Perhaps the most obvious follow-up to this work is to exhibit a version of our superspace deformation with differing amounts of supersymmetry. For instance, it should be possible to define deformations of SUSY-QM theories, which are related by dimensional reduction to the supercurrent-squared deformations of theories with $(0,2)$ or $(2,2)$ supersymmetry \cite{Chang:2019kiu,Jiang:2019hux}. The case of an $\mathcal{N}= 4$ SUSY-QM theory, which descends from a $\mathcal{N} = (2, 2)$ field theory is perhaps more interesting since such field theories are especially well-studied.

It may be that such an analysis is more amenable to a different technique for obtaining the supercurrents than the one we have used here. In the $2d$ case, such supercurrents for theories with $\mathcal{N} = (1, 1)$ supersymmetry were straightforward to compute using either a Noether procedure \cite{Chang:2018dge} or via coupling to supergravity \cite{Baggio:2018rpv}. However, in the case of $2d \,, \, \mathcal{N} = (2, 2)$ theories, it was more convenient to couple to the appropriate supergravity rather than employing a Noether approach \cite{Chang:2019kiu}. From this intuition, one might expect that the computation of supercharges for deformations of $\mathcal{N} = 4$ SUSY-QM theories might likewise be easier to perform by coupling to worldline supergravity.

It would also be interesting to understand $T\overline{T}$-type deformations in theories with even more supersymmetry, like $\mathcal{N} = 8$ or maximal SUSY.\footnote{Other deformations of QM theories with more supersymmetry, albeit not related to $T\overline{T}$, have been considered in \cite{Ivanov:2018czk,Ivanov:2019gxo,Sidorov:2014lvo,Fedoruk:2017efi}. See also \cite{Galajinsky:2020hsy,Kozyrev:2021agn,Kozyrev:2021icm} for discussions of the super-Schwarzian with more SUSY.} Such an endeavor is complicated by the absence of a conventional superspace, which makes all of the supersymmetries manifest. One could, of course, work with a reduced superspace like $\mathcal{N} = 2$ or $\mathcal{N} = 4$, which geometrizes a subset of the supersymmetry transformations, but the action of the non-manifest SUSY generators will then be corrected order-by-order in $\lambda$ after turning on a $T\overline{T}$-like deformation.

\emph{Connections to Supersymmetric BF Gauge Theory}

Another direction concerns the holographic interpretation of these results. We have emphasized that part of the motivation for considering deformations of $(0+1)$-dimensional theories whose Lagrangians take the purely kinetic form:
\begin{align}\label{purely_kinetic}
    S = \int \, dt \, g_{ij} ( X ) \dot{X}^i \dot{X}^j \,,
\end{align}
where the $X^i$ are coordinates on a Lie group $G$, is that such theories are dual to BF gauge theories with gauge group $G$. This relationship holds with or without supersymmetry; in the SUSY case, the dual is a SUSY-BF theory, and the quantum mechanics theory admits an interpretation as a particle moving on a supergroup. In the special case that the gauge group is an extension of $\mathrm{SL}(2, \mathbb{R})$, the dual BF theory is also related to JT gravity \cite{Iliesiu:2019xuh} and to other interesting theories such as SYK.

There have been some interpretations offered for the holographic interpretations of the $T\overline{T}$-like deformation of quantum mechanics in these various dual theories, at least in the non-manifestly-supersymmetric context. For instance, connections to cutoff JT gravity and the Schwarzian have been discussed in \cite{Gross:2019ach,Gross:2019uxi,Iliesiu:2020zld,Stanford:2020qhm,Griguolo:2021wgy}, related analyses of the dual matrix models have been carried out in \cite{Rosso:2020wir,Ebert:2022gyn}, and a connection to modified boundary conditions in BF gauge theory is discussed in chapter \ref{ch:BF}. 

It would be very interesting to extend these holographic interpretations to the case with manifest supersymmetry. In the undeformed case, the correspondence between the quantum mechanical theory of a particle moving on an $\mathrm{SL}(2, \mathbb{R})$ group manifold and the BF gauge theory with gauge group $\mathrm{SL}(2, \mathbb{R})$ is lifted to the supersymmetric setting by promoting the gauge group to either $\mathrm{OSp}( 1 \, \vert \, 2)$ for $\mathcal{N} = 1$ SUSY or $\mathrm{OSp} ( 2 \, \vert \, 2)$ for $\mathcal{N} = 2$ SUSY, as is nicely reviewed in Section 4.2 of \cite{Mertens:2018fds}. Here $\mathrm{OSp} ( N \, \vert \, 2p )$ is the orthosymplectic supergroup, a particular sub-supergroup of $\mathrm{GL}(N \, \vert \, 2p)$, which is the supergroup version of the general linear group $\mathrm{GL}(N)$.\footnote{In particular, $\mathrm{OSp} ( N \, \vert \, 2p )$ is the sub-supergroup of $\mathrm{GL}(N \, \vert \, 2p)$ which preserves a symmetric bilinear form on the bosonic elements (analogous to the orthogonal group) and preserves a symplectic form on the fermionic elements (analogous to the symplectic group); hence the name ``orthosymplectic.''} We focus on the $\mathcal{N} = 2$ case which was the main focus of this thesis.\footnote{ Various aspects of the $\mathcal{N} = 1$ version of this theory, including its relationship to the super-Schwarzian and the properties of boundary-anchored Wilson lines, have been studied in \cite{Fan:2021wsb}.} This $\mathcal{N} = 2$ supersymmetric BF theory was analyzed in \cite{Astorino:2002bj,Livine:2007dx}, and its action can be written as
\begin{align}
    S_{\text{BF}}^{\mathcal{N} =2} = \int_{M} \mathrm{STr} \left( \Phi F \right) - \frac{1}{2} \oint_{\partial M} \mathrm{STr} ( \Phi \mathscr{A}_t )\,,
\end{align}
where $\mathrm{STr}$ is the supertrace, $\Phi$ is the supersymmetric analogue of the scalar $\phi$ appearing in the usual BF Lagrangian $\mathcal{L}_{\text{BF}} = \mathrm{tr} ( \phi F )$, and $F = d \mathscr{A} + \mathscr{A} \wedge \mathscr{A}$ is the field strength of a supersymmetric gauge connection $\mathscr{A}$. In this $\mathcal{N} = 2$ case, each of $\mathscr{A}$ and $\Phi$ admit an expansion in terms of the $8$ generators of the $\mathfrak{osp} ( 2 \, \vert \, 2)$ Lie superalgebra; these consist of the usual $3$ generators of $\mathfrak{sl} (2, \mathbb{R})$, along with four fermionic generators, and one additional bosonic $\mathfrak{u} ( 1 )$ generator required by supersymmetry.

One would like to understand what modification of the bulk super-BF theory corresponds to turning on the $f(\mathcal{Q}, \bar{\mathcal{Q}})$ operator in the boundary SUSY-QM theory. We can view this question as the dimensional reduction of a related query: what has happened to a bulk $\mathrm{AdS}_3$ supergravity theory, written in Chern-Simons variables when the dual supersymmetric field theory is deformed by supercurrent-squared? The standard intuition from $\mathrm{AdS}/\mathrm{CFT}$ is that the addition of a double-trace operator in the field theory corresponds to a modification of the boundary conditions for the bulk fields \cite{Klebanov:1999tb,Witten:2001ua}, although it is not clear that this intuition should generically apply for irrelevant double-trace deformations as opposed to relevant (or marginal) operators. In the non-supersymmetric context, it has been argued that this expectation is indeed correct and that activating $T\overline{T}$ in a $2d$ CFT corresponds to a rotation of the sources and expectation values in the dual $\mathrm{SL} ( 2, \mathbb{R} ) \times \mathrm{SL} ( 2, \mathbb{R} )$ Chern-Simons theory \cite{Llabres:2019jtx}. A similar rotation of boundary conditions appears in the non-supersymmetric setting of a $2d$ BF gauge theory, which is dual to a boundary $(0+1)$-dimensional theory in chapter \ref{ch:BF}. It would be interesting to see whether the deformed super-BF theory, dual to a quantum mechanics theory deformed by $f(\mathcal{Q}, \bar{\mathcal{Q}})$ likewise admits such an interpretation, perhaps involving a linear mixing of the coefficient functions multiplying the $8$ generators of $\mathfrak{osp} ( 2 \, \vert \, 2 )$ in the expansions of $\Phi$ and $\mathcal{A}$.

\emph{Deformations of Multiple Scalars; Target Space Geometry}

Another avenue for investigation is seeking solutions to the $f(\mathcal{Q},\bar{\mathcal{Q}})$ flow equations for theories with multiple scalars. In this work, we have only managed to find a closed-form result (\ref{susy_qm_1_scalar_soln}) for the deformed theory in the case of a single scalar, and even then, we have only found an expression that is on-shell equivalent to the full solution since we have imposed one implication of the superspace equations of motion. But of course, the most interesting examples are the deformed theories of a particle moving on a higher-dimensional manifold, such as the $3$-dimensional $\mathrm{SL}(2, \mathbb{R})$ group manifold relevant for the Schwarzian theory. We have already mentioned in the analysis of the corresponding question for $2d$ field theories around equation (\ref{finite_lambda_2d_schematic}) that solving the flow in this context is much more difficult because one expects a system of coupled PDEs for the functions multiplying the various two-fermion, four-fermion, etc. terms in the superspace Lagrangian. However, if one could find a partial or approximate solution with multiple scalars -- perhaps after going partly on-shell, as we have done here -- the result could be interesting.

For example, given such a solution, we could ask whether the resulting deformed theory still admits an interpretation as a point particle moving on some deformed target-space geometry. One might think not since our intuition is that the ordinary $T\overline{T}$ flow in $2d$ generates theories that are no longer local QFTs. Analogously, one might expect that $f(\mathcal{Q}, \bar{\mathcal{Q}})$ deformed SUSY-QM theories exhibit some signature of non-locality. For instance, the particle whose position is described by the $X^i$ in the undeformed theory could become delocalized into a ``fuzzy particle'' over a length scale controlled by $\lambda$. It would be interesting to ask whether other properties of the target manifold can be probed in this case or if the target manifold itself is changed.

On the other hand, in the undeformed theory, the Witten index of the theory is controlled by the Euler characteristic of the target space. Since our $f(\mathcal{Q}, \bar{\mathcal{Q}})$ flow is the supersymmetric extension of an $f(H)$ deformation -- which does not affect the energy eigenstates but merely modifies their energy eigenvalues -- it seems that this index remains unchanged under our deformation, which suggests that the target space topology is also unmodified.

There is some evidence that other indices cannot flow under $T\overline{T}$-like deformations. For instance, related indices like the elliptic genus have been shown not to flow under the usual $T\overline{T}$ in two dimensions if the seed theory is conformal \cite{Datta:2018thy}, and the same conclusion seems likely to hold if the undeformed theory is integrable but not conformal \cite{Ebert:2020tuy}. Nonetheless, it would be worthwhile to make this intuition precise in the SUSY-QM case and perhaps look for other index-like quantities that do flow under $f(\mathcal{Q}, \bar{\mathcal{Q}})$ and which may admit an interpretation via target space geometry.

\emph{Relation to Supersymmetric SYK}

We mention one final future direction, along the lines of the previously mentioned question about the relationship of this deformation with super-BF theory, but which is also related to the issue of defining our $f ( \mathcal{Q}, \bar{\mathcal{Q}} )$ deformation in quantum mechanics with a potential.

It is well-known that the Schwarzian or particle-on-a-group theory is also related to the SYK model of Majorana fermions with random all-to-all interactions \cite{kitaev_talk,PhysRevLett.70.3339}. The SYK model has a supersymmetric extension \cite{Fu:2016vas,Murugan:2017eto,Peng:2017spg}; for an (incomplete) collection of related works on the SYK model and supersymmetry, see \cite{Kanazawa:2017dpd,Yoon:2017gut,Hunter-Jones:2017raw,Hunter-Jones:2017crg,Narayan:2017hvh,Forste:2017apw,Garcia-Garcia:2018ruf,Kato:2018kop,Sun:2019yqp,Behrends:2019sbd,Berkooz:2020xne,Peng:2020euz,Gates:2021jdm,Peng:2016mxj,Li:2017hdt} and references therein.

The application of $T\overline{T}$-like deformations in quantum mechanics to the non-supersymmetric SYK model was carried out in \cite{Gross:2019uxi} (see also \cite{He:2021dhr}). In that case, after shifting the ground state energy of the model by a constant $E_0$, it was pointed out that there are two choices for how to perform the deformation:
\begin{enumerate}
    \item First, perform the average over disorder in the undeformed model and then deform the Hamiltonian by the desired $T\overline{T}$ or $f(H)$ operator.
    
    \item Begin by deforming the Hamiltonian by some $f(H)$ operator and then perform the disorder average in the deformed theory.
\end{enumerate}
The authors of \cite{Gross:2019uxi} point out that it is easier to do the former since if one first deforms the Hamiltonian then this procedure will introduce higher powers of the disorder, which makes the resulting disorder average difficult. Although the latter provides a microscopic picture of physics.

It would be interesting to carry out a version of this analysis in the supersymmetric setting using the techniques developed in the present work. To do this, one should use a presentation of the supersymmetric SYK action which is written directly in $\mathcal{N} = 1$ or $\mathcal{N} = 2$ superspace, such as those developed in \cite{Fu:2016vas,Bulycheva:2018qcp}. For concreteness, let us focus on the $\mathcal{N} = 2$ case. The degrees of freedom for the $\mathcal{N} = 2$ super-SYK model are packaged into chiral superfields $\Psi, \bar{\Psi}$ which obey the constraints
\begin{align}
    \bar{D} \Psi = D \bar{\Psi} = 0 \,.
\end{align}
The Lagrangian is a sum of a kinetic $F$-term plus a holomorphic superpotential:
\begin{gather}
    L = \int \, d \bar{\theta} \, \mathcal{A}_{\text{kin}} + \left( \int d \theta \, \mathcal{A}_{\text{potential}} + \text{c.c} \right) \,, \nonumber \\
    \mathcal{A}_{\text{kin}} = \bar{\Psi} D_i \Psi \, \qquad \mathcal{A}_{\text{potential}} = C_{i_1 \cdots i_k} \Psi_{i_1} \cdots \Psi_{i_k} \,.
\end{gather}
To study the appropriate $T\overline{T}$-type deformation of such a superspace Lagrangian, one would, therefore, need to generalize the analysis presented in this work to allow for fermionic superfields and potentials. One could also attempt to understand deformations of SYK via the dimensional reductions of appropriate two-dimensional field theories \cite{Turiaci:2017zwd,Murugan:2017eto}, or to investigate $T\overline{T}$-like deformations in related disordered supersymmetric models \cite{Chang:2018sve,Chang:2021fmd,Chang:2021wbx}.
 \chapter{$T\overline{T}$-Deformed Free Energy of the Airy Model}
\label{ch:Airy}
Sharpening the correspondence of JT gravity and its matrix model description under the $T\overline{T}$ deformation is of interest. To proceed, we simplify the problem by considering the Airy model and deform Airy correlators in the same way as in $T\overline{T}$-deformed JT gravity. We use those correlators to numerically compute the annealed and quenched free energies for $\lambda > 0$ and $\lambda < 0$ from an integral representation of the replica trick. At the leading order in $\lambda$ and low temperatures, we confirm that the genus-zero quenched free energy monotonically decreases as a function of temperature when perturbation theory is valid. We then study the all-genus quenched free energy at low temperatures, where we discover and discuss subtleties due to non-perturbative effects in the Airy model, as well as the contributions from the non-perturbative branch under the $T\overline{T}$ deformation.
\section{Introduction}
An application of the $T\overline{T}$ deformation relevant to this chapter is probing the low-temperature limit of JT gravity and its Airy model description.\footnote{Hereafter, we will refer to the Airy limit of Gaussian matrix models as the ``Airy model.''} To further motivate the Airy model, we list a few reasons why it is interesting and useful to study. Firstly, the genus expansion can be summed via \cite{Okounkov:2001usa} allowing one to make definite statements on non-perturbative corrections. Secondly, JT gravity at low temperatures, or more precisely the 't Hooft limit:
\begin{equation}
\label{eq:'t Hooft}
    \hbar \rightarrow 0\,, \quad \beta \rightarrow \infty\,, \quad \operatorname{for ~ fixed}~ \hbar \beta\,,
\end{equation}
is dual to the Airy model at all genera known from \cite{Okuyama:2020ncd}. The authors of \cite{Okuyama:2020ncd} showed all the non-trivial information of the spectral curve of JT gravity is still preserved under the 't Hooft limit \eqref{eq:'t Hooft}. Additionally, there is an important caveat as \cite{Saad:2019lba} finds: the exact eigenvalue density has an exponential leakage when $E < 0$, which is denoted as the ``classically forbidden'' region causing the system unstable. This non-perturbative instability is not special for the Airy model as the same phenomena occur for the matrix model dual of JT gravity.

However, a recent proposal by \cite{Johnson:2019eik} improves the non-perturbative behavior of JT gravity by removing the non-perturbative instabilities. Therefore, one should only expect this relation between JT gravity and the Airy model to confidently hold perturbatively. There are subtleties for the Airy model's quenched free energy not monotonically decreasing as a function of temperature using directly the replica trick as done by \cite{Engelhardt:2020qpv}. 

Fortunately, Okuyama \cite{Okuyama:2021pkf} showed this failure of monotonicity arises due to analytical continuation issues in the correlators $\langle Z^n\rangle$ to $\langle Z^{n=0}\rangle$ and proposed an alternative formulation of the replica trick to correctly give a monotonically decreasing quenched free energy in the low-temperature limit after summing over all genus. With these motivations of the Airy model, we wish to investigate how some of these features change under the $T\overline{T}$ deformation with a holographic picture in mind to understand JT gravity and its matrix model dual.

Through the lens of the $T\overline{T}$ deformation, it is interesting to probe observables between JT gravity and its double-scaled matrix model dual description at a finite cutoff governed by $\lambda$, as done in \cite{Gross:2019ach,Gross:2019uxi,Iliesiu:2020zld,Stanford:2020qhm,Rosso:2020wir,Griguolo:2021wgy}. The purpose of this chapter is to further sharpen the correspondence between JT gravity at low temperatures and its Airy model description by computing the correlators and quenched free energy. We now comment on the deformed energy spectrum in JT gravity.

\begin{figure}[h]
     \begin{subfigure}[b]{0.4\textwidth}
         \centering
         \includegraphics[width=\textwidth]{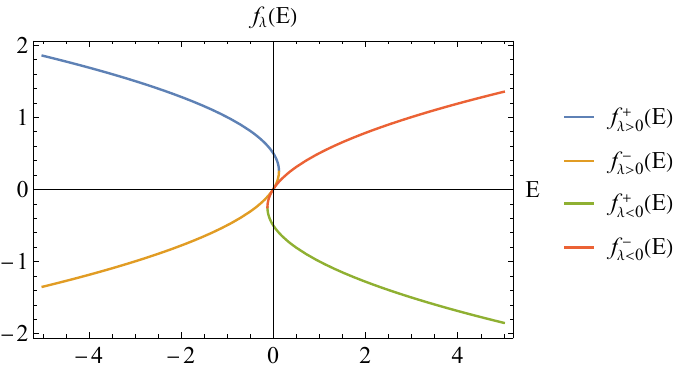}
         \caption{$|\lambda| = 1$}
         \label{fig:ncT1a}
     \end{subfigure}
     \begin{subfigure}[b]{0.4\textwidth}
         \centering
         \includegraphics[width=\textwidth]{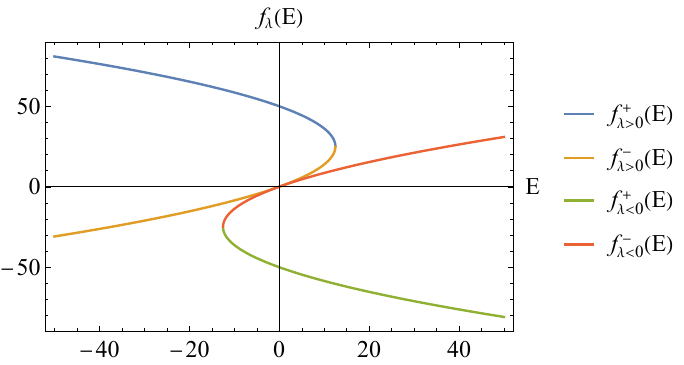}
         \caption{$|\lambda| = 0.01$}
     \end{subfigure}
     \caption{We plot the entire real deformed energy spectrum as a function of the undeformed energy $E$ when $|\lambda| = 1$ and $|\lambda| = 0.01$. Here $f^{+}_{\lambda > 0}(E)$ (blue curve) and $f^{+}_{\lambda < 0}(E)$ (green curve) describe the non-perturbative branch. While $f^-_{\lambda > 0}(E)$ (orange curve) and $f^-_{\lambda < 0}(E)$ (red curve) describe the perturbative branch.}
     \label{NPP}
\end{figure}

Throughout several instances in this thesis, we established and used the following differential equation for the one-dimensional stress scalar $T^\tau_\tau = f^{\pm}_\lambda (E)$:
\begin{equation}
\label{f(E)}
   \frac{\partial f^\pm_\lambda (E)}{\partial \lambda} = \frac{f^\pm_\lambda (E)^2}{\frac{1}{2}-2\lambda f^\pm_\lambda(E)}
\end{equation}
with two solutions
\begin{equation}
\label{energyspectra}
    f^{\pm}_\lambda(E) = \frac{1\pm \sqrt{1-8\lambda E}}{4\lambda}\,.
\end{equation}
As can be seen in the deformed energy spectrum \eqref{energyspectra} and figure \ref{NPP}, the sign of $\lambda > 0$ violates unitarity when the undeformed energy is $E > \frac{1}{8\lambda}$. The authors of \cite{Iliesiu:2020zld} carefully dealt with this violation of unitarity and restored it in their rigorous non-perturbative treatment for the deformed partition function from a Wheeler-de Witt wavefunction perspective. In short, one writes a linear combination of the wavefunction of the two branches \eqref{energyspectra} such that the density of states stays real for all energies. 

Other treatments of the deformation parameter $\lambda$'s sign are addressed by \cite{Rosso:2020wir} from a double-scaled matrix model perspective of JT gravity with an attempt to define the dual deformed matrix model description at a finite cutoff. Unfortunately, the analysis of \cite{Rosso:2020wir} was unable to match the $T\overline{T}$-deformed correlators between JT gravity and the matrix model description, but did provide several alternative methods on how one could properly make this correspondence well-defined. We will discuss the importance of \cite{Iliesiu:2020zld} and \cite{Rosso:2020wir} more throughout various places of this chapter.\footnote{A recent proposal in \cite{He:2022bbb} successfully matched the $T\overline{T}$-deformed partition functions of $\mathcal{N} = 1$ type 0A and type 0B JT supergravity with the associated matrix models. Additional evidence of the duality from \cite{He:2022bbb} was calculating the deformed $T\overline{T}$-deformed matrix model correlators via topological recursion relations.}

It is helpful to point out in this chapter, that we will think about JT gravity and its $T\overline{T}$ deformation from the point of view of the matrix model description. As shown in \cite{Saad:2019lba}, higher topology contributions in JT gravity are captured entirely by a double-scaled matrix model. The partition function of JT gravity at any genus and with any number of boundaries can be computed from the correlators in its matrix model description. From the point of view of boundary theory, we only know that JT gravity on a disk is dual to Schwarzian theory on the boundary. Also, the result of \cite{Saad:2019lba} showed the connected $n$-point function in JT gravity -- namely, the partition function on a connected surface with $n$ boundaries -- does not factorize. The boundary theory is an ensemble of theories, not a single theory. Since the replica trick is essential to compute quenched free energy and requires knowledge of $n$-point correlators, we will think from the point of view of the matrix model and its correlators rather than directly from the boundary theory.

In section \ref{review}, we first explain how the $T\overline{T}$-deformed partition function of various topologies in JT gravity is found through an integral transformation of the undeformed partition function. Additionally, in section \ref{review}, we then extend this integral transformation to find a relation between the deformed and undeformed correlators via: 
\begin{equation}
    \langle Z(\beta_1) \cdots Z(\beta_n)\rangle_\lambda = \int_{-\infty}^{\infty} dE_1 \cdots dE_n \, \rho(E_1,E_2,\cdots,E_n) e^{-\beta_1 f^-_\lambda(E_1)} \cdots e^{-\beta_n f^-_\lambda(E_n)}\,,
\end{equation}
where $\rho(E_1,\cdots,E_n)$ is defined by ensuring
\begin{equation}
    \langle Z(\beta_1) \cdots Z(\beta_n)\rangle_0 = \int_{-\infty}^{\infty} dE_1 \cdots dE_n \, \rho(E_1,E_2,\cdots,E_n) e^{-\beta_1 E_1} \cdots e^{-\beta_n E_n}\,.
\end{equation}
Only when $\lambda < 0$, we can use the integration kernel $K(\beta,\beta') = \frac{\beta}{\sqrt{-8 \pi \lambda} \beta'^{\frac{3}{2}}} e^{\frac{(\beta - \beta')^2}{8\lambda \beta'}}$ to be reviewed in section \ref{review} to conveniently compute the deformed correlators:
\begin{equation}
\label{eq:2kernels}
\langle Z (\beta_1) \cdots Z (\beta_n) \rangle_\lambda =  \int^\infty_0  d\beta_1' K(\beta_1, \beta_1')  \cdots \int^\infty_0  d\beta_n' K(\beta_n, \beta_n') \langle Z(\beta_1') \cdots Z(\beta_n') \rangle_0\,,
\end{equation}
where
\begin{equation} 
\label{eq:trans}
\int_0^\infty d\beta' K(\beta,\beta')e^{-\beta' E} = e^{-\beta f_\lambda^-(E)}\,.
\end{equation}
This is the case when we look at any particular order in the genus expansion. However, when we look at the exact solution of the Airy model, non-perturbative effects there extend the integration range to $E_i \in (-\infty,\infty)$, reflecting the translational invariance of the underlying effective Hamiltonian \cite{Johnson:2019eik}. As we later encounter the non-perturbative effect of the $T\overline{T}$ deformation, hereafter, we refer to this as the \textit{non-perturbative instability} to distinguish the two. 

With the density of states for the Airy model, the undeformed one-point function is given by
\begin{equation} \langle Z(\beta) \rangle_{\text{Airy}} = \int_{-\infty}^{\infty} dE\, \rho_{\text{Airy}}(E)e^{-\beta E}\,,
\end{equation}
where, from \cite{Forrester:1993vtx,Ginsparg:1993is}, the density of states is
\begin{equation} 
\label{eq:}
\rho_{\text{Airy}}(E) = \text{Ai}'(-E)^2 + E \,\text{Ai}(-E)^2\,,
\end{equation}
and $\text{Ai}(x)$ is the Airy function of the first kind.

In this Airy model, we should use the following 
\begin{equation}\label{airy1pt} \langle Z(\beta) \rangle_{\text{Airy},\lambda} = \int_{-\infty}^{\infty} dE\, \rho_{\text{Airy}}(E) e^{-\beta f^-_\lambda(E)}\,,
\end{equation}
instead of \eqref{eq:2kernels}, which is invalid because the integral transformation \eqref{eq:trans} of $e^{-\beta E}$ diverges when $E<\frac{1}{8\lambda}$.

From determining the deformed correlators in section \ref{review}, an immediate application is to calculate the annealed and quenched free energies in JT gravity at low temperatures where certain approximations are feasible. In section \ref{Airy}, we review a new way of computing the quenched free energy from an integral representation of replica trick due to Okuyama \cite{Okuyama:2021pkf}:
\begin{equation}
\label{okuyama1}
 \lim _{n \rightarrow 0} \frac{\left\langle Z^{n}\right\rangle-1}{n} =  \ln \langle Z\rangle-\int_{0}^{\infty} \frac{d x}{x}\left[\left\langle e^{-Z x}\right\rangle-e^{-\langle Z\rangle x}\right]\,,
\end{equation}
where $\left\langle e^{-Z x}\right\rangle$ is fully determined by connected correlators $\langle Z^n\rangle_c$. We elaborate later in section \ref{Airy} why we use this integral representation \eqref{okuyama1} instead of directly using the replica trick in the low-temperature regime, and we comment more on the non-perturbative contributions from the $T\overline{T}$ deformation in section \ref{sec:comments on non-perturbative}. 

However, it is generally difficult to evaluate the correlators of JT gravity under the $T\overline{T}$ deformation let alone sum over all the connected correlators to use Okuyama's formula \eqref{okuyama1}. Since JT gravity is known to have a matrix model description, we will simplify the problem by studying a simpler matrix model, the Airy model, to make progress. Here, it is important to notice that although the $T\overline{T}$ deformation of random matrix models has been studied in \cite{Rosso:2020wir}, the correlators there do not match those of $T\overline{T}$-deformed JT gravity studied in \cite{Iliesiu:2020zld}. 

The goal is to understand JT gravity and its matrix model description at a finite cutoff. We will not follow the $T\overline{T}$ deformation of the matrix model defined in \cite{Rosso:2020wir}. Instead, we require the correlators of the Airy model to transform in the same way as the correlators in JT gravity do under the $T\overline{T}$ deformation, and we will take this as a working definition for our version of $T\overline{T}$ deformation applied to the Airy model. This chapter will not attempt to completely explore this version of deformation for matrix models. 

It is also important to point out one caveat of our simplification. Na\"ively, the double-scaled matrix model dual to JT gravity shares the same non-perturbative instability as the Airy model we considered; for instance, see \cite{Saad:2019lba,Johnson:2019eik}. However, there have been studies on how to improve the non-perturbative behavior of JT gravity and remove this undesired feature \cite{Johnson:2019eik}.

In section \ref{sec:Airy}, we use \eqref{okuyama1} to numerically evaluate the quenched free energy in the Airy model for both $\lambda > 0$ and $\lambda < 0$. As a warm-up, we first compute the quenched free energy $F_{q,\lambda}(T)$ at genus-zero at the leading order of $\lambda$ in perturbation theory. We confirm that $F_{q,\lambda}(T)$ is a monotonic function in $T$ at low temperature when a leading order approximation is valid. Additionally, we find that $\lambda < 0$ deformation decreases the quenched free energy while the $\lambda > 0$ increases it. Intuitively, this sign of the deformation parameter $\lambda > 0$ corresponds to JT gravity in a finite box with Dirichlet boundary conditions at $r_c =\frac{\pi \lambda}{4G}$. The increase of the quenched free energy for the $\lambda > 0$ theory is related to the fact that the $T\overline{T}$ deformation cuts off the spectrum. 

On the gravity side, one can think of this phenomenon as a gravitational redshift. For an object in a gravitational potential, the energy measured at the conformal boundary (which is the undeformed case) is reduced compared to the energy measured at the particle location. For example, in the extreme case where a particle is sent towards a black hole's horizon, the energy at infinity vanishes and is negative when the particle is inside the event horizon. Here, in the cutoff gravitational theory, we are measuring this local energy at the finite cutoff boundary, which is closer to the particle compared to the conformal boundary. Thus, the amount of redshift decreases, and the energy we measure increases.

Next, we use the low-temperature approximation $\langle Z(\beta)^n \rangle_c \simeq \langle Z(n\beta) \rangle$ of the Airy model to compute the quenched free energy for finite $\lambda$. Notice that this computation includes contributions from all genera as well as non-perturbative effects in the Airy model, which makes $\rho_{\text{Airy}}(E < 0) \neq 0$.

For the $\lambda > 0$, as studied in \cite{Iliesiu:2020zld,Griguolo:2021wgy}, there are contributions from the non-perturbative branch. We compute the quenched free energy with this branch included and excluded. In both cases, we find the quenched free energy to be divergent.

For $\lambda < 0$, the issue of the complex-valued energy always arises regardless of the value of $\lambda$ since $\rho_{\text{Airy}}(E)$ has support on the entire real axis. Although the issue of complex energy for $\lambda < 0$ has been noticed before in \cite{Zamolodchikov:2004ce} for $2d$ CFT since the ground state energy is $-\frac{c}{12}$, we emphasize this is different from our case, as the energy spectrum is still bounded below by $-\frac{c}{12}$. 

However, for the Airy model, the issue of complex energy will always arise regardless of the value of $\lambda$. This has not been noticed before for $\lambda < 0$. A simple solution is imposing a hard cutoff in the energy $E$, namely, we simply remove these states with complex-valued energy. We will see in section \ref{goodsign} that this option would lead to a violation of the $T\overline{T}$ flow equation by a boundary term in the integral. 

Another option would be to include those states with complex-valued energy to demand the partition function to be real-valued. One must include the other branch as well. By a careful choice of coefficient, one can make sure that the boundary terms cancel each other properly so the flow equation is satisfied. We then compute the quenched free energy for both cases: If we exclude the non-perturbative contribution, we find the quenched free energy to be finite and monotonic at low temperatures and decrease under the deformation. If we include the non-perturbative contribution, we find the quenched free energy diverges.

\section{$T\overline{T}$ deformation in JT gravity and matrix models}
\subsection{$T\overline{T}$ deformed correlation functions in JT gravity}
\label{review}

We review how one computes the deformed partition function in JT gravity via an integral transformation of the partition function in the undeformed theory. As already alluded to in the introduction, the deformed energy spectrum is given by \eqref{energyspectra}, and given the density of states, one can immediately compute the deformed partition function systematically as
\begin{equation}
\label{DeformedZ}
    Z_\lambda(\beta) = \int ^\infty_{-\infty} dE ~ \rho(E) e^{-\beta f^-_\lambda (E)}\,.
\end{equation}
For the moment, we will only consider the contribution from the perturbative branch $f_\lambda^-(E)$. We will return to the potential contribution from the non-perturbative branch $f_\lambda^+(E)$ later in section\ref{sec:comments on non-perturbative}.

The partition function $Z_\lambda (\beta)$ satisfies a flow equation derived by \cite{Iliesiu:2020zld}:
\begin{equation}
\label{flow}
    \left[ 4 \lambda \partial_\lambda \partial_\beta +2 \beta \partial^2_\beta + \left( 1 - \frac{4\lambda}{\beta}\right) \partial_\lambda \right] Z_\lambda (\beta) = 0\,,
\end{equation}
which is closely related to the usual inviscid Burgers' equation from the $2d$ $T\overline{T}$-deformed energy spectra \eqref{f(E)} (e.g. see \cite{Jiang:2019epa}). The deformed partition function \eqref{DeformedZ} may be written in terms of an integral transformation involving a kernel and the undeformed partition function \cite{Gross:2019ach} when $\rho (E < \frac{1}{8\lambda} ) = 0$
\begin{equation}
\begin{aligned}
\label{Direct}
  Z_\lambda(\beta) &=  \frac{1}{2\pi i} \int^\infty_0 d \beta' Z_0(\beta') \int^{i\infty}_{-i\infty} dE e^{-\beta f^-_\lambda (E) + \beta' E} \\&= \int^\infty_0 d\beta' K(\beta, \beta') Z_0(\beta')\,,
  \end{aligned}
\end{equation}
where $K(\beta, \beta')$ is a kernel determined from the deformed energy spectrum \eqref{energyspectra}:
\begin{equation}\label{kernel}
\begin{aligned}
    K(\beta, \beta') &= \frac{1}{2\pi i} \int^{i\infty}_{-i\infty} dE e^{-\beta f^-_\lambda (E) + \beta' E} = \frac{\beta}{\sqrt{-8 \pi \lambda} \beta'^{\frac{3}{2}}} e^{\frac{(\beta - \beta')^2}{8\lambda \beta'}}\,,
\end{aligned}
\end{equation}
namely, the inverse Laplace transform of the Boltzmann factor after deformation for $\lambda<0$.

With the integral transform \eqref{Direct} and its kernel \eqref{kernel} at hand, one can proceed to compute the partition function of the deformed JT gravity on disk, trumpet, and other topologies for $\lambda < 0$. For example, the map between the undeformed and deformed disk and trumpet partition functions, respectively, are 
\begin{equation}
    \begin{aligned}
    &Z_0 (\beta')_\text{D}=\frac{e^{\frac{\pi^2}{\beta'}}}{4\sqrt{\pi} \beta'^{\frac{3}{2}}} \implies  Z_\lambda (\beta)_{\text{D}} = \frac{\beta }{\sqrt{-8\lambda}\pi} \frac{e^{- \frac{\beta}{4\lambda}}}{\beta^2 + 8\pi^2 \lambda} K_2 \left( - \frac{\sqrt{\beta^2 + 8 \pi^2 \lambda}}{4\lambda} \right)\,, \\  &Z_0(b ,\beta)_{\text{T}} = \frac{e^{- \frac{b^2}{4\beta}}}{2\sqrt{\pi \beta}} \implies Z_\lambda(b ,\beta)_{\text{T}} = \frac{\beta}{2\pi \sqrt{-2\lambda}} \frac{e^{- \frac{\beta}{4\lambda}}}{\sqrt{\beta^2 - 2b^2 \lambda}} K_1 \left( - \frac{\sqrt{\beta^2 - 2b^2 \lambda}}{4\lambda} \right)\,,
    \end{aligned}
\end{equation}
where we have used an identity for the modified Bessel functions of the second kind 
\begin{equation}
\label{eq:BesselK}
    \int^\infty_0 d \beta' (\beta')^{-m-\frac{3}{2}} e^{\frac{a}{\beta'}} e^{ \frac{(\beta - \beta')^2}{8 \beta' \lambda}} =\frac{ 2 e^{-\frac{\beta}{4\lambda}}}{(\beta^2 + 8 a \lambda)^{\frac{2m+1}{4}}} K_{\frac{2m+1}{2}} \left( - \frac{\sqrt{\beta^2 + 8 a \lambda}}{4\lambda} \right)\,,\quad m\in\mathbb{R}\,,
\end{equation}
when $\lambda<0$ and $8a \lambda + \beta^2 > 0$. 

As in \cite{Iliesiu:2020zld}, knowing $Z_\lambda(\beta)_{\text{D}}$ and $Z_\lambda(b,\beta)_{\text{T}}$ allows us to build general correlation functions in the deformed JT gravity. The connected correlators on a hyperbolic Riemann surface with $n$ boundary components and genus $g$ are
\begin{equation}
\label{connected cor}
    \left\langle Z(\beta)^{n}\right\rangle_{\mathrm{conn}, \lambda}=\sum_{g=0}^{\infty} e^{-S_{0}(2 g+n-2)} Z_{g, n; \lambda}(\beta)\,,
\end{equation}
where $S_0$ is the two-dimensional Einstein-Hilbert action, and $Z_{g,n;\lambda} (\beta)$ are defined from \cite{Griguolo:2021wgy} as follows:
\begin{equation}
\begin{aligned}
\label{eq:ingredientsZ}
Z_{0,1; \lambda}(\beta)&= Z_\lambda(\beta)_{\mathrm{D}}\,, \\
Z_{0,2;\lambda}(\beta_1,\beta_2)&= \int_{0}^{\infty} d b~ b Z_\lambda(b, \beta_1)_{\text{T}}Z_\lambda(b, \beta_2)_{\text{T}}\,, \\
Z_{g, n; \lambda}(\beta_1,\cdots,\beta_n)&= \int_{0}^{\infty}\Bigg(\prod_{j=1}^{n} d b_{j}~ b_{j} Z_\lambda\left(b_{j}, \beta_i\right)_{\text {T}}\Bigg) V_{g, n}\left(b_{1}, \ldots, b_{n}\right)\,,
\end{aligned}
\end{equation}
with the Weil-Petersson volume $V_{g, n}\left(b_{1}, \ldots, b_{n}\right)$ of a Riemann surface $\Sigma_{g,n}$ (i.e. with genus $g$ and $n$ distinct marked points $p_i$) defined in \cite{Mirzakhani:2006eta} as \footnote{A few of $Z_{g, n; \lambda}(\beta)$ has already been computed in \cite{Griguolo:2021wgy}. Also, see \cite{Dijkgraaf:2018vnm,Saad:2019lba} for a review of Weil-Petersson volumes in $2d$ topological gravity and matrix models.} 
\begin{equation}
    V_{g, n}\left(b_{1}, \ldots, b_{n}\right)= \frac{1}{(2\pi^2)^{3g-3+n}} \int_{\overline{\mathcal{M}}_{g, n}} \exp \left(\omega+\frac{1}{2} \sum_{i=1}^{n} \psi_{i} b_{i}^{2}\right)\,.
\end{equation}
Here $\overline{\mathcal{M}}_{g,n}$ is the Deligne-Mumford compactification of the moduli space $\mathcal{M}_{g,n}$ of $\Sigma_{g,n}$ of complex dimension $(3g-3+n)$, $\psi_i\equiv c_1(\mathcal{L}_i)$ is the first Chern class of the tautological line bundle $\mathcal{L}_i$ over $\overline{\mathcal{M}}_{g,n}$ whose fiber at the point $(x_1, \cdots, x_n)\in\overline{\mathcal{M}}_{g,n}$ is the cotangent line to the curve $C$ at $x_i$, $\omega$ is the Weil-Petersson symplectic form on $\overline{\mathcal{M}}_{g,n}$, and $b_i$ is the length of the $i$th geodesic boundary component of $\Sigma_{g,n}$.

From the definition in \eqref{eq:ingredientsZ}, one might wonder if the Weil-Petersson volumes should flow under the $T\overline{T}$ deformation. One possible way to see why Weil-Petersson volumes do not flow under the deformation\footnote{However, in the context of topological recursion, both the resolvent $R_{g,n;\lambda}$ and function $W_{g,n;\lambda}$ is deformed \cite{Griguolo:2021wgy}, while their relation to each other $W_{g,n;\lambda}(z_1,\cdots,z_n)\equiv(-2)^n z_1\cdots z_n R_{g,n;\lambda}(-z_1^2,\cdots,-z_n^2)$ is intact. So in the deformed theory, $V_{g,n}$ is no longer the Laplace transform of $W_{g,n;\lambda}$. The topological recursion formula in terms of $W_{g,n;\lambda}$ \cite{Eynard:2007fi} is covariant under the deformation, and retains the same form.} is that the flow equation \eqref{flow} should be satisfied on each \textit{asymptotic} boundary component with proper legnth $\beta_i$. But by definition, the flow equation only contains derivatives with respect to $\lambda$ and $\beta$, not $b_i$, the length of \textit{geodesic} boundary component to be glued together. This fact was also adopted in \cite{Rosso:2020wir,Griguolo:2021wgy}.

Then for generic $Z_{g,n;\lambda}(\beta)$, we write the deformed partition functions as 
\begin{equation}
    Z_{g,n;\lambda}(\beta_1,\cdots,\beta_n) = \int_0^\infty dE_1 \cdots dE_n \, \rho_{g,n}(E_1,\cdots,E_n) e^{-\beta_1 f_\lambda^-(E_1)} \cdots e^{-\beta_n f_\lambda^-(E_n)}\,,
\end{equation}
where
\begin{equation}
    \rho_{g,n}(E_1,\cdots,E_n) = \int_0^\infty \bigg(\prod_{j=1}^n db_j \, b_j \, \rho_{\text{T}}(b_j,E_j)\bigg) V_{g,n}(b_1,\cdots,b_n)\,,
\end{equation}
and 
\begin{equation}
    \rho_{\text{T}}(b,E) = \frac{\cos(b\sqrt{E})}{2\pi\sqrt{E}}\quad \text{such that} \quad \int_0^\infty dE \, \rho_{\text{T}}(b,E) e^{-\beta E} = Z_0(b,\beta)_{\text{T}}\,.
\end{equation}
To conclude this section, we recall the differential operator presentation of the $T\overline{T}$ deformation introduced in \eqref{eq:derivative}, which is sufficient for computing perturbative expansions in $\lambda$ and will be used in section\ref{genus0sec}, by rewriting the exponential
\begin{equation}
    \begin{aligned}
    \label{diffop} 
     e^{-\beta f^-_\lambda(E)} &= e^{-\beta \sum^\infty_{m=1} c_m \lambda^m E^{m+1}} e^{-\beta E} \\\
&= \mathscr{D}_{y;\lambda}|_{y=\beta} \, e^{-y E}\,.
    \end{aligned}
\end{equation}

It is then straightforward to compute partition functions of generic topologies using this differential operator. However, for multiple boundary components, one starts from the undeformed partition function with different inverse temperatures $\beta_i$ on each boundary component and then applies the differential operator $\mathscr{D}_{y_i;\lambda}|_{y_i = \beta_i}$ to each boundary component separately
\begin{equation}
    \langle Z(\beta_1)\cdots Z(\beta_n) \rangle_\lambda = \left( \prod_{i=1}^n \mathscr{D}_{y_i;\lambda}|_{y_i = \beta_i} \right) \langle Z(y_1) \cdots Z(y_n) \rangle_0\,,
\end{equation}
which, in particular, yields
\begin{equation}
    \langle Z(\beta)^n \rangle_\lambda = \left( \prod_{i=1}^n \mathscr{D}_{y_i;\lambda}|_{y_i = \beta} \right) \langle Z(y_1) \cdots Z(y_n) \rangle_0\,.
\end{equation}
We will use this differential operator presentation to perform perturbation calculation in the leading order of $\lambda$ in section \ref{genus0sec}.

\subsection{Quenched free energy}
\label{Airy}
We briefly review a method of performing the replica trick
\begin{equation}
\label{ReplicaTrick}
\langle \ln Z \rangle  =  \lim_{n \rightarrow 0}  \frac{\left \langle Z^n \right \rangle - 1}{n}\,,
\end{equation}
following \cite{Okuyama:2021pkf}. The replica trick \eqref{ReplicaTrick} has been shown in  \cite{Okuyama:2021pkf} to be written as a rather convenient integral representation
\begin{equation}
\label{OkuyamaF}
   \langle\ln  Z\rangle=  \ln \langle Z\rangle - \int_{0}^{\infty} \frac{d x}{x}\left[\left\langle e^{-Z x}\right\rangle-e^{-\langle Z\rangle x}\right]
\end{equation}
such that the analytical continuation from $\langle Z^n\rangle$ to $\langle Z^{n=0}\rangle$ remains unambiguous. From \eqref{OkuyamaF}, the first term is the annealed free energy while the second term encodes the contribution from Euclidean replica wormholes with the interpretation that the operator $e^{-Z x}$ creates spacetime boundary components, so-called ``spacetime D-brane'' or ``SD-brane'' introduced in the context of baby universes \cite{Marolf:2020xie}. Additionally, the term $ \langle e^{-Zx}  \rangle$ can be rewritten as $e^{-\mathcal{Z}(x)}$ in terms of the following generating function of connected correlators
\begin{equation}
\label{calZdef}
    \mathcal{Z}(x) = \sum^{\infty}_{n=1} \frac{(-x)^{n+1}}{n!} \left \langle Z^n \right \rangle_c\,.
\end{equation}
This will turn out to be an important quantity when we compute the $T\overline{T}$-deformed annealed and quenched free energies of the Airy model in the upcoming sections. 

A motivation to study a simple observable, such as free energy, is to see how Euclidean replica wormholes contribute to the Euclidean gravitational path integral
\begin{equation}
\label{factorization}
    \langle Z(B) \rangle = \int_{B} ~[dg]~ e^{-I[g]}
\end{equation}
with spacetime boundary $B$, metric measure $[dg]$ and Euclidean JT gravity action $I[g]$. An obvious way to determine the presence of Euclidean wormholes is to see whether correlation functions of the partition function cease to factorize
\begin{equation}
\label{failure}
    \langle Z(B)^n \rangle \overset{?}{=}\langle Z(B) \rangle^n
\end{equation}
among $n$ boundary components. 

It turns out not to be the case due to the factorization failure as shown in \eqref{OkuyamaF}, and this fact can be confirmed by directly computing the annealed and quenched free energies\footnote{In \cite{Alishahiha:2020jko}, the quenched and annealed free energies in JT gravity with conical deficit angles were computed following the formalism developed by \cite{Mertens:2019tcm,Maxfield:2020ale,Witten:2020ert,Witten:2020wvy} and observed the same pathology of monotonicity failing at low temperatures as \cite{Engelhardt:2020qpv} finds.} as done in \cite{Engelhardt:2020qpv}:
\begin{equation}
    \begin{aligned}
        F_a(\beta) =- \beta^{-1}  \ln \langle Z \rangle\,, \quad F_q(\beta)= - \beta^{-1} \langle \ln Z \rangle\,,
    \end{aligned}
\end{equation}
at inverse temperature $\beta$. They are shown not to be the same indeed, clearly indicating the factorization failure already hinted at by \eqref{OkuyamaF}. 

Unfortunately, the authors of \cite{Engelhardt:2020qpv} computed $F_q(\beta)$ with direct usage of the replica trick \eqref{ReplicaTrick}, and their analysis at low temperature found that $F_q(\beta)$ is \emph{not} monotonically decreasing as a function of temperature. This is fundamentally due to the non-uniqueness of analytically continuing $\langle Z^{n} \rangle$ to $\langle Z^{n = 0}\rangle$. Given this conundrum at low temperature, the correct analytical continuation was performed by Okuyama \cite{Okuyama:2021pkf} without directly using the replica trick \eqref{ReplicaTrick}, and there $F_q(\beta)$ was shown to be a monotonically decreasing function of temperature. 

We now review how this is done following the proof in \cite{Okuyama:2021pkf}. The correlator $\langle Z^n \rangle$ can be expanded in terms of connected correlators $\langle Z^k \rangle_c$ as \cite{Okuyama:2019xvg}: 
\begin{equation}
\langle Z^n\rangle=\langle Z\rangle^n\left[1+\frac{1}{2}n(n-1)\frac{\langle Z^2\rangle_c}{\langle Z\rangle^2}+\cdots\right],
\end{equation}
so now the analytic continuation of  $\langle Z^n \rangle$ is unambiguous due to being rewritten as a polynomial in $n$ up to an overall $\langle Z \rangle^n$. To further find the integral representation \eqref{OkuyamaF}, we generalize the above expansion to
\begin{equation}
\label{polynomial}
    \langle Z^n \rangle = \langle Z \rangle^n \sum_{j_i \geq 0} \frac{n!}{\left( n - \sum_{\ell \geq 2} \ell j_\ell \right)!} \prod_{k \geq 2} \frac{1}{j_k!} \left( \frac{1}{k!} \frac{\langle Z^k \rangle_c}{\langle Z \rangle^k} \right)^{j_k}\,, 
\end{equation}
where  $i \geq 2$ and the $j_k$'s constitute an integer partition
\begin{equation}
    \sum_{k=1}^{n} k j_{k}=n\,.
\end{equation}
Now, using the standard prescription for analytical continuation
\begin{equation}
    \lim_{n \rightarrow 0} \frac{1}{n} \frac{n!}{(n-m)!}= (-1)^{m-1} (m-1)!
\end{equation}
and then the identity
\begin{equation}
\label{eq:trick}
    \int^\infty_0 dy ~ y^{k-1} e^{-y} = (k-1)!,
\end{equation}
the quenched free energy $F_q(\beta)$ can now be written as the integral representation \eqref{OkuyamaF}, whose integration range inherits that of \eqref{eq:trick}. 

\subsection{Comments on non-perturbative contributions}
\label{sec:comments on non-perturbative}
We clarify more of the \emph{non-perturbative} features arising from the $T\overline{T}$ deformation appearing in this chapter.\footnote{Another non-perturbative effect we will encounter is in the Airy model, whose density of states $\rho_{\text{Airy}}(E < 0) \neq 0$. This is already present in the undeformed theory and is unrelated to \cite{Griguolo:2021wgy}. Yet another non-perturbative effect takes place upon summing over all genus. We believe the non-perturbative effects in summing over genus results in non-perturbative instability. We will often refer to this effect as the non-perturbative instability to distinguish the two. But based on context, the readers should have no trouble in distinguishing the two.} The perturbative branch is denoted by the negative branch in the energy spectrum \eqref{energyspectra} since $\lim_{\lambda \rightarrow 0} f^-_\lambda(E) = E$. In contrast, the $\lambda \rightarrow 0$ limit of the positive branch for \eqref{energyspectra} diverges as expected. Most papers omit this branch in their perturbative analysis and only consider the negative branch. Unfortunately, when $\lambda > 0$, the spectrum along the flow becomes complex-valued for large enough energies. To resolve this issue, as explicitly shown in \cite{Iliesiu:2020zld}, one is forced to include the non-perturbative contribution such that the partition function 
\begin{equation}
\label{eq:nonpert}
    Z_{\lambda > 0} (\beta) = \int^\infty_{-\infty} dE\, \rho_+ (E) e^{-\beta f^+_\lambda (E)} + \int^\infty_{0} dE \,\rho_- (E) e^{-\beta f^-_\lambda (E)} 
\end{equation}
is real with appropriate constraints on the density of states $\rho_\pm (E)$ for JT gravity and is a solution to the flow equation \eqref{flow}. The appropriate constraints on $\rho_\pm (E)$ are explained more in-depth by \cite{Iliesiu:2020zld}.
An alternative approach to naturally incorporate non-perturbative effects is through a resurgent analysis. In \cite{Griguolo:2021wgy}, the disk and trumpet deformed JT partition functions written as power series in $\lambda$ are Borel resummed to obtain non-perturbative results, which are used to study how the partition functions summed over topologies (i.e. topological recursion) and spectral form factor are modified under the $T\overline{T}$ deformation. 

\section{Quenched deformed free energy for Airy model}
\label{sec:Airy}
In this section, we compute the $T\overline{T}$-deformed annealed and quenched free energies in the Airy model. It is important to point out that our deformation of the double-scaled matrix model is different from the ones considered in \cite{Rosso:2020wir}, whose $T\overline{T}$ deformation of the double-scaled matrix model dual to JT gravity does not completely match the correlators with the $T\overline{T}$-deformed JT gravity. 

Eventually, our ultimate goal is to understand the quenched free energy in $T\overline{T}$-deformed JT gravity. However, this is a rather difficult problem since even for the one-point function $\langle Z(\beta)\rangle_{\operatorname{JT}}$ of the undeformed theory, there is no analytical expression that includes all-genus contributions, let alone non-perturbative effects. In contrast, there has been much progress with numerical calculations \cite{Johnson:2019eik,Johnson:2020exp,Johnson:2020heh,Johnson:2020lns,Johnson:2021rsh}. Therefore, we simplify and study the Airy model instead, which is known to be the low-energy approximation of JT gravity. Since it is a double-scaled matrix model, one could apply the $T\overline{T}$ deformation defined in \cite{Rosso:2020wir}. We already know that the $T\overline{T}$ deformation defined and studied there does not provide an exact match between JT gravity and its matrix model dual in terms of general correlators. This is not a good choice if the goal is to understand the quenched free energy in $T\overline{T}$-deformed JT gravity. 

To investigate the Airy model with the hope of retaining some essential features of the $T\overline{T}$-deformed JT gravity, we will have to work with a different deformation from the one considered in \cite{Rosso:2020wir}. Since one can construct $T\overline{T}$-deformed JT correlators with any number of boundary components and genera from basic observables like the deformed disk and trumpet partition functions by gluing together pants decomposition \cite{Iliesiu:2020zld}, we can recast the $T\overline{T}$ deformation of the correlators of JT gravity in various ways (e.g. using the differential operator $\mathscr{D}_{y;\lambda}|_{y=\beta}$) as reviewed in section \ref{review}. To match the $T\overline{T}$-deformed matrix model dual to JT gravity, correlators on both sides must deform the same way. We can then apply the same recipe of deforming the matrix model correlators to the Airy model instead of using the one defined in \cite{Rosso:2020wir}. This is the deformation we will adopt and study in this chapter. 

To be more specific, we will take the deformed $n$-point functions in the Airy model as
\begin{equation} 
\label{eq:deformednpt}
\langle Z(\beta_1) \cdots Z(\beta_n) \rangle_{\text{Airy},\lambda} = \int dE_1 \cdots  dE_n \, \rho_{\text{Airy},n}(E_1,\cdots,E_n) e^{-\beta_1 f_\lambda^-(E_1)}\cdots e^{-\beta_n f_\lambda^-(E_n)}
\end{equation}
where $\rho_{\text{Airy},n}(E_1,\cdots, E_n)$ is such that
\begin{equation} 
\label{eq:undeformednpt}
\langle Z(\beta_1) \cdots Z(\beta_n) \rangle_{\text{Airy}, 0} = \int dE_1 \cdots dE_n \, \rho_{\text{Airy},n}(E_1,\cdots,E_n) e^{-\beta_1 E_1} \cdots e^{-\beta_n E_n}\,.
\end{equation}
It is important to notice that if we are interested in computing the deformed correlators at any given genus, there will not be non-perturbative effects (from the Airy model itself before $T\overline{T}$ deforming) so the integration range of $E_i$ is $[0,\infty)$, and we can safely apply the integral transformation \eqref{Direct} on \eqref{eq:undeformednpt} to obtain \eqref{eq:deformednpt} for convenience. However, if we are interested in the exact result, for instance, $\langle Z(\beta)\rangle_{\text{Airy},\lambda}$, then we cannot use the integral transformation, no matter if $\lambda$ is positive or negative, because even for $\lambda<0$, $\rho_{\text{Airy}}$ has support over the entire real axis of $E$. 

There are a few other important caveats to mention. 

Firstly, carrying over the $T\overline{T}$ deformation of JT gravity correlators to matrix model correlators determines the perturbative branch of the $T\overline{T}$ deformation. There will be non-perturbative contributions of the $T\overline{T}$ deformation if we are interested in results with \emph{finite} $\lambda$. We will analyze those non-perturbative effects case-by-case as we encounter them by matching the deformed Airy correlators with either the genus-zero result or the flow equation. A possible systematical treatment would be adapting the resurgent analysis performed in \cite{Griguolo:2021wgy}. However, in our study, eventually, we will have to sum over not only all orders in $\lambda$ but all genus $g$ as well. As hinted in this chapter's introduction, it is well-known that there is a non-perturbative instability in the exact spectral density in the Airy model that is related to the genus expansion.  

Secondly, related to the non-perturbative instability of the Airy model, the same non-perturbative instability appears in the na\"ive matrix model dual of the JT gravity as well \cite{Saad:2019lba,Johnson:2019eik}. There have been works to improve the non-perturbative feature of the JT gravity and remove this undesired feature \cite{Saad:2019lba,Johnson:2019eik,Stanford:2019vob}. There are possibilities that the non-perturbative instability can qualitatively affect the behavior of the quenched free energy in the deformed theory and is important to understand if this is the case. A possible proposal is to extend our study to the JT gravity at general temperatures with the non-perturbative instability taken care of.

Thirdly, it should be emphasized that we do not provide a complete description of our version of the $T\overline{T}$ deformation for the matrix model. We require that in the deformed theory, every correlator has to transform to match the gravity side. One can take this as the working definition of our version of the $T\overline{T}$ deformation for matrix models. To further illustrate the difference between our version and the one in \cite{Rosso:2020wir}, let us begin with the standard Hermitian matrix integral
\begin{equation}\label{MatrixIntegral}
    Z = \int [dM] e^{-\text{Tr}V(M)} = \int d^N x \prod_{1\leq i < j \leq N} (x_i - x_j)^2 e^{-\sum^N_{i=1} V(x_i)}\,,
\end{equation}
where $x_i$ are the eigenvalues of the $N \times N$ Hermitian matrix $M$ and the Vandermonde determinant $\prod_{1\leq i < j \leq N} (x_i - x_j)^2$ appears from diagonalizing $M$. 

The undeformed correlation functions are determined by integrating operators against \eqref{MatrixIntegral}:
\begin{equation}
    \langle O_1(M) \cdots O_n(M) \rangle_0 = \int [dM] e^{-\text{Tr}V(M)} O_1(M) \cdots O_n(M)\,.
\end{equation}
In \cite{Rosso:2020wir}, it is assumed that the deformed matrix model still takes the form of \eqref{MatrixIntegral}, and the only physical quantity that changes is the matrix model potential which shifts from $V(M)$ to $V_{\lambda}(M) = c_{\lambda} V\left(M-2 \lambda M^{2}\right)$. As a somewhat unfortunate consequence of this assumption, the $n$-point functions in the deformed matrix model still do not match the $n$-point functions in the deformed JT gravity. In our case, we start by requiring that the $n$-point functions in the deformed matrix model match the results in deformed JT gravity leading to a different deformation on the matrix model side. In particular, na\"ively the integration measure $[dM]$ of matrices will receive corrections from the $T\overline{T}$ deformation as well if we want to keep the potential $\operatorname{Tr}V(M)$ as a single trace operator as implicitly assumed in \cite{Rosso:2020wir}. To see this, consider the correlators 
\begin{equation}
\begin{aligned}
  &\left  \langle \text{Tr}\, e^{-\beta_1 M} \cdots \text{Tr} \,e^{-\beta_n M} \right \rangle_0 \\
  &\quad\quad\quad\quad\quad\quad=\int d^Nx \prod_{1\leq i \leq j \leq N} (x_i - x_j)^2 e^{-\sum^N_{i=1} V(x_i)} \bigg(\sum^N_{i_1 = 1} e^{-\beta_1 x_{i_1}}\bigg) \cdots \bigg(\sum^N_{i_n=1} e^{-\beta_n x_{i_n}}\bigg)
  \end{aligned}
\end{equation}
in the undeformed matrix model. 

To match the gravitational result, its deformation should take the following form:
\begin{equation}
\begin{aligned}
\label{half-diff1}
  &\left  \langle \text{Tr}\, e^{-\beta_1 f_\lambda^-(M)} \cdots \text{Tr}\, e^{-\beta_n f_\lambda^-(M)} \right \rangle_\lambda \\
  & \quad\quad\,=\int d^Nx \prod_{1\leq i < j\leq N}(x_i-x_j)^2 e^{-\sum^N_{i=1} V(x_i)} \bigg(\sum^N_{i_1 = 1}e^{-\beta_1 f^-_\lambda (x_{i_1})}\bigg) \cdots \bigg(\sum^N_{i_n =1} e^{-\beta_n f^-_\lambda (x_{i_n})}\bigg)\,.
  \end{aligned}
\end{equation}
To determine the deformed matrix integral, we consider a change of variables for $x_i$, such that $f^-_\lambda (x_i) = x_i'$, so that we are computing the same correlators
\begin{equation}
\label{half-diff2}
\left\langle \text{Tr}\,e^{-\beta_1 M'} \cdots \text{Tr} \,e^{-\beta_n M'} \right\rangle_\lambda
\end{equation}
in terms of the new variable $M'$. This will not only change the potential $V(M)$, but the integration measures $[dM]$ as well if we want $e^{-\text{Tr}V(M)}$ to contain only single trace operators.

To further illustrate this fact, we consider the change of variables $y_i = f_\lambda^-(x_i)$ as well as neglect for the moment issues with the branch cut from the square root in $f^-_\lambda (x_i)$ and potential non-perturbative subtleties related to the change of variable for $M$.\footnote{This resembles the ``half-diffeomorphism,'' which appears in the formulation of $T\overline{T}$ deformation by coupling the 2D field theory to topological gravity in \cite{Coleman:2019dvf} where either the metric or coordinates change under the deformation, but not both.} Then $x_i = y_i(1-2\lambda y_i)$ and this implies
\begin{equation}
\begin{aligned}
   \left[ dM \right] e^{-\sum_{i=1}^N V(x_i)} = (d^N y)\prod_{1\leq i < j \leq N} (y_i - y_j)^2 e^{-\operatorname{Tr}V(y_i(1-2\lambda y_i))} \prod_{i,j=1,\cdots,N}(1-2\lambda(y_i+y_j))
\end{aligned}
\end{equation}
Since only the product of the matrix measure $[dM]$ and exponential $e^{-\operatorname{Tr}V(M)}$ are unambiguously defined, one could consider absorbing the extra piece
\begin{equation}
    \prod_{i,j = 1,\cdots,N}(1-2\lambda(y_i+y_j))
\end{equation}
into the definition of $e^{-\operatorname{Tr}V(M)}$ to retain the form of the measure $[dM]$. However, this will lead to an infinite sum over double-trace operators as follows:
\begin{equation}
\begin{aligned}
   \label{eq:MMstartingpoint}
    & \prod_{i,j = 1,\cdots, N}(1-2\lambda(y_i + y_j)) \nonumber  \\ &= \exp \Bigg( \sum_{i,j = 1,\cdots,N}\log(1-2\lambda(y_i + y_j)) \Bigg) \nonumber \\ 
    &= \exp \Bigg(\sum_{i,j = 1,\cdots,N}\sum_{m=1}^\infty \frac{(2\lambda)^m}{m} \sum_{p=0}^m C_m^p y_i^p y_j^{m-p} \Bigg) \nonumber  \\ 
    &= \exp \Bigg(\sum_{m=1}^\infty \frac{(2\lambda)^m}{m}\sum_{p=0}^m \operatorname{Tr}(M^p) \operatorname{Tr}(M^{m-p}) \Bigg)\,. 
\end{aligned}
\end{equation}
Either way, the deformation presented here violates the implicit assumptions in \cite{Rosso:2020wir} and would be the starting point on the matrix model side. However, it is unclear whether one should first $T\overline{T}$ deform and take the double-scaling limit or vice versa. Further analysis is beyond the scope of this thesis.

\subsection{Genus-zero quenched free energy at leading order of $\lambda$}\label{genus0sec}
We begin our analysis on the deformed quenched free energy $F_{q,\lambda}(T)$ at genus-zero perturbatively. To determine $F_{q,\lambda}(T)$, we first compute the deformed genus-zero multi-boundary connected correlators $\langle Z(\beta_1)\cdots Z(\beta_n)\rangle_{c,\lambda}^{g=0}$. The undeformed connected correlators at genus-zero have been computed in \cite{Okuyama:2020ncd} using the genus-zero Korteweg–De Vries (KdV) flow:
\begin{equation}
    \langle Z(\beta_1) \cdots Z(\beta_n) \rangle_{c,0}^{g=0} = g_s^{n-2} \bigg(\sum^n_{i=1} \beta_i\bigg)^{n-3} \prod^n_{i=1} \left(\frac{\beta_i}{2\pi}\right)^{\frac{1}{2}}\,,
\end{equation}
where $g_s\equiv \sqrt{2}$ is the genus-counting parameter.\footnote{Here in fact $g_s = \sqrt{2} \hbar$ and we will set $\hbar = 1$. Adopting the conventions by \cite{Okuyama:2021pkf}, $\hbar$ is the genus-counting parameter in the Airy limit of matrix models while $g_s$ is the natural genus-counting parameter in 2D topological gravity. Also refer to earlier discussions in \cite{Okuyama:2019xbv,Okuyama:2020ncd}.}

The deformed correlators can be computed directly using the integral transformation \eqref{Direct} and, by construction, solve the flow equation \eqref{flow}. For instance, for $\lambda < 0$, the deformed one-point, two-point, and $n$-point correlators are given by 
\begin{equation}
\begin{aligned}
\label{eq:Zcorrs}
	\langle Z(\beta)\rangle^{g=0}_{\lambda} &= g_s^{-1} \frac{e^{-\frac{\beta}{4\lambda}}}{2\pi\beta\sqrt{-\lambda}} K_2\bigg(-\frac{\beta}{4\lambda}\bigg)\,,\\
    \langle Z(\beta)^2 \rangle^{g=0}_{c,\lambda} &= - \frac{\beta_1 \beta_2}{16 \pi^2 \lambda} \int^\infty_0 d\beta_1' \int^\infty_0 d\beta_2' \, \frac{\sqrt{\beta_1' \beta_2'}}{\beta_1' + \beta_2'} \, \frac{1}{(\beta_1' \beta_2')^{\frac{3}{2}}}e^{\frac{(\beta_1 - \beta_1')^2}{8 \lambda \beta_1'} + \frac{(\beta_2 - \beta_2')^2}{8 \lambda \beta_2'} }\,, \\
    \langle Z(\beta)^n \rangle^{g=0}_{c,\lambda} &=g_s^{n-2} \frac{e^{-\frac{n\beta}{4\lambda}}}{\beta^3} \bigg(\frac{\beta^2}{2\pi \sqrt{-\lambda}}\bigg)^n \, \sum_{i_j\geq 0}^{\sum_j i_j = n-3} \frac{(n-3)!}{i_1! \cdots i_n!} \prod_{j=1}^n K_{i_j}\left(-\frac{\beta}{4\lambda}\right)\,, \quad n\geq 3\,.
\end{aligned}	
\end{equation}

\begin{figure}[h]
    \centering
    \includegraphics[scale = 0.1]{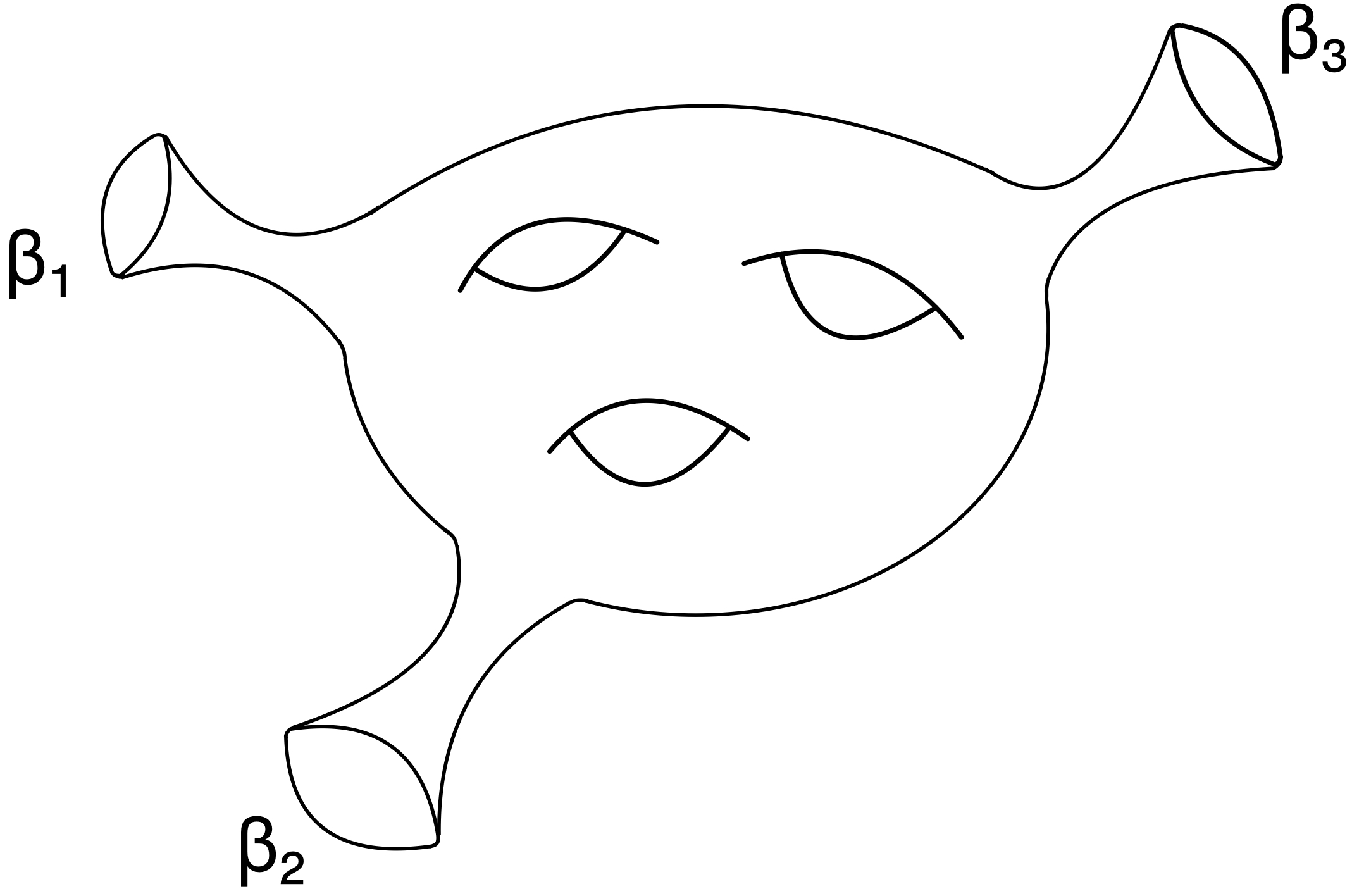}
    \caption{As an example, we have drawn a genus-three Riemann surface with three boundary components, each of which is a circle of circumference $\beta_i$.}
    \label{fig:surface}
\end{figure}
Generally, how one evaluates the above sums of products of modified Bessel functions remains unclear, let alone compute
\begin{equation}
    \mathcal{Z}^{g=0}_\lambda (x)= \sum_{n=1}^\infty \frac{(-x)^n}{n!}\langle Z(\beta)^n \rangle_{c,\lambda}^{g=0}
\end{equation}
in \eqref{calZdef} as in \cite{Okuyama:2021pkf}. 

Therefore, instead, we consider working with perturbation theory and keep the leading order in $\lambda$. For this purpose, it is convenient (in comparison with our ``working definition'' introduced at the beginning of section \ref{sec:Airy}) to express the $T\overline{T}$ deformation in terms of differential operators and only keep the leading order term in $\lambda$ so that the deformation acts as 
\begin{equation} 
\label{eq:diff}
\langle Z(\beta_1) \cdots Z(\beta_n)\rangle_\lambda \rightarrow \bigg(1 - 2 \lambda \sum^n_{i=1} \beta_i \partial_{\beta_i}^2\bigg) \langle Z(\beta_1) \cdots Z(\beta_n)\rangle_0 + O(\lambda^2)\,.
\end{equation}

Then, we find 
\begin{equation}
    \langle Z(\beta)^n \rangle_{c,\lambda}^{g=0} = \langle Z(\beta)^n \rangle_{c}^{g=0} - \frac{\lambda}{\beta^4} n^{n-4} \left(\frac{\beta^3}{\pi}\right)^{\frac{n}{2}}\left(\frac{7}{4}n^2 - 10n + 12\right) + O(\lambda^2)\,.
\end{equation}
On the other hand, 
\begin{equation}
    \mathcal{Z}_\lambda^{g=0}(x) = - \sum_{n=1}^\infty \frac{(-x)^n}{n!} \langle Z(\beta)^n \rangle_{c,\lambda}^{g=0}
\end{equation}
can be evaluated using a similar trick as in \cite{Okuyama:2021pkf}. Specifically, focusing on the leading order piece in $\lambda$, we have the following sum:
\begin{equation} \mathcal{Z}^{g=0}_\lambda(x) - \mathcal{Z}^{g=0}_0(x) = -\frac{\lambda}{\beta^4} \sum_{n=1}^\infty \frac{(-1)^{n-1}n^{n-4}}{n!} \left(x\sqrt{\frac{\beta^3}{\pi}}\right)^n \left(\frac{7}{4}n^2 - 10n +12\right)\,.
\end{equation}
Let $z = x \sqrt{\frac{\beta^3}{\pi}}$, and the sum can be decomposed into three separate pieces:
\begin{equation} 
\label{LambWDiff} 
\begin{aligned}
&A(W(z)) = \sum_{n=1}^\infty \frac{(-1)^{n-1} n^{n-2}}{n!} z^n\,,\quad B(W(z)) = \sum_{n=1}^\infty \frac{(-1)^{n-1} n^{n-3}}{n!} z^n\,,\\ &C(W(z)) = \sum_{n=1}^\infty \frac{(-1)^{n-1} n^{n-4}}{n!} z^n\,,
\end{aligned}
\end{equation}
satisfying
\begin{equation} (z\partial_z) A(W(z)) = W(z), \quad (z\partial_z)^2 B(W(z)) = W(z), \quad (z\partial_z)^3 C(W(z)) = W(z)\,,
\end{equation}
where $W(z)$ is the Lambert function defined by the following Taylor series expansion:
\begin{equation} W(z) \equiv \sum_{n=1}^\infty \frac{(-1)^{n-1}n^{n-1}}{n!}z^n\,.
\end{equation}
The above differential equations \eqref{LambWDiff} can be solved by making the ansatz
\begin{equation} A = A_2 W ^2 + A_1 W\,, \quad B = B_3 W^3 + B_2 W^2 + B_1 W\,, \quad C = C_4 W^4 + C_3 W^3 + C_2 W^2 + C_1 W\,,
\end{equation}
and using the property of the Lambert function
\begin{equation} z\partial_z W(z) = \frac{W(z)}{1+W(z)}\,.
\end{equation}
We then find
\begin{equation} A = \frac{1}{2}W^2 + W, \quad B = \frac{1}{6}W^3 + \frac{3}{4}W^2 + W, \quad C = \frac{1}{24}W^4 + \frac{11}{36}W^3 + \frac{7}{8}W^2 + W
\end{equation}
and
\begin{equation} \mathcal{Z}_\lambda^{g = 0}(z) = \frac{B(W(z))}{2} + \frac{c_1\lambda}{2\beta} \left(\frac{7}{4}A(W(z)) -10 B(W(z)) + 12 C(W(z)) \right)\,.
\end{equation}
Then using \eqref{OkuyamaF}, we compute the $\mathcal{O}(\lambda)$ corrections
\begin{equation}
\begin{aligned}
\label{intg0}
   & \langle \log Z\rangle^{g=0}_\lambda \\
  =& \log\langle Z\rangle^{g=0}_\lambda - \int_0^\infty dx \frac{e^{-\beta^{-3}\mathcal{Z}^{g=0}_\lambda(x)} - e^{-x\langle Z\rangle^{g=0}_\lambda}}{x} \\
=& \log \left[\frac{1}{\sqrt{4\pi\beta^3}}\bigg(1-\frac{15\lambda}{2\beta}\bigg)\right]  \\
&-\int_0^\infty dW\,\frac{1+W}{W}\left[e^{-\frac{1}{\beta^3}\left(\frac{B(W)}{2} - \frac{\lambda}{\beta}\left(\frac{7}{4}A(W) - 10B(W) + 12 C(W)\right)\right)} - e^{-\frac{1}{2\beta^3}We^W\big(1-\frac{15\lambda}{2\beta}\big)}\right] \\&+ \mathcal{O}(\lambda^2) \\
=&\log\bigg(\frac{1}{\sqrt{4\pi\beta^3}}\bigg) - \frac{15\lambda}{2\beta} - \int_0^\infty \frac{dW}{W}(1+W)\bigg(e^{-\frac{B(W)}{2\beta^3}} - e^{-\frac{1}{2\beta^3} We^W}\bigg) \\
&- \frac{\lambda}{8\beta^4}\int_0^\infty dW (1+W)\bigg(e^{-\frac{1}{24\beta^3}W(12+9W+2W^2)}(30+31W+16W^2+4W^3)\\
& - 30 e^{W\left(1-\frac{1}{2\beta^3}e^{W}\right)}\bigg) + \mathcal{O}(\lambda^2)\,,
\end{aligned}
\end{equation}
where we have changed the integration variable from $z$ to $W(z)$ and used
\begin{equation} 
    \langle Z(\beta) \rangle_{\lambda}^{g=0} = \frac{1}{\sqrt{4\pi\beta^3}}\bigg(1-\frac{15\lambda}{2\beta} + \mathcal{O}(\lambda^2)\bigg)\,.
\end{equation}
Notice that in the last line of \eqref{intg0}, we expanded in $\lambda$ for the logarithm and exponential since our result is only valid in the leading order of $\lambda$.

This integral \eqref{intg0} can be evaluated numerically, and, below we plot the quenched free energy $F_q(\beta)$ at genus-zero with $T\overline{T}$ deformation coupling $\lambda = -\frac{1}{20}$ against temperature $T$.

\begin{figure}[h]
    \centering
    \includegraphics[scale = 0.65]{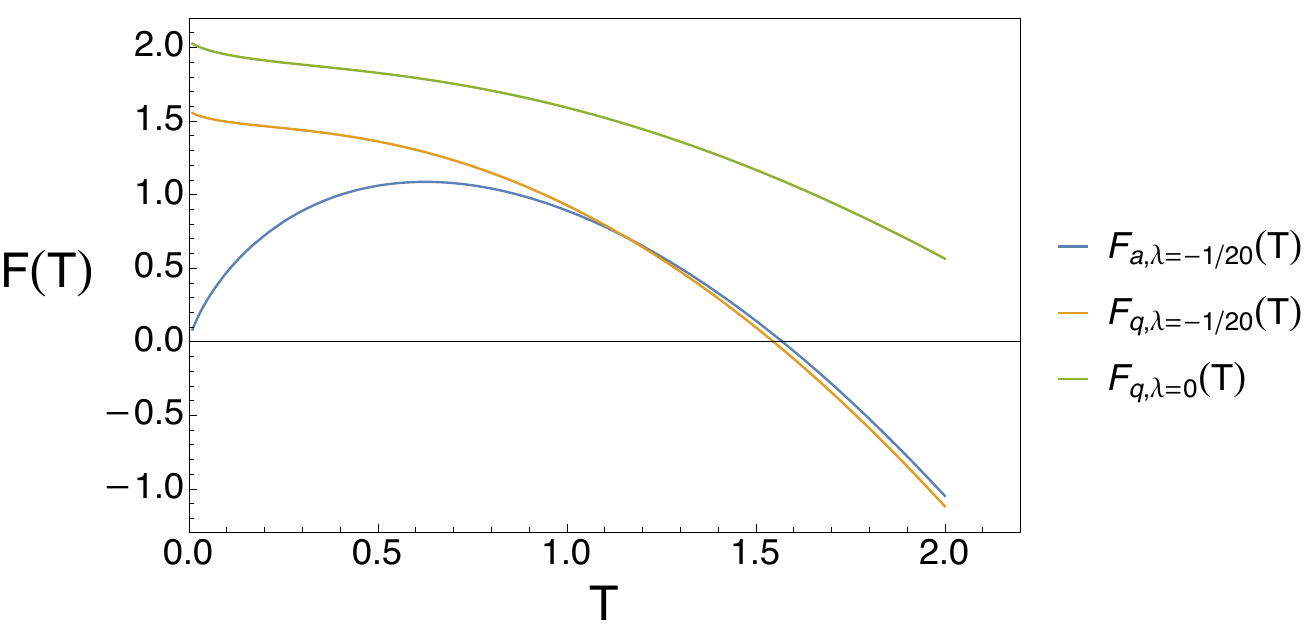}
    \caption{The annealed and quenched free energies for the deformed Airy model as a function of temperature $T$ are plotted. The blue and orange curves are quenched and annealed free energies, respectively, for the deformed theory at $\lambda = -\frac{1}{20}$ and the green curve is the quenched free energy for the undeformed Airy model. Here $\lambda <0$ lowers the quenched free energy.}
    \label{fig:genus0}
\end{figure}

\begin{figure}[h]
    \centering
    \includegraphics[scale = 0.68]{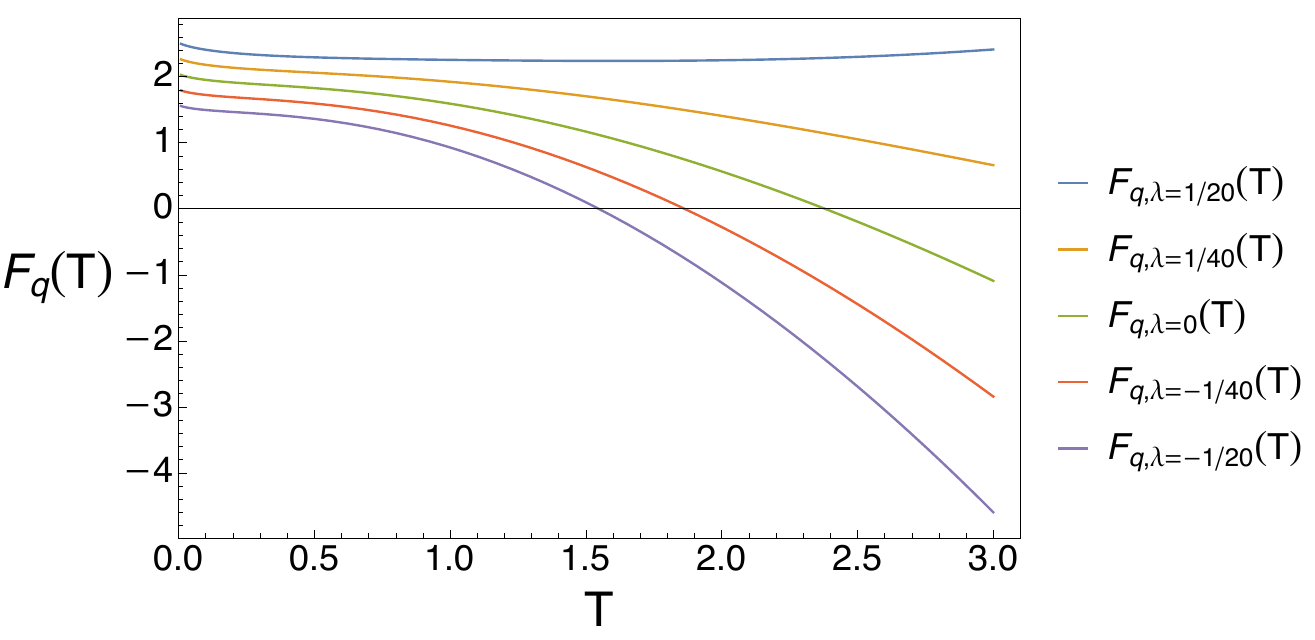}
    \caption{The quenched free energy $F_{q,\lambda}(T)$ for the deformed Airy model as a function of temperature $T$ is plotted. We notice that a deformation with $\lambda < 0$ lowers the quenched free energy while the one with $\lambda > 0$ increases $F_q(T)$. For both signs of $\lambda$, $F_{q,\lambda}(T)$ monotonically decreases as $T$ increases for $T<1$; however, for $\lambda < 0$, this breaks down for $\lambda =\frac{1}{20}$ and at around $T = 1$. Notice that the perturbative expansion of $\lambda$ always appears as $\lambda T$. For the first-order approximation to hold, $\lambda T$ should be small. Therefore, this breakdown of monotonicity when $T$ becomes large is likely because we go beyond the validity of perturbation theory.}
    \label{fig:badlambdas}
\end{figure}

We also plot the genus-zero quenched free energy for various signs of $\lambda$, again using differential operators as in \eqref{eq:diff}, see figure \ref{fig:badlambdas}. Notice in the perturbative expansion of $\lambda$, that $\lambda$ also always appears as $\lambda T$. Hence, for the leading order approximation to hold, $\lambda T$ must be small. In the numerical calculation, we find $F_{q,\lambda}(T)$ monotonically decreases as a function of $T$ for $\lambda > 0$. However, monotonicity can break down when $\lambda T$ is too large for  $\lambda > 0$ and this is likely due to $\lambda T$ exceeding the range of validity of leading order approximation in $\lambda$.

\subsection{All genus quenched free energy in low temperature limit}
Now, using the low-temperature approximation as in \cite{Okuyama:2021pkf}, we compute the quenched free energy starting from the relation in the undeformed theory \cite{Okuyama:2019xvg,Okuyama:2021pkf}: 
\begin{equation} 
\label{lowTapproximation} \langle Z(\beta)^n \rangle_c \simeq \langle Z(n\beta) \rangle, \quad T \lesssim 1\,.
\end{equation}
Under the low-temperature approximation, the deformation of $\langle Z(\beta)^n \rangle_c \simeq \langle Z(n\beta) \rangle $ is easily computable. This is seen from expressing the undeformed correlator for $n$ boundary components $\left \langle Z(\sum_{i=1}^n \beta_i) \right \rangle_0 $ as
\begin{equation} \left \langle Z \left (\sum_{i=1}^n \beta_i \right) \right \rangle_0 = \int dE \, \rho(E) e^{-\sum_{i=1}^n \beta_i E}\,,
\end{equation}
the term dependent on $\beta$ essentially factorizes. Thus, the deformed $n$-point function in low temperatures is
\begin{equation} \left \langle Z(n\beta) \right \rangle_\lambda  = \int_{-\infty}^\infty dE \, \rho(E) e^{-n\beta f^-_\lambda (E)}\,.
\end{equation}
However, we must show that the change from the $T\overline{T}$ deformation for $\langle Z(n\beta) \rangle_{c,\lambda} - \langle Z(n\beta) \rangle_{c,0}$ is of lower order compared to the correction $\langle Z(\beta)^n \rangle_c - \langle Z(n\beta) \rangle$ with \eqref{lowTapproximation}. This is the case because the correction to the approximation $\langle Z(\beta)^{n}\rangle \simeq Z(n\beta)\rangle$ is exponentially suppressed as $e^{-c_0 \beta^3}$ in the low temperature limit $\beta\rightarrow \infty$, where $c_0$ is some positive constant. One can explicitly check this approximation for $n = 2, 3$, where the exact expression of the partition functions can be conveniently found in \cite{Okounkov:2001usa}. For instance, with $n = 2$, we have
\begin{equation} \label{n=2} \langle Z(\beta)^2 \rangle_{c,0} = \langle Z(2\beta) \rangle_0 \,\text{erf}\left(\sqrt{\frac{\beta^3}{2}}\right)\,. \end{equation}
Using the asymptotic expansion of the error function:
\begin{equation} 
\label{eq:error}
\text{erf}(x) = 1 - \frac{e^{-x^2}}{x\sqrt{\pi}}\sum_{n=0}^\infty (-1)^n \frac{(2n-1)!!}{(2x^2)^n}, \quad \text{as\,\,}x\rightarrow \infty\,, \end{equation}
it is clear that the correction is suppressed by $e^{-\frac{\beta^3}{2}}$ in the low temperature limit for \eqref{n=2}. 

Similarly, for $n=3$:
\begin{equation} \langle Z(\beta)^3 \rangle_{c,0} = \langle Z(3\beta) \rangle_0 \left(1 - 12 T\left(\sqrt{3\beta^3},\frac{1}{\sqrt{3}}\right)\right)\,.\end{equation}
From the definition of Owen's $T$ function:
\begin{equation}
\label{eq:owen}
T(h,a) \equiv \frac{1}{2\pi} \int_0^a \frac{e^{-\frac{1}{2}h^2(1+x^2)}}{1+x^2} dx\,, 
\end{equation}
one can see in the large $h$ limit (i.e. large $\beta$ limit), it is indeed suppressed as $e^{-\frac{1}{2}h^2} = e^{-\frac{3}{2}\beta^3}$.

The connected correlator has a well-known integral representation given by \cite{Okounkov:2001usa} and is known to be a closed form only for $n=1, 2, 3$. We can easily check that  $\langle Z(\beta)^n \rangle_c \simeq\langle Z(n \beta)\rangle$ when $n=1,2,3$ for small temperatures, but proving this for $n >3$ is numerically difficult. We will content ourselves and assume $\langle Z(\beta)^n \rangle_c \simeq\langle Z(n \beta)\rangle$ is true for all $n$.

Meanwhile, even at the leading order of $\lambda$, the $T\overline{T}$ deformation will give polynomial corrections in $\beta$ for low temperatures. Hence, the corrections in the approximation $\langle Z(\beta)^n \rangle_0 \simeq \langle Z(n\beta)\rangle_0$ can still be neglected even when we consider the $T\overline{T}$ deformation.

\begin{figure}[h]
    \centering
    \includegraphics[scale = 1.5]{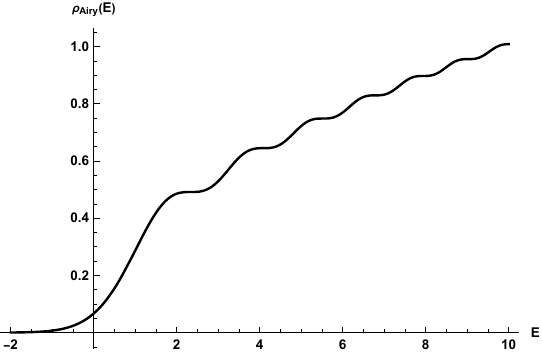}
    \caption{We plot the Airy density of states \eqref{eq:airydensityofstates1}. As one can see, the density of states is supported along the entire real line and exponentially suppressed when $E < 0$.}
    \label{fig:airy1111}
\end{figure}

There is a caveat: the all-genus density of states does not vanish for $E<0$. Instead, it is exponentially suppressed when $E<0$: 
%
%
%
\begin{equation} \label{eq:airydensityofstates1} \rho_{\text{Airy}}(E) =\text{Ai}'\left(-E\right)^2 +  E\, \text{Ai}\left(-E\right)^2
\end{equation}
and is plotted in figure \ref{fig:airy1111}.

This means for either sign of $\lambda$, there will be states with complex-valued energy. We discuss the two cases separately, and in each case, there will be a plausible non-perturbative contribution from the $f^+_\lambda (E)$ branch defined in \eqref{f(E)}.

The deformed quenched free energy is defined and computed as
\begin{equation}
\label{eq:deformfree}
    F_{q,\lambda}(\beta)=-T\langle \log Z\rangle_\lambda=T\int_0^{\infty}\frac{dx}{x}\left[e^{-\mathcal{Z}_\lambda(x)}-e^{-x\langle Z\rangle_\lambda}\right]-T\log\langle Z\rangle_\lambda \,.
\end{equation}

\subsection{Quenched free energy for $\lambda > 0$}
\label{sec:bad}
In this subsection, we study the quenched free energy of the deformed Airy model with $\lambda > 0$. In the deformed theory's spectrum, as explained in this chapter's introduction, the deformed energy $f^-_\lambda (E  > \frac{1}{8\lambda} )$ is complex-valued. Therefore, a cutoff in $E$ at $\frac{1}{8\lambda}$ is needed to remove the complex-valued energy states. Furthermore, there could be contributions arising from non-perturbative states with energy given by the other branch $f^+_\lambda(E)$. This kind of non-perturbative effects have been rigorously studied in \cite{Griguolo:2021wgy, Iliesiu:2020zld}. Their analyses lead one to conjecture what the non-perturbative contribution could be in the deformed Airy theory for $\lambda > 0$.

We study the quenched free energy with and without the non-perturbative contributions. In both cases, we find the quenched free energy $F_{q,\lambda}(T)$ diverges for every $\lambda>0$ and $T$ in our low-temperature approximation. It is unclear to us whether the low-temperature approximation which causes $F_{q,\lambda}(T)$ to diverge or this is the feature of the deformation itself. Additionally, we do not know if repeating the same calculation in JT gravity will lead to the same problem. 

For $\lambda > 0$, the issue with complex-valued energy is fixed by imposing a cutoff in the energy $E$ such that the deformed energy spectrum is real-valued. Furthermore, the resurgent analysis in section 4.2 of \cite{Griguolo:2021wgy} indicates that there could be non-perturbative contributions to the partition function coming from the other branch $f^+_\lambda(E)$. More specifically, the deformed genus-zero partition function of JT gravity is given by 
\begin{equation}
\begin{aligned}
\label{eq:JTdisk}
\langle Z(\beta) \rangle^{g=0}_{\operatorname{JT},\lambda} &= \int_0^{\frac{1}{8\lambda}} dE \, \rho_{\operatorname{JT}}^{g=0}(E) \bigg(e^{-\beta f^-_\lambda (E)} - e^{-\beta f^+_\lambda (E)}\bigg)\,,
\end{aligned}
\end{equation}
where 
\begin{equation} \rho_{\operatorname{JT}}^{g=0}(E) = \frac{\sinh{\left( 2\pi \sqrt{E}\right)}}{4\pi^2}
\end{equation}
is just the usual JT gravity density of states for the disk. After a change of variables 
\begin{equation}
\mathcal{E}=\frac{1}{4\lambda}\left(1\pm\sqrt{1-8\lambda E}\right)
\end{equation}
for the two terms in \eqref{eq:JTdisk} respectively, they combine into a single integral (see, for example, \cite{Griguolo:2021wgy})
\begin{equation}
\begin{aligned}
\langle Z(\beta) \rangle^{g=0}_{\operatorname{JT},\lambda}=\int_0^{\frac{1}{2\lambda}} d\mathcal{E} \, \rho_{\operatorname{JT},\lambda}^{g=0}(\mathcal{E}) e^{-\beta \mathcal{E}}\,,
\end{aligned}
\end{equation}
where
\begin{equation}
    \begin{aligned}
\rho_{\operatorname{JT},\lambda}^{g=0}(E) &=  \frac{d\mathcal{E}^{-1}}{dE}\, \rho_{\operatorname{JT},0}^{g=0} \nonumber\left(\mathcal{E}^{-1}\right) \\&=  (1-4\lambda E)\, \rho_{\operatorname{JT}}^{g=0}(E-2\lambda E^2)\,.
\end{aligned}
\end{equation}
Motivated by this, it is natural to expect that for the Airy model deformed by $\lambda > 0$ we should have the one-point function upon an identical change of variables as before 
\begin{equation}
    \begin{aligned}
    \label{eq:Airyguess}
\langle Z(\beta) \rangle_{\text{Airy},\lambda} &= \int_{-\infty}^{\frac{1}{8\lambda}} dE \, \rho_{\text{Airy}}(E) \, \bigg(e^{-\beta f^-_\lambda (E)} - e^{-\beta f^+_\lambda (E)}\bigg) \\
&= \int_{-\infty}^{\infty} d\mathcal{E} \, \rho_{\text{Airy},\lambda}(\mathcal{E}) e^{-\beta \mathcal{E}}\,,
    \end{aligned}
\end{equation}
where
\begin{equation} \rho_{\text{Airy},\lambda}(E) = (1-4\lambda E) \rho_{\text{Airy}, 0}(E-2\lambda E^2)\,,
\end{equation}
and the second term in the first line of \eqref{eq:Airyguess} signals non-perturbative effects. Note that here, instead of using the integration transformation for $\lambda < 0$, we must resort to our working definition introduced at the beginning of this section as the new prescription for the $T\overline{T}$ deformation, which agrees with \eqref{eq:nonpert}. 

\begin{figure}[h]
    \centering
    \includegraphics[scale = 0.7]{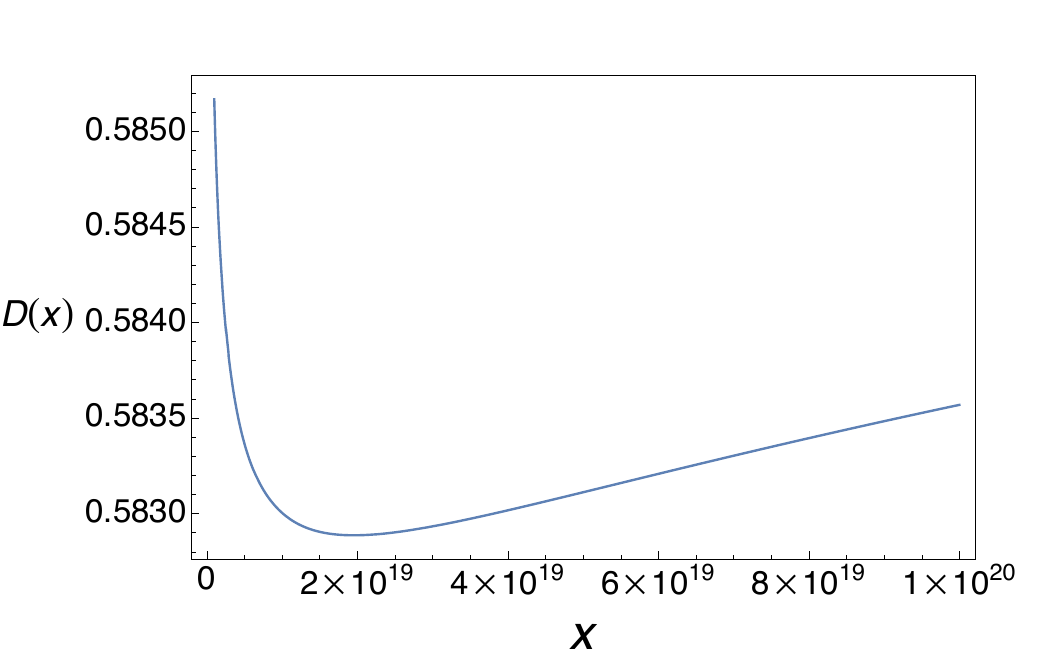}
    \caption{We plot $D(x)$ at $\lambda  = \frac{1}{15}$ and $T = \frac{1}{12}$ without the non-perturbative branch. We see the turning point is at $x\sim 2\times 10^{19}$.}
    \label{fig:diverge1}
\end{figure}

One can use this result with the approximation $\langle Z(\beta)^n \rangle \simeq \langle Z(n\beta) \rangle$ to compute the quenched free energy. However, in this case, we find that the quenched free energy diverges. To see this, we look at the difference between the quenched free energy $F_q(\beta)$ and the annealed free energy $F_a(\beta)$
\begin{equation} F_{q,\lambda} - F_{a,\lambda} = \int_0^\infty \frac{dx}{x} \left(e^{-\mathcal{Z}_\lambda(x)} -e^{-x \langle Z \rangle_\lambda}\right)\,,
\end{equation}
where 
\begin{equation}
\label{eq:zcalbad}
\begin{aligned}
    \mathcal{Z}_\lambda(x) &=-\sum_{n=1}^{\infty}\frac{(-x)^n}{n!}\langle Z(n\beta)\rangle_{\text{Airy},\lambda} \\ 
    &=-\sum_{n=1}^{\infty}\frac{(-x)^n}{n!}\int_{-\infty}^{\frac{1}{8\lambda}}dE\, \rho_{\text{Airy}}(E)\left(e^{-n\beta f^-_\lambda (E)}  - e^{-n\beta f^+_\lambda (E)}\right) \\
    &= \int_{-\infty}^{\frac{1}{8\lambda}} dE \, \rho_{\text{Airy}}(E)\left[e^{-x e^{-\beta f^+_\lambda (E)}} - e^{-x e^{-\beta f^-_\lambda (E)}}\right]\,.
\end{aligned}
\end{equation}
In order for the $x$-integral to converge, the difference $D(x) \equiv e^{-\mathcal{Z}_\lambda(x)}-e^{-x \langle Z \rangle_\lambda}$ must at least go to zero as $x\rightarrow \infty$ but at least faster than $\frac{1}{\ln x}$. We numerically plot $D(x)$ for $\lambda = \frac{1}{15}$ and $T = \frac{1}{12}$ in figure \ref{fig:diverge1}. As one can see, the function $D(x)$ is monotonically decreasing with $x$ at the beginning, but turns around and starts monotonically increasing at a very large value $x\sim 2\times 10^{19}$.

One might wonder if the non-perturbative branch with the energy $f^+_\lambda (E)$ causes the integral to diverge. We can certainly only include the perturbative branch when computing the quenched free energy. One can numerically show this divergence from $D(x)$ asymptotic to some finite number in the $x\rightarrow \infty$ limit.

\begin{figure}[h]
    \centering
    \includegraphics[scale = 0.7]{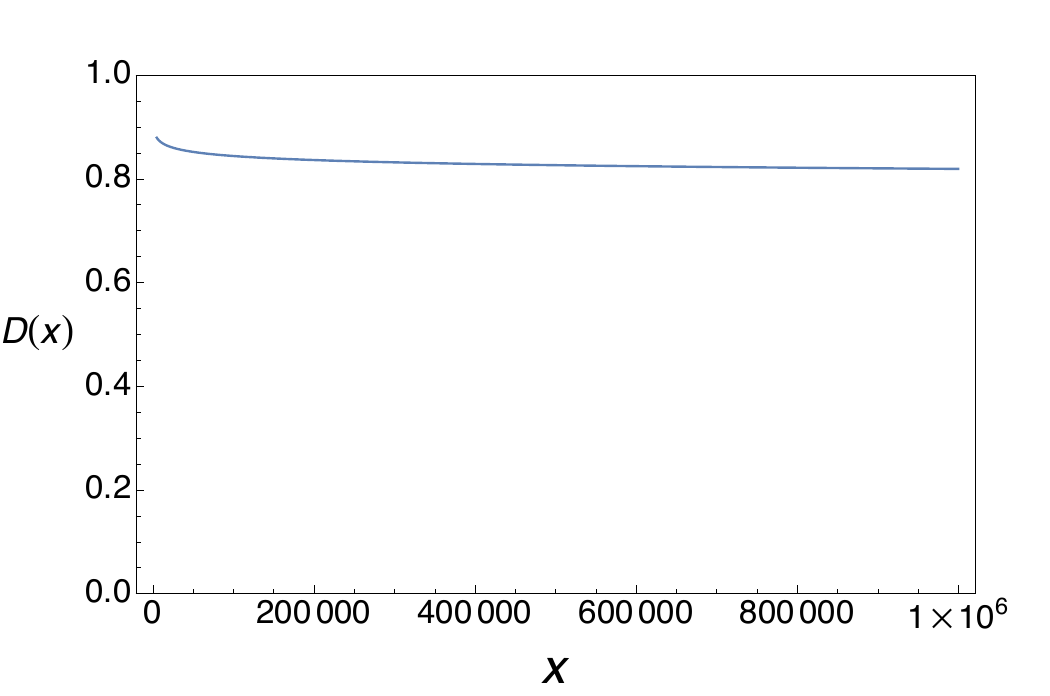}
    \caption{We plot $D(x)$ at $\lambda = \frac{1}{10}$ and $T = \frac{1}{10}$ without the non-perturbative branch. Though not as dramatic as figure \ref{fig:diverge1}, as one can see $D(x)$ still does not go to zero as $x\rightarrow \infty$.}
    \label{fig:diverge2}
\end{figure}

It is unclear to us if such divergence is intrinsically physical or due to any of our approximations. There are several possibilities. For instance, Okuyama's formula \eqref{OkuyamaF} may fail for the deformed theory in general. The derivation of \eqref{OkuyamaF} in \cite{Okuyama:2021pkf} requires one to exchange the integral with an infinite sum which is not convergent. This may lead to the failure of \eqref{OkuyamaF} in the deformed theory. Another possibility could be that the divergence is due to the non-perturbative instability of the Airy model. A reliable way to rule out some of these possibilities is to extend our work to JT gravity with the proper improvement of its non-perturbative behavior.

\subsection{Quenched free energy for $\lambda < 0$ }\label{goodsign}
In this subsection, we compute the quenched free energy for $\lambda < 0$. Since non-perturbatively, the density of states of the Airy model extends to $E = -\infty$, the deformed theory will have unitarity issues caused by complex-valued energy, which seems not to be discussed before in the previous literature of deformed JT gravity. All the densities of states in the literature have lower bounds (i.e. $\rho(E  \leq E_0 ) = 0$). Thus, by considering $\tilde{\rho}(E) = \rho(E-E_0)$, we can also have a well-defined spectrum for all $\lambda < 0$. The non-perturbative effect makes $\rho(E) \neq 0$ for all $E\in \mathbb{R}$. Therefore, the deformed energy spectrum will be complex-valued for $E < \frac{1}{8\lambda}$. 

One such treatment is by imposing a cutoff in the deformed energy spectrum up to when it becomes complex-valued. This cutoff resolves the unitarity issues, however, but this leads to a violation of the flow equation \eqref{flow} of $\langle Z(\beta) \rangle_\lambda$ as a boundary term arises at the cutoff $E = \frac{1}{8\lambda}$. Alternatively, we may include these states with complex-valued energy, but we must then include their corresponding states from the non-perturbative sectors to ensure that the partition function is real-valued. We will refer to this part as non-perturbative. 

By carefully choosing coefficients, the contributions in the boundary piece of this term cancel the boundary term \eqref{eq:bdry} in the first option. Thus, the flow equation of the one-point function $\langle Z(\beta) \rangle_\lambda$ will be satisfied in this case. We study the quenched energy with and without the non-perturbative contribution. Excluding the non-perturbative contribution, we numerically confirm the quenched free energy is monotonically decreasing with temperature $T$ at a given $\lambda < 0$. We also find the quenched free energy monotonically decreases as we increase the absolute value of $\lambda$. Including the non-perturbative branch, unfortunately, we find that the quenched free energy computed from Okuyama's formula \eqref{OkuyamaF} diverges in general, and we illustrate this subtlety numerically in this subsection. 

We start with how to treat the complex-valued energy states in the deformed spectrum. One may expect that the correct answer is given by the exact recipe for the $\lambda > 0$. We cut off the spectrum below $E<\frac{1}{8\lambda}$ where the deformed energy becomes complex-valued and includes the other branch for the remaining spectrum. However, there are two objections. The first objection is that if we consider the genus expansion, the spectrum at genus-zero does not extend to $E<0$. As a result, the deformed spectrum remains unchanged, and $\lambda < 0$ is well-defined making no additional branch required. If we included the other branch through the genus expansion, the genus-zero partition function will receive corrections from the other branch as well, which leads to inconsistencies. The second objection is that the Boltzmann weight $e^{-\beta f^+_\lambda (E)}$ diverges as $e^{\beta \sqrt{ \frac{E}{2 |\lambda |}} }$ when $E\rightarrow \infty$. Hence, the contribution from the other branch is divergent. 

These two reasons suggest we should not include the contribution from the other branch for the real-valued energy region. Thus, one might conclude the deformed partition function for $\lambda < 0$ is simply given by truncating the spectrum with complex-valued energy:
\begin{equation} \langle Z(\beta) \rangle_{\lambda,\text{guess}} = \int_{\frac{1}{8\lambda}}^\infty dE \, \rho_{\text{Airy}}(E) e^{-\beta f_\lambda^-(E)} = \int_{\frac{1}{4\lambda}}^\infty dE\, \rho_{\text{Airy},\lambda}(E) e^{-\beta E}\,,
\end{equation}
where 
\begin{equation} \rho_{\text{Airy},\lambda}(E) = (1-4\lambda E)\rho_{\text{Airy}}(E(1-2\lambda E))\,.
\end{equation}
There is a caveat: the deformed partition function should satisfy the differential equation \eqref{flow} derived in \cite{Iliesiu:2020zld}: 
\begin{equation} 
\label{eq:flow}
\left[4\lambda \partial_\lambda \partial_\beta + 2\beta \partial_\beta^2 - \bigg(\frac{4\lambda}{\beta} - 1\bigg)\partial_\lambda\right] \langle Z(\beta) \rangle_\lambda = 0\,.
\end{equation}
For convenience, we introduce the differential operator
\begin{equation}
    \mathcal{F} \equiv 4\lambda \partial_\lambda \partial_\beta + 2\beta \partial_\beta^2 - \left(\frac{4\lambda}{\beta} - 1\right)\partial_\lambda\,.
\end{equation}
Next, consider a change of variables $E = \tilde{E} + \frac{1}{8\lambda}$ so that the bound of the integral does not depend on $\lambda$:
\begin{equation} \langle Z(\beta) \rangle_{\lambda, \text{guess}} = \int_0^\infty d\tilde{E} \, \rho_{\text{Airy}}\left(\tilde{E} + \frac{1}{8\lambda}\right) e^{-\beta f_\lambda^-\left(\tilde{E} + \frac{1}{8\lambda} \right)}\,.
\end{equation}
As one can show, $\mathcal{F}$ acting on the integrand leads to a total derivative
\begin{equation} 
\begin{aligned}
&\mathcal{F}\left[\rho_{\text{Airy}}\left(\tilde{E} + \frac{1}{8\lambda}\right) e^{-\beta f_\lambda^-\left(\tilde{E} + \frac{1}{8\lambda}\right)}\right] \\
&= \frac{d}{d\tilde{E}}\left[e^{-\beta f_\lambda^- \left(\tilde{E}+\frac{1}{8\lambda}\right)} \rho_{\text{Airy}}\left(\tilde{E}+\frac{1}{8\lambda}\right) \frac{4\lambda - \beta\sqrt{-8\lambda \tilde{E}}}{8\beta\lambda^2}\right]\,.
\end{aligned}
\end{equation}
Therefore
\begin{equation}
    \begin{aligned}
    \label{eq:bdry}
\mathcal{F}\left[\langle Z(\beta) \rangle_{\lambda,\text{guess}}\right] &= \left[e^{-\beta f_\lambda^-(\tilde{E} + \frac{1}{8\lambda})} \rho_{\text{Airy}} \left(\tilde{E}+\frac{1}{8\lambda}\right)\frac{4\lambda - \beta\sqrt{-8\lambda \tilde{E}}}{8\beta\lambda^2}\right]\Bigg|_{\tilde{E}=0}^{\tilde{E}=\infty} \\
&= -\frac{e^{-\beta f_\lambda^-(\frac{1}{8\lambda})}}{2\beta\lambda}\rho_{\text{Airy}}\left(\frac{1}{8\lambda}\right)\,.
    \end{aligned}
\end{equation}
%

Now we see the problem: our guess $\langle Z(\beta) \rangle_{\lambda,\text{guess}}$ violates the flow equation \eqref{eq:flow} due to the appearance of the boundary term \eqref{eq:bdry}. This is just another manifestation of the non-perturbative effect of the Airy model. If $\rho_{\text{Airy}}(E)$ had been supported on $[E_0,\infty)$, we can shift the ground state energy such that $\tilde{\rho}_{\text{Airy}}(E) = \rho_{\text{Airy}}(E - E_0)$ to remove the complex-valued energy and make $\tilde{\rho}_{\text{Airy}}(\frac{1}{8\lambda}) = 0$ such that the flow equation is satisfied. However, this is not possible due to the non-perturbative effects as $\rho_{\text{Airy}}(E)$ has support on the entire real axis.

Thus, to make sure that $\langle Z(\beta) \rangle_{\lambda, \text{guess}}$ satisfies the flow equation \eqref{eq:flow} while keeping it finite, we must include the complex-valued energy region where $E \in (-\infty, \frac{1}{8\lambda})$ to cancel the unwanted boundary term. To make sure the deformed partition function is real, we must also add its complex conjugate, i.e. the contribution from the other branch. Therefore, the partition function $\langle Z(\beta) \rangle_\lambda$ for $\lambda < 0$ should be given by
\begin{equation} 
\label{eq:good1pt}
\langle Z(\beta) \rangle_{\text{Airy},\lambda} = \int_{\frac{1}{8\lambda}}^\infty dE\, \rho_{\text{Airy}}(E) e^{-\beta f_\lambda^-(E)} + \int_{-\infty}^{\frac{1}{8\lambda}} dE\, \rho_{\text{Airy}}(E) \frac{e^{-\beta f_\lambda^-(E)} + e^{-\beta f_\lambda^+(E)}}{2}\,,
\end{equation}
where the sum of exponentials in the second integrand can be rewritten in terms of cosine as
\begin{equation}
\label{eq:cosine1}
e^{-\frac{1}{4\lambda T}}\int_{-\infty}^{\frac{1}{8\lambda}} dE\, \rho_{\text{Airy}}(E)\cos \left( \frac{\sqrt{8\lambda E-1}}{4\lambda T}\right)
\end{equation}
and is a highly oscillatory integral when $\lambda$ or $T$ is small, but can still fairly easily be numerically evaluated to be finite. As one can check, the boundary terms cancel with each other, and the flow equation \eqref{eq:flow} is satisfied.

One can then use Okuyama's formula \eqref{OkuyamaF} to numerically compute the quenched free energy. In this case, we can express $\mathcal{Z}_\lambda(x)$ using \eqref{eq:good1pt} as the following:
\begin{equation}
    \begin{aligned}
    \label{eq:calZ}
\mathcal{Z}_\lambda(x) &\simeq  - \sum_{n=1}^\infty \frac{(-x)^n}{n!} \langle Z(n\beta)\rangle_{\text{Airy},\lambda} \\
&= -\int_{\frac{1}{8 \lambda}}^\infty dE\, \rho_{\text{Airy}}(E) \sum_{n=1}^\infty \frac{(-x)^n}{n!}e^{-n\beta f_\lambda^-(E)} \\
&- \frac{1}{2}\int_{-\infty}^{ \frac{1}{8 \lambda} } dE\, \rho_{\text{Airy}}(E) \sum_{n=1}^\infty\frac{(-x)^n}{n!} \left(e^{-n\beta f_\lambda^-(E)} + e^{-n\beta f_\lambda^+(E)}\right) \\
&= \int_{\frac{1}{8 \lambda}}^\infty dE\, \rho_{\text{Airy}}(E) \left(1 - e^{-x e^{-\beta f_\lambda^-(E)}}\right)\\
&+ \frac{1}{2}\int_{-\infty}^{\frac{1}{8 \lambda}} dE\, \rho_{\text{Airy}}(E)\left[2 - e^{-x e^{-\beta f_\lambda^+(E)}} - e^{-x e^{-\beta f_\lambda^-(E)}}\right]\,.
    \end{aligned}
\end{equation}
Since the second integral of \eqref{eq:calZ} can be rewritten as
\begin{equation}
\label{eq:cosine2}
    \int_{-\infty}^{\frac{1}{8\lambda}} dE\, \rho_{\text{Airy}}(E)\Bigg[1-e^{-xe^{-\frac{1}{4 \lambda T}} \cos\frac{\sqrt{8\lambda E-1}}{4 \lambda T}} \cos \left(xe^{-\frac{1}{4\lambda T}}\sin\frac{\sqrt{8 \lambda E -1}}{4 \lambda T}\right)\Bigg]\,.
\end{equation}

We can turn off the non-perturbative effect by including only the first term in \eqref{eq:calZ}. In the next subsection, we study the quenched free energy first with and then without the non-perturbative contribution.

\subsubsection{Without the non-perturbative contribution}
We first study the quenched free energy without the contribution from the non-perturbative part. In this case, the numerical calculation is straightforward without subtleties. We numerically confirm that the deformed quenched free energy $F_{q,\lambda}(T)$ monotonically decreases as $T$ increases. Furthermore, we find the quenched free energy $F_{q,\lambda}(T)$ monotonically decreases as the absolute value of $\lambda$ increases and present our numerical results below in figures \ref{fig:ncl} and \ref{fig:ncT}. 

\begin{figure}[h!]
     \centering
     \begin{subfigure}[b]{0.4\textwidth}
         \centering
         \includegraphics[width=\textwidth]{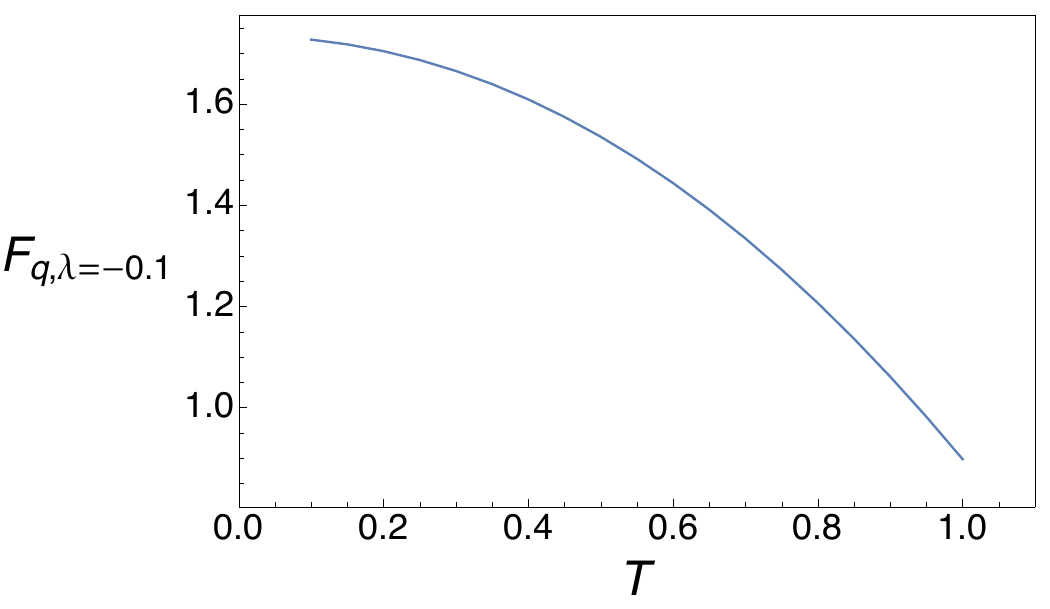}
         \caption{$\lambda = - 0.1$}
         \label{fig:ncl1}
     \end{subfigure}
     \hfill
     \begin{subfigure}[b]{0.4\textwidth}
         \centering
         \includegraphics[width=\textwidth]{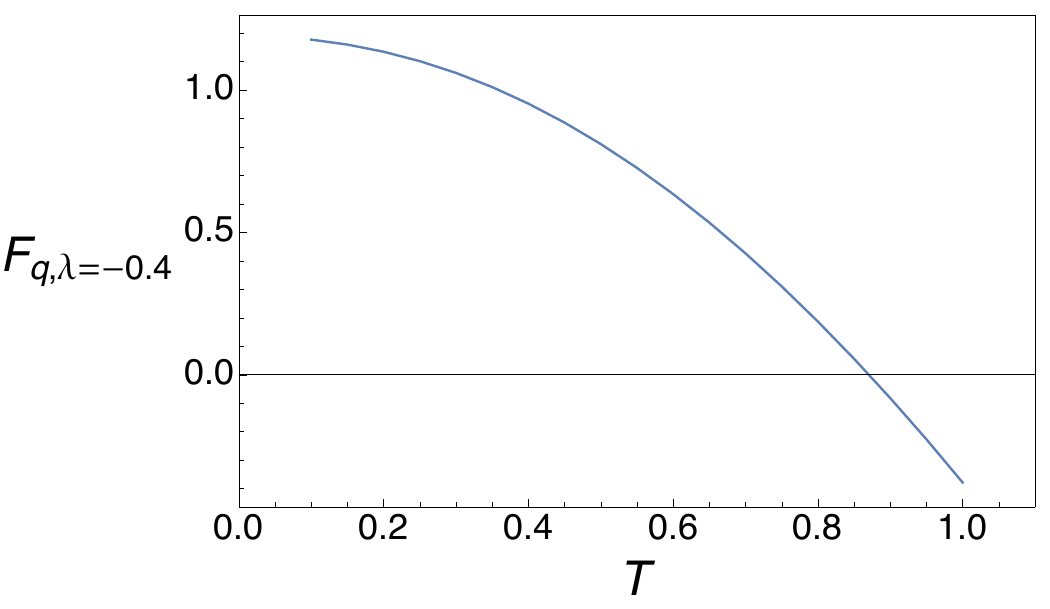}
         \caption{$\lambda = - 0.4$}
         \label{fig:ncl2}
     \end{subfigure}
     \hfill
     \begin{subfigure}[b]{0.4\textwidth}
         \centering
         \includegraphics[width=\textwidth]{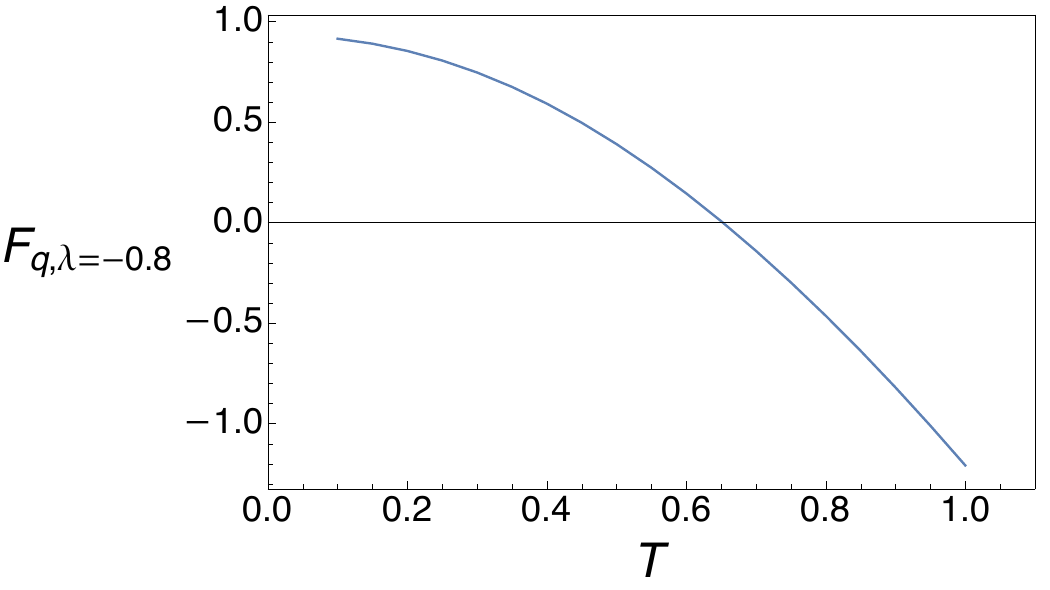}
         \caption{$\lambda = - 0.8$}
         \label{fig:ncl8}
     \end{subfigure}
     \hfill
     \begin{subfigure}[b]{0.4\textwidth}
         \centering
         \includegraphics[width=\textwidth]{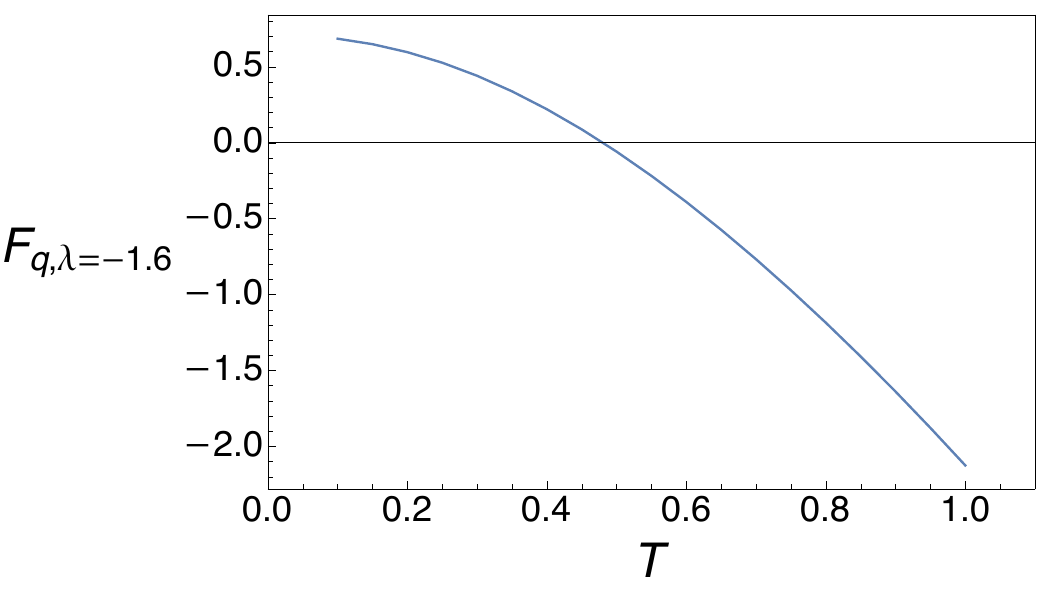}
         \caption{$\lambda = - 1.6$}
         \label{fig:ncl16}
     \end{subfigure}
     \hfill
     \begin{subfigure}[b]{0.4\textwidth}
         \centering
         \includegraphics[width=\textwidth]{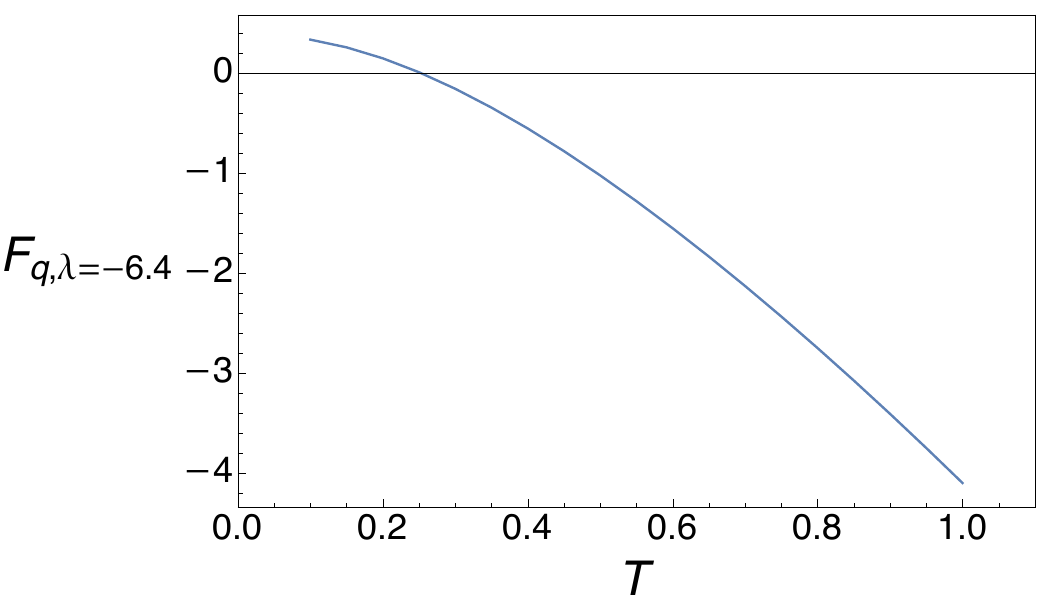}
         \caption{$\lambda = - 6.4$}
         \label{fig:ncl64}
     \end{subfigure}
     \hfill
     \begin{subfigure}[b]{0.4\textwidth}
         \centering
         \includegraphics[width=\textwidth]{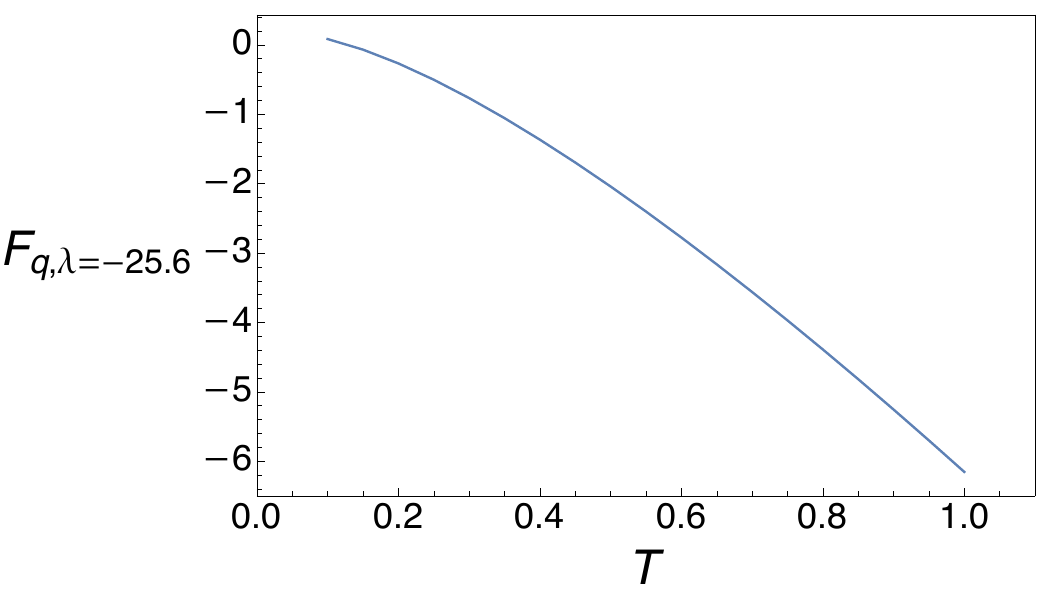}
         \caption{$\lambda = - 25.6$}
         \label{fig:ncl256}
     \end{subfigure}
     \caption{We plot the quenched free energy $F_{q,\lambda}(T)$ without the contribution from the non-perturbative branch as a function of $T$ for $\lambda = -0.1,-0.4,-0.8,-1.6,-6.4,-25.6$. As shown, $F_{q,\lambda}(T)$ is a monotonically decreasing function of $T$ in the deformed theory.} 
     \label{fig:ncl}
\end{figure}

\begin{figure}[h]
     \centering
     \begin{subfigure}[b]{0.4\textwidth}
         \centering
         \includegraphics[width=\textwidth]{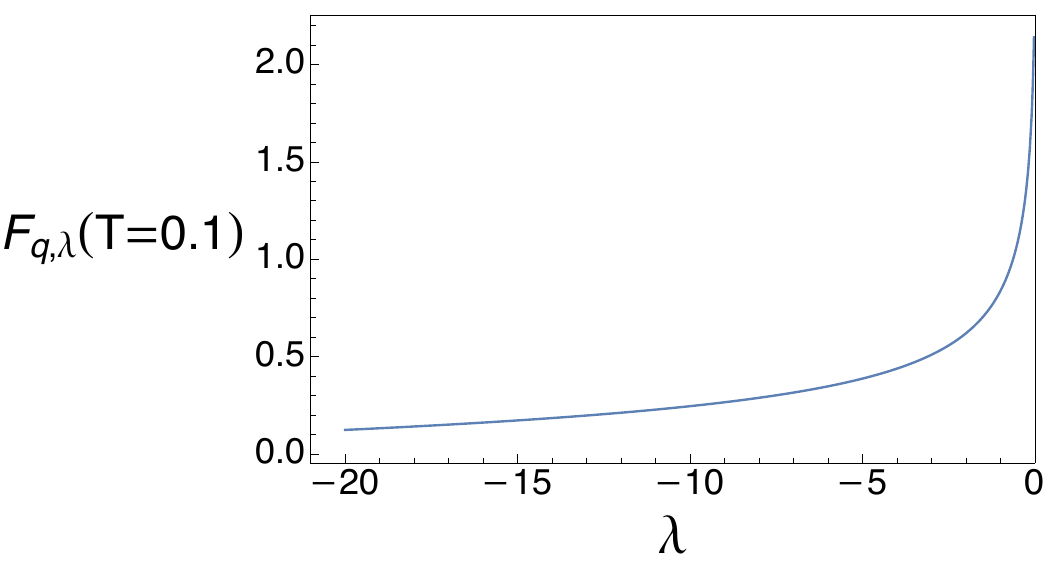}
         \caption{$T = 0.1$}
         \label{fig:ncT1}
     \end{subfigure}
     \hfill
     \begin{subfigure}[b]{0.4\textwidth}
         \centering
         \includegraphics[width=\textwidth]{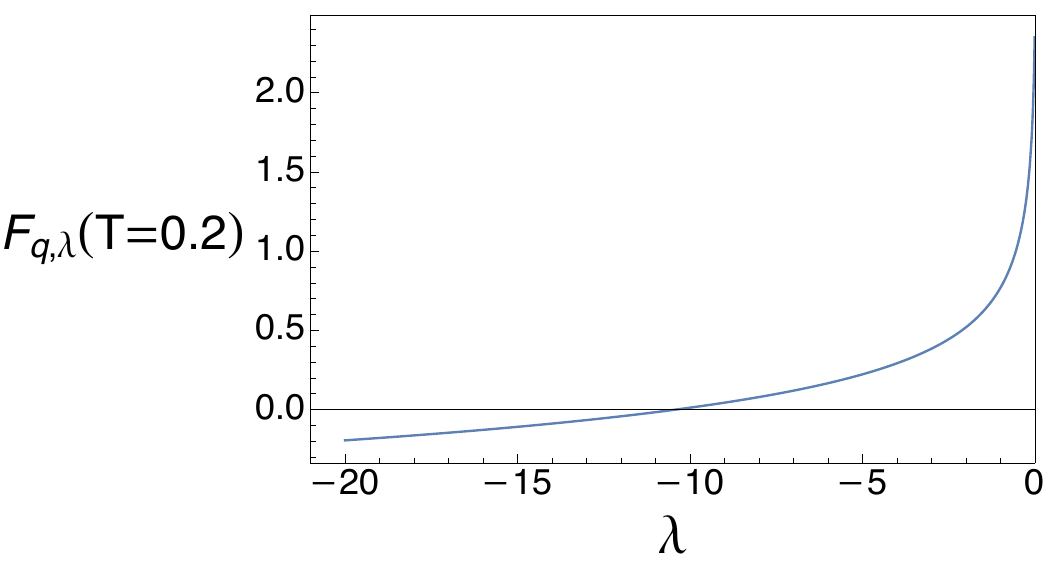}
         \caption{$T = 0.2$}
         \label{fig:ncT2}
     \end{subfigure}
     \hfill
     \begin{subfigure}[b]{0.4\textwidth}
         \centering
         \includegraphics[width=\textwidth]{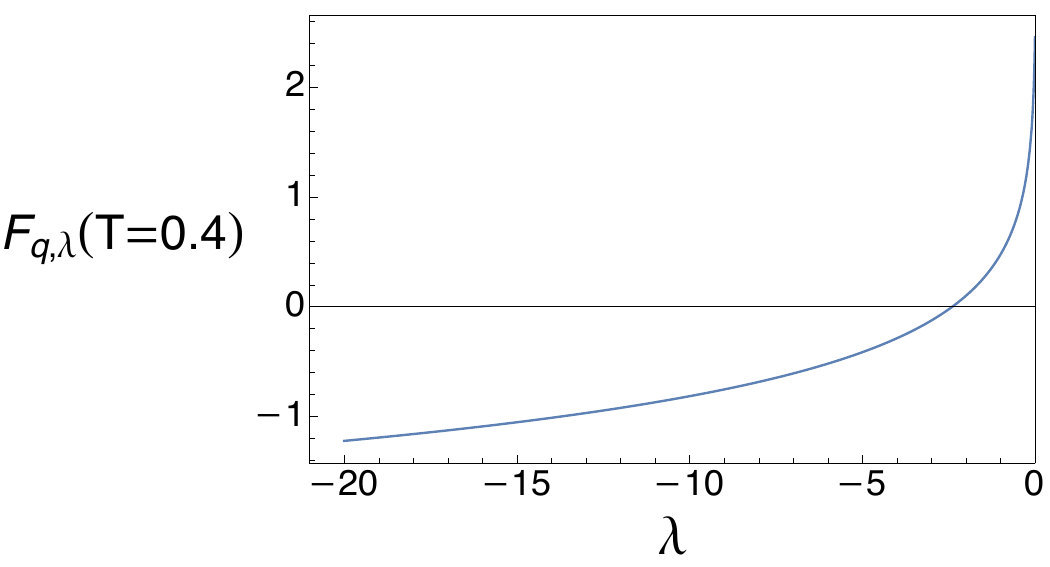}
         \caption{$T = 0.4$}
         \label{fig:ncT4}
     \end{subfigure}
     \hfill
     \begin{subfigure}[b]{0.4\textwidth}
         \centering
         \includegraphics[width=\textwidth]{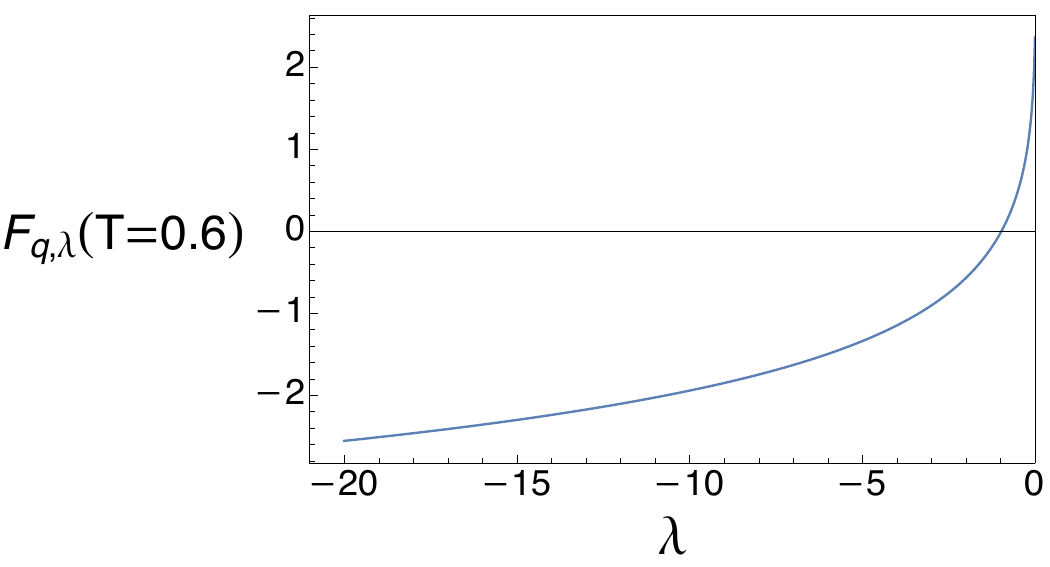}
         \caption{$T = 0.6$}
         \label{fig:ncT6}
     \end{subfigure}
     \caption{We plot the quenched free energy $F_{q,\lambda}(T)$ without the contribution from the non-perturbative branch as a function $\lambda$ for fixed $T = 0.1,0.2,0.4,0.6$. As shown, $F_{q,\lambda}(T)$ is monotonically decreasing as $|\lambda|$ increases.}
     \label{fig:ncT}
\end{figure}

\newpage
\subsubsection{With the non-perturbative contribution}
\begin{figure}[H]
    \centering
    \includegraphics[scale = 0.7]{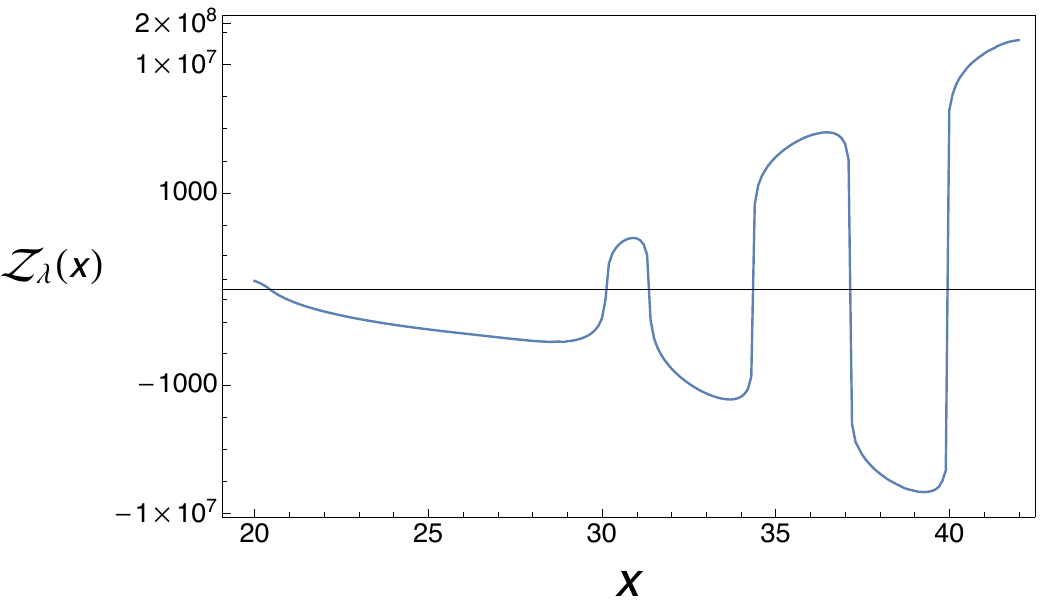}
    \caption{We plot $\mathcal{Z}_\lambda(x)$ against $x$ at $T = 0.3$ and $\lambda = -2$. One can see that the valleys' depths deepening towards the right makes $e^{-\mathcal{Z}_\lambda(x)}$ unbounded in the large $x$ limit.}
    \label{fig:cosine0}
\end{figure}

Similarly, in section \ref{sec:bad} (e.g., figures \ref{fig:diverge1} and \ref{fig:diverge2}), we numerically demonstrate the integral diverges by showing the difference $D(x) \equiv e^{-\mathcal{Z}_\lambda(x)}-e^{-x \langle Z \rangle_\lambda}$ does not vanish in the large $x$ limit. Notice that $e^{-x \langle Z\rangle_\lambda}\rightarrow 0$ in the large $x$ limit, and we can show that $D(x)$ diverges as $x\rightarrow\infty$ by proving $\mathcal{Z}_\lambda(x)$ is oscillating with its amplitude rapidly increasing with $x$. Instead, we numerically plot figure \ref{fig:cosine0} to illustrate this fact by showing the depth of the three valleys deepening as $x$ increases, which shows the divergence of $D(x)$. 

\section{Conclusion}
\label{Conclusion}
In this chapter, we studied the $T\overline{T}$-deformed correlators for JT gravity and its matrix model description. Additionally, we computed the quenched free energy of the Airy model under the $T\overline{T}$ deformation for both signs of $\lambda$ and their non-perturbative features. We briefly summarize our numerical results. At genus-zero and the leading order of perturbation theory in $\lambda$, we confirmed the quenched free energy $F_{q,\lambda}(T)$ is a monotonic function in $T$ for a given $\lambda$ within the validity domain of the leading order approximation. We also find $\lambda<0$ decreases $F_{q,\lambda}(T)$ while $\lambda > 0$ increases $F_{q,\lambda}(T)$. For all genera and in the low-temperature approximation, we computed the quenched free energy $F_{q,\lambda}(T)$ using \eqref{OkuyamaF}, which diverges regardless of whether we include the non-perturbative contribution of the $T\overline{T}$ deformation or not when $\lambda > 0$. For $\lambda < 0$, we can numerically compute $F_{q,\lambda}(T)$ without including the potential non-perturbative contributions and confirm the monotonicity of $F_{q,\lambda}(T)$ at low temperature. Additionally, we find $F_{q,\lambda}(T)$ decreases by the deformation and matches the result in perturbation theory at genus zero. When including the possible contributions from the non-perturbative branch, we find $F_{q,\lambda}(T)$ computed from \eqref{OkuyamaF} diverges again.

\chapter{Root-$T\overline{T}$ Deformed Boundary Conditions in Holography}
\label{ch:root-TT}
In this chapter, we develop the holographic dictionary for pure $\mathrm{AdS}_3$ gravity where the Lagrangian of the dual $2d$ conformal field theory has been deformed by an arbitrary function of the energy-momentum tensor. In addition to the $T\overline{T}$ deformation, examples of such functions include a class of marginal stress tensor deformations, which are special because they leave the generating functional of connected correlators unchanged up to a redefinition of the source and expectation value. Within this marginal class, we identify the unique deformation that commutes with the $T\overline{T}$ flow, which is the root-$T\overline{T}$ operator, and write down the modified boundary conditions corresponding to this root-$T\overline{T}$ deformation. We also identify the unique marginal stress tensor flow for the cylinder spectrum of the dual CFT, which commutes with the inviscid Burgers' flow driven by $T\overline{T}$, and we propose this unique flow as a candidate root-$T\overline{T}$ deformation of the energy levels. We study BTZ black holes in $\mathrm{AdS}_3$ subject to root-$T\overline{T}$ deformed boundary conditions and find that their masses flow in a way that is identical to that of our candidate root-$T\overline{T}$ energy flow equation, which offers evidence that this flow is the correct one. Finally, we obtain the root-$T\overline{T}$ deformed boundary conditions for the gauge field in the Chern-Simons formulation of $\mathrm{AdS}_3$ gravity.
\section{Introduction} \label{intro}

A promising strategy for learning more about holography is to begin with a relatively well-understood holographic correspondence and then deform it in some controlled way. We will focus on the case of an asymptotically $\mathrm{AdS}_3$ bulk, which is dual to a two-dimensional conformal field theory. Given such a holographic boundary theory, we can view the $\mathrm{CFT}_2$ as essentially defining the $3d$ gravitational theory. More precisely, the $\mathrm{CFT}_2$ defines the boundary conditions that the fields of the bulk gravity theory should obey at infinity.

To consider a concrete example, we recall that every translation-invariant quantum field theory admits a conserved stress tensor operator $T_{\alpha \beta}$. In the holographic dictionary, this boundary stress tensor operator is dual to the asymptotic bulk metric. One way to see this is to vary the action $S$ of the $3d$ gravitational theory, including both the Einstein-Hilbert term and appropriate boundary terms and put this varied quantity on-shell using the bulk equations of motion. The resulting expression can be written as a boundary integral
\begin{align}\label{3d_bulk_variation}
    \delta S \Big\vert_{\text{on-shell}} = \frac{1}{2} \int_{\partial M_3} d^2 x \, \sqrt{\gamma} \, T_{\alpha \beta} \, \delta \gamma^{\alpha \beta} \,,
\end{align}

For the on-shell variation of the action to vanish, we require $\delta \gamma^{\alpha \beta} = 0$ on $\partial M_3$, which means that we impose Dirichlet boundary conditions on the metric near infinity. The quantity $T_{\alpha \beta}$ which appears in (\ref{3d_bulk_variation}) is then identified with the expectation value of the stress tensor operator of the boundary theory; the procedure described above furnishes an explicit expression for $T_{\alpha \beta}$ in terms of functions appearing in the Fefferman-Graham expansion of the metric near infinity. We interpret this by saying that the asymptotic metric $\gamma^{\alpha \beta}$ is a source for the stress tensor operator of the dual $\mathrm{CFT}_2$.

Now consider a deformation of the boundary conformal field theory. One familiar way to perform such a deformation is to add an integrated local operator to the action defining the $2d$ theory, so that
\begin{align}\label{general_O_deformation}
    S_0 \longrightarrow S_0 + \delta S = S_0 + \mu \int d^2 x \, \sqrt{\gamma} \, \mathcal{O} ( x ) \,, 
\end{align}
where $\mathcal{O} ( x )$ is a local operator and $\mu$ is a real parameter. Because the $\mathrm{CFT}_2$ defines the boundary conditions that the bulk fields obey at infinity, it is natural to expect that such a deformation would change these boundary conditions. This is the case for many such multi-trace deformations \cite{Witten:2001ua}, at least subject to the usual caveats that one should restrict attention to the effects on light single-trace operators at large-$N$.

For instance, one much-studied example is a double-trace deformation, where the object $\mathcal{O} ( x ) $ appearing in (\ref{general_O_deformation}) is the square of an operator which is dual to a fundamental field in the gravity theory. In this work, we will focus on deformations constructed from the stress-energy tensor $T_{\alpha \beta}$; because this operator is present in \emph{any} translation-invariant quantum field theory, such deformations are in a sense universal. An operator $\mathcal{O} ( x )$ which is constructed from products of components $T_{\alpha \beta}$ is a double-trace operator, by the definition given above, because the stress-energy tensor is dual to the bulk metric, which is a fundamental field of the gravity theory. One particularly nice Lorentz-invariant double-trace combination of components $T_{\alpha \beta}$ is
\begin{align}\label{OTT_def}
    \mathcal{O}_{T\overline{T}} = T^{\alpha \beta} T_{\alpha \beta} - \left( T^\alpha{}_\alpha \right)^2 \,.
\end{align}

This combination defines the so-called $T\overline{T}$ operator, which has generated considerable research interest in recent years. For the moment, let us focus on the properties of this operator purely as an object in the $2d$ boundary theory (and postponing its bulk interpretation). By infinitesimally adding this operator $\mathcal{O}_{T\overline{T}}$ at each step along a flow, one can define a one-parameter family of theories that obeys the differential equation
\begin{align}\label{TT_flow}
    \frac{\partial S^{(\lambda)}}{\partial \lambda} = - \frac{1}{2} \int d^2 x \, \sqrt{\gamma} \, \mathcal{O}^{(\lambda)}_{T\overline{T}} ( x ) \,,
\end{align}
where the superscript $\lambda$ is meant to emphasize that we must re-compute the operator $\mathcal{O}_{T\overline{T}}^{(\lambda)}$ using the deformed stress tensor $T_{\alpha \beta}^{(\lambda)}$ at each point along the flow.\footnote{\label{notation_footnote}Throughout this chapter, we always use the symbol $\lambda$ to denote the parameter of a $T\overline{T}$ flow, while we use the symbol $\mu$ either for the parameter of a generic deformation of a boundary field theory or for the parameter of the root-$T\overline{T}$ flow, which we introduce shortly. Note that $\mu$ is never a spacetime index in this chapter and appendix \ref{sec:AdS3TTApp}.}

We make three sets of observations.
\begin{enumerate}
    \item\label{TT_one} First note that $\mathcal{O}_{T\overline{T}}$ is a dimension-four operator, which means that it is irrelevant in the Wilsonian sense. As a consequence, the flow equation (\ref{TT_flow}) is quite unusual from the perspective of the renormalization group. Ordinarily, one imagines beginning with a conformal field theory and then adding an integrated \emph{relevant} operator in the spectrum of the theory, which triggers a flow to the infrared. In a loose sense, the $T\overline{T}$ flow is the inverse of this familiar paradigm, as we add an integrated \emph{irrelevant} operator, which modifies the definition of the theory in the ultraviolet.
    
    \item\label{TT_two} The quantity $\mathcal{O}_{T\overline{T}}$ defined in (\ref{OTT_def}) involves products of stress tensor operators. As products of coincident local operators are generally divergent in quantum field theory, it is far from obvious that the combination $\mathcal{O}_{T\overline{T}}$ defines a local operator at all. However, it has been shown that one can begin with a point-split quantity 
    \begin{align}
        \mathcal{O}_{T\overline{T}} ( x, y ) = T^{\alpha \beta} ( x ) \, T_{\alpha \beta} ( y ) - T^\alpha{}_\alpha ( x ) \, T^\beta{}_\beta ( y ) \,, 
    \end{align}
    and then take a coincident point-limit $\lim\limits_{y \to x} \mathcal{O}_{T\overline{T}} ( x, y )$. Surprisingly, this procedure \emph{does} define a sensible local operator, up to certain total derivative ambiguities which can be ignored \cite{Zamolodchikov:2004ce,Smirnov:2016lqw}.
    
    \item\label{TT_three} This deformation is ``nice'' in the sense that it preserves many desirable properties of the undeformed theory, such as integrability \cite{Smirnov:2016lqw,Cavaglia:2016oda,Conti:2018jho,Chen:2021aid} and supersymmetry \cite{Baggio:2018rpv,Chang:2018dge,Jiang:2019hux,Chang:2019kiu, Ferko:2019oyv,Ferko:2021loo,Ebert:2022xfh}. Relatedly, observables in the deformed theory can often be described with simple closed-form expressions; a few examples include the finite-volume spectrum \cite{Smirnov:2016lqw,Cavaglia:2016oda}, $S$-matrix \cite{Dubovsky:2017cnj}, and torus partition function \cite{Cardy:2018sdv,Datta:2018thy,Aharony:2018bad}.
\end{enumerate}

Because the operator $\mathcal{O}_{T\overline{T}}$ appears to be rather special from the field theory perspective, one might suspect that this deformation corresponds to some fairly natural modification of the asymptotic boundary conditions for the bulk fields in the holographic dual. This turns out to be the case \cite{Guica:2019nzm}. To see this, one first defines the $\lambda$-dependent quantities
\begin{align} \label{TT_deformed_gamma_T}\begin{split}
    \gamma^{(\lambda)}_{\alpha \beta} &= \gamma^{(0)}_{\alpha \beta} - 2 \lambda \widehat{T}_{\alpha \beta}^{(0)} + \lambda^2 \widehat{T}^{(0)}_{\alpha \rho} \,  \widehat{T}^{(0)}_{\sigma \beta} \, \gamma^{(0) \rho \sigma} \,, \\
    \widehat{T}^{(\lambda)}_{\alpha \beta} &= \widehat{T}^{(0)}_{\alpha \beta} - \lambda \widehat{T}^{(0)}_{\alpha \rho} \, \widehat{T}^{(0)}_{\sigma \beta} \gamma^{(0) \rho \sigma} \,,
\end{split}\end{align}
where $\widehat{T}_{\alpha \beta} = T_{\alpha \beta} - \gamma_{\alpha \beta} T^\alpha{}_\alpha$ is the trace-reversed stress tensor. In terms of these quantities, the boundary action that solves the $T\overline{T}$ flow equation (\ref{TT_flow}) has the property that its variation can be written as
\begin{align}\label{deformed_delta_S}
    \delta S = \frac{1}{2} \int d^2 x \, \sqrt{\gamma^{(\lambda)} } \, T_{\alpha \beta}^{(\lambda)} \, \delta \gamma^{(\lambda) \alpha \beta} \,.
\end{align}
This is \emph{exactly} of the same form as the usual on-shell bulk variation, (\ref{3d_bulk_variation}), except written in terms of the $\lambda$-dependent metric and stress tensor. For the variation of the action to vanish, we now require that $\delta \gamma^{(\lambda) \alpha \beta} = 0$, which means that we impose Dirichlet boundary conditions on the deformed metric $\gamma^{(\lambda) \alpha \beta}$ at infinity. In terms of the original variables, this looks like a certain choice of mixed boundary conditions on the metric at infinity since we now hold fixed a combination of the original metric $\gamma^{(0)}_{\alpha \beta}$ and its radial derivative, which is related to $T^{(0)}_{\alpha \beta}$.

One might ask whether there are other universal deformations constructed from stress tensors that admit interpretations as particularly simple modified boundary conditions. Another candidate is the recently-proposed root-$T\overline{T}$ operator \cite{Ferko:2022cix}, which is defined as
\begin{align}\label{root_TT_def}
    \mathcal{R} = \sqrt{ \frac{1}{2} T^{\alpha \beta} T_{\alpha \beta} - \frac{1}{4} \left( T^\alpha{}_\alpha \right)^2 } \,.
\end{align}
By way of comparison, let us revisit the three points \ref{TT_one} - \ref{TT_three} which we made concerning the $T\overline{T}$ operator and consider the analogous statements for root-$T\overline{T}$. 

\begin{enumerate}

    \item Whereas $T\overline{T}$ is an irrelevant operator, the root-$T\overline{T}$ operator is classically \emph{marginal}. For instance, it has been checked in a large class of examples that the stress tensor of a root-$T\overline{T}$ deformed CFT still has a vanishing trace. As a consequence, the coupling constant $\mu$ parameterizing the root-$T\overline{T}$ flow is dimensionless.
    
    \item Although $T\overline{T}$ is quantum-mechanically well-defined, it is not known whether the root-$T\overline{T}$ operator can be defined at the quantum level by point-splitting. Understanding the quantum properties of this operator remains an important open problem.
    
    \item The root-$T\overline{T}$ deformation shares some of the ``niceness'' properties of the ordinary $T\overline{T}$ deformation. For instance, flow equations for the root-$T\overline{T}$-deformed Lagrangian can often be solved in closed form \cite{Ferko:2022cix}, and the root-$T\overline{T}$ deformation preserves classical integrability in many examples \cite{Borsato:2022tmu}. However, formulas for root-$T\overline{T}$ deformed spectra, $S$-matrices, and partition functions have not been obtained.
\end{enumerate}

Although much less is known about the root-$T\overline{T}$ operator, there are many hints that this deformation might lead to an interesting class of models. One is the relation to the ModMax theory \cite{Bandos:2020jsw,Bandos:2020hgy,Bandos:2021rqy,Lechner:2022qhb} in four dimensions. This theory and its Born-Infeld extension obey $4d$ analogs of the root-$T\overline{T}$ and $T\overline{T}$ flow equations, respectively \cite{Babaei-Aghbolagh:2022uij,Conti:2018jho}, and both flows can be supersymmetrized \cite{Ferko:2022iru,Ferko:2023ruw}. The root-$T\overline{T}$ operator also appears in a flow equation that generates the $3d$ Born-Infeld Lagrangian or its supersymmetric extension \cite{Ferko:2023sps}. Further, the dimensional reduction of the ModMax theory is identical to the theory obtained by root-$T\overline{T}$ deforming a collection of $2d$ free scalars \cite{Babaei-Aghbolagh:2022leo,Conti:2022egv}. A $(0+1)$-dimensional version of the root-$T\overline{T}$ deformation was studied in \cite{Garcia:2022wad}, which also preserves integrability. This operator has been connected to ultra/non-relativistic limits and the BMS group in three dimensions \cite{Rodriguez:2021tcz,Bagchi:2022nvj}, and to nonlinear automorphisms of the conformal algebra \cite{Tempo:2022ndz}. See also \cite{Hou:2022csf} for an analysis of $T\overline{T}$ and root-$T\overline{T}$-like deformations using characteristic flows.

Given the interest in the root-$T\overline{T}$ operator from the field theory perspective, it is natural to ask whether there are modified boundary conditions for the bulk metric that implement this deformation, as (\ref{TT_deformed_gamma_T}) do in the $T\overline{T}$ case. In this work, we will argue that the answer to this question is yes, and the analogous expressions are
\begin{align}
\begin{split}\label{root_TT_deformed_bcs}
    \gamma_{\alpha \beta}^{(\mu)} &= \cosh ( \mu ) \gamma_{\alpha \beta}^{(0)} + \frac{\sinh ( \mu )}{\mathcal{R}^{(0)}} \widetilde{T}_{\alpha \beta}^{(0)} \,, \\
    \widetilde{T}_{\alpha \beta}^{(\mu)} &= \cosh ( \mu )   \widetilde{T}_{\alpha \beta}^{(0)} + \sinh ( \mu ) \mathcal{R}^{(0)} \gamma^{(0)}_{\alpha \beta} \,,
\end{split}
\end{align}
where we have defined $\widetilde{T}_{\alpha \beta} = T_{\alpha \beta} - \frac{1}{2} \gamma_{\alpha \beta} T^\alpha{}_\alpha$, which is the traceless part of the stress tensor (not to be confused with the \emph{trace-reversed} stress tensor $\hat{T}_{\alpha \beta}$), and 
\begin{align}
    \mathcal{R}^{(0)} &= \sqrt{ \frac{1}{2} T^{(0) \alpha \beta} T_{\alpha \beta}^{(0)} - \frac{1}{4} \left(T^\alpha{}_\alpha \right)^2 } \nonumber \\
    &= \sqrt{ - \det \left(   \widetilde{T}_{\alpha \beta}^{(0)} \right) } \,, 
\end{align}
is the root-$T\overline{T}$ operator as before.

This means that -- from the viewpoint of holography -- the root-$T\overline{T}$ deformation plays a similar role as the $T\overline{T}$ deformation (or other $f(T)$ deformations), insofar as it imposes certain mixed boundary conditions where some function of the metric $\gamma_{\alpha \beta}$ and stress tensor $T_{\alpha \beta}$ is held fixed. However, the mixed boundary conditions which appear in the root-$T\overline{T}$ case are considerably more exotic because they involve the expression $\mathcal{R}^{(0)}$ which is non-analytic in the stress tensor $T^{(0)}_{\alpha \beta}$. Despite this unusual feature, we will show that the root-$T\overline{T}$ deformed boundary conditions have several surprisingly nice properties: for instance, various combinations of deformed quantities, like $T_{\alpha \beta}^{(\mu)} \, \delta \gamma^{(\mu) \alpha \beta}$ and $\det ( \gamma_{\alpha \beta}^{(\mu)} )$, are equal to their undeformed values and the root-$T\overline{T}$ deformed boundary conditions commute with the $T\overline{T}$-deformed boundary conditions, in a sense which we will make precise below. These unexpectedly simple relations, along with the pressing need to more deeply understand theories of root-$T\overline{T}$ type, motivate us to undertake a detailed study of the boundary conditions (\ref{root_TT_deformed_bcs}) in the remainder of the present work.

The layout of this chapter is as follows. In section \ref{sec:dictionary}, we review the holographic dictionary under general multi-trace deformations and apply these results to stress tensor deformations of $\mathrm{AdS}_3 / \mathrm{CFT}_2$. In section \ref{sec:consistency}, we use consistency conditions, such as commutativity between $T\overline{T}$ and root-$T\overline{T}$, to identify the root-$T\overline{T}$ deformed boundary conditions and the flow equation for the finite-volume spectrum of the field theory under a root-$T\overline{T}$ deformation. In section \ref{sec:AdS3RootTT}, we study $\mathrm{AdS}_3$ gravity with these root-$T\overline{T}$ deformed boundary conditions in both the metric and Chern-Simons formalisms and perform a holographic computation of the deformed spacetime mass which agrees with our flow equation for the root-$T\overline{T}$ deformed spectrum. In section \ref{sec:conclusion}, we conclude and identify directions for future research.

\section{Holographic dictionary for stress tensor deformations}\label{sec:dictionary}

The connection between deformations of a field theory by local operators and modified boundary conditions for the gravity dual was pointed out in the early days of the $\mathrm{AdS}$/$\mathrm{CFT}$ duality. For double-trace deformations, the effect on the CFT partition function was discussed in \cite{Gubser:2002vv} and its relation to modified boundary conditions was explored in \cite{Klebanov:1999tb,Witten:2001ua,Berkooz:2002ug,Mueck:2002gm,Diaz:2007an,Hartman:2006dy,Gubser:2002zh}. A generalization to multi-trace deformations, which we will follow in section \ref{subsec:multi-trace}, was laid out in \cite{Papadimitriou:2007sj}. Although earlier work focused on relevant or marginal deformations, the analysis of irrelevant $T\overline{T}$ and $J \bar{T}$ deformations is described in \cite{Bzowski:2018pcy,Guica:2019nzm} and the lecture notes \cite{monica_notes}. See also \cite{Nguyen:2021pdz} for a recent discussion of the generating functional of connected stress tensor correlators in holography (without $T\overline{T}$-like deformations).

In this section, we will review some of this well-known material to apply it to more general stress tensor deformations in $\mathrm{AdS}_3$/$\mathrm{CFT}_2$. An arbitrary scalar constructed from the stress tensor $T_{\alpha \beta}$ for a two-dimensional field theory can be written as a function of two independent invariants,
\begin{align}\label{two_invariants}
    f \left( T_{\alpha \beta} \right) = f \left( T^\alpha{}_\alpha, T^{\alpha \beta} T_{\alpha \beta} \right) \,, 
\end{align}
since all higher traces of $T_{\alpha \beta}$ are related to these two by trace identities. At the classical level, any such function can be used to generate a deformation of a quantum field theory. The usual $T\overline{T}$ deformation corresponds to
\begin{align}
    f = T^{\alpha \beta} T_{\alpha \beta} - \left( T^\alpha{}_\alpha \right)^2 = \mathcal{O}_{T\overline{T}} \,, 
\end{align}
whereas the root-$T\overline{T}$ deformation is
\begin{align}
    f = \sqrt{ \frac{1}{2} T^{\alpha \beta} T_{\alpha \beta} - \frac{1}{4} \left( T^\alpha{}_\alpha \right)^2 } = \mathcal{R} \,.
\end{align}
As we will explain, for any operator $f$ which is chosen as a deformation of the two-dimensional field theory, one can find the modified generating functional in the large-$N$ limit by a path integral argument. For certain choices of $f$, it is then possible to explicitly solve for the modified boundary conditions in the $3d$ bulk gravity theory.

The surprising feature of deformation by a \emph{marginal} combination of stress tensors, such as the root-$T\overline{T}$ operator, is that the additive shift in the generating functional of connected CFT correlators vanishes to leading order in $\frac{1}{N}$. Although such a deformation still has non-trivial effects on observables, this feature means that we will not be able to find the corresponding modified boundary conditions in the usual way. We will instead need to use a different argument, which will be the subject of section \ref{sec:consistency}.

\subsection{Multi-trace deformations}\label{subsec:multi-trace}
We first review the reasoning, which is used to find the change in the generating functional under a general multi-trace deformation of the $\mathrm{CFT}$. We follow \cite{Papadimitriou:2007sj} except for the mild generalization that we allow deformations by general scalar quantities constructed from operators carrying arbitrary indices, which allows us to include the case of stress tensor deformations. The analysis of this subsection applies in any dimension, so we will temporarily work in general spacetime dimension $d$ before specializing to $d = 3$ in later subsections. In this section, we will also explicitly retain factors of $N$ to make the role of the large-$N$ limit more transparent. Although in later sections, we will always implicitly work in a large-$N$ or large-$c$ limit to have a classical bulk gravity dual, we will typically not emphasize the central charge dependence of quantities appearing in path integrals.

Consider a $\mathrm{CFT}_d$ dual to a bulk $\mathrm{AdS}_{d+1}$ gravity theory. Let $\mathcal{O}_A$ be a collection of local operators in the conformal field theory, which are single traces in the sense that each is dual to a fundamental field of the bulk gravity theory. For instance, one can imagine each $\mathcal{O}_A$ as being dual to a light scalar field in the $3d$ bulk, in which case $A$ is an internal index. When we specialize to stress tensor deformations in $\mathrm{AdS}_3$/$\mathrm{CFT}_2$, we will instead think of $\mathcal{O}_A$ as some component $T_{\alpha \beta}$ of the energy-momentum tensor, in which case $A$ is a multi-index of spacetime indices. For now, we will treat both cases uniformly by using an abstract index $A$, which may transform under the action of some unspecified Lie group $G$.

We will deform the action by adding $N^2 \mu \int d^d x \, \sqrt{\gamma} \, f ( \mathcal{O} )$. Here $f$ is a scalar function of $\mathcal{O}_A$, in the sense that it is invariant under the action of $G$, $\mu$ is a coupling constant with the appropriate dimension, and $\gamma_{\alpha \beta}$ is the boundary metric. We will assume that $f ( 0 ) = 0$ but make no further assumptions about the function $f$. For simplicity, in the remainder of this subsection we will assume $\gamma_{\alpha \beta} = \eta_{\alpha \beta}$ and thus omit factors of $\sqrt{\gamma}$. Quantities in the deformed theory will be decorated by a $\mu$ superscript or subscript, whereas quantities in the undeformed theory will carry a $(0)$ label. Our goal will be to find a relationship between the generating functionals of connected  $\mathcal{O}_A$  correlators in the deformed and undeformed theories, which we write as $W^{(\mu)} [ J^{(\mu)} ]$ and $W^{(0)} [ J^{(0)} ]$, respectively, and which are defined by the path integrals
\begin{align}\begin{split}\label{two_generating_functionals}
    e^{- W^{(0)} [ J^{(0)} ] } &= \int \mathcal{D} \psi \, e^{ - S_0 - N^2 \int d^d x \, J^{(0) A} ( x ) \mathcal{O}_A ( x ) }\,, \\
    e^{- W^{(\mu)} [ J^{(\mu)} ] } &= \int \mathcal{D} \psi \, e^{- S_0 - N^2 \int d^d x \left( \mu f ( \mathcal{O} ) + J^{(\mu) A} (x) \mathcal{O}_A ( x ) \right) } \,.
\end{split}\end{align}
Here $J^{(0) A}$ and $J^{(\mu) A}$ are sources that are linearly coupled to the operators $\mathcal{O}_A$ in the undeformed and deformed theories, respectively. For simplicity, we suppress the $A$ indices on the sources $J^{(0) A}$, $J^{(\mu) A}$ when they appear as arguments in generating functionals. Correlators of the operators $\mathcal{O}_A$ are obtained from functional derivatives with respect to the source; for instance, the one-point function in the undeformed theory is given by
\begin{align}\label{OA_one_point}
    \langle \mathcal{O}_A \rangle_0 = \frac{1}{N^2} \frac{\delta W [ J^{(0)} ]}{\delta J^{(0) A}} \equiv \sigma_A ( x ) \,, 
\end{align}
where we introduce the shorthand $\sigma_A$ for convenience.

In the large-$N$ limit, all multi-point functions of operators $\mathcal{O}_A$ factorize into products of one-point functions of the form (\ref{OA_one_point}). This fact implies a simple relation between the two generating functionals in (\ref{two_generating_functionals}). To see this, we begin by changing variables in the path integral expression for $\exp \left( - W^{(\mu)} [ J^{(\mu)} ] \right)$, defining
\begin{align}\label{source_shift}
    \widetilde{J}_A = J^{(\mu)}_A + \mu \frac{\partial f ( \mathcal{O} ) }{\partial \mathcal{O}^A} \,.
\end{align}
This shift is performed because a general function $f ( \mathcal{O} )$ will have a term linear in $\mathcal{O}$ in its Taylor series expansion. Such a linear term in the effective action obstructs us from directly applying the results of large-$N$ factorization.\footnote{One way of understanding this, which is nicely explained in chapter 8 of \cite{coleman}, is to consider diagrammatics. For an effective action with a term linear in $\mathcal{O}$, there are infinitely many tree graphs that can be constructed with two external lines, since any lines may end on linear vertices. This complicates the large-$N$ analysis, which usually proceeds by noting that the leading contribution at large-$N$ comes from tree graphs with a minimal number of external lines (of which there should be finitely many). Performing the shift (\ref{source_shift}) removes the linear vertex and repairs this undesirable feature.} After implementing this shift to remove the linear term, the generating functional becomes
\begin{equation}
\begin{aligned}
\label{generating_functional_multi_trace}
   e^{- W^{(\mu)} \left( J^{(\mu)} \right) } = \int \mathcal{D} \psi \, e^{- S_{0} - N^2 \int d^d x \, \left( \mu f ( \mathcal{O} ) + \widetilde{J}^{A} \mathcal{O}_A - \mu  \mathcal{O}_A \partial^A f \right) }\,, 
\end{aligned}
\end{equation}
where we introduce the shorthand $\partial^A f = \frac{\partial f ( \mathcal{O} ) }{\partial \mathcal{O}_A}$. We may now relate this expression to the undeformed generating functional evaluated on $\widetilde{J}$ as
\begin{align}\label{generating_functional_intermediate}
    e^{- W^{(\mu)} \left[ J^{(\mu)} \right] } 
    &= \int \mathcal{D} \psi \, e^{ - S_{0} - N^2 \int d^d x \, \left( \widetilde{J}^{A} \mathcal{O}_A \right) } e^{- \mu N^2 \int d^2 x \, \left( f ( \mathcal{O} ) - \mathcal{O}_A \partial^A f \right) } \, \nonumber \\
    &= e^{- W^{(0)} [ \widetilde{J} ] } e^{- \mu N^2 \int d^d x \, \left( f ( \sigma ) - \sigma_A \partial^A f ( \sigma ) \right) } + O \left( \frac{1}{N} \right) \,.
\end{align}
The key observation is that the path integral on the second line of (\ref{generating_functional_intermediate}) defines a certain expectation value, namely of the second exponential factor, but in the large-$N$ limit, we may use factorization to evaluate this expectation value by replacing all instances of $\mathcal{O}_A$ with its one-point function $\sigma_A$. When $\mu = 0$, the argument of the second exponential factor vanishes, and the two generating functions are equal, as expected.

The upshot of this manipulation is that, by taking logarithms of the first and last expressions of (\ref{generating_functional_intermediate}) and discarding subleading terms as $N \to \infty$, we conclude
\begin{align}
    - W^{(\mu)} \left[ J^{(\mu)} \right] = - W^{(0)} \big[ \,  \widetilde{J} \, \big] - \mu N^2 \int d^d x \, \left( f ( \sigma ) - \sigma_A \partial^a f ( \sigma ) \right) \,, 
\end{align}
or in terms of the rescaled generating functionals $w [ J ] = \frac{1}{N^2} W [ J ]$,
\begin{align}\label{final_generating_functional}
    w^{(\mu)} \left[ J^{(\mu)} \right] = w^{(0)} \big[ \,  \widetilde{J} \, \big] + \mu \int d^d x \, \left( f ( \sigma ) - \sigma_A \partial^A f ( \sigma ) \right) \,,
\end{align}
where now $\sigma ( x ) = \frac{\delta w^{(0)} [ \tilde{J} ]}{\delta \tilde{J} (x)}$.

Note (\ref{final_generating_functional}) is the main result which allows us to find the change in the generating functional under an arbitrary multi-trace deformation, including by non-analytic operators like root-$T\overline{T}$. The deformation by any such operator has two separate effects. First, the generating functional $w^{(\mu)}$ is shifted by a term involving an integral of $f(\sigma) - \sigma_A \partial^A f ( \sigma )$. Second, the effective source $J^{(\mu) A}$ which is used for computing one-point functions is shifted by a term proportional to the derivative of $f$, as in (\ref{source_shift}).

It will sometimes be convenient to use a varied form of (\ref{final_generating_functional}). The variations of the two generating functionals are defined by varying the sources and holding the corresponding one-point functions fixed:
\begin{align}\begin{split}\label{defn_variations}
    \delta W^{(0)} \left[ J^{(0)} \right] &= \int d^d x \, \langle O_A \rangle_0 \, \delta J^{(0) A} \,, \\
    \delta W^{(\mu)} \left[ J^{(\mu)} \right] &= \int d^d x \, \langle O_A \rangle_\mu \, \delta J^{(\mu) A } \,.
\end{split}\end{align}
Varying (\ref{final_generating_functional}) and substituting for $\delta w^{(0)}$ then gives
\begin{align}\label{generating_functional_varied}
    \delta w^{(\mu)} \left[ J^{(\mu)} \right] &= \delta w^{(0)} \big[ \,  \widetilde{J} \, \big] + \mu \int d^d x \, \delta \left( f ( \sigma ) - \sigma_A \partial^A f ( \sigma ) \right) \, \nonumber \\
    &= \int d^d x \, \left( \langle O_A \rangle_0 \, \delta \widetilde{J}^{A} + \mu \delta \left( f ( \sigma ) - \sigma_A \partial^A f ( \sigma ) \right) \right) \,.
\end{align}
Finally, equating this result with the expression for $\delta w^{(\mu)}$ in terms of $\delta J^{(\mu) A}$ gives
\begin{align}\label{varied_generating_funcational_matching}
    \int d^d x \, \langle O_A \rangle_\mu \, \delta J^{(\mu) A } = \int d^d x \, \left( \langle O_A \rangle_0 \, \delta \widetilde{J}^{A} + \mu \delta \left( f ( \sigma ) - \sigma_A \partial^A f ( \sigma ) \right) \right) \,.
\end{align}
Note (\ref{final_generating_functional}) is the main result that allows us to find the change in the generating functional under an arbitrary multi-trace deformation, including by non-analytic operators like root-$T\overline{T}$. The deformation by any such operator has two separate effects. First, the generating functional $w^{(\mu)}$ is shifted by a term involving an integral of $f(\sigma) - \sigma_A \partial^A f ( \sigma )$. Second, the effective source $J^{(\mu) A}$, which is used for computing one-point functions is shifted by a term proportional to the derivative of $f$, as in (\ref{source_shift}).

It will sometimes be convenient to use a varied form of (\ref{final_generating_functional}). The variations of the two generating functionals are defined by varying the sources and holding the corresponding one-point functions fixed:
\begin{align}\begin{split}\label{defn_variations}
    \delta W^{(0)} \left[ J^{(0)} \right] &= \int d^d x \, \langle O_A \rangle_0 \, \delta J^{(0) A} \,, \\
    \delta W^{(\mu)} \left[ J^{(\mu)} \right] &= \int d^d x \, \langle O_A \rangle_\mu \, \delta J^{(\mu) A } \,.
\end{split}\end{align}
Varying (\ref{final_generating_functional}) and substituting for $\delta w^{(0)}$ then gives
\begin{align}\label{generating_functional_varied}
    \delta w^{(\mu)} \left[ J^{(\mu)} \right] &= \delta w^{(0)} \big[ \,  \widetilde{J} \, \big] + \mu \int d^d x \, \delta \left( f ( \sigma ) - \sigma_A \partial^A f ( \sigma ) \right) \, \nonumber \\
    &= \int d^d x \, \left( \langle O_A \rangle_0 \, \delta \widetilde{J}^{A} + \mu \delta \left( f ( \sigma ) - \sigma_A \partial^A f ( \sigma ) \right) \right) \,.
\end{align}
Finally, equating this result with the expression for $\delta w^{(\mu)}$ in terms of $\delta J^{(\mu) A}$ gives
\begin{align}\label{varied_generating_funcational_matching}
    \int d^d x \, \langle O_A \rangle_\mu \, \delta J^{(\mu) A } = \int d^d x \, \left( \langle O_A \rangle_0 \, \delta \widetilde{J}^{A} + \mu \delta \left( f ( \sigma ) - \sigma_A \partial^A f ( \sigma ) \right) \right) \,.
\end{align}
Here (\ref{varied_generating_funcational_matching}) will be useful for finding the modified boundary conditions for bulk fields after deforming the boundary field theory by some operator $f$. In particular, for a given deformation, one can match the coefficients of independent variations in (\ref{varied_generating_funcational_matching}) to obtain differential equations whose solution gives the deformed boundary conditions. This is especially helpful for studying more general deforming operators $f$, which depend both on the operators $\mathcal{O}_A$ and their sources $J^A$. Deformations by scalars constructed from the stress tensor $T_{\alpha \beta}$ are of this more complicated form since they involve contractions with the boundary metric $\gamma^{\alpha \beta}$ which plays the role of the source for $T_{\alpha \beta}$. It is shown in \cite{Papadimitriou:2007sj} that an analysis of the varied (\ref{varied_generating_funcational_matching}) yields the correct modification to the stress tensor $T_{\alpha \beta}$ after a multi-trace deformation, which is convenient because this analysis is more straightforward than a direct computation from the deformed generating functional.

As a sanity check, it is useful to consider the case of a double-trace deformation,
\begin{align}
    \mu f ( \sigma ) = \frac{1}{2} \mu^{AB} \sigma_A \sigma_B \,, 
\end{align}
where $\mu^{AB}$ is a field-independent symmetric tensor. In this case,
\begin{align}
    \mu \partial^A f = \mu^{AB} \sigma_B \,.
\end{align}
This means that the source $J_\mu^{(A)}$ satisfies
\begin{align}\label{double_trace_source_shift}
    J^{(\mu) A} = \widetilde{J}^A - \mu^{AB} \sigma_B \,, 
\end{align}
and, thus, the source has been shifted by a term linear in the corresponding expectation value. The deformed and undeformed generating functionals are related by
\begin{align}\label{double_trace_generating_shift}
    w^{(\mu)} \left[ J^{(\mu)} \right] = w^{(0)} \big[ \,  \widetilde{J} \, \big] - \frac{1}{2} \int d^d x \, \mu^{AB} \sigma_A \sigma_B \,.
\end{align}
Therefore, we see that a double-trace deformation is especially simple: although we deformed the action by \emph{adding} an integrated quantity proportional to $\int d^d x \, \mu^{AB} \sigma_A \sigma_B$, the generating functional has been deformed by \emph{subtracting} such a quantity.

\subsection{Compatibility with Hubbard-Stratonovich}\label{subsec:hub-strat}
In the case of a double-trace deformation, the general analysis of section \ref{subsec:multi-trace} is equivalent to another common technique for deriving the modified holographic dictionary, namely the Hubbard-Stratonovich transformation. This method exploits the fact that a double-trace deformation is quadratic in fields and, therefore, can be decoupled by integrating in an appropriate auxiliary field. The Hubard-Stratonovich technique has a long history and was already used in \cite{Gubser:2002vv} to study the effect of a double-trace deformation on the dual CFT, which is nicely reviewed in \cite{Bzowski:2018pcy,Guica:2019nzm}. We note that a similar strategy was used in \cite{Cardy:2018sdv} to replace the $T\overline{T}$ operator with a coupling to a metric-like field $h_{\alpha \beta}$ and interpret the deformation as random geometry. However, this decoupling procedure does not straightforwardly apply to more general multi-trace deformations, such as the square root-type deformation by $\mathcal{R}$. For completeness, we now briefly review this alternative derivation and confirm that the resulting modification to the generating functional is identical.

We again work in general spacetime dimension $d$ and focus on a deformation of the CFT action, which takes the form:
\begin{align}
    S_0 \longrightarrow S_0 + \frac{N^2}{2} \int d^d x \, \mu^{AB} \mathcal{O}_A \mathcal{O}_B \,, 
\end{align}
where the $\mathcal{O}_A$ are single-trace operators as before, and consider the deformed generating functional defined in (\ref{two_generating_functionals}),
\begin{align}\label{generating_functional_double_trace}
    e^{- W^{(\mu)} [ J^{(\mu)} ] } &= \int \mathcal{D} \psi \, e^{ - S_0 - N^2 \int d^d x \left( \frac{\mu^{AB}}{2} \mathcal{O}_A \mathcal{O}_B + J^{(\mu) A} (x) \mathcal{O}_A ( x ) \right) } \,.
\end{align}
We now integrate in an auxiliary field. To emphasize the similarity with the random geometry analysis of \cite{Cardy:2018sdv}, we will use the notation $h_A$ for this Hubbard-Stratonovich field. One has the path integral identity
\begin{align}\label{HS_identity}
    1 = \mathcal{N} \int \mathcal{D} h \, e^{ \frac{N^2}{2} \int d^2 x \, h^A \left( \mu^{-1} \right)_{AB} h^B} \,. 
\end{align}
Here, the quantity $\mathcal{N}$ is a normalization factor, which is defined by the property that it normalizes the path integral on the right side of (\ref{HS_identity}) to one. It can also be formally written as $\mathcal{N} = \frac{1}{\sqrt{ \det \left( \frac{\mu}{N^2} \right) } }$, although we will use the shorter expression $\mathcal{N}$ to avoid cluttering formulas. Inserting this identity into the expression (\ref{generating_functional_double_trace}) for the generating functional,
\begin{equation}
\begin{aligned}
      e^{- W^{(\mu)} \left( J^{(\mu)} \right)}  
   & = \mathcal{N} \int \mathcal{D} \psi \, \mathcal{D} h \, e^{ - S_{0} - N^2 \int d^d x \, \left( J^{(\mu) A} \mathcal{O}_A  + \frac{\mu^{AB}}{2} \mathcal{O}_A \, \mathcal{O}_B - \frac{1}{2} \, h^A \left( \mu^{-1} \right)_{AB} h^B \right)}\,  \\
    &= \mathcal{N} \int \mathcal{D} \psi \, \mathcal{D} h \, e^{- S_{0} - N^2 \int d^d x \, \left( J^{(\mu) A} + \widehat{h}^A \right) \mathcal{O}_A - \frac{1}{2} \widehat{h}^A \mu^{-1}_{AB} \widehat{h}^B} \,,
\end{aligned}
\end{equation}
where in the last step, we have completed the square in the integrand by writing quantities in terms of a shifted auxiliary field
\begin{align}
    \widehat{h}^A = h^A + \mu^{AB} \mathcal{O}_B \,.
\end{align}
Seeing that the combination $\widehat{h}^A + J^{(\mu) A}$ now acts as the source for $\mathcal{O}_A$, we perform a second change of variables to
\begin{align}
    \widetilde{h}^A = \widehat{h}^A + J^{(\mu) A}
\end{align}
to find

\begin{equation}
\begin{aligned}
   \label{HS_intermediate}
    e^{ - W^{(\mu)} \left( J^{(\mu)} \right) }  
    &=  \mathcal{N} \int \mathcal{D} \psi \, \mathcal{D} h \, e^{- S_{0} - N^2 \int d^d x \, \left( \widetilde{h}^A \mathcal{O}_A + \frac{1}{2} \left( \widetilde{h}^A - J^{(\mu) A} \right) \mu^{-1}_{AB} \left( \widetilde{h}^B - J^{(\mu) B} \right) \right) } \\
    &=  \mathcal{N} \int \mathcal{D} h \, e^{- W^{(0)} [ \widetilde{h} ] } \, e^{- N^2 \int d^d x \, \left( \frac{1}{2} \left( \widetilde{h}^A - J^{(\mu) A} \right) \mu^{-1}_{AB} \left( \widetilde{h}^B - J^{(\mu) B} \right) \right) } \,.
\end{aligned}
\end{equation}
In the second step, we have noted that performing the path integral including the first two terms in the exponential, $S_0$ and the coupling $\widetilde{h}^A \mathcal{O}_A$, defines the undeformed generating functional $\exp \left( - W^{(0)} \big[ \, \widetilde{h} \, \big] \right)$ since $\widetilde{h}^A$ acts as the source for $\mathcal{O}_A$. We have also implicitly used large-$N$ factorization since the third term in the exponential also depends on the operators $\mathcal{O}_A$. In the last line of (\ref{HS_intermediate}), all implicit instances of such operators are understood to be replaced with the corresponding one-point functions.

In the large-$N$ limit, the remaining path integral over $h$ can be performed using the saddle point approximation. The saddle occurs at the point $\widetilde{h}^A$ which satisfies
\begin{align}\label{htilde_saddle}
    - \frac{\delta W^{(0)} [ \widetilde{h} ] }{\delta \widetilde{h}^A} - \mu^{-1}_{AB} \left( \widetilde{h}^B - J^{(\mu) B} \right) = 0 \,.
\end{align}
On the other hand, the quantity $- \frac{\delta W^{(0)} [ \widetilde{h} ] }{\delta h^A}$ defines the one-point function $\langle \mathcal{O}_A \rangle_0 \equiv \sigma_A$, where we again introduce the shorthand $\sigma_A$ for the undeformed expectation value of $\mathcal{O}_A$.

Because we are modifying the field theory, the local operator $\mathcal{O}_A^{(0)}$ in the undeformed theory could correspond to some different operator $\mathcal{O}_A^{(\mu)}$ in the deformed theory. Therefore, in principle, we should distinguish between deformed and undeformed operators, in addition to distinguishing between deformed and undeformed \emph{expectation values}, which we have written as $\langle \; \cdot \; \rangle_\mu$ and $\langle \; \cdot \;  \rangle_0$, and which differ in that they are computed using path integrals weighted by different actions. However, we will see that for both the $T\overline{T}$ deformation and the root-$T\overline{T}$ deformation, the deformed and undeformed operators agree:
\begin{align}\label{same_operators}
    \mathcal{O}_A^{(0)} = \mathcal{O}_A^{(\mu)} \,.
\end{align}
More precisely, we will see that the derivatives of the operators $\mathcal{O}^{(\lambda)}_{T\overline{T}}$ and $\mathcal{R}^{(\mu)}$ with respect to the appropriate flow parameters $\lambda$ and $\mu$, respectively, both vanish. Thus, we will simply assume that (\ref{same_operators}) holds in the present analysis; one can view this as an extra condition one might impose to single out a preferred class of deforming operators.

Solving (\ref{htilde_saddle}) then yields $\widetilde{h}^A = J^{(\mu) B} + \mu^{AB} \sigma_B$. We, therefore, find that
\begin{align}\label{HS_saddle}
   e^{- W^{(\mu)} [ J^{(\mu) } ] } \sim e^{- W^{(0)} \left[ J^{(\mu) B} + \mu^{AB} \sigma_B \right] - \frac{N^2}{2} \int d^d x \, \mu^{AB} \sigma_A \sigma_B } \,. 
\end{align}
Here, we write $\sim$ to indicate both overall proportionality, since the saddle point integral introduces an additional prefactor which we will not track, and also the approximation to leading order in $\frac{1}{N}$. Taking logarithms and discarding the normalization, we conclude
\begin{align}
    W^{(\mu)} [ J^{(\mu)} ] &= W^{(0)} \big[ \, \widetilde{J} \, \big] - \frac{N^2}{2} \int d^d x \, \mu^{AB} \sigma_A \sigma_B \,, \nonumber \\
    \widetilde{J}^A &= J^{(\mu) A} + \mu^{AB} \sigma_B \,.
\end{align}
We see that this exactly reproduces equations (\ref{double_trace_source_shift}) and (\ref{double_trace_generating_shift}), after shifting to the re-scaled generating functionals $w$ by dividing through by $N^2$.

Therefore, the two approaches that we have described are equivalent to the case of double-trace deformations. Both computations use the assumption of large-$N$ in a key way. In the first method, we used large-$N$ factorization in (\ref{generating_functional_intermediate}), and in the Hubbard-Stratonovich approach, we used both large-$N$ factorization in (\ref{HS_intermediate}) and a saddle-point approximation in (\ref{HS_saddle}).

However, it is important to emphasize that the first method is more general since it applies to arbitrary multi-trace deformations. The Hubbard-Stratonovich technique crucially relies on the path integral identity (\ref{HS_identity}), which is a Gaussian integral and can, therefore, only introduce a quadratic dependence on $h^A$. Such a quadratic auxiliary field term is sufficient to decouple a double-trace deformation like the usual $T\overline{T}$, but for more general operators such as root-$T\overline{T}$, we will instead resort to the multi-trace analysis.

\subsection{Stress tensor deformations of $\mathrm{AdS}_3 / \mathrm{CFT}_2$}\label{sec:stress_tensor_AdS/CFT}

In the remainder of this work, we will focus on deformations that are constructed from the energy-momentum tensor rather than from general operators $\mathcal{O}_A$. It is worth pointing out that such deformations are qualitatively different in three bulk spacetime dimensions, which is our primary case of interest. In $\mathrm{AdS}_3$, the bulk metric has no local propagating degrees of freedom. As a result, we do not need to impose the usual restrictions that a deforming operator built from $\mathcal{O}_A$ be relevant or marginal to retain analytic control.

An irrelevant deformation built from an operator $\mathcal{O}_A$, which is dual to a dynamical field, such as a light scalar, would generically backreact on the metric and, therefore, become difficult to study. In contrast, an irrelevant deformation constructed from the $2d$ stress tensor $T_{\alpha \beta}$, such as the $T\overline{T}$ deformation, does not lead to any backreaction because the dual field is the (non-dynamical) bulk metric $g_{\alpha \beta}$. This means that we are free to consider deformations by \emph{any} scalar function $f ( T )$ of the stress tensor, even those with arbitrarily large dimensions, and study the resulting mixed boundary conditions in the $\mathrm{AdS}_3$ bulk.

As we mentioned around (\ref{two_invariants}), the most general Lorentz scalar, which can be constructed from a $2d$ stress tensor $T_{\alpha \beta}$ is
\begin{align}
    f \left( T_{\alpha \beta} \right) = f \left( x_1, x_2 \right) \,, \qquad  x_1 = T^\alpha{}_\alpha \,, \qquad x_2 = T^{\alpha \beta} T_{\alpha \beta}  \,,
\end{align}
where we introduce $x_1 = \text{Tr} ( T )$ and $x_2 = \text{Tr} ( T^2 )$.

In the notation of section \ref{subsec:multi-trace}, this corresponds to $\mathcal{O}_A = T_{\alpha \beta}$ and $\sigma_A = \langle T_{\alpha \beta} \rangle_0$, where $A$ is a multi-index of two boundary spacetime indices. We note that
\begin{align}
    \frac{\partial f}{\partial T_{\alpha \beta}} = \frac{\partial f}{\partial x_1} \gamma^{\alpha \beta} + 2 \frac{\partial f}{\partial x_2} T^{\alpha \beta} \,, 
\end{align}
where $\gamma_{\alpha \beta}$ is the $2d$ metric.

We may now import the general results for the shift in the generating functional under a multi-trace deformation defined by
\begin{align}
    \frac{\partial S^{(\mu)}}{\partial \mu} = \int d^2 x \, \sqrt{\gamma} \, f ( x_1, x_2 ) \,.
\end{align}
Because the boundary metric $\gamma_{\alpha \beta}$ now plays a more important role, we restore factors of $\sqrt{\gamma}$ in integrals which were omitted in sections \ref{subsec:multi-trace} and \ref{subsec:hub-strat}.

Using (\ref{final_generating_functional}), we find
\begin{align}\label{fT_generating_functional}
    w^{(\mu)} \left[ J^{(\mu)} \right] = w^{(0)} \big[ \,  \widetilde{J} \, \big] + \mu \int d^2 x \, \sqrt{\gamma} \, \left( f ( x_1, x_2 ) - \left( x_1 \frac{\partial f}{\partial x_1} + 2 x_2 \frac{\partial f}{\partial x_2} \right) \right) \,,
\end{align}
where now we use $x_1, x_2$ interchangeably for the operators $T^\alpha{}_\alpha$, $T^{\alpha \beta} T_{\alpha \beta}$ and the expectation values $\langle T^\alpha{}_\alpha \rangle $, $\langle T^{\alpha \beta} \rangle \langle T_{\alpha \beta} \rangle$, as justified by large-$N$ factorization.

In these formulas, the source $J^{(\mu)}$ which couples to the deformed stress tensor $T_{\alpha \beta}^{(\mu)}$ is the deformed metric $\gamma_{\alpha \beta}^{(\mu)}$. This means that the deformation by $f$ involves both single-trace operators and their sources, which makes the behavior of this deformation more complicated. While a deformation by a function that depends only on operators $\mathcal{O}_A$ (but not their sources $J^A$) shifts the sources and leaves the expectation values $\langle \mathcal{O}_A \rangle$ unchanged, a deformation which depends on both $\mathcal{O}_A$ and $J^A$ will shift both the sources and the one-point functions. In this case, as we discussed above, it is more convenient to use the varied (\ref{varied_generating_funcational_matching}), which allows us (in principle, at least) to find expressions for both the deformed sources and the deformed expectation values. In this context, the appropriate varied equation for a stress tensor deformation after taking the limit as $\mu \to 0$ is
\begin{align}\label{general_variational_method}
    \frac{\partial}{\partial \mu} \int d^d x \, \sqrt{\gamma^{(\mu)}} \, T_{\alpha \beta}^{(\mu)} \, \delta \gamma^{(\mu) \alpha \beta} = \int d^d x \,  \delta \left[ \sqrt{\gamma^{(\mu)}} \left( f ( x_1, x_2 ) - \left( x_1 \frac{\partial f}{\partial x_1} + 2 x_2 \frac{\partial f}{\partial x_2} \right) \right) \right] \,.
\end{align}
The operator $\delta$ appearing on the right side acts on both $T_{\alpha \beta}$ and $\gamma_{\alpha \beta}$. Given a particular choice of deformation $f(x_1, x_2)$, one can then attempt to match the $\delta \gamma_{\alpha \beta}$ and $\delta T_{\alpha \beta}$ terms on both sides of (\ref{general_variational_method}) and solve the resulting coupled differential equations in $\mu$ to obtain solutions for the deformed quantities $T_{\alpha \beta}^{(\mu)}$ and $\gamma_{\alpha \beta}^{(\mu)}$.

The known results for the $T\overline{T}$ deformation can be recovered by setting
\begin{align}
    f ( x_1, x_2 ) = - \frac{1}{2} \left( x_2 - x_1^2 \right) = - \frac{1}{2} \mathcal{O}_{T\overline{T}} \,,
\end{align}
where the factor of $- \frac{1}{2}$ is a choice of normalization for the operator. Substituting this deformation $f$ into (\ref{general_variational_method}) and stripping off the integrals gives the condition
\begin{align}\label{TT_varied_flow}
    \partial_\lambda \left( \sqrt{\gamma^{(\lambda)}} T_{\alpha \beta}^{(\lambda)} \, \delta \gamma^{(\lambda) \alpha \beta} \right) = \delta \left( \sqrt{\gamma^{(\lambda)}} \left( T^{(\lambda) \alpha \beta} T_{\alpha \beta}^{(\lambda)} - \left( T^{(\lambda) \alpha}{}_\alpha\right)^2 \right) \right) \,,
\end{align}
where we have changed the label for the deformation parameter from $\mu$ to $\lambda$ to emphasize that this flow is associated with the $T\overline{T}$ deformation (see footnote \ref{notation_footnote}). The indices in (\ref{TT_varied_flow}) are raised and lowered with the deformed metric $\gamma^{(\lambda)}_{\alpha \beta}$. One can solve this equation with the initial conditions $\gamma^{(\lambda)}_{\alpha \beta} \to \gamma^{(0)}_{\alpha \beta}$, $T_{\alpha \beta}^{(\lambda)} \to T_{\alpha \beta}^{(0)}$ as $\lambda \to 0$, as described in \cite{Guica:2019nzm} and reviewed in appendix \ref{app:TT_bc_pde_soln}. The solution to this differential equation can be expressed in terms of the trace-reversed stress tensor, $\widehat{T}_{\alpha \beta} = T_{\alpha \beta} - \gamma_{\alpha \beta} T^\rho{}_\rho$, in terms of which one finds
\begin{align}\begin{split}\label{TT_def_bc_later}
    \gamma^{(\lambda)}_{\alpha \beta} &= \gamma^{(0)}_{\alpha \beta} - 2 \lambda \widehat{T}_{\alpha \beta}^{(0)} + \lambda^2 \widehat{T}^{(0)}_{\alpha \rho} \,  \widehat{T}^{(0)}_{\sigma \beta} \, \gamma^{(0) \rho \sigma} \,, \\
    \widehat{T}^{(\lambda)}_{\alpha \beta} &= \widehat{T}^{(0)}_{\alpha \beta} - \lambda \widehat{T}^{(0)}_{\alpha \rho} \, \widehat{T}^{(0)}_{\sigma \beta} \gamma^{(0) \rho \sigma} \,,
\end{split}\end{align}
which reproduces (\ref{TT_deformed_gamma_T}) for the $T\overline{T}$-deformed boundary conditions which we quoted in the introduction.

One might ask whether there are other choices for the deforming operator $f$ which are distinguished in some sense. For instance, it is natural to ask whether there is any choice of $f$ for which the shift in the generating functional appearing in (\ref{fT_generating_functional}) vanishes. Such a function $f$ satisfies the differential equation:
\begin{align}
    f ( x_1, x_2 ) = x_1 \frac{\partial f}{\partial x_1} + 2 x_2 \frac{\partial f}{\partial x_2} \,, 
\end{align}
which has a general solution
\begin{align}
    f ( x_1, x_2 ) = x_1 g \left( \frac{x_2}{x_1^2} \right)\,,
\end{align}
where $g$ is an arbitrary function. We demand that this deformation is well-defined if the seed theory is a CFT, for which $x_1 = T^\alpha{}_\alpha = 0$. The only way for the argument of the function $g$ to remain finite when $x_1 = 0$ is if $g ( y ) = \sqrt{c_1 y}$, in which case
\begin{align}
    f ( x_1, x_2 ) = x_1 \sqrt{ c_1 \frac{x_2}{x_1^2} } = \sqrt{ c_1 x_2 } \,.
\end{align}
Choosing the normalization factor $c_1 = \frac{1}{2}$, we find
\begin{align}
    f ( x_1, x_2 ) = \sqrt{ \frac{1}{2} x_2 } = \sqrt{ \frac{1}{2} T^{\alpha \beta} T_{\alpha \beta} } = \mathcal{R} \, \big\vert_{T^\alpha{}_\alpha = 0}  \,.
\end{align}
Therefore, the only physical sensible stress tensor deformation of a CFT with a vanishing shift in (\ref{fT_generating_functional}) is, up to proportionality, the root-$T\overline{T}$ operator $\mathcal{R}$ defined in (\ref{root_TT_def}). Note that this argument fixes the dependence of $f$ on $x_2$ but not on $x_1$ since we have restricted it to the case of a conformal theory for which $x_1 = 0$. We will determine the dependence on $x_1$ by demanding that this deformation commute with the $T\overline{T}$ deformation in section \ref{sec:consistency}.

Suppose that we wish to identify the deformed metric $\gamma_{\alpha \beta}^{(\mu)}$ and stress tensor $T_{\alpha \beta}^{(\mu)}$ associated with a deformation by this operator $\mathcal{R}$. One immediately encounters the subtlety that the differential (\ref{general_variational_method}) reduces to
\begin{align}\label{rtt_variational_pde}
    \frac{\partial}{\partial \mu} \left( \sqrt{\gamma^{(\mu)}} T_{\alpha \beta}^{(\mu)} \, \delta \gamma^{(\mu) \alpha \beta } \right) = 0 \,.
\end{align} 
This means that the operator $\mathcal{R}$ is in the kernel of the map, which sends deformations to sources on the right side of the differential (\ref{general_variational_method}). There are multiple solutions to (\ref{rtt_variational_pde}). The most obvious one is the trivial solution $\gamma^{(\mu)}_{\alpha \beta} = \gamma^{(0)}_{\alpha \beta}$ and $T^{(\mu)}_{\alpha \beta} = T^{(0)}_{\alpha \beta}$. Another less obvious solution can be conveniently written in terms of the traceless part of the stress tensor, $\widetilde{T}_{\alpha \beta} = T_{\alpha \beta} - \frac{1}{2} \gamma_{\alpha \beta} T^\rho{}_\rho$. That solution is
\begin{align}\begin{split}\label{root_TT_deformed_bcs_later}
    \gamma_{\alpha \beta}^{(\mu)} &= \cosh ( \mu ) \gamma_{\alpha \beta}^{(0)} + \frac{\sinh ( \mu )}{\mathcal{R}^{(0)}} \widetilde{T}_{\alpha \beta}^{(0)} \,, \\
    \widetilde{T}_{\alpha \beta}^{(\mu)} &= \cosh ( \mu ) \widetilde{T}_{\alpha \beta}^{(0)} + \sinh ( \mu )\mathcal{R}^{(0)} \gamma^{(0)}_{\alpha \beta} \,,
\end{split}\end{align}
where $\mathcal{R}^{(0)}$ is the root-$T\overline{T}$ operator constructed from the undeformed metric and stress tensor. One can verify that the expressions (\ref{root_TT_deformed_bcs_later}) solve the differential (\ref{rtt_variational_pde}). Several quantities of interest remain individually undeformed along this flow:
\begin{align}
    \det \left( \gamma_{\alpha \beta}^{(\mu)} \right)  = \det \left( \gamma_{\alpha \beta}^{(0)} \right) \,, \qquad T_{\alpha \beta}^{(\mu)} \, \delta \gamma^{(\mu) \alpha \beta } = T_{\alpha \beta}^{(0)} \, \delta \gamma^{(0) \alpha \beta } \,, \qquad \mathcal{R}^{(\mu)} = \mathcal{R}^{(0)} \,.
\end{align}
If the only condition we impose is that our deformed metric and stress tensor satisfies (\ref{rtt_variational_pde}), then there is no way to distinguish between the trivial solution and the $\mu$-dependent solution (\ref{root_TT_deformed_bcs_later}). Furthermore, it is not immediately obvious; whether there are other solutions $\gamma^{(\mu)}_{\alpha \beta}$, $T^{(\mu)}_{\alpha \beta}$ which also satisfy this flow. For this reason, from the perspective of the deformed generating functional, we cannot uniquely identify a single solution for the deformed metric and stress tensor, which corresponds to the root-$T\overline{T}$ deformation.

To circumvent this ambiguity, we will pursue a complementary analysis that does not rely on the deformed generating functional. Instead, we will stipulate a set of consistency conditions in which we expect the root-$T\overline{T}$ deformed metric and stress tensor to satisfy and demonstrate that (\ref{root_TT_deformed_bcs_later}) is the only solution with these properties. This gives an independent piece of evidence that these deformed boundary conditions are the correct ones, which correspond to a root-$T\overline{T}$ deformation of the boundary theory. We turn to this argument in the next section.

\section{Root-\texorpdfstring{$T\overline{T}$}{TT} from consistency conditions}\label{sec:consistency}

We have seen that the root-$T\overline{T}$ deformation is subtle to treat holographically because it belongs to a class of deformations for which the combination:
\begin{align}\label{root_TT_class_deformations}
    \int d^2 x \, \sqrt{\gamma^{(\mu)}} T_{\alpha \beta}^{(\mu)} \, \delta \gamma^{(\mu) \alpha \beta }
\end{align}
is independent of $\mu$. This class also includes trivial deformations, such as boundary diffeomorphisms or scale transformations, which leave the theory unchanged.

However, we expect that the root-$T\overline{T}$ deformation is \emph{not} such a trivial deformation and should modify the behavior of the theory in some way. One piece of evidence for this is that the $2d$ root-$T\overline{T}$ deformation of a collection of bosons is the dimensional reduction of the $4d$ root-$T\overline{T}$ deformation of the free Maxwell theory \cite{Conti:2022egv}, which gives rise to the ModMax theory. This ModMax theory represents a genuine modification of the Maxwell theory in that physical observables are modified; one example is that the ModMax theory exhibits birefringence, whereas the Maxwell theory does not.

We would, therefore, like to distinguish the root-$T\overline{T}$ deformed theory from other deformations in the same class that obeys (\ref{root_TT_class_deformations}). To do this, we will enumerate a list of properties that we expect the root-$T\overline{T}$ deformed theory to obey and search for the most general deformation that satisfies these properties. This will allow us to identify both a candidate set of deformed boundary conditions $\gamma^{(\mu)}_{\alpha \beta}$, $T^{(\mu)}_{\alpha \beta}$ and a proposal for the deformed finite-volume spectrum of a root-$T\overline{T}$ deformed CFT on a cylinder.

An important ingredient in this analysis is the assumption that the root-$T\overline{T}$ deformation commutes with the ordinary $T\overline{T}$ deformation, in a sense which we will make precise. This expectation is motivated by the observation that classical $T\overline{T}$ and root-$T\overline{T}$ flows for the Lagrangian exhibit this property in many examples \cite{Ferko:2022cix,Borsato:2022tmu}. The property of commuting with $T\overline{T}$ is \emph{not} shared by generic marginal stress tensor deformations. A simple example is the marginal deformation generated by the trace of the stress tensor,
\begin{align}\label{trace_deformation}
    \frac{\partial S}{\partial \mu} = \int d^2 x \, \sqrt{\gamma} \, T^a{}_a \,.
\end{align}
The flow (\ref{trace_deformation}) simply generates scale transformations, so a conformal field theory is invariant under such a deformation. However, a $T\overline{T}$-deformed field theory is not scale-invariant because the theory has a dimensionful scale set by $\lambda$. Thus, scale transformations do not commute with $T\overline{T}$. Deforming a CFT first by (\ref{trace_deformation}) and then $T\overline{T}$-deforming with parameter $\lambda$ is the same as only performing the $T\overline{T}$ step, whereas first deforming the CFT by $T\overline{T}$ and then performing the scale transformation (\ref{trace_deformation}) is \emph{not} the same as $T\overline{T}$-deforming by $\lambda$.

\subsection{Derivation of deformed boundary conditions}\label{sec:boundary_condition_derivation}

We aim to find a one-parameter family of modified boundary conditions $\gamma^{(\mu)}_{\alpha \beta}$, $T_{\alpha \beta}^{(\mu)}$ with the following properties:

\begin{enumerate}

    \item\label{trace_assumption} The deformed boundary conditions should correspond to a classically marginal deformation of the dual field theory. This means that the parameter $\mu$ is dimensionless and that, if the undeformed stress tensor is traceless so that
    \begin{align}
        \gamma^{(0) \alpha \beta} T^{(0)}_{\alpha \beta} = 0 \,,
    \end{align}
    then the deformed stress tensor is also traceless with respect to the deformed metric,
    \begin{align}
        \gamma^{(\mu) \alpha \beta} T^{(\mu)}_{\alpha \beta} = 0 \,.
    \end{align}

    \item\label{group_assumption} The deformations of the metric and stress tensor form a group. In particular, deformations compose. If we deform an initial configuration by $\mu_1$,
    \begin{align}
        \gamma^{(0)}_{\alpha \beta} \,, T^{(0)}_{\alpha \beta} \overset{\mu_1}{\longrightarrow} \gamma^{(\mu_1)}_{\alpha \beta} \,, T^{(\mu_1)}_{\alpha \beta} \,,
    \end{align}
    and then use these quantities as the initial condition for a second deformation by $\mu_2$,
    \begin{align}
        \gamma^{(\mu_1)}_{\alpha \beta} \,, T^{(\mu_1)}_{\alpha \beta} \overset{\mu_2}{\longrightarrow} \gamma^{(\mu_1 + \mu_2)}_{\alpha \beta} \,, T^{(\mu_1 + \mu_2)}_{\alpha \beta} \,,
    \end{align}
    then the doubly-deformed quantities are identical to those obtained from doing a single deformation by the total parameter $\mu_1 + \mu_2$,
    \begin{align}
        \gamma^{(0)}_{\alpha \beta} \,, T^{(0)}_{\alpha \beta} \xrightarrow{\mu_1 + \mu_2} \gamma^{(\mu_1 + \mu_2)}_{\alpha \beta} \,, T^{(\mu_1 + \mu_2)}_{\alpha \beta} \,.
    \end{align}
    Here we assume that $\mu = 0$ is the identity element, so that the deformed boundary conditions reduce to the undeformed boundary conditions as the deformation parameter is taken to zero. We further assume the group to be non-trivial, so a deformation by $\mu \neq 0$ must be different from the identity.
    
    \item\label{bc_commute_assumption} The root-$T\overline{T}$ deformation commutes with the ordinary $T\overline{T}$ deformation, in the following sense. If we first deform the metric and stress tensor using the $T\overline{T}$ deformed boundary conditions and flow by parameter $\lambda$, and then use these deformed quantities as the initial condition for a root-$T\overline{T}$ flow by parameter $\mu$, then the result is identical to first deforming by root-$T\overline{T}$ with parameter $\mu$ and then by $T\overline{T}$ with parameter $\lambda$.
    
\begin{align}\label{}
\includegraphics[width=.95\linewidth]{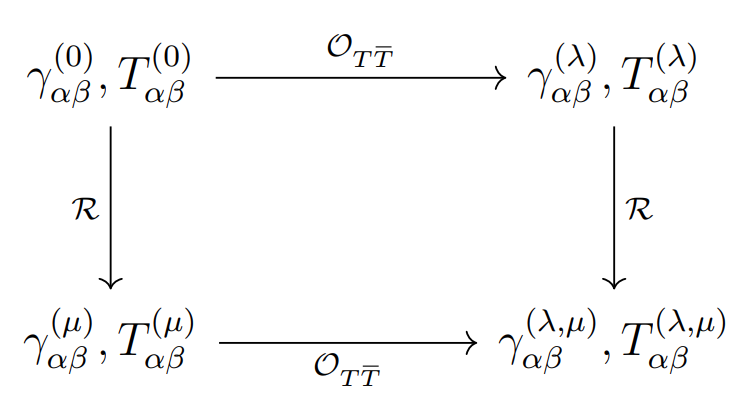}
\end{align}
    
\end{enumerate}

We will first use assumptions \ref{trace_assumption} and \ref{group_assumption} to determine the nature of the modified boundary conditions when the seed theory is conformal, and then use the third assumption to extend this procedure to the case when the undeformed theory is non-conformal.

\emph{Conformal seed theory}

For a conformal seed theory satisfying $\gamma^{(0) \alpha \beta} T^{(0)}_{\alpha \beta} = 0$, the only independent dimensionful Lorentz scalar quantity in the problem is $T^{(0) \alpha \beta} T^{(0)}_{\alpha \beta}$. Ordinarily, two independent scalars can be constructed from a general $2 \times 2$ matrix $M$ -- for instance, $\text{Tr}( M )$ and $\text{Tr} ( M^2 )$ -- but we have assumed that the trace of the stress tensor vanishes. We can equivalently say that any Lorentz scalar built from a traceless stress tensor is a function of
\begin{align}
    \mathcal{R}^{(0)} = \sqrt{ \frac{1}{2} T^{(0) \alpha \beta} T_{\alpha \beta}^{(0)} } \,.
\end{align}
On the other hand, there are also only two functionally independent symmetric $2$-tensors available in this problem, namely $\gamma_{\alpha \beta}^{(0)}$ and $T_{\alpha \beta}^{(0)}$. Again, one could attempt to form a new independent $2$-tensor by taking products of the form:
\begin{align}
    \left( T^2 \right)_{\alpha \beta} = T^{(0)}_{\alpha \gamma} T^{(0) \gamma}{}_\beta  \,, 
\end{align}
but because of the tracelessness condition, one has the identity
\begin{align}\label{traceless_simp}
    \left( T^2 \right)_{\alpha \beta} = \left( \mathcal{R}^{(0)} \right)^2 \gamma_{\alpha \beta} \,, 
\end{align}
so this combination does not give an independent tensor structure. All higher powers of the stress tensor will also be proportional to either $\gamma_{\alpha \beta}^{(0)}$ or $T_{\alpha \beta}^{(0)}$ with coefficients that are functions of $\mathcal{R}^{(0)}$.

We, therefore, find that the most general ansatz for deformed symmetric tensors $\gamma_{\alpha \beta}^{(\mu)}$ and $T_{\alpha \beta}^{(\mu)}$ with the correct scaling dimensions is
\begin{align}\label{root_TT_cft_seed_ansatz}
    \gamma^{(\mu)}_{\alpha \beta} &= f_1 ( \mu ) \gamma_{\alpha \beta}^{(0)} + \frac{f_2 ( \mu )}{\mathcal{R}^{(0)}} T_{\alpha \beta}^{(0)} \,, \nonumber \\
    T^{(\mu)}_{\alpha \beta} &= f_3 ( \mu ) T_{\alpha \beta}^{(0)} + f_4 ( \mu ) \mathcal{R}^{(0)} \gamma^{(0)}_{\alpha \beta} \,.
\end{align}
All that remains is to fix the four functions $f_i ( \mu )$. First, we will use the assumption that the deformed stress tensor remains traceless with respect to the deformed metric so that
\begin{align}
    \gamma^{(\mu) \alpha \beta} T^{(\mu)}_{\alpha \beta} = 0 \,.
\end{align}
This condition is satisfied if and only if
\begin{align}\label{f4_constraint}
    f_4 ( \mu ) = \frac{f_2 ( \mu ) f_3 ( \mu )}{f_1 ( \mu )} \,,
\end{align}
which fixes one of the functions. 

Next, we impose that subsequent deformations form a group, which is listed as assumption \ref{group_assumption} above. On the one hand, we can first deform the metric and stress tensor by parameter $\mu_1$ to obtain
\begin{align}\label{root_TT_cft_sequential}
    \gamma^{(\mu_1)}_{\alpha \beta} &= f_1 ( \mu_1 ) \gamma_{\alpha \beta}^{(0)} + \frac{f_2 ( \mu_1 )}{\mathcal{R}^{(0)}} T_{\alpha \beta}^{(0)} \,, \nonumber \\
    T^{(\mu_1)}_{\alpha \beta} &= f_3 ( \mu_1 ) T_{\alpha \beta}^{(0)} + \frac{f_2 ( \mu_1 ) f_3 ( \mu_1 )}{f_1 ( \mu_1 )} \mathcal{R}^{(0)} \gamma^{(0)}_{\alpha \beta} \,,
\end{align}
where we have used (\ref{f4_constraint}), and then use (\ref{root_TT_cft_sequential}) as the initial condition for a second deformation by parameter $\mu_2$. This gives one set of deformed quantities, $\gamma^{(\mu_1 + \mu_2)}_{\alpha \beta}$ and $T_{\alpha \beta}^{(\mu_1 + \mu_2)}$. On the other hand, we can deform all at once by a combined parameter $\mu_1 + \mu_2$, to yield
\begin{align}\begin{split}\label{root_TT_two_step}
    \gamma^{\prime (\mu_1 + \mu_2)}_{\alpha \beta} &= f_1 ( \mu_1 + \mu_2 ) \gamma_{\alpha \beta}^{(0)} + \frac{f_2 ( \mu_1 + \mu_2 )}{\mathcal{R}^{(0)}} T_{\alpha \beta}^{(0)} \,, \\
    T^{\prime (\mu_1 + \mu_2)}_{\alpha \beta} &= f_3 ( \mu_1 + \mu_2 ) T_{\alpha \beta}^{(0)} + \frac{f_2 ( \mu_1 + \mu_2 ) f_3 ( \mu_1 + \mu_2 )}{f_1 ( \mu_1 + \mu_2 )} \mathcal{R}^{(0)} \gamma^{(0)}_{\alpha \beta} \,.
\end{split}\end{align}
We have decorated the quantities in (\ref{root_TT_two_step}) with primes to emphasize that they may differ from the results $\gamma^{(\mu_1 + \mu_2)}_{\alpha \beta}$ and $T_{\alpha \beta}^{(\mu_1 + \mu_2)}$ of performing the two deformations sequentially. We then impose the constraints
\begin{align}\label{group_constraints}
    \gamma^{(\mu_1 + \mu_2)}_{\alpha \beta} = \gamma^{\prime (\mu_1 + \mu_2)}_{\alpha \beta} \,, \qquad T^{ (\mu_1 + \mu_2)}_{\alpha \beta} = T_{\alpha \beta}^{\prime (\mu_1 + \mu_2)} \,.
\end{align}
In performing the algebra to find the implications of equations (\ref{group_constraints}), we will assume that $f_1 > 0$ and $f_3 > 0$, which is convenient for simplifying expressions like $\sqrt{ f_1^2 }$ which appear in intermediate steps. This sign choice is reasonable because we are interested in deformations for which $f_1 ( 0 ) = f_3 ( 0 ) = 1$, so these functions should remain positive, at least for sufficiently small deformation parameters.

After making this assumption, one finds that these equations hold if and only if
\begin{align}\begin{split}\label{function_groups_constraints}
    f_1 ( \mu_1 + \mu_2 ) &= f_1 ( \mu_1 ) f_1 ( \mu_2 ) + f_2 ( \mu_1 ) f_2 ( \mu_2 ) \,, \\
    f_2 ( \mu_1 + \mu_2 ) &= f_1 ( \mu_2 ) f_2 ( \mu_1 ) + f_1 ( \mu_1 ) f_2 ( \mu_2 ) \,, \\
    f_3 \left( \mu_1 + \mu_2 \right) &= f_3 ( \mu_1 ) f_3 ( \mu_2 ) \left( 1 + \frac{f_2 ( \mu_1 ) f_2 ( \mu_2 )}{f_1 ( \mu_1 ) f_1 ( \mu_2 ) } \right) \,.
\end{split}\end{align}
We can turn the first two conditions in (\ref{function_groups_constraints}) into differential equations for $f_1$ and $f_2$ with the initial condition that $f_1(0) = 1$ and $f_2 ( 0 ) = 0$, which is required so that the deformation reproduces the undeformed theory as $\mu \to 0$. For instance, taking a derivative of the first line of (\ref{function_groups_constraints}) with respect to $\mu_2$ and then taking $\mu_2 = 0$ yields:
\begin{align}
    f_1' ( \mu_1 ) = f_1' ( 0 ) f_1 ( \mu_1 ) + f_2' ( 0 ) f_2 ( \mu_1 ) \,. 
\end{align}
To ease notation, let $f_1' ( 0 ) = a$ and $f_2' ( 0 ) = b$. Differentiating the second line of (\ref{function_groups_constraints}) with respect to $\mu_1$ and then taking $\mu_1$ to zero gives $f_2' ( \mu_2 ) = b f_1 ( \mu_2 ) + a f_2 ( \mu_2 )$. Thus, we have a system of differential equations
\begin{align}
    f_1' ( x ) = a f_1 ( x ) + b f_2 ( x ) \,, \qquad f_2 ' ( x ) = b f_1 ( x ) + a f_2 ( x ) \,, 
\end{align}
whose general solution with the initial conditions $f_1 ( 0 ) = 1$, $f_2 ( 0 ) = 0$ is
\begin{align}
    f_1 ( x ) = e^{a x} \cosh ( b x ) \,, \qquad f_2 ( x ) = e^{a x} \sinh ( b x ) \,.
\end{align}
For this class of solutions, the constraint in the third line of (\ref{function_groups_constraints}) then imposes
\begin{align}
    f_3 \left( \mu_1 + \mu_2 \right) &= f_3 ( \mu_1 ) f_3 ( \mu_2 ) \left( 1 + \tanh ( b \mu_1 ) \tanh ( b \mu_2 ) \right) \,,
\end{align}
which can be turned into a differential equation with the initial condition $f_3 ( 0 ) = 1$ as above. The result of this procedure is $f_3 ( x ) = e^{c x} \cosh ( b x )$, where $c$ is another constant.

Therefore, the most general $\mu$-dependent deformation of the metric and stress tensor consistent with our assumptions is
\begin{align}\begin{split}\label{root_TT_cft_a_b}
    \gamma^{(\mu)}_{\alpha \beta} &= e^{a \mu} \left( \cosh ( b \mu ) \gamma_{\alpha \beta}^{(0)} + \frac{\sinh ( b \mu ) }{\mathcal{R}^{(0)}} T_{\alpha \beta}^{(0)} \right) \,, \\
    T^{(\mu)}_{\alpha \beta} &= e^{c \mu} \left( \cosh ( b \mu ) T_{\alpha \beta}^{(0)} + \sinh ( b \mu ) \mathcal{R}^{(0)} \gamma^{(0)}_{\alpha \beta} \right) \,,
\end{split}\end{align}
where $a$, $b$, $c$ are arbitrary constants.

Some comments are in order. First, the deformations associated with the parameters $a$ and $c$ are simply the freedom to re-scale the metric or stress tensor by a constant $\mu$-dependent factor, which is expected since such a scaling respects conformal symmetry and forms a group. However, any such change in coordinates can be undone by a diffeomorphism, along with a redefinition of the stress tensor by a multiplicative factor (which does not affect conservation). Therefore, we will set $a = c = 0$ in what follows.

Second, the choice of the parameter $b$ corresponds to the scaling of the dimensionless flow parameter $\mu$, or equivalently to our choice of normalization for the operator $\mathcal{R}$. If $b = 0$, then there is no change in the metric or stress tensor (up to diffeomorphisms) for any value of $\mu$, and the group structure of our deformation is the trivial group. This violates our assumption \ref{group_assumption}, where we demand that the deformations form a non-trivial group, so $b = 0$ is forbidden. For simplicity, we will choose $b = 1$. With these choices, our modified boundary conditions for the case of a conformal seed theory are
\begin{align}\begin{split}\label{root_TT_cft_final}
    \gamma^{(\mu)}_{\alpha \beta} &= \cosh ( \mu ) \gamma_{\alpha \beta}^{(0)} + \frac{\sinh ( \mu ) }{\mathcal{R}^{(0)}} T_{\alpha \beta}^{(0)} \,, \\
    T^{(\mu)}_{\alpha \beta} &= \cosh ( \mu ) T_{\alpha \beta}^{(0)} + \sinh ( \mu ) \mathcal{R}^{(0)} \gamma^{(0)}_{\alpha \beta}  \,.
\end{split}\end{align}
We conclude that, up to diffeomorphisms and normalization, the unique choice of modified $\mathrm{AdS}_3$ boundary conditions which implement a marginal deformation of a $\mathrm{CFT}_2$ satisfying our assumptions are (\ref{root_TT_cft_final}). This is perhaps not too surprising since there is only a single Lorentz invariant that be constructed from a traceless stress tensor, so we had only one choice of scalar $\mathcal{R}^{(0)}$ which could appear in the modified boundary conditions. However, when both $T^{\alpha \beta} T_{\alpha \beta}$ and $T^\alpha{}_\alpha$ are nonzero, there is more freedom in the deformation, and we will require additional input to uniquely identify the appropriate analog of (\ref{root_TT_cft_final}).

\emph{Non-conformal seed theory}

Typically, one would not be interested in seed theories for which $T^{(0)^\alpha}{}_\alpha \neq 0$, since a generic non-conformal $2d$ QFT will not have any $\mathrm{AdS}_3$ dual. An important exception is if the seed theory \emph{itself} was obtained through applying an irrelevant stress tensor deformation, such as the $T\overline{T}$ deformation, to a conformal seed theory. Such a $T\overline{T}$-deformed CFT has a stress tensor with a non-vanishing trace,\footnote{\label{trace_flow_footnote} To wit, the trace satisfies the trace flow equation $T^\alpha{}_\alpha = - 2 \lambda \mathcal{O}_{T\overline{T}}$.} and yet it is dual to an $\mathrm{AdS}_3$ bulk with modified boundary conditions as we have described. One might, therefore, ask what happens if we use such a $T\overline{T}$-deformed theory as the input for a second deformation by root-$T\overline{T}$.

First we consider the most general expression for the modified boundary conditions $\gamma^{(\mu)}_{\alpha \beta}$ and $T^{(\mu)}_{\alpha \beta}$, which depend both on $\mu$ and the undeformed quantities $\gamma^{(0)}_{\alpha \beta}$ and $T^{(0)}_{\alpha \beta}$, with the property that these expressions reduce to (\ref{root_TT_cft_final}) in the special case where $T^{(0) \alpha}{}_\alpha = 0$. Because the stress tensor is no longer traceless, its square will not be proportional to the metric. As a result, one might believe that there are now three independent tensor structures in the problem, namely
\begin{align}
    \gamma_{\alpha \beta}^{(0)} \,, \quad T_{\alpha \beta}^{(0)} \,, \, \text{and} \quad \left( T^{(0)} \right)^2_{\alpha \beta} = T_{\alpha \rho}^{(0)} \gamma^{(0) \rho \sigma} T_{\sigma \beta}^{(0)} \,, 
\end{align}
and that the most general deformed metric $\gamma^{(\mu)}_{\alpha \beta}$ and stress tensor $T^{(\mu)}_{\alpha \beta}$ will each be a linear combination of three different tensor structures, with appropriate coefficients.

However, this is not the case, and there are still only two tensor structures. One can see this by writing quantities in terms of the traceless part of the stress tensor,
\begin{align}
    \widetilde{T}_{\alpha \beta} = T_{\alpha \beta} - \frac{1}{2} T^\rho{}_\rho \gamma_{\alpha \beta} \,.
\end{align}
Then $T_{\alpha \beta} = \widetilde{T}_{\alpha \beta} + \frac{1}{2} \gamma_{\alpha \beta} T$, where we write $T = T^\alpha{}_\alpha$ for the trace of the stress tensor to lighten notation. We also suppress the $(0)$ superscripts for the moment. Then the putative new tensor structure arising from the square of the stress tensor is
\begin{align}
    T^2_{\alpha \beta} &= T_{\alpha \sigma} \gamma^{\sigma \rho} T_{\rho \beta} \nonumber \\
    &= \left( \widetilde{T}_{\alpha \sigma} + \frac{1}{2} \gamma_{\alpha \sigma} T \right) \gamma^{\sigma \rho} \left( \widetilde{T}_{\rho \beta} + \frac{1}{2} \gamma_{\rho \beta} T \right) \nonumber \\
    &= \widetilde{T}^2_{\alpha \beta} + \widetilde{T}_{\alpha \beta} T + \frac{1}{4} T^2 \gamma_{\alpha \beta} \,.
\end{align}
We have already seen in (\ref{traceless_simp}) that $\widetilde{T}^2_{\alpha \beta} = \mathcal{R}^2 \gamma_{\alpha \beta}$. We conclude that there are still only two independent tensor structures $\gamma_{\alpha \beta}$ and $\widetilde{T}_{\alpha \beta}$, and that a generic candidate expression for a deformed symmetric tensor like $\gamma^{(\mu)}_{\alpha \beta}$ or $T^{(\mu)}_{\alpha \beta}$ must still be a linear combination of these two structures with appropriate scalar coefficients.

However, what \emph{has} changed is that there are now two Lorentz scalars that can be constructed from $T^{(0)}_{\alpha \beta}$ rather than just one. One way of parameterizing the two functionally independent scalars is $x_1 = T^{(0)\alpha}{}_\alpha$, $x_2 = T^{(0) \alpha \beta} T^{(0)}_{\alpha \beta}$, as we have done above. It will be more useful to instead use $x_1$ and $\mathcal{R}^{(0)} = \sqrt{ \frac{1}{2} x_2 - \frac{1}{4} x_1^2 }$. Clearly any function of $x_1$ and $x_2$ can also be expressed as a function of $x_1$ and $\mathcal{R}^{(0)}$.

A convenient way to write most general deformed boundary conditions is
\begin{align}\begin{split}\label{root_TT_noncft_general}
    \gamma^{(\mu)}_{\alpha \beta} &= f_1 ( \mu , y ) \cosh ( \mu ) \gamma_{\alpha \beta}^{(0)} + f_2 ( \mu , y ) \frac{\sinh ( \mu ) }{\mathcal{R}^{(0)}}  \widetilde{T}_{\alpha \beta}^{(0)} \,, \\
    \widetilde{T}^{(\mu)}_{\alpha \beta} &= f_3 ( \mu , y )  \cosh ( \mu ) \widetilde{T}_{\alpha \beta}^{(0)} + f_4 ( \mu , y ) \sinh ( \mu )  \mathcal{R}^{(0)} \gamma^{(0)}_{\alpha \beta} \,.
\end{split}\end{align}
The functions $f_i$ may depend on $\mu$ and on the dimensionless combination 
\begin{align}
    y \equiv \frac{x_1}{\mathcal{R}^{(0)}} \,.
\end{align}
The expressions (\ref{root_TT_noncft_general}) give the most general deformed boundary conditions that can be constructed from a non-conformal seed theory. To reduce to the earlier results (\ref{root_TT_cft_final}) in the case of a CFT seed, we impose
\begin{align}
    f_i ( \mu, 0 ) = 1 \,.
\end{align}
We now expect that there should be many solutions for the functions $f_i$, which correspond to boundary deformations by different marginal operators. For instance, one could deform by some operator of the form
\begin{align}
    \mathcal{O} = \sqrt{ c_1 T^{(0) \alpha \beta} T^{(0)}_{\alpha \beta} + c_2 \left( T^{(0)\alpha}{}_\alpha \right)^2 } + c_3 T^{(0)\alpha}{}_\alpha \,, 
\end{align}
for any choice of $c_i$. There should exist some choice of modified boundary conditions corresponding to any such operator, and we expect that any such deformation will satisfy the group property described in assumption \ref{group_assumption} above.

Rather than perform a systematic investigation of all such allowed deformed boundary conditions, we will attempt to single out a unique deformation within this family by imposing our assumption \ref{bc_commute_assumption}, namely that this deformation commutes with $T\overline{T}$. More precisely, we can first substitute a metric $\gamma_{\alpha \beta}^{(0)}$ and traceless stress tensor $T_{\alpha \beta}^{(0)}$ into the expressions (\ref{TT_def_bc_later}) to obtain the $T\overline{T}$-deformed boundary conditions $\gamma_{\alpha \beta}^{(\lambda)}$ and $T_{\alpha \beta}^{(\lambda)}$, and then substitute these two expressions into (\ref{root_TT_noncft_general}) to obtain
\begin{align}
    \gamma_{\alpha \beta}^{(\lambda, \mu)} \,, \quad T_{\alpha \beta}^{(\lambda, \mu)} \,.
\end{align}
On the other hand, we could instead first substitute $\gamma_{\alpha \beta}^{(0)}$, $T_{\alpha \beta}^{(0)}$ into (\ref{TT_def_bc_later}) to obtain $\gamma_{\alpha \beta}^{(\mu)}$ and $T_{\alpha \beta}^{(\mu)}$, and then substitute  these into (\ref{TT_def_bc_later}) to find
\begin{align}
    \gamma_{\alpha \beta}^{(\mu, \lambda)} \,, \quad T_{\alpha \beta}^{(\mu, \lambda)} \,.
\end{align}
We then impose the two constraints
\begin{align}\label{commute_constraint_algebraic}
    \gamma_{\alpha \beta}^{(\mu, \lambda)} = \gamma_{\alpha \beta}^{(\lambda, \mu)} \,, \qquad T_{\alpha \beta}^{(\mu, \lambda)} = T_{\alpha \beta}^{(\lambda, \mu)} \,.
\end{align}
This equation can be analyzed explicitly in components, by beginning with a general metric with entries $\gamma_{zz}$, $\gamma_{\bar{z} \bar{z}}$, $\gamma_{z \bar{z}} = \gamma_{\bar{z} z}$ and a general stress tensor compatible with the tracelessness constraints, and then evaluating both sides of (\ref{commute_constraint_algebraic}). We will omit the general expressions resulting from this procedure, which are not especially enlightening, and proceed to the implications of (\ref{commute_constraint_algebraic}). The constraint arising from demanding that $\gamma^{(\mu, \lambda)}_{z \bar{z}} = \gamma^{(\lambda, \mu)}_{z \bar{z}}$ is
\begin{align}
    f_2 ( \mu , y ) = 1 + \frac{4 + y^2}{4 y} \coth ( \mu ) \left( 1 - f_1 \right) \,.
\end{align}
Substituting this result and demanding that $\gamma^{(\mu, \lambda)}_{z z} = \gamma^{(\lambda, \mu)}_{z z}$ then yields
\begin{align}
    f_1 = 1 \,.
\end{align}
Therefore, $f_1 = f_2 = 1$. Using these constraints and requiring that $T^{(\mu,\lambda)}_{z \bar{z}} = T^{(\lambda, \mu)}_{z \bar{z}}$ gives
\begin{align}
    f_4 = 1 + \frac{4 y \coth ( \mu ) }{4 + y^2} \left( 1 - f_3 \right) \,,
\end{align}
and substituting this back into the equation $T^{(\mu,\lambda)}_{z z} = T^{(\lambda, \mu)}_{z z}$ yields
\begin{align}
    f_3 = 1 \,.
\end{align}
Therefore, all four of the undetermined functions must satisfy $f_i = 1$ to commute with $T\overline{T}$. We conclude that the only expressions for modified boundary conditions which are consistent with our assumptions are
\begin{align}
\begin{split}\label{root_TT_deformed_metric_bcs_later}
    \gamma_{\alpha \beta}^{(\mu)} &= \cosh ( \mu ) \gamma_{\alpha \beta}^{(0)} + \frac{\sinh ( \mu )}{\mathcal{R}^{(0)}} \widetilde{T}_{\alpha \beta}^{(0)} \,, \\
   \widetilde{T}_{\alpha \beta}^{(\mu)} &= \cosh ( \mu ) \widetilde{T}_{\alpha \beta}^{(0)} + \sinh ( \mu ) \mathcal{R}^{(0)} \gamma^{(0)}_{\alpha \beta} \,,
\end{split}
\end{align}
which are exactly the ones that we claim correspond to the root-$T\overline{T}$ deformation.

\subsection{Derivation of deformed energy levels}\label{sec:Energy_Levels}

The $T\overline{T}$ deformation of a QFT on a cylinder of radius $R$ has a well-known effect on the spectrum of the theory \cite{Zamolodchikov:2004ce,Smirnov:2016lqw,Cavaglia:2016oda}. For an energy eigenstate $|n\rangle$ with energy $E_n ( R, \lambda )$ and momentum $P_n$, the deformed energy satisfies the inviscid Burgers' equation,
\begin{align}\label{inviscid_burgers}
    \frac{\partial E_n}{\partial \lambda} = E_n \frac{\partial E_n}{\partial R} + \frac{P_n^2}{R} \,, 
\end{align}
and the momentum $P_n$ remains unchanged. If the undeformed theory is a CFT so that all of the undeformed energy levels are of the form $E_n^{(0)} \equiv E_n ( R, 0 ) = \frac{a_n}{R}$ for constants $a_n$, then (\ref{inviscid_burgers}) can be solved in closed form to obtain
\begin{align}
    E_n ( R , \lambda ) = \frac{R}{2 \lambda} \left( \sqrt{ 1 + \frac{4 \lambda E_n^{(0)} }{R} + \frac{4 \lambda^2 P_n^2}{R^2} } - 1  \right) \,.
\end{align}
The flow equation (\ref{inviscid_burgers}) can be derived by using the point-splitting definition of the local $T\overline{T}$ operator in any translation-invariant QFT and then expressing the components of the energy-momentum tensor in terms of $E_n$, $R$, $\frac{\partial E_n}{\partial R}$, and $P_n$. 

In the case of the root-$T\overline{T}$ deformation, it is not known whether one can define a local operator $\mathcal{R}$ by point-splitting. Therefore, we cannot give a rigorous derivation of a flow equation like (\ref{inviscid_burgers}) for a quantum field theory deformed by root-$T\overline{T}$. However, in the spirit of the preceding subsection, we can attempt to list the properties that such a flow equation would necessarily possess and then see whether there exists a unique differential equation satisfying these properties. We stress that this type of argument does not constitute proof that a root-$T\overline{T}$ deformed QFT exists and has a particular spectrum. It would merely show that, \emph{assuming} that the root-$T\overline{T}$ deformation is well-defined quantum-mechanically and behaves in the expected way, then there is only one possible flow equation that the spectrum could satisfy.\footnote{Note that such a differential equation for the cylinder spectrum is distinct from a flow equation for the classical Hamiltonian density, which has been studied in \cite{Tempo:2022ndz}.}

Before enumerating the desired properties of such a flow equation, it is useful to obtain a rough expectation for what the result might look like. Suppose, for the sake of argument, that there exists a local operator $\mathcal{R} ( x )$ in the spectrum of a QFT with the property:
\begin{align}\label{root_TT_optimistic}
    \langle \mathcal{R} \rangle = \sqrt{ \frac{1}{2} \langle T^{\alpha \beta} \rangle \langle T_{\alpha \beta} \rangle - \frac{1}{4} \langle T^\alpha{}_\alpha \rangle^2 } \,,
\end{align}
and consider a deformation of the action given by $\frac{\partial S}{\partial \mu} = \int d^2 x \, \mathcal{R}$. By expressing the components of the stress tensor for the theory on a cylinder of radius $R$ in terms of energies and momenta, exactly as in the derivation of the inviscid Burgers' equation for $T\overline{T}$, one would arrive at a putative root-$T\overline{T}$ flow equation
\begin{align}
    \frac{\partial E_n}{\partial \mu} = \sqrt{ \frac{1}{4} \left( E_n - R \frac{\partial E_n}{\partial R} \right)^2 - P_n^2 } \,, 
\end{align}
or equivalently
\begin{align}\label{root_TT_energy_flow}
    \left( \frac{\partial E_n}{\partial \mu} \right)^2 - \frac{1}{4} \left( E_n - R \frac{\partial E_n}{\partial R} \right)^2 + P_n^2 = 0  \,. 
\end{align}
If the initial condition for this flow is a CFT, so $E_n \sim \frac{1}{R}$ and $P_n \sim \frac{1}{R}$, then the solution is
\begin{align}\label{root_TT_deformed_CFT_energies}
    E_n ( R, \mu ) = \cosh ( \mu ) E_n ( R, 0 ) + \sinh ( \mu ) \sqrt{ \left( E_n^{(0)} \right)^2 - P_n^2 } \,, 
\end{align}
where $E_n^{(0)} = E_n ( R , 0 )$, and again the momenta $P_n$ are unaffected.

Much like the root-$T\overline{T}$ flow equation for the Lagrangian, this candidate deformation of the energy levels forms a two-parameter family of commuting deformations along with the $T\overline{T}$ flow. Beginning with a CFT, the solution for the doubly-deformed spectrum is
\begin{align}\label{doubly_deformed}
   & E_n ( R, \mu , \lambda ) \nonumber \\&= \frac{R}{2 \lambda} \left( \sqrt{ 1 + 4 \lambda \left( \cosh ( \mu ) E_n^{(0)} + \sinh ( \mu ) \sqrt{ \left( E_n^{(0)} \right)^{2} - P_n^2 } \right) + \frac{4 \lambda^2}{R^2} P_n^2 } -  1 \right) \,, 
\end{align}
where $E_n^{(0)} = E_n ( R, 0, 0)$. The spectrum (\ref{doubly_deformed}) satisfies the two commuting flow equations
\begin{align}
    \left( \frac{\partial E_n}{\partial \mu} \right)^2 - \frac{1}{4} \left( E_n - R \frac{\partial E_n}{\partial R} \right)^2 + P_n^2 = 0 \,, \qquad \frac{\partial E_n}{\partial \lambda} - E_n \frac{\partial E_n}{\partial R} - \frac{P_n^2}{R} = 0 \,, 
\end{align}
corresponding to the root-$T\overline{T}$ and $T\overline{T}$ deformations, respectively.

In section \ref{sec:AdS3RootTT}, we will give an argument that (\ref{root_TT_energy_flow}) is the correct flow equation using holography. For now, we would like to argue that this partial differential equation is the only reasonable possibility. To that end, we would like to look for the most general flow equation for a spectrum with the following properties:

\begin{enumerate}

    \item\label{marginal_flow_assumption} The flow is generated by a marginal stress tensor deformation. This means that it is a partial differential equation for $\frac{\partial E_n}{\partial \mu}$, where $\mu$ is a dimensionless parameter, which arises from a deformation of the Euclidean action by a Lorentz scalar constructed from the stress-energy tensor.

    \item The momentum $P_n$ is undeformed along the flow, so $P_n ( \mu ) = P_n ( 0 )$.

    \item\label{commuting_energy_assumption} The flow equation forms a two-parameter family of commuting flows with the inviscid Burgers' equation associated with the $T\overline{T}$ deformation.
\end{enumerate}
We will show that the only flow equation consistent with \ref{marginal_flow_assumption} - \ref{commuting_energy_assumption} is (\ref{root_TT_energy_flow}). First, it will be useful to express the possible Lorentz scalars constructed from $T_{\alpha \beta}$ in terms of energies and momenta. We work in Euclidean signature with coordinates $(x, y)$, where $x \sim x + R$ is the compact direction of the cylinder and $y$ is the Euclidean time direction. In an energy eigenstate $| n \rangle$, the components of the stress tensor are related to the energy $E_n$ and momentum $P_n$ of the state as follows:
\begin{align}
    T_{yy} = - \frac{1}{R} E_n ( R ) \,, \qquad T_{xx} = - \frac{\partial E_n ( R )}{\partial R} \,, \qquad T_{xy} = \frac{i}{R} P_n ( R ) \,.
\end{align}
Furthermore, as we have described above, any Lorentz scalar constructed from $T_{\alpha \beta}$ is a function of the two independent invariants
\begin{align}
    x_1 = T^\alpha{}_\alpha = - \frac{2 E_n}{R} - 2 \frac{\partial E_n}{\partial R} \,, \qquad x_2 = T^{\alpha \beta} T_{\alpha \beta} = \left( \frac{\partial E_n}{\partial R} \right)^2 + \frac{E_n^2}{R^2} - \frac{2 P_n^2}{R^2} \,.
\end{align}
We are interested in a deformation of the form $\partial_\mu S_E = \int d^2 x \, f ( x_1, x_2 )$. The Euclidean Lagrangian density is the Hamiltonian density, whose integral over a spatial slice gives the energy $E_n$ of a state. Therefore, this deformation of the Euclidean action can be written as
\begin{align}
    \partial_\mu E_n = \int_{0}^{R} \, dx \, f ( x_1, x_2 ) = R f ( x_1, x_2 ) \,.
\end{align}
We have assumed that this flow is generated by a marginal deformation, which means that $f ( x_1, x_2 )$ has mass dimension $2$. The function $f$ must therefore be homogeneous of degree $\frac{1}{2}$ in $x_2$ and degree $1$ in $x_1$. This allows us to scale the factor of $R$ into the arguments of $f$:
\begin{align}
    \frac{\partial E_n}{\partial \mu} = f \left( R x_1, R^2 x_2 \right) \,.
\end{align}
Next, we use the second assumption, that the momenta $P_n$ are undeformed along the flow. This means that the theory is connected to some conformal field theory by a flow along which the momenta are constant, and, therefore, the dependence of momenta on the radius is fixed to be $P_n \sim \frac{1}{R}$ as in a CFT. It is convenient to define dimensionless momenta $p_n = R P_n$. We will also re-scale $x_1$ by a factor of $-\frac{1}{2}$ for convenience and write
\begin{align}\begin{split}
    \frac{\partial E_n}{\partial \mu} &= f \left( \tilde{x}_1, \tilde{x}_2 \right) \,, \\
    \tilde{x}_1 = E_n + R \frac{\partial E_n}{\partial R} \,, &\qquad \tilde{x}_2 = R^2 \left( \frac{\partial E_n}{\partial R^2} \right)^2 + E_n^2 - \frac{2 p_n^2}{R^2} \,.
\end{split}\end{align}
This is the most general ansatz for a flow equation consistent with our first two assumptions. We will now fix the dependence of $f$ on $\tilde{x}_1, \tilde{x}_2$ using the third assumption.

Consider a two-parameter family of theories with energies $E_n ( R, \lambda, \mu )$ which satisfy the simultaneous partial differential equations
\begin{align}
    \frac{\partial E_n}{\partial \mu} = f \left( \tilde{x}_1, \tilde{x}_2 \right) \,, \qquad \frac{\partial E_n}{\partial \lambda} = E_n \frac{\partial E_n}{\partial R} + \frac{p_n^2}{R^3} \,.
\end{align}
Differentiating the $\mu$ flow equation with respect to $R$ gives
\begin{align}\label{d2_mu_R}
   \frac{\partial^2 E_n}{\partial \mu \, \partial R} \nonumber &= \frac{\partial f}{\partial \tilde{x}_1} \left( 2 \frac{\partial E_n}{\partial R} + R \frac{\partial^2 E_n}{\partial R^2} \right) \\&+ \frac{2}{R^3} \frac{\partial f}{\partial \tilde{x}_2} \left( 2 p_n^2 + R^3 \frac{\partial E_n}{\partial R} \left( E_n + R \frac{\partial E_n}{\partial R} + R^2 \frac{\partial^2 E_n}{\partial R^2 } \right) \right) \,,
\end{align}
while the derivative of the $\lambda$ flow equation with respect to $R$ is
\begin{align}
    \frac{\partial^2 E_n}{\partial \lambda \, \partial R} = E_n \frac{\partial^2 E_n}{\partial R^2} + \left( \frac{\partial E_n}{\partial R} \right)^2 - \frac{3 p_n^2}{R^4} \,.
\end{align}
We may compute the mixed second partial derivative $\frac{\partial^2 E_n}{\partial \mu \, \partial \lambda}$ in two ways. Taking a $\mu$ derivative of the expression for $\frac{\partial E_n}{\partial \lambda}$ and simplifying using (\ref{d2_mu_R}) yields
\begin{align}\label{lambda_then_mu}
    \frac{\partial^2 E_n}{\partial \mu \, \partial \lambda} &= f ( \tilde{x}_1 , \tilde{x}_2 ) \frac{\partial E_n}{\partial R} + E_n \frac{\partial f}{\partial \tilde{x}_1} \left( 2 \frac{\partial E_n}{\partial R} + R \frac{\partial^2 E_n}{\partial R^2 } \right) \nonumber \\
    & + \frac{2 E_n}{R^3} \frac{\partial f}{\partial \tilde{x}_2} \left( 2 p_n^2 + R^3 \frac{\partial E_n}{\partial R} \left( E_n + R \frac{\partial E_n}{\partial R} + R^2 \frac{\partial^2 E_n}{\partial R^2 } \right) \right) \,.
\end{align}
On the other hand, the $\lambda$ derivative of $\frac{\partial E_n}{\partial \mu}$ is
\begin{align}\label{mu_then_lambda}
    \frac{\partial^2 E_n}{\partial \lambda \, \partial \mu} &= \frac{1}{R^3} \frac{\partial f}{\partial \tilde{x}_1} \left( - 2 p_n^2 + R^4 \left( \frac{\partial E_n}{\partial R} \right)^2 + R^3 E_n \frac{\partial E_n}{\partial R} + R^4 \frac{\partial^2 E_n}{\partial R^2} \right) \nonumber \\
    &+ \frac{\partial f}{\partial \tilde{x}_2} \bigg[ 2 E_n \left( \frac{p_n^2}{R^3} + E_n \frac{\partial E_n}{\partial R} \right) \nonumber \\&+ \frac{2}{R^2} \frac{\partial E_n}{\partial R} \left( - 3 p_n^2 + R^4 \left( \frac{\partial E_n}{\partial R} \right)^2 + R^4 E_n \frac{\partial^2 E_n}{\partial R^2 } \right) \bigg] \,.
\end{align}
By our third assumption, the $\lambda$-flow must commute with the $\mu$-flow, and hence the two mixed second partial derivatives must be equal. We set (\ref{lambda_then_mu}) equal to (\ref{mu_then_lambda}) and eliminate the variables $p_n$ and $\frac{\partial E_n}{\partial R}$ in favor of $y_1, y_2$. This leads to the differential equation:
\begin{align}\label{mixed_second_pde}
    0 = \tilde{x}_1 f + \tilde{x}_1^3 \frac{\partial f}{\partial \tilde{x}_2} - 3 \tilde{x}_1 \tilde{x}_2 \frac{\partial f}{\partial \tilde{x}_2} - \tilde{x}_2 \frac{\partial f}{\partial \tilde{x}_1} + E_n \left( \tilde{x}_1 \frac{\partial f}{\partial \tilde{x}_1} + 2 \tilde{x}_2 \frac{\partial f}{\partial \tilde{x}_2} - f \right) \,.
\end{align}
The function $f$ can depend on the variables $\tilde{x}_1$ and $\tilde{x}_2$ but not on the function $E_n ( R, \lambda, \mu )$ directly. Thus in order for (\ref{mixed_second_pde}) to be consistent, the $E_n$-dependent and $E_n$-independent terms must vanish separately:
\begin{align}\begin{split}\label{two_pieces_pde}
    0 &= \tilde{x}_1 \frac{\partial f}{\partial \tilde{x}_1} + 2 \tilde{x}_2 \frac{\partial f}{\partial \tilde{x}_2} - f \,, \\
    0 &= \tilde{x}_1 f + \tilde{x}_1^3 \frac{\partial f}{\partial \tilde{x}_2} - 3 \tilde{x}_1 \tilde{x}_2 \frac{\partial f}{\partial \tilde{x}_2} - \tilde{x}_2 \frac{\partial f}{\partial \tilde{x}_1} \,.
\end{split}\end{align}
The solution to the first line of (\ref{two_pieces_pde}) is
\begin{align}
    f = \tilde{x}_1 g \left( \frac{\tilde{x}_2}{\tilde{x}_1^2} \right) \,, 
\end{align}
where $g$ is an arbitrary function. Letting $X = \frac{\tilde{x}_2}{\tilde{x}_1^2}$ and substituting this ansatz into the second line of (\ref{two_pieces_pde}) gives
\begin{align}
    \left( X - 1 \right) \left( g ( X ) + \left( 1 - 2 X \right) g' ( X ) \right) = 0 \,.
\end{align}
There are two possibilities. The first factor vanishes if $X = 1$, which is the case when
\begin{align}
    T^{\alpha \beta} T_{\alpha \beta} = \frac{1}{4} \left( T^\alpha{}_\alpha \right)^2 \,.
\end{align}
This is a trivial case in which the stress tensor for the theory is degenerate, and the two trace structures become dependent. We will discard this solution and require that $X$ is not identically equal to $1$. This leaves us with the second possibility, $g ( X ) + \left( 1 - 2 X \right) g' ( X ) = 0$, which has the solution
\begin{align}
    g ( X ) = c_1 \sqrt{ 2 X - 1 } \,, 
\end{align}
where $c_1$ is an arbitrary constant. Tracing back through the changes of variables, this corresponds to a deformation of the form:
\begin{align}
    f ( x_1, x_2 ) = c_1 \sqrt{ 2 x_2 - x_1^2 } \,.
\end{align}
Choosing the normalization to be $c_1 = \frac{1}{2}$, we conclude that the function $f$ is
\begin{align}
    f \left( x_1, x_2 \right) = \sqrt{ \frac{1}{2} x_2 - \frac{1}{4} x_1^2 } = \mathcal{R} \,, 
\end{align}
which is precisely the root-$T\overline{T}$ operator. The flow equation for the energies is
\begin{align}
    \frac{\partial E_n}{\partial \mu} = \sqrt{ \frac{1}{4} \left( E_n - R \frac{\partial E_n}{\partial R} \right)^2 - P_n^2 } \,, 
\end{align}
and taking the square of this equation recovers (\ref{root_TT_energy_flow}).

We conclude that there is only a single marginal deformation of the cylinder spectrum for a $2d$ quantum field theory, which is constructed from the energy-momentum tensor and commutes with the irrelevant $T\overline{T}$ flow. This unique deformation is the one that corresponds to the combination of stress tensors, which appears in the classical root-$T\overline{T}$ deformation. We reiterate that this does not represent proof that the root-$T\overline{T}$ operator is necessarily well-defined at the quantum level. However, if there exists \emph{any} deformation of the quantum theory with the properties that we listed, it must lead to exactly the flow equation that one would have na\"ively guessed would correspond to the root-$T\overline{T}$ deformation, as we did around (\ref{root_TT_optimistic}).

\section{AdS$_3$ gravity with root-$T\overline{T}$ deformed boundary conditions}
\label{sec:AdS3RootTT}

In this section, we aim to show that the root-$T\overline{T}$ deformed boundary conditions derived in section \ref{sec:boundary_condition_derivation} are compatible with our proposed flow equation for the spectrum in section \ref{sec:Energy_Levels}. To do this, we will compute the mass (or energy) of a spacetime with root-$T\overline{T}$ deformed boundary conditions and compare this deformed mass to its undeformed value.

It is well-known that the notion of mass is subtle in a generally covariant theory, and there are many definitions of the total mass of spacetime that are applicable in different contexts. In our case, since we are interested in asymptotically $\mathrm{AdS}_3$ spacetimes, it will be most convenient to define the spacetime mass as the integral of the quasi-local Brown-York stress tensor \cite{PhysRevD.47.1407}. In a $d$-dimensional spacetime $M_d$, this mass integral is given by
\begin{align}\label{spacetime_mass}
    M = \int_{\Sigma} d^{d-1} x \, \sqrt{\sigma} \, u^\mu T_{\mu \nu} \xi^\nu \,, 
\end{align}
where $\Sigma$ is a spacelike surface in the boundary $\partial M_d$ with metric $\sigma_{\alpha \beta}$, $u^\mu$ is a timelike unit normal, and $\xi^\nu$ is the Killing vector associated with time translations. Our strategy will be to compute the mass (\ref{spacetime_mass}) by choosing a convenient coordinate system generated by a field-dependent change of variables that implements our root-$T\overline{T}$ deformed boundary conditions. Such field-dependent diffeomorphisms have also appeared in various works in the context of the ordinary $T\overline{T}$ deformation \cite{Conti:2018tca,Guica:2019nzm,Conti:2022egv}. 

It would be interesting to study the mass of $\mathrm{AdS}_3$ spacetimes subject to modified boundary conditions using a more general prescription such as the covariant phase space formalism \cite{ Iyer:1994ys,Wald:1999wa}. The result of a mass calculation in this formalism is guaranteed to agree with (\ref{spacetime_mass}), but because this machinery maintains covariance, it may be possible to obtain mass flow equations associated with $T\overline{T}$ and root-$T\overline{T}$ deformations (or even more general stress tensor deformations) without resorting to a field-dependent diffeomorphism. 

We will also obtain the corresponding root-$T\overline{T}$ deformed boundary conditions in the Chern-Simons description of AdS$_3$ gravity. In this formalism, the definition of the deformed spacetime mass is not immediately obvious. As we will review around (\ref{eq:boundaryundeformedCS}), in the undeformed theory with conventional boundary conditions, it is straightforward to show that the mass (\ref{spacetime_mass}) is equal to the value of the Chern-Simons boundary term which imposes the appropriate boundary conditions. We will see by explicit computation that this remains true when this boundary term is modified to the one that imposes the root-$T\overline{T}$ deformed boundary conditions. This provides evidence that the Chern-Simons boundary term continues to yield the spacetime mass even with modified boundary conditions, which one might attempt to prove more generally by a computation in the canonical formulation.

\subsection{Metric formalism}

First, we will briefly review the salient details in AdS$_3$ gravity to set the stage for the deformed energy spectrum computation.\footnote{For useful reviews on the AdS$_3$/CFT$_2$ correspondence, see \cite{Kraus:2006wn,Donnay:2016iyk,Compere:2018aar}.} Pure three-dimensional general relativity contains no local degrees of freedom, but is nontrivial enough to have black hole solutions and is a useful arena to study interesting phenomena in a controllable manner. A general solution of AdS$_3$ gravity can be written in the Fefferman-Graham expansion \cite{fefferman1985conformal}
\begin{equation}
\label{eq:AdS3Background}
ds^2 = \frac{\ell^2 d\rho^2}{4 \rho^2} + \left( \frac{ g^{(0)}_{\alpha \beta} (x^\alpha)}{\rho} + g^{(2)}_{\alpha \beta} (x^\alpha) + \rho g^{(4)}_{\alpha \beta} (x^\alpha)  \right) dx^\alpha \, dx^\beta\,,
\end{equation}
which terminates at second order \cite{Skenderis:1999nb} and where $\rho = 0$ is the AdS$_3$ boundary. The AdS$_3$ radius is $\ell$ and the three-dimensional Einstein equations determine $g^{(4)}_{\alpha \beta}$ in terms of the other two Fefferman-Graham expansion coefficients as
\begin{equation}
\label{eq:g4}
    g^{(4)}_{\alpha \beta} = \frac{1}{4} g^{(2)}_{\alpha \gamma} g^{(0) \gamma \delta} g^{(2)}_{\delta \beta}\,.
\end{equation}
Asymptotically AdS$_3$ solutions realize two copies of the Virasoro algebra, which are generated by Brown-Henneaux diffeomorphisms \cite{Brown:1986nw} that preserve the leading asymptotics of the metric (\ref{eq:AdS3Background}). Such diffeomorphisms correspond to conformal transformations in the $2d$ boundary theory. From the AdS/CFT dictionary \cite{Balasubramanian:1999re,deHaro:2000vlm}, the Fefferman-Graham quantity $g^{(2)}_{\alpha \beta}$ is proportional to the expectation value of the boundary CFT stress tensor:
\begin{equation}
g^{(2)}_{\alpha \beta} = 8 \pi G \ell \left( T_{\alpha \beta} - g^{(0)}_{\alpha \beta} T^{ \alpha}_\alpha \right) \equiv 8 \pi G \ell \widehat{T}_{\alpha \beta} \,, 
\end{equation}
and $g^{(0)}_{\alpha \beta}$ is the metric on the boundary where the dual CFT lives. To derive the energy spectrum of this background \eqref{eq:AdS3Background} with the root-$TT$ deformed boundary conditions, we borrow some of the key methods developed to study holographic aspects of the double-trace $T\overline{T}$ deformation in the metric formalism \cite{Guica:2019nzm} and Chern-Simons formalism \cite{He:2020hhm} at large-$N$. See appendix \ref{sec:AdS3TTApp} for a review of these methods. As a consequence of our analysis, we will also find that the bulk spacetime exhibits superluminal propagation for one sign of the root-$T\overline{T}$ deformation parameter $\mu$, which is also the case for the bad-sign $T\overline{T}$ deformation.

\emph{Root-$T\overline{T}$ deformed theory}

In section \ref{sec:boundary_condition_derivation}, we argued that the root-$T\overline{T}$ deformed boundary metric and stress tensor are
\begin{align}\begin{split}\label{root_TT_deformed_bcs_repeat}
    \gamma_{\alpha \beta}^{(\mu)} = \cosh ( \mu ) \gamma_{\alpha \beta}^{(0)} + \frac{\sinh ( \mu )}{\mathcal{R}^{(0)}} \widetilde{T}_{\alpha \beta}^{(0)} \,, \quad
    \widetilde{T}_{\alpha \beta}^{(\mu)} = \cosh ( \mu )  \widetilde{T}_{\alpha \beta}^{(0)} + \sinh ( \mu ) \mathcal{R}^{(0)} \gamma^{(0)}_{\alpha \beta} \,.
\end{split}\end{align}
The strategy we follow here is motivated by the analysis of $T\overline{T}$-deformed AdS$_3$/CFT$_2$ in \cite{Guica:2019nzm}. As in the holographic analysis of the $T\overline{T}$ deformation, we will make two assumptions about the root-$T\overline{T}$ flow. The first is that this deformation is smooth and, therefore, preserves the boundary theory's degeneracy of states; for a black hole solution, this corresponds to the statement that the black hole horizon area is unchanged. The second assumption is that the deformation does not affect the momentum quantum number $P_n$ in the boundary field theory, which is quantized in units of $\frac{1}{R}$, where $R$ is the cylinder radius. 
Thus, we will equate the deformed and undeformed areas and angular momenta. These assumptions imply the root-deformed energy spectrum, which was derived by consistency conditions in section \ref{sec:Energy_Levels}. Unlike the $T\overline{T}$ deformation, the trace of the root-$T\overline{T}$ theory does not flow, as expected for a classically marginal deformation. 

We will now focus on the case of a Ba\~nados geometry  \cite{Banados:1998gg}, which is parameterized by two quantities $\mathcal{L} ( u )$ and $\bar{\mathcal{L}} ( v )$. A Ba\~nados geometry's Fefferman-Graham quantities are defined:
\begin{equation}
    \begin{aligned}
    \label{eq:FG1}
    g^{(0)}_{\alpha \beta} dx^\alpha \, dx^\beta &= du \, dv\,, \\
        g^{(2)}_{\alpha \beta} dx^\alpha dx^\beta &= \mathcal{L}(u) du^2 + \bar{\mathcal{L}}(v) dv^2\,, \\
        g^{(4)}_{\alpha \beta} dx^\alpha dx^\beta &= \mathcal{L}(u) \bar{\mathcal{L}}(v) dudv
    \end{aligned}
\end{equation}
implying that the metric \eqref{eq:AdS3Background} becomes
\begin{align}
\label{eq:BanadosMetric}
    ds^2 &= \frac{\ell^2 d\rho^2}{4\rho^2} + \frac{du \, dv}{\rho} + \mathcal{L}(u) du^2 + \bar{\mathcal{L}}(v) dv^2 + \rho \mathcal{L}(u) \bar{\mathcal{L}}(v) \, du \, dv \,. 
\end{align}
The root-$T\overline{T}$ deformed boundary metric and stress tensor given in \eqref{root_TT_deformed_bcs_repeat} are therefore
\begin{equation}
\begin{aligned}
\label{eq:root-TTstress}
\gamma^{(\mu)}_{\alpha \beta} &= (\cosh  \mu ) g_{\alpha \beta}^{(0)} + \frac{\sinh \mu }{2\sqrt{\mathcal{L}(u) \bar{\mathcal{L}}(v) }}  g^{(2)}_{\alpha \beta} \\& =   \frac{1}{2} \left( \begin{array}{cc}
  \sqrt{\frac{\mathcal{L}(u)}{\bar{\mathcal{L}}(v)}} \sinh \mu  & \quad  \cosh \mu  \\
 \cosh \mu  & \quad   \sqrt{\frac{\bar{\mathcal{L}}(v)}{\mathcal{L}(u)}} \sinh \mu
    \end{array}\right)\,, 
\\ \widetilde{T}^{(\mu)}_{\alpha \beta} &=
 \frac{\cosh \mu}{2} g^{(2)}_{\alpha \beta} +\sqrt{\mathcal{L}(u) \bar{\mathcal{L}}(v)  } (\sinh \mu) g^{(0)}_{\alpha \beta} \\&=  \frac{1}{2} \left( \begin{array}{cc}
      \mathcal{L}(u) \cosh \mu   & \quad \sqrt{\mathcal{L}(u) \bar{\mathcal{L}}(v)}  \sinh \mu \\
    \sqrt{\mathcal{L}(u) \bar{\mathcal{L}}(v)} \sinh \mu     & \quad \bar{\mathcal{L}}(v) \cosh \mu
    \end{array} \right)\,,
 \end{aligned}
\end{equation}
where we work in conventions such that $4\pi G \ell = 1$ and substituted \eqref{eq:FG1} into \eqref{root_TT_deformed_bcs_repeat}. We also used $g^{(2)}_{\alpha \beta} =2 T_{\alpha \beta}$ and computed the operator $\displaystyle{\mathcal{R} = \sqrt{\mathcal{L}(u) \bar{\mathcal{L}}(v)}}$. We define new coordinates $(U, V)$ by
\begin{equation}
\begin{aligned}
\label{eq:rootTTcoordtransformation}
dU &= \left( \cosh \frac{\mu}{2} \right) du + \sqrt{\frac{\bar{\mathcal{L}}(v)}{\mathcal{L}(u)}} \left( \sinh \frac{\mu}{2} \right) dv\,, \\ dV &= \left( \cosh \frac{\mu}{2} \right) dv + \sqrt{\frac{\mathcal{L}(u)}{\bar{\mathcal{L}}(v)}} \left( \sinh \frac{\mu}{2} \right) du\,.  
\end{aligned}
\end{equation}
which has the property that the metric, when written in these new variables, returns to the standard form:
\begin{equation}
\gamma^{(\mu)}_{\alpha \beta} dx^\alpha \, dx^\beta = dU \, dV \,.
\end{equation}
Expressing \eqref{eq:rootTTcoordtransformation} in matrix notation, we may write the field-dependent coordinate transformation and its inverse as
\begin{equation}
\begin{aligned}
\label{eq:inverserootTT}
\left( \begin{array} {c}
dU\\dV
\end{array}
\right) &= \left( \begin{array}{cc}
    \cosh \frac{\mu}{2}   & \quad \sqrt{\frac{\bar{\mathcal{L}} (v) }{\mathcal{L}(u)  }} \sinh \frac{\mu}{2}  \\
   \sqrt{\frac{\mathcal{L}(u) }{\bar{\mathcal{L}}(v)  }} \sinh \frac{\mu}{2}   & \quad \cosh \frac{\mu}{2} 
    \end{array} \right) \left( \begin{array} {c}
du\\dv
\end{array}
\right)\,, \\ 
\left( \begin{array} {c}
du\\dv
\end{array}
\right) &= \left( \begin{array}{cc}
    \cosh \frac{\mu}{2}   & \quad -\sqrt{\frac{\bar{\mathcal{L}}(v) }{\mathcal{L}(u) }} \sinh \frac{\mu}{2}  \\
  - \sqrt{\frac{\mathcal{L}(u)}{\bar{\mathcal{L}}(v)}} \sinh \frac{\mu}{2}   & \quad \cosh \frac{\mu}{2} 
    \end{array} \right) \left( \begin{array} {c}
dU\\dV
\end{array}
\right).
\end{aligned}
\end{equation}
For black hole solutions with constant $(\mathcal{L}(u), \bar{\mathcal{L}}(v)) \equiv \left( \mathcal{L}_\mu, \bar{\mathcal{L}}_\mu\right)$, the Fefferman-Graham quantities in the $(U, V)$ coordinates are
\begin{equation}
\begin{aligned}
\label{eq:FGRTT}
g^{(0)}_{\alpha \beta} dx^\alpha \, dx^\beta&= du \, dv
\\&= -\frac{1}{2} \sinh \mu \left( \sqrt{\frac{\mathcal{L}_\mu}{\bar{\mathcal{L}}_\mu}} dU^2 + \sqrt{\frac{\bar{\mathcal{L}}_\mu}{\mathcal{L}_\mu}} dV^2 \right) + \cosh \mu dU \, dV\,,
\\ 
g^{(2)}_{\alpha \beta} dx^\alpha \, dx^\beta&= \mathcal{L}_\mu du^2 + \bar{\mathcal{L}}_\mu dv^2 \\&= \cosh \mu \left( \mathcal{L}_\mu dU^2 + \bar{\mathcal{L}}_\mu dV^2 \right) - 2 \sqrt{\mathcal{L}_\mu \bar{\mathcal{L}}_\mu} \sinh \mu dU \, dV\,,
\\
g^{(4)}_{\alpha \beta} dx^\alpha \, dx^\beta &= \mathcal{L}_\mu \bar{\mathcal{L}}_\mu  du \, dv \\&= \mathcal{L}_\mu \bar{\mathcal{L}}_\mu \left( -\frac{1}{2} \sinh \mu \left( \sqrt{\frac{\mathcal{L}_\mu}{\bar{\mathcal{L}}_\mu}} dU^2 + \sqrt{\frac{\bar{\mathcal{L}}_\mu}{\mathcal{L}_\mu}} dV^2 \right) + \cosh \mu dU \, dV \right)\,,
\end{aligned}
\end{equation}
which yields the deformed metric from substituting \eqref{eq:FGRTT} into \eqref{eq:AdS3Background}
\begin{equation}
\begin{aligned}
\label{eq:deformedrootFGmetric}
ds^2 &= \frac{\ell^2 d\rho^2}{4\rho^2} + \left( - \frac{1}{2 \rho} \sqrt{ \frac{\mathcal{L}_\mu}{\bar{\mathcal{L}}_\mu}   } \sinh \mu + \mathcal{L}_\mu \cosh \mu - \frac{\rho}{2} \mathcal{L}_\mu \sqrt{\mathcal{L}_\mu \bar{\mathcal{L}}_\mu} \sinh \mu \right) dU^2 \\&+ \left(  - \frac{1}{2 \rho} \sqrt{\frac{\bar{\mathcal{L}}_\mu}{\mathcal{L}_\mu}} \sinh \mu + \bar{\mathcal{L}}_\mu \cosh \mu - \frac{\rho}{2} \bar{\mathcal{L}}_\mu \sqrt{\mathcal{L}_\mu \bar{\mathcal{L}}_\mu} \sinh \mu\right) dV^2 \\&+ \left( \frac{1}{\rho} \cosh \mu -2 \sqrt{\mathcal{L}_\mu \bar{\mathcal{L}}_\mu} \sinh \mu + \rho \mathcal{L}_\mu \bar{\mathcal{L}}_\mu \cosh \mu \right) dU \, dV \,.
\end{aligned}
\end{equation}
%
The metric \eqref{eq:AdS3Background} in terms of these Fefferman-Graham quantities \eqref{eq:FGRTT} at the event horizon $\rho_h = \left(\mathcal{L}_\mu \bar{\mathcal{L}}_\mu \right)^{-\frac{1}{2}}$ is
\begin{equation}
\begin{aligned}
\hspace{-10pt}
\label{eq:metricdeformedrootTT}
ds^2 \big\vert_{\rho = \rho_h} &= \frac{\ell^2 \mathcal{L}_\mu \bar{\mathcal{L}}_\mu}{4} d\rho^2 \\&+ e^{-\mu} \left( \left( \sqrt{\mathcal{L}_\mu} + \sqrt{\bar{\mathcal{L}}_\mu } \right)^2 d\phi^2 + \left( \sqrt{\mathcal{L}_\mu} - \sqrt{\bar{\mathcal{L}}_\mu } \right)^2 dT^2 + 2 \left( \mathcal{L}_\mu - \bar{\mathcal{L}}_\mu \right) dT \, d\phi \right)\,,
\end{aligned}
\end{equation}
where $(U, V) = (\phi + T, \phi - T)$. The undeformed and deformed event horizon areas are read off from \eqref{eq:metricdeformedrootTT}
\begin{equation}
\begin{aligned}
A^{(0)} &= \int^R_0 d\phi\sqrt{g_{\phi \phi}} = R\left( \sqrt{\mathcal{L}_0} + \sqrt{\bar{\mathcal{L}}_0} \right)\,, \\ A^{(\mu)} &= \int^R_0 d\phi\sqrt{g_{\phi \phi}}|_{\rho_h} = R e^{-\frac{\mu}{2}} \left( \sqrt{\mathcal{L}_\mu} + \sqrt{\bar{\mathcal{L}}_\mu} \right)\,. 
\end{aligned}
\end{equation}
Now, to extract the deformed energy and angular momentum. Using \eqref{eq:root-TTstress} and \eqref{eq:FGRTT}, we write the components of the stress tensor
\begin{equation}
\label{eq:ttrootcomponentsenergymomentum}
T^{(\mu)}_{\alpha \beta} \, dx^\alpha \, dx^\beta = \frac{1}{2} (\mathcal{L}_\mu dU^2 + \bar{\mathcal{L}}_\mu dV^2) 
= \frac{1}{2} ( \mathcal{L}_\mu +\bar{\mathcal{L}}_\mu) (dT^2 + d\phi^2) + (\mathcal{L}_\mu - \bar{\mathcal{L}}_\mu) \, d\phi \, dT \,.
\end{equation}
Restoring factors of $4 \pi G \ell$, the deformed energy and angular momentum from \eqref{eq:ttrootcomponentsenergymomentum} are
\begin{equation}
\label{eq:E&J}
E_\mu = \int^R_0 d\phi \, T^{(\mu)}_{TT} = \frac{R}{8\pi G \ell} (\mathcal{L}_\mu + \bar{\mathcal{L}}_\mu), \quad J_\mu = \int^R_0 d\phi T^{(\mu)}_{T\phi} = \frac{R}{8 \pi G \ell}(\mathcal{L}_\mu - \bar{\mathcal{L}}_\mu)\,.
\end{equation}
The root-$T\overline{T}$ deformed energy \eqref{eq:E&J} is a simple sum $\mathcal{L}_\mu +\bar{\mathcal{L}}_\mu$, reminiscent of a CFT's energy, which is a sign that the root-$T\overline{T}$ deformed theory remains a CFT. This simplicity of energy ceases to exist for the $T\overline{T}$ deformation due to the deformed theory being non-conformal. In the $T\overline{T}$ deformation of AdS$_3$/CFT$_2$, the energy is
\begin{equation}
\label{eq:EnergyTT}
E_\lambda = \frac{R}{8\pi G \ell} \frac{\mathcal{L}_\lambda + \bar{\mathcal{L}}_\lambda - 2 \rho_c \mathcal{L}_\lambda \bar{\mathcal{L}}_\lambda }{1-\rho_c^2 \mathcal{L}_\lambda \bar{\mathcal{L}}_\lambda  } \,, 
\end{equation}
with $(\mathcal{L}_\lambda, \bar{\mathcal{L}}_\lambda)$ defined in \cite{Guica:2019nzm} and \eqref{eq:solsTT1}. The next ingredients are the areas and angular momenta, which obey
\begin{equation}
\label{eq:areaangularmomentumTTroot}
\sqrt{\mathcal{L}_0} + \sqrt{\bar{\mathcal{L}}_0} = e^{- \frac{\mu}{2}} \left( \sqrt{\mathcal{L}_\mu} + \sqrt{\bar{\mathcal{L}}_\mu} \right), \quad \mathcal{L}_0 - \bar{\mathcal{L}}_0 = \mathcal{L}_\mu - \bar{\mathcal{L}}_\mu \,, 
\end{equation}
and the solutions of \eqref{eq:areaangularmomentumTTroot} are
\begin{equation}
\label{eq:rootTTLLb}
\mathcal{L}_\mu = \left( \sqrt{\mathcal{L}_0} \cosh \frac{\mu}{2} + \sqrt{\bar{\mathcal{L}}_0} \sinh \frac{\mu}{2} \right)^2\,, \quad \bar{\mathcal{L}}_\mu = \left( \sqrt{\bar{\mathcal{L}}_0} \cosh \frac{\mu}{2} + \sqrt{\mathcal{L}_0} \sinh \frac{\mu}{2} \right)^2\,.
\end{equation}
One can invert \eqref{eq:rootTTLLb} to find that
\begin{equation}
\mathcal{L}_0 = \left( \sqrt{\mathcal{L}_\mu} \cosh \frac{\mu}{2} - \sqrt{\bar{\mathcal{L}}_\mu} \sinh \frac{\mu}{2} \right)^2\,, \quad \bar{\mathcal{L}}_0 = \left( \sqrt{\bar{\mathcal{L}}_\mu} \cosh \frac{\mu}{2} -\sqrt{\mathcal{L}_\mu}   \sinh \frac{\mu}{2} \right)^2\,.
\end{equation}
Using the relations (\ref{eq:rootTTLLb}) to express the deformed energy in terms of $\mathcal{L}_0$, $\bar{\mathcal{L}}_0$ then yields
\begin{equation}
\begin{aligned}
\label{eq:energyrootTT}
E_\mu = \frac{R}{8 \pi G \ell} \left( (\mathcal{L}_0 + \bar{\mathcal{L}}_0 ) \cosh \mu+ 2 \sqrt{\mathcal{L}_0 \bar{\mathcal{L}}_0 }\sinh \mu \right)\,.
\end{aligned}
\end{equation}
We can rewrite \eqref{eq:energyrootTT} in terms of the undeformed energy and angular momentum by recalling that 
\begin{equation}
\begin{aligned}
E_0 &= \frac{R}{8 \pi G \ell} \left( \mathcal{L}_0 + \bar{\mathcal{L}}_0 \right)\,, \quad J_0 = \frac{R}{8 \pi G \ell} \left( \mathcal{L}_0 - \bar{\mathcal{L}}_0 \right) \,, \\
\mathcal{L}_0&= \frac{4\pi G \ell}{R} \left( E_0 +J_0 \right)\,, \quad \bar{\mathcal{L}}_0= \frac{4\pi G \ell}{R} \left( E_0 -J_0 \right) \,.
\end{aligned}
\end{equation}
After identifying the bulk angular momentum $J_0$ with the CFT momentum $P_0$, this gives the same energy spectrum (\ref{root_TT_deformed_CFT_energies}) we found from consistency conditions, namely
\begin{equation}
\label{eq:root-TT-energyspectrum}
E_\mu = E_0 \cosh \mu+  \sqrt{E_0^2 - P_0^2 }\sinh \mu\,.
\end{equation}

\emph{Propagation speed}

In the case of the usual $T\overline{T}$ deformation, there is a sharp distinction between the two signs of the deformation parameter $\lambda$. In our conventions, $\lambda > 0$ corresponds to the ``good sign'' of the flow. With this choice of sign, so long as $\lambda$ is not too large, all of the energy eigenvalues in a $T\overline{T}$-deformed CFT remain real. For the ``bad sign'' $\lambda < 0$, however, all but finitely many of the energies in the deformed theory become complex.\footnote{In some cases, these complex energies can be removed by performing multiple $T\overline{T}$ deformations in a row. For instance, one can deform a pair of CFTs by the bad sign of $\lambda$ and then subsequently deform the tensor product of these theories by a good-sign flow to cure the spectrum \cite{Ferko:2022dpg}.} This signals a pathology in the bad-sign-deformed theory, which appears robust to the type of $T\overline{T}$-deformation one uses. For instance, a single-trace $T\overline{T}$ deformation with the bad sign corresponds to a bulk dual with closed timelike curves \cite{Chakraborty:2019mdf}. The conventional double-trace $T\overline{T}$ deformation, which is the version considered in this work, with the bad choice of sign is dual to a bulk spacetime which exhibits superluminal propagation \cite{McGough:2016lol,Hartman:2018tkw}.

It is natural to ask whether the root-$T\overline{T}$ deformation has a similar pathology for one choice of the sign. Such a pathology would not be visible at the level of the formula (\ref{eq:root-TT-energyspectrum}) for the root-$T\overline{T}$ deformed spectrum, which appears to yield real energies for either sign of $\mu$. However, we will now show that the sign choice $\mu < 0$ leads to a bulk spacetime, which allows superluminal propagation. This suggests that, as with $T\overline{T}$, only the positive sign of the root-$T\overline{T}$ flow parameter may lead to a sensible deformed theory.

To demonstrate this superluminal propagation, we begin with a diagonal stress tensor $\widetilde{T}_{\alpha \beta} (0) = \operatorname{diag}(\widetilde{T}_{tt}(0), \widetilde{T}_{xx}(0))$ on a two-dimensional space equipped with $(t, x)$ coordinates and flat metric $\eta_{\alpha \beta} = \operatorname{diag} (-1, 1)$. The boundary deformed metric \eqref{root_TT_deformed_bcs_repeat} in this setting is 
\begin{equation}
\begin{aligned}
ds^2 &= \left( - dt^2 + dx^2 \right) \cosh \mu + \frac{\sinh \mu}{\mathcal{R}^{(0)}} \left( \widetilde{T}_{tt}(0) dt^2 + \widetilde{T}_{xx}(0)dx^2 \right)
\\&=-  e^{-\mu } dt^2 + e^\mu dx^2\,,
\end{aligned}
\end{equation}
where we have used $ \widetilde{T}_{tt} = \widetilde{T}_{xx} = \mathcal{R}^{(0)}$ and is straightforward to show from
\begin{equation}
    \begin{aligned}
        \widetilde{T}_{tt} &= T_{tt} - \frac{1}{2} \eta_{tt} T^\rho{}_\rho =   T_{tt} + \frac{1}{2} (-T_{tt} + T_{xx}) = \frac{1}{2} (T_{tt} + T_{xx}), \\
        \widetilde{T}_{xx} &= T_{xx} - \frac{1}{2} \eta_{xx} T^\rho_\rho = T_{xx} - \frac{1}{2} (-T_{tt} + T_{xx}) = \frac{1}{2} ( T_{tt} + T_{xx}),\\
        \widetilde{T}_{tt} + \widetilde{T}_{xx} &= T_{tt} + T_{xx} \,,  \\
       \mathcal{R}^{(0)} &= \sqrt{ \frac{1}{2} T^{(0) \alpha \beta} T^{(0)}_{\alpha \beta} - \frac{1}{4} (T^{(0)\alpha}{}_\alpha)^2 } \\&=  \sqrt{ \frac{1}{2} (T_{tt}^2 + T_{xx}^2) - \frac{1}{4} (-T_{tt} + T_{xx})^2 }\\&= \frac{T_{tt} + T_{xx}}{2}\,.
   \end{aligned}
\end{equation}

Null geodesics obey $ds^2 = 0$ which have the following propagation speed 
\begin{equation}
v=  e^{-\mu}\,, 
\end{equation}
and in particular we see that $v > 1$ if $\mu <0$ while $0 < v < 1$ if $\mu >0$. This confirms that the bulk supports superluminal propagation for the negative sign of the root-$T\overline{T}$ deformation parameter.

This result might have been anticipated because the root-$T\overline{T}$ deformation is closely connected to the Modified Maxwell or ModMax theory of electrodynamics in four spacetime dimensions. In particular, the $4d$ root-$T\overline{T}$ deformation of the free Maxwell theory yields the ModMax theory \cite{Babaei-Aghbolagh:2022uij, Ferko:2023ruw}, and the dimensional reduction of this theory to two spacetime dimensions is the Modified Scalar theory which is obtained from a root-$T\overline{T}$ flow of free scalars \cite{Conti:2022egv}. It was already pointed out in \cite{Bandos:2020jsw} that the $4d$ ModMax theory also allows for superluminal propagation when $\gamma < 0$, which corresponds to $\mu < 0$ in our notation. This gives another reason to suspect that the root-$T\overline{T}$ deformation may be ill-behaved for $\mu < 0$.

\subsection{Chern-Simons formalism}

The connections associated to the  Ba\~{n}ados geometry \eqref{eq:BanadosMetric} are
\begin{equation}
\begin{aligned}
\label{eq:connectionsundeformed}
A& =-\frac{1}{2\rho} L_0 d\rho + \frac{1}{\ell} \left(- \sqrt{\rho} \mathcal{L}_0 L_{-1} + \frac{1}{\sqrt{\rho}} L_{1}  \right)du =    \left(\begin{array}{ll}
-\frac{d\rho}{4\rho} & \quad -\frac{\sqrt{\rho} \mathcal{L}_0 du }{\ell }\\
-\frac{du}{\ell \sqrt{\rho}  } & \quad \frac{d\rho}{4\rho}
\end{array}\right)   \,,\\
\bar{A} &= \frac{1}{2\rho} L_0 d\rho + \frac{1}{\ell} \left( \frac{1}{\sqrt{\rho}} L_{-1}-\sqrt{\rho} \bar{\mathcal{L}}_0 L_1 \right)dv =    \left(\begin{array}{ll}
\frac{d\rho}{4\rho} & \quad \frac{dv}{\ell \sqrt{\rho} } \\
 \frac{\sqrt{\rho} \bar{\mathcal{L}}_0  dv}{\ell } & \quad -\frac{d\rho}{4\rho}
\end{array}\right) \,.
\end{aligned}
\end{equation}

In radial gauge, we may extract the radial dependence from the bulk connections as
\begin{equation}
\begin{aligned}
A(\rho, u) &= b^{-1}(\rho) (d + a(u)) b(\rho), \quad \bar{A}(\rho, v) = b (d + \bar{a}(v)) b^{-1}(\rho)\,, \\ b(\rho) &= e^{- \frac{1}{2} L_0 \ln \rho} =  \left( \begin{array}{cc}
 \rho^{-\frac{1}{4}}& \quad 0 \\
0 & \quad \rho^{\frac{1}{4}}
\end{array}
\right) \,. 
\end{aligned}
\end{equation}
where the boundary connections are

\begin{equation}\label{CS_boundary_banados}
    a(u) =\frac{1}{\ell} \left(-  \mathcal{L}_0(u) L_{-1} + L_{1}  \right)du, \quad \bar{a}(v) =  \frac{1}{\ell} \left( L_{-1} -  \bar{\mathcal{L}}_0(v) L_1 \right)dv \,.
\end{equation}
To compare more easily with our metric formalism analysis, we work in the same temporal and periodic coordinates\footnote{We distinguish between the undeformed coordinates $(t, \varphi)$ and the deformed coordinates $(T, \phi)$.}
\begin{equation}
\label{eq:tphivars}
t= \frac{1}{2} \left( u+v \right)\,, \quad \varphi = \frac{1}{2} \left( u-v \right)\,, \quad \varphi \sim \varphi + R\,.
\end{equation}
In these variables \eqref{eq:tphivars}, the chiral boundary conditions are $A_t = A_\varphi\,, \bar{A}_t = - \bar{A}_\varphi$. To have a variational principle that realizes these chiral boundary conditions, we add the following boundary term to the total Chern-Simons action:
\begin{equation}
\label{eq:boundaryundeformedCS}
S =  S_{\text{CS}}[A] - S_{\text{CS}}[\bar{A}] + \frac{\ell}{16 \pi G} \int_{\partial M} dt \, d\varphi \, \operatorname{Tr} \left(A_\varphi^2 + \bar{A}_\varphi^2\right)\,.
\end{equation}
In the undeformed theory with conventional boundary conditions, it can be shown that the boundary term in \eqref{eq:boundaryundeformedCS} that imposes the chiral boundary conditions is related to the mass of the bulk spacetime, which can be defined via other means in the metric formalism \cite{Regge:1974zd, Coussaert:1995zp}. In our case, this mass is simply the black hole's total energy. We can see this explicitly by substituting \eqref{eq:connectionsundeformed} into the boundary action, which yields:
\begin{equation}
S_{\text{bdry}} = \frac{\ell}{16 \pi G} \int_{\partial M_3} dt \,  d\varphi \, \operatorname{Tr} \left(A_\varphi^2 + \bar{A}_\varphi^2\right) =  \frac{1}{8\pi G \ell} \int_{\partial M_3} dt \, d\varphi \left( \mathcal{L}_0 + \bar{\mathcal{L}}_0 \right) \,, 
\end{equation}
since the undeformed energy is
\begin{equation}
E_0 = \frac{R}{8 \pi G \ell} \left( \mathcal{L}_0 + \bar{\mathcal{L}}_0 \right)\,.
\end{equation}
It is not obvious, without performing a computation in the canonical formulation, that the Chern-Simons boundary action will continue to yield the mass of the bulk spacetime in the presence of a boundary deformation. However, we will find that this is indeed the case when the boundary is deformed by the root-$T\overline{T}$ operator.

Next, we will understand the mixed boundary conditions imposed by the root-$T\overline{T}$ deformation in Chern-Simons variables. There are two equivalent approaches that one might use to find the deformed boundary conditions. One strategy to find the deformed Chern-Simons connections is to use the field-dependent coordinate transformation \eqref{eq:inverserootTT}. In describing this approach, we will work with an explicit choice of coordinate system.

The second method is to work with a covariant expansion of the boundary connections in terms of vielbeins $e_i^a$ and their dual expectation values $f_i^a$, which are related to the stress tensor. One can then work out the mixing of sources and expectation values in Chern-Simons variables, either by imposing consistency conditions of the kind discussed in section \ref{sec:boundary_condition_derivation}, or by taking the results for $\gamma_{\alpha \beta}^{(\mu)}$ and $\widetilde{T}_{\alpha \beta}^{(\mu)}$ in (\ref{root_TT_deformed_metric_bcs_later}) as given and then finding the modification in Chern-Simons variables which reproduce these results. 

\emph{Root-$T\overline{T}$ deformed Chern-Simons: coordinate approach}

The first way to find the deformed Chern-Simons connections is to use the field-dependent coordinate transformation \eqref{eq:inverserootTT}. In describing this approach, we will work directly with boundary coordinates $U, V$ for the deformed theory and $u, v$ for the undeformed theory. Transforming the connections $A$ and $\bar{A}$ using this change of coordinates yields
\begin{equation}
\begin{aligned}
A(\mu)&= -\frac{1}{2\rho} L_0 d\rho + \frac{1}{\ell} \left( -\sqrt{\rho} \mathcal{L}_\mu  L_{-1} + \frac{1}{\sqrt{\rho}} L_1 \right) \left( \cosh \frac{\mu}{2} dU - \sqrt{\frac{  \bar{\mathcal{L}}_\mu  }{\mathcal{L}_\mu}} \sinh \frac{\mu}{2} dV \right)\,,\\
\bar{A}(\mu) &= \frac{1}{2\rho} L_0 d\rho + \frac{1}{\ell} \left( \frac{1}{\sqrt{\rho}}  L_{-1} - \sqrt{\rho} \bar{\mathcal{L}}_\mu L_1 \right) \left( - \sqrt{\frac{\mathcal{L}_\mu}{\bar{\mathcal{L}}_\mu}} \sinh \frac{\mu}{2}  dU + \cosh \frac{\mu}{2} dV \right)\,.
\end{aligned}
\end{equation}
It is straightforward to see the mixed boundary conditions in this root-$T\overline{T}$ deformed setting:
\begin{equation}
\begin{aligned}
\sqrt{\frac{\bar{\mathcal{L}}_\mu}{\mathcal{L}_\mu}} \left(\sinh \frac{\mu}{2} \right) A_U(\mu) + \left(\cosh \frac{\mu}{2} \right) A_V(\mu) &= 0\,, \\ \sqrt{\frac{\bar{\mathcal{L}}_\mu}{\mathcal{L}_\mu}} \left( \cosh \frac{\mu}{2} \right) \bar{A}_U(\mu) + \left( \sinh \frac{\mu}{2} \right)   \bar{A}_V (\mu) &= 0\,.
\end{aligned}
\end{equation}

Moreover, we can extract the deformed boundary Chern-Simons connections:
\begin{equation}\label{deformed_cs_coordinate_approach}
\begin{aligned}
a(\mu)&= \frac{L_1-\mathcal{L}_\mu  L_{-1} }{\ell}  \left( \left( \cosh \frac{\mu}{2} - \sqrt{\frac{\bar{\mathcal{L}}_\mu}{\mathcal{L}_\mu} }  \sinh \frac{\mu}{2} \right) d\phi  + \left( \cosh \frac{\mu}{2} + \sqrt{\frac{\bar{\mathcal{L}}_\mu}{\mathcal{L}_\mu}} \sinh \frac{\mu}{2} \right) dT
  \right)\\
  \bar{a}(\mu) &= \frac{ L_{-1} -  \bar{\mathcal{L}}_\mu L_1}{\ell}\left( \left( \cosh \frac{\mu}{2} -\sqrt{\frac{\mathcal{L}_\mu}{\bar{\mathcal{L}}_\mu}} \sinh \frac{\mu}{2}\right) d\phi - \left(\cosh \frac{\mu}{2}  +\sqrt{\frac{\mathcal{L}_\mu}{\bar{\mathcal{L}}_\mu}} \sinh \frac{\mu}{2}    \right) dT \right)   
  \end{aligned}
\end{equation}
which obey 
\begin{equation}
\label{eq:BCsroot-TT}
a_T(\mu) = \frac{\cosh \frac{\mu}{2} +   \sqrt{\frac{\bar{\mathcal{L}}_\mu}{\mathcal{L}_\mu}} \sinh \frac{\mu}{2} }{ \cosh \frac{\mu}{2} - \sqrt{\frac{\bar{\mathcal{L}}_\mu}{\mathcal{L}_\mu}} \sinh \frac{\mu}{2}   } a_\phi (\mu)\,, \quad \bar{a}_{T}(\mu) = - \frac{ \cosh \frac{\mu}{2} + \sqrt{\frac{\mathcal{L}_\mu}{\bar{\mathcal{L}}_\mu}} \sinh \frac{\mu}{2}   }{\cosh \frac{\mu}{2}    - \sqrt{ \frac{\mathcal{L}_\mu}{\bar{\mathcal{L}}_\mu}  } \sinh \frac{\mu}{2} } \bar{a}_\phi(\mu)\,.
\end{equation}
To make contact with our discussion of the horizon area in the metric formalism, we note that one may compute the BTZ black hole's Bekenstein-Hawking entropy (and thus its horizon area) directly in the Chern-Simons formalism. Following \cite{deBoer:2013gz}, the black hole entropy is given in terms of Chern-Simons quantities as
\begin{equation}
\label{eq:Entropy}
    S = C \operatorname{Tr} \left( (\lambda_\phi - \bar{\lambda}_\phi)L_0 \right)\,,
\end{equation}
where $C$ is a constant that depends on the central charge $c$, but whose precise value is not important for this discussion, $\lambda_\phi$ and $\bar{\lambda}_\phi$ are diagonal traceless matrices containing the eigenvalues of $a_\phi$ and $\bar{a}_\phi$.

Note (\ref{eq:Entropy}) was derived in \cite{deBoer:2013gz} using a particular boundary term that is appropriate for the Drinfeld-Sokolov form of the connections, which in our case corresponds to a Ba\~nados type solution. We note that the root-$T\overline{T}$ deformed connections are \emph{not} of this form when written in terms of the original coordinates, which will become clear when we obtain covariant expressions for the deformed in (\ref{TT_CS_mixed_bcs}). However, when we write the deformed connections in new coordinates $(T, \phi)$, the connections are of Ba\~nados type, albeit characterized by deformed parameters $\mathcal{L}_\mu$ and $\bar{\mathcal{L}}_\mu$. Therefore it is justified to use the expression (\ref{eq:Entropy}) so long as we work in the transformed coordinates.

Diagonalizing the connections given in \eqref{deformed_cs_coordinate_approach}, one finds
\begin{equation}\label{root_TT_lambda_matrices}
\begin{aligned}
    \lambda_\phi &= \frac{1}{\ell} \left( \begin{array}{cc}
       \sqrt{\mathcal{L}_\mu} \cosh \frac{\mu}{2} - \sqrt{\bar{\mathcal{L}}_\mu} \sinh \frac{\mu}{2}  & \quad   0\\
   0      & \quad  - \sqrt{\mathcal{L}_\mu} \cosh \frac{\mu}{2} + \sqrt{\bar{\mathcal{L}}_\mu} \sinh \frac{\mu}{2} 
    \end{array} \right)\,, \\ 
    \bar{\lambda}_\phi  &= \frac{1}{\ell}  \left( \begin{array}{cc}
     -  \sqrt{\bar{\mathcal{L}}_\mu} \cosh \frac{\mu}{2} + \sqrt{\mathcal{L}_\mu} \sinh \frac{\mu}{2}  & \quad   0\\
   0      & \quad    \sqrt{\bar{\mathcal{L}}_\mu} \cosh \frac{\mu}{2} - \sqrt{\mathcal{L}_\mu} \sinh \frac{\mu}{2}
    \end{array} \right)\,.
    \end{aligned}
\end{equation}
Therefore
\begin{equation}\label{CS_two_entropies}
    S^{(0)} = \frac{C}{\ell} \left(\sqrt{\mathcal{L}_0} + \sqrt{\bar{\mathcal{L}}_0}\right)\,, \quad S^{(\mu)} = \frac{C}{\ell} e^{- \frac{\mu}{2}}  \left( \sqrt{\mathcal{L}_\mu  }  + \sqrt{\bar{\mathcal{L}}_\mu} \right) \,.
\end{equation}
Equating the two entropies in (\ref{CS_two_entropies}) then gives the same area equation which we found in \eqref{eq:areaangularmomentumTTroot} using a metric-formalism analysis.

The corresponding boundary term which we must add to have a well-defined variational principle with respect to these root-$T\overline{T}$ deformed boundary conditions is

\begin{align}
\label{eq:0okjnbgfr456789}
\delta S_{\text{bdry}} =\frac{\ell}{8\pi G} \int_{\partial M_3} dT \, d\phi \, &\Bigg( \operatorname{Tr} \left[  \frac{\cosh \frac{\mu}{2} +   \sqrt{\frac{\bar{\mathcal{L}}_\mu}{\mathcal{L}_\mu}} \sinh \frac{\mu}{2} }{ \cosh \frac{\mu}{2} - \sqrt{\frac{\bar{\mathcal{L}}_\mu}{\mathcal{L}_\mu}} \sinh \frac{\mu}{2}   } a_\phi(\mu)\ \delta a_\phi (\mu) 
 \right] \nonumber \\
&\qquad + \operatorname{Tr} \left[ \frac{ \cosh \frac{\mu}{2} + \sqrt{\frac{\mathcal{L}_\mu}{\bar{\mathcal{L}}_\mu}} \sinh \frac{\mu}{2}   }{\cosh \frac{\mu}{2}    - \sqrt{ \frac{\mathcal{L}_\mu}{\bar{\mathcal{L}}_\mu}  } \sinh \frac{\mu}{2} } \bar{a}_\phi(\mu)\, \delta  \bar{a}_\phi (\mu) \right] \Bigg) \,.
\end{align}
We calculate the boundary connections variations in \eqref{eq:BCsroot-TT} and evaluate relevant traces:
\begin{equation}
\begin{aligned}
\operatorname{Tr} \left( a_\phi (\mu) \delta a_\phi (\mu) \right) &= \frac{  \sqrt{\frac{\mathcal{L}_\mu}{\bar{\mathcal{L}}_\mu}}  \cosh \frac{\mu}{2} -  \sinh \frac{\mu}{2} }{\ell^2}  \left( \sqrt{ \frac{\bar{\mathcal{L}}_\mu}{\mathcal{L}_\mu} } \left(\cosh \frac{\mu}{2} \right) \delta \mathcal{L}_\mu - \left(\sinh \frac{\mu}{2} \right) \delta \bar{\mathcal{L}}_\mu \right) \\
\operatorname{Tr} \left( \bar{a}_\phi (\mu) \delta \bar{a}_\phi (\mu) \right) &= \frac{\sqrt{\frac{\bar{\mathcal{L}}_\mu}{\mathcal{L}_\mu}}  \cosh \frac{\mu}{2} -  \sinh \frac{\mu}{2} }{\ell^2}  \left(   \sqrt{ \frac{\mathcal{L}_\mu}{\bar{\mathcal{L}}_\mu}   } \left( \cosh \frac{\mu}{2} \right) \delta \bar{\mathcal{L}}_\mu - \left(\sinh \frac{\mu}{2} \right) \delta \mathcal{L}_\mu\right)  
\end{aligned}
\end{equation}
to simplify \eqref{eq:0okjnbgfr456789} into
\begin{equation}
\delta S_{\text{bdry}} = \frac{1}{8\pi G \ell} \int_{\partial M} dT \, d\phi \left( \delta \mathcal{L}_\mu + \delta \bar{\mathcal{L}}_\mu \right) \,, 
\end{equation}
from which $S_{\text{bdry}}$ is easily read off
\begin{equation}
\label{eq:boundaryactionroot-tt}
S_{\text{bdry}}= \frac{1}{8 \pi G \ell} \int_{\partial M} dT \, d\phi \, \left(  \mathcal{L}_\mu + \bar{\mathcal{L}}_\mu \right)\,.
\end{equation}
In summary, we have shown that the final expression (\ref{eq:boundaryactionroot-tt}) for the deformed boundary action in Chern-Simons variables is identical to that of the root-$T\overline{T}$ deformed energy, given in (\ref{eq:E&J}), of the spacetime computed in the metric formalism.\footnote{In \cite{He:2020hhm} it was shown that the corresponding Chern-Simons boundary action for $T\overline{T}$-deformed boundary conditions also matches the $T\overline{T}$-deformed spacetime energy \eqref{eq:EnergyTT}.}

Note that we have not given any \emph{a priori} justification that the deformed Chern-Simons boundary action yields the spacetime mass when the boundary theory is deformed by a general multi-trace operator. Although it is easy to show that this is true in the undeformed theory, a general proof that the Chern-Simons boundary action computes the spacetime mass in the presence of modified boundary conditions would require a computation of the Hamiltonian using an analysis of the canonical structure. We will not pursue such an analysis here. However, the fact that the deformed boundary action (\ref{eq:boundaryactionroot-tt}) does agree with the energy computed in the metric formulation may be viewed as an \emph{a posteriori} argument that such an analysis in the canonical formulation would conclude that the boundary term equals the spacetime energy in the case of root-$T\overline{T}$ deformed boundary conditions.

\emph{Root-$T\overline{T}$ deformed Chern-Simons: covariant approach}

We now describe the second approach. To make the sources and expectation values in Chern-Simons variables explicit, it is convenient to expand the boundary gauge fields as
\begin{align}\label{llabres_expansion}
    a_i &= 2 e_i^+ L_{1} - f_i^- L_{-1} + \omega_i L_0 \,, \nonumber \\
    \bar{a}_i &= f_i^+ L_{1} - 2 e_i^- L_{-1} + \omega_i L_0 \,.
\end{align}
In the case of a Ba\~nados-type geometry, this expansion reduces to the one given in (\ref{CS_boundary_banados}). In these expansions, $e_i^a$ plays the role of the boundary vielbein, where we use middle Latin letters $i, j, k$ for curved boundary indices and early Latin letters $a, b, c$ for flat boundary indices. We have chosen the numerical factors appearing in the expansions (\ref{llabres_expansion}) to simplify our final results, but they will lead to some unfamiliar factors of $2$ in certain expressions. For instance, the boundary metric in these conventions is
\begin{align}\label{metric_factors_2}
    \gamma_{ij} = 2 e_i^a \eta_{ab} e_j^b \,, 
\end{align}
which has an additional factor of $2$ compared to the standard definition. We also define
\begin{align}
    e = \det ( e_j^b ) \,, 
\end{align}
so that $\det ( \gamma_{ij} ) = - 4 e^2$, and the Levi-Civita symbols with flat and curved indices are
\begin{align}
    \epsilon_{a b} = \begin{bmatrix} 0 & 1 \\ -1 & 0 \end{bmatrix}_{ab} \,, \qquad \epsilon^{ij} = \frac{1}{2 e} \begin{bmatrix} 0 & 1 \\ -1 & 0 \end{bmatrix}^{ij} \,, 
\end{align}
These satisfy various identities such as $g^{ij} = - \epsilon^{i k} \epsilon^{j l} g_{kl}$, $\epsilon^{ab} = 2 \epsilon^{ij} e_i^a e_j^b$, and so on, with factors of $2$ that can be traced back to the definition (\ref{metric_factors_2}). Flat indices are raised and lowered with $\eta_{ab}$, where we take $\eta_{+-} = \eta_{-+} = - 1$ in this subsection. We refer the reader to section 2 of \cite{Ebert:2022ehb}, or to \cite{Llabres:2019jtx}, for more details on these notational conventions.

In the holographic dictionary, this vielbein $e_i^a$ is the source while the other expansion coefficients $f_i^a$ are the dual expectation values, which are related to the boundary stress tensor with one flat and one curved index according to the relation:
\begin{align}
    T^i_a = \frac{1}{4 \pi G} \epsilon_{ab} \epsilon^{ij} f_j^b \,.
\end{align}
We will assume that the boundary spin connection $\omega$ vanishes in the undeformed theory, which is appropriate for a flat boundary.

We expect, based on the general analysis of section \ref{sec:dictionary}, that the addition of a multi-trace boundary term in Chern-Simons variables will impose modified boundary conditions in which some deformed source $e_i^a ( \mu )$ is now held fixed in the variational principle, rather than the undeformed source $e_i^a ( 0 )$. In the case of a boundary $T\overline{T}$ deformation, we recall from \cite{Ebert:2022ehb,Llabres:2019jtx} that the resulting modification of the sources and expectation values is simply
\begin{align}\label{TT_CS_mixed_bcs}
    e_i^a ( \lambda ) = e_i^a ( 0 ) + \frac{\lambda}{4 \pi G} f_i^a \,, \qquad f_i^a ( \lambda ) = f_i^a ( 0 ) \,.
\end{align}
One can determine the analog of (\ref{TT_CS_mixed_bcs}), which corresponds to a boundary root-$T\overline{T}$ deformation by following the procedure of section \ref{sec:boundary_condition_derivation}. That is, we first write down the most general expression for deformed quantities $e_i^a ( \mu )$ and $f_i^a ( \mu )$ which depend on a dimensionless parameter $\mu$, preserve tracelessness for a conformal seed theory, and commute with the $T\overline{T}$-deformed boundary conditions (\ref{TT_CS_mixed_bcs}). We will not carry out these steps explicitly since they are identical to section \ref{sec:boundary_condition_derivation} after changing from metric variables to Chern-Simons variables. Instead, we simply quote the result, which for a CFT seed is
\begin{align}\label{CS_rTT_deformed_bcs}
    e_i^a ( \mu ) &= \cosh \left( \frac{\mu}{2} \right) e_i^a ( 0 ) + \frac{\sinh \left( \frac{\mu}{2} \right)}{ \mathcal{R}^{(0)}} f_i^a ( 0 ) \nonumber\,, \\ f_i^a ( \mu ) &= \cosh \left( \frac{\mu}{2} \right) f_i^a ( 0 ) + \sinh \left( \frac{\mu}{2} \right) \mathcal{R}^{(0)} e_i^a ( 0 ) \,. 
\end{align}
Here $\mathcal{R}^{(0)}$ is the usual root-$T\overline{T}$ operator, which can be expressed purely in Chern-Simons variables. Again, these expressions will have some unusual numerical factors introduced by (\ref{metric_factors_2}). For instance, we can reproduce a general stress tensor on a standard flat metric via
\begin{align}
   e^i_a = \frac{1}{\sqrt{2}} \begin{bmatrix} 0 & 1 \\ 1 & 0 \end{bmatrix}^i_a \,, \qquad f^i_a = \frac{4 \pi G}{\sqrt{2}} \begin{bmatrix} T_{zz} & - T_{z \bar{z}} \\ - T_{z \bar{z}} & T_{\bar{z} \bar{z}} \end{bmatrix}^i_a \,, 
\end{align}
and then the stress tensor with two curved indices is
\begin{align}
    T_{ij} = T^k{}_a e^a_j g_{k i} = \begin{bmatrix} T_{zz} & T_{z \bar{z}} \\ T_{z \bar{z}} & T_{\bar{z} \bar{z}} \end{bmatrix}_{ij} \,, 
\end{align}
and its trace is $g^{ij} T_{ij} = - 2 T_{z \bar{z}}$, while the invariant $T^{ij} T_{ij}$ is
\begin{align}
    T^{ij} T_{ij} = 2 \left( T_{zz} T_{\bar{z} \bar{z}} + T_{z \bar{z}}^2 \right) \,.
\end{align}
It is straightforward to covariantize these statements and obtain expressions for $\mathcal{R}^{(0)}$. For instance, in the case of a conformal seed theory with a traceless stress tensor, we find
\begin{align}
    \mathcal{R}^{(0)} = \frac{1}{4 \pi G} \sqrt{ - f_i^a f_j^b \epsilon_{ab} \epsilon^{ij} } \,.
\end{align}
In the general case where the undeformed stress tensor is not traceless, we can define a traceless part of $f_i^a$, which is the analog of $\widetilde{T}_{\alpha \beta}$, as
\begin{align}
    \widetilde{f}_i^a = f_i^a - 4 \pi G \, e_i^a \left( e_j^b T^j_b \right) \,,  
\end{align}
and then express the root-$T\overline{T}$ operator as
\begin{align}
    \mathcal{R}^{(0)} = \frac{1}{4 \pi G} \sqrt{ - \widetilde{f}_i^a \widetilde{f}_j^b \epsilon_{ab} \epsilon^{ij} } \,.
\end{align}
One can then check that, after transforming from Chern-Simons variables to metric variables, the deformed quantities (\ref{CS_rTT_deformed_bcs}) reproduce the metric and stress tensor (\ref{TT_CS_mixed_bcs}) in the case of a conformal seed theory (or for a general seed, if we replace $f_i^a$ with $\widetilde{f}_i^a$).

In particular, the deformed connections computed with the $e_i^a ( \mu )$ in (\ref{CS_rTT_deformed_bcs}) agree with those in (\ref{deformed_cs_coordinate_approach}). One can see this by expressing the $f_i^a$ in terms of $\mathcal{L}$ and $\bar{\mathcal{L}}$ and choosing coordinates $(\phi, T)$. Indeed, it must have been the case that these agree since the coordinate transformation that was used to obtain (\ref{deformed_cs_coordinate_approach}) is precisely the one that generates the root-$T\overline{T}$ deformed metric and stress tensor in the metric formalism, and the deformed vielbein (\ref{CS_rTT_deformed_bcs}) reproduces these quantities. Therefore, the two methods are equivalent.

\section{Conclusion}\label{sec:conclusion}

In this chapter, we have investigated several properties of the root-$T\overline{T}$ operator in holography. Among our main results is the proposal (\ref{root_TT_energy_flow}) for the flow of the finite-volume spectrum of a root-$T\overline{T}$ deformed CFT. We have explicitly verified that this flow equation matches the deformed spacetime mass for a class of Ba\~nados-type $\mathrm{AdS}_3$ solutions subject to root-$T\overline{T}$ deformed boundary conditions. This represents the first calculation, which may shed light on quantum aspects of the root-$T\overline{T}$ deformation. Although a quantum definition of the root-$T\overline{T}$ operator itself is still not known in the field theory, we have sidestepped this issue by working in a large-$N$ limit and performing a holographic calculation in the bulk dual. Besides the question of a quantum definition of the root-$T\overline{T}$ operator, there remain many other avenues for future research, two of which we outline below. We believe that a better understanding of these issues will offer new insights into non-analytic root-$T\overline{T}$-like (or ModMax-like) theories.

\emph{Correlation functions}

An immediate, and important, next step would be to study correlation functions in a root-$T\overline{T}$ deformed $\mathrm{CFT}_2$. Due to the awkwardness of the square root of an operator, calculating a root-$T\overline{T}$ deformed correlation function in perturbation theory seems difficult and ambiguous. However, because the root-$T\overline{T}$ operator exhibits some of the special properties of the $T\overline{T}$ operator, there might be hope that a perturbative calculation is feasible. In particular, we have seen that demanding commutativity of the $T\overline{T}$ and root-$T\overline{T}$ flows is a powerful constraint that allowed us to uniquely fix the deformed boundary conditions and flow equation for the spectrum. One might conjecture that perturbative corrections to correlation functions may also be fixed by imposing commutativity of the following diagram:
  \begin{align*}\label{}
\includegraphics[width=.5\linewidth]{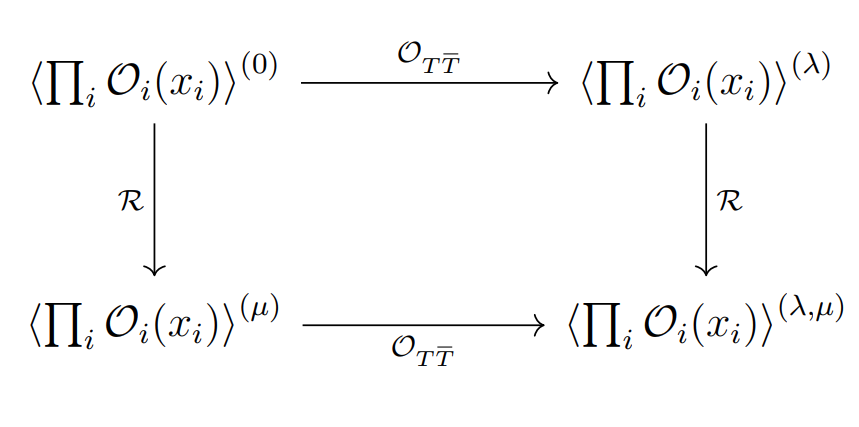}
\end{align*}
To be more concrete, the $T\overline{T}$-deformed two-point planar stress tensor correlators take the following form from dimensional analysis, translational and rotational symmetry \cite{Kraus:2018xrn,Ebert:2022cle}
\begin{equation}
    \begin{aligned}
    \label{eq:uhreh}
        \langle T_{zz} (x) T_{zz} (0) \rangle^{(\lambda)} &= \frac{1}{z^4} f_1 (y), \quad
            \langle T_{zz} (x) T_{z\bar{z}} (0) \rangle^{(\lambda)} = \frac{1}{z^3 \bar{z}} f_2 (y)\,,\\
                    \langle T_{zz} (x) T_{\bar{z} \bar{z}} (0) \rangle^{(\lambda)} &= \frac{1}{z^2 \bar{z}^2} f_3 (y)\,,\quad 
                                     \langle T_{z\bar{z}} (x) T_{z \bar{z}} (0) \rangle^{(\lambda)}= \frac{1}{z^2 \bar{z}^2} f_4 (y)\,,
    \end{aligned}
\end{equation}
where $y = z \bar{z}/\lambda$ and the functions $f_i(y)$ are fixed by stress tensor conservation $\partial^\alpha T_{\alpha \beta} = 0$ and the trace flow equation $T_{z\bar{z}} = -\pi \lambda T\overline{T}$ giving $T_{z\bar{z}} = - \pi \lambda T_{zz} T_{\bar{z} \bar{z}} + \mathcal{O}(\lambda^2)$. Using the trace flow equation, we can easily determine $f_4(y)$ at $O(\lambda^2)$:
\begin{equation}
\begin{aligned}
\langle T_{z\bar{z}}(x) T_{z\bar{z}}(0)  \rangle^{(\lambda)} &= \langle (-\pi \lambda T_{zz}(x) T_{\bar{z} \bar{z} } (x))(-\pi \lambda T_{zz}(0) T_{\bar{z} \bar{z} } (0)) \rangle^{(0)} + \cdots
\\&= \pi^2 \lambda^2 \langle T_{zz} (x) T_{zz}(0) \rangle^{(0)} \langle T_{\bar{z}\bar{z}} (x) T_{\bar{z}\bar{z}}(0) \rangle^{(0)}  + \cdots
\\&= \frac{\pi^2 \lambda^2 c^2}{4 z^4 \bar{z}^4} + \cdots \nonumber \\
\implies f_4(y) &= \frac{\pi^2 c^2}{4y^2}  + \cdots\,.
\end{aligned}
\end{equation}
The correlators also obey $\partial^\alpha \langle T_{\alpha \beta} (x) T_{\rho \sigma}(0) \rangle^{(\lambda)} = 0$ which give three conservation equations:
\begin{equation}
\begin{aligned}
\label{conservation equations}
\beta = \rho = \sigma = z:  & \partial_{\bar{z}} \langle T_{zz} (x) T_{zz} (0) \rangle^{(\lambda)} + \partial_{z} \langle T_{\bar{z} z} (x) T_{zz}(0) \rangle^{(\lambda)}\\& =  \partial_{\bar{z}}\left( \frac{f_1(y)}{z^4} \right) +   \partial_{z} \left( \frac{f_2(y)}{z^3 \bar{z}} \right) = 0\,, \\
 \beta =\bar{z}, \rho = \sigma = z:& \partial_{\bar{z}} \langle T_{z \bar{z}} (x) T_{zz} (0) \rangle^{(\lambda)} + \partial_{z} \langle T_{\bar{z} \bar{z}} (x) T_{zz} (0) \rangle^{(\lambda)} \\&=  \partial_{\bar{z}} \left( \frac{f_2(y)}{z^3 \bar{z}}\right) +  \partial_{z}\left(  \frac{f_3(y)}{z^2 \bar{z}^2}\right) = 0\,, \\ \beta =z, \rho = z, \sigma = \bar{z}: &\partial_{\bar{z}} \langle T_{z \bar{z}} (x) T_{\bar{z} z}(0) \rangle^{(\lambda)} + \partial_{z} \langle T_{\bar{z} z} (x) T_{z \bar{z}}(0) \rangle^{(\lambda)} \\&=  \partial_{\bar{z}}\left( \frac{f_2(y)}{z^3 \bar{z}} \right) +  \partial_{z} \left( \frac{f_4(y)}{z^2 \bar{z}^2} \right) = 0\,.
\end{aligned}
\end{equation}
Since $f_4(y)$ at $O(\lambda^2)$ is known, we can determine the other $f_i(y)$ from solving \eqref{conservation equations} with initial conditions that the seed theory's correlators are recovered when $\lambda = 0$:
\begin{equation}\hspace{-25pt}
    \begin{aligned}
        f_1 (y) &= \frac{c}{2} + \frac{5 \pi^2 \lambda^2 c^2}{6 z^2 \bar{z}^2} + \cdots \,, \; f_2 (y) = - \frac{\pi^2 \lambda^2 c^2}{3 z^2 \bar{z}^2} + \cdots \,, \\ f_3(y) &= \frac{\pi^2 \lambda^2 c^2}{4 z^2 \bar{z}^2} + \cdots \,, \;  f_4(y) =\frac{\pi^2 \lambda^2 c^2}{4 z^2 \bar{z}^2} + \cdots \,.
    \end{aligned}
\end{equation}
For the root-$T\overline{T}$ case, one should follow a similar logic as in the above $T\overline{T}$ example:
\begin{equation}
    \begin{aligned}
    \label{eq:uhreh1}
        \langle T_{zz} (x) T_{zz} (0) \rangle^{(\mu)} &= \frac{1}{z^4} g_1 (u), \quad \quad \; 
            \langle T_{zz} (x) T_{z\bar{z}} (0) \rangle^{(\mu)} = \frac{1}{z^3 \bar{z}} g_2 (u)\,,\\
                    \langle T_{zz} (x) T_{\bar{z} \bar{z}} (0) \rangle^{(\mu)} &= \frac{1}{z^2 \bar{z}^2} g_3 (u)\,,\quad 
                                     \langle T_{z\bar{z}} (x) T_{z \bar{z}} (0) \rangle^{(\mu)} = \frac{1}{z^2 \bar{z}^2} g_4 (u)\,,
    \end{aligned}
\end{equation}
    where the $g_i(u)$ obey the same stress tensor conservation equations \eqref{conservation equations}. It would be interesting to see whether one or more of the $g_i(u)$ can be fixed from demanding commutativity of the $T\overline{T}$ and root-$T\overline{T}$ flows. For example, perhaps commutativity may fix one of the $g_i ( u )$, and then conservation of the stress tensor may fix the others. It would also be interesting to understand this commutativity and correlators in the context of quantum corrections, such as the two-loop corrected $T\overline{T}$-deformed planar stress tensor correlators found in \cite{Ebert:2022cle}.

\emph{The fate of conformal symmetry}

The root-$T\overline{T}$ operator is classically marginal and thus preserves conformal invariance at the classical level. It is an important open question to determine the fate of conformal symmetry in the quantum theory, assuming that a quantum definition of the root-$T\overline{T}$ operator exists. Quantum corrections might make this operator marginally relevant or marginally irrelevant, which would mean that conformal invariance is broken at the quantum level. 

One way to probe this question is to investigate the high-energy density of states. In any two-dimensional CFT, the degeneracy of states for large energy and high temperature is described by the Cardy formula \cite{Cardy:1986ie}, which fixes the asymptotic scaling to be
\begin{equation}
\label{eq:Cardy}
    \rho (E_0) \sim \exp\left(2\pi \sqrt{\frac{cE_0}{3}}\right) \,, \quad S(E_0) \sim 2\pi \sqrt{\frac{cE_0}{3}} \,.
\end{equation}
Therefore, to investigate whether a root-$T\overline{T}$ deformed CFT remains a CFT, one might ask whether its high-energy behavior agrees with (\ref{eq:Cardy}). A sketch of an argument in support of this claim might proceed as follows. First, since the root-$T\overline{T}$ deformed energy spectrum
depends on both the energy $E_n^{(0)}$ and momentum $P_n$ of the corresponding state in the undeformed theory, we cannot immediately use the na\"ive Cardy formula (\ref{eq:Cardy}), which has already coarse-grained over all states with energy near $E_n$ but with any momentum $P_n$. However, we may use a generalization of the Cardy formula, which accounts for the spin of a CFT state \cite{Hartman:2014oaa,Pal:2019zzr}. In terms of left-moving and right-moving energies, this formula reads:
\begin{equation}
    \rho(E_L,E_R) \sim \exp \left( 2\pi \sqrt{\frac{cE_L}{6}} + 2\pi \sqrt{\frac{cE_R}{6}} \right) \,.
\end{equation}
One can express our deformed spectrum (\ref{root_TT_deformed_CFT_energies}) in terms of the left-moving and right-moving energies, such that $E_\mu = \left( E_L \right)_\mu + \left( E_R \right)_\mu$ and $P_\mu = \left( E_L \right)_\mu - \left( E_R \right)_\mu = P_0$, which satisfy
\begin{equation}
\begin{aligned}
    \sqrt{(E_L)_0} &= \sqrt{(E_L)_\mu} \cosh \left(\frac{\mu}{2}\right) - \sqrt{(E_R)_\mu} \sinh \left(\frac{\mu}{2}\right) \,, \; \\\sqrt{(E_R)_0} &= \sqrt{(E_R)_\mu} \cosh \left(\frac{\mu}{2} \right) - \sqrt{(E_L)_\mu} \sinh \left(\frac{\mu}{2} \right) \,.
\end{aligned}
\end{equation}
The deformed density of states $\rho_\mu ( E_L, E_R)$ is then obtained by expressing the density of states of the undeformed theory in terms of the deformed left-moving and right-moving energies. Up to a factor which is unimportant for the leading exponential behavior, we find
\begin{equation}
    \rho_\mu(E_L,E_R) \sim \exp \left( 2\pi\sqrt{\frac{c E_L}{6}} e^{-\mu/2} +2\pi \sqrt{\frac{c E_R}{6}} e^{-\mu/2} \right) \,.
\end{equation}
It, therefore, appears that the high-energy density of states for the deformed theory still has the (generalized) Cardy behavior appropriate for a conformal field theory, although with a new effective central charge $c_{\text{eff}} = c e^{-\mu}$. In particular, this gives one hint that the root-$T\overline{T}$ deformation (if it indeed is well-defined quantum mechanically) may be marginally \emph{relevant} since the central charge appears to decrease along the flow for positive $\mu$.\footnote{One can also see this by considering the behavior of the spectrum (\ref{root_TT_deformed_CFT_energies}) as $\mu \to \infty$. In this limit, it appears that all negative-energy states in the undeformed theory approach zero deformed energy, while all undeformed positive-energy states have deformed energies that grow without bound. This suggests that the large-$\mu$ root-$T\overline{T}$ deformed theory becomes a gapped system with only a finite number of states.}

Although suggestive, some subtleties prevent this argument from being fully rigorous. One is that we have not, strictly speaking, demonstrated that the root-$T\overline{T}$ flow equation holds for an arbitrary state in the deformed theory. Our gravitational calculation only demonstrates that this flow equation holds for holographic states which are dual to Ba\~nados-type geometries, and only in the large-$N$ regime. A robust quantum definition of the root-$T\overline{T}$ operator might allow one to analyze the high-energy behavior of the deformed theory and determine whether it still exhibits Cardy behavior.

Another way of probing the fate of conformal invariance is to investigate modular properties of the root-$T\overline{T}$ deformed torus partition function. If one could show that the deformed partition function remains modular invariant, this would offer further evidence that the theory is conformal. One strategy for doing this would be to derive a differential equation that the deformed partition function satisfies. In the case of the $T\overline{T}$ deformation, it is known \cite{Cardy:2018sdv,Aharony:2018bad} that the torus partition function obeys the flow equation
\begin{align}
    \partial_\lambda Z_\lambda ( \tau, \bar{\tau} ) = \left( \tau_2 \partial_{\tau} \partial_{\bar{\tau}} + \frac{1}{2} \left( \partial_{\tau_2} - \frac{1}{\tau_2} \right) \lambda \partial_\lambda \right) Z_\lambda ( \tau, \bar{\tau} ) \,, 
\end{align}
where the torus' modulus is $\tau = \tau_1 + i \tau_2$ and that $Z_\lambda$ is invariant under a modular transformation if the $T\overline{T}$ parameter $\lambda$ also transforms. It appears that a root-$T\overline{T}$ deformed theory obeys an analogous flow equation,
\begin{align}\label{Z_flow_root_TT}
    \partial_\mu^2 Z_\mu ( \tau , \bar{\tau } )  =  \left( \tau_2^2 \left( \partial_\tau \partial_{\bar{\tau}} \right) + \tau_2 \partial_{\tau_2} \right) Z_\mu ( \tau , \bar{\tau } ) \,,
\end{align}
which suggests that the root-$T\overline{T}$ deformed theory may be modular invariant.
\chapter{Flows in the Space of Interacting Chiral Boson Theories}

\label{ch:ChiralBoson}

We study interacting theories of $N$ left-moving and $\bar{N}$ right-moving Floreanini-Jackiw bosons in two dimensions. A parameterized family of such theories is shown to enjoy (non-manifest) Lorentz invariance if and only if its Lagrangian obeys a flow equation driven by a function of the energy-momentum tensor. We discuss the canonical quantization of such theories along classical stress tensor flows, focusing on the case of the root-$T\overline{T}$ deformation, where we obtain perturbative results for the deformed spectrum in a certain large-momentum limit. In the special case $N = \bar{N}$, we consider the quantum effective action for the root-$T\overline{T}$-deformed theory by expanding around a general classical background, and we find that the one-loop contribution vanishes for backgrounds with constant scalar gradients. Our analysis can also be interpreted via dual $U(1)$ Chern-Simons theories in three dimensions, which might be used to describe deformations of charged $\mathrm{AdS}_3$ black holes or quantum Hall systems.

\section{Introduction} 
In physics, we are frequently interested in parameterized families of classical or quantum field theories. The tangent vectors to these families often have an interpretation as operators within a given theory. One familiar example appears in the study of conformal field theories, which may possess certain exactly marginal operators. Deforming a CFT by a marginal operator generates motion on the conformal manifold, which is one such family of theories.

Another simple one-parameter family generated from any quantum field theory is the well-known renormalization group flow. We can interpret this as a curve of theories labeled by an energy scale $\mu$.
For a CFT, this curve degenerates to a single point, but for other QFTs, one finds an infinite family of theories connecting two RG fixed points at the UV and IR ends of this flow. The operator that plays the role of the tangent vector to this curve is the trace of the energy-momentum tensor, which generates scale transformations.

The renormalization group example is especially useful because it is \emph{universal}: any translation-invariant field theory admits an energy-momentum tensor $T_{\mu \nu}$, so we may always deform by the trace $T^\mu{}_\mu$ to flow toward the infrared. It is natural to explore other deformations constructed from the stress tensor, which are also universal. These stress tensor deformations generate a larger class of flows, which includes the renormalization group flow as a special case, but which also includes other famous examples such as the $T\overline{T}$ deformation of two-dimensional quantum field theories \cite{Zamolodchikov:2004ce, Smirnov:2016lqw,Cavaglia:2016oda}.

Even at the classical level, stress tensor flows often give rise to interesting parameterized families of theories. For instance, consider classical theories of a single Abelian gauge field $A_\mu$ whose Lagrangians depend on the field strength $F_{\mu \nu}$ but not its derivatives. Construct the parameterized family which contains the Maxwell theory, $\mathcal{L} = - \frac{1}{4} F_{\mu \nu} F^{\mu \nu}$, and all other theories that can be reached from the Maxwell theory by deformations involving the energy-momentum tensor. This family is precisely the collection of theories of non-linear electrodynamics which are invariant under electric-magnetic duality rotations \cite{Ferko:2023wyi}, which is of interest in its own right.\footnote{Strictly speaking, there are some isolated points in this space such as the Bialynicki-Birula theory \cite{Bialynicki-Birula:1992rcm} which are not connected to Maxwell, so to be precise we should say that the family generated in this way gives one connected component in the space of duality-invariant theories.}

Another example concerns theories of a two-form gauge potential $A_{\mu \nu}$ with a self-dual three-form field strength $F_{\mu \nu \rho}$ in six spacetime dimensions. Any family of such theories -- e.g., the collection of interacting chiral tensor theories which describe the worldvolume theory on an M5-brane, labeled by a parameter $T$ that controls the tension of the brane -- also obeys a stress tensor flow equation \cite{Ferko:2024zth}. We say that both $4d$ theories of duality-invariant electrodynamics and $6d$ chiral tensor theories are closed under stress tensor flows, in the sense that deforming any member of one of these classes of theories by a Lorentz scalar constructed from $T_{\mu \nu}$ produces another member of the same class.

In this chapter, we will investigate another space of theories, which is also closed under deformations involving the energy-momentum tensor. The theories that we consider here describe the dynamics of a collection of $N$ chiral and $\bar{N}$ anti-chiral bosonic fields in two spacetime dimensions. The simplest member of this class, $N = 1$ and $\bar{N} = 0$, is the theory of a single chiral boson which is described by the Floreanini-Jackiw action \cite{Floreanini:1987as}, namely
\begin{align}\label{fj_action}
    S_{\text{FJ}} = \frac{1}{2} \int d^2 x \, \left( \partial_t \phi \partial_\theta \phi - \partial_\theta \phi \partial_\theta \phi \right) \, .
\end{align}
Here we work in a $2d$ spacetime with coordinates $(t, \theta)$. As is well-known, it is not straightforward to write a manifestly Lorentz-invariant Lagrangian for a field that obeys a chirality (or self-duality) constraint. One approach, which we will follow in this work, is to sacrifice manifest Lorentz invariance and work with actions of the form (\ref{fj_action}) that explicitly single out a preferred time coordinate $t$; we will then need to impose that the theory enjoy a non-manifest Lorentz symmetry. Another strategy is to introduce one or more auxiliary fields to restore manifest Lorentz invariance, which is the tactic used to describe chiral tensor theories in six dimensions using, e.g., the Pasti-Sorokin-Tonin (PST) formulation \cite{Pasti:1995tn,Pasti:1996vs,Pasti:1997gx} (and later extended to higher dimensions \cite{Buratti:2019guq}). A related technique was used to present a manifestly Lorentz invariant description of the Floreanini-Jackiw action in \cite{Townsend:2019koy}.

For a single chiral (or anti-chiral) boson, it is known that no Lorentz-invariant self-interactions are possible \cite{Buratti:2019guq,Bandos:2020hgy}, so (\ref{fj_action}) is the only allowed theory with $N = 1$. In this work, we will give a new interpretation of this fact: all Lorentz-invariant interacting chiral boson theories are generated from stress tensor deformations, but (\ref{fj_action}) is a fixed point of all such flows, and, therefore, there is no way to deform it to include interactions. However, for a theory with $N \geq 1$ chiral and $\bar{N} \geq 1$ anti-chiral bosons, such self-interactions are possible, and it is natural to describe them with an interaction function $V ( \partial_\theta \phi^i , \partial_\theta \bar{\phi}^{\overline{i}} )$ that depends on the spatial derivatives of the fields:
\begin{align}\label{interaction_function}
    S_{\text{int}} = \frac{1}{2} \int d^2 x \, \left( \partial_t \phi^i \partial_\theta \phi^i - \partial_t \bar{\phi}^{\overline{i}} \partial_\theta \bar{\phi}^{\overline{i}} - V ( \partial_\theta \phi^i , \partial_\theta \bar{\phi}^{\overline{i}} ) \right) \, .
\end{align}
In this expression, $i = 1 , \ldots , N$ runs over the chiral fields and $\overline{i} = 1 , \ldots , \bar{N}$ labels the anti-chiral fields. We will be primarily interested in theories that are invariant under the $O ( N ) \times O ( \bar{N} )$ symmetry rotating the chiral and anti-chiral bosons among themselves, although we will give some results that do not make this assumption; we will see that it is also possible to promote (\ref{interaction_function}) to include a target-space metric $G_{ij}$, $\bar{G}_{\overline{i} \overline{j}}$ for the bosons, or couplings to an antisymmetric tensor field $B_{ij}$, $\bar{B}_{\overline{i} \overline{j}}$, (which in general breaks $O ( N ) \times O ( \bar{N} )$) without significantly changing our analysis. Because the Lagrangian appearing in (\ref{interaction_function}) is first-order in time derivatives, the function $V$ can also be interpreted as the Hamiltonian of the model. This structure is similar to that of the PST description of a $6d$ chiral tensor theory, after gauge-fixing the auxiliary field $v_\mu$ of this formalism to the value $v_\mu =\delta^0_\mu$, whose action is
\begin{align}
    S_{\text{PST, gauge-fixed}} = \int d^6 x \, \left( \frac{1}{4} B_{ij} \partial_0 A^{ij} - \mathcal{H} ( s, p ) \right) \, .
\end{align}
Here $s = \frac{1}{4} B^{ij} B^{kl} \delta_{ik} \delta_{jl}$ and $p = \sqrt{p^i p_i}$, where $p_i = \frac{1}{8} \varepsilon_{i j k l m} B^{jk} B^{lm}$, are two $SO(5)$-invariant quantities constructed from the ``magnetic field'' $B_{ij}$, where $E^{ij}$ and $B^{ij}$ are related to the fundamental field $F_3 = d A_2$ as $E^{ij} = F^{i j 0}$, $B^{ij} = \widetilde{F}^{i j 0}$, and $\widetilde{F}$ denotes the Hodge dual of $F$. This gauge-fixed form of the PST action is closely related to the Perry-Schwarz formalism \cite{Perry:1996mk}. In our two-dimensional example, the role of the magnetic components $B^{ij}$ of the three-form field strength is played by the spatial derivatives $\partial_\theta \phi^i$ and $\partial_\theta \bar{\phi}^{\overline{i}}$ of the bosons.

Although we will not consider other formulations of chiral boson theories in this work, let us briefly mention that several other approaches have been used to describe such models. One presentation, due to Sen \cite{Sen:2015nph,Sen:2019qit}, introduces an additional ``spectator'' field which decouples from the dynamics; $T\overline{T}$ flows within this formalism have been studied in \cite{Chakrabarti:2020dhv,Chakrabarti:2022lnn,Chakrabarti:2023czz}.\footnote{The latter analysis also illuminates a surprising connection between the solvability of $T\overline{T}$-like deformations and that of another deformation of quantum mechanics involving a $\cosh ( p )$ kinetic term \cite{Grassi:2018bci}.} Another presentation introduced by Mkrtchyan includes an additional auxiliary scalar field $R$ and reduces to the PST form of the chiral boson action after integrating out $R$ \cite{Mkrtchyan:2019opf}. See \cite{Arvanitakis:2022bnr} for a comparison of some of these formulations and the realization of chiral bosons via a $3d$ Chern-Simons theory. Finally, a notable presentation by Siegel \cite{Siegel:1983es} expresses the chiral boson action in terms of a symmetric and traceless auxiliary tensor field $\lambda^{\alpha \beta}$:
\begin{equation}
\begin{aligned}\label{eq:Siegel}
 S_{\text{Siegel}}&=-\frac{1}{4} \int d^2x \bigg[ \partial_\alpha \phi \partial^\alpha \phi + \lambda^{\alpha \beta} \left( \partial_\alpha \phi - \epsilon_{\alpha \sigma} \partial^\sigma \phi \right) \left( \partial_\beta \phi - \epsilon_{\beta \rho} \partial^\rho \phi \right)\bigg]
 \\&=\int dt d\theta \bigg[ \frac{1}{4} \left(\partial_t \phi \partial_t \phi  - \partial_\theta \phi \partial_\theta \phi \right) + \frac{\lambda^{01} - \lambda^{00}}{2} \left(\partial_t \phi - \partial_\theta \phi \right)^2\bigg]\,.
\end{aligned}
\end{equation}
Siegel's action \eqref{eq:Siegel} is \emph{classically} equivalent to the Floreanini-Jackiw action \eqref{fj_action} assuming one can gauge the two independent components of $\lambda^{\alpha \beta}$ to $(\lambda^{00}, \lambda^{01}) =( \frac{1}{4}, - \frac{1}{4})$  \cite{Bernstein:1988zd}. For applications extending Siegel's action to gravity and string theory, see \cite{Gates:1987sy,Bellucci:1988uv,Kuzenko:1990mk,Bellucci:1991id,Kuzenko:1991ew,Gates:1991am}. The study of chiral bosons and other self-dual fields has a long history, and we refer the reader to an incomplete sampling \cite{Deser:1976iy,Marcus:1982yu,Henneaux:1987hz,Henneaux:1988gg,Schwarz:1993vs,Mezincescu:2022hnb,Evnin:2022kqn} of earlier work, and references therein, for other results.

Our motivation for studying this class of interacting chiral boson theories in this work is twofold. The first reason is purely classical: we would like to characterize the space of all such interacting theories, each of which is determined by an interaction function $V$, which enjoy non-manifest Lorentz invariance. As we will see, this condition will require that the function $V$ satisfy a certain partial differential equation which is very similar to those that appear in the cases of $4d$ duality-invariant electrodynamics \cite{Ferko:2023wyi} and $6d$ chiral tensor theories \cite{Ferko:2024zth}. The space of solutions to this partial differential equation is intimately connected to stress tensor flows. More precisely, given any parameterized family of Lorentz-invariant theories with interaction functions $V^{(\lambda)}$ labeled by a parameter $\lambda$, we will show that $\partial_\lambda V^{(\lambda)}$ can always be written as a function of the stress tensor $T_{\mu \nu}^{(\lambda)}$ of the theory at the same value of the parameter $\lambda$. Conversely, any flow equation of the form
\begin{align}\label{classical_flow}
    \partial_\lambda V^{(\lambda)} = f \left( T_{\mu \nu}^{(\lambda)} , \lambda \right) \, , \qquad \lim_{\lambda \to 0} V^{(\lambda)} = V^{(0)} \, ,
\end{align}
along with a Lorentz-invariant initial condition $V^{(0)}$, gives rise to a one-parameter family of Lorentz-invariant theories. Therefore, families of Lorentz-invariant interacting chiral boson theories are in one-to-one correspondence with stress tensor flows. These statements are the precise $2d$ analogs of the $4d$ and $6d$ results in \cite{Ferko:2023wyi} and \cite{Ferko:2024zth}.

The second motivation for this study concerns quantization. The general form (\ref{interaction_function}) of an interacting theory is convenient for canonical quantization, since the dependence on time derivatives is fixed and thus the definition of the conjugate momenta is unaffected by the interaction function. One can study the quantization of theories in this class in a uniform way, at least for cases that admit a controlled perturbative expansion which makes calculations tractable. When we consider the quantization of a one-parameter family of theories defined by interaction functions $V^{(\lambda)}$ that satisfy a differential equation of the form (\ref{classical_flow}), we will say that we are studying ``quantization along the classical flow.''

We will be especially interested in quantization along the flow driven by the function 
\begin{align}\label{root_TT_flow}
    \partial_\gamma V^{(\gamma)} = \mathcal{R} \left[ T_{\mu \nu} \right] = \frac{1}{\sqrt{2}} \sqrt{ T^{\mu \nu} T_{\mu \nu} - \frac{1}{2} \left( T^\mu{}_\mu \right)^2 } \, , 
\end{align}
where we suppress the dependence of $T_{\mu \nu}$ on the flow parameter $\gamma$. This non-analytic combination $\mathcal{R}$ is the two-dimensional root-$T\overline{T}$ operator \cite{Ferko:2022cix}, which is the unique marginal combination of stress tensors that defines a classical flow equation which commutes with the irrelevant $T\overline{T}$ flow in $2d$. The root-$T\overline{T}$ deformation shares some of the remarkable properties of the $T\overline{T}$ deformation, such as preserving classical integrability in many examples \cite{Borsato:2022tmu} and admitting a holographic interpretation in terms of modified boundary conditions for $\mathrm{AdS}_3$ gravity \cite{Ebert:2023tih}. It also plays a role in classical flows for $3d$ gauge theories \cite{Ferko:2023sps} and has connections to $\text{BMS}_3$ symmetry and ultra/non-relativistic limits of $2d$ CFTs \cite{Rodriguez:2021tcz,Bagchi:2022nvj,Tempo:2022ndz}.

Another motivation for studying this operator is that the corresponding commuting $T\overline{T}$-like and root-$T\overline{T}$-like flows in four spacetime dimensions, with the initial condition given by the free Maxwell Lagrangian, were shown in \cite{Conti:2018jho,Babaei-Aghbolagh:2022uij,Ferko:2022iru,Conti:2022egv,Ferko:2023ruw} to produce an interesting family of gauge theories referred to as ModMax-Born-Infeld, which was first written down in \cite{Bandos:2020hgy}. This family depends on two parameters $\lambda$ and $\gamma$. When $\gamma$ is taken to zero, the theory reduces to the $4d$ Born-Infeld model which gives an effective description of the gauge dynamics on a D3-brane. As $\lambda \to 0$, one recovers the so-called Modified Maxwell or ModMax theory, which is the unique conformally invariant and electric-magnetic duality-invariant extension of the Maxwell theory \cite{Bandos:2020jsw}. This theory can be supersymmetrized \cite{Bandos:2021rqy,Kuzenko:2021cvx,Kuzenko:2021pqm}
and the entire class of ModMax-Born-Infeld theories can be lifted to a similar family of $6d$ chiral tensor theories \cite{Bandos:2020hgy} which also satisfies commuting stress tensor flow equations \cite{Ferko:2024zth}. For an introduction to these and other theories of non-linear electrodynamics, see the lecture notes \cite{Sorokin:2021tge}.

Although several classical aspects of the ModMax theory (and its ModMax-Born-Infeld extension) have been studied \cite{Dassy:2021ulu,Nastase:2021uvc,Escobar:2021mpx,Lechner:2022qhb,Neves:2022jqq}, the quantization of this model appears to be more subtle because the Lagrangian is non-analytic around $F_{\mu \nu} = 0$. One strategy is to perform perturbative quantization of this theory around a non-zero background for the field strength \cite{cianThesis}.\footnote{Another approach would be to use heat kernel techniques. We are grateful to Sergei Kuzenko and Dmitri Sorokin for discussions on this topic and for informing us of their unpublished results. See also \cite{pinelliThesis} for a Master's thesis which computes the one-loop effective action for ModMax using such techniques.}
Another approach is to look for lower-dimensional analogs of the ModMax theory, which one might hope are simpler to quantize. The most extreme case is to dimensionally reduce the Modified Maxwell theory all the way down to $(0+1)$ spacetime dimensions, which yields a theory of particle mechanics known as the ModMax oscillator \cite{Garcia:2022wad,Ferko:2023ozb} that can be quantized exactly \cite{Ferko:2023iha}. An intermediate case is to reduce ModMax from $4d$ to $2d$, which was done in \cite{Conti:2022egv}, and this reduction yields precisely the same theory that one obtains by deforming a collection of free scalars by the $2d$ root-$T\overline{T}$ flow \cite{Ferko:2022cix,Babaei-Aghbolagh:2022leo}. This ``Modified Scalar'' theory is the model whose quantization we consider in the present work.

For one non-chiral boson, or one left-moving and one-right moving chiral boson, the Modified Scalar theory collapses to a free massless scalar with a re-scaled kinetic term, but for multiple scalars, the theory is non-trivial. As we will see later, the Modified Scalar theory with a general number of scalars may also be related to a free theory by a series of more complicated, non-local field redefinitions; similar field redefinitions, and related non-local ``dressed'' operators, have also played a role in the study of $T\overline{T}$ flows \cite{Kruthoff:2020hsi,Guica:2020uhm,Kraus:2021cwf,Ebert:2022cle,Kraus:2022mnu}.

One of our goals in studying the quantization of this model is to test a flow equation for certain energies in a root-$T\overline{T}$ deformed CFT, which was obtained via a holographic analysis in \cite{Ebert:2023tih}. Under some assumptions, this equation predicts that the deformed energy $E_\gamma$ associated with a seed CFT state that has undeformed energy $E_0$ and momentum $P_0$ is
\begin{align}\label{zero_mode_formula}
    E_\gamma = E_0 \cosh ( \gamma ) + \sqrt{ E_0^2 - P_0^2 } \sinh ( \gamma ) \, .
\end{align}
This formula was derived for states dual to BTZ black holes in $\mathrm{AdS}_3$ with mass $M \geq 0$ and spin $| J | \leq M$, which correspond to \emph{constant} stress tensor backgrounds. We will, therefore, refer to the flow equation (\ref{zero_mode_formula}) as the ``zero-mode energy formula''  since it applies to states of a CFT on a cylinder whose stress tensors are constant along the circular direction (that is, the formula applies to the zero mode of the stress tensor). It would be quite unusual if this energy formula held \emph{universally}, even for states whose stress tensors are spatially varying. And indeed, we will see explicitly in this work that the zero-mode energy formula fails for states with such spatial gradients. One might therefore think of (\ref{zero_mode_formula}) as the first term in a gradient expansion, which is corrected by terms that depend on derivatives $\partial T$.\footnote{The idea of performing such a gradient expansion is philosophically similar to the strategy adopted in hydrodynamics or the fluid-gravity correspondence \cite{Bhattacharyya:2007vjd} (see \cite{Rangamani:2009xk} for a review).}

The key ingredient in our check of the energy formula (\ref{zero_mode_formula}), which allows us to resolve the square root and perform a perturbative analysis, is to consider a certain large-momentum limit and expand in powers of $\frac{1}{p}$. Although this approach involves a specific choice of background around which to expand, one could expand around \emph{any} field configuration for which the gradients of the scalars are non-vanishing, since the combination of stress tensors (\ref{root_TT_flow}) which appears in the classical Lagrangian for the Modified Scalar theory is only non-analytic around zero-energy configurations. We will also present a related analysis which involves expanding around a general classical background for $N = \bar{N}$, in which case the equal number of chiral and anti-chiral bosons can be assembled into a manifestly Lorentz invariant theory of $N$ non-chiral bosons, and compute loop corrections to the Modified Scalar action. This offers further insight into the perturbative quantization of this model.

This chapter is organized as follows. In section \ref{sec:classical}, we compute the stress tensor for a generic interacting chiral boson theory and study classical properties of flows driven by functions of $T_{\mu \nu}$, such as preservation of the Lorentz invariance condition. We then give a complementary perspective on such chiral boson theories in section \ref{sec:cs}, interpreting them as the boundary duals to $U(1)$ Chern-Simons gauge theories, and we show that deformations such as root-$T\overline{T}$ can be implemented using certain modified boundary conditions for the bulk gauge fields. In section \ref{sec:quantization}, we review general machinery for the canonical quantization of first-order systems like (\ref{interaction_function}) along classical stress tensor flows using a mode expansion; we then specialize to quantization along the root-$T\overline{T}$ flow and study the cases of $\left( N , \bar{N} \right) = ( 1, 1 )$ and $\left( N , \bar{N} \right) = ( 2, 1 )$ in detail. In section \ref{sec:cian}, we perform a diagrammatic analysis of quantum corrections along the root-$T\overline{T}$ flow for a deformed theory of $N = \bar{N}$ \emph{non-chiral} bosons, using the background field method. Finally, section \ref{sec:Conclusion&Outlook1} summarizes our results and outlines some interesting future directions. An order-by-order analysis for more general stress tensor flows is presented in appendix \ref{app:TTn}, and the computational steps used to evaluate certain Feynman diagrams in dimensional regularization have been relegated to appendix \ref{app:feynman}.

\section{Classical Stress Tensor Flows for Chiral Boson Theories}\label{sec:classical}

In this section, we will discuss some generalities about classical deformations of interacting chiral boson theories which are driven by functions of the energy-momentum tensor. Quite generally, we refer to any differential equation for the Lagrangian which takes the form
\begin{align}\label{stress_tensor_flow_defn}
    \frac{\partial \mathcal{L}^{(\lambda)}}{\partial \lambda} = f \left( T_{\alpha \beta}^{(\lambda)} , \lambda \right) \, , 
\end{align}
along with an initial condition $\mathcal{L}^{(\lambda = 0)} = \mathcal{L}^{(0)}$, as a stress tensor flow. We emphasize that the function $f$ is a Lorentz scalar constructed from the Hilbert stress tensor associated with the Lagrangian $\mathcal{L}^{(\lambda)}$, and not with the undeformed theory $\mathcal{L}^{(0)}$. For theories that can be coupled to gravity using only the metric tensor $g_{\alpha \beta}$, the stress tensor is given by
\begin{align}\label{hilbert_defn_standard}
    T_{\alpha \beta} = - \frac{2}{\sqrt{-g}} \frac{\delta S}{\delta g^{\alpha \beta}} = - 2 \frac{\partial \mathcal{L}}{\partial g^{\alpha \beta}} + g_{\alpha \beta} \mathcal{L} \, .
\end{align}
However, for theories involving fermions or the chiral bosons of interest in this work, the standard definition (\ref{hilbert_defn_standard}) is not sufficient. We will instead need to work in a tetrad formalism, introducing vielbein fields (or frame fields) $E^a{}_\alpha$ so that
\begin{align}
    g_{\alpha \beta} =E^a{}_\alpha E^b{}_\beta \eta_{ab} \, .
\end{align}
We will use Greek symbols such as $\alpha$ and $\beta$ to refer to curved\footnote{We use the term ``curved'' for spacetime indices, even when we set the spacetime metric $g_{\alpha \beta}$ to the flat Minkowski metric $\eta_{\alpha \beta}$, to distinguish them from ``flat'' tangent space indices like those on $\eta_{ab}$.} indices in the two-dimensional spacetime with metric $g_{\alpha \beta}$ on which our fields are defined, in contrast with early Latin letters like $a$ and $b$ which refer to the flat tangent-space indices that are raised and lowered with the Minkowski metric $\eta_{ab}$. These are not to be confused with the lowercase middle Latin symbols like $i$ which are used to index the chiral scalars $\phi^i$, or their antichiral variants $\bar{i}$ which are decorated with a bar and label the anti-chiral scalars $\bar{\phi}^{\overline{i}}$.

We also define $E = \det \left( E^a{}_\alpha \right) = \sqrt{|g|}$. Because this determinant is non-vanishing, the matrix $E^a{}_\alpha$ has an inverse, which we refer to as the inverse vielbein and write as $E^\alpha{}_a$. This inverse frame field obeys
\begin{align}
 E^a{}_\alpha E^\alpha{}_b = \delta^a{}_b  \, , \qquad E^\alpha{}_a E^a{}_\beta = \delta^\alpha{}_\beta \, ,
\end{align}
and similarly
\begin{align}
E^\alpha{}_a E^\beta{}_b g_{\alpha \beta} = \eta_{ab} \, .
\end{align}
Within the tetrad formalism, the appropriate generalization of the Hilbert stress tensor with one curved and one flat index is defined by
\begin{align}\label{stress_tensor_flat_curved}
    T_\beta{}^a = - \frac{1}{E} \frac{\delta S}{\delta E^\beta{}_a} \, .
\end{align}
All tangent space indices can be converted to spacetime indices, or vice-versa, by contracting with vielbeins or inverse vielbeins as needed. For instance, the conventional stress tensor with two curved indices is then
\begin{align}
    T_{\alpha \beta} = T_\alpha{}^a E^\gamma{}_a g_{\gamma \beta} \, .
\end{align}
The tetrad formalism will allow us to compute the energy-momentum tensor and define stress tensor flows for an arbitrary interacting chiral boson theory of the type in equation (\ref{interaction_function}). We will perform the coupling to vielbeins in such a way that the stress tensor is automatically symmetric, $T_{\alpha \beta} = T_{\beta \alpha}$, but this is not sufficient to guarantee that the theory is invariant under boosts; for a generic choice of the interaction function $V$, the theory is \emph{not} Lorentz-invariant. In this work, we will be primarily interested in theories which \emph{do} enjoy Lorentz invariance, although this Lorentz symmetry will not be manifest within this formalism. Therefore, we will now pause to discuss the non-manifest Lorentz invariance of these models, including the conditions this imposes upon the interaction function $V$ and the connection between Lorentz symmetry and stress tensor flows.

\subsection{Lorentz invariance}\label{sec:lorentz}

We begin by reviewing one way to see the non-manifest Lorentz invariance of the simplest theory within the class of interest, the Floreanini-Jackiw action describing a single chiral boson. Although this is a well-known story, the discussion will fix our notation and set the stage for the analysis of Lorentz invariance with more general interaction functions.

\emph{One free chiral boson}

Much like the electric-magnetic duality invariance of the $4d$ Maxwell theory, which is a symmetry of the equations of motion but not of the action itself, the Lorentz symmetry of the chiral boson theories we study here will be easier to understand at the level of the equations of motion. We illustrate this simple principle beginning with the action (\ref{fj_action}), which we rewrite for convenience:
\begin{equation}
\label{eq:FJR}
    S= \frac{1}{2} \int d^2x \left( \dot{\phi} \phi' - \phi^{\prime 2} \right) \, .
\end{equation}
Here, we have defined
\begin{align}
    \dot{\phi} = \partial_t \phi = \frac{\partial \phi}{\partial x^0} \, ,  \quad \phi' = \partial_\theta \phi = \frac{\partial \phi}{\partial x^1} \, , 
\end{align}
to ease notation. Now consider an infinitesimal Lorentz boost $\Lambda^\alpha{}_\beta = \delta^\alpha{}_\beta + \omega^\alpha{}_\beta $ with parameter $\omega_{0 1} = - \omega_{10} = \epsilon$. In this section, we work in Lorentzian signature with spacetime metric $\eta_{\alpha \beta} = \left[ \begin{smallmatrix} -1 & 0 \\ 0 & 1 \end{smallmatrix} \right]$. The change in the components of $\partial^\alpha \phi$ is
\begin{align}
    \delta \left( \partial^\alpha \phi \right) = \omega^\alpha{}_\beta \partial^\beta \phi \, ,
\end{align}
and thus the components of the covector $\partial_\alpha \phi = ( \dot{\phi} , \phi' )$ transform as
\begin{equation}
\label{eq:Iboost}
    \dot{\phi} \rightarrow \dot{\phi} + \epsilon \phi'\,, \quad \phi' \rightarrow \phi' + \epsilon \dot{\phi} \, .
\end{equation}
The change in the action \eqref{eq:FJR} is therefore

\begin{equation}
    \delta S = \frac{\epsilon}{2} \int d^2x \left( \dot{\phi} - \phi' \right)^2 + \mathcal{O}(\epsilon^2) \, .
\end{equation}
This is not an off-shell total derivative, so it is not manifest that this transformation is a symmetry of the theory. However, this property is more transparent if we work directly with the equations of motion. The Euler-Lagrange equation associated with (\ref{eq:FJR}) is
\begin{equation}
    \dot{\phi}' - \phi'' = 0 \, ,
\end{equation}
where $\dot{\phi}' = \partial_t \partial_\theta \phi$. This equation of motion can be expressed as $\partial_\theta \left( \dot{\phi} - \phi' \right) = 0$, which means that the quantity $\dot{\phi} - \phi'$ is independent of the spatial coordinate $\theta$:
\begin{align}\label{eom_no_gauge}
    \dot{\phi} - \phi' = f ( t ) \, .
\end{align}
The time-dependent function $f ( t )$ can be thought of as a choice of gauge, which is not physically meaningful. Indeed, suppose that we transform the function $\phi$ by
\begin{align}\label{delta_phi_gauge}
    \delta \phi = h ( t ) 
\end{align}
for a general time-dependent function $h$. Then $\delta \dot{\phi} = \dot{h}$ and $\delta \phi' = 0$, so the change in the Floreanini-Jackiw action is
\begin{align}
    \delta S = \frac{1}{2} \int d^2 x \, \left( \dot{h} \phi' \right) = \frac{1}{2} \int d^2 x \, \partial_\theta \left( \dot{h} \phi \right) \, , 
\end{align}
which is an integral of a total spatial derivative, and thus the action is unchanged. Therefore, given any solution to the equations of motion which takes the form (\ref{eom_no_gauge}), we are always free to perform a gauge transformation (\ref{delta_phi_gauge}) with
\begin{align}
    h ( t ) = \int^{t} f ( t' ) \, dt' \, ,  \qquad \dot{h} ( t ) = f ( t ) \, , 
\end{align}
which has the effect of eliminating the function $f ( t )$ on the right side of (\ref{eom_no_gauge}), and thus brings the equation of motion to the form
\begin{align}\label{eom_gauge_choice}
    \dot{\phi} - \phi' = 0 \, .
\end{align}
We will always work in the gauge (\ref{eom_gauge_choice}) in what follows. If we write equation (\ref{eom_gauge_choice}) as
\begin{align}
    \mathcal{E} ( \dot{\phi} , \phi' ) = 0 \, , \qquad \mathcal{E} = \dot{\phi} - \phi' \, , 
\end{align}
then acting with a Lorentz transformation on this quantity $\mathcal{E}$ gives
\begin{align}
    \delta \mathcal{E} = \delta \left( \dot{\phi} - \phi' \right) = - \epsilon \left( \dot{\phi} - \phi' \right) = - \epsilon \mathcal{E} \, .
\end{align}
That is, the variation of the equation of motion is proportional to the equation of motion itself. This means that, on the mass shell, the equations of motion are invariant under Lorentz transformations, which we write as
\begin{align}
    \delta \mathcal{E} \simeq 0 \, , 
\end{align}
where the symbol $\simeq$ means ``equal when the fields satisfy their equations of motion.'' This is sufficient for the theory to enjoy Lorentz invariance.

From this simple exercise, we see that the Floreanini-Jackiw theory of 
a single chiral boson does indeed exhibit non-manifest Lorentz invariance. This discussion also motivates a couple of definitions. We say that any function $\mathcal{O}$ of the fields and their derivatives is a \emph{Lorentz-invariant function} if $\delta \mathcal{O} \simeq 0$, that is, if the quantity $\mathcal{O}$ is invariant under Lorentz transformations when the fields satisfy their equations of motion. Likewise, we say that a Lagrangian $\mathcal{L}$ defines a \emph{Lorentz-invariant theory} if the Euler-Lagrange equations associated with $\mathcal{L}$ can be written as $\mathcal{E} = 0$ where $\mathcal{E}$ is a Lorentz-invariant function. 

\emph{Multiple interacting bosons}

We now promote the action to depend on $N$ chiral bosons and $\bar{N}$ anti-chiral bosons. A general theory with interactions that depend on spatial derivatives of the fields is\footnote{In this chapter, we do not consider higher-derivative interactions.}
\begin{equation}\label{interacting_again}
    S = \int d^2x \left( \frac{1}{2} (\dot{\phi}^i \phi^{\prime \, i} - \dot{\bar{\phi}} {}^{\overline{i}} \, \bar{\phi}^{\prime \, \overline{i}}  ) - V ( \phi', \bar{\phi}' ) \right) \, ,
\end{equation}
where we suppress indices on the fields in the argument of the interaction function $V$. Following the notation of the $N = 1$ analysis above, we can write the equations of motion for this model as a collection of equations $\mathcal{E}^i = 0$ and $\bar{\mathcal{E}}^{\overline{i}} = 0$, where
\begin{align}\label{interacting_eom}
    \mathcal{E}^i = \dot{\phi}^i - \frac{\partial V}{\partial \phi^{\prime \, i}} \, , \qquad \bar{\mathcal{E}}^{\overline{i}} = \dot{\bar{\phi}}{}^{\overline{i}} + \frac{\partial V}{\partial \bar{\phi}^{\overline{i}}} \, .
\end{align}
Note that we do not distinguish between upstairs and downstairs $i, j$ and $\overline{i}$, $\overline{j}$ indices on the scalars, instead choosing index placement for typographical convenience. In expressing the equations of motion as the vanishing of the quantities (\ref{interacting_eom}), we have also implicitly chosen the analog of the gauge $h(t) = 0$, as in the discussion around equation (\ref{eom_gauge_choice}) for the case of one chiral boson.

Let us again consider a Lorentz boost parameterized by $\omega_{01} = - \omega_{10} = \epsilon$. All of the fields transform in the same way as before:
\begin{equation}
    \dot{\phi}^i \rightarrow \dot{\phi}^i + \epsilon \phi^{\prime \, i} \, , \quad \phi^{\prime \, i} \rightarrow \phi^{\prime \, i} + \epsilon \dot{\phi}^i \, , \quad \dot{\bar{\phi}}{}^{\overline{i}} \rightarrow \dot{\bar{\phi}}{}^{\overline{i}} + \epsilon \bar{\phi}^{\prime \, \overline{i}} \,, \quad \bar{\phi}^{\prime \, \overline{i}} \rightarrow \bar{\phi}^{\prime \, \overline{i}} + \epsilon \dot{\bar{\phi}}{}^{\overline{i}} \, .
\end{equation}
We now ask: under what conditions on the interaction function $V$ will the action (\ref{interacting_again}) define a Lorentz-invariant theory, which means that $\delta \mathcal{E}^i \simeq 0$ and $\delta \bar{\mathcal{E}}^{\overline{i}} \simeq 0$ under this Lorentz transformation? The variation of the chiral equations of motion is
\begin{align}
    \delta \mathcal{E}^i &= \delta \dot{\phi}^i - \frac{\partial^2 V}{\partial \phi^{\prime \, i} \, \partial \phi^{\prime \, j}} \, \delta \phi^{\prime \, j } - \frac{\partial^2 V}{\partial \phi^{\prime \, i} \, \partial \bar{\phi}^{\prime \, \overline{j}}} \, \delta \bar{\phi}^{\prime \, \overline{j} } \nonumber \\
    &= \epsilon \phi^{\prime \, i} - \epsilon V_{i j} \dot{\phi}^j - \epsilon V_{i \overline{j}}  \dot{\bar{\phi}}{}^{\overline{j}} \nonumber \\
    &\simeq \epsilon \left[ \phi^{\prime \, i} - V_{ij} V_j + V_{i \overline{j}} V_{\overline{j}} \right] \, ,
\end{align}
where in the second step we have introduced the notation
\begin{align}
    V_i = \frac{\partial V}{\partial \phi^{\prime \, i}} \, , \quad V_{ij} = \frac{\partial^2 V}{\partial \phi^{\prime \, i} \, \partial \phi^{\prime \, j}} \, , \quad V_{i \overline{j}} = \frac{\partial^2 V}{\partial \phi^{\prime \, i} \, \partial \bar{\phi}^{\prime \, \overline{j}}} \, , 
\end{align}
and so on, and in the third line, we have replaced the time derivatives $\dot{\phi}^i$, $\dot{\bar{\phi}}{}^{\overline{j}}$ using the equations of motion and therefore used the on-shell equality symbol $\simeq$. An identical calculation for the anti-chiral equations of motion gives
\begin{align}
    \delta \bar{\mathcal{E}}^{\overline{i}} \simeq \epsilon \left[ \bar{\phi}^{\prime \, \overline{i}} - V_{\overline{i} \overline{j}} V_{\overline{j}} + V_{\overline{i} j} V_j \right] \, .
\end{align}
Therefore, for the quantities $\mathcal{E}^i$ and $\bar{\mathcal{E}}^{\overline{i}}$ to be Lorentz-invariant functions, we must impose the two conditions
\begin{align}
    \phi^{\prime \, i } + V_{i \overline{j}} V_{\overline{j}} = V_{ij} V_j \, , \qquad \bar{\phi}^{\prime \, \overline{i}} + V_{\overline{i} j} V_j = V_{\overline{i} \overline{j}} V_{\overline{j}} \, .
\end{align}
It is convenient to write these two equations in terms of the derivatives of products,
\begin{align}\label{lorentz_intermediate}
    \phi^{\prime \, i } + \frac{1}{2} \partial_i \left( V_{\overline{j}} V_{\overline{j}} \right) = \frac{1}{2} \partial_i \left( V_j V_j \right) \, , \qquad \bar{\phi}^{\prime \, \overline{i}} + \frac{1}{2} \partial_{\overline{i}} \left( V_j V_j \right) = \frac{1}{2} \partial_{\overline{i}} \left( V_{\overline{j}} V_{\overline{j}} \right) \, ,
\end{align}
where the repeated $j$, $\overline{j}$ indices are summed and where $\partial_i = \frac{\partial}{\partial \phi^{\prime \, i}}$, $\partial_{\overline{i}} = \frac{\partial}{\partial \bar{\phi}^{\prime \, \overline{i}}}$.

We can now integrate the first of the equations (\ref{lorentz_intermediate}) with respect to $\phi^{\prime \, i}$ and the second with respect to $\bar{\phi}^{\prime \, \overline{i}}$ to find
\begin{align}\label{lorentz_intermediate_two}
    \left( \phi^{\prime \, i} \right)^2 + V_{\overline{j}} V_{\overline{j}} = V_{j} V_{j} + C^i ( \phi^{\prime \, k \neq i} , \bar{\phi}^{\prime \, \overline{k}} ) \, , \qquad \left( \bar{\phi}^{\prime \, \overline{i}} \right)^2 + V_j V_j = V_{\overline{j}} V_{\overline{j}} + \bar{C}^{\overline{i}} ( \phi^{\prime \, k} , \bar{\phi}^{\prime \, \overline{k} \neq \overline{i}} ) \, .
\end{align}
Here we have introduced two integration constants, $C^i$ which is independent of $\phi^{\prime \, i}$ and $\bar{C}^{\overline{i}}$ which is independent of $\bar{\phi}^{\prime \, \overline{i}}$. Also note that equation (\ref{lorentz_intermediate_two}) holds separately for each fixed $i$ and $\overline{i}$; the quantity $\left( \phi^{\prime \, i} \right)^2$ is the square of one such fixed $\phi^{\prime \, i}$, and is not summed on $i$. We can fix these integration constants by noting that the choice of interaction function
\begin{align}\label{free_interaction}
    V ( \phi , \bar{\phi} ) = \frac{1}{2} \left( \phi^{\prime \, j} \phi^{\prime \, j} + \bar{\phi}^{\prime \, \overline{j}} \bar{\phi}^{\prime \, \overline{j}} \right) \, ,
\end{align}
which is just a sum of non-interacting chiral and anti-chiral bosons, must necessarily satisfy the Lorentz-invariance condition. This will be true if we choose
\begin{align}
    C^i = \bar{\phi}^{\prime \, \overline{j}} \bar{\phi}^{\prime \, \overline{j}} \, + \, \sum_{k \neq i} \phi^{\prime \, k} \phi^{\prime \, k} , \qquad \bar{C}^{\overline{i}} = \phi^{\prime \, j} \phi^{\prime \, j} \, + \, \sum_{\overline{k} \neq \overline{i}} \bar{\phi}^{\prime \, \overline{k}} \bar{\phi}^{\prime \, \overline{k}} \, ,
\end{align}
which means that the two equations in (\ref{lorentz_intermediate_two}) are proportional to one another, and we are left with the single condition
\begin{align}\label{lorentz_intermediate_three}
    \phi^{\prime \, j} \phi^{\prime \, j} - \bar{\phi}^{\prime \, \overline{j}} \bar{\phi}^{\prime \, \overline{j}} = V_j V_j - V_{\overline{j}} V_{\overline{j}} \, , 
\end{align}
for Lorentz invariance. Suppose that we now further assume that the interaction function is invariant under $O ( N )$ rotations of the $N$ chiral fields and $O ( \bar{N} )$ rotations of the $\bar{N}$ anti-chiral fields. This means that we can parameterize $V$ as a function of the two invariants\footnote{The invariant $S$ should not be confused with the action $S = \int d^2 x \, \mathcal{L}$; we trust that the reader can distinguish between the two based on context.}
\begin{align}\label{S_and_P_invariants}
    S = \frac{1}{2} \left( \phi^{\prime \, j} \phi^{\prime \, j} + \bar{\phi}^{\prime \, \overline{j}} \bar{\phi}^{\prime \, \overline{j}} \right) \, , \qquad P = \frac{1}{2} \left( \phi^{\prime \, j} \phi^{\prime \, j} - \bar{\phi}^{\prime \, \overline{j}} \bar{\phi}^{\prime \, \overline{j}} \right) \, .
\end{align}
Note that, for the theory defined by the free interaction function (\ref{free_interaction}), the quantities $S$ and $P$ represent the total Hamiltonian density and momentum density, respectively. In terms of these variables, the condition (\ref{lorentz_intermediate_three}) can be written as
\begin{align}\label{lorentz_pde_final}
    V_S^2 + \frac{2 S}{P} V_S V_P + V_P^2 = 1 \, .
\end{align}
Partial differential equations of the schematic form (\ref{lorentz_pde_final}) have appeared in many contexts. Most directly relevant for this analysis, precisely the same differential equation appears as the condition for Lorentz invariance of the phase space actions for theories of self-dual electrodynamics in $d = 4$ or for chiral tensor theories in $d = 6$; see, for instance, sections 2.2 and 2.3 of \cite{Bandos:2020hgy} for these two cases, respectively. Our condition (\ref{lorentz_pde_final}) is merely the $2d$ version of these results, in the case where one considers arbitrary numbers of chiral and anti-chiral bosons. Note that, in the case $\bar{N} = 0$ which describes only chiral bosons, the two invariants (\ref{S_and_P_invariants}) collapse to
\begin{align}
    S = P \, , 
\end{align}
so that $V$ is a function of one variable, and the constraint (\ref{lorentz_intermediate_three}) simplifies to
\begin{align}
    \phi^{\prime \, j} \phi^{\prime \, j} = V_j V_j \, , 
\end{align}
or in terms of the variable $S = \frac{1}{2} \phi^{\prime \, j} \phi^{\prime \, j}$,
\begin{align}\label{only_S}
    V_S = 1 \, .
\end{align}
This means that the only solution is the free case, $V = S = \frac{1}{2} \phi^{\prime \, j} \phi^{\prime \, j}$, in accordance with known results. The same conclusion holds for only anti-chiral bosons, $N = 0$ but $\bar{N} > 0$.

A similar partial differential equation, which differs only by signs, occurs as the condition for a Lagrangian for $4d$ non-linear electrodynamics to have equations of motion that are invariant under electric-magnetic duality rotations. In this case, the appropriate PDE reads
\begin{align}\label{duality_pde}
    \mathcal{L}_S^2 - \frac{2 S}{P} \mathcal{L}_S \mathcal{L}_P - \mathcal{L}_P^2 = 1 \, , 
\end{align}
where now $S = - \frac{1}{4} F_{\mu \nu} F^{\mu \nu}$ and $P = - \frac{1}{4} F_{\mu \nu} \widetilde{F}^{\mu \nu}$ are the two independent Lorentz scalars that can be constructed from the field strength $F_{\mu \nu}$, and $\widetilde{F}_{\mu \nu}$ denotes the Hodge dual of $F_{\mu \nu}$. This version of the differential equation, with the signs as in (\ref{duality_pde}), also appears as the condition for a certain class of non-linear sigma models in $d = 2$ to have equations of motion which are equivalent to the flatness of a Lax connection which takes a prescribed form \cite{Borsato:2022tmu} (see equations (7.3) - (7.5) of \cite{Ferko:2023wyi} for the definitions of $S$ and $P$ in this case).

In either presentation, with the choice of signs in (\ref{lorentz_pde_final}) or the one in (\ref{duality_pde}), this differential equation has many solutions besides the free one. For instance, equation (\ref{lorentz_pde_final}) admits the two-parameter family of solutions
\begin{align}\label{scalar_modified_nambu_goto}
    V ( S, P ; \gamma, \lambda ) = \frac{1}{\lambda} \left( \sqrt{ 1 + 2 \lambda \left( \cosh ( \gamma ) S + \sinh ( \gamma ) \sqrt{ S^2 - P^2 } \right) + \lambda^2 P^2 } - 1 \right) \, .
\end{align}
This family of interaction functions is the $2d$ chiral boson analog of the two-parameter family of $4d$ ModMax-Born-Infeld gauge theories, which we mentioned in the introduction. As in the $4d$ case, the function $V$ of equation (\ref{scalar_modified_nambu_goto}) satisfies two commuting flow equations which relate $\partial_\lambda V$ and $\partial_\gamma V$ to an irrelevant $T\overline{T}$-like and a marginal root-$T\overline{T}$-like operator built from the energy-momentum tensor of the model, respectively:
\begin{align}\label{two_nice_flows}
    \frac{\partial V}{\partial \lambda} &= - \mathcal{O}_{TT} = - \frac{1}{4} \left( T^{\alpha \beta} T_{\alpha \beta} - \left( T^\alpha{}_\alpha \right)^2 \right) \, , \nonumber \\ \frac{\partial V}{\partial \gamma} &= - \mathcal{R} = - \frac{1}{\sqrt{2}} \sqrt{ T^{\alpha \beta} T_{\alpha \beta} - \frac{1}{2} \left( T^\alpha{}_\alpha \right)^2 } \, .
\end{align}
This example illustrates that, at least in this case, solutions to the differential equation (\ref{lorentz_pde_final}) can be obtained by deforming the interaction function by Lorentz-invariant quantities, such as Lorentz scalars constructed from $T_{\mu \nu}$. This statement applies quite generally to any deformation of $V$ by a Lorentz-invariant function, as we describe next.

\emph{Lorentz-invariant functions}

In the preceding discussion, we derived a condition on the function $V$ (equation (\ref{lorentz_pde_final})) which guarantees that this interaction function describes a Lorentz-invariant theory. By definition, this means that the equations of motion $\mathcal{E}^i$, $\bar{\mathcal{E}}^i$ are Lorentz-invariant functions. One might ask, more generally, given an arbitrary function $\mathcal{O} ( S, P )$ which depends on the two combinations $S$ and $P$ defined in (\ref{S_and_P_invariants}), under what conditions is $\mathcal{O}$ a Lorentz-invariant function? That is, for which operators $\mathcal{O}$ is $\delta \mathcal{O} \simeq 0$, where $\delta$ is a Lorentz transformation?

This question can be answered using a similar calculation as the one above. One has
\begin{align}
    \delta \mathcal{O} ( S, P ) &= \mathcal{O}_S \delta S + \mathcal{O}_P \delta P \nonumber \\
    &= \mathcal{O}_S \left( \phi^{\prime \, j} \, \delta \phi^{\prime \, j}  + \bar{\phi}^{\prime \, \overline{j}} \, \delta \bar{\phi}^{\prime \, \overline{j}} \right) + \mathcal{O}_P \left( \phi^{\prime \, j} \delta \phi^{\prime \, j}  - \bar{\phi}^{\prime \, \overline{j}} \, \delta \bar{\phi}^{\prime \, \overline{j}} \right) \, ,
\end{align}
where subscripts represent partial derivatives with respect to the argument. On-shell, one has the variations
\begin{align}
    \delta \phi^{\prime \, j} = \epsilon \dot{\phi}^{j} \simeq \epsilon V_j \, , \qquad \delta \bar{\phi}^{\prime \, j} = \epsilon \dot{\bar{\phi}}{}^{\overline{j}} \simeq - V_{\overline{j}} \, , 
\end{align}
and thus one finds
\begin{align}
    \delta \mathcal{O} \simeq \epsilon \mathcal{O}_S \left( \phi^{\prime \, j} V_j - \bar{\phi}^{\prime \, \overline{j}} V_{\overline{j}} \right) + \epsilon \mathcal{O}_P \left( \phi^{\prime \, j} V_j + \bar{\phi}^{\prime \, \overline{j}} V_{\overline{j}} \right) \, .
\end{align}
Expressing the derivatives of $V$ in terms of $V_S$ and $V_P$ using
\begin{align}
    V_j = \left( V_S + V_P \right) \phi^{\prime \, j} \, , \qquad V_{\overline{j}} = \left( V_S - V_P \right) \bar{\phi}^{\prime \, \overline{j}} \, , 
\end{align}
we conclude that $\delta \mathcal{O} \simeq 0$ if and only if
\begin{align}\label{lorentz_invariant_function}
    V_S \mathcal{O}_S + \frac{S}{P} \left( V_S \mathcal{O}_P + V_P \mathcal{O}_S \right) + V_P \mathcal{O}_P = 0 \, .
\end{align}
It is easy to see that the condition (\ref{lorentz_invariant_function}) is identical to the constraint that one finds by expanding the Lorentz-invariance condition (\ref{lorentz_pde_final}) for a perturbed interaction function
\begin{align}
    V ( S, P ) \to V ( S, P ) + \lambda \mathcal{O} ( S , P ) \, , 
\end{align}
assuming that $V$ itself satisfies the Lorentz-invariance condition, and then demanding that the deformed interaction function preserve this condition (\ref{lorentz_pde_final}) to leading order in $\lambda$.

We conclude that linearized Lorentz-preserving deformations of a boost-invariant theory of chiral bosons, described by an interaction function $V$, are in one-to-one correspondence with Lorentz-invariant functions $\mathcal{O}$ within this same theory defined by $V$. Again, this result is the $2d$ analog of the corresponding statements about linearized deformations which preserve electric-magnetic duality invariance in $4d$ \cite{Ferko:2023wyi} or PST gauge invariance in $6d$ \cite{Ferko:2024zth}. As in those contexts, this extends to an all-orders result: given a one-parameter family of interaction functions $V 
( \lambda )$ with an initial condition $V_0 = V ( \lambda = 0 )$ which satisfies (\ref{lorentz_pde_final}), the entire family of functions $V ( \lambda )$ satisfies the Lorentz invariance condition if and only if
\begin{align}\label{all_orders_flow}
    \frac{\partial V(\lambda)}{\partial \lambda} = \mathcal{O}^{(\lambda)} \, , 
\end{align}
where at each value of $\lambda$, the function $\mathcal{O}^{(\lambda)}$ obeys the constraint (\ref{lorentz_invariant_function}) with respect to the interaction function $V ( \lambda )$ at the same value of $\lambda$. 

There are several ways to prove this claim, which we will not present in detail since they are similar to the $4d$ and $6d$ cases. One strategy is to first argue that any such family of Lorentz-invariant functions $\mathcal{O}^{(\lambda)}$ can be expressed in terms of Lorentz scalars constructed from $T_{\mu \nu}^{(\lambda)}$, as we will show shortly, and then to show that an all-orders flow of the form (\ref{all_orders_flow}) driven by a function of the stress tensor preserves the Lorentz-invariance condition, by following an inductive argument like that in appendix A.1 of \cite{Ferko:2023wyi}.

\subsection{Stress tensor for general interacting theory}

We now turn to the computation of the energy-momentum tensor for a generic member of our class of chiral boson theories. Contractions built from this stress tensor, such as $T^\mu{}_\mu$ and $T^{\mu \nu} T_{\mu \nu}$, are canonical examples of the Lorentz-invariant functions which yield Lorentz-preserving deformations (\ref{all_orders_flow}) of the interaction function -- and, in fact, \emph{any} such deformation can be expressed in terms of such stress tensor scalars, as we will see.

In order to calculate the stress tensor defined in (\ref{stress_tensor_flat_curved}), we will couple a general theory of chiral bosons to gravity in the vielbein formulation following the approach of \cite{Bastianelli:1989cu}, which demonstrated how to perform this coupling for the standard Floreanini-Jackiw boson with interaction function $V ( S, P ) = S$. In the case of a general interaction function, the corresponding Lagrangian including the vielbein couplings takes the form
\begin{align}\label{general_class_vielbeins}
    \mathcal{L} = \frac{1}{2} \left( G_{ij} \dot{\phi}^i \phi^{\prime \, j} - \bar{G}^{\overline{i} \overline{j}} \dot{\bar{\phi}}{}^{\overline{i}} \overline{\phi}^{\prime j} \right) - \left( E_\theta^- E_t^+ + E_t^- E_\theta^+ \right)  P - E V ( S, P ) + \mathcal{L}_{\text{top}} \, , 
\end{align}
where now $S$ and $P$ are coupled to the frame fields as
\begin{align}
    S = - \frac{1}{4 E_\theta^- E_\theta^+} \left( G_{ij} \phi^{\prime \, i} \phi^{\prime \, j} + \bar{G}_{\overline{i} \overline{j}} \bar{\phi}^{\prime \overline{i}} \bar{\phi}^{\prime \overline{j}} \right) \, , \quad P = - \frac{1}{4 E_\theta^- E_\theta^+} \left( G_{ij} \phi^{\prime \, i} \phi^{\prime \, j} - \bar{G}_{\overline{i} \overline{j}} \bar{\phi}^{\prime \overline{i}} \bar{\phi}^{\prime \overline{j}} \right) \, .
\end{align}
A few remarks are in order. We work in light-cone coordinates $x^a = x^{\pm}$ for the tangent space indices, so the vielbeins and inverse vielbeins carry one $( + , - )$ index and one $(t, \theta)$ index. After varying with respect to the vielbeins, we will set them to their flat-space values 
\begin{align}\label{flat_vielbeins}
    E^+_t = - E^+_\theta = E^-_\theta = E^-_t = \frac{1}{\sqrt{2}} \, ,
\end{align}
at the end of the calculation, which is appropriate for the light-cone tangent space metric $\eta_{ab} = \left[ \begin{smallmatrix} 0 & - 1 \\ - 1 & 0 \end{smallmatrix} \right]$. We have also introduced general target-space metrics $G_{ij} ( \phi )$ and $\bar{G}_{\overline{i} \overline{j}} ( \bar{\phi} )$ for the chiral and anti-chiral bosons, which does not affect the computation of the stress tensor. In equation (\ref{general_class_vielbeins}), we have allowed for the inclusion of a general topological term $\mathcal{L}_{\text{top}}$, which does not couple to the frame fields and which therefore drops out of the computation of $T_{\mu \nu}$. An example of such a topological term is a coupling to a target-space antisymmetric tensor field $B_{ij}$, $\bar{B}_{\overline{i} \overline{j}}$. In manifestly Lorentz-invariant notation, which is perhaps more familiar, such a coupling would be expressed as $B_{ij} \epsilon^{\alpha \beta} \partial_\alpha \phi^i \partial_\beta \phi^j$, and is independent of the metric.

Note that, in the special case $G_{ij} = \delta_{ij}$, $\bar{G}_{\overline{i} \overline{j}} = \delta_{\overline{i} \overline{j}}$, $\mathcal{L}_{\text{top}} = 0 $, and with the vielbeins equal to their flat-space values (\ref{flat_vielbeins}), the Lagrangian (\ref{general_class_vielbeins}) reduces to
\begin{equation}\label{interacting_again_again}
    \mathcal{L} = \frac{1}{2} (\dot{\phi}^i \phi^{\prime \, i} - \dot{\bar{\phi}} {}^{\overline{i}} \, \bar{\phi}^{\prime \, \overline{i}}  ) - V ( S, P  )  \, ,
\end{equation}
which agrees with (\ref{interacting_again_again}), and the quantities $S$ and $P$ become
\begin{align}\label{S_and_P_invariants_again}
    S = \frac{1}{2} \left( \phi^{\prime \, j} \phi^{\prime \, j} + \bar{\phi}^{\prime \, \overline{j}} \bar{\phi}^{\prime \, \overline{j}} \right) \, , \qquad P = \frac{1}{2} \left( \phi^{\prime \, j} \phi^{\prime \, j} - \bar{\phi}^{\prime \, \overline{j}} \bar{\phi}^{\prime \, \overline{j}} \right) \, ,
\end{align}
which agrees with (\ref{S_and_P_invariants}). 

It may come as a surprise that the kinetic terms in (\ref{general_class_vielbeins}), which involve $\dot{\phi}^i \phi^{\prime \, j}$ and $\dot{\bar{\phi}}{}^{\overline{i}} \overline{\phi} {}^{\prime \overline{j}} $, are independent of the vielbeins, and do not even include a factor of $E$ which plays the role of $\sqrt{ g }$ that usually accompanies any scalar within a spacetime integral. This is a consequence of the specific method for coupling the chiral boson to gravity developed in \cite{Bastianelli:1989cu}, which first introduces an unconstrained bosonic field and then incorporates auxiliary fields $P$ and $b$ which enforce the chirality constraint. This combined system is then coupled to gravity, and then integrating out the auxiliary fields $P$ and $b$ has the effect of eliminating the factor of $E$ that normally multiplies the kinetic term. We will see in section \ref{sec:cs} that the absence of vielbein dependence in these terms has a natural interpretation in the dual Chern-Simons description of chiral boson theories.

We can now explicitly perform the variation with respect to the vielbeins to compute the stress tensor $T_a{}^\beta$, as defined in equation (\ref{stress_tensor_flat_curved}), or more usefully, the version $T_{\alpha \beta}$ with two spacetime indices:
\begin{align}\label{general_stress_tensor}
    T_{tt} &= V ( S, P ) \, , \nonumber \\
    T_{t \theta} &= - P = T_{\theta t} \, , \nonumber \\
    T_{\theta \theta} &= - V + 2 \left( S V_S + P V_P \right) \, .
\end{align}
Note that the off-diagonal terms of $T_{\alpha \beta}$ are therefore identical and both proportional to $P$, which has the interpretation of the momentum along the $\theta$ circle. This is a consequence of the way we have coupled to the vielbeins in the second term of (\ref{general_class_vielbeins}), which is proportional to $P$ but which vanishes in the flat-space limit.

In principle, one could consider more general couplings of these chiral boson theories to vielbeins, which would lead to stress tensors that may not be symmetric and which are related to (\ref{general_stress_tensor}) by an improvement transformation. However, we find the choice of coupling that we have made here to be physically motivated for the problem of studying flow equations of the form (\ref{all_orders_flow}) which are connected to the free interaction function (\ref{free_interaction}). For instance, in the quantum theory, the momentum along a circle of radius $R$ is quantized in units of $\frac{1}{R}$, and, therefore, cannot flow with any deformation parameterized by a continuous $\lambda$. The coupling to frame fields which leads to (\ref{general_stress_tensor}) makes this manifest, even at the level of the classical stress tensor, since for any interaction function $V ( S, P )$, the linear momentum along the circle is fixed to its value $T_{t \theta} = - P$ in the free theory.

The trace of the stress tensor,
\begin{align}\label{trT}
    \operatorname{Tr}( T ) = T^\alpha{}_\alpha = - 2 \left( V - S V_S - P V_P \right) \, , 
\end{align}
vanishes if the interaction function $V$ is a homogeneous function of degree $1$ in its arguments, which is equivalent to the scale invariance of the theory as expected. The other Lorentz invariant that one can construct from the stress tensor is
\begin{align}\label{trT2}
    \operatorname{Tr} ( T^2 ) = T^{\mu \nu} T_{\mu \nu} = V^2 - 2 P^2 + \left( V - 2 \left( S V_S + P V_P \right) \right)^2 \, .
\end{align}
One can check by explicit computation that the two invariants (\ref{trT}) and (\ref{trT2}) each satisfy the condition (\ref{lorentz_invariant_function}), assuming that the interaction function $V$ itself obeys the condition (\ref{lorentz_pde_final}). In fact, more is true: given either of these two Lorentz-invariant functions $T^\mu{}_\mu$ and $T^{\mu \nu} T_{\mu \nu}$, locally and away from exceptional points, we can implicitly express any other Lorentz-invariant function $f$ in terms of this stress tensor invariant. To see this, let $f ( S, P )$ and $g ( S, P )$ be any two functions that satisfy the Lorentz-invariance condition (\ref{lorentz_invariant_function}). Consider the Jacobian for the change of variables from $(S, P)$ to $(f, g)$, namely
\begin{align}
    J = \begin{bmatrix} f_S & f_P \\ g_S & g_P \end{bmatrix} \, ,
\end{align}
and, in particular, its determinant,
\begin{align}\label{jac_det}
    \det ( J ) = f_S g_P - f_P g_S \, .
\end{align}
Since $f$ and $g$ each satisfy equation (\ref{lorentz_invariant_function}), we can solve this equation to express one of the partial derivatives of each function in terms of the other. For instance, we can choose
\begin{align}
    f_S = - \frac{f_P \left( P V_P + S V_S \right) }{S V_P + P V_S} \, , \qquad g_S = - \frac{g_P \left( P V_P + S V_S \right) }{S V_P + P V_S} \, .
\end{align}
Substituting these into the determinant (\ref{jac_det}), we find
\begin{align}
    \det ( J ) =  - \frac{f_P g_P \left( P V_P + S V_S \right) }{S V_P + P V_S} + \frac{f_P g_P \left( P V_P + S V_S \right) }{S V_P + P V_S} = 0 \, . 
\end{align}
Because $\det ( J ) = 0$, this change of variables is singular, which means that there exists a functional relation of the form $F ( f, g ) = 0$. By the implicit function theorem, under some assumptions on the derivatives of $F$, we can locally express $f ( S, P )$ as a function of $g ( S, P )$, or vice-versa. Thus, ignoring exceptional points, any pair of Lorentz-invariant functions are functionally dependent. Since the quantities $T^{\mu \nu} T_{\mu \nu}$ and $T^\mu{}_\mu$ are examples of such invariant quantities, it follows that any other Lorentz-invariant function -- again, away from singular points, and excluding trivial examples such as the case where one of the functions is a constant -- can be expressed as a function of the stress tensor.

Combining this conclusion with the previous statement around equation (\ref{all_orders_flow}), it also follows that, given any parameterized family of interaction functions $V ( \lambda )$ for Lorentz-invariant theories, one can write
\begin{align}
    \frac{\partial V ( \lambda )}{\partial \lambda} = \mathcal{O}^{(\lambda)} \equiv f ( T_{\mu \nu}^{(\lambda)} , \lambda ) \, ,
\end{align}
where in the last step we have used that $\mathcal{O}^{(\lambda)}$ can be implicitly expressed as a function of the stress tensor, given that this $\mathcal{O}^{(\lambda)}$ satisfies the Lorentz-invariance condition (\ref{lorentz_invariant_function}).

Therefore, the stress tensor flows that we have introduced in equation (\ref{stress_tensor_flow_defn}) are quite generic: any family of Lorentz-invariant interaction functions obeys a differential equation of this form, and conversely, any such flow equation (along with a Lorentz-invariant initial condition) defines a family of Lorentz-invariant interacting chiral boson theories.

Interesting examples of such flows are the ones defined in equation (\ref{two_nice_flows}), which are driven by the operators $\mathcal{O}_{T\overline{T}}$ and $\mathcal{R}$. We can express these two operators in terms of the interaction function $V$ and its derivatives using the general results (\ref{trT}) and (\ref{trT2}):
\begin{align}\label{simplified_TT_and_R}
    \mathcal{O}_{T\overline{T}} &= V \left( S V_S + P V_P \right) - \frac{1}{2} \left( V^2 + P^2 \right) \nonumber \, , \\ \mathcal{R} &= \sqrt{ \left( S V_S + P V_P + P \right) \left( S V_S + P V_P - P \right) } \, .
\end{align}
One can check directly that the two-parameter family of interaction functions (\ref{scalar_modified_nambu_goto}) solves the flow equations driven by the two operators given in (\ref{simplified_TT_and_R}).\footnote{When $\gamma = 0$, one recovers the theory of $T\overline{T}$-deformed Floreanini-Jackiw bosons, which also appears in the boundary graviton action for AdS$_3$ gravity at a finite radial cutoff; see equation (3.70) of \cite{Ebert:2022cle}.} The root-$T\overline{T}$ flow equation can also be solved in more generality. Suppose we begin from the flow equation
\begin{align}\label{root_TT_flow_V}
    \frac{\partial V ( \gamma )}{\partial \gamma} = - \mathcal{R} = - \sqrt{ \left( S V_S + P V_P + P \right) \left( S V_S + P V_P - P \right) } \, , 
\end{align}
and we furthermore assume that the function $V$ satisfies the Lorentz-invariance condition (\ref{lorentz_pde_final}) everywhere along the flow (which it is guaranteed to do, assuming the initial condition is Lorentz-invariant). Then the general solution to the differential equation (\ref{root_TT_flow_V}) with initial condition $V ( \gamma = 0 , S , P ) = V_0 ( S, P )$ is
\begin{align}\label{general_solution_root_TT}
    V ( \gamma, S , P ) = V_0 \left( \cosh ( \gamma ) S + \sinh ( \gamma ) \sqrt{ S^2 - P^2 } , P \right) \, .
\end{align}
That is, we simply replace all occurrences of the variable $S$ in the initial condition $V_0 ( S, P )$ with the combination $\cosh ( \gamma ) S + \sinh ( \gamma ) \sqrt{ S^2 - P^2 }$, while leaving all occurrences of $P$ unchanged. The result is a solution to (\ref{root_TT_flow_V}) with the correct initial condition at $\gamma = 0$.

Let us point out that the formulas (\ref{general_stress_tensor}) for the stress tensor components are valid when $N \geq 1$ and $\bar{N} \geq 1$. In the case of all chiral bosons ($\bar{N} = 0$), or all anti-chiral bosons ($N = 0$), the two invariants $S$ and $P$ become proportional to one another, so some of the structures in the Lagrangian collapse. For instance, for a theory of all chiral bosons, we have $S = P$ and the components of the stress tensor are
\begin{align}
    T_{tt} &= V ( S ) \, , \nonumber \\
    T_{t \theta} &= - S = T_{\theta t} \, , \nonumber \\
    T_{\theta \theta} &= - V ( S ) + 2 S V' ( S ) \, .
\end{align}
We have seen that the only solution to the Lorentz-invariance condition (\ref{lorentz_pde_final}) for all chiral bosons is $V = S$, and the stress tensor for this theory is
\begin{align}
    T_{\alpha \beta} = \frac{1}{2} \phi^{\prime j} \phi^{\prime j} \begin{bmatrix} 1 & -1 \\ -1 & 1 \end{bmatrix} \, .
\end{align}
Here one has $T^\alpha{}_\alpha = 0$ and $T^{\alpha \beta} T_{\alpha \beta} = 0$. The same conclusion holds for all anti-chiral bosons, where we have $S = -P$ rather than $S = P$, but again one finds $\operatorname{Tr} ( T ) = 0 = \operatorname{Tr} ( T^2 )$. For either of these scenarios, since both Lorentz scalars constructed from the stress tensor are vanishing, the theory is a fixed point of all Lorentz-preserving stress tensor deformations.\footnote{Another way to see this is by considering complex coordinates $(w, \bar{w})$, with $T = T_{ww}$ and $\bar{T} = T_{\bar{w} \bar{w}}$. A theory of all chiral bosons has $\bar{T} = 0$ and a theory of all anti-chiral bosons has $T = 0$. In either case, the product $T\overline{T}$ vanishes, and the trace vanishes by conformal invariance, so any Lorentz-preserving stress tensor flow is trivial. Of course, one could generate non-trivial interacting models by breaking Lorentz invariance and studying, for example, $f(T)$ (or $f(\bar{T})$) flows, but we will not pursue this option here.}

This gives another way to understand the fact that no Lorentz-invariant interactions are possible between only chiral bosons, or only anti-chiral bosons. Indeed, if a family of such interacting theories did exist, they would necessarily satisfy a stress tensor flow equation. But no such flow can exist which includes the free theory $V = S$, as this theory is left invariant by any stress tensor deformation. Since a theory of only chiral bosons has the Hamiltonian $\mathcal{H} = S = P$, one can view it as a $2d$ version of the $4d$ theory of Bialynicki-Birula electrodynamics, which is also a fixed point of all stress tensor flows.

\subsection{Self-duality and chirality}\label{sec:sd_and_chiral}
To conclude this section, we will point out one additional feature of the chiral boson models considered here. Although this property is trivially satisfied for any interacting chiral boson theory, regardless of the interaction function $V ( S, P )$, the analogous property for theories in the dual Chern-Simons description will play an important role in the next section.

Suppose that we begin with a general action of the form that we have been considering, which we repeat here for convenience:
\begin{equation}\label{interacting_start_selfduality}
    S = \int d^2x \left( \frac{1}{2} (\dot{\phi}^i \phi^{\prime \, i} - \dot{\bar{\phi}} {}^{\overline{i}} \, \bar{\phi}^{\prime \, \overline{i}}  ) - V ( S, P ) \right) \, .
\end{equation}
We would like to exchange the gradients $\partial_\alpha \phi^i = ( \dot{\phi}^i , \phi^{\prime i} )$ of the scalar fields for a vector field $A_\alpha = ( A_0, A_1 )$, and likewise for the anti-chiral scalars. To do this, we introduce a collection of Lagrange multiplier fields $\lambda^{i \alpha}$ and $\bar{\lambda}^{\overline{i} \alpha}$, and write the equivalent action
\begin{equation}
\begin{aligned}
\label{equivalent_interacting}
    S = \int d^2x &\bigg( \frac{1}{2} \left( A_0^i A_1^i - \bar{A}_0^{\overline{i}} \bar{A}_1^{\overline{i}} \right) - V ( S_A , P_A ) \\&+ \frac{1}{2} \lambda^{\alpha i} ( A_\alpha^i - \partial_\alpha \phi^i ) - \frac{1}{2} \bar{\lambda}^{\alpha \overline{i}} ( \bar{A}_\alpha^{\overline{i}} - \partial_\alpha \bar{\phi}^{\overline{i}} ) \bigg) \, .
\end{aligned}
\end{equation}
Here the variables $S_A$ and $P_A$ are defined by replacing instances of $(\phi^{\prime i}, \bar{\phi}^{\prime \overline{i}})$ with $(A_1^i, \bar{A}^{\overline{i}}_1)$:
\begin{align}
    S_A = \frac{1}{2} \left( A_1^i A_1^i + \bar{A}_1^{\overline{i}} \bar{A}_1^{\overline{i}} \right) \, , \qquad P_A = \frac{1}{2} \left( A_1^i A_1^i - \bar{A}_1^{\overline{i}} \bar{A}_1^{\overline{i}} \right) \, .
\end{align}
If one integrates out the auxiliary fields $\lambda^{\alpha i}$ and $\bar{\lambda}^{\alpha \overline{i}}$ in the action (\ref{equivalent_interacting}), these fields simply act as Lagrange multipliers which set $A_\alpha^i = \partial_\alpha \phi^i$ and $\bar{A}_\alpha^{\overline{i}} = \partial_\alpha \bar{\phi}^{\overline{i}}$, and the action then reduces to (\ref{interacting_start_selfduality}). 

However, suppose that we wish to proceed in the opposite direction, instead integrating out the fields $A_\alpha^i$ and $\bar{A}_\alpha^{\overline{i}}$. To do this, we vary the action with respect to the fields $(A_\alpha^i, \bar{A}_\alpha^{\overline{i}})$ and  to obtain their equations of motion, whose solutions take the form
\begin{align}\label{A_eom_soln}
    A_0^i &= - \lambda^{1 i} - 2 ( V_{S_A} + V_{P_A} ) \lambda^{0 i}  \, , \qquad A_1^i = - \lambda^{0 i} \, , \nonumber \\
    \bar{A}_0^{\overline{i}} &= - \bar{\lambda}^{1 \overline{i}} + 2 ( V_{S_A} - V_{P_A} ) \bar{\lambda}^{0 \overline{i}} \, , \qquad \bar{A}_1^{\overline{i}} = - \bar{\lambda}^{0 \overline{i}}   \, .
\end{align}
Integrating out $A_\alpha^i$ and $\bar{A}_\alpha^{\overline{i}}$ by replacing them with their on-shell values (\ref{A_eom_soln}) then gives
\begin{align}
    S = \int d^2 x \, \left( \frac{1}{2} \left( \bar{\lambda}^{0 \overline{i}} \bar{\lambda}^{1 \overline{i}} - \lambda^{0 i} \lambda^{1 i} \right) - V ( S_\lambda, P_\lambda ) + \frac{1}{2} \left( \phi^i \partial_\alpha \lambda^{\alpha i} - \bar{\phi}^{\overline{i}} \partial_\alpha \bar{\lambda}^{\alpha \overline{i}} \right) \right)  \, ,
\end{align}
 where we have integrated by parts to move the derivatives on the final two terms, and where now $S_\lambda$ and $P_\lambda$ are defined as
\begin{align}\label{S_lambda_P_lambda}
    S_\lambda = \frac{1}{2} \left( \lambda^{0 i} \lambda^{0 i} + \bar{\lambda}^{0 \overline{i}} \bar{\lambda}^{0 \overline{i}} \right) \, , \qquad P_\lambda = \frac{1}{2} \left( \lambda^{0 i} \lambda^{0 i} - \bar{\lambda}^{0 \overline{i}} \bar{\lambda}^{0 \overline{i}} \right) \, .
\end{align}
Note that (\ref{S_lambda_P_lambda}) involve the \emph{time} components of the $\lambda$ fields, rather than the spatial components. We see that the fields $\phi^i$ and $\bar{\phi}^{\overline{i}}$ act as Lagrange multipliers to enforce the constraints
\begin{align}
    \partial_\alpha \lambda^{\alpha i} = 0 = \partial_\alpha \bar{\lambda}^{\alpha \overline{i}} \, , 
\end{align}
which admit the general solutions
\begin{align}
    \lambda^{\alpha i} = \epsilon^{\alpha \beta} \partial_\beta \psi^i \, , \qquad \bar{\lambda}^{\alpha \overline{i}} = \epsilon^{\alpha \beta} \partial_\beta \bar{\psi}^{\overline{i}} \, ,
\end{align}
for some scalar fields $\psi^i$, $\bar{\psi}^{\overline{i}}$. Here we use the conventions $\epsilon^{0 1} = 1$, so
\begin{align}\label{lambda_to_psi}
    \lambda^{0 i} = \partial_x \psi^i = \psi^{\prime i} \, , \quad \lambda^{1 i} = - \partial_t \psi^i = - \dot{\psi}^i \, , \quad \bar{\lambda}^{0 \overline{i}} = \partial_x \bar{\psi}^{\overline{i}} = \bar{\psi}^{\prime \overline{i}} \, , \quad \bar{\lambda}^{1 \overline{i}} = - \partial_t \bar{\psi}^{\overline{i}} = - \dot{\bar{\psi}}{}^{\overline{i}} \, .
\end{align}
After integrating out $\phi^i$ and $\bar{\phi}^{\overline{i}}$ and replacing $\lambda^{\alpha i}$, $\bar{\lambda}^{\alpha \overline{i}}$ in favor of $\psi^i$, $\bar{\psi}^{\overline{i}}$, we arrive at the dual form of the action
\begin{align}\label{psi_action}
    S = \int d^2x \left( \frac{1}{2} (\dot{\psi}^i \psi^{\prime \, i} - \dot{\bar{\psi}} {}^{\overline{i}} \, \bar{\psi}^{\prime \, \overline{i}}  ) - V ( S_\psi , P_\psi ) \right) \, ,
\end{align}
where, according to the map in equation (\ref{lambda_to_psi}), the dualization has replaced time components with space components in the definition of the $S$ and $P$ variables,
\begin{align}\label{S_and_P_invariants_psi}
    S_\psi = \frac{1}{2} \left( \psi^{\prime \, i} \psi^{\prime \, i} + \bar{\psi}^{\prime \, \overline{i}} \bar{\psi}^{\prime \, \overline{i}} \right) \, , \qquad P_\psi = \frac{1}{2} \left( \psi^{\prime \, i} \psi^{\prime \, i} - \bar{\psi}^{\prime \, \overline{i}} \bar{\psi}^{\prime \, \overline{i}} \right) \, .
\end{align}
The result (\ref{psi_action}) is in fact identical to our starting point (\ref{interacting_start_selfduality}). Therefore, any interacting chiral boson theory is ``self-dual'' in the sense that the theory is left invariant under the process of introducing auxiliary fields and then integrating out to express the theory in terms of the ``dual'' $\psi$ variables rather than the original $\phi$ variables.

Versions of this simple argument are well-known in various contexts. The observation that the standard Floreanini-Jackiw action with $V ( S, P ) = S$ exhibits this self-duality appeared in \cite{Miao:1999pr}, which we have simply generalized to the interacting case. Similar manipulations also appear, for instance, when discussing T-duality in string theory from the worldsheet point of view.

However, we would like to emphasize two aspects of this observation. The first is that, unlike Lorentz invariance -- which only holds for interaction functions which satisfy the differential equation (\ref{lorentz_pde_final}) -- this self-duality holds for \emph{any} system of interacting chiral bosons, regardless of the form of $V ( S, P )$. We will therefore take the view that the property of self-duality should be part of the \emph{definition} of a theory of chiral bosons. Since we have seen that any chiral boson theory enjoys self-duality in the sense described here when presented in the Floreanini-Jackiw formulation, we will demand that any other description of chiral bosons should also have a corresponding self-duality property. That is, we will take self-duality as a necessary condition for a theory to describe chiral degrees of freedom.

The second observation is that, if one rewrites the action (\ref{equivalent_interacting}) as
\begin{align}
    S &= \int d^2x \left( \mathcal{L}_A + \frac{1}{2} \lambda^{\alpha i} ( A_\alpha^i - \partial_\alpha \phi^i ) - \frac{1}{2} \bar{\lambda}^{\alpha \overline{i}} ( \bar{A}_\alpha^{\overline{i}} - \partial_\alpha \bar{\phi}^{\overline{i}} ) \right) \, , \nonumber \\
    \mathcal{L}_A &= \frac{1}{2} \left( A_0^i A_1^i - \bar{A}_0^{\overline{i}} \bar{A}_1^{\overline{i}} \right) - V ( S_A , P_A ) \, , 
\end{align}
then the equations of motion for the fields $A_\alpha^i$ and $\bar{A}_\alpha^{\overline{i}}$ are
\begin{align}\label{generalized_legendre}
    \lambda^{\alpha i} = - 2 \frac{\partial \mathcal{L}_A}{\partial A_\alpha^i} \, , \qquad \bar{\lambda}^{\alpha \overline{i}} = 2 \frac{\partial \mathcal{L}_A}{\partial \bar{A}_\alpha^{\overline{i}}} \, .
\end{align}
Therefore, in a sense, one can think of the fields $\lambda$, $\bar{\lambda}$ as the duals (or conjugates) of the fields $A$ and $- \bar{A}$. Since the fields $A_\alpha^i = \partial_\alpha \phi^i$ and $\bar{A}_\alpha^{\overline{i}} = \partial_\alpha \bar{\phi}^{\overline{i}}$ are given by derivatives of a scalar field on-shell, one can also view the relations (\ref{generalized_legendre}) as a sort of Legendre transform. From this perspective, the self-duality of chiral boson models is the statement that such theories are invariant under a Legendre transform, or that one is free to rotate the fields $A_\alpha$, $\bar{A}_\alpha$ into their duals $\lambda_\alpha$ and $- \bar{\lambda}_\alpha$. This is very similar to the structure of theories of duality-invariant nonlinear electrodynamics in four dimensions, which are invariant under rotations mixing the field strength $F_{\mu \nu}$ with a certain dual field strength tensor $G_{\mu \nu}$. We will review this structure in more detail around equation (\ref{duality_rotations}) in the next section.

\section{Deformations of Dual Chern-Simons Theories}\label{sec:cs}

The chiral boson theories that we have considered in section \ref{sec:classical} often arise as the edge modes, or boundary duals, associated with the dynamics of Chern-Simons gauge fields in $3d$ bulk theories \cite{Witten:1996hc,ELITZUR1989108,Belov:2006jd}. For instance, physical descriptions of a quantum Hall droplet often involve a gauge field defined on a disk whose circular boundary supports edge modes modeled by chiral bosons \cite{PhysRevB.25.2185,PhysRevLett.64.220,PhysRevLett.64.216}.
Another example is found in $\mathrm{AdS}_3$ holography, where a collection of $U(1)$ Chern-Simons gauge fields in the bulk are dual to a corresponding collection of chiral currents in the $2d$ boundary. The addition of such bulk Chern-Simons terms to the action for $\mathrm{AdS}_3$ gravity allows BTZ black hole solutions to carry $U(1)$ charges \cite{Kraus:2006nb,Kraus:2006wn,Moussa:2008sj}.

In this section, we will show how stress tensor deformations of $2d$ chiral boson theories can be interpreted from the perspective of $3d$ bulk Chern-Simons gauge theories. We will begin by making some preliminary observations about the behavior of such $3d$ Chern-Simons theories in the presence of general boundary terms.

\subsection{$U(1)$ Chern-Simons theories with general boundary terms}\label{sec:cs_bdry}

Throughout this section, we will consider gauge theories defined on a bulk spacetime manifold $\mathcal{M}_3$ with boundary $\partial \mathcal{M}_3$. We will not specify whether $\partial \mathcal{M}_3$ is a true physical boundary or a conformal boundary, since our results apply uniformly in both cases.

Let us give a concrete example for each of these two cases to keep in mind as applications. In the former case, with a physical boundary, an example is furnished by the spacetime manifold $\mathcal{M}_3 = \mathcal{H}_2^+ \times \mathbb{R}_t$, where
\begin{align}
    \mathcal{H}_2^+ = \left\{ ( x, y ) \mid x, y \in \mathbb{R} \, , \, y \geq 0 \right\}
\end{align}
is the upper half-plane, viewed as a spatial manifold, and the factor of $\mathbb{R}_t$ represents a non-compact time direction. In this case, the boundary is $\partial \mathcal{M}_3 = \mathbb{R}_x \times \mathbb{R}_t$, where $\mathbb{R}_x$ is the spatial boundary $\partial \mathcal{H}_2^+ = \left\{ ( x, 0 ) \mid x \in \mathbb{R} \right\}$ and $\mathbb{R}_t$ is again the time direction.

An example of the latter case, with a conformal boundary, is a three-dimensional negatively curved bulk manifold $\mathcal{M}_3$, which asymptotically approaches an $\mathrm{AdS}_3$ spacetime that is characterized by a length scale $\ell$. The metric on $\mathcal{M}_3$ plays almost no role in this example, since the bulk Chern-Simons action is topological, but it is convenient to use the structure of the metric to characterize the conformal boundary $\partial \mathcal{M}_3$. The most general asymptotically $\mathrm{AdS}_3$ metric can be written in the form of a Fefferman-Graham expansion
\begin{align}\label{fg_expansion}
    ds^2 = \frac{\ell^2}{4 \rho^2} \, d \rho^2 + \left( \frac{g_{\alpha \beta}^{(0)} ( x^\gamma ) }{\rho} + g_{\alpha \beta}^{(2)} ( x^\gamma ) + \rho g_{\alpha \beta}^{(4)} ( x^\gamma ) \right) \, dx^\alpha \, d x^\beta \, .
\end{align}
The important point about this asymptotic form is that it determines a conformal boundary $\partial \mathcal{M}_3$ for our spacetime, located near $\rho = 0$, which has a boundary metric $g_{\alpha \beta}^{(0)} ( x^\gamma )$ determined by the leading term in the expansion (\ref{fg_expansion}). Here $\rho$ has the interpretation of a bulk radial coordinate whereas $x^\alpha$ label the two coordinates on the conformal boundary.

From now onwards, we will not distinguish between the two qualitatively different cases above, using the notation $\partial \mathcal{M}_3$ for either type of boundary. We will describe the $2d$ boundary in Euclidean signature using coordinates $x^\alpha = (w, \bar{w})$ and the flat metric
\begin{align}
    ds^2 = g_{\alpha \beta} \, dx^\alpha \, dx^\beta = dw \, d \bar{w} \, .
\end{align}
Although this signature and coordinate choice differs from the ones used in section \ref{sec:classical}, it allows for easier comparison with the holographic analysis of the root-$T\overline{T}$ deformation in \cite{Ebert:2023tih}. We will also use the convention that
\begin{align}
    \sqrt{ g } = \sqrt{ \det \left( \begin{bmatrix} 0 & \frac{1}{2} \\ \frac{1}{2} & 0 \end{bmatrix} \right) } = \frac{i}{2} \, , 
\end{align}
which will introduce some unfamiliar factors of $i$ in various places.

Our primary interest is to study the dynamics of Abelian gauge fields defined on the bulk manifold $\mathcal{M}_3$. Consider a collection of $U(1)$ gauge fields $(A_i, \bar{A}_{\overline{i}})$. Of course, the standard kinetic term for such gauge fields is the Maxwell term $F_{\alpha \beta}^i F^{\alpha \beta}_i$ where $F^i = d A^i$ is the field strength associated with the gauge field $F^i$. However, as we are in three spacetime dimensions, it is also possible to write down a Chern-Simons term which takes the form $A_i \wedge d A^i$ for the gauge fields $A^i$. The Maxwell term involves two derivatives and two factors of $A_i$, whereas the Chern-Simons term has only a single derivative and two factors of $A_i$. Therefore, by power counting, we see that the infrared behavior of the theory will be dominated by the Chern-Simons terms.

This motivates us to study the gauge theory with purely Chern-Simons couplings for the gauge fields $A_i$ and $\bar{A}_{\overline{i}}$, which we parameterize as
\begin{align}\label{eq:EuclideanU(1)AdS3}
    I_{\text{CS}} = \frac{i}{8 \pi} \int \left( k^{ij} A_i \wedge dA_j - \bar{k}^{\overline{i} \overline{j}} \bar{A}_{\overline{i}} \wedge d \bar{A}_{\overline{j}} \right) \, , 
\end{align}
where $k^{ij}$ and $\bar{k}^{\overline{i} \overline{j}}$ are constant matrices which we assume are symmetric and have positive eigenvalues.\footnote{Throughout this section we will use the symbol $I$ rather than $S$ for actions to emphasize that we are in Euclidean signature.} These matrices will play the role of the metrics $G_{ij}$ and $\bar{G}_{\overline{i} \overline{j}}$ of section \ref{sec:classical}. 

In addition to the Chern-Simons term (\ref{eq:EuclideanU(1)AdS3}), one can add a boundary term of the form
\begin{align}\label{boundary_term}
    I_{\text{bdry}} = - \frac{1}{8 \pi} \int_{\partial \mathcal{M}_3} d^2 x \, \sqrt{g} \, \mathcal{L}_{\text{bdry}} \left( A_{i \alpha} , \bar{A}_{\overline{i} \alpha} \right) \,,
\end{align}
where $\mathcal{L}_{\text{bdry}}$ is a Lorentz scalar constructed from the quantities $A_{i \alpha}$, $\bar{A}_{\overline{i} \alpha}$, which are the restrictions of the three-dimensional gauge fields to the boundary $\partial \mathcal{M}_3$. The full description of the theory is then given by the combined action
\begin{align}
    I = I_{\text{CS}} + I_{\text{bdry}} \, .
\end{align}
The standard choice of boundary term is the one which corresponds to the free interaction function $V ( \phi , \bar{\phi} )$ given in equation (\ref{free_interaction}), and is written as
\begin{equation}\label{free_CS_bdry}
    I_{\text{bdry}} = - \frac{1}{16 \pi} \int_{\partial \mathcal{M}_3} d^2 x \, \sqrt{g} \, g^{\alpha \beta} \left( k^{ij} A_{i \alpha} A_{j \beta} + \bar{k}^{\overline{i} \overline{j}} \bar{A}_{\overline{i} \alpha} \bar{A}_{\overline{j} \beta} \right) \, .
\end{equation}
However, in this section we will be interested in studying more general choices of boundary term, especially those which arise by deformations of the conventional boundary term (\ref{free_CS_bdry}).

It may seem strange that one can write down a general boundary term (\ref{boundary_term}) which is an arbitrary function of the variables $A_{i \alpha}$ and $\bar{A}_{\overline{i} \alpha}$, or after assuming Lorentz invariance and $O ( N ) \times O ( \bar{N} )$ symmetry under rotations of the gauge fields, an arbitrary function of the two combinations
\begin{align}
    S = \frac{1}{2} \left( k^{ij} A_i^\alpha A^{j}_{\alpha} + \bar{k}^{\overline{i} \overline{j}} \bar{A}_{\overline{i}}^\alpha \bar{A}^{\overline{j}}_{\alpha} \right) \, , \qquad P = \frac{1}{2} \left(  k^{ij}  A_i^\alpha A^{j}_{\alpha} - \bar{k}^{\overline{i} \overline{j}} \bar{A}_{\overline{i}}^\alpha \bar{A}^{\overline{j}}_{\alpha} \right) \, .
\end{align}
Any such boundary term $\mathcal{L}_{\text{bdry}} ( S , P )$ is manifestly compatible with boundary Lorentz invariance. This is in contrast with the analysis of section \ref{sec:classical}, where only interaction functions $V ( S, P )$ which obey the differential equation (\ref{lorentz_pde_final}) yield Lorentz-invariant theories.

The resolution to this tension is that the Floreanini-Jackiw and Chern-Simons descriptions of Lorentz-invariant chiral boson theories make different aspects of the models manifest. In the Floreanini-Jackiw description of section \ref{sec:classical}, it is manifest that the bosons $\phi^i$ are chiral since the theory is automatically self-dual (which we take as part of the definition of chirality) as we saw in section \ref{sec:sd_and_chiral}. However, it is not manifest that the Floreanini-Jackiw equations of motion respect Lorentz invariance, and requiring boost symmetry imposes a condition on $V ( S, P )$. Conversely, in the Chern-Simons description, it is manifest that the boundary theory enjoys Lorentz invariance since $\mathcal{L}_{\text{bdry}}$ is a Lorentz scalar. However, it is not manifest that the theory describes \emph{chiral} edge modes, which in particular requires that the theory be self-dual under the appropriate notion of duality transformation. Demanding chirality, or self-duality, will yield a constraint on $\mathcal{L}_{\text{bdry}}$, to be given in equation (\ref{CS_chirality_constraint}).

An analogy with electrodynamics is apt. Suppose that one wishes to describe a theory of an Abelian gauge field $A_\mu$ in four spacetime dimensions, whose Lagrangian $\mathcal{L}$ depends on the field strength $F_{\mu \nu}$ but not its derivatives. We assume that the equations of motion of this theory are invariant under both Lorentz transformations and under electric-magnetic duality rotations $\delta_\theta$ which act as
\begin{align}\label{duality_rotations}
    \delta_\theta F_{\mu \nu} = \theta G_{\mu \nu} ( F ) \, ,
\end{align}
where $G_{\mu\nu} = - \frac{1}{2} \varepsilon_{\mu\nu\rho\tau} \widetilde{G}^{\rho\tau}$ is the Hodge dual of $\widetilde{G}_{\mu \nu}$, which is itself defined as
\begin{align}
\widetilde{G}_{\mu \nu} = 2 \frac{\partial \mathcal{L}}{\partial F^{\mu \nu}} \, .
\end{align}
One option for describing such a theory is by giving the Lagrangian $\mathcal{L}$ itself. As the Lagrangian is a Lorentz scalar, this description makes Lorentz invariance manifest. However, invariance under duality rotations (\ref{duality_rotations}) is not automatic, and requires that the Lagrangian satisfy the differential equation (\ref{duality_pde}). 

Another option is to describe the theory in terms of its Hamiltonian $\mathcal{H} ( \vec{D}, \vec{B} )$, where $\vec{D} = \frac{\partial \mathcal{L}}{\partial \vec{E}}$ is the electric displacement. In these variables, the duality transformation (\ref{duality_rotations}) acts as an $SO(2)$ rotation which mixes the vectors $\vec{D}$ and $\vec{B}$. The most general duality-invariant Hamiltonian can be written as a function of the two variables
\begin{align}
    s = \frac{1}{2} \left( | \vec{D} |^2 + | \vec{B} |^2 \right) \, , \qquad p = | \vec{D} \times \vec{B} | \, .
\end{align}
These quantities $s$ and $p$ are invariant under $SO(3)$ rotations of the spatial coordinates and under duality rotations, so any Hamiltonian $\mathcal{H} ( s, p )$ is manifestly duality-invariant. However, because the canonical formulation has singled out a time direction as special, Lorentz invariance is no longer manifest. Imposing boost symmetry requires that the Hamiltonian satisfy the differential equation
\begin{align}
    \mathcal{H}_s^2 + \frac{2 s}{p} \mathcal{H}_s \mathcal{H}_p + \mathcal{H}_p^2 = 1 \, .
\end{align}
The upshot is that, in the electrodynamics example, either Lorentz invariance or duality invariance can be made manifest, and then imposing a partial differential equation will ensure that the remaining non-manifest symmetry will be respected.

In the chiral boson version of this story, the Floreanini-Jackiw formulation is analogous to the Hamiltonian presentation of $4d$ duality-invariant electrodynamics, since any theory of Floreanini-Jackiw bosons is automatically self-dual although Lorentz invariance is not manifest. The Chern-Simons presentation, on the other hand, is analogous to the Lagrangian description, since Lorentz invariance is manifest but self-duality is not guaranteed.

To understand the condition which must be imposed upon the Chern-Simons boundary term to ensure self-duality, which is the subject of section \ref{sec:cs_self_duality}, it will first be useful to study the currents obtained from varying the boundary gauge fields.

\emph{Boundary currents}

Quite generically, we expect that gauge fields couple to conserved currents. In the case of the $3d$ Chern-Simons theory, although we have not coupled the bulk gauge fields to any sources in $\mathcal{M}_3$, the variation of the on-shell action localizes to a boundary term, so we can therefore define boundary currents that live in $\partial \mathcal{M}_3$. We normalize these currents as
\begin{align}\label{currents_defn}
    J^\alpha_i = - \frac{2 \pi i}{\sqrt{g}} \frac{\delta I}{\delta A_\alpha^i} \Big\vert_{\text{on-shell}} \, , \qquad \bar{J}^\alpha_{\overline{i}} = - \frac{2 \pi i}{\sqrt{g}}  \frac{\delta I}{\delta \bar{A}_\alpha^{\overline{i}}} \Big\vert_{\text{on-shell}} \, .
\end{align}
We would like to compute these currents in a Chern-Simons theory with a general boundary term that is an arbitrary function of the $O(N) \times O ( \bar{N} )$ invariant combinations $S$ and $P$. To do this, we consider a general variation of the action. The Chern-Simons term varies as
\begin{align}
    \delta I_{\text{CS}} &= \frac{i}{8 \pi} \int_{\mathcal{M}_3} \left( k^{ij} \left( \delta A_i \wedge d A_j + A_i \wedge d \delta A_j \right) - \bar{k}^{\overline{i} \overline{j}} \left( \delta \bar{A}_{\overline{i}} \wedge d \bar{A}_{\overline{j}} + \bar{A}_{\overline{i}} \wedge d \delta \bar{A}_{\overline{j}} \right) \right) \nonumber \\
    &= \frac{i}{4 \pi} \int_{\mathcal{M}_3} \left( k^{ij}  \delta A_i \wedge d A_j - \bar{k}^{\overline{i} \overline{j}} \delta \bar{A}_{\overline{i}} \wedge d \bar{A}_{\overline{j}}\right) \\&- \frac{i}{8 \pi} \int_{\mathcal{M}_3} d \bigg( k^{ij} A_i \wedge \delta A_j - \bar{k}^{\overline{i} \overline{j}} \bar{A}_{\overline{i}} \wedge \delta \bar{A}_{\overline{j}} \bigg) \, .
\end{align}
The first term vanishes after imposing the bulk equations of motion $d A_j = 0 = d \bar{A}_{\overline{j}}$, while the second term localizes to a boundary contribution,
\begin{align}
    \delta I_{\text{CS}} \Big\vert_{\text{on-shell}} = - \frac{i}{8 \pi} \int_{\partial \mathcal{M}_3} \left( k^{ij} A_\alpha^i \, \delta A_\beta^j - \bar{k}^{\overline{i} \overline{j}} \bar{A}^{\overline{i}}_{\alpha} \, \delta \bar{A}^{\overline{j}}_{\beta} \right) \, dx^\alpha \wedge dx^\beta \, .
\end{align}
Since we are assuming that $\mathcal{L}_{\text{bdry}}$ takes the form
\begin{align}
    \mathcal{L}_{\text{bdry}} = f ( S, P ) \, ,
\end{align}
the variation of the boundary term can be written as
\begin{align}
    \delta I_{\text{bdry}} = - \frac{1}{8 \pi} \int_{\partial \mathcal{M}_3} \sqrt{g} \left( \left( f_S + f_P \right) k^{ij} A_i^\alpha \delta A_{j \alpha} + \left( f_S - f_P \right) \bar{k}^{\overline{i} \overline{j}} \bar{A}_{\overline{i}}^\alpha \delta \bar{A}_{\overline{j} \alpha} \right) \, .
\end{align}
In coordinates $(w, \bar{w})$, after raising the indices using $A^w_i = 2 A_{\bar{w} i}$ and $A^{\bar{w}}_i = 2 A_{w i}$, the variation of the combined action is then
\begin{align}\label{CS_variation_total_action}\hspace{-10pt}
    \delta I \Big\vert_{\text{on-shell}} &= - \frac{i}{8 \pi} \int_{\partial \mathcal{M}_3} \left( k^{ij} \left( A_w^i \, \delta A_{\bar{w}}^j - A_{\bar{w}}^i \, \delta A_w^j  \right) - \bar{k}^{\overline{i} \overline{j}} \left( \bar{A}^{\overline{i}}_{w} \, \delta \bar{A}^{\overline{j}}_{\bar{w}} - \bar{A}^{\overline{i}}_{\bar{w}} \, \delta \bar{A}^{\overline{j}}_{w} \right) \right) \nonumber \\
    &- \frac{1}{4 \pi} \int_{\partial \mathcal{M}_3} \sqrt{g} \bigg( k^{ij} \left( f_S + f_P \right) \left( A_{\bar{w}}^{i} \delta A^{j}_{w} + A_{w}^{i} \delta A^{j}_{\bar{w}}  \right) \nonumber \\&+ \bar{k}^{\overline{i} \overline{j}} \left( f_S - f_P \right) \left( \bar{A}^{\overline{i}}_{\bar{w}} \delta \bar{A}^{\overline{j}}_{w} + \bar{A}^{\overline{i}}_{w} \delta \bar{A}^{\overline{j}}_{\bar{w}} \right) \bigg) \, .
\end{align}
Using $\sqrt{g} = \frac{i}{2}$, we can therefore read off the currents (\ref{currents_defn}),
\begin{align}\label{currents_wwbar}
    J^w_i = \frac{i}{2} k^{ij} \left( f_S + f_P - 1 \right) A_{\bar{w}}^j \, , \qquad J^{\bar{w}}_i = \frac{i}{2} k^{ij} \left( f_S + f_P + 1 \right) A_w^j \, , \nonumber \\
    \bar{J}^w_{\overline{i}} = \frac{i}{2} \bar{k}^{\overline{i} \overline{j}} \left( f_S - f_P + 1 \right) \bar{A}_{\bar{w}}^{\overline{j}} \, , \qquad \bar{J}^{\bar{w}}_{\overline{i}} = \frac{i}{2} \bar{k}^{\overline{i} \overline{j}} \left( f_S - f_P - 1 \right) \bar{A}_w^{\overline{j}} \, .
\end{align}
These can also be written more covariantly as
\begin{align}
    J_{\alpha i} &= \frac{i}{4} k^{ij} \left( g_{\alpha \beta} ( f_S + f_P ) + \frac{1}{2} \epsilon_{\alpha \beta} \right) A^\beta_j \, , \nonumber \\
    \bar{J}^\alpha_{\overline{i}} &= \frac{i}{4} \bar{k}^{\overline{i} \overline{j}} \left( g_{\alpha \beta} ( f_S - f_P ) - \frac{1}{2} \epsilon_{\alpha \beta} \right) \bar{A}^\beta_{\overline{j}} \, ,
\end{align}
which agrees with (\ref{currents_wwbar}) for $g_{w \bar{w}} = \frac{1}{2}$, $\epsilon_{w \bar{w}} = 1 = - \epsilon_{\bar{w} w}$.

We note that variation of the total on-shell action has two qualitatively different contributions. The terms in the first line of (\ref{CS_variation_total_action}) are ``universal'' in the sense that they are present for any Chern-Simons theory and do not depend on the details of the boundary term $f ( S, P )$. These universal terms are also independent of the boundary metric, since they come from the integral of a $2$-form. In contrast, the terms on the second line of (\ref{CS_variation_total_action}) are ``model-dependent'' as they make explicit reference to the choice of boundary term $f ( S, P )$. Furthermore, these terms are metric-dependent and include an overall factor of $\sqrt{ g }$.

These two types of terms are analogous to those in the Lagrangian (\ref{general_class_vielbeins}) which couples a generic chiral boson theory to gravity. In that setting, the role of the ``universal'' and metric-independent contributions is played by the kinetic terms $G_{ij} \dot{\phi}^i \phi^{\prime \, j}$ and $\bar{G}_{\overline{i} \overline{j}} \dot{\bar{\phi}}{}^{\overline{i}} \overline{\phi}^{\prime j}$, which as we explained below equation (\ref{S_and_P_invariants_again}), do not include a factor of $E$. The Chern-Simons perspective gives another way to understand the metric-independence of these terms, since they may be viewed as the duals of contributions which arise from a topological bulk term. Similarly, the remaining metric-dependent and interaction-function-dependent terms in (\ref{general_class_vielbeins}) can be viewed as the analogs of the second line of (\ref{CS_variation_total_action}).

The expressions for the $J_i^\alpha$ and $\bar{J}_{\overline{i}}^\alpha$ also determines the boundary conditions on the gauge fields which we impose in order to have a well-defined variational principle. In general, the on-shell variation of the action can be written as
\begin{align}\label{delta_I_os}
    \delta I \Big\vert_{\text{on-shell}} \sim \int_{\partial \mathcal{M}_3} \left( J_i^\alpha \, \delta A^{i}_{\alpha} + \bar{J}_{\overline{i}}^{\alpha} \, \delta \bar{A}^{\overline{i}}_{\alpha} \right) \, .
\end{align}
We must ensure that the quantity (\ref{delta_I_os}) vanishes to have a good variational principle. To do this, we impose boundary conditions which hold fixed some particular combination of the boundary gauge fields $A^{i}_{\alpha}$ and $\bar{A}^{\overline{i}}_{\alpha}$. Schematically, this constraint takes the form
\begin{align}\label{F_bdry_conditions}
    F \left( A^{i}_{\alpha} \right) = 0 \, , \qquad \bar{F} \left( \bar{A}^{\overline{i}}_{\alpha} \right) = 0 \, ,
\end{align}
where the precise form of the functions $F$ and $\bar{F}$ depend on the case under consideration. In particular, this means that the allowed variations of the gauge fields must be constrained to satisfy the equations
\begin{align}\label{boundary_delta_A_constraints}
    \frac{\partial F}{\partial A^i_\alpha} \delta A^i_\alpha = 0 \, , \qquad \frac{\partial \bar{F}}{\partial \bar{A}^{\overline{i}}_\alpha} \delta \bar{A}^{\overline{i}}_{\alpha} = 0 \, .
\end{align}
For instance, if both of the boundary variations $\delta A^i_w$ and $\delta A^i_{\bar{w}}$ are non-zero, the constraints (\ref{boundary_delta_A_constraints}) can in principle be inverted to express one of these two boundary variations in terms of the other. This means that only one combination of the boundary gauge fields is free to fluctuate, while the other is held fixed. This is in agreement with the general expectation that imposing Dirichlet boundary conditions on \emph{both} components of the gauge field is too strong, and one would not find smooth solutions to the equations of motion for generic choices of the fixed gauge fields.

We also note that these boundary conditions will restrict the class of bulk gauge transformations that are permissible. A general gauge transformation $A^i \to A^i + d \Lambda^i$, $\bar{A}^{\overline{i}} \to \bar{A}^{\overline{i}} + d \bar{\Lambda}^{\overline{i}}$ in the bulk leads to a variation of the Chern-Simons term which takes the form
\begin{align}
    \delta I_{\text{CS}} = \frac{i}{8 \pi} \int_{\partial \mathcal{M}_3} \left( k^{ij} d A^i \wedge \Lambda^j - \bar{k}^{\overline{i} \overline{j}} d \bar{A}^{\overline{i}} \wedge \bar{\Lambda}^{\overline{j}} \right) \, ,
\end{align}
which, for general choices of the gauge parameters, will not be compatible with our choice of boundary conditions. We must therefore allow only a subclass of bulk gauge transformations which preserve the desired boundary conditions. Physically, one can think of this restriction as giving rise to physical degrees of freedom on the boundary.

To give a specific example illustrating the general observations above, let us consider the standard boundary term $f = S$. In this case, evaluating the currents (\ref{currents_wwbar}) gives
\begin{align}\label{holo_currents}
    J_{\bar{w}}^{i} &= 0 \, , \quad J_{w}^i = \frac{i}{2} k^{ij} A^j_w \, , \quad \bar{J}_{\bar{w}}^{\overline{i}} = \frac{i}{2} \bar{k}^{\overline{i} \overline{j}} \bar{A}^{\overline{j}}_{\bar{w}} \, , \quad \bar{J}_{w}^{\overline{i}} = 0 \, .
\end{align}
Therefore, with the conventional boundary term, the currents $J_\alpha^i$ are purely holomorphic and the currents $\bar{J}^{\overline{i}}_\alpha$ are purely anti-holomorphic. The variation of the on-shell action is
\begin{align}\label{variation_std_term}
    \delta I \Big\vert_{\text{on-shell}} \sim \int_{\partial \mathcal{M}_3} \left( J^i_w \, \delta A^i_{\bar{w}} + \bar{J}^{\overline{i}}_{\bar{w}} \, \delta \bar{A}^{\overline{i}}_{w} \right) \, .
\end{align}
The variation (\ref{variation_std_term}) vanishes if we require that $\delta A^i_{\bar{w}} = 0$ and $\delta \bar{A}^{\overline{i}}_w = 0$, which is equivalent to imposing Dirichlet boundary conditions on the components $A^i_{\bar{w}}$ and $\bar{A}^{\overline{i}}_w$ at the boundary $\partial \mathcal{M}_3$. For instance, one can demand that these components are both set to zero, which corresponds to the choice of functions $F$, $\bar{F}$ in (\ref{F_bdry_conditions}) given by
\begin{align}
    F \left( A^{i}_{\alpha} \right) = A^i_{\bar{w}} =  0 \, , \qquad \bar{F} \left( \bar{A}^{\overline{i}}_{\alpha} \right) = \bar{A}^{\overline{i}}_{w} = 0 \, .
\end{align}
We must then allow only bulk gauge transformations which do not change the values of $A^i_{\bar{w}}$ and $\bar{A}^{\overline{i}}_{w}$ on the boundary, and this restriction gives rise to boundary degrees of freedom. To see why these degrees of freedom are chiral, it is convenient to think of the holomorphic currents as $J_w^i = \partial \varphi^i$ and the anti-holomorphic currents as $\bar{J}_{\bar{w}}^{\overline{i}} = \bar{\partial} \varphi^i$, where the $\varphi^i$ are $c = 1$ free bosons. Then it is clear that the $J_w^i$ play the role of the left-moving chiral half of a non-chiral boson, and the $\bar{J}_{\bar{w}}^{\overline{i}}$ act as the right-moving anti-chiral parts.

\subsection{Self-duality condition for Chern-Simons theories}\label{sec:cs_self_duality}

Let us now consider the question of self-duality for Chern-Simons theories. As we argued in section \ref{sec:sd_and_chiral}, self-duality should be viewed as a necessary condition to impose on the theory so that it describes chiral degrees of freedom. In the Floreanini-Jackiw description, self-duality meant that we could express the action either in terms of the original variables $A_\alpha^{i} = \partial_\alpha \phi^{i}$, or in terms of the dual variables $\lambda_\alpha^i = \epsilon_{\alpha \beta} \partial^\beta \psi^i$. 
The relationship between $A_\alpha$ and $\lambda_\alpha$, as expressed around equation (\ref{generalized_legendre}) is very similar to the relationship between the boundary Chern-Simons gauge field $A_\alpha$ and the corresponding current. Let us compare them side-by-side. In section \ref{sec:sd_and_chiral}, we had the relations
\begin{align}\label{lambda_to_J}
    \lambda^{\alpha i} = - 2 \frac{\partial \mathcal{L}}{\partial A_\alpha^i} \, , \qquad \bar{\lambda}^{\alpha \overline{i}} = 2 \frac{\partial \mathcal{L}}{\partial \bar{A}_\alpha^{\overline{i}}} \, ,
\end{align}
where in this formula the symbol $A_\alpha$ refers to the vector field appearing in the action (\ref{equivalent_interacting}). In the Chern-Simons setting, we instead have the schematic relations 
\begin{align}\label{J_CS_analogy}
    J^{\alpha i} &= - \frac{2 \pi i}{\sqrt{g}} \frac{\delta I}{\delta A_\alpha^i} \Big\vert_{\text{on-shell}} = - \frac{2 \pi i}{\sqrt{g}} \frac{\partial \mathcal{L}_{\text{on-shell}}}{\partial A_\alpha^i} \, , \nonumber \\
    \bar{J}^{\alpha \overline{i}} &=  - \frac{2 \pi i}{\sqrt{g}}  \frac{\delta I}{\delta \bar{A}_\alpha^{\overline{i}}} \Big\vert_{\text{on-shell}} = - \frac{2 \pi i}{\sqrt{g}}  \frac{\partial \mathcal{L}_{\text{on-shell}}}{\partial \bar{A}_\alpha^{\overline{i}}} \, ,
\end{align}
where now the symbol $A_\alpha$ refers to the boundary Chern-Simons gauge field.\footnote{In equation (\ref{J_CS_analogy}), the partial derivatives of the Lagrangian $\mathcal{L}_{\text{on-shell}}$ are understood to be defined as the integrands of corresponding variations of the on-shell action in the middle expression of each line.} Insofar as the gauge field acts as a good proxy for the gradient of the Floreanini-Jackiw bosons, this suggests that the role of the dual variable $\lambda_\alpha^i$ is now played by
\begin{align}
    \lambda_\alpha^i = - \frac{1}{2} J_\alpha^i \, , \qquad \bar{\lambda}_\alpha^{\overline{i}} = \frac{1}{2} \bar{J}_\alpha^{\overline{i}} \, , 
\end{align}
where the sign difference is due to the relative sign in (\ref{generalized_legendre}), which itself originates from the difference in signs between the kinetic terms for chiral and anti-chiral bosons.

This analogy leads us to propose a notion of self-duality for Chern-Simons theories. We will phrase this condition via an infinitesimal transformation, rather than a finite one. That is, in section \ref{sec:sd_and_chiral}, the duality transformation was a $\mathbb{Z}_2$ action which replaced the fields $A_\alpha$ with the fields $\lambda_\alpha$. In the present context, we will instead propose a continuous transformation which infinitesimally rotates the fields $A_\alpha^i$, $\bar{A}_\alpha^{\overline{i}}$ into their duals $J_\alpha^i$, $- \bar{J}_\alpha^{\overline{i}}$.

We say that a Chern-Simons theory with boundary term $f ( S, P )$ is \emph{self-dual} if the on-shell variation of the action identically vanishes under the transformation
\begin{align}\label{cs_sd_trans}
    \delta A_\alpha^i = \epsilon J_\alpha^i \, , \qquad \delta \bar{A}_\alpha^{\overline{i}} = - \epsilon \bar{J}_\alpha^{\overline{i}} \, .
\end{align}
To see why this is the right notion of self-duality, let us find the condition on the boundary term $f ( S, P )$ under which the transformation (\ref{cs_sd_trans}) is a symmetry. By equation (\ref{delta_I_os}), under this variation the change in the on-shell action is
\begin{align}
    \delta I \Big\vert_{\text{on-shell}} &\sim \int_{\partial \mathcal{M}_3} \left( J_i^\alpha \, \delta A^{i}_{\alpha} + \bar{J}_{\overline{i}}^{\alpha} \, \delta \bar{A}^{\overline{i}}_{\alpha} \right) \, \nonumber \\
    &= \epsilon \int_{\partial \mathcal{M}_3} \left( J_i^\alpha \, J^{i}_{\alpha} - \bar{J}_{\overline{i}}^{\alpha} \, \bar{J}^{\overline{i}}_{\alpha} \right) \, ,
\end{align}
so the rotation (\ref{cs_sd_trans}) is a symmetry if and only if
\begin{align}\label{JJ_is_JbJb}
    J^i_w J^i_{\bar{w}} - \bar{J}^{\overline{i}}_w \bar{J}^{\overline{i}}_{\bar{w}} = 0 \, .
\end{align}
Using the general expression (\ref{currents_wwbar}) for the currents, and expressing the condition in terms of $S$ and $P$, we find that (\ref{JJ_is_JbJb}) is equivalent to the condition
\begin{align}\label{CS_chirality_constraint}
    f_S^2 + \frac{2 S}{P} f_S f_P + f_P^2 = 1 \, .
\end{align}
Remarkably, the Chern-Simons boundary term is self-dual if and only if it satisfies precisely the same differential equation (\ref{lorentz_pde_final}) which the Floreanini-Jackiw interaction function $V(S, P)$ must satisfy in order to guarantee Lorentz invariance. Because of the identical structure of the constraints on $f ( S , P)$ and $V(S,P)$, some of our observations from section \ref{sec:classical} can be immediately translated to analogous statements in the Chern-Simons setting.

For instance, if $\bar{N} = 0$ so that the theory features only a collection of unbarred gauge fields $A_\alpha^i$ but no barred fields $\bar{A}_{\alpha}^{\overline{i}}$, the two invariants collapse as $S = P$ and the only solution to the constraint (\ref{CS_chirality_constraint}) is $f ( S, P ) = S$. This is consistent with the comments around equation (\ref{only_S}) in the Floreanini-Jackiw formulation, namely that no Lorentz-invariant interactions are possible for a system of purely chiral (or purely anti-chiral) bosons. Here we are seeing the Chern-Simons counterpart of this statement: although we can write down any boundary term $f(S)$ that we like, and still respect boundary Lorentz invariance, only the choice $f(S) = S$ will respect chirality (or self-duality) of the boundary theory.

In the remainder of this section, we will view the differential equation (\ref{CS_chirality_constraint}) as a consistency condition which a boundary term $f ( S, P )$ must satisfy to describe chiral bosons. One can also understand this constraint as an analog of electric-magnetic duality invariance for $3d$ Chern-Simons theories. Of course, the conventional form of electric-magnetic duality is inapplicable for $3d$ gauge theories, since the Hodge dual of a two-form field strength $F_2$ in three spacetime dimensions is a one-form, which is interpreted as the field strength of a dual scalar rather than a dual $1$-form. However, demanding invariance under the duality rotation (\ref{cs_sd_trans}) is closely related to imposing invariance under the rotations (\ref{duality_rotations}); in both cases, the symmetry exchanges the field appearing in the Lagrangian with a certain ``dual'' that is defined via the derivative of the Lagrangian with respect to this field.

\emph{ Linear and non-linear self-duality constraints for currents}

One typically describes a free chiral $p$-form field in $2p$ dimensions, where $p$ is odd, as a form which satisfies a linear Hodge self-duality constraint. For instance, a free chiral $3$-form field $F_3$ in six dimensions obeys $\ast F_3 = F_3$. Likewise, the Floreanini-Jackiw bosons $\phi^i$, $\bar{\phi}^{\overline{i}}$ with free interaction function $V ( S, P ) = S$ are self-dual and anti-self-dual, respectively. Introducing interactions for such $p$-forms then modifies this constraint to a non-linear self-duality condition, which can be viewed as determining the self-dual part of the $p$-form as a function of the anti-self-dual part, or vice-versa.

We would now like to see how these self-duality constraints can be understood from the Chern-Simons description of chiral bosons. Since we are working in a two-dimensional Euclidean spacetime, the appropriate self-duality conditions for a one-form are \emph{imaginary} self-duality or anti-self-duality, since the definition of the Hodge star,
\begin{align}
    \left( \ast V \right)_\beta = \sqrt{ g } V^\alpha \epsilon_{\alpha \beta} \, ,
\end{align}
includes a factor of $\frac{i}{2}$ from the measure $\sqrt{ g }$. With these conventions, the dual of a general one-form $V_\alpha$ with components $V_w$, $V_{\bar{w}}$ is
\begin{align}
    \left( \ast V \right)_\alpha = \left( - i V_w , i V_{\bar{w}} \right) \, .
\end{align}
Thus a holomorphic one-form $V_\alpha = ( V_w, 0 )$ obeys an imaginary anti-self-duality condition
\begin{align}
    \ast V = - i V \, , 
\end{align}
whereas a purely anti-holomorphic one-form $V_\alpha = ( 0, V_{\bar{w}} )$ is imaginary-self-dual,
\begin{align}
    \ast V = i V \, .
\end{align}
We therefore see that all of the currents $J_i$ and $\bar{J}_i$ of equation (\ref{holo_currents}), which correspond to the standard boundary term $f(S,P) = S$,  satisfy $\ast J_i = - i J_i$ and $\ast \bar{J}_i = + i \bar{J}_i$. This can also be expressed by defining the projectors onto imaginary-self-dual and imaginary-anti-self-dual parts of a one-form,
\begin{align}
    P_{\pm} = \frac{1}{2} \left( 1 \mp i \ast \right) \, .
\end{align}
In terms of these projectors, the fact that the $J^\alpha_i$ are purely holomorphic can be expressed as $P_- J^\alpha_i = J^\alpha_i$, and the fact that the $\bar{J}^{\alpha}_{\overline{i}}$ are purely anti-holomorphic is equivalent to the statement that $P_+ \bar{J}^{\alpha}_{\overline{i}} = \bar{J}^{\alpha}_{\overline{i}}$. Therefore, by adding the boundary term $f ( S, P ) = S$ to the Chern-Simons action, we obtain chiral currents which obey a \emph{linear} self-duality condition. This is the image of the usual statement that free chiral $p$ forms in $2p$ dimensions, for $p$ odd, obey linear self-duality constraints.

Next we would like to understand how a more general boundary term gives rise to a \emph{non-linear} self-duality constraint, which corresponds to an interacting system of boundary chiral bosons. In this case, rather than obeying the standard chirality constraints
\begin{align}
    P_- J_i^\alpha = J_i^\alpha \, , \qquad P_+ \bar{J}_{\overline{i}}^\alpha = \bar{J}_{\overline{i}}^\alpha \, ,
\end{align}
which correspond to (linear) Hodge imaginary-self-duality or imaginary-anti-self-duality,
\begin{align}\label{linear_sd}
    \ast J_i = - i J_i \, , \qquad \ast \bar{J}_{\overline{i}} = i \bar{J}_{\overline{i}} \, ,
\end{align}
the currents will satisfy more general, non-linear or twisted self-duality conditions, each characterized by an operator $\mathcal{T}^{(i)}$ or $\bar{\mathcal{T}}^{(i)}$:
\begin{align}
    \ast J_i = \mathcal{T}^{(i)} J_i \, , \qquad \ast \bar{J}_{\overline{i}} = \bar{\mathcal{T}}^{(\overline{i})} \bar{J}_{\overline{i}} \, .
\end{align}
In the case where $\mathcal{T}^{(i)} = - i \, \mathbb{I}$ and $\bar{\mathcal{T}}^{(\overline{i})} = i \, \mathbb{I}$ are both proportional to the identity operator $\mathbb{I}$, this reduces to the standard chirality condition (\ref{linear_sd}). In the more general case we allow $\mathcal{T}^{(i)}$, $\bar{\mathcal{T}}^{(i)}$ to be non-trivial operators which can depend on the fields.

Twisted self-dual boundary conditions characterized by operators of this form have been considered in \cite{Severa:2016prq,Arvanitakis:2022bnr}, primarily in the setting of non-Abelian Chern-Simons theories. In the Abelian case, which is the focus of this work, no non-trivial operator $\mathcal{T}$ exists for a system obeying (\ref{CS_chirality_constraint}) with either $N = 0$ or  $\bar{N} = 0$ (i.e. a self-dual theory which only describes fields $\bar{A}_{\overline{i}}^\alpha$ but no $A_i^\alpha$, or with only  $A_i^\alpha$  but none of the $\bar{A}_{\overline{i}}^\alpha$, respectively). This is again related to the statement, which we have seen in section \ref{sec:lorentz}, that there are no possible Lorentz-invariant interactions for a system of purely chiral (or purely anti-chiral) bosons.\footnote{Alternatively, this is because there are no solutions to the self-duality equation (\ref{CS_chirality_constraint}) besides the trivial solution $f ( S, P ) = S$ when either $N = 0$ or $\bar{N} = 0$.} However, in a theory which features both chiral and anti-chiral bosons -- or both $A_i^\alpha$ and $\overline{A}_{\overline{i}}^\alpha$, from the Chern-Simons perspective -- such Lorentz-invariant interactions are possible, which manifests as the existence of allowable operators $\mathcal{T}$ besides the identity.

It is easy to see that, for a general boundary term $\mathcal{L}_{\text{bdry}} =f( S, P )$, the currents
\begin{align}
    J_w^i = \frac{i}{4} k^{ij} \left( f_S + f_P + 1 \right) A_w^j \, , \qquad J_{\bar{w}}^i = \frac{i}{4} k^{ij} \left( f_S + f_P - 1 \right) A_{\bar{w}}^j \, ,
\end{align}
satisfy the non-linear self-duality condition
\begin{align}
    \left( \ast J^i \right)_\alpha &= \left( \mathcal{T}^{(i)} \right)_{\alpha}{}^\beta J^i_\beta \, \nonumber , \\
    \mathcal{T}^{(i)} &= - i \begin{bmatrix} 1 & 0 \\ - \frac{2 k^i{}_j A_{\bar{w}}^j }{k^i{}_k A_w{}^k} \frac{f_S + f_P - 1}{f_S + f_P + 1} & 1 \end{bmatrix} \, .
\end{align}
This expression gives the components of the matrix $\mathcal{T}^{(i)}$ with respect to its Lorentz indices $\alpha, \beta = w, \bar{w}$, where $i$ is a fixed internal index. When $f_S = 1$ and $f_P = 0$, we see that $\mathcal{T}^{(i)}$ reduces to $-i \, \mathbb{I}$, which expresses the usual imaginary-anti-self-duality constraint. 

Similarly, the general currents
\begin{align}
    \bar{J}^{\overline{i}}_w = \frac{i}{4} \bar{k}^{\overline{i} \overline{j}} \left( f_S - f_P - 1 \right) \bar{A}_w^{\overline{j}} \, , \qquad \bar{J}^{\overline{i}}_{\bar{w}} = \frac{i}{4} \bar{k}^{\overline{i} \overline{j}} \left( f_S - f_P + 1 \right) \bar{A}_{\bar{w}}^{\overline{j}} \, ,
\end{align}
satisfy the non-linear self-duality condition
\begin{align}
    \left( \ast \bar{J} \right)^{\overline{i}}_\alpha &= \left( \bar{\mathcal{T}}^{(\overline{i})} \right)_\alpha{}^\beta \bar{J}^{\overline{i}}_\beta \, , \nonumber \\
    \bar{\mathcal{T}}^{(\overline{i})} &= i \begin{bmatrix} 1 & - \frac{2 \overline{k}^{\overline{i}}{}_{\overline{j}} \bar{A}_w^{\overline{j}} ( 1 + f_P - f_S ) }{ \overline{k}^{\overline{i}}{}_{\overline{k}} \bar{A}_{\bar{w}}^{\overline{k}} ( -1 + f_P - f_S ) } \\ 0 & 1 \end{bmatrix} \, .
\end{align}
Likewise, when $f_S = 1$ and $f_P = 0$, we see that $\bar{\mathcal{T}}^{(\overline{i})} = i \, \mathbb{I}$ so this reduces to the usual imaginary-self-duality condition $\ast \bar{J}^{\overline{i}} = i \bar{J}^{\overline{i}}$.

We should point out that, in other studies of twisted self-duality in Chern-Simons theories such as \cite{Severa:2016prq,Arvanitakis:2022bnr}, the twisting operator $\mathcal{T}$ commutes with the Hodge star operation. As a result, acting with the Hodge star operator on each side of the twisted self-duality constraint $\ast J = \mathcal{T} J$, one has
\begin{align}
    \ast \ast J = \ast \mathcal{T} J = \mathcal{T} \ast J = \mathcal{T}^2 J \, .
\end{align}
Since the Hodge star is an anti-involution, $\ast \ast = - \mathbb{I}$, in two Euclidean dimensions, one therefore arrives at the constraint
\begin{align}\label{anti_involution}
    \mathcal{T}^2 = - \mathbb{I} \, .
\end{align}
In Lorentzian signature, this would instead give the constraint $\mathcal{T}^2 = \mathbb{I}$.

However, in our case the twisting operators $\mathcal{T}^{(i)}$ and $\bar{\mathcal{T}}^{(\overline{i})}$ have non-trivial structure in their Lorentz indices and therefore do not commute with the Hodge star. This is why, in our case, these twisting operators do not satisfy an anti-involutive constraint like (\ref{anti_involution}).

One can now proceed as in the linear case and define projection operators
\begin{align}\label{four_projection_operators}
    P_+^{(i)} &= \begin{bmatrix} 0 & 0 \\ - \frac{k^i_j A_{\bar{w}}^j }{k^i_k A_w^k} \frac{f_S + f_P - 1}{f_S + f_P + 1} & 1 \end{bmatrix} \, , \qquad \bar{P}_{+}^{(\overline{i})} = \begin{bmatrix} 0 & \frac{ \overline{k}^{\overline{i}}_{\overline{j}} \bar{A}_w^{\overline{j}} ( 1 + f_P - f_S ) }{\overline{k}^{\overline{i}}_{\overline{k}} \bar{A}_{\bar{w}}^{\overline{k}} ( -1 + f_P - f_S ) } \\ 0 & 1 \end{bmatrix}  \, , \nonumber \\
    P_-^{(i)} &=  \begin{bmatrix} 1 & 0 \\ \frac{k^i_j A_{\bar{w}}^j }{k^i_k A_w^k} \frac{f_S + f_P - 1}{f_S + f_P + 1} & 0 \end{bmatrix} \, , \qquad \bar{P}_-^{(\overline{i})} = \begin{bmatrix} 1 & - \frac{\overline{k}^{\overline{i}}_{\overline{j}} \bar{A}_w^{\overline{j}} ( 1 + f_P - f_S ) }{\overline{k}^{\overline{i}}{}_{\overline{k}} \bar{A}_{\bar{w}}^{\overline{k}} ( -1 + f_P - f_S ) } \\ 0 & 0 \end{bmatrix} \, ,
\end{align}
which satisfy the expected properties of orthogonal projectors,
\begin{align}
    \left( P_\pm^{(i)} \right)^2 = P_\pm^{(i)}  \, , \quad \left( \bar{P}_{\pm}^{(\overline{i})} \right)^2 = \bar{P}_{\pm}^{(\overline{i})} \, , \quad P_\pm^{(i)} P_\mp^{(i)} = 0 = \bar{P}_{\pm}^{(\overline{i})} \bar{P}_{\mp}^{(\overline{i})}   \, , 
\end{align}
along with the chirality conditions
\begin{align}
    P_-^{(i)} J^i = J^i \, , \quad P_+^{(i)} J^i = 0 \, , \quad \bar{P}_{+}^{(\overline{i})} \bar{J}^{\overline{i}} = \bar{J}^{\overline{i}} \, , \quad \bar{P}_{-}^{(\overline{i})} \bar{J}^{\overline{i}} = 0 \, .
\end{align}
Therefore, even in the interacting case, one can view the currents as satisfying an appropriate non-linear self-duality constraint. This expresses, in Chern-Simons language, the equations of motion (\ref{interacting_eom}) for interacting Floreanini-Jackiw bosons.

We should point out that this construction has now produced two separate pairs of projection operators $P_{\pm}^{(i)}$, $\bar{P}_{\pm}^{(\overline{i})}$ for each fixed choice of indices $i, \overline{i}$, or equivalently, two separate twist operators $\mathcal{T}^{(i)}$ and $\bar{\mathcal{T}}^{(i)}$. This is in contrast with the linear-self duality constraint, which is described by only two projectors $P_{\pm} = \frac{1}{2} \left( 1 \mp i \ast \right)$, where
\begin{align}\label{linear_projectors}
    P_+ = \bar{P}_+ \, \qquad P_- = \bar{P}_- \, .
\end{align}
In the linear case, there are relations that cause these four operators to collapse to just two independent projectors, and it is clear that these operators project onto one-dimensional eigenspaces which represent physically opposite chiralities.

In the non-linear case, there are also relations (albeit more complicated ones) between the two twist operators. For instance, one can see that $\mathcal{T}^{(i)}$ can be obtained from $\bar{\mathcal{T}}^{(i)}$ by simultaneously transposing the matrix in its Lorentz indices and interchanging all barred and unbarred quantities. That is, one exchanges
\begin{align}
    k^{ij} \longleftrightarrow \bar{k}^{\overline{i} \overline{j}} \, , \quad A^i \longleftrightarrow \bar{A}^{\overline{i}} \, \quad w \longleftrightarrow \bar{w} \, ,
\end{align}
which also has the effect of sending $P \to -P$ (and thus $f_P \to - f_P$). This relation holds regardless of the choice of boundary term. When the function $f ( S, P )$ satisfies the self-duality condition (\ref{CS_chirality_constraint}) necessary to describe chiral modes, there are further constraints between the twist operators. To see one such constraint, we can rewrite (\ref{CS_chirality_constraint}) as
\begin{align}
    J^i \wedge \ast J^i = \bar{J}^{\overline{i}} \wedge \ast \bar{J}^{\overline{i}} \, .
\end{align}
Since $\ast J^i = \mathcal{T}^{(i)} J^i$ and $\ast \bar{J}^{\overline{i}} = \bar{\mathcal{T}}^{(\overline{i})} \bar{J}^{\overline{i}}$, this relation can also be expressed as
\begin{align}\label{wedge_constraint}
    J^i \wedge \mathcal{T}^{(i)} J^i =  \bar{J}^{\overline{i}} \wedge \bar{\mathcal{T}}^{(\overline{i})} \bar{J}^{\overline{i}} \, .
\end{align}
Equation (\ref{wedge_constraint}) is a consequence of the fact that, when the boundary term obeys the self-duality constraint, the chiral and anti-chiral twist operators are ``compatible'' in a sense which generalizes the statements that $\mathcal{T}^{(i)} = - \bar{\mathcal{T}}^{(\overline{i})}$, or that the projection operators satisfy (\ref{linear_projectors}), in the linear case.

\subsection{Current deformations of boundary terms}\label{sec:cs_defs}

We will now consider flow equations which modify the boundary term $\mathcal{L}_{\text{bdry}}$ of a bulk Chern-Simons theory.\footnote{Although we focus on $U(1)$ Chern-Simons theories in this work, stress tensor deformations of the boundary term for $SL(2) \times SL(2)$ Chern-Simons have been considered in \cite{Llabres:2019jtx,Ouyang:2020rpq,Ebert:2022ehb}.} In particular, we are interested in differential equations for $\mathcal{L}_{\text{bdry}}$ which are driven by conserved quantities. We will refer to any such flow equation as a ``current deformation'' regardless of whether the conserved currents driving the flow are the objects $J_\alpha^i$ and $\bar{J}_\alpha^{\overline{i}}$ defined in equation (\ref{currents_defn}), or the energy-momentum tensor $T_{\alpha \beta}$, which is another type of conserved current in the theory.

Let us first study deformations which involve the spin-$1$ currents $J_\alpha^i$ and $\bar{J}_\alpha^{\overline{i}}$. A general flow equation in this class takes the form
\begin{align}
    \frac{\partial \mathcal{L}_{\text{bdry}}}{\partial \lambda} = \mathcal{O} \left( J_\alpha^i , \bar{J}_\alpha^{\overline{i}} , \lambda \right) \, ,
\end{align}
where $\mathcal{O}$ is a Lorentz scalar and $O ( N ) \times O ( \bar{N} )$ singlet constructed from the currents. Within this class, there are fewer interesting possibilities. The most natural deformation to consider is to begin with the conventional boundary term $\mathcal{L}_{\text{bdry}} = S$ and deform by a marginal combination of the form
\begin{align}
    \mathcal{O} = k_{ij} J_\alpha^i J^{ \alpha j} \, , \; \text{ or } \quad \mathcal{O} = \bar{k}_{\overline{i} \overline{j}} \bar{J}_\alpha^{\overline{i}} \bar{J}^{\alpha \overline{j}} \, .
\end{align}
However, by virtue of the chirality of the currents given in equation (\ref{holo_currents}), both of these operators vanish. One might instead construct a deforming operator which mixes the currents on the two sides, such as
\begin{align}\label{JJbar_defn}
    \mathcal{O} = C_{i \overline{j}} J_\alpha^i  \bar{J}^{\alpha \overline{j}} \, ,
\end{align}
where $C_{i \overline{j}}$ is a constant tensor with mixed indices. For instance, in the case $N = \bar{N}$, we do not need to distinguish between barred and unbarred indices, and can choose $C_{i \overline{j}} = \delta_{i \overline{j}} \equiv \delta_{ij}$.\footnote{Of course, when $N \neq \bar{N}$, a deformation of this form does not preserve $O(N) \times O ( \bar{N} )$ symmetry. For instance, a deformation by $\sum_{i=1}^{M} J^i_\alpha \bar{J}^{i \alpha}$, where $M = \min ( N, \bar{N} )$, treats the currents asymmetrically.} Let us consider the effect of this deformation with the simplifying assumption $k^{ij} = \bar{k}^{ij} = \delta^{ij}$. In this case, at leading order in the deformation parameter, one finds a deformed boundary term
\begin{align}\label{JJ_deformed}
    \mathcal{L}_{\text{bdry}}^{(1)} = \frac{1}{2} \left( A_i^\alpha A^{i}_{\alpha} +  \bar{A}_{i}^\alpha \bar{A}^{i}_{\alpha} \right) + \lambda A_i^\alpha \bar{A}^i_{\alpha} \, ,
\end{align}
up to the normalization of $\lambda$. That is, such an operator has introduced an off-diagonal mixing between the barred and unbarred gauge fields. Ignoring possible subtleties about quantization of the Chern-Simons levels, such a quadratic mixing can always be undone by performing a Bogoliubov-like field redefinition. Indeed, note that beginning with the undeformed boundary term
\begin{align}
    \mathcal{L}_{\text{bdry}}^{(0)} = \frac{1}{2} \left( A_i^\alpha A^{i}_{\alpha} +  \bar{A}_{i}^\alpha \bar{A}^{i}_{\alpha} \right)
\end{align}
and then performing a change of variables to
\begin{align}\label{CS_bogoliubov}
    A^i_\alpha = \cosh ( \mu ) B^i_\alpha + \sinh ( \mu ) \bar{B}^i_\alpha \, , \qquad \bar{A}^{i}_{\alpha}  = \cosh ( \mu ) \bar{B}^i_\alpha + \sinh ( \mu ) B^i_\alpha\, ,
\end{align}
gives the transformed boundary term
\begin{align}
    \mathcal{L}_{\text{bdry}}^{(0)} = \cosh ( 2 \mu )  \left[ \frac{1}{2} \left( B_i^\alpha B^{i}_{\alpha} +  \bar{B}_{i}^\alpha \bar{B}^{i}_{\alpha} \right) + \tanh ( 2 \mu ) B_i^\alpha \bar{B}^i_\alpha \right] \, .
\end{align}
Up to an overall rescaling, this is equivalent to the deformed boundary term (\ref{JJ_deformed}) if we identify $\tanh ( 2 \mu ) = \lambda$. Therefore, the marginal $J \bar{J}$ deformation of equation (\ref{JJbar_defn}) can be viewed as inducing a rotation between the fields $A_\alpha^i$ and $\bar{A}_\alpha^{\overline{i}}$. We will see later that the root-$T\overline{T}$ deformation, in the case $N = \bar{N} = 1$, is qualitatively similar to this $J \bar{J}$ deformation.

In principle, one could consider more general operators constructed from the currents $J$ and $\bar{J}$, such as powers of the form $\mathcal{O} = \left( J_\alpha^i  \bar{J}^{\alpha i} \right)^n$ or other structures such as $\mathcal{O} = \left( J_\alpha^i J_\beta^i  \bar{J}^{\alpha \overline{j}}\bar{J}^{\beta \overline{j}} \right)^m$, both of which preserve $O(N) \times O ( \bar{N})$ symmetry. These operators are irrelevant for $n > 1$ and $m > \frac{1}{2}$, respectively. However, we will now instead turn our attention to deformations which are constructed from the energy-momentum tensor, 
\begin{align}
    \frac{\partial \mathcal{L}_{\text{bdry}}}{\partial \lambda} = \mathcal{O} \left( T_{\alpha \beta}^{(\lambda)} , \lambda \right) \, .
\end{align}
The first choice that one must make in defining such a flow is \emph{which} stress tensor to use. There are generally many definitions of the energy-momentum tensor which are all conserved but which differ by improvement transformations. One natural choice is the Hilbert stress tensor defined by varying the metric. Of course, neither the Chern-Simons action (\ref{eq:EuclideanU(1)AdS3}) nor the boundary action (\ref{boundary_term}) depend on the \emph{bulk} metric, but the term $I_{\text{bdry}}$ does depend on the boundary metric. One can therefore define a boundary stress tensor,
\begin{align}\label{hilbert_cs}
    T_{\alpha \beta} = - \frac{2}{\sqrt{g}} \frac{\delta I}{\delta g^{\alpha \beta}} = - \frac{2}{\sqrt{g}} \frac{\delta I_{\text{bdry}}}{\delta g^{\alpha \beta}} \, .
\end{align}
However, this stress tensor is qualitatively different from the one obtained in equation (\ref{general_stress_tensor}) by coupling a chiral boson theory to vielbeins. In that context, the coupling to vielbeins treated chiral and anti-chiral modes differently, and as a result the stress tensor component $T_{t \theta} = - P$ is sensitive to the difference between chiral and anti-chiral fields. Exchanging the fields $\phi$ with $\bar{\phi}$, and vice-versa, reverses the sign of $P$ and therefore changes $T_{t \theta}$.

In contrast, since both the barred gauge fields and unbarred gauge fields couple to the boundary metric in the same way, the Hilbert stress tensor (\ref{hilbert_cs}) treats the fields $(A_\alpha^i, \bar{A}_{\overline{i}}^\alpha)$ on equal footing. Unlike (\ref{general_stress_tensor}), the Hilbert stress tensor associated with the standard boundary Lagrangian $\mathcal{L}_{\text{bdry}} = S$ is unchanged under the process of exchanging barred and unbarred gauge fields. To make this point explicit, let us write this boundary term as 
\begin{align}\label{Lbdry_A}
    \mathcal{L}_{\text{bdry}} = \frac{1}{2} S^\alpha{}_\alpha \, , \qquad S_{\alpha \beta} = k^{ij} A_{i \alpha} A_{j \beta} + \bar{k}^{\overline{i} \overline{j}} \bar{A}_{\overline{i} \alpha} \bar{A}_{\overline{j} \beta} \, .
\end{align}
With this definition, one has $S^\alpha_\alpha = 2 S$. The Hilbert stress tensor computed from (\ref{Lbdry_A}), after rescaling to eliminate the overall prefactor of $- \frac{1}{16 \pi}$ in $I_{\text{bdry}}$, is
\begin{align}\label{std_lbdry_hilbert}
    T_{\alpha \beta} = - S_{\alpha \beta} + g_{\alpha \beta} S \, .
\end{align}
Deforming the standard boundary term by a generic function of the stress tensor (\ref{std_lbdry_hilbert}), which necessarily involves the single independent non-vanishing Lorentz invariant $T^{\alpha \beta} T_{\alpha \beta}$, will introduce dependence on the new variable
\begin{align}
    S_2 = S_{\alpha \beta} S^{\alpha \beta} \, .
\end{align}
Note that $S_2$ is functionally independent from the invariant $P = \frac{1}{2} \left(  k^{ij}  A_i^\alpha A^{j}_{\alpha} - \bar{k}^{\overline{i} \overline{j}} \bar{A}_{\overline{i}}^\alpha \overline{A}^{\overline{j}}_{\alpha} \right)$. Therefore, the class of boundary terms that can be described by functions $f ( S, P )$ is not closed under deformations by the Hilbert stress tensor. Instead, to describe flows driven by this choice of stress tensor, we should instead parameterize the boundary term as a function of different invariants:
\begin{align}\label{boson_like_class}
    \mathcal{L}_{\text{bdry}} = f ( S_1 , S_2 ) \, ,
\end{align}
where
\begin{align}
    S_1 = \operatorname{Tr} ( S ) = S_\alpha{}^\alpha = 2 S \, , \qquad S_2 = \operatorname{Tr} ( S^2 ) = S_{\alpha \beta} S^{\alpha \beta} \, .
\end{align}
The structure of Hilbert stress tensor deformations of the class of functions (\ref{boson_like_class}) is identical to the structure of such flows for a collection of \emph{non-chiral} bosons. Indeed, as was worked out in \cite{Ferko:2022cix}, a general Lagrangian for a collection of $N$ non-chiral bosons $\varphi^i$ with target space metric $G_{ij}$ is a function of the matrix
\begin{align}
   X_\alpha{}^\beta = G_{ij} \partial_\alpha \varphi^i \partial^\beta \varphi^j \, ,
\end{align}
which has two independent traces, 
\begin{align}
    x_1 = \operatorname{Tr} ( X ) = X_\alpha{}^\alpha \, , \qquad x_2 = \operatorname{Tr} ( X^2 ) = X_\alpha{}^\beta X_\beta{}^\alpha \, .
\end{align}
All higher traces can be expressed in terms of $x_1$ and $x_2$ using identities derived from the Cayley-Hamilton theorem for $2 \times 2$ matrices. Precisely the same results apply in the Chern-Simons context, except replacing the matrix $X_\alpha{}^\beta$ with $S_\alpha{}^\beta$ and thus replacing the invariants $x_1$, $x_2$ with $S_1$, $S_2$. For instance, the Hilbert stress tensor associated with a general boundary term (\ref{boson_like_class}) is
\begin{align}
    T_{\alpha \beta} = - 2 \frac{\partial f}{\partial S_1} S_{\alpha \beta} - 4 \frac{\partial f}{\partial S_2} S_{\alpha \gamma} S^\gamma{}_\beta  + g_{\alpha \beta} f \, .
\end{align}
One can then construct deformations of the boundary term which depend on the two independent traces of the stress tensor, which can be written as
\begin{align}
    T^{\alpha \beta} T_{\alpha \beta} &= 2 \left( f + 2 S_1^2 \frac{\partial f}{\partial S_2} \right) \left( f - 2 S_1 \left( \frac{\partial f}{\partial S_1} + S_1 \frac{\partial f}{\partial S_2} \right) \right) + 8 S_2^2 \left( \frac{\partial f}{\partial S_2} \right)^2 \nonumber \\
    &\quad + 4 S_2 \left( \left( \frac{\partial f}{\partial S_1} \right)^2 + 6 S_1 \frac{\partial f}{\partial S_1} \frac{\partial f}{\partial S_2} - 2 \frac{\partial f}{\partial S_2} \left( f - 2 S_1^2 \frac{\partial f}{\partial S_2} \right) \right)  \, , \\
    T^\alpha{}_\alpha &= - 2 S_1 \frac{\partial f}{\partial S_1} - 4 S_2 \frac{\partial f}{\partial S_2} + 2 f \, .
\end{align}
All of the results concerning stress tensor flows for non-chiral bosons in two dimensions (see, for instance, \cite{Ferko:2022cix} and section 4 of \cite{Ferko:2023ruw}) therefore immediately apply to deformations of Chern-Simons boundary terms which take the form (\ref{boson_like_class}).

One way to think about this class of deformations, using the parameterization (\ref{boson_like_class}) and the Hilbert stress tensor, is the following. In the case $N = \bar{N}$ -- when the unbarred gauge fields $A^i_\alpha$ and barred gauge fields $\bar{A}^i_\alpha$ are dual to equal numbers of chiral bosons $\phi^i$ and anti-chiral bosons $\bar{\phi}^i$, respectively -- one can collect these fields into a collection of non-chiral bosons $\varphi^i$ as
\begin{align}
    \varphi^i = \frac{1}{\sqrt{2}} \left( \phi^i + \bar{\phi}^{i} \right) \, .
\end{align}
We will revisit the quantization of the boundary theory after performing this repackaging of the field content into non-chiral fields in section \ref{sec:cian}. We claim that deformations using the Hilbert stress tensor and the parameterization (\ref{boson_like_class}) are appropriate for understanding flows in which the bosons are assembled into non-chiral fields in this way. This is why such flows are naturally studied using the invariants $(S_1, S_2)$, which have the same structure as the ones appearing in $T\overline{T}$-like deformations of non-chiral bosons, rather than the invariants $(S, P)$, which we have used in section \ref{sec:classical} to understand stress tensor flows for chiral bosons. 

One might ask whether there is a different presentation of stress tensor deformations for the boundary term whose structure is more similar to that of flows in the Floreanini-Jackiw description of section \ref{sec:classical}. This brings us to the second natural choice of stress tensor, besides the Hilbert definition in equation (\ref{hilbert_cs}). Rather than coupling the boundary theory to a metric on $\partial \mathcal{M}_3$, one could instead couple to vielbeins in the same way as we did in equation (\ref{general_class_vielbeins}) for chiral boson theories. To do this, we again introduce frame fields $E^a{}_\alpha$, although now the flat indices will be raised or lowered with the \emph{Euclidean} tangent-space metric $\eta_{ab} = \left[ \begin{smallmatrix} 0 & 1 \\ 1 & 0 \end{smallmatrix} \right]$. In this case, the appropriate flat-space values for the vielbeins are
\begin{align}\label{euclidean_flat}
    E^+_w = E^-_{\bar{w}} = \frac{1}{\sqrt{2}} \, , \qquad E^+_{\bar{w}} = E^-_{w} = 0 \, ,
\end{align}
whose inverses produce the desired spacetime metric $ds^2 = dw \, d \bar{w}$,
\begin{align}
    E^a_\alpha E^b_\beta \eta_{a b} = g_{\alpha \beta} = \begin{bmatrix} 0 & \frac{1}{2} \\ \frac{1}{2} & 0 \end{bmatrix} \, .
\end{align}
One can then couple the Chern-Simons boundary term $I_{\text{bdry}}$ to vielbeins as 
\begin{equation}\label{CS_bdry_coupled}
    I_{\text{bdry}} = - \frac{i}{16 \pi} \int_{\partial \mathcal{M}_3} d^2 x \, \left(  2 \left( E^+_w E^-_w - E^+_{\bar{w}} E^-_{\bar{w}} \right) P + 2 E f ( S, P ) \right) \, ,
\end{equation}
where we include factors of $2$ since, in the conventions of this section, $E = \frac{1}{2}$. Likewise, the overall factor of $i$ in (\ref{CS_bdry_coupled}) arises because $\sqrt{g} = \frac{i}{2}$ but $E = \frac{1}{2}$. To compare with equation (\ref{general_class_vielbeins}), note that in the conventions of section \ref{sec:classical}, we instead had $E = 1$. Now $S$ and $P$ are coupled to vielbeins as
\begin{align}
    S &= \frac{1}{4 \left( E^+_w E^-_{\bar{w}} + E^-_w E^+_{\bar{w}} + E^+_w E^-_w + E^+_{\bar{w}} E^-_{\bar{w}} \right) } \left( k^{ij} A_{i w} A^{j}_{j \bar{w}} + \bar{k}^{\overline{i} \overline{j}} \bar{A}_{\overline{i} w} \bar{A}^{\overline{j}}_{\bar{w}} \right) \, , \nonumber \\
    P &= \frac{1}{4 \left( E^+_w E^-_{\bar{w}} + E^-_w E^+_{\bar{w}} + E^+_w E^-_w + E^+_{\bar{w}} E^-_{\bar{w}} \right)} \left(  k^{ij}  A_{i w} A^{j}_{\bar{w}} - \bar{k}^{\overline{i} \overline{j}} \bar{A}_{\overline{i} w} \bar{A}^{\overline{j}}_{\bar{w}} \right) \, ,
\end{align}
in such a way that they reduce to their flat-space values when the vielbeins are given by (\ref{euclidean_flat}). Because these expressions are written with explicit $(w, \bar{w})$ indices, the resulting coupling to gravity is not manifestly Lorentz-invariant. However, this is to be expected since we are performing the equivalent of the procedure used in equation (\ref{general_class_vielbeins}) for coupling Floreanini-Jackiw bosons to gravity, which is also not manifestly Lorentz-invariant.

We now compute the stress tensor (\ref{stress_tensor_flat_curved}) using this coupling to the frame fields. In order to make comparison with the results of section \ref{sec:classical} easier, we will re-scale the stress tensor by an overall factor to absorb the multiplicative constant of $- \frac{i}{16 \pi}$ in the boundary term (\ref{CS_bdry_coupled}), as well as the relative factor of $2$ due to the conventions for the vielbein in this section. Therefore we instead compute
\begin{align}\label{stress_tensor_flat_curved_rescaled}
    T_\beta{}^a = - \frac{ 8 \pi i}{E} \frac{\delta S}{\delta E^\beta{}_a} \, ,
\end{align}
and convert to spacetime indices to find
\begin{align}
    T_{ww} &= - \frac{1}{4} \left( 2 S V_S + 2 P \left( 1 + V_P \right) \right) \, , \nonumber \\
    T_{w \bar{w}} &= T_{\bar{w} w} = \frac{1}{2} \left( V - S V_S - P V_P \right) \, , \nonumber \\
    T_{\bar{w} \bar{w}} &= \frac{1}{2} \left( P - P V_P - S V_S \right) \, .
\end{align}
The two Lorentz scalars that we use for constructing flows are therefore
\begin{align}
    T^\alpha{}_\alpha &= 2 \left( V - S V_S - P V_P \right) \, , \nonumber \\
    T^{\alpha \beta} T_{\alpha \beta} &= V^2 - 2 P^2 + \left( V - 2 \left( S V_S + P V_P \right) \right)^2 \, ,
\end{align}
which exactly matches equations (\ref{trT}) and (\ref{trT2}).

It now follows that all of our comments about stress tensor flows in section \ref{sec:classical} immediately apply to deformations of Chern-Simons boundary terms which are constructed using the stress tensor (\ref{stress_tensor_flat_curved_rescaled}) obtained from coupling to vielbeins, as opposed to the standard Hilbert stress tensor. For instance, any deformation by a function of the vielbein stress tensor (\ref{stress_tensor_flat_curved_rescaled}) necessarily preserves the condition (\ref{CS_chirality_constraint}). This means that, if one begins with a seed Chern-Simons boundary term which is invariant under the symmetry (\ref{cs_sd_trans}) that guarantees the chirality (or self-duality) of the theory, and then deforms this seed by any function of the energy-momentum tensor, the resulting deformed boundary term will also be invariant under the same symmetry. Furthermore, any one-parameter family of Chern-Simons boundary terms which are all invariant under the duality rotation (\ref{cs_sd_trans}) must satisfy a differential equation driven by a function of the vielbein stress tensor.

It also follows that the closed-form solutions to flow equations driven by functions of the stress tensor discussed in section \ref{sec:classical} -- such as the two-parameter family of solutions (\ref{scalar_modified_nambu_goto}) to the commuting $T\overline{T}$ and root-$T\overline{T}$ flow equations -- also have obvious analogs for deformations of Chern-Simons boundary terms. Besides solving these differential equations directly, a complementary way to analyze stress tensor deformations is by performing a perturbative expansion which computes the deformed action order-by-order in the flow parameter. This approach is discussed in appendix \ref{app:TTn} for deformations by various functions of the energy-momentum tensor, using the version of $T_{\alpha \beta}$ defined by coupling to vielbeins.

To conclude this section, let us summarize and mention some applications. We have seen that the boundary term of a bulk $U(1)$ Chern-Simons theory can be deformed either by functions of the Hilbert stress tensor or by functions of the vielbein stress tensor (\ref{stress_tensor_flat_curved_rescaled}). The former deformations lead to a class of modified boundary terms $\mathcal{L}_{\text{bdry}} ( S_1, S_2 )$ with the same properties as Lagrangians obtained by stress tensor deformations of non-chiral boson theories. Conversely, the latter flows generate a family of boundary terms $\mathcal{L}_{\text{bdry}} ( S, P )$ with the same structure as the Lagrangians in section \ref{sec:classical} arising from stress tensor deformations of chiral boson theories. We have thus described two complementary ways to view deformations of Chern-Simons boundary terms by functions of the energy-momentum tensor.

These results provide a general framework for studying three-dimensional $U(1)$ Chern-Simons theories subject to boundary deformations. Throughout our discussion, we have been agnostic as to the specific setting in which such Chern-Simons terms arise, but let us briefly mention two specific applications of the formalism we have developed. One context in which these results could be useful is when considering $\mathrm{AdS}_3/\mathrm{CFT}_2$ holography with $U(1)$ gauge fields. One could use our machinery to derive flow equations for various observables under stress tensor deformations, just as \cite{Ebert:2022ehb} found expressions for $T\overline{T}$-deformed Wilson lines and loops, and \cite{Ebert:2023tih} obtained formulas for the masses of BTZ black holes under a boundary root-$T\overline{T}$ deformation. For instance, one could use the results of this section to analyze the dependence of the $U(1)$ charges of charged BTZ black holes as a function of the deformation parameter for boundary $T\overline{T}$ or root-$T\overline{T}$ deformations. Another possible application of these results is to study quantum Hall systems subject to boundary deformations, which we will briefly describe in the conclusion of this chapter.

\section{Quantization Along Classical Flows}\label{sec:quantization}

In this section, we will consider the quantization of a member of the general class of interacting chiral boson models. We will work purely within the Floreanini-Jackiw description, described by an action of the form (\ref{interacting_again}), rather than in the Chern-Simons formulation of section \ref{sec:cs}. We will also work in Lorentzian signature with spacetime coordinates $(t, \theta)$. Although in the preceding discussion we have been agnostic as to the spacetime topology, within this section we will assume that $\theta$ is compact and subject to the identification $\theta \sim \theta + 2 \pi$. We focus on the case of a compact spatial manifold because our primary observable of interest is the finite-volume spectrum of energy levels $E_n$, and in particular how these energies depend on a deformation parameter along a stress tensor flow.

The most well-studied example of a stress tensor deformation for which the deformed cylinder spectrum can be determined is the $T\overline{T}$ deformation. Under the $T\overline{T}$ flow, the energy levels of the deformed theory obey the inviscid Burgers' equation,
\begin{align}\label{burgers}
    \frac{\partial E_n}{\partial \lambda} = E_n \frac{\partial E_n}{\partial R} + \frac{P_n^2}{R} \, ,
\end{align}
where $R$ is the radius of the cylinder and $E_n, P_n$ are the energy and momentum of the eigenstate under consideration \cite{Zamolodchikov:2004ce, Smirnov:2016lqw,Cavaglia:2016oda}.\footnote{One can also study various generalizations of this flow for the spectrum, such as the energy levels of tensor product theories where the factors are sequentially deformed by multiple $T\overline{T}$ flows \cite{Ferko:2022dpg}.}

This example is remarkable because the flow equation (\ref{burgers}) can be proven directly at the quantum level using the properties of the local $T\overline{T}$ operator, which is defined by
\begin{align}\label{TT_point_split}
    \mathcal{O}_{T\overline{T}} ( x ) = \lim_{y \to x} \left( T^{\alpha \beta} ( x ) T_{\alpha \beta} ( y ) - T^\alpha{}_\alpha ( x ) T^\beta{}_\beta  ( y ) \right) \, .
\end{align}
It was demonstrated in \cite{Zamolodchikov:2004ce} that the coincident point limit on the right side of (\ref{TT_point_split}) actually gives rise to a well-defined local operator, up to total derivative ambiguities which can be ignored. One can therefore prove results about a $T\overline{T}$-deformed quantum field theory at the quantum level using the properties of this operator; for instance, an argument involving a certain factorization property of $\mathcal{O}_{T\overline{T}}$ and the interpretation of the components of the stress tensor in terms of energy and momentum lead to the flow equation (\ref{burgers}). 

This is in contrast with a different method for attempting to learn about the quantum mechanical properties of a stress tensor deformation, which we refer to as \emph{quantization along a classical flow}. In this case, one first finds the solution to a differential equation of the form (\ref{stress_tensor_flow_defn}) for the Lagrangian of a deformed theory, and then attempts to quantize this deformed Lagrangian directly.

Assuming that a given classical deformation can be rigorously defined at the quantum level, we do not expect that quantization along the classical flow will give accurate information about \emph{all} aspects of the deformed quantum field theory. Indeed, this is already true for the $T\overline{T}$ deformation. For instance, it can be shown that the S-matrix of a $T\overline{T}$-deformed quantum field theory is equal to the S-matrix of the undeformed theory multiplied by a certain momentum-dependent phase known as a CDD factor \cite{Dubovsky:2012wk,Dubovsky:2013ira}. However, if one studies scattering using quantization along the classical $T\overline{T}$ flow, one finds that this CDD factor is not reproduced unless one adds specific counter-terms which are engineered to obtain the expected scattering behavior \cite{Rosenhaus:2019utc,Dey:2021jyl,Chakrabarti:2022lnn}. Therefore, quantization along the classical flow is not sufficient to fully characterize the properties of the $T\overline{T}$-deformed theory without additional input from the quantum definition.\footnote{Another argument for this conclusion is that quantization of theories with fermions along the classical $T\overline{T}$ flow can give different Hilbert spaces depending on which definition of the stress tensor one uses \cite{Lee:2021iut,Lee:2023uxj}.}

Despite this, one may hope that quantization of a classical deformed Lagrangian will still give \emph{some} information about the corresponding deformation at the quantum level, at least in particular limiting cases. For instance, the solution to the classical $T\overline{T}$ flow equation beginning from a seed theory of free scalars is the Nambu-Goto action of string theory, and one generically expects that string theories exhibit a high-energy density of states which is Hagedorn rather than Cardy. This predicted Hagedorn scaling agrees with an analysis of the high-energy behavior of a $T\overline{T}$-deformed CFT at the quantum level, which can be seen either from the energies \cite{Cavaglia:2016oda} or the partition function \cite{Cardy:2018sdv,Datta:2018thy}. Thus certain limiting features of the quantum theory can still be inferred from the $T\overline{T}$-deformed classical Lagrangian.

For other stress tensor deformations, like the root-$T\overline{T}$ flow, it is not yet known whether one can give a rigorous definition of the deforming operator at the quantum level. Therefore, we do not yet have any exact data about the deformed quantum theory against which to compare results obtained by other methods. However, extrapolating from the $T\overline{T}$ case, one might perform quantization along a classical root-$T\overline{T}$ flow in the hope that this procedure will still give useful information in certain limits. Our goal in this section is to carry out this quantization procedure for root-$T\overline{T}$-deformed theories of chiral bosons and examine the behavior of the deformed spectrum in such limiting cases.

One regime for which we have additional data about the root-$T\overline{T}$-deformed spectrum is the limit of a large-$c$ holographic CFT which admits a bulk $\mathrm{AdS}_3$ dual. When restricting to states for which the stress tensor is approximately constant (which are dual to BTZ black holes), one obtains the formula (\ref{zero_mode_formula}) for the root-$T\overline{T}$ deformed spectrum \cite{Ebert:2023tih}. We will see that our analysis using quantization along the classical flow agrees with this ``zero mode formula'' for states that correspond to constant stress tensor backgrounds. However, we will also be able to probe other limits of a root-$T\overline{T}$-deformed theory, such as a large-momentum limit which is \emph{not} close to a constant stress tensor configuration for which the zero mode formula is expected to apply. This result may therefore give novel information about the behavior of a putative root-$T\overline{T}$-deformed field theory in a different regime.

\label{sec:Quantization of first-order deformed}

\subsection{Generalities on quantization}\label{sec:general_quantization}

Let us now study the quantum mechanics of interacting chiral boson models such as (\ref{interacting_again}). This Floreanini-Jackiw form of the Lagrangian, although it is not manifestly Lorentz-invariant, is nonetheless convenient for quantization because it is first-order in time derivatives. This allows us to perform canonical quantization in a uniform way which does not depend on the details of the interaction function $V ( S, P )$.

We begin by reviewing some basic features of quantization of first-order systems in the simpler setting of $(0+1)$-dimensional theories, i.e. particle mechanics.

\emph{Quantization of first-order particle mechanics}

We will first consider a collection of $(0 + 1)$-dimensional fields $q_i ( t )$, whose time derivatives will be denoted $\dot{q}_i ( t )$. A general first-order Lagrangian for such a system takes the form
\begin{align}\label{eq:o0iuh1fsd}
    L = \frac{1}{2} C^{ij} q_i \dot{q}_j - V ( q ) \, ,
\end{align}
where $C^{ij}$ is a non-singular constant matrix. Without loss of generality, we may assume that $C^{ij}$ is antisymmetric. Indeed, if we instead split $C^{ij} = C^{[ij]} + C^{(i j)}$ into symmetric and anti-symmetric parts, the Lagrangian would be
\begin{align}
    L = \frac{1}{2} C^{[ij]} q_i \dot{q}_j + \frac{1}{2} C^{(ij)} \frac{d}{dt} \left( q_j q_i \right) - V ( q ) \, ,
\end{align}
where the second term is a total time derivative that can be ignored. 

The canonical momentum which is conjugate to $q_j ( t )$ is
\begin{align}\label{first_order_momentum}
    p^j= \frac{\partial L}{\partial \dot{q}_j} = \frac{1}{2} C^{ij} q_i \, ,
\end{align}
and thus the Hamiltonian associated with (\ref{eq:o0iuh1fsd}) is
\begin{align}\label{first_order_hamiltonian}
    H ( q, p ) = \frac{\partial L}{\partial \dot{q}_i} \dot{q}_i - L = V ( q ) \, .
\end{align}
The Hamiltonian (\ref{first_order_hamiltonian}) appears to depend only on the position variables but not on the momenta, but this is misleading, since equation (\ref{first_order_momentum}) implies that some combinations of the $q_i$ \emph{are} momenta. The Euler-Lagrange equations arising from the Lagrangian (\ref{eq:o0iuh1fsd}) are
\begin{align}
    C^{ij} \dot{q}_i = \frac{\partial H}{\partial q_j} \, ,
\end{align}
where we now use the symbols $V$ and $H$ interchangeably. Alternatively, by defining $C_{ij}$ to be the inverse matrix $\left( C^{-1} \right)_{ij}$ of $C^{ij}$, the equations of motion can be written as
\begin{align}\label{first_order_EL}
    \dot{q}_i = C_{ij} \frac{\partial H}{\partial q_j} \, .
\end{align}
Next we consider the quantization of this model. Ordinarily, for Lagrangians which are quadratic in time derivatives, one would impose the canonical commutation relations
\begin{align}\label{ccr}
    [ x_i, p_j ] = i \delta_{ij} \, .
\end{align}
However, imposing the relations (\ref{ccr}) for a first-order system like (\ref{eq:o0iuh1fsd}) gives results that differ from the correct commutation relations by a factor of $2$. To arrive at the correct relations, we follow the prescription outlined in appendix A of \cite{Floreanini:1987as}, and further justified in \cite{Faddeev:1988qp}, which is to define commutators so that the Heisenberg-picture time evolution of operators in the quantum theory takes the same form as the classical Euler-Lagrange equations.\footnote{In conventional quantum systems with second-order Lagrangians, the fact that these two equations should take the same form is the content of the Ehrenfest theorem. We demand that the same is true here.}

In general, the Heisenberg equation of motion for an operator $\mathcal{O}$ reads $\dot{\mathcal{O}} = i [ H , \mathcal{O} ]$. In the case of the operator $\mathcal{O} = q_i$, we have
\begin{align}\label{heisenberg_eom}
    \dot{q}_i = i [ H , q_i ] = i \frac{\partial H}{\partial q_j} [ q_j, q_i ] \, .
\end{align}
Comparing (\ref{heisenberg_eom}) to (\ref{first_order_EL}), we find that the two take the same form if we identify
\begin{align}\label{correct_commutation}
    [ q_i, q_j ] = i C_{ij} \, .
\end{align}
As we mentioned, since $p^j = \frac{1}{2} C^{ij} q_i$, this differs from the canonical prescription (\ref{ccr}) which would give $[ q_i, q_j ] = 2 i C_{ij}$. The errant factor of $2$ is due to the fact that, in a first-order system, there is a constraint on the phase space.

\emph{Quantization of first-order field theories}

Having reviewed the quantum mechanics of first-order $(0+1)$-dimensional systems, we now turn to the quantization of first-order $(1+1)$-dimensional field theories, and in particular the theories of chiral bosons which are the focus of this work.

As a simple example to set the stage, we will first consider a single chiral boson described by the Floreanini-Jackiw Lagrangian (\ref{fj_action}) which we repeat here:
\begin{equation}\label{FJ_later}
    \mathcal{L} = \frac{1}{2}  \left(\phi' \dot{\phi} - \phi' \phi' \right) \, .
\end{equation}
As usual, we write $\dot{\phi}$ for the time derivative of $\phi$ and $\phi'$ for the spatial derivative of $\phi$. The quantization of this system in infinite volume, i.e. with a spatial coordinate $x \in \mathbb{R}$, was first studied in \cite{Floreanini:1987as}. In short, one can view $x$ as a continuous generalization of the discrete labels $i$, $j$ in \eqref{eq:o0iuh1fsd} and rewrite the first term as
\begin{equation}
\begin{aligned}
    \frac{1}{2} \int dx\, \partial_x \phi(x,t) \dot{\phi}(x,t) &= \frac{1}{2} \int dx \int dy \, \delta(x-y) \partial_x \phi(x,t) \dot{\phi}(y,t) \\
    &= - \frac{1}{2} \int dx \int dy \, [\partial_x \delta(x-y)] \phi(x,t) \dot{\phi}(y,t) \, .
\end{aligned}
\end{equation}
The role of the constant antisymmetric matrix $C^{ij}$ in the particle mechanics example is now played by the function 
\begin{align}
    C(x-y) = -\partial_x \delta(x-y) \, ,
\end{align}
and the role of the inverse matrix $C_{ij}$ is played by the Green's function of $C ( x - y )$. This suggests that we impose the commutation relations
\begin{equation}
    [\phi(x), \phi(y)] =  - \frac{i}{2}\text{sgn}(x-y)~,
\end{equation}
which is the field theory analog of (\ref{correct_commutation}) and which matches the result in \cite{Floreanini:1987as}. It is then straightforward to use the above equal-time commutation relations to confirm the Heisenberg equations of motion are indeed equivalent to the Euler-Lagrange equations of the Lagrangian (\ref{FJ_later}), which describe a chiral boson:
\begin{equation}
    \frac{\partial \mathcal{O}}{\partial t} = -i [\mathcal{O},H] \quad \implies \quad \dot{\phi} = \phi' \, .
\end{equation}
Next we will study this theory in finite volume. We now replace the spatial coordinate $x \in \mathbb{R}$ with an angular coordinate $\theta$ labeling a position on $S^1$, and subject to the identification $\theta \sim \theta + 2 \pi$. We will also assume that the \emph{target space} is compact, which means that $\phi$ likewise takes values in a circle so that $\phi \sim \phi + 2 \pi$. As we will see, the structure of this theory on a cylinder is closely related to the particle mechanics example considered above.

First let us write the function $\phi(t,\theta)$ using a mode expansion:
\begin{equation}\label{phi_mode_expansion}
    \phi(t,\theta) = \frac{1}{\pi} x(t) + p(t) \theta + \frac{1}{\sqrt{2\pi}} \sum_{n=1}^\infty \frac{1}{\sqrt{n}} \left( a_{n}(t) e^{in\theta}+ a^\dagger_n(t) e^{-in\theta} \right) \, .
\end{equation}
We have included a zero-mode term $x(t)$ in addition to a momentum contribution which is linear in $\theta$; the latter is permissible, despite not being periodic in $\theta$, since both $\theta \sim \theta + 2 \pi$ and $\phi \sim \phi + 2 \pi$, so such a term is compatible with our identifications if $p \in \mathbb{Z}$. The remaining sum is the standard Fourier expansion of the periodic part of $\phi$ in the $\theta$ direction.

It is now necessary to distinguish between the Lagrangian density $\mathcal{L}$ and the Lagrangian $L = \int d \theta \, \mathcal{L}$. Substituting the mode expansion (\ref{phi_mode_expansion}) into the Lagrangian density (\ref{FJ_later}) and performing the integral over the $\theta$ coordinate gives
\begin{equation}
    L = \int_{0}^{2\pi} d\theta \, \mathcal{L} =  p \dot{x} - \pi p^2 +  \frac{i}{2} \left(\sum_{n=1}^\infty  (\dot{a}^\dagger_n a_n - \dot{a}_n a_n^\dagger) \right) - \left(\frac{1}{2} \sum_{n =1} ^\infty n (a_n a^\dagger_n + a^\dagger_n a_n)  \right) ~,
\end{equation}
where we have dropped a term that is a total derivative in time. Because $p$ is integer-quantized, as we mentioned above, the first term describes the well-known quantum system which is a particle on a ring. The Hilbert space is generated by states $|p\rangle$ labeled by integer $p \in \mathbb{Z}$ with energy $E_p = \pi p^2$. The remaining terms are nothing but the familiar first-order particle mechanics system discussed previously. To make this analogy clearer, it is convenient to define
\begin{align}
    a_{-n} = a_n^\dagger \, ,
\end{align}
so that the Lagrangian can be written as
\begin{equation}\label{fj_rewritten_modes}
    L =  p \dot{x} - \pi p^2 +  \frac{i}{2} \left(\sum_{n=1}^\infty  (\dot{a}_{-n} a_n - \dot{a}_n a_{-n}) \right) - \left(\frac{1}{2} \sum_{n =1} ^\infty n (a_n a_{-n} + a_{-n} a_n)  \right) \, .
\end{equation}
The $a_n$'s now play the role of $q_i$'s, except the modes are labeled by $n \in \mathbb{Z}$ so the phase space is infinite-dimensional. Comparing the two sums in the Lagrangian (\ref{fj_rewritten_modes}) with the general form \eqref{eq:o0iuh1fsd}, we find that the two agree if we identify
\begin{equation}
    C^{n,m} = i \, \text{sgn}(n) \delta_{n,-m} \, .
\end{equation}
Therefore, when we promote the $a_n$ from functions appearing in the expansion of the classical field $\phi$ to quantum operators, the appropriate commutation relations (\ref{correct_commutation}) are
\begin{align}
    [a_n,a_m] = \text{sgn}(n) \delta_{n,-m} \, .
\end{align}
When expressed in terms of $a^\dagger_m$, this is the familiar commutation relation of ladder operators:
\begin{equation}\label{standard_ladders}
    [a_n, a_m^\dagger] = \delta_{n,m} \, .
\end{equation}
It is perhaps surprising that, if we had worked with the Fourier modes $a_n$ of the field $\phi$ from the beginning (rather than with the field $\phi$ itself), then imposing the standard commutation relations (\ref{standard_ladders}) gives the correct result, without the errant factor of $2$ which we mentioned around equation (\ref{correct_commutation}) that occurs due to the phase space constraint on first-order systems. The reason for this is that, after performing the mode expansion, the positive Fourier modes $a_n$ with $n > 0$ act as the position variables and the negative modes $a_n$ with $n < 0$ (or equivalently $a^\dagger_n$) act as the conjugate momentum variables. Therefore, in Fourier space, the separation between coordinates and momenta is automatic, and we need not impose phase space constraints or consider commutation relations like (\ref{correct_commutation}) which na\"ively appear to involve two position variables.\footnote{See section 6.1.3 of \cite{Tong:2016kpv} for a pedagogical review of the quantization of the chiral boson from this momentum-space perspective, and later sections of this reference for applications to quantum Hall physics.}

The Hamiltonian obtained from the Legendre transform of the Lagrangian (\ref{fj_rewritten_modes}), written in terms of $a_n^\dagger$ rather than $a_{-n}$, is
\begin{align}
    H &= \pi p^2 + \frac{1}{2} \sum_{n=1}^\infty n (a_n a_n^\dagger + a_n^\dagger a_n) \nonumber \\
    &= -\frac{1}{24} + \pi p^2 +  \sum_{n=1}^\infty n a^\dagger_n a_n \, ,
\end{align}
where we have used $a_n a_n^\dagger = a_n^\dagger a_n + 1$ and the well-known $\zeta$-function regularization
\begin{equation}
    \sum_{n=1}^\infty n = - \frac{1}{12} \, .
\end{equation}
It is straightforward to generalize the above discussion to the case of multiple chiral and anti-chiral bosons. We work with a Lagrangian density for $N$ chiral bosons $\phi^i$, $i = 1, \ldots, n$, and $\bar{N}$ anti-chiral bosons $\bar{\phi}^{\overline{i}}$, of the form (\ref{interaction_function}) which we have been considering in section \ref{sec:classical}. For simplicity we take trivial target-space metrics for the bosons, $G_{ij} = \delta_{ij}$ and $\bar{G}_{\overline{i} \overline{j}} = \delta_{\overline{i} \overline{j}}$. The Lagrangian density for this system is then
\begin{equation}
    \mathcal{L} = \frac{1}{2}  \left(\phi'_i \dot{\phi}^i - \bar{\phi}_{\overline{i}}' \dot{\bar{\phi}} {}^{\overline{i}} \right) -V ( \phi_i', \bar{\phi}_{\overline{i}}^{\prime} ) ~.
\end{equation}
We expand both the chiral and anit-chiral fields in modes as
\begin{equation}\label{eq:phiphibarexpan}
\begin{aligned}
    \phi_i(t,\theta) &= \frac{1}{\pi} x_i(t) + p_i(t) \theta + \frac{1}{\sqrt{2\pi}} \sum_{n = 1}^{\infty} \frac{1}{\sqrt{n}} \left(a_{i,n}(t) e^{in\theta} + a_{i,n}^\dagger(t) e^{-in\theta} \right)~, \\
    \bar{\phi}_{\overline{i}} (t,\theta) &= - \frac{1}{\pi} \bar{x}_{\overline{i}} (t) + \bar{p}_{\overline{i}} (t) \theta + \frac{1}{\sqrt{2\pi}} \sum^\infty_{n=1}\frac{1}{\sqrt{n}}\left( b_{\overline{i} , n}(t) e^{-in\theta} + b_{\overline{i} ,n}^\dagger(t) e^{in\theta}\right)~.
\end{aligned}
\end{equation}
The non-zero commutation relations between the various expansion coefficients are
\begin{equation}\label{general_bosons_commutators}
    [x_i, p_j] = i \delta_{ij}, \quad [ \bar{x}_{\overline{i}}, \bar{p}_{\overline{j}} ] = i \delta_{\overline{i} \overline{j}}, \quad [a_{i,n}, a_{j,m}^\dagger] = \delta_{i j} \delta_{n m}, \quad [b_{\overline{i},n}, b_{\overline{j},m}^\dagger] = \delta_{\overline{i} \overline{j}} \delta_{n m} \, , 
\end{equation}
with all other commutators vanishing.

Note that here we take all $\phi_i$ and $\bar{\phi}_{\overline{i}}$ to be compact with radius $2\pi$. Therefore, the eigenvalues of $p_j$ and $\bar{p}_{\overline{j}}$ must be integers. The Hamiltonian is given by
\begin{equation}
    H = \int_0^{2\pi} d\theta \, V(\phi'_i, \bar{\phi}'_{\overline{j}} ) ~.
\end{equation}
The commutation relations (\ref{general_bosons_commutators}) allow us to build the Hilbert space of the quantum theory for any potential $V$. In the next subsection, we will use this to study the spectrum of the ``Modified Scalar'' theory, that is, the theory obtained by applying a root-$T\overline{T}$ deformation to a seed theory of free chiral and anti-chiral bosons.

\subsection{Root-$T\overline{T}$-deformed spectrum}
\label{sec:quantumspectrum}

We will now use the formalism reviewed in section \ref{sec:general_quantization} to study root-$T\overline{T}$-deformed free boson theories. In principle, this can be done for any numbers $(N, \bar{N})$ of chiral and anti-chiral bosons, respectively. However, there is a sharp distinction between the case $N = \bar{N} = 1$, for which the deformation is comparatively simple and can be interpreted as a re-scaling of the target space radius for the boson, and all other cases with $N \geq 1$ and $\bar{N} \geq 1$, where the deformation is more non-trivial.\footnote{Note that, if either $N = 0$ or $\bar{N} = 0$, then the theory is a fixed point of stress tensor flows so the root-$T\overline{T}$ deformation is trivial.} We will therefore first discuss the simpler case $N = \bar{N} = 1$ in detail, and then as an illustrative example of the latter class, we will study the example with $N = 2$ and $\bar{N} = 1$. We expect that the qualitative features of the deformed $(N, \bar{N}) = (2, 1)$ model will be similar to those of theories with larger $N$ and $\bar{N}$.

\emph{One compact boson}

Let us begin by studying the root-$T\overline{T}$ deformation of a single (non-chiral) $c = 1$ compact boson, or equivalently, a pair of $N = 1$ left-moving and $\bar{N} = 1$ right-moving chiral bosons. It was already mentioned in the initial work \cite{Ferko:2022cix} that, in this case, the root-$T\overline{T}$ flow simply rescales the kinetic term for the boson, which corresponds to a change in the radius if the scalar is compact. We will revisit this claim by describing the deformed model in terms of chiral bosons and determining the quantum spectrum exactly to confirm that the root-$T\overline{T}$ deformation of a compact boson is just a change of radius. 

This formalism also provides a way to realize a compact boson at an arbitrary radius -- even at irrational points where the theory does not factorize into the chiral part and anti-chiral part -- using a Lagrangian for one chiral and one anti-chiral boson with a quadratic mixing term. Furthermore, treating this example in detail will allow us to test the ``zero mode formula'' given in equation (\ref{zero_mode_formula}) that is expected, due to evidence from holography \cite{Ebert:2023tih}, to describe the energies of states in root-$T\overline{T}$-deformed CFTs for which the energy-momentum tensor is constant in space. We will see explicitly that this zero mode formula fails to give the energies of deformed states for which this assumption is violated.

The Lagrangian for a root-$T\overline{T}$-deformed seed theory of one left-moving and one right-moving chiral boson takes the form (\ref{interaction_function}) with an interaction function $V ( S, P , \gamma )$ given by the $\lambda \to 0$ limit of equation (\ref{scalar_modified_nambu_goto}). To be pedantic, the resulting Lagrangian is technically
\begin{equation}
    \mathcal{L}^{(\gamma)} = \frac{1}{2} \left( \phi' \dot{\phi} - \bar{\phi}' \dot{\bar{\phi}} \right) - \frac{\cosh ( \gamma )}{2} \left( \phi^{\prime 2} + \bar{\phi}^{\prime 2} \right) - \sinh ( \gamma ) \sqrt{ \left( \phi' \right)^2 \left( \bar{\phi}' \right)^2 }  \, .
\end{equation}
That is, because $\phi'$ and $\dot{\phi}$ can take both positive and negative values the final term is really proportional to $| \phi' | \cdot | \bar{\phi}' |$. However, we will ignore this subtlety and simply replace $\sqrt{ \left( \phi' \right)^2 \left( \bar{\phi}' \right)^2 }$ with $\phi' \bar{\phi}'$. This can be justified, for instance, by restricting attention to small fluctuations of the fields around a background for which the gradients are large and positive, so that both $\phi'$ and $\bar{\phi}'$ have fixed positive sign. This corresponds to a solution with large positive values of $p_i$ and $\bar{p}_i$ in the expansion of equation (\ref{eq:phiphibarexpan}). We will take a similar large-momentum limit in the analysis with several bosons below, again resolving the square root, which is more non-trivial in that setting because of an additional term under the root.

After making this simplification, the Lagrangian we wish to study becomes
\begin{equation}\label{eq:1bL}
    \mathcal{L}^{(\gamma)} = \frac{1}{2} \left( \phi' \dot{\phi} - \bar{\phi}' \dot{\bar{\phi}} \right) - \frac{\cosh ( \gamma )}{2} \left( \phi^{\prime 2} + \bar{\phi}^{\prime 2} \right)  - \sinh ( \gamma )  \phi' \bar{\phi}'  \,.
\end{equation}
As discussed previously, the Hilbert space factorizes into two parts: the particles on a ring and the infinite tower of harmonic oscillators. Due to the special form of \eqref{eq:1bL}, the Hamiltonian does not mix the two parts. Therefore, we can study them separately.

Let us first consider the sector of the Hilbert space which describes the particles on a ring. We write the states in this Hilbert space as $|p,\overline{p}\rangle$, which are labeled by two quantized momenta $p,\overline{p} \in \mathbb{Z}$. The corresponding Hamiltonian and the momentum operator are
\begin{equation}
\label{eq:0iouv1f}
    H_{\text{PR}}^{(\gamma)} = \pi (p^2 + \overline{p}^2) \cosh ( \gamma  ) + 2\pi p \overline{p} \sinh ( \gamma ) \,, \quad P^{(\gamma)}_{\text{PR}} = \pi (p^2 - \overline{p}^2 ) = P^{(0)}_{\text{PR}} \, ,
\end{equation}
where we use the subscript PR to denote \underline{p}articles on a \underline{r}ing.

Because the corresponding undeformed states at $\gamma = 0$ have energies
\begin{align}
    H_{\text{PR}}^{(0)} = \pi (p^2 + \overline{p}^2) \, ,
\end{align}
we see that the prediction for the deformed energies from the zero mode formula (\ref{zero_mode_formula}) is
\begin{align}\label{zero_mode_works_winding}
    E_{\text{PR}}^{(\gamma)} &= H_{\text{PR}}^{(0)} \cosh ( \gamma ) + \sqrt{ \left( H_{\text{PR}}^{(0)} \right)^2 - \left( P^{(0)}_{\text{PR}} \right)^2 } \sinh ( \gamma ) \nonumber \\
    &= \pi (p^2 + \overline{p}^2) \cosh ( \gamma  ) + \sqrt{ \left( \pi (p^2 + \overline{p}^2) \right)^2 - \left( \pi (p^2 - \overline{p}^2 ) \right)^2 } \sinh ( \gamma ) \nonumber \\
    &= \pi (p^2 + \overline{p}^2) \cosh ( \gamma  ) + 2\pi p \overline{p} \sinh ( \gamma ) \, ,
\end{align}
which indeed agrees with the true deformed energies $H_{\text{PR}}^{(\gamma)}$ of equation (\ref{eq:0iouv1f}), subject to the usual caveat that we have used the assumption $\sqrt{ p^2 \bar{p}^2 } = p \bar{p}$.

It is not too surprising that these states have deformed energies which agree with the zero-mode formula, since the corresponding saddle points have constant stress-energy tensors, and this is the assumption under which the formula (\ref{zero_mode_formula}) was derived in holography. 

To see this explicitly, we look for solutions to the equations of motion associated with the deformed Lagrangian $\mathcal{L}^{(\gamma)}$ in equation (\ref{eq:1bL}), which are
\begin{equation}
\begin{aligned}
\label{eq:0i9ou111efs@}
  \dot{\phi}' - \phi'' \cosh ( \gamma ) - \bar{\phi}'' \sinh ( \gamma ) = 0\,, \quad 
  \dot{\bar{\phi}} {}^{\prime} + \bar{\phi}'' \cosh ( \gamma ) + \phi'' \sinh ( \gamma )  = 0 \, .
\end{aligned}
\end{equation}
One can integrate these equations with respect to the spatial coordinate $\theta$, up to an undetermined integration constant $h ( t )$ which is an arbitrary function of $t$. As in the discussion around equation (\ref{eom_gauge_choice}), one can always set $h ( t ) = 0$ by a gauge transformation. Specializing to this $h = 0$ gauge, the equations of motion become
\begin{equation}\label{winding_h=0_eom}
    \dot{\phi} - \phi' \cosh ( \gamma ) - \bar{\phi}' \sinh ( \gamma ) = 0, \quad \dot{\bar{\phi}} + \bar{\phi}' \cosh ( \gamma ) + \phi' \sinh ( \gamma ) = 0 \, .
\end{equation}
We wish to solve the equations of motion (\ref{winding_h=0_eom}) subject to the boundary conditions
\begin{equation}
\label{eq:deformedperiodicity}
    \phi(\theta + 2\pi, t) - \phi(\theta, t) = 2\pi p\,, \quad \bar{\phi} ( \theta + 2\pi, t ) - \bar{\phi} (\theta, t) = 2 \pi \bar{p}\,,
\end{equation}
where $p, \bar{p} \in \mathbb{Z}$. The desired solutions with such periodic boundary conditions are
\begin{equation}
\begin{aligned}
 \phi_p^{(\gamma)}&= p \theta + \left(p \cosh ( \gamma ) + \bar{p} \sinh ( \gamma )  \right) t \, , \quad
\bar{\phi}_{\bar{p}}^{(\gamma)} = \bar{p} \theta -\left(\bar{p} \cosh ( \gamma ) + p \sinh ( \gamma ) \right) t\,.
\end{aligned}
\end{equation}
Since these solutions $\phi_{p}^{(\gamma)}$ and $\bar{\phi}_{\overline{p}}^{(\gamma)}$ depend on $t,\theta$ linearly, the corresponding stress-energy tensor is constant. Therefore, it is reasonable that the energies of these states are indeed governed by the energy formula derived via AdS$_3$/CFT$_2$ holography for constant stress tensor backgrounds, as we found around equation (\ref{zero_mode_works_winding}).\footnote{Strictly speaking, the derivation of this zero mode formula also assumes that the boundary theory is a large-$c$ holographic CFT for which we can trust semiclassical bulk gravity. However, this assumption does not seem strictly necessary for the zero mode formula to hold, since the theory we study here has $c = 1$.}

We would also like to point out that the energies of these states agree with the energies of momentum states for a compact boson with a different radius. To see this, it is convenient to change variables as
\begin{equation}
   w= \sqrt{\pi} \left( p+ \bar{p} \right)\,, \quad \bar{w} =\sqrt{\pi} \left(p- \bar{p}\right)\,, \quad R = \exp \left( - \frac{\gamma}{2} \right) \, ,
\end{equation}
so that the deformed Hamiltonian \eqref{eq:0iouv1f} can be written as
\begin{equation}
\label{eq:okjbgfty2!}
    H^{(\gamma)}_{\text{PR}}  = \frac{1}{2} \left( \frac{w^2}{R^2} + R^2 \bar{w}^2 \right) \, .
\end{equation}
This supports the claim that the root-$T\overline{T}$ deformation, in this case, corresponds to a rescaling of the target-space radius for the compact boson. However, to verify this conclusion, we should also study the effect of the deformation in the other sector of the Hilbert space, which describes an infinite tower of harmonic oscillators.

We turn to this task now. Expanding the field $\phi$ and $\bar{\phi}$ as in \eqref{eq:phiphibarexpan}, we find the Hamiltonian operator and the momentum operator for this oscillator sector are given by
\begin{equation}
\begin{aligned}
    H^{(\gamma)}_{\text{OS}} &= \sum_{n=1}^{\infty} n  (a_n^\dagger a_n + b^\dagger_n b_n)\cosh (\gamma) + \sum_{n=1}^\infty n (a_n^\dagger b_n^\dagger + a_n b_n)\sinh(\gamma ) - \frac{1}{12} \cosh (\gamma)\,, \\
    P^{(\gamma)}_{\text{OS}} &= \sum_{n=1}^{\infty} n(a_n^\dagger a_n - b_n^\dagger b_n) \, , 
\end{aligned}
\end{equation}
where we have performed normal ordering as before and where OS stands for oscillators. This Hamiltonian has exactly the same spectrum as its undeformed counterpart, which can be made manifest by the following Bogoliubov transformation:
\begin{equation}\label{bogoliubov}
\begin{aligned}
    a_n=\tilde{a}_n \cosh \left( \frac{\gamma}{2} \right) - \tilde{b}^\dagger_n \sinh \left( \frac{\gamma}{2} \right) \, , \quad
    b_n= \tilde{b}_n \cosh \left( \frac{\gamma}{2} \right) - \tilde{a}^{\dagger}_n \sinh \left( \frac{\gamma}{2} \right) \, .
\end{aligned}
\end{equation}
We note that this has the same structure as the change of variables which diagonalized the mixing term between the two Chern-Simons gauge fields in equation (\ref{CS_bogoliubov}). This transformation preserves the commutation relation, i.e.
\begin{equation}
    [\tilde{a}_n, \tilde{a}^\dagger_n] = [\tilde{b}_n, \tilde{b}^\dagger_n] = 1~.
\end{equation}
In terms of the new oscillators, the Hamiltonian then reduces to the undeformed one,
\begin{equation}
\label{eq:plkjngtyui12}
    H^{(\gamma)}_{\text{OS}} = - \frac{1}{12} + \sum^\infty_{n=1} n \left(\tilde{a}^{\dagger}_n \tilde{a}_n + \tilde{b}^{\dagger}_n \tilde{b}_n \right)\,,
\end{equation}
while the momentum operator is unchanged,
\begin{equation}
    P_{\text{OS}}^{(\gamma)} = \sum_{n=1}^\infty n \left(\tilde{a}_n^\dagger \tilde{a}_n - \tilde{b}_n^\dagger \tilde{b}_n\right)\,.
\end{equation}
Hence, we conclude that the energies in the oscillator sector of the Hilbert space do not flow under the root-$T\overline{T}$ deformation. This agrees with the effect of changing the radius for a compact boson, which likewise does not change the energies of oscillator excitations.

Therefore, combining this result with the flow of $H_{\text{PR}}^{(\gamma)}$, we conclude that indeed the root-$T\overline{T}$ deformation corresponds to a change of radius for a single compact boson. 

We have also verified that the zero-mode energy formula (\ref{zero_mode_formula}) proposed in \cite{Ebert:2023tih} does not apply to generic states in a root-$T\overline{T}$-deformed CFT. For instance, any state with $p = \overline{p} = 0$ but with oscillator excitations will have an energy that is unchanged by the root-$T\overline{T}$ flow, whereas the formula (\ref{zero_mode_formula}) would predict that the energy flows with $\gamma$. This is because such oscillator states have non-constant stress tensors and therefore violate the assumptions under which the zero-mode formula was derived. However, we reiterate that the states which \emph{do} have constant stress tensors -- namely, states with general $p$ and $\bar{p}$ but no oscillator excitations -- indeed have energies which flow according to the zero mode formula.

\emph{Multiple compact bosons}

Next we aim to study the spectrum for the theory of root-$T\overline{T}$-deformed free bosons when there are more fields, rather than just a single left-mover and a single right-mover. All of these cases are qualitatively similar, in the sense that the argument of the square root appearing in the Lagrangian is no longer a perfect square, and thus cannot be resolved to a simple product of fields as in the $N = \bar{N} = 1$ case above. For simplicity, we will therefore focus on the first non-trivial case, which has $N = 2$ left-movers and $\bar{N} = 1$ right-movers (the case with $N = 1$ and $\bar{N} = 2$ is identical, after exchanging chiral and anti-chiral fields).

The Hamiltonian for the deformed $(N, \bar{N}) = (2, 1)$ theory is
\begin{equation}
\begin{aligned}
\label{eq:(2,1)Ham}
    H^{(\gamma)} &=  \int d\theta \left[ \frac{1}{2} \left( \phi^{\prime 2}_1+\phi^{\prime 2}_2 + \bar{\phi}^{\prime 2}_1  \right) \cosh ( \gamma ) + \sqrt{\phi^{\prime 2}_1 + \phi^{\prime 2}_2} \bar{\phi}'_1 \sinh ( \gamma )\right] \, .
\end{aligned}
\end{equation}
To resolve the square root, our strategy will be to expand in large positive momenta and compute the energies perturbatively. The mode expansion for the fields takes the form
\begin{equation}
\begin{aligned}
\hspace{-10pt}\label{expansion_(2,1)}
    \phi_j &= p_j \theta + \frac{1}{\sqrt{2\pi}} \sum_{n=1}^\infty \frac{1}{\sqrt{n}} \left( a_{j, n} e^{in \theta} +a_{j, n}^{\dagger} e^{-in \theta} \right)\,, \\ \bar{\phi}_1 &= \bar{p}_1 \theta  +\frac{1}{\sqrt{2\pi}}  \sum_{n=1}^\infty \frac{1}{\sqrt{n}} \left( b_{1, n}^{\dagger} e^{in \theta} +b_{1, n} e^{-in \theta} \right)\,,
\end{aligned}
\end{equation}
where $j = 1, 2$ and periodicity requires $\bar{p}_1$, $p_1$, $p_2 \in \mathbb{Z}$. Substituting the expansion (\ref{expansion_(2,1)}) into our Hamiltonian (\ref{eq:(2,1)Ham}) and expanding in large $p_1$ and $\bar{p}_1$, to leading order we find
\begin{equation}
\begin{aligned}
\label{eq:ipou1;p;}
    H^{(\gamma)} &=  \left( \pi ( p_1^2 + p_2^2 + \bar{p}_1^2) - \frac{1}{8} + \sum^\infty_{n=1} n \left( N_{1, n} + N_{2, n} + \bar{N}_{1, n} \right) \right) \cosh ( \gamma) \\ 
    &\qquad + \left(2\pi p_1 \bar{p}_1  +  \sum^\infty_{n=1} n \left( a_{1, n} b_{1, n} + a^\dagger_{1, n} b^\dagger_{1, n} \right) \right) \sinh ( \gamma )  + \cdots \, ,
\end{aligned}
\end{equation}
where $N_{i, n} = a_{i, n}^\dagger a_{i, n}$ and $\bar{N}_{1, n}= b_{1, n}^\dagger b_{1, n}$ are number operators at level $n$ for left- and right-movers respectively.

We would now like to compare the spectrum of the true large-momentum Hamiltonian (\ref{eq:ipou1;p;}) to the zero-mode formula (\ref{zero_mode_formula}) predicted from holography for states with constant stress tensors. The undeformed Hamiltonian and momentum are
\begin{equation}\label{undeformed_H_P_(2,1)}
\begin{aligned}
    H^{(0)} &= \pi \left( p_1^2 + p_2^2 + \bar{p}_1^2 \right) - \frac{1}{8} + \sum^\infty_{n=1} n \left( N_{1, n} + N_{2, n} + \bar{N}_{1, n} \right) \,, \\
    P^{(0)} &= \pi \left( p_1^2 + p_2^2 - \bar{p}_1^2\right) - \frac{1}{24} + \sum^\infty_{n=1} n \left( N_{1, n} + N_{2, n} - \bar{N}_{1, n} \right) \, .
\end{aligned}
\end{equation}
Let us restrict to an eigenstate of both the momentum operators $p_1$, $p_2$, $\bar{p}_1$ and the number operators $N_{1, n}$, $N_{2, n}$, $\bar{N}_{1, n}$, in the undeformed theory. The energy and momentum of such a state are also given by the expressions (\ref{undeformed_H_P_(2,1)}), if we simply re-interpret each symbol representing an operator as instead representing the corresponding eigenvalue.\footnote{We have chosen not to denote operators by decorating them with hats, which would distinguish between operators $\widehat{N}_1$ and their corresponding eigenvalues $N_1$, to avoid cluttering the formulas.}

Substituting the energy and momentum eigenvalues for this state into the zero-mode formula (\ref{zero_mode_formula}) then gives a predicted value for a deformed energy:
\begin{align}
\label{eq:iopihjvg2}
    E^{(\gamma)}_{\text{zero mode}} &= \left( \pi \left(p_1^2 + p_2^2 + \bar{p}_1^2 \right)  - \frac{1}{8} + \sum^\infty_{n=1} n \left( N_{1, n} + N_{2, n} + \bar{N}_{1, n} \right) \right) \cosh ( \gamma ) \nonumber \\
    &\qquad + 2\pi p_1 \bar{p}_1 \sinh ( \gamma ) + \cdots \, .
\end{align}
We should stress that equation (\ref{zero_mode_formula}) is a prediction for the deformed \emph{spectrum} and not for the deformed \emph{eigenstates}. Therefore, even if equation (\ref{eq:iopihjvg2}) were correct, this would simply mean that there exists \emph{some} state in the deformed theory whose energy is $E^{(\gamma)}_{\text{zero mode}}$. 

However, with this caveat aside, it is now easy to see why the formula (\ref{eq:iopihjvg2}) is incorrect, and what effect it fails to take into account. Were it not for the final term in the true Hamiltonian (\ref{eq:ipou1;p;}), which involves $\sum^\infty_{n=1} n \left( a_{1, n} b_{1, n} + a^\dagger_{1, n} b^\dagger_{1, n} \right)$, then any eigenstate of the undeformed theory would remain an eigenstate of the deformed theory at this order in the momentum expansion, and its energy would indeed be given by (\ref{eq:iopihjvg2}). This is simply because the first several terms of the true Hamiltonian (\ref{eq:ipou1;p;}) agree with the zero-mode prediction (\ref{eq:iopihjvg2}), after replacing operators with their eigenvalues. However, the presence of this final term in (\ref{eq:ipou1;p;}) means that an eigenstate of the undeformed Hamiltonian will \emph{not} remain an eigenstate in the deformed theory, since terms like $a_{1, n} b_{1, n}$ will mix such a state into other states with different oscillator numbers. We conclude that the zero-mode energy formula (\ref{eq:iopihjvg2}) is not correct for the deformed spectrum, even in this large-momentum limit.

\emph{Possible interpretation of root-$T\overline{T}$ deformation for higher $N$, $\bar{N}$}

We have seen that, in the special case $N = \bar{N} = 1$, the root-$T\overline{T}$ deformations of chiral bosons admits a simple interpretation as a rescaling of the target-space radius. This can also be understood from the observation that, for this case, the oscillator sector of the deformed theory is equivalent to that of the undeformed theory due to the Bogoliubov transformation (\ref{bogoliubov}). To conclude this section, we would like to make some speculative remarks about possible generalizations of this interpretation to cases with higher $N$ and $\bar{N}$, which seem considerably more complicated.

First let us point out that, for the case $(N, \bar{N}) = (1, 1)$, the Bogoliubov transformation which returns the oscillator sector of the root-$T\overline{T}$ deformed theory to its undeformed form also has an analog at the level of the Lagrangian and Hamiltonian densities. Indeed, for the quadratic theory \eqref{eq:1bL}, one can write the Lagrangian and Hamiltonian densities as
 \begin{equation}
 \begin{aligned}\label{big_phi_L_H}
 \mathcal{L}&=  \frac{1}{2} \left( \Phi' \dot{\Phi} - \bar{\Phi}' \dot{\bar{\Phi}} \right) - \frac{1}{2} \left( \Phi^{\prime 2} + \bar{\Phi}^{\prime 2} \right) \,,\\
     \mathcal{H} &= \frac{1}{2} \left( \Phi^{\prime 2} + \bar{\Phi}^{\prime 2} \right)\,,
\end{aligned}
 \end{equation}
where we have made a field redefinition
\begin{equation}
\begin{aligned}
\hspace{-15pt} \label{eq:rotationphis}
 \left( \begin{array}{c}
\Phi\\ \bar{\Phi}
 \end{array} \right) &= \left( \begin{array}{cc}
    \cosh \left( \frac{\gamma}{2} \right)  &  \quad \sinh \left( \frac{\gamma}{2} \right)  \\
   \sinh \frac{\gamma}{2} & \quad  \cosh \left(\frac{\gamma}{2} \right)
 \end{array} \right)  \left( \begin{array}{c}
\phi\\ \bar{\phi}
 \end{array} \right)\,, \\  \left( \begin{array}{c}
\phi\\ \bar{\phi}
 \end{array} \right) &= \left( \begin{array}{cc}
    \cosh \left( \frac{\gamma}{2} \right)  &  \quad - \sinh \left( \frac{\gamma}{2} \right)  \\
 -  \sinh \left( \frac{\gamma}{2} \right) & \quad  \cosh \left( \frac{\gamma}{2} \right)
 \end{array} \right)  \left( \begin{array}{c}
\Phi\\ \bar{\Phi}
 \end{array} \right) \, .
\end{aligned}
\end{equation}
The deformed equations of motion, written in terms of the new fields $\Phi$ and $\bar{\Phi}$, are
\begin{equation}
\label{eq:0i9uy232}
    \Phi'' = \dot{\Phi}', \quad \bar{\Phi}'' = - \dot{\bar{\Phi}}' \, ,
\end{equation}
which take the same form as those in the undeformed theory. Again, this is analogous to the field redefinition (\ref{CS_bogoliubov}) in the Chern-Simons setting, which undoes a similar quadratic mixing between the barred and unbarred fields induced by a $J \bar{J}$ deformation.

Next let us consider how this observation might extend to multiple bosons. We focus on the case of $N = \bar{N}$ for simplicity. The deformed Hamiltonian density for an equal number of left- and right-moving chiral bosons is
\begin{equation}\label{ham_density_field_redef}
\mathcal{H}^{(\mu)} = \frac{1}{2}\left( \phi^{\prime}_j \phi^{\prime}_j + \bar{\phi}^{\prime}_{\bar{j}} \bar{\phi}^{\prime }_{\bar{j}}  \right) \cosh ( \gamma ) + \sqrt{ \phi^{\prime}_j \phi^{\prime}_j  \bar{\phi}^{\prime}_{\bar{j}} \bar{\phi}^{\prime}_{\bar{j}}   } \sinh ( \gamma ) \, . 
\end{equation}
We now ask whether some more complicated field redefinition might return this Hamiltonian to a quadratic one, as in the case of (\ref{big_phi_L_H}). When $N=\bar{N}=2$, at least formally, one can attempt to perform a change of variables that resembles a transformation to polar coordinates in a $2d$ target space:
\begin{equation}
\begin{aligned}
\label{eq:fieldred2}
\left( \begin{array}{c}
\phi^{\prime}_1(\theta, t)\\\phi^{\prime}_2 (\theta, t)
\end{array} \right) & = \left( \begin{array}{cc}
   r'(\theta, t) \cos \left( \Theta'(\theta, t) \right)   \\   r'(\theta, t)  \sin \left( \Theta'(\theta, t) \right)
\end{array} \right)\,, \\  \left( \begin{array}{c}
\bar{\phi}^{\prime}_1(\theta, t)\\\bar{\phi}^{\prime}_2 (\theta, t)
\end{array} \right)  &= \left( \begin{array}{cc}
 \bar{r}'(\theta, t)\cos \left( \bar{\Theta}'(\theta, t) \right)  \\  \bar{r}'(\theta, t) \sin \left( \bar{\Theta}'(\theta, t) \right)
\end{array} \right)\,.
\end{aligned}
\end{equation}
Here we interpret $\Theta'(\theta, t)$ and $r'(\theta, t)$ as spatial derivatives of new fields which depend on the derivatives $\phi' ( \theta, t )$ in a nonlinear way. In terms of these quantities, the Hamiltonian density (\ref{ham_density_field_redef}) with $N = \bar{N} = 2$ takes the form
\begin{equation}
\begin{aligned}\label{eq:H1rdef}
    \mathcal{H}^{(\mu)}&= \frac{1}{2} \left( \phi^{\prime 2}_1 + \phi^{\prime 2}_2 + \bar{\phi}^{\prime 2}_1 + \bar{\phi}^{\prime 2}_2 \right) \cosh \left( \gamma \right) +  \sqrt{\left(\phi^{\prime 2}_1 + \phi^{\prime 2}_2 \right) \left( \bar{\phi}^{\prime 2}_1 + \bar{\phi}^{\prime 2}_2 \right)} \sinh \left( \gamma \right)\\ &= \frac{1}{2} \left( r^{\prime 2} + \bar{r}^{\prime 2} \right) \cosh \left( \gamma \right) + r^\prime \bar{r}^\prime \sinh \left( \gamma \right)\,,
\end{aligned}
\end{equation}
where we assumed $r'\bar{r}'>0$ in order to simplify the square root. Now we perform a second field redefinition, just as in (\ref{eq:rotationphis}), to a new field $\rho$: 
\begin{equation}\label{eq:rotationphisrho}
    \left( \begin{array}{c}
r(\theta, t)\\ \bar{r}(\theta, t)
    \end{array} \right) = \left(\begin{array}{cc}
     \cosh \left( \frac{\gamma}{2} \right)    & \quad - \sinh \left( \frac{\gamma}{2} \right)  \\
      - \sinh \left( \frac{\gamma}{2} \right)   & \quad  \cosh \left( \frac{\gamma}{2} \right)
    \end{array}\right)  \left( \begin{array}{c}
\rho(\theta, t)\\ \bar{\rho}(\theta, t)
    \end{array} \right)\,.
\end{equation}
Expressing the Hamiltonian density \eqref{eq:H1rdef} in terms of the $\rho$ variables rather than the $r$ variables, we conclude
\begin{equation}\label{eq:simpleH2-2}
    \mathcal{H}^{(\mu)} = \frac{1}{2} \left( \rho^{\prime 2} + \bar{\rho}^{\prime 2} \right)\,.
\end{equation}
Therefore, again at a formal classical level, it appears that this series of field redefinitions has returned the Hamiltonian density to that of the free theory. Furthermore, the latter change of variables (\ref{eq:rotationphisrho}) can be interpreted as rescaling the overall target space radius $r$, much as in the $(N, \bar{N}) = (1, 1)$ case. For a larger number of bosons $N = \bar{N} > 2$, one can perform a similar series of manipulations using higher-dimensional spherical coordinates.

Several technical issues preclude us from taking this series of field redefinitions seriously, at least without further investigation. First, the change of variables (\ref{eq:fieldred2}) was at the level of \emph{derivatives} of the fields, and it is not clear that this corresponds to a sensible change of variables for the fields themselves. Second, all of these manipulations have been purely classical, and it is not guaranteed that one could make sense of these field redefinitions within a path integral (which would produce Jacobian factors from each change of variables). And third, we have not been careful about the identifications that each field is subject to. For instance, if indeed the field $\Theta$ can be interpreted as a target-space angle in polar coordinates, then it should be subject to the identification $\Theta \sim \Theta + 2 \pi$.

Nonetheless, it would be very interesting if an argument of this form could be used to endow the root-$T\overline{T}$ deformation of $N$ chiral and anti-chiral bosons with a geometrical target-space interpretation.

\section{Perturbative Quantization Using Background Field Method} \label{sec:cian}

In the preceding sections, we have considered interacting theories with arbitrary numbers $N ,\bar{N}$ of chiral and anti-chiral bosons, respectively, and sacrificed manifest Lorentz invariance in order to use a first-order formulation which is convenient for canonical quantization. In the special case $N = \bar{N}$, however, we also have the option of assembling the field content of our theory into $N$ \emph{non-chiral} bosons by summing the left-movers and right-movers:
\begin{align}
    \varphi^i = \frac{1}{\sqrt{2}} \left( \phi^i + \bar{\phi}^{i} \right) \, .
\end{align}
Here we now use the same index $i = 1 , \ldots, N$ for both the chiral and anti-chiral fields, rather than distinct indices $i$ and $\overline{i}$. As this change of variables is merely a field redefinition, stress tensor deformations of such a theory of $N$ bosons must be equivalent, regardless of whether the theory is presented in terms of left-movers and right-movers $\phi^i$, $\bar{\phi}^i$, or in terms of their non-chiral counterparts $\varphi^i$. Indeed, for the case of the $T\overline{T}$ deformation of a free seed theory, this equivalence was checked explicitly in \cite{Ouyang:2020rpq}.

In this section, we will provide a complementary analysis of the perturbative quantization of the Modified Scalar theory using this presentation in terms of non-chiral fields $\varphi^i$. For concreteness, we will focus on the case where both the fields $\varphi^i$ and the Lorentzian spacetime coordinates $(t, x)$ are non-compact, and we will use middle Greek letters like $\mu$, $\nu$ (rather than early Greek letters like $\alpha$, $\beta$, which were used in sections \ref{sec:classical} and \ref{sec:cs}) for spacetime indices in this section. We will write $g_{\mu \nu}$ for the (Minkowski) spacetime metric.

In terms of the non-chiral fields $\varphi^i$, the Lagrangian for the Modified Scalar theory can be written in the manifestly Lorentz-invariant form
\begin{align}\label{modscalar_lorentz_invariant_lagrangian}
    \mathcal{L} = \frac{1}{2} \left( \cosh \left( \gamma \right) \partial_{\mu} \varphi^{i}\partial^{\mu} \varphi^{i} + \sinh \left( \gamma \right) \sqrt{2 \left( \partial_\mu \varphi^{i} \partial^{\nu} \varphi^{i} \right) \left( \partial_\nu \varphi^{j} \partial^{\mu} \varphi^{j} \right) - \left( \partial_\mu \varphi^{i} \partial^{\mu} \varphi^{i} \right)^2 } \right) \, .
\end{align}
The advantage of this representation is that one can more easily apply standard diagrammatic techniques to compute loop corrections in the quantum theory. Of course, the second term in the Lagrangian (\ref{modscalar_lorentz_invariant_lagrangian}) is still non-analytic around the vacuum of the theory, or around any field configuration for which
\begin{align}
    \partial_\mu \varphi^i = 0 \, .
\end{align}
We will circumvent this issue by working in a background field expansion around a field configuration $\varphi^i$ for the scalars which we assume has non-zero gradients and which satisfies the classical equations of motion for the theory, but which is otherwise arbitrary. 

\subsection{Background field expansion and Feynman rules}

Throughout this section, we will use the notation
\begin{align}
    \varphi^i = C^i + Q^i \, , 
\end{align}
where $C^i$ is a classical (background) field configuration around which we perform our expansion, and $Q^i$ is a quantum field which is allowed to fluctuate within the path integral. This classical background $C^i$ is the analog of the large-momentum configuration around which we performed our expansion in section \ref{sec:quantumspectrum}. Our goal will be to investigate the terms which contribute to the quantum effective action, as a function of the background $C^i$.

To avoid cluttering the formulas, it will also be convenient to adopt the following shorthand for spacetime derivatives of the various fields:
\begin{align}
  \varphi_\mu{}^i = \partial_\mu \varphi^i \, , \quad C_\mu{}^i = \partial_\mu C^i \, , \quad Q_\mu{}^i = \partial_\mu Q^i \, .
\end{align}
In our analysis of chiral boson theories, we introduced two useful quantities $S$ and $P$ in equation (\ref{S_and_P_invariants}) which were independent combinations of derivatives of the scalar fields. In the present non-chiral analysis, let us similarly introduce the quantities
\begin{align}\label{non_chiral_S_P_defn}
    S = \varphi_{\mu}{}^{i} \varphi^{\mu i} \, , \qquad P^2 = \varphi_{\mu}{}^{i} \varphi^{\nu i} \varphi_{\nu}{}^{j} \varphi^{\mu j} \, .
\end{align}
We note that these are not the precise analogs of $S$ and $P$ in the chiral setting; for instance, the role of the combination $S^2 - P^2$ in section \ref{sec:classical} is now played by $2 P^2 - S^2$. Therefore, in terms of these quantities (\ref{non_chiral_S_P_defn}), the Modified Scalar Lagrangian (\ref{modscalar_lorentz_invariant_lagrangian}) can be written as
\begin{align}\label{modscal_S_P_form}
    \mathcal{L} = \frac{1}{2} \left( \cosh ( \gamma ) S + \sinh ( \gamma ) \sqrt{ 2 P^2 - S^2 } \right) \, .
\end{align}
We decompose $S$ into a classical piece $S_C$ and a quantum piece $S_Q$, along with a cross term:
\begin{align}
    S &= \left( C_\mu{}^i + Q_\mu{}^i \right) \left( C{\mu i} + Q^{\mu i} \right)  \nonumber \\
    &= \underbrace{C_{\mu i} C^{\mu i}}_{S_C} + 2 C_\mu{}^i Q^{\mu i} + \underbrace{Q_\mu{}^iQ_i{}^\mu }_{S_Q} \, .
\end{align}

Next we will consider the splitting of $S^2$ and $P^2$ into classical and quantum pieces. Because we assume that the field configuration $C^i$ is a solution to the classical equations of motion, by definition the action is stationary to linear order when expanding around such a solution. This means that the effective action cannot contain any terms which are linear in the fluctuation field $Q^i$, because the sum of all such contributions must conspire to form an on-shell total derivative. We will therefore label all terms linear in $Q^{\mu i}$ as ``on-shell deriv.'' and ignore them in what follows, although with the caveat that \emph{individual} terms of this form need not separately drop out; we are only guaranteed that the \emph{combined} effect of all such terms is to form an on-shell total derivative.

With this in mind, the quantity $S^2$ can be expanded as
\begin{align}
    S^2 &= S_C^2 + \underbrace{4S_C C_{\mu}^{i} Q^{\mu i}}_{\text{on-shell deriv.}} + 2 S_C S_Q  + 4C_{\mu}{}^{i}Q^{\mu i} C_\nu{}^{j} Q^{\nu j}  + \underbrace{4S_Q C_{\mu}{}^{i} Q^{\mu i}}_{\mathcal{O}\left( Q^3 \right) }+ \underbrace{S_Q^2}_{\mathcal{O}\left( Q^{4} \right) } \nonumber \\
    &\simeq S_C^2 + 2 S_C S_Q  + 4C_{\mu}{}^{i} Q^{\mu i} C_\nu^{j} Q^{\nu j} \, ,
\end{align}
where the symbol $\simeq$ means equal modulo all terms that are either linear in $Q^i$ (which will form on-shell total derivatives) or that are of cubic order or higher in $Q^i$ (which do not contribute to the one loop effective action). A similar computation for $P^2$ gives
\begin{align}
     P^2 &= C_{\mu}{}^{i} C^{\mu j} C_{\nu}{}^{i} C^{\nu j} + \underbrace{4C_{\mu}{}^{i} C^{\mu j} C^{\nu i} Q_{\nu}{}^{j}}_{\text{on-shell deriv.}} + 2C_{\mu}{}^{i} C^{\nu i} Q_{\nu j} Q^{\mu j} + 2Q_{\mu i} C^{\nu i} Q_{\nu j} C^{\mu j} \nonumber \\
    &\qquad + 2Q_{\mu}{}^{i} C^{\nu i} C_{\nu}{}^{j} Q^{\mu j} + \mathcal{O}\left( Q^3 \right) \, \nonumber \\
    &\simeq  C_{\mu}{}^{i} C^{\mu j} C_{\nu}{}^{i} C^{\nu j}  + 2C_{\mu}{}^{i} C^{\nu i} Q_{\nu}{}^{j} Q^{\mu j} + 2Q_{\mu}{}^{i} C^{\nu i} Q_{\nu}{}^{j} C^{\mu j}  + 2Q_{\mu}{}^{i} C^{\nu i} C_{\nu}{}^{j} Q^{\mu j} \, .
\end{align}
Therefore, the combination $2 P^2 - S^2$ under the square root in (\ref{modscal_S_P_form}) has an expansion
\begin{align}\label{2psq_sq_quadratic}
\begin{aligned}
2 P^2-S^2 & \simeq 2 P_C^2-S_C^2-2 S_C S_Q-4 C_\mu{ }^i Q^{\mu i} C_\nu{ }^j Q^{\nu j}+4 C_\mu{ }^i C^{\nu i} Q_\nu{ }^j Q^{\mu j} \\
& +4 Q_\mu{ }^i C^{\nu i} Q_\nu{ }^j C^{\mu j}+4 Q_\mu{ }^i C^{\nu i} C_\nu{ }^j Q^{\mu j} \\
& \equiv 2 P_C^2-S_C^2+2 \mathcal{Q}_1 \,.
\end{aligned}
\end{align}
Here we introduce the shorthand $\mathcal{Q}_1$ which is proportional to the correction to the classical part of (\ref{2psq_sq_quadratic}) up to quadratic order in fluctuations,
\begin{equation}
\begin{aligned}
\mathcal{Q}_1 & =-S_C S_Q-2 C_\mu{ }^i Q^{\mu i} C_\nu{ }^j Q^{\nu j}+2 C_\mu{ }^i C^{\nu i} Q_\nu{ }^j Q^{\mu j} \\
& +2 Q_\mu{ }^i C^{\nu i} Q_\nu{ }^j C^{\mu j}+2 Q_\mu{ }^i C^{\nu i} C_\nu{ }^j Q^{\mu j}\,,
\end{aligned}
\end{equation}
which is not to be confused with $Q^i$ or $Q_\mu{}^i = \partial_\mu Q^i$. Let us also define $\mathcal{Q}_2 \simeq \mathcal{Q}_1^2$ to be the square of this quantity, retaining terms only up to second order in $Q^i$, so that
\begin{align}
    \mathcal{Q}_2 = S_C^2 C_{\mu}{}^{i} Q^{\mu i} C_{\nu}{}^{j} Q^{\nu j} - 8 S_C C_{\mu}{}^{i} Q^{\mu i} C_{\nu}{}^{j} C^{\nu k} C^{\rho j} Q_{\rho}{}^{k} + 16 \left( C_{\nu}{}^{j} C^{\nu k} C^{\rho j} Q_{\rho}{}^{k} \right)^2 \, .
\end{align}
In terms of these combinations, we can expand the square root appearing in (\ref{modscal_S_P_form}) as
\begin{align}\label{sqrt_expansion_Q}
    \sqrt{ 2 P^2 - S^2 } \simeq \sqrt{ 2 P_C^2 - S_C^2 } + \frac{\mathcal{Q}_1}{ \sqrt{ 2 P_C^2 - S_C^2  } } - \frac{\mathcal{Q}_2}{2 \left( 2 P_C^2 - S_C^2  \right)^{3/2}} \, .
\end{align}
Finally, we can express the Modified Scalar Lagrangian expanded to quadratic order in fluctuations around a given classical solution as
\begin{align}
    \mathcal{L} \simeq \mathcal{L}_C + \frac{1}{2} \left( \cosh ( \gamma ) S_Q + \sinh ( \gamma ) \left( \frac{\mathcal{Q}_1}{ \sqrt{ 2 P_C^2 - S_C^2  } } - \frac{\mathcal{Q}_2}{2 \left( 2 P_C^2 - S_C^2  \right)^{3/2}} \right) \right) \, ,
\end{align}
where $\mathcal{L}_C$ represents the Lagrangian evaluated on the background solution $C^i$, i.e.
\begin{align}
    \mathcal{L}_C &\equiv \frac{\cosh ( \gamma ) }{2}S_C  + \frac{\sinh ( \gamma ) }{2}\sqrt{2P_C^2 - S_C^2} \, .
\end{align}
It is also convenient to write the Lagrangian for the quantum field $Q^i$ in terms of a bilinear form. Defining the tensor
\begin{equation}
 \begin{aligned}
       \label{P_tensor}
& P_{\mu \nu}{ }^{i j}=-\left(\frac{-S_C g_{\mu \nu} \delta^{i j}-2 C_\mu{ }^i C_\nu{ }^j+2 C_\mu{ }^k C_\nu{ }^k \delta^{i j}+2 C_\mu{ }^j C_\nu{ }^i+2 C_\rho{ }^i C^{\rho j} g_{\mu \nu}}{2 \sqrt{2 P_C^2-S_C^2}}\right. \\
& \left.-\frac{S_C^2 C_\mu{ }^i C_\nu{ }^j-8 S_C C_\mu{ }^i C_\rho{ }^k C^{\rho j} C_\nu{ }^k+16 C_\rho{ }^k C^{\rho i} C_\mu{ }^k C_\tau{ }^m C^{\tau j} C_\nu{ }^m}{4\left(2 P_C^2-S_C^2\right)^{\frac{3}{2}}}\right), \\
&
\end{aligned}
\end{equation}
we can write the Lagrangian $\mathcal{L}_Q$ for the fluctuating field as
\begin{align}
    \mathcal{L}_Q &= Q^{\mu i}\left( \frac{\cosh ( \gamma ) }{2}g_{\mu \nu} \delta^{ij} + \sinh ( \gamma ) P_{\mu \nu}{}^{ij}  \right) Q^{\nu j} \, , 
\end{align}
or after integrating by parts to move the derivative acting on $Q^{\mu i} = \partial^\mu Q^i$, as
\begin{align}
    \mathcal{L}_Q &=  - Q^{i} \left( \frac{\cosh ( \gamma ) }{2} \delta^{ij} \partial^2 +  \sinh ( \gamma ) \left( \partial^{\mu} P_{\mu \nu}{}^{ij} \right) \partial^\nu + \sinh ( \gamma )  P_{\mu \nu}{}^{ij} \partial^\mu \partial^\nu\right)  Q^{j} \label{eq:SFQuantum_Lagrangian} \, .
\end{align}
The first term in (\ref{eq:SFQuantum_Lagrangian}) is proportional to a conventional free kinetic term for the fields $Q^i$. The second and third terms, involving $P_{\mu \nu}{}^{ij}$ and its derivative, encode the interactions which are induced by expanding around the classical field configuration $C^i$.

\emph{Feynman rules}

Now that we have obtained the Lagrangian (\ref{eq:SFQuantum_Lagrangian}), we may read off the Feynman rules which we will need for computing diagrams. The propagator for the quantum field is
\begin{align}
       D^{ij}  &= -\frac{i}{\cosh ( \gamma ) } \frac{\delta^{ij}}{k^2 } \, . \label{eq:SFpropagator}
\end{align}
Next we must work out the vertex associated to the interaction between $Q^i$ and the classical field via the combination $P_{\mu \nu}{}^{m n}$. We will draw quantum fields as solid lines and the cumulative effect of the background fields as a single coiled line. Consider the trivalent interaction between a field $Q^i$ with momentum $p$, a field $Q^j$ with momentum $q$, and an insertion of the background $P_{\mu \nu}{}^{m n}$ with momentum $r$. This vertex can be visualized as
\begin{align}\label{interaction_vertex}
    \raisebox{-0.5\height}
    {\includegraphics[width=0.38\linewidth]{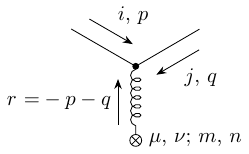}} \; .
\end{align}
Let the vertex factor for this interaction be $g_{ij}$.\footnote{The vertex factor $g_{ij}$ should not be confused with the target-space metric $G_{ij} ( \phi )$ for the bosons which appears in equation (\ref{general_class_vielbeins}). We also note that the value $r$ of the classical field momentum must be integrated over in this trivalent interaction, but we do not include this integral in the expression (\ref{vertex_value}) for $g_{ij}$.} There are four ways that we can get a contribution to this factor from the Lagrangian (\ref{eq:SFQuantum_Lagrangian}). First, there is a piece arising from the term $\sinh ( \gamma ) \left( \partial^{\mu} P_{\mu \nu}{}^{m n} \right) \partial^\nu$ when $m = j$ and $n = i$, which gives a term proportional to $r^\mu q^\nu$ because of the first derivative $\partial^\mu$ acting on $P_{\mu \nu}{}^{m n}$ and the second derivative $\partial^\nu$ acting on $Q^j$. There is another term of the same form when $m = i$ and $n = j$. Then there are two more contributions from the term $\sinh ( \gamma )  P_{\mu \nu}{}^{m n} \partial^\mu \partial^\nu$, when either $m = j$ and $n = i$, or when $m = i$ and $n = j$, which both come with a factor of $q^\mu q^\nu$ from the two derivatives acting on $Q^j$. Altogether, the value of this vertex is
\begin{align}\label{vertex_value}
    g_{ij} &= i \sinh ( \gamma ) \bigg( \delta^{m j} \delta^{ni} P_{\mu \nu}^{mn} r^\mu q^\nu + \delta^{m i} \delta^{nj} P_{\mu \nu}^{mn} r^\mu q^\nu \nonumber \\&+ \delta^{m j} \delta^{ni} P_{\mu \nu}^{mn} q^\mu q^\nu + \delta^{m i} \delta^{nj} P_{\mu \nu}^{mn} q^\mu q^\nu \bigg) \nonumber \\
    &= - i \sinh ( \gamma ) \left( \delta^{m j} \delta^{ni} P_{\mu \nu}^{mn} p^\mu q^\nu + \delta^{m i} \delta^{nj} P_{\mu \nu}^{mn} p^\mu q^\nu \right) \, \nonumber \\
    &= - i \sinh ( \gamma ) \left( P^{ij}_{\mu \nu} + P^{ji}_{\mu \nu} \right)  p^\mu q^\nu \, ,
\end{align}
where in the second step we have used $r^\mu = - q^\mu - p^\mu$ to cancel terms. This gives the desired value of the trivalent vertex $g_{ij}$ between $Q^i$, $Q^j$, and the classical background. However, in the calculations that follow, it will be convenient to factor out the dependence on $P_{\mu \nu}^{mn}$ and use an ``uncontracted'' vertex factor $\widetilde{g}$ defined by
\begin{align}\label{uncontracted_vertex_factor}
    g_{ij} &= P_{\mu \nu}^{mn} \left( \widetilde{g}^{mn}_{ij} \right)^{\mu \nu} \, , \nonumber \\
    \left( \tilde{g}^{mn}_{ij} \right)^{\mu \nu} &= - i \sinh ( \gamma ) \left( \delta^{m j} \delta^{ni} p^\mu q^\nu + \delta^{m i} \delta^{nj} p^\mu q^\nu \right) \, .
\end{align}
Let us emphasize that $\left( \tilde{g}^{mn}_{ij} \right)^{\mu \nu}$ is \emph{not} the full value of the interaction vertex, but rather a useful intermediate quantity which has removed all factors of $P_{\mu \nu}^{mn}$. After computing Feynman diagrams using ``uncontracted'' vertices $\tilde{g}$, we must contract the final result with one factor of $P_{\mu \nu}^{mn}$ for each vertex in order to recover the true value of the diagram.

\subsection{Quantum effective action}

We are now ready to compute the leading quantum corrections to the Modified Scalar Lagrangian. Most of our discussion will focus on the one-loop effective action, defined by the first term beyond the classical contribution in the expansion
\begin{align}\label{effective_action}
    \Gamma [ C^i ] = S [ C^i ] + \frac{i}{2} \operatorname{Tr} \left[ \log \left( \frac{\delta^2 S}{\delta \varphi^i \, \delta \varphi^j } \right) \Bigg\vert_{\varphi^k = C^k} \right] + \cdots \, . 
\end{align}
Although we are primarily interested in the one-loop contribution to $\Gamma$, we will also present some partial results concerning corrections at higher loop order.

There are several techniques for computing the one-loop effective action $\Gamma$. One way is to use heat kernel methods; we will not pursue this strategy here, but we refer the reader to the thesis \cite{pinelliThesis} for a discussion of this approach in the related context of the $4d$ ModMax theory. Rather, we will compute contributions to the effective action perturbatively, using the Feynman rules derived in the preceding subsection. This amount to a diagrammatic evaluation of the one-loop determinant of the operator $\frac{\delta^2 S}{\delta \varphi^i \, \delta \varphi^j }$, which is the operator appearing in $\mathcal{L}_Q$ that we have computed in equation (\ref{eq:SFQuantum_Lagrangian}).

In particular, our goal is to evaluate divergent Feynman diagrams in the Modified Scalar theory using dimensional regularization, as a function of the background configuration $C^i$. Each such divergent contribution necessitates the addition of an appropriate counterterm to cancel the divergence. The collection of all such counterterms which must be added to the classical Lagrangian therefore reproduces the additional terms that appear in the quantum effective action, giving a characterization of the corrections in the expansion (\ref{effective_action}).

\emph{Constant background, one-loop diagrams}

Let us begin by considering the simpler case in which the background field configuration $C^i$ is linear in the spacetime coordinates, which means that the classical field has constant gradients. That is, we assume that $C_\mu{}^i = \partial_\mu C^i$ is constant for such backgrounds, so that $\partial_\mu C_{\nu}{}^i = 0$ for all $\mu, \nu, i$. In this case, no momentum can flow through the classical fields in the interaction vertex (\ref{interaction_vertex}), which implies that $r = 0$ and thus $p = - q$.

To obtain the one-loop effective action $\Gamma$, we must evaluate all Feynman diagrams built from the quantum field propagator and interaction vertex (\ref{interaction_vertex}) which contain at most one loop. This corresponds to an infinite series of diagrams given by
\begin{align}\label{infinite_series_constant_bg}
    \Gamma = \raisebox{-0.5\height}{\includegraphics[width=0.2\linewidth]{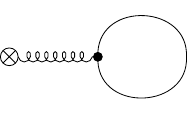}} \; + \; \raisebox{-0.5\height}{\includegraphics[width=0.25\linewidth]{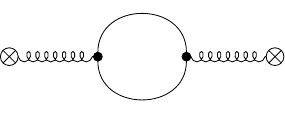}} \; + \; \raisebox{-0.5\height}{\includegraphics[width=0.2\linewidth]{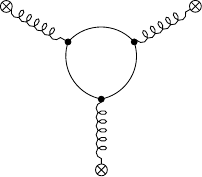}} \; + \; \cdots \, .
\end{align}
Let $\mathcal{D}_n$ represent the value of the diagram in the series (\ref{infinite_series_constant_bg}) which has $n$ insertions of the classical background. The first diagram in this infinite series is
\begin{align}
    \mathcal{D}_1 = \raisebox{-0.5\height}
 {\includegraphics[width=0.25\linewidth]{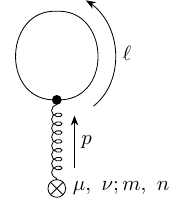}} \, .
\end{align}
Following the comments around equation (\ref{uncontracted_vertex_factor}), we will evaluate this diagram -- and the others in this section -- by using the Feynman rule associated with the uncontracted vertex factor $\tilde{g}$, and then contracting with $P_{\mu \nu}^{m n}$. Doing this and simplifying the resulting sum of Kronecker delta functions using symmetry, one finds
\begin{align}
    \mathcal{D}_1 &= P_{\mu \nu}^{m n} \sinh ( \gamma ) \delta^{im} \delta^{jn} \int \frac{d^d\ell}{\left( 2\pi \right)^{d}} \left( 2i \ell^{\mu} \ell^{\nu} \right) D^{ij} \, .
\end{align}
A term in the integrand which is proportional to $\ell^\mu \ell^\nu$ will produce a result which scales like $\ell^2$ and which is a symmetric tensor in $\mu$ and $\nu$. The only constant symmetric $2$-tensor in the problem is the spacetime metric $g^{\mu \nu}$, so the integral of such a term must be proportional to $\ell^2 g^{\mu \nu}$. By taking the trace, one can fix the dimensionless constant to be $\frac{1}{d}$. Thus, within the integral, we can make the replacement
\begin{align}\label{sym_replacements_1}
    \ell^{\mu} \ell^{\nu} &\to \frac{1}{d} \ell^2 g^{\mu \nu} \, .
\end{align}
Using this replacement and the propagator (\ref{eq:SFpropagator}), we find
\begin{align}
    \mathcal{D}_1 &= - \frac{2\tanh ( \gamma )}{d}  P_{\mu \nu}^{m n} \delta^{m n} \int \frac{d^d \ell}{\left( 2\pi \right)^{d}} g^{\mu \nu} \, .
\end{align}
The integrand is now independent of $\ell$. Although this integral diverges as $\Lambda^{d}$ with a na\"ive cutoff at momentum $\Lambda$, within dimensional regularization it is exactly zero \cite{Anselmi:2019pdm}. 

This result relies only on the momentum dependence of the integral. However, note that the insertions of additional vertices appearing in the higher one-loop diagrams $\mathcal{D}_n$ will not change the momentum dependence of the integral. In general, we will have $n$ propagators $D_{ij}$ of the form (\ref{eq:SFpropagator}), each of which is proportional to $\frac{1}{\ell^2}$, and $n$ copies of the vertex factor (\ref{uncontracted_vertex_factor}). Because the vertex factor contains products of momenta like $\ell^\mu \ell^\nu$, the integrand of $\mathcal{D}_n$ will involve a product of $2n$ momenta. We can replace such factors using a generalization of the argument which led to the replacement rule (\ref{sym_replacements_1}). That is, any integral involving a totally symmetric product of $2n$ momenta must yield a result which is proportional to $\ell^{2n}$ multiplied by a totally symmetrized combination of $n$ metric tensors, since the metric is the only symmetric tensor in the problem. This leads to the replacement
\begin{align}
    \prod_{i=1}^{n} \ell^{\mu_{2i- 1 }} \ell^{\mu_{2i}} \to \frac{\ell^{2n} \left( d -2 \right)!!}{\left( d-2 + 2n \right)!!} g^{(\mu_{1}\mu_2} \cdots g^{\mu_{2n - 1} \mu_{2n})} \, ,\label{eq:symmetrization}
\end{align}
where we have used the double factorial $n!! = n \cdot (n-2) \ldots 4 \cdot 2$.
We thus find an overall factor of $\ell^{2n}$ from the vertex factors, in addition to a compensating factor of $\frac{1}{\ell^{2n}}$ from the $n$ copies of the propagator, each of which scales like $\frac{1}{\ell^2}$. Note that all of these momenta are equal due to momentum conservation around the loop, as we assumed that no momentum can be carried by the classical fields, so the powers of loop momentum precisely cancel. Therefore, every diagram $\mathcal{D}_n$ involves an integrand which is independent of momentum, and thus vanishes in dimensional regularization just as $\mathcal{D}_1$ does.

We conclude that the perturbative one-loop effective action $\Gamma [ C^i ]$, with constant background field strength $C_\mu{}^i$, vanishes in dimensional regularization. This implies that under these assumptions, there are no 1-loop corrections to the classical theory.

\emph{Constant background, multi-loop diagrams}

Proceeding to higher loops, more vertices in the perturbative expansion become accessible, beginning at two loops with a vertex cubic in the quantum field. The first of such diagrams that is not a tadpole, shown in equation (\ref{not_tadpole}), emerges at order $\mathcal{O}(\gamma^2)$, and one can show that it nontrivially vanishes within dimensional regularization.
\begin{align}\label{not_tadpole}
    \raisebox{-0.5\height}{\includegraphics[width=0.35\linewidth]{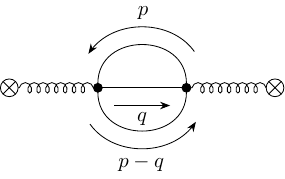}} \, .
\end{align}
The introduction of multiple loop momenta prevents the simple argument in the one-loop case from generalizing immediately. However, since with constant backgrounds there cannot be any external momenta and there is no characteristic scale present in these integrals, it will always be possible to iteratively symmetrize using (\ref{eq:symmetrization}) and integrate over each loop momentum, leaving a symmetrizable integral that will vanish in dimensional regularization. Therefore we expect that the argument presented above generalizes to all loops, implying that the full effective action $\Gamma [ C^i ]$ admits no corrections for constant background field strengths $C_\mu{}^i$.

\emph{Background-varying, one-loop diagrams}

We now study the more general case in which we do not assume that $\partial_\mu C_{\nu}^{i}= 0$, instead allowing the background field to vary. Besides requiring that the field configuration $C^i$ is a solution to the classical equations of motion, we make no further assumptions.

For this general background analysis, let us use the same notation $\mathcal{D}_n$ for the diagrams appearing in the infinite sum (\ref{infinite_series_constant_bg}). The first diagram in this series, $\mathcal{D}_1$, is unchanged from the constant background case, and thus it identically vanishes in dimensional regularization. 

The first nontrivial diagram is
\begin{align}\label{D2_diagram}
    \mathcal{D}_2 = \raisebox{-0.5\height}{\includegraphics[width=0.35\linewidth]{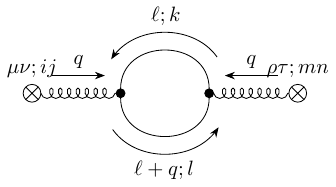}} \, .
\end{align}
As usual, it will be convenient to strip off factors of $P_{\mu \nu}{}^{ij}$ when computing the value of this diagram. This corresponds to evaluating the diagram using the ``uncontracted'' vertex $\tilde{g}$ of (\ref{uncontracted_vertex_factor}) and contracting the result with factors of $P_{\mu \nu}{}^{ij}$. To this end, let us write the value of the diagram as
\begin{align}\label{D2_stripped}
   & \mathcal{D}_2= \frac{\tanh^2 ( \gamma )}{2} \int \frac{d^dq}{\left( 2\pi \right)^{d}} P_{\mu \nu}{}^{ij} \left( -q \right) \nonumber \\& \Bigg(  \int \frac{d^d\ell}{\left( 2\pi \right)^{d}} \frac{\delta^{ik} \delta^{jl} \ell^{\mu} \left( \ell + q \right)^{\nu} + \delta^{il} \delta^{jk} \left( \ell + q \right)^{\mu} \ell^{\nu} }{\ell^2 \left( \ell + q \right)^2} \nonumber \\
    &\hspace{90pt} \cdot \left( \delta^{km} \delta^{ln} \ell^{\rho} \left( \ell + q \right)^{\tau} + \delta^{kn} \delta^{lm} \left( \ell + q \right)^{\rho} \ell^{\tau} \right) \Bigg) P_{\rho \tau}{}^{m n}\left( q \right)  \, .
\end{align}
Using the symmetry property $P_{\mu \nu}^{ij} = P_{\nu \mu}^{ji}$, this can also be expressed as
\begin{align}\label{I2_to_D2}
    \mathcal{D}_2 &= 2 \tanh^2 ( \gamma ) \int \frac{d^dq}{\left( 2\pi \right)^{d}} P_{\mu \nu}^{ij} \left( -q \right) {\mathcal{I}}^{(\mu \nu) (\rho \tau)}_2 P_{\rho \tau}^{i j}\left( q \right) \, ,
\end{align}
where we have defined the simpler integral
\begin{align}\label{simpler_integral_body}
    {\mathcal{I}}^{\mu \nu \rho \tau}_2 = \int \frac{d^d \ell}{\left( 2\pi \right)^{d}} \frac{( \ell + q )^{\nu}\ell^{\mu} ( \ell + q )^{\tau} \ell^{\rho} }{\ell^2 \left( \ell + q \right)^2} \, ,
\end{align}
and where our conventions for symmetrization are $T^{( \mu \nu ) } = \frac{1}{2} \left( T^{\mu \nu} + T^{\nu \mu} \right)$.

To study the divergence structure of the diagram $\mathcal{D}_2$, it suffices to evaluate the quantity ${\mathcal{I}}^{\mu \nu \rho \tau}_2$ in dimensional regularization, which is performed in appendix \ref{sec:scalar_field_1_loop_2_vertex_calc}. The resulting divergent contribution is
\begin{align}\hspace{-10pt}\label{divergent_final_body}
    {\mathcal{I}}^{\mu \nu \rho \tau}_2 &= \left( \frac{1}{\epsilon} \right) \frac{-i}{24\left( 4\pi \right)} \bigg[ q^{2}  \left( g^{\mu \nu} g^{\rho \tau}  + g^{\mu \rho} g^{\nu \tau} + g^{\mu \tau} g^{\nu \rho} \right) \nonumber \\
    &\quad+2 \left( q^{\nu} q^{\mu} g^{\tau \rho} + g^{\mu \tau} q^{\nu} q^{\rho}+ g^{\nu \mu} q^{\tau} q^{\rho} + g^{\nu \rho} q^{\mu} q^{\tau} \right)  +4\left( g^{\mu \rho} q^{\nu} q^{\tau} + g^{\nu \tau} q^{\mu} q^{\rho} \right) \bigg] \, .
\end{align}
In order to cancel this $\frac{1}{\epsilon}$ divergence, one would introduce a counter-term which involves two factors of $P_{\mu \nu}{}^{ij}$ in the Lagrangian. Therefore, in the background-varying case, there is a non-trivial contribution to the quantum effective action at one loop. Because the higher diagrams $\mathcal{D}_n$ will involve higher powers of $\gamma$, the result (\ref{divergent_final_body}) represents the complete one-loop effective action at $\mathcal{O} ( \gamma^2 )$.

With the two-vertex diagram evaluated, to complete the computation of the one-loop effective action, we seek to evaluate all remaining diagrams containing one loop. Fortunately, there is only one diagram $\mathcal{D}_n$ for each number of vertices $n$. The details of the evaluation of this diagram are presented in appendix \ref{app:one_loop_n_vertex}. Here we merely summarize the results. The value of $\mathcal{I}_n$ can be written as 
\begin{align}
    \left( {\mathcal{I}}_{n} \right)^{\mu_1 \ldots \mu_{2n}} &= \left( n - 1 \right)!\int_0^{1} \left( \prod_{i=0}^{n-1} d^dx_{i} \right) \delta \left( \sum_{i=0}^{n-1} x_{i} - 1  \right)  \left( \mathbf{C}_{2n}^{\mu_1 \ldots \mu_{2n}} + \mathbf{D}_{2n}^{\mu_1 \ldots \mu_{2n}} \right) \, ,
\end{align}
where we have defined

\begin{align}
    \mathbf{C}_{2n}^{\mu_1 \ldots \mu_{2n}} &= \frac{i\left( d - 2 \right)!!}{\left( d - 2 + 2n \right)!!} g^{\mu_1 \cdots \mu_{2n}}  \frac{\Gamma \left( n + \frac{d}{2} \right)  }{\left( 4\pi \right)^{\frac{d}{2}} \Gamma \left( n \right) \Gamma \left( \frac{d}{2} \right)} \Gamma \left( -\frac{d}{2} \right)\Delta^{d} \, \, \nonumber \\
    \mathbf{D}_{2n}^{\mu_1 \ldots \mu_{2n}} &= \frac{i\left( d - 2 \right)!!}{\left( d - 4 + 2n \right)!!}\sum_{a=1}^{2n} \sum_{b>a}^{2n} g^{\{ \mu \neq \mu_a, \mu_b \}}   f^{\mu_a} \left( x,q,a \right)  f^{\mu_{b}} \left( x,q,b \right) \nonumber \\
    &\qquad \cdot \frac{\Gamma \left( n-1+\frac{d}{2} \right) }{\left( 4\pi \right)^{\frac{d}{2}} \Gamma \left( n \right) \Gamma \left( \frac{d}{2} \right)  } \Gamma \left( 1 - \frac{d}{2} \right) \Delta^{d - 2} \, .
\end{align}
The notation $g^{\mu_1 \cdots \mu_{2n}}$ refers to a symmetrized product of metric tensor factors, which is defined in equation (\ref{shorthand_metric}). Similarly, $g^{\{ \mu \neq \mu_a, \mu_b \}}$ is shorthand for such a symmetrized product of metrics which which omits the two indices $\mu_a$ and $\mu_b$, which is explained in more detail around equation (\ref{vertex_intermediate}). Finally, the function $f^\mu ( x, q, a )$ is defined in equation (\ref{cursed_defn}).

In dimensional regularization, with $d = 2 ( 1 + \epsilon )$ and as $\epsilon \to 0$, the overall momentum dependence and divergence structure of these terms is
\begin{align}\label{one_loop_n_vertex_final}
    \mathbf{C}_{2n}^{\mu_1 \ldots \mu_{2n}} &\sim \frac{1}{\epsilon} q^2 g^{\mu_1 \cdots \mu_{2n}}  \, , \nonumber \\
    \mathbf{D}_{2n}^{\mu_1 \ldots \mu_{2n}} &\sim \frac{1}{\epsilon} \sum_{a=1}^{2n} \sum_{b>a}^{2n} q^{\mu_a} q^{\mu_b} g^{\{ \mu \neq \mu_a, \mu_b \}}  \, ,
\end{align}
which is of the same qualitative form as the one-loop, two-vertex contribution (\ref{divergent_final_body}).

Therefore, the full one-loop effective action for the Modified Scalar theory is obtained by introducing counterterms that cancel the divergent contributions which we have described in equations (\ref{divergent_final_body}) and (\ref{one_loop_n_vertex_final}). Because, after Fourier transforms, only two derivatives arise acting on the external background vertices, the counterterms are invariant under classical conformal transformations.

\emph{Background varying, two vertex, $m$-loop diagrams}

One could imagine computing the quantum effective action (\ref{effective_action}) using a double expansion in both the number $n$ of vertices and the number $m$ of loops. The preceding subsections have discussed the contributions at one loop but for any number of vertices. We have also argued that higher loop corrections vanish when expanding around constant backgrounds.

It is then natural to ask what one can say about the higher-loop contributions in the general case of varying backgrounds. Although the structure of the problem quickly becomes quite complicated, we can make some general remarks by restricting to two vertices but any number of loops. For instance, we can consider a diagram with $m + 1$ internal quantum field lines, each of which runs between two interaction vertices with a classical background field, thus forming $m$ loops:

\begin{align}\label{two_vertex_n_loop}
    \raisebox{-0.5\height}{\includegraphics[width=0.5\linewidth]{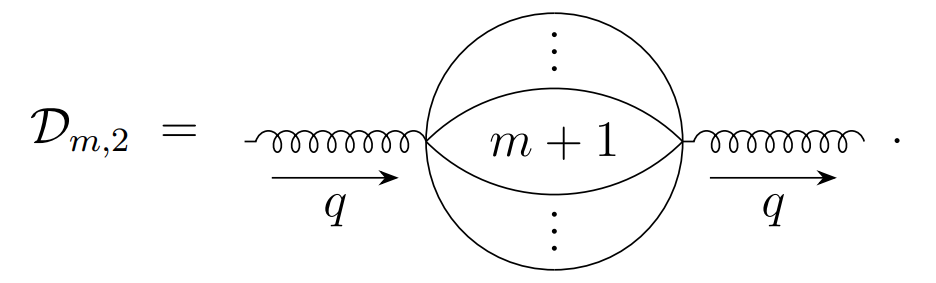}} 
\end{align}
We use the notation $\mathcal{D}_{m, n}$ for a diagram which has $m$ loops and $n$ vertices. In this notation, the one-loop diagrams which we called $\mathcal{D}_n$ in the preceding subsections would be denoted $\mathcal{D}_{1, n}$. For example, the diagram $\mathcal{D}_2$ of equation (\ref{D2_diagram}) would be written as $\mathcal{D}_{1, 2}$, since it is of the form in equation (\ref{two_vertex_n_loop}) with $m = 1$ because it has $2 = 1 + 1$ internal lines between two vertices and thus one loop. Similarly, a diagram with $4$ internal lines between two vertices would have three loops and be denoted $\mathcal{D}_{3, 2}$.

In order to study the diagrams $\mathcal{D}_{m, 2}$, we will need to derive a new Feynman rule for the $(m + 2)$-valent vertex involving $( m + 1)$ quantum fields lines and one insertion of the classical background. These higher vertex factors will come from further terms in the expansion of the square root in equation (\ref{sqrt_expansion_Q}),
\begin{align}\label{implicit_m_vertex}
    \sqrt{ 2 P^2 - S^2 } &= \sqrt{ 2 P_C^2 - S_C^2 } + \sum_{N=1}^{\infty} \binom{\frac{1}{2}}{N} \frac{ 2^N \mathcal{Q}_N }{\left( 2P_C^2 - S_C^2 \right)^{N - \frac{1}{2}}} \nonumber \\
    &= \sqrt{ 2 P_C^2 - S_C^2 } + \sum_{M=2}^{\infty} P^{i_{1} \cdots i_{M}}_{\mu_1 \cdots \mu_{M}} \prod_{k=1}^{M} \partial^{\mu_{k}} Q_{i_k} \, .
\end{align}
In the first line, the factor of $2^N$ is a choice of normalization which is needed to match our conventions for $\mathcal{Q}_1$ and $\mathcal{Q}_2$ above. We will not compute the higher terms $\mathcal{Q}_N$ explicitly, but we instead schematically denote the collection of all contributions from these terms which involve a product of $M$ derivatives of the quantum fields by writing the tensor $P^{i_{1} \cdots i_{M}}{}_{\mu_1 \cdots \mu_{M}}$. When $M = 2$, this is precisely the tensor $P_{\mu \nu}{}^{ij}$ of equation (\ref{P_tensor}). We have changed the summation variable to $M$ in the second line to emphasize that one must collect contributions from several $\mathcal{Q}_N$ at each fixed order in $M$. There are no linear vertices in $Q^i$, so the $M = 1$ term is absent, but both the $N = 1$ term $\mathcal{Q}_1$ and the $N = 2$ term $\mathcal{Q}_2$ of the first sum contributes to the quadratic $M = 2$ interaction of the second sum, and so on.

In terms of the tensors $P^{i_{1} \cdots i_{M}}{}_{\mu_1 \cdots \mu_{M}}$ which are defined implicitly through the expansion in equation (\ref{implicit_m_vertex}), the Feynman rule for an $(M+1)$-valent interaction with one classical field insertion is
\begin{align} 
    \raisebox{-0.5\height}{\includegraphics[width=0.5\linewidth]{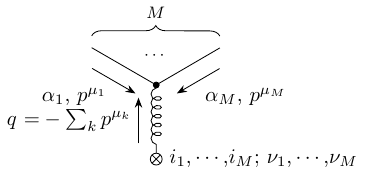}}
    &= \int d^dq P^{i_{1} \cdots i_M}{}_{\nu_1 \cdots \nu_{M}} \left( q \right) \prod_{k=1}^{M}  p^{\nu_k}_{i_k}  \, . \label{n_vertex_rule}
\end{align}
Using this Feynman rule, we can compute the value of the diagram $\mathcal{D}_{m, 2}$ in equation (\ref{two_vertex_n_loop}). Such a diagram has two vertices of the form (\ref{n_vertex_rule}), each with $M = m + 1$, along with $m$ loop momenta $\ell_i$. The contribution from this diagram is given by the integral
\begin{align}\hspace{-25pt}
    \mathcal{D}_{m, 2} &= \frac{\sinh^2 ( \gamma )}{\cosh^{m + 1} ( \gamma ) }\int \frac{d^dq}{( 2 \pi )^d} \left( \int \prod_{i=1}^{m}  \frac{d^d \ell_{i} }{ ( 2 \pi  )^d } \right) \left( \frac{1}{\prod_{j=1}^{m + 1} p^\mu_j p_{\mu j} } \right) \nonumber \\
    &\qquad \qquad \qquad \cdot P^{ i_1  \cdots  i_{m + 1} }{}_{ \nu_1  \cdots \nu_{m + 1}} \left( q \right) \left( \prod_{k=1}^{m + 1}  p_{i_k}^{\nu_k} p_{j_k}^{\mu_k} \right)  P^{j_1 \cdots j_{m + 1}}{}_{\mu_1 \cdots \mu_{m + 1}} \left( -q \right) \, .
\end{align}
Here the momenta of the internal lines are chosen to be $p_1 = q - \ell_1$, $p_{i} = \ell_{i-1} - \ell_{i}$ for $1 < i < m + 1$, and $p_{m+1} = \ell_{m}$, so that the total momentum satisfies
\begin{align}
    \sum_{i= 1}^{m+1} p_{i} &= q \, .
\end{align}
Besides the diagrams $\mathcal{D}_{m, 2}$ drawn in equation (\ref{two_vertex_n_loop}), one might ask whether we should account for additional diagrams where a loop begins and ends on the same vertex. However, diagrams of this form do not contribute, as they vanish in dimensional regularization. We can see this by noting that the momentum $\ell$ running in such a loop will appear in the vertex factor only in the combination $\ell^{\mu} \ell^{\nu}$, and in the propagator in the form $\frac{1}{\ell^2}$. Therefore, the value of any diagram will be proportional to
\begin{align}
    \int \frac{d^d \ell}{\left( 2\pi \right)^{d}} \frac{\ell^{\mu} \ell^{\nu}}{\ell^2} &= \frac{g^{\mu \nu}}{d}\int \frac{d^d\ell}{\left( 2\pi \right)^{d}} 1 \ , 
\end{align}
which we have seen vanishes in dimensional regularization in the limit $d \to 2$, as desired.

Next let us consider the divergence structure of the diagram $\mathcal{D}_{m, 2}$. It is convenient to isolate the part of the integrand which depends on the loop momenta and evaluate it separately. To do this, let us define
\begin{align}\label{frak_L_defn}
    \left( \mathfrak{L}_{m, 2} \right)_{\{ i j \}}^{\{ \mu \nu \}} &= \int \, \left(  \prod_{i=1}^{m}  \frac{d^d \ell_i}{ ( 2 \pi  )^d } \right) \left( \frac{1}{\prod_{j=1}^{m + 1} p^\mu_j p_{\mu j} } \right) \left( \prod_{k=1}^{m + 1}  p_{i_k}^{\nu_k} p_{j_k}^{\mu_k} \right) \, .
\end{align}
Here we use $\{ i j \}$ as a shorthand for the multi-index $\{ i_1 \ldots i_{m+1} j_1 \ldots j_{m+1} \}$ and $\{ \mu \nu \}$ for $\{ \mu_1 \ldots \mu_{m+1} \nu_1 \ldots \nu_{m+1} \}$. We will sometimes suppress these multi-indices in writing $\mathfrak{L}_{m, 2}$ for convenience. The quantity $\mathfrak{L}_{m, 2}$ determines the value of the diagram $\mathcal{D}_{m, 2}$ as
\begin{align}
    \mathcal{D}_{m, 2} &= \frac{\sinh^2 ( \gamma )}{\cosh^{m + 1} ( \gamma ) }\int \frac{d^dq}{( 2 \pi )^d} P^{ i_1 \cdots  i_{m + 1}}{}_{\nu_1  \cdots \nu_{m + 1}}(q) \left( \mathfrak{L}_{m, 2} \right)_{\{ i j \} \{ \mu \nu \}} P^{j_1 \cdots j_{m + 1}}{}_{\mu_1 \cdots \mu_{m + 1}} \left( -q \right) \, ,
\end{align}
so to understand the divergences in $\mathfrak{D}_{m, 2}$, it suffices to understand those in $\mathfrak{L}_{m, 2}$.

One can evaluate $\mathfrak{L}_{m, 2}$ by performing the integral over each loop momentum in succession. The details of one such integration, namely the integral over the final variable $\ell_m$, are presented in appendix \ref{sec:scalar_field_n_loop_2_vertex_calc}. After evaluating this single integral over $\ell_m$, one obtains a result proportional to $\Gamma \left( -\frac{d}{2} \right) \ell^{d}_{m - 1}$. One can then apply the same argument recursively to conclude that performing all $m$ of the integrals generates $m$ factors of this form. After evaluating all $m$ integrals, the final dependence on the momentum $q$ takes the form
\begin{align}\label{final_n_loop_2_vertex_dimreg}
    \left( \mathfrak{L}_{m, 2} \right)_{\{ i j \}}^{\{ \mu \nu \}} \sim \Gamma\left( -\frac{d}{2} \right)^{m} q^{d m} \, ,
\end{align}
where we show only the dependence on $q$ and $d$ but suppress the tensor structure in the $i, j, \mu, \nu$ indices.\footnote{Each integral yields $6$ different symmetrizations of the external indices. Thus the exact form of an $m$-loop diagram contains many different index structures and is challenging to write explicitly in general.}

It is also useful to translate the divergence structure of equation (\ref{final_n_loop_2_vertex_dimreg}) in dimensional regularization to an equivalent dependence on a momentum cutoff $\Lambda$. For $d = 2 ( 1 + \epsilon )$ we have the limiting behavior $\Gamma \left( - \frac{d}{2} \right) \sim \frac{1}{\epsilon}$, and a divergence proportional to $\frac{1}{\epsilon}$ in dimensional regularization corresponds to a logarithmic divergence of the form $\log ( \Lambda )$. Therefore, the $m$-loop, $2$-vertex contributions from (\ref{final_n_loop_2_vertex_dimreg}) yield divergences of the form
\begin{align}
    \left( \mathfrak{L}_{m, 2} \right)_{\{ i j \}}^{\{ \mu \nu \}} \sim \left( \frac{1}{\epsilon} \right)^{m} q^{2m} \sim \left( \log \Lambda \right)^m \, .
\end{align}
This is a different divergence structure than the one which we have seen in our study of the $1$-loop effective action, which would necessitate the addition of different counterterms. Additionally, each of these counterterms is classically conformal and has a different higher derivative dependence on the external classical field momenta.

We conclude this section with some further comments. Even though our analysis for
non-constant backgrounds is very preliminary, no clear organizational principle seems to
emerge in this hierarchy of divergences and necessary counterterms. Though this might be
a feature of our perturbative approach, it begins to suggest that this non-analytic model is
non-renormalizable, which might also spoil the quantum conformal invariance of the model.
Ultimately, the theory might retain a sensible interpretation only as an effective field theory.
Yet, it remains a very interesting fact that there are no quantum corrections for constant background fields $C_\mu^i$. We leave other open questions for further future investigations

\section{Conclusion} \label{sec:Conclusion&Outlook1}

In this work, we have explored the space of interacting chiral boson theories from several perspectives. We showed that, when written in a Floreanini-Jackiw representation, the property of non-manifest Lorentz invariance is closely related to stress tensor deformations: indeed, every parameterized family of Lorentz-invariant chiral boson theories can be interpreted as a deformation by some function of the energy-momentum tensor. In the dual description using $U(1)$ gauge fields with a Chern-Simons action, Lorentz invariance is manifest but chirality (or self-duality) is not, and in this setting we find that every family of \emph{self-dual} Chern-Simons boundary terms likewise obeys a flow equation driven by a function of the stress tensor. We have also explained how a general boundary term for such a bulk $U(1)$ Chern-Simons theory imposes modified boundary conditions on the gauge fields which lead to a non-linear self-duality condition for the currents; this mirrors the analogous non-linear self-duality constraints obeyed by interacting Floreanini-Jackiw bosons.

We then studied the quantization of interacting chiral boson models, focusing on a root-$T\overline{T}$-deformed system of free bosons. We characterized the finite-volume spectrum both for one left-moving and one right-moving boson, where the root-$T\overline{T}$ deformation acts as a rescaling of the target space radius, and also for two left-moving bosons and one right-moving boson, where the deformation is more complicated but can be analyzed perturbatively in a large-momentum expansion.  In doing so, we confirmed that the zero-mode formula (\ref{zero_mode_formula}) derived via holography does not apply to generic states, but does apply in certain states with constant stress tensors. 
We also gave a classical/heuristic argument on how a set of field redefinitions might turn all these models into free ones.
Finally, we have studied the quantum effective action for the theory of root-$T\overline{T}$-deformed bosons with equal numbers of left- and right-movers. Intriguingly, we find that the one-loop effective action vanishes around classical backgrounds which are linear in the spacetime coordinates.

There are several interesting directions for future research, some of which we summarize in what follows. Understanding more about these issues, and in particular developing a clearer picture of field theories with non-analytic interaction terms such as the Modified Scalar theory, may teach us new lessons about previously unexplored models within the space of quantum field theories.

\emph{Supersymmetry}

There has been a great deal of work on supersymmetric extensions of deformations constructed from the energy-momentum tensor \cite{Baggio:2018rpv,Chang:2018dge,Jiang:2019hux,Chang:2019kiu, Coleman:2019dvf,Ferko:2019oyv,He:2019ahx,Ebert:2020tuy,Ferko:2021loo} and other conserved currents \cite{Jiang:2019trm}, including analogous deformations of $1d$ theories by conserved charges \cite{Gross:2019ach,Gross:2019uxi,Ebert:2022xfh,Ebert:2022ehb,Ferko:2023ozb}.

A natural direction for further investigation is to seek such a supersymmetric generalization of the results in this work. This would involve coupling a supersymmetric theory of interacting chiral bosons and their fermionic superpartners to supergravity, which would give expressions for the fields in the stress tensor supermultiplet.

In the case of a single free chiral boson and its fermionic partner, the procedure for performing this coupling to supergravity was explained in \cite{Bastianelli:1989cu}, building on earlier results for the supergravity couplings of non-chiral fields \cite{Brink:1976sc}. The bosonic truncation of this supergravity coupling reproduces the coupling to vielbeins which we have used in this work. It would be interesting to generalize this technique and couple an arbitrary number of chiral and anti-chiral bosons, and their fermionic counterparts, to supergravity, and then consider flows in the space of such supersymmetric interacting theories, much as we have done here. In principle, one could perform this analysis either using component fields -- which was the strategy adopted in \cite{Bastianelli:1989cu} -- or using a superspace formulation, such as the one employed in \cite{Gates:1987sy,Bellucci:1987mj}. One might also hope to interpret these theories using a bulk description involving a supersymmetric Chern-Simons theory, which would give a supersymmetric generalization of the results in section \ref{sec:cs}.

\emph{Quantum Hall physics}

A famous application of $U(1)$ Chern-Simons theories, and the chiral bosons which describe their edge modes, occurs in the study of the quantum Hall effect. The essential reason for this, as we mentioned in section \ref{sec:cs}, is that the Chern-Simons term is more relevant at low energies than the Maxwell term. Therefore, in an effectively $(2 + 1)$-dimensional system -- such as a flat slab of material subject to a background magnetic field -- one expects that the low-energy effective action $S_{\text{eff}} [ A ]$ will be controlled by the Chern-Simons term $S_{\text{CS}} [ A ]$. Computing the associated current which we defined in equation (\ref{currents_defn}), 
\begin{align}\label{currents_conclusion}
    J_i \sim \frac{\delta S_{\text{CS}}}{\delta A_i} \, ,
\end{align}
therefore gives predictions for the behavior of the system. For instance, in the integer quantum Hall effect, this current $J_i$ agrees with the Hall conductivity of an integer number of filled Landau levels, if this integer $\nu \in \mathbb{Z}$ is related to the Chern-Simons level appropriately.

We have seen that a Chern-Simons theory on a manifold with boundary supports chiral bosons on the edge. In the quantum Hall setting, these chiral edge modes describe propagating fluctuations in the charge density at the edge of the physical sample. Remarkably, the quantum mechanics of this chiral boson theory contains a great deal of information about the interior of the sample. For instance, by carrying out the quantization of a single Floreanini-Jackiw boson as we described in section \ref{sec:general_quantization}, one finds a Hamiltonian which correctly predicts the spectrum (including degeneracies) of excited modes for the Laughlin wavefunction which describes the fractional quantum Hall effect.\footnote{See the reviews \cite{Tong:2016kpv,RevModPhys.75.1449}, or the incomplete sampling of some of the original works \cite{PhysRevB.41.12838,PhysRevB.43.11025,1992IJMPB...6.1711W}, for further discussion on this subject.}

One might ask whether the modified Chern-Simons boundary terms which we have considered in this work could be used to model some variant of a conventional quantum Hall system. For instance, it would be very interesting if an experimentally realizable modification of a quantum Hall droplet would subject the system to a boundary term like the one which is generated by the root-$T\overline{T}$ deformation. If so, this could offer a way to study the effective dynamics of the Modified Scalar theory -- and other theories obtained via stress tensor deformations -- in the laboratory.

\emph{Non-perturbative analysis}

All of the results concerning the quantum theory of root-$T\overline{T}$-deformed bosons presented in this work have been obtained in perturbation theory, by expanding around a classical background. For instance, we have attempted a perturbative analysis of the effective action and noticed that a hierarchy of counterterms emerged in the Modified Scalar theory. However, it seems likely that the most interesting features of root-$T\overline{T}$-deformed theories at the quantum level -- assuming that they exist -- will only be visible non-perturbatively. It is therefore important to find a way to study the quantization of such root-$T\overline{T}$ deformed theories beyond perturbation theory, which will likely require a new perspective.

One way to re-frame these deformed theories, which may be useful for a non-perturbative analysis, is via geometry. In the case of the related $T\overline{T}$ deformation, many insights have resulted from presentations of the flow in terms of coupling to gravity \cite{Dubovsky:2017cnj,Dubovsky:2018bmo} or random geometry \cite{Cardy:2018sdv}, or realizing the deformation via a field-dependent change of variables \cite{Conti:2018tca,Conti:2022egv,Morone:2024ffm}. A similar geometrical interpretation may be possible for the root-$T\overline{T}$ deformation. For instance, the Modified Scalar Lagrangian (\ref{modscalar_lorentz_invariant_lagrangian}) can be rewritten as
\begin{align}\label{rtt_background_metric}
    \mathcal{L} &= \frac{1}{2} g^{\mu \nu} \partial_\mu \varphi^i \partial_\nu \varphi^i \, , \nonumber \\
    g^{\mu \nu} &= \cosh ( \gamma ) \eta^{\mu \nu} + \sinh ( \gamma ) \left( \frac{2 \partial^\mu \varphi^{j} \partial^\nu \varphi^{j} - \eta^{\mu \nu} \partial_\rho \varphi^{j} \partial^\rho \varphi^{j} }{\sqrt{ 2 \partial_\sigma \varphi^i \partial^\tau \varphi^i \partial_\tau \varphi^k \partial^\sigma \varphi^k  - \left( \partial_\sigma \varphi^i \partial^\sigma \varphi^i \right)^2 } } \right) \, ,
\end{align}
which is equivalent to a theory of \emph{free} scalar fields coupled to a field-dependent metric. Even at the perturbative level, such a rewriting of the deformation may be useful -- for instance, it may be possible to adapt existing heat kernel techniques\footnote{See \cite{Vassilevich:2003xt} and references therein for a review.} which compute the quantum effective actions for theories on background metrics to handle field-dependent metrics such as (\ref{rtt_background_metric}), which could reproduce results like those in section \ref{sec:cian} from a different point of view. However, it would be even more useful if such a geometrical presentation of the root-$T\overline{T}$ flow could furnish us with a non-perturbative definition of the quantum theory.

Another potential way to approach the study of renormalisation of the Modified Scalar theory, and analyse its quantum conformal symmetry, is by using non-perturbative functional renormalisation group approaches. An attempt to use such techniques for $T\overline{T}$ deformed scalar theories has been made in \cite{Liu:2023omp}. It would be intriguing to reattempt this analysis for non-analytic models and generic $T\overline{T}$-like deformations, including root-$T\overline{T}$.

A third strategy is to bypass the classical Lagrangian (\ref{modscalar_lorentz_invariant_lagrangian}) and attempt to define the quantum Modified Scalar theory directly by characterizing the set of local operators in the theory along with their correlation functions. For instance, one could proceed under the assumption that the theory in question is a CFT, and see whether this leads to a contradiction.\footnote{An example of such a contradiction would be finding an operator which can be neither a primary nor a descendant, which is used to demonstrate that the Maxwell theory is not conformal except in four dimensions \cite{El-Showk:2011xbs}. Alternatively, one could use the more formal machinery of algebraic/axiomatic QFT.} Here there appears to be an interesting tension. Standard lore suggests that, in any $\mathrm{CFT}_2$ with a conserved vector current $J$, its Hodge dual $\ast J$ must also be conserved. For a putative theory of root-$T\overline{T}$-deformed $\varphi^i$, it appears that the operators $J^i_\mu = \partial_\mu \varphi^i$ should not be conserved at finite $\gamma$ due to the source terms in the equations of motion, although their duals $\widetilde{J}^i_\mu = \epsilon_{\mu \nu} \partial^\nu \varphi^i$ \emph{are} conserved (at least for non-compact scalars).\footnote{The analogous tension for the ModMax theory can be phrased in terms of generalized global symmetries: if a $4d$ CFT has a $U(1)_1$ magnetic one-form global symmetry, then it must also have the corresponding $U(1)_1$ electric one-form global symmetry, and vice-versa \cite{Hofman:2018lfz}. A $4d$ ModMax CFT would appear to have the magnetic $1$-form symmetry of the Maxwell theory but not the electric one, since $\partial_\mu \widetilde{F}^{\mu \nu} = 0$ but $\partial_\mu F^{\mu \nu} \neq 0$.} If the quantum Modified Scalar theory does exist, it would be very interesting to see how this tension is resolved. Perhaps the quantum theory is not a CFT, or perhaps it is not even a local quantum field theory, much like a $T\overline{T}$-deformed CFT is believed to become non-local due to its Hagedorn density of states at high energies.

\appendix
\chapter{Effective Field Theory Details of the $T^2$ Operator}
\label{app:EFT}
\section{$T^2$ flow equation}
\label{app:EFT1}
Here, we provide the details of the EFT interpretation of the $T^2$ operator. We first prove \eqref{eq:hartmanFlow}. Let's start on the right-hand side of \eqref{eq:hartmanFlow}. Note that
\begin{equation}
\begin{aligned}
    &\left(\widetilde{T}_{i j}+a_d \widetilde{C}_{i j}\right)^2 \\&= \left( \frac{1}{8 \pi G} \left( K_{ij} - K g^{0}_{ij} + (d-1) g^0_{ij} \right) - a_d \widetilde{C}_{ij} + a_d \widetilde{C}_{ij} \right)^2
    \\&= \frac{1}{64 \pi^2 G^2} \left(  K_{ij} - K g^{0}_{ij} + (d-1) g^0_{ij}  \right)\left(  K^{ij} - K g^{0ij} + (d-1) g^{0ij}  \right)
    \\&=  \frac{1}{64 \pi^2 G^2}  \bigg( K_{ij} K^{ij}- K g^{0ij}K_{ij} + (d-1) g^{0ij} K_{ij} - K g^{0}_{ij} K^{ij} + K^2 g^{0ij} g_{0ij}\\& - (d-1) K g^0_{ij} g^{0ij} + (d-1) g^0_{ij} K^{ij} - (d-1) K g^0_{ij} g^{0ij} +(d-1)^2 g^{0ij} g^0_{ij}   \bigg)
    \\&= \frac{1}{64 \pi^2 G^2}  \bigg( K_{ij} K^{ij}- K^2  + (d-1) K - K^2 + K^2 d - d(d-1) K  \\&+ (d-1)K - d(d-1) K  +d(d-1)^2   \bigg)
    \\&=\frac{1}{64 \pi^2 G^2}  \bigg( K_{ij} K^{ij} +(d-2)K^2 -2(d-1)^2K  +d(d-1)^2   \bigg)
    \\&= \frac{1}{64 \pi^2 G^2}  \bigg( K_{ij} K^{ij}+(d-K-1) (d^2 +2K - d(K+1))   \bigg)\,.
\end{aligned}
\end{equation}
The second term is 
\begin{equation}
    \begin{aligned}
   & \left(\widetilde{T}_i^i+a_d \widetilde{C}_i^i\right)^2 \\&= \left(\frac{1}{8 \pi G} \left( g^{0ij} K_{ij} - K g^{0ij} g^{0}_{ij} + (d-1)g^{0ij}  g^0_{ij} \right) - a_d g^{0ij} \tilde{C}_{ij}   +a_d \widetilde{C}_i^i\right)^2
    \\&= \left(\frac{1}{8 \pi G} \left( g^{0ij} K_{ij} - K g^{0ij} g^{0}_{ij} + (d-1)g^{0ij}  g^0_{ij} \right)\right)^2
    \\&=\left(\frac{1}{8 \pi G} \left( K - Kd  + d(d-1) \right)\right)^2
    \\&= \frac{1}{64 \pi^2 G^2} \left( d-1 \right)^2 (d-K)^2\,.
    \end{aligned}
\end{equation}
Therefore
\begin{equation}
    \begin{aligned}
    \label{eq:ffffs}
  &\left(\widetilde{T}_{i j}+a_d \widetilde{C}_{i j}\right)^2-\frac{1}{d-1}\left(\widetilde{T}_i^i+a_d \widetilde{C}_i^i\right)^2 \\&= \frac{1}{64 \pi^2 G^2}  \bigg( K_{ij} K^{ij}+(d-K-1) (d^2 +2K - d(K+1))   \bigg) \\&- \frac{1}{64 \pi^2 G^2 (d-1)} \left( d-1 \right)^2 (d-K)^2
  \\&= \frac{1}{64 \pi^2 G^2} \left( K_{ij} K^{ij} + d(1-d+ 2K) - K(K+2)\right).
    \end{aligned}
\end{equation}
We can trade out $K_{ij}K^{ij}$ by using the Hamiltonian constraint for the radial slicing in the bulk
\begin{equation}
\label{eq:Einsteinradial}
    K^2 - K_{ij} K^{ij} - d(d-1) - \widetilde{R} + 16 \pi G \tilde{t}^r_r = 0\,.
\end{equation}
Substituting \eqref{eq:Einsteinradial} into \eqref{eq:ffffs}, we find 
\begin{equation}
    \begin{aligned}
  &\left(\widetilde{T}_{i j}+a_d \widetilde{C}_{i j}\right)^2-\frac{1}{d-1}\left(\widetilde{T}_i^i+a_d \widetilde{C}_i^i\right)^2 \\&= \frac{1}{64 \pi^2 G^2} \left(  K^2  - d(d-1) - \widetilde{R} + 16 \pi G \tilde{t}^r_r + d(1-d+ 2K) - K(K+2)\right)
  \\&= \frac{1}{64 \pi^2 G^2} \left(-2(d-1) (d-K)   - \widetilde{R} + 16 \pi G \tilde{t}^r_r \right)\,.
    \end{aligned}
\end{equation}
Hence
\begin{equation}
    \begin{aligned}
    \label{eq:RHS11}
      &  -4 \pi G\left[\left(\widetilde{T}_{i j}+a_d \widetilde{C}_{i j}\right)^2-\frac{1}{d-1}\left(\widetilde{T}_i^i+a_d \widetilde{C}_i^i\right)^2\right]-\frac{\widetilde{R}}{16 \pi G}+\widetilde{t}_r^r \\&=  -4 \pi G\left[\frac{1}{64 \pi^2 G^2} \left(-2(d-1) (d-K)   - \widetilde{R} + 16 \pi G \tilde{t}^r_r \right)\right]-\frac{\widetilde{R}}{16 \pi G}+\widetilde{t}_r^r
      \\&= -\frac{1}{16 \pi G} \left(-2 (d-1) (d-K) \right) + \frac{\widetilde{R}}{16 \pi G} - \Tilde{t}^r_r -\frac{\widetilde{R}}{16 \pi G} + \Tilde{t}^r_r 
      \\&=  \frac{1}{8\pi G} (d-1) (d-K)\,.
    \end{aligned}
\end{equation}
The left-hand side of \eqref{eq:hartmanFlow} is simply
\begin{equation}
    \begin{aligned}
   \widetilde{T}_i^i+a_d \widetilde{C}_i^i &= \frac{1}{8 \pi G} \left( K -d K  + d(d-1)  \right) - a_d \widetilde{C}^i_{i} +a_d \widetilde{C}_i^i
   \\&=\frac{1}{8 \pi G} (d-1) (d-K)
    \end{aligned}
\end{equation}
which equals to \eqref{eq:RHS11}. Therefore, we have proven  \eqref{eq:hartmanFlow}.

Finally, let's prove \eqref{eq:X}. Recall that we  can write the deformation of the classical action in terms of a local operator $ \mathbb{X}$ as
\begin{equation}
\label{eq:gee1}
    \frac{\partial S_{\text{EFT}}}{\partial \lambda} = \int d^dx \sqrt{\gamma} \mathbb{X}
\end{equation}
where $\lambda$ is a  dimensionful scale related to the size of the deformation and cutoff. In a theory with a single dimensionful scale $\lambda$, invariance under a change of units implies the effective action is modified by the trace
\begin{equation}
\label{eq:gee2}
    \lambda \frac{\partial W}{\partial \lambda} = \frac{1}{\operatorname{dim} (\lambda)} \int d^dx \sqrt{\gamma} T^i_i  = - \frac{1}{d}  \int d^dx \sqrt{\gamma} T^i_i  = \int d^dx \sqrt{\gamma}  \mathbb{X}
\end{equation}
implying that
\begin{equation}
    \mathbb{X} =-\frac{1}{d \lambda } T^i_i\,.
\end{equation}
All we need to do is determine $T^i_i$ from $\widetilde{T}^i_i$ to find $ \mathbb{X}$. We use the dictionary from \cite{Hartman:2018tkw} to help, which is
\begin{equation}
    \begin{aligned}
\text{Bulk coordinates}&: \quad (r, x)\\
\text{Bulk spacetime metric}&: \quad g_{\mu \nu} \\
\text{Induced metric at $r=r_c$}&: \quad g^0_{ij} (x) = g_{ij} (r_c, x) \\
\text{Boundary metric}&: \quad \gamma_{ij} = r^{-2}_c g^{0}_{ij} \\
\text{Bulk scalar field}&: \quad \phi \\
\text{Boundary value}&: \quad \phi_0 (x) = \phi(r_c, x)\\
\text{CFT source} &: \quad J = r_c^{d - \Delta} \phi_0 \\
\text{Bulk on-shell action}&: \quad W[g^0, \phi_0]\\
\text{Bulk Brown-York tensor}&: \quad \tilde{T}_{ij} \\
\text{Boundary stress tensor}&: \quad T_{ij} = r^{d-2}_c \Tilde{T}_{ij}\,.
    \end{aligned}
\end{equation}
Proceeding to calculate $T^i_i$ from \eqref{eq:hartmanFlow}
\begin{equation}
\begin{aligned}
   & \widetilde{T}_i^i+a_d \widetilde{C}_i^i=-4 \pi G\left[\left(\widetilde{T}_{i j}+a_d \widetilde{C}_{i j}\right)^2-\frac{1}{d-1}\left(\widetilde{T}_i^i+a_d \widetilde{C}_i^i\right)^2\right]-\frac{\widetilde{R}}{16 \pi G}+\widetilde{t}_r^r
   \\ &\implies  \widetilde{T}_i^i=-4 \pi G\left[\left(\widetilde{T}_{i j}+a_d \widetilde{C}_{i j}\right)^2-\frac{1}{d-1}\left(\widetilde{T}_i^i+a_d \widetilde{C}_i^i\right)^2\right]-\frac{\widetilde{R}}{16 \pi G}+\widetilde{t}_r^r - a_d \widetilde{C}_i^i
   \\& \implies r_c^{-d} T_i^i=-4 \pi G\left[\left(\widetilde{T}_{i j}+a_d \widetilde{C}_{i j}\right)^2-\frac{1}{d-1}\left(\widetilde{T}_i^i+a_d \widetilde{C}_i^i\right)^2\right]-\frac{\widetilde{R}}{16 \pi G}+\widetilde{t}_r^r - a_d \widetilde{C}_i^i
      \\& \implies  T_i^i=-4 \pi G r_c^d \left[\left(\widetilde{T}_{i j}+a_d \widetilde{C}_{i j}\right)^2-\frac{1}{d-1}\left(\widetilde{T}_i^i+a_d \widetilde{C}_i^i\right)^2\right]\\&+ r_c^d \left( -\frac{\widetilde{R}}{16 \pi G}+\widetilde{t}_r^r - a_d \widetilde{C}_i^i\right)\,.
    \end{aligned}
\end{equation}
Therefore
\begin{equation}
    \begin{aligned}
        \mathbb{X} &=  - \frac{1}{d\lambda} T^i_i
        \\&= \frac{4 \pi G r_c^d}{d\lambda} \left[\left(\widetilde{T}_{i j}+a_d \widetilde{C}_{i j}\right)^2-\frac{1}{d-1}\left(\widetilde{T}_i^i+a_d \widetilde{C}_i^i\right)^2\right]- \frac{r_c^d}{d\lambda} \left( -\frac{\widetilde{R}}{16 \pi G}+\widetilde{t}_r^r - a_d \widetilde{C}_i^i\right)
        \\&= \frac{4 \pi G r_c^d}{d\lambda} \left[\left(\widetilde{T}_{i j}+a_d \widetilde{C}_{i j}\right)^2-\frac{1}{d-1}\left(\widetilde{T}_i^i+a_d \widetilde{C}_i^i\right)^2\right]- \frac{r_c^d}{d\lambda} \left( -\frac{\widetilde{R}}{16 \pi G}+\widetilde{t}_r^r - a_d \widetilde{C}_i^i\right)\,.
    \end{aligned}
\end{equation}
Note that
\begin{equation}
    \begin{aligned}
        \left(\widetilde{T}_{i j}+a_d \widetilde{C}_{i j}\right)^2 &=  g^{0ik} g^{0jm} \left(\widetilde{T}_{i j}+a_d \widetilde{C}_{i j}\right)\left(\widetilde{T}_{km}+a_d \widetilde{C}_{km}\right)
        \\&= g^{0ik} g^{0jm}  r_c^{4-2d} \left(T_{ij} + r_c^{d-2}a_d \widetilde{C}_{i j}\right) \left( T_{km}+r_c^{d-2}a_d \widetilde{C}_{km}\right)
        \\&=r_c^{-2}\gamma^{0ik} r_c^{-2} \gamma^{0jm} r_c^{4-2d} \left( T_{ij} + a_d r_c^{d-2} \widetilde{C}_{i j}\right) \left(T_{km}+a_d r_c^{d-2}\widetilde{C}_{km}\right)
        \\&=r_c^{-2d}\gamma^{0ik} \gamma^{0jm}  \left( T_{ij} + a_d r_c^{d-2} \widetilde{C}_{i j}\right) \left(T_{km}+a_d r_c^{d-2}\widetilde{C}_{km}\right)
        \\&=r_c^{-2d} \left( T_{ij} + a_d r_c^{d-2} \widetilde{C}_{i j}\right)^2
    \end{aligned}
\end{equation}
and 
\begin{equation}
  \begin{aligned}
      \left(\widetilde{T}_i^i+a_d \widetilde{C}_i^i\right)^2 &= \left(g^{0ij} \tilde{T}_{ij} + a_d r_c^{d-2} g^{0ij} \widetilde{C}_{ij}\right)^2 \\&=\left( r_c^{-2}\gamma^{0ij} r_{c}^{2-d}\tilde{T}_{ij} + a_d r_c^{-2}\gamma^{0ij} \widetilde{C}_{ij}\right)^2
      \\&=\left(r_{c}^{-d} \gamma^{0ij} \tilde{T}_{ij} + a_d r_c^{-2}\gamma^{0ij} \widetilde{C}_{ij}\right)^2
      \\&= r_c^{-2d} \left( T_{i}^i + a_d r_c^{d-2} \widetilde{C}_{i}^i\right)^2
  \end{aligned}
\end{equation}
yielding
\begin{equation}
    \begin{aligned}
      &  \mathbb{X} =  \frac{4 \pi G}{d\lambda r_c^d} \left[\left(\widetilde{T}_{i j}+a_d \widetilde{C}_{i j}\right)^2-\frac{1}{d-1}\left(\widetilde{T}_i^i+a_d \widetilde{C}_i^i\right)^2\right]- \frac{r_c^d}{d\lambda} \left( -\frac{\widetilde{R}}{16 \pi G}+\widetilde{t}_r^r - a_d \widetilde{C}_i^i\right)
        \\& \implies  \mathbb{X}=  \left[\left(T_{i j}+a_d \widetilde{C}_{i j}\right)^2-\frac{1}{d-1}\left(T_i^i+a_d \widetilde{C}_i^i\right)^2\right]- \frac{r_c^d}{d\lambda} \left( -\frac{\widetilde{R}}{16 \pi G}+\widetilde{t}_r^r - a_d \widetilde{C}_i^i\right)\,,
    \end{aligned}
\end{equation}
where we used \eqref{eq:TTlambdadef}. We have proven \eqref{eq:X}.

\subsection{Details of the $T^2$-deformed energy spectrum}
Here, we explain in more detail what went behind in deriving the $T^2$-deformed flow equation for the energy spectrum on the square $(d-1)$-dimensional torus in \eqref{eq:pppsa}. Thus
\begin{equation}
T_{i j} T^{i j} = 
\varepsilon^2  + (d-1) \left(\frac{1}{L^{d-2}} \frac{d\left(\varepsilon L^{d-1}\right)}{d L} \right)^2\,, 
\end{equation}
where
\begin{equation}
\label{eq:12e3w3rr3wr4r4w4444444}
  T_{\tau \tau}=\varepsilon, \quad T_{a a} =\frac{1}{L^{d-2}} \frac{d\left(\varepsilon L^{d-1}\right)}{d L}  = (d-1) \varepsilon + L \frac{d \varepsilon}{dL}\,.
\end{equation}
One can easily see \eqref{eq:12e3w3rr3wr4r4w4444444} is true from thermodynamics on the torus
\begin{equation}
    dE =P_a dV \implies P_a = \frac{dE_n}{dV_a} =\frac{dE_n}{A dL_a} = \frac{1}{\prod_{a \neq b} L_a} \frac{d(\varepsilon \prod_{i=1}^{d-1} L_i)}{d L_a}\,.
\end{equation}
For a square torus, the pressure
\begin{equation}
    P = \frac{1}{L^{d-2}} \frac{d}{dL} \left( \varepsilon L^{d-1} \right)\,.
\end{equation}
The trace squared piece is
\begin{equation}
   (T^i_i)^2 = \left( \varepsilon +\frac{ d-1}{L^{d-2}} \frac{d\left(\varepsilon L^{d-1}\right)}{d L} \right) ^2 = \varepsilon^2 + \frac{2\varepsilon (d-1)}{L^{d-2}} \frac{d\left(\varepsilon L^{d-1}\right)}{d L} + \left( \frac{ d-1}{L^{d-2}} \frac{d\left(\varepsilon L^{d-1}\right)}{d L} \right)^2\,,
\end{equation}
where the $(d-1)$ comes from tracing over the $d-1$ dimensional torus indices. Thus
\begin{equation}
\begin{aligned}
    T_{i j} T^{i j}-\frac{1}{d-1}\left( T_i^i\right)^2  &=    \varepsilon^2  +  (d-1) \left(\frac{1}{L^{d-2}} \frac{d\left(\varepsilon L^{d-1}\right)}{d L} \right)^2 \\& - \frac{1}{d-1} \bigg(  \varepsilon^2 + \frac{2\varepsilon  (d-1)}{L^{d-2}} \frac{d\left(\varepsilon L^{d-1}\right)}{d L} + \left( \frac{ d-1}{L^{d-2}} \frac{d\left(\varepsilon L^{d-1}\right)}{d L} \right)^2 \bigg)
    \\&= \frac{d-2}{d-1} \varepsilon^2 - \frac{2 \varepsilon}{d-1} \frac{1}{L^{d-2}} \frac{d (\varepsilon L^{d-1})}{d L}\,.
    \end{aligned}
\end{equation}
We have  proven \eqref{eq:pppsa}.
  \section{Scalar theory and Hamilton-Jacobi equation}
  As an example, let's consider gravity decoupled, and the bulk contains a scalar field $\phi$. The bulk path integral is computed by the on-shell action, $W[r_c; \phi_0 (x)]$. The flow of this functional is governed by the Hamilton-Jacobi equation
\begin{equation}
\label{eq:flow}
    \frac{\partial}{\partial r_c} W[r_c; \phi_0] = - H\left[\phi_0, \frac{\delta W}{\delta \phi_0}\right]\,.
\end{equation}
Here $H$ is the scalar Hamiltonian with evolution in the $r$ direction. To derive the EFT at a finite cutoff, we first write $Z_{\text{gravity}} = e^{-W}$ and apply the flow equation  \eqref{eq:flow} to this dictionary in the generalized holographic principle
\begin{equation}
\label{eq:generalizedholographicprinciple}
    Z_{\text{EFT}} [r_c; \gamma_{ij}, J] = Z_{\text{gravity}} [g^0_{ij} = r_c^2 \gamma_{ij}, \phi_0 = r_c^{\Delta - d} J]
\end{equation}
then translate back to the field theory. In other words
\begin{equation}
    W = - \ln Z_{\text{gravity}}
\end{equation}

so the flow equation equation implies that
\begin{equation}
\begin{aligned}
    \frac{\partial}{\partial r_c} W &= - \frac{1}{Z_{\text{gravity}}} \frac{d}{dr_c} Z_{\text{gravity}}
\end{aligned}
\end{equation}
and using the Hamilton-Jacobi equation
\begin{equation}
    \frac{d}{dr_c} Z_{\text{gravity}}= H Z_{\text{gravity}}\,.
\end{equation}
From the generalized holographic principle \eqref{eq:generalizedholographicprinciple}, we have the EFT relation
\begin{equation}
\label{eq:rereg}
    \frac{d}{dr_c} Z_{\text{EFT}} = H Z_{\text{EFT}} = \int D \varphi H e^{ - S_{\text{EFT}} + \int d^dx \sqrt{\gamma} \mathcal{O} \phi_0 r_c^{d-\Delta} }\,.
\end{equation}
Let us look at the far left side of \eqref{eq:rereg} and take an $r_c$ derivative
\begin{equation}
\begin{aligned}
    \frac{d}{dr_c} Z_{\text{EFT}} &= \int D \varphi \frac{d}{dr_c} e^{-S_{\text{EFT}} + \int d^dx \sqrt{\gamma} \mathcal{O} \phi_0 r_c^{d-\Delta}}
    \\&= \int D \varphi \left( - \frac{d}{dr_c} S_{\text{EFT}} + \frac{d-\Delta}{r_c} \int d^dx \sqrt{\gamma} \mathcal{O} \phi_0 r_c^{d-\Delta} \right) e^{-S_{\text{EFT}} + \int d^dx \sqrt{\gamma} \mathcal{O} \phi_0 r_c^{d-\Delta}} \,.
\end{aligned}
\end{equation}
We arrive at
\begin{equation}
    -\frac{d}{dr_c} S_{\text{EFT}}  + \frac{d-\Delta}{r_c} \int d^dx \sqrt{\gamma} \mathcal{O} \phi_0 r_c^{d-\Delta} =  H
\end{equation}
implying that our flow equation is
\begin{equation}
\label{eq:rejper}
\frac{d}{dr_c} S_{\text{EFT}} = - H + \frac{d- \Delta}{r_c} \int d^dx \sqrt{\gamma} \mathcal{O} \phi_0 r_c^{d-\Delta}\,.   
\end{equation}
Furthermore, define
\begin{equation}
    \hat{S}_{\text{EFT}} = S_{\text{EFT}} - \int d^dx \sqrt{\gamma}  \mathcal{O} \phi_0 r_c^{d-\Delta} 
\end{equation}
so that \eqref{eq:rejper} becomes
\begin{equation}
\frac{d}{dr_c} \hat{S}_{\text{EFT}} (r_c, J(r_c)) = - H\,.
\end{equation}
Instead of writing the above flow equation with a $\frac{d}{dr_c}$, let's use $\frac{\partial}{\partial r_c}$ via chain rule
\begin{equation}
\begin{aligned}
\label{eq:ergjipe3w1}
    \frac{d}{dr_c} \hat{S}_{\text{EFT}} (r_c, J(r_c)) &= \frac{\partial \hat{S}_{\text{EFT}} }{\partial r_c} + \frac{\partial \hat{S}_{\text{EFT}}}{\partial J} \frac{\partial J}{\partial r_c}
    \\&= \frac{\partial \hat{S}_{\text{EFT}} }{\partial r_c} + \frac{d-\Delta}{r_c}  \frac{\partial \hat{S}_{\text{EFT}}}{\partial J}J(r_c)\,,
    \end{aligned}
\end{equation}
where the source is 
\begin{equation}
    J(r_c) = \phi_0 r_c^{d-\Delta} \implies \frac{\partial}{\partial r_c} J(r_c)= \frac{d-\Delta}{r_c} J(r_c)
\end{equation}
and the dual operator is defined by taking a derivative of the EFT action with respect to the source
\begin{equation}
\label{eq:erg213f}
    \mathcal{O} = - \frac{1}{\sqrt{\gamma}} \frac{\partial \hat{S}_{\text{EFT}}}{\partial J}\,.
\end{equation}
Therefore, using \eqref{eq:erg213f},  \eqref{eq:ergjipe3w1} becomes
\begin{equation}
\begin{aligned}
  \frac{\partial \hat{S}_{\text{EFT}} }{\partial r_c}  =    \frac{d}{dr_c} \hat{S}_{\text{EFT}} (r_c, J(r_c))+ \frac{d-\Delta}{r_c}  \int d^dx \sqrt{\gamma} \mathcal{O}  J\,.
    \end{aligned}
\end{equation}
We can simplify more because we know that $\frac{d}{dr_c} \hat{S}_{\text{EFT}} = -H$. Thus
\begin{equation}
\label{eq:master1}
\frac{\partial \hat{S}_{\text{EFT}} (r_c, J)}{\partial r_c} = - H [r_c^{\Delta - d} J, - r_c^{d-\Delta} \sqrt{\gamma} \mathcal{O}] + \frac{d-\Delta}{r_c} \int d^dx \sqrt{\gamma} J \mathcal{O}\,.
\end{equation}

\chapter{AdS$_3$ Gravity}
\label{app:ads3}

\section{Gravitational trace flow equation}
\label{eq:flowAdS3dS3}
We write the three-dimensional line element in a coordinate system such that
\begin{equation}
\label{eq:ooijre222}
    ds^2 = d\rho^2 + h_{ij}(x, \rho) dx^i dx^j
\end{equation}
so that the extrinsic curvature is 
\begin{equation}
    K_{ij} = \frac{1}{2} \partial_\rho h_{ij}\,.
\end{equation}
The action for pure AdS$_3$ gravity after integration by parts is
\begin{equation}
I=-\frac{1}{16 \pi G} \int_{M_3} d^3 x \sqrt{g}\left(R+2\right)-\frac{1}{8 \pi G} \int_{\partial M_3} d^2 x \sqrt{h}\left(K-1\right)\,,
 \end{equation}
 where we choose our boundary action to have a well-defined variational principle to have Dirichlet boundary conditions on the metric.

Hence, the vacuum Einstein equations 
\begin{equation}
    R_{\mu \nu} -\frac{1}{2} R g_{\mu \nu} - g_{\mu \nu} = 0
\end{equation}
in the coordinate system \eqref{eq:ooijre222} become
\begin{equation}
\begin{aligned}
\label{eq:EinsteinIJRHORHORHO}
& E_j^i=-\partial_\rho\left(K_j^i-\delta_j^i K\right)-K K_j^i+\frac{1}{2} \delta_j^i\left[K^{m n} K_{m n}+K^2\right]-\delta_j^i=0\,, \\
& E_j^\rho=\nabla^i\left(K_{i j}-K h_{i j}\right)=0\,, \\
& E_\rho^\rho=-\frac{1}{2} R^{(2)}+\frac{1}{2}\left[K^2-K^{i j} K_{i j}\right]-1=0\,.
\end{aligned}
\end{equation}
The on-shell variation of the gravitational action with respect to the boundary metric is the stress tensor
\begin{equation}
\begin{aligned}
\label{eq:AdS3StressTensor11111}
    T_{ij} &= \frac{4\pi}{\sqrt{h}} \frac{\delta I_{\text{on-shell}}}{\delta h^{ij}} 
    \\&= \frac{1}{4G} \left( K_{ij} - K h_{ij} +  h_{ij} \right)\,.
\end{aligned}
\end{equation}
The trace is
\begin{equation}
    \begin{aligned}
    T^i_i &= h^{ij} T_{ij} 
    \\&= \frac{1}{4G} \left( h^{ij} K_{ij} - h^{ij} h_{ij} K + h_{ij} h^{ij} \right)
    \\&= \frac{1}{4G} \left( K - 2 K + 2 \right)
    \\&= \frac{1}{4G} \left( 2 - K\right)
    \end{aligned}
\end{equation}
and the $T\overline{T}$ operator is
\begin{equation}
    T\overline{T} = \frac{1}{8} \left( T^{ij} T_{ij} - (T^i_i)^2 \right)\,,
\end{equation}
where
\begin{equation}
    \begin{aligned}
    \label{eq:TTAppendixAAAf}
    T^{ij} T_{ij} &= \frac{1}{16 G^2} \left( K^{ij} - K h^{ij} + h^{ij} \right) \left( K_{ij} - K h_{ij} + h_{ij} \right) \\&=\frac{1}{16 G^2}\bigg(K^{i j} K_{i j}-K K^{i j} h_{i j}+K^{i j} g_{i j}-K h^{i j} K_{i j}\\&+K^2 h^{i j} g_{i j}-K h^{i j} h_{i j}+h^{i j} K_{i j}-K h^{i j} h_{i j}+h^{i j} h_{i j}\bigg) \\
& =\frac{1}{16 G^2}\left(K^{i j} K_{i j}-K^2+K-K^2+2 K^2-2 K+K-2 K+2\right) \\
& =\frac{1}{16 G^2}\left(K^{i j} K_{i j}-2 K+2\right)\,.
    \end{aligned}
\end{equation}
From the radial Einstein equation $E^\rho_\rho = 0$ in \eqref{eq:EinsteinIJRHORHORHO}, we have
\begin{equation}
   E^\rho_\rho: \quad -\frac{1}{2} R^{(2)}+\frac{1}{2}\left[K^2-K^{i j} K_{i j}\right]-1=0 \implies  K^{ij} K_{ij} = K^2 - R^{(2)} - 2 
\end{equation}
so \eqref{eq:TTAppendixAAAf} becomes
\begin{equation}
\begin{aligned}
    T^{ij} T_{ij} &= \frac{1}{16 G^2} \left( K^2 - R^{(2)} - 2  - 2K +2  \right) \\&=\frac{1}{16 G^2} \left( K^2 - R^{(2)}   - 2K  \right)\,.
\end{aligned}
\end{equation}
Finally, the trace squared is
\begin{equation}
    (T^i_i)^2 = \frac{1}{16 G^2} \left(K^2 -4K +4 \right)\,.
\end{equation}
The AdS$_3$ $T\overline{T}$ operator is defined as
\begin{equation}
    \begin{aligned}
    T\overline{T} &= \frac{1}{8} \left( \left[\frac{1}{16 G^2} \left( K^2 - R^{(2)}   - 2K   \right) \right]- \left[ \frac{1}{16 G^2} \left(K^2 -4K +4 \right) \right] \right) 
    \\&= -\frac{R^{(2)}}{128 G^2} + \frac{K - 2}{64 G^2} \\&= - \frac{R^{(2)}}{128G^2} - \frac{1}{16 G} T^i_i\,.
    \end{aligned}
\end{equation}
Thus, the AdS$_3$ trace flow equation is
\begin{equation}
\begin{aligned}
\label{eq:flowAdS3111}
    T^i_i &= - 16 G T\overline{T} - \frac{1}{8G} R^{(2)} \\&=-4\pi \lambda T\overline{T} - \frac{c}{12} R^{(2)}\,.
    \end{aligned}
\end{equation}

For a flat boundary metric $R^{(2)} = 0$
\begin{equation}
    T^i_i = - 16 G T\overline{T}
\end{equation}
and comparing to Zamolodchikov's trace flow equation for deforming a CFT, $T^i_i = - 4\pi \lambda T \bar{T}$, we find 
\begin{equation}
    \lambda = \frac{4G}{\pi}
\end{equation}
in this set of conventions.

As a quick check, if our metric is conformal $h_{ij} = e^{2\rho} h_{ij}$ then
\begin{equation}
    K_{ij} = g_{ij}, \quad K^{ij} K_{ij} = h^{ij} h_{ij} =2, \quad K =h^{ij} K_{ij} = 2, \quad R^{(2)} = 0 
\end{equation}
implying that $T^i_i= -16 G T\overline{T} =0$, which is expected that the trace of the stress tensor vanishes.

\subsection{dS$_3$ trace flow}
The action for pure dS$_3$ gravity is
\begin{equation}
I=-\frac{1}{16 \pi G} \int_M d^3 x \sqrt{g}\left(R-2\right)-\frac{1}{8 \pi G} \int_{\partial M} d^2 x \sqrt{h}\left(K-1\right)
 \end{equation}
and the stress tensor is the same as AdS$_3$ \eqref{eq:AdS3StressTensor11111}. The only part of this analysis that changes is the vacuum Einstein equations
\begin{equation}
    R_{\mu \nu} -\frac{1}{2} R g_{\mu \nu} + g_{\mu \nu} = 0
\end{equation}
in the coordinates \eqref{eq:ooijre222}
\begin{equation}
\begin{aligned}
& E_j^i=-\partial_\rho\left(K_j^i-\delta_j^i K\right)-K K_j^i+\frac{1}{2} \delta_j^i\left[K^{m n} K_{m n}+K^2\right]+\delta_j^i=0\,, \\
& E_j^\rho=\nabla^i\left(K_{i j}-K h_{i j}\right)=0\,, \\
& E_\rho^\rho=-\frac{1}{2} R^{(2)}+\frac{1}{2}\left[K^2-K^{i j} K_{i j}\right]+1=0\,.
\end{aligned}
\end{equation}
As we saw from AdS$_3$, we will need the $E^\rho_\rho$ equation
\begin{equation}
   E_\rho^\rho: \quad -\frac{1}{2} R^{(2)}+\frac{1}{2}\left[K^2-K^{i j} K_{i j}\right]+1=0 \implies  K^{ij} K_{ij} = K^2 - R^{(2)} + 2\,. 
\end{equation}
Therefore
\begin{equation}
\begin{aligned}
    T^{ij} T_{ij} &= \frac{1}{16 G^2} \left( K^2 - R^{(2)} + 2  - 2K +2  \right) \\&= \frac{1}{16 G^2} \left( K^2 - R^{(2)} - 2K +4  \right)\,.
\end{aligned}
\end{equation}
Therefore, the dS$_3$ $T\overline{T}$ operator is defined as
\begin{equation}
    \begin{aligned}
    T\overline{T} &= \frac{1}{8} \left( \left[\frac{1}{16 G^2} \left( K^2 - R^{(2)} - 2K +4    \right) \right]- \left[ \frac{1}{16 G^2} \left(K^2 -4K +4 \right) \right] \right) 
    \\&= -\frac{R^{(2)}}{128 G^2} + \frac{K}{64 G^2}
    \\&=-\frac{R^{(2)}}{128 G^2} + \frac{1-2 GT^i_i}{32 G^2}
    \\&= -\frac{R^{(2)}}{128 G^2} -  \frac{1}{16 G} T^i_i  + \frac{1}{32 G^2}\,.
    \end{aligned}
\end{equation}
Therefore, the dS$_3$ trace flow equation is
\begin{equation}
\begin{aligned}
T^i_i &= - 16 G T\overline{T} - \frac{1}{8G} R^{(2)} + \frac{1}{2G} \\&= -4\pi \lambda T\overline{T} - \frac{c}{12} R^{(2)} + \frac{2}{\pi \lambda}\,,
\end{aligned}
\end{equation}
which differs from the AdS$_3$ flow equation \eqref{eq:flowAdS3111} by an additive constant $ \frac{2}{\pi \lambda}$.

\section{ The Chern-Simons action with a cutoff spherical boundary and the Weyl anomaly}\label{sphere}

The maximally symmetric solution to the $3d$ Einstein equations with a negative cosmological constant in the Euclidean signature has the topology of a solid sphere. Its metric can be written as
\begin{equation}
\label{eq:dSsphere}
    ds^2 = d\eta^2 + \sinh^2 \eta \, d\Omega_2^2\,,
\end{equation}
where $\eta  \geq 0$ and $d\Omega_2^2 = d\theta^2 + \sin^2 \theta \, d\varphi^2$ is the metric of the $S^2$ conformal boundary.
In this section, we calculate the classical action of this geometry using both the metric and Chern-Simons language and show how the Weyl anomaly emerges. Our analysis differs from that of \cite{Cotler:2018zff}, where the Weyl anomaly appeared as a logarithmic divergence of the boundary action near the poles.
Instead, we work with two coordinate patches and see the Weyl anomaly that appears from the nontrivial relation between the gauge connections on each patch.

\subsection{Metric calculation}

The on-shell value of the Einstein-Hilbert action \eqref{eq:ereerer435456} can be calculated using $R = 6\Lambda = -6$
\begin{align}
\label{eq:IMetric}
    I_{\text{EH}} &= -\frac1{8G} \int_{0}^{\eta_c} d\eta \, (-4 \sinh^2 \eta)
    = -\frac1{4G} (2\eta_c - \sinh 2\eta_c)\,.
\end{align}
To calculate the boundary action in \eqref{dd}, we first relate $\eta$ to the Fefferman--Graham coordinate $r$ defined in \eqref{db} by $2\eta = -\ln r$. Observing that $K = \tfrac12 g^{ij} \partial_\eta g_{ij}$, and $\sqrt{\det g_{ij}} R(g_{ij}) = 2 \sin \theta$, we obtain
\begin{align}
\label{eq:IMetricBdy}
    I_{\text{bndy}} &= - \frac1{2G} \left[ \sinh^2 \eta_c (2 \coth \eta_c - 1) \right] + \frac{\eta_c}{2 G}\,.
\end{align}
As expected, both the exponential and the linear divergences in $\eta_c$ cancel with $I_\text{EH}$ 
\begin{align}
    I = I_{\text{EH}} + I_{\text{bndy}}
    = -\frac1{4G} (1 - e^{-2 \eta_c})\,.
\end{align}
The term linear in $\eta_c$ is logarithmic in $r_c$ and cannot be canceled by adding to the action a covariant local boundary term.
Instead, we use the second term in \eqref{dd}, which is proportional to $\eta_c$ times the Ricci curvature of the boundary.
This term is not covariant since it depends explicitly on the coordinate value $\eta_c$. Indeed, this term signals the presence of a Weyl anomaly in the CFT and manifests itself on the gravity side as the absence of diffeomorphism invariance.

\subsection{Chern-Simons calculation}

In the previous section, it was not necessary to choose explicit coordinates on the boundary two-sphere to do this calculation. Indeed, the action only depended on its overall area. The fact that $S^2$ cannot be covered in a single coordinate patch did not pose any problems. We need to face this issue to do the analogous Chern-Simons calculation.

\subsubsection{Stereographic projection}
It is possible to cover the whole sphere except for one point using stereographic projection. We define the complex coordinate
\begin{equation}
    z_{\text{S}} = \cot \left( \frac{\theta}{2} \right) e^{i\varphi}\,,
\end{equation}
which is regular everywhere but the north pole at $\theta = 0$ and in terms of which two-sphere metric is
\begin{equation}
    d \Omega_2^2 = \frac{4 d z_{\text{S}} \, d \overline z_{\text{S}}}{(1 + z_{\text{S}} \overline z_{\text{S}})^2}\,.
\end{equation}
Similarly, we can cover all but the south pole using
\begin{equation}
	z_{\text{N}} = z_{\text{S}}^{-1} = \tan \left( \frac{\theta}{2} \right) e^{-i \varphi}\,,
\end{equation}
which gives the same metric as before and is related to $z_{\text{S}}$ by a rotation of the sphere that maps the north to the south pole: $(\theta, \varphi) \to (\pi - \theta, -\varphi)$.
We can choose a local Lorentz frame for which the associated zweibein and spin connection, which has only a single component in two dimensions, are
\footnote{
In Euclidean signature, the flatness condition \eqref{aamda} contains additional minus signs,
\begin{align}
    d e^+ + \omega \wedge e^+ = d e^- - \omega \wedge e^- = 0
    \,.
\end{align}
This can be traced back to the minus sign in the Lorentzian identity $\epsilon_{\mu\nu\rho} \epsilon^{\mu\sigma\tau} = - \delta_{\nu\rho}^{\sigma\tau}$, whereas that minus sign is absent in Euclidean signature.
}
\begin{align}
	e_{\text{S,N}}^+ &= -\frac{i \, d z}{1 + z \overline z}
	\,, &
	e_{\text{S,N}}^- &=- \frac{i \, d \overline z}{1 + z \overline z}
	\,, &
	\omega_{\text{S,N}} &= -\frac{z \, d \overline z - \overline z \, d z}{1 + z \overline z}
	\,,
\end{align}
where $z$ is either $z_{\text{S}}$ or $z_{\text{N}}$.
In terms of the original variables, this gives
\begin{align}
	e_{\text{S}}^+ &= \frac12 e^{i\varphi} (i d \theta + \sin \theta \, d \varphi)
	\,, &
	e_{\text{N}}^+ &= -\frac12 e^{-i\varphi} (i d \theta + \sin \theta \, d \varphi)
	\,, \nonumber \\
	e_{\text{S}}^- &= \frac12 e^{-i\varphi} (i d \theta - \sin \theta \, d \varphi)
	\,, &
	e_{\text{N}}^- &= -\frac12 e^{i\varphi} (i d \theta - \sin \theta \, d \varphi)
	\,, \nonumber \\
	\omega_{\text{S}} &= 2i \cos^2 \left(\tfrac\theta2 \right) d \varphi
	\,, &
	\omega_{\text{N}} &= -2i \sin^2 \left (\tfrac\theta2 \right) d \varphi
	\,.
\end{align}

Using these coordinates, the 3-dimensional Chern-Simons gauge connections on each patch are
\begin{align}
\label{eq:ANS}
	A &= (\omega + d \eta) L_0 + e^\eta e^+ L_1 - e^{-\eta} e^- L_{-1}
	\,, \nonumber \\
	\overline A &= (\omega - d \eta) L_0 + e^{-\eta} e^+ L_1 - e^\eta e^- L_{-1}\,.
\end{align}
\subsubsection{Action}

We are now ready to calculate the on-shell action in the Chern-Simons language. The total action consists of two terms, the Einstein-Hilbert bulk action \eqref{eq:ereerer435456} and the boundary contribution \eqref{dd}.

\begin{figure}[htbp]
	\centering
	\includegraphics{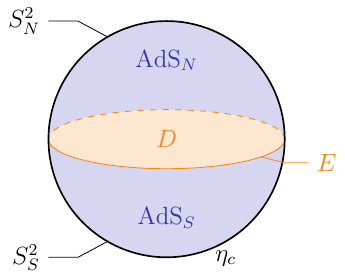}
	\caption{Global Euclidean AdS$_3$ and its different components and boundaries. The bulk of AdS$_3$ is composed of two patches, AdS$_{\text{S}}$ and AdS$_{\text{N}}$, whose boundaries at $\eta_c$ are the southern and a northern hemisphere $S^2_{\text{S}}$ and $S^2_{\text{N}}$, respectively. The bulk components are separated by an equatorial disk $D$, while the boundary hemispheres touch the equator $E$.
	}
\label{fig:sphere}
\end{figure}

Starting with the Einstein-Hilbert action, we can rewrite it in terms of Chern-Simons gauge connections as follows
\footnote{%
There is an additional factor of $i$ in the relation between these actions because we now work in Euclidean signature. 
We will not change the gauge group with respect to the main text but include factors of $i$ in the gauge connections.
}
\begin{align}
\label{eq:EH2CS}
	I_{\text{EH}}
	&= -\frac1{16\pi G} \int_{M_3} d^3 x \sqrt{g} (R - 2\Lambda)
	\\
	&= -\frac{ik}{4\pi} \int_{M_3}  \operatorname{Tr}(A \wedge d A + \tfrac23 A^3) + \frac{ik}{4\pi} \int_{M_3}  \operatorname{Tr}(\overline A \wedge d \overline A + \tfrac23 \overline A^3) + \frac{ik}{4\pi} \int_{M_3} d(  \operatorname{Tr} A \wedge \overline A)
	\ , \nonumber
\end{align}
where $k = 1 / 4 G$.
We will split up the integral over the manifold $M_3$ into a contribution from AdS$_{\text{N}}$ and AdS$_{\text{S}}$ as depicted in figure \ref{fig:sphere}.
To evaluate the first term, we can use the explicit form of the gauge potentials \eqref{eq:ANS}
\begin{align}
	I_{\text{CS}}[A]
	&= -\frac{ik}{8\pi} \int_0^{\eta_c} d\eta \left( \int_{\text{AdS}_{\text{S}}} d\omega_{\text{S}} + \int_{\text{AdS}_\text{N}} d\omega_{\text{N}} \right)
	= -\frac{k}2 \eta_c
	\,.
\end{align}
One can check that the result does not depend on the location of the disk $D$, which separates the two patches, as long as it does not cross either of the poles.
The second term in \eqref{eq:EH2CS} yields the same contribution, $-I_{\text{CS}}[\overline A] = - \frac{k }{2}\eta_c$.
The total derivative in the third term of \eqref{eq:EH2CS} will contribute not only on the cutoff boundary $S^2 = S^2_{\text{S}} \cup S^2_{\text{N}}$ at $\eta_c$ but also on the internal boundary $D$ that separates northern from the southern hemisphere,
\begin{align}
\label{eq:AbA}
	\frac{ik}{4\pi} \int_{M_3} d(  \operatorname{Tr} A \wedge \overline A)
	&= \frac{ik}{4\pi} \int_{S^2}  \operatorname{Tr} A \wedge \overline A - \frac{ik}{4\pi} \int_{D}  \operatorname{Tr}( A_N \wedge \overline A_N - A_S \wedge \overline A_S )
	\,.
\end{align}
The signs are fixed by comparing the volume form in the bulk, which we took $\propto d\eta \wedge d\theta \wedge d\varphi$, with the one on $S^2$ that we choose $\propto d\theta \wedge d\varphi$ and on the disk $D$ which we fix to be $\propto d\eta \wedge d\varphi$.
There is an additional sign for $A_{\operatorname{S}} \wedge \overline A_{\operatorname{S}}$ coming from the outward-pointing normal $n_i d x^i = - d\theta$.
Calculate from \eqref{eq:ANS} that $ \operatorname{Tr} A \wedge \overline A = d\eta \wedge \omega + 2 \sinh(2\eta) e^+ \wedge e^-$ and pulling this back to each of the boundaries, we find
\begin{align}
	\frac{ik}{4\pi} \int_{M_3} d(  \operatorname{Tr} A \wedge \overline A)
	&= -\frac{ik}{2\pi} \sinh(2\eta_c) \int_{S^2} \frac{i}2 \sin \theta \, d\theta \wedge d\phi + \frac{ik}{4\pi} \int_D 2 i d\eta \wedge d\varphi
	\nonumber \\
	&= k \sinh(2\eta_c) - k \, \eta_c
	\,.
\end{align}
Altogether, we find
\begin{align}
	I_{\text{EH}} = -k (2\eta_c - \sinh(2\eta_c))
	\ ,
\end{align}
which agrees with the metric calculation \eqref{eq:IMetric}.

What remains to be calculated is the boundary action $I_{\text{bndy}}$ \eqref{dd} on the $S^2$ boundary of $M$.
We explain in appendix \ref{sec:GHY_CS} how to express the extrinsic curvature in terms of the Chern-Simons gauge connections. In Euclidean signature, the result reads
\begin{align}
	I_{\text{bndy}}
	&= -\frac{k}{2\pi} \int_{S^2} d^2 x \sqrt{\det g_{ij}} (\partial_a n^a - 1) - \frac{ik}{2\pi} \int_{S^2}  \operatorname{Tr}(A \wedge \overline A)
	\,.
\end{align}
In the case of interest, $\partial_a n^a$ vanishes, and we choose a gauge for which the $L_0$ component of the gauge connections is normal to the boundary, and the other components are parallel.
The action then simplified to
\begin{align}
\label{eq:IbndyAbA}
	I_{\text{bndy}} &= -\frac{ik}{4\pi} \int_{S^2}  \operatorname{Tr} \left[ 2A \wedge \overline A - L_0 (A - \overline A) \wedge (A - \overline A) \right]
	\ .
\end{align}
The first term was already calculated in \eqref{eq:AbA}. Indeed, it adds up with \eqref{eq:IbndyAbA} to give the boundary contribution given in \eqref{eq:ConsistentAction}.
The second one only depends on the boundary frame field,
\begin{align}
I_{\text{bndy}}
	&= -2k [\sinh(2\eta_c) - \sinh^2(\eta_c)]\,,
\end{align}
which agrees with \eqref{eq:IMetricBdy}.

\subsubsection{General one-loop comments}
\label{sec:sphere1-loop}
In this thesis, we considered planar boundaries at a finite cutoff so it is natural for one to consider curved boundaries such as a sphere, however, this is complicated as one is faced with solving non-asymptotic curved boundary and patching conditions. We outline the steps one would follow. We have 6 Gauss parameters $(\lambda, \Psi, F, \bar{\lambda}, \bar{\Psi}, \bar{F})$ for each patch and boundary. There are 4 boundary conditions leaving one with $6-4=2$ free functions $(F, \bar{F})$. One would then want to find an action $I=I_{\text{Sphere}}+I_{\text{Disk}}$ in terms of these free functions and then study loop corrections. For one-loop, we can find linear fluctuations from the spherical boundary and patching conditions by expanding these free functions. 

\section{Chern-Simons action as a boundary term}\label{app:ActionAsBoundaryTerm}

To reduce the action to a boundary term, we start by implementing the space-time split \eqref{eq:CSActionSplit} and making use of the constraints $\tilde{\mathcal{F}} = \tilde{\overline{\mathcal{F}}} = 0$. Here
\begin{equation}\label{eq:BulkTerm0}
\begin{split}
S[A,\bar{A}] =&{k\over 4\pi} \int_{M_3} \operatorname{Tr}\left[
 \tilde{A}\wedge dt\wedge \partial_t \tilde{A}-  \tilde{\bar{A}}\wedge dt\wedge \partial_t \tilde{\bar{A}}
    \right] \\
    &+{k\over 4\pi} \int_{M_3}  \operatorname{Tr}\left[
 - \tilde{d}\left(    \tilde{A}\wedge A_t dt   \right)+ \tilde{d}\left(    \tilde{\bar{A}}\wedge \bar{A}_t dt   \right)
    \right]
    + I_{\text{bndy}}\,.
    \end{split}
\end{equation}
The second line of \eqref{eq:BulkTerm0} involves boundary terms that can easily be evaluated in terms of $(g, \overline{g})$. The first line is a bulk term which, when evaluated on the flat connections \eqref{eq:ConstrainedConnections}, reads
\begin{equation}\label{eq:BulkTerm1}
{k\over 4\pi} \int_{M_3}  \operatorname{Tr}\left[
 \tilde{A}\wedge dt\wedge \partial_t \tilde{A}
    \right]
    =
    {k\over 4\pi} \int_{M_3}  \operatorname{Tr}\left[
-{1\over 3} \left( g^{-1}dg\right)^3 - \tilde{d}\left(  g^{-1}\partial_t g\wedge g^{-1}\tilde{d}g   \right)
    \right] \,,
\end{equation}
and similarly for the barred connections. The second term on the right-hand side of \eqref{eq:BulkTerm1} is already a boundary term. The first term is a Wess-Zumino term, which becomes a boundary term once an explicit parametrization for the group element $g$ is chosen. For the Gauss parametrization \eqref{zm}, one finds
\begin{equation}\label{eq:BulkTerm2}
    {k\over 4\pi} \int_{M_3}  \operatorname{Tr}\left[
-{1\over 3} \left( g^{-1}dg\right)^3
    \right]
     =  - {k\over 4\pi} \int_{\partial M_3} \lambda^2 \wedge d\Psi\wedge dF
    \,.
\end{equation}
Combining equations \eqref{eq:BulkTerm2}, \eqref{eq:BulkTerm1}, and \eqref{eq:BulkTerm0} yields an expression for the full action written as a boundary term. Its expression as a functional of the Gauss parameters has been written in the main text in equation \eqref{zs}.
\section{Relation between Chern-Simons theory at finite cutoff and coupling to topological gravity}\label{app:AdSasTopo}

The objective of this appendix is to connect the ideas of AdS$_3$ gravity with a finite cutoff and the $T\overline{T}$ deformation of a conformal field theory as described by coupling to topological gravity;
see \cite{Dubovsky:2017cnj,Dubovsky:2018bmo,Tolley:2019nmm,Caputa:2020lpa} for relevant background.  In this appendix, we follow the conventions in \cite{Caputa:2020lpa}, with $g_{ij} = \delta_{ab} e^a_i e^b_j$. 

The topological gravity formulation is based on the observation that the  $T\overline{T}$ flow equation for the deformed action 
\begin{equation}
{d I_{\lambda_{T\overline{T}} }\over d\lambda_{T\overline{T}}} = -{1\over 4} \int d^2 x\, \sqrt{g}  \text{det}T^{i}_{j}
\end{equation}
can be solved by defining an action with auxiliary fields that will be integrated out. In particular, we define 
\begin{equation}\label{eq:DeformedActionTopological}
I_{\lambda_{T\overline{T}}} = I_{\text{grav}}[\tilde{e^a}_i,e^a_i] + I_0[e^a_i ,\psi]\,,
\end{equation}
where $\psi$ and $\tilde{e}^a_i$ are the original matter fields and veilbein in the undeformed theory, and $e^a_i$  is the vielbein of the deformed theory. The action $I_0$ is the undeformed action, while the topological gravity action reads
\begin{equation}
I_{\text{grav}} = {1\over 4\pi^2 \lambda_{T\overline{T}}} \int_{\partial M_3} d^2 x\, \epsilon^{ij}\epsilon_{ab} (e - \tilde{e})_{i}^a(e - \tilde{e})_{j}^b\,.
\end{equation}
In this appendix we take $|\epsilon^{ij}|=|\epsilon_{ab}|= 1$ and also write $e=\det e^a_i$. 
To obtain the action for the fields $\psi$ coupled to the background vielbein  $e$, the prescription is to path integrate over  $\tilde{e}^a_i$, which in the classical limit can be performed by extremizing \eqref{eq:DeformedActionTopological} with respect to  $\tilde{e}$.

An important ingredient is the stress tensor of the deformed theory, 
\begin{equation}\label{eq:StressTensorDefinitionTopologicalGravity}
T^{i}_a = {2\pi \over e} {\delta I_{\lambda_{T\overline{T}}}\over \delta e^a_{i}} = -{1\over \pi \lambda_{T\overline{T}}}  \epsilon^{ij}\epsilon_{ab} (\tilde{e}^b_{j}-e^b_{j}) \,.
\end{equation}
As we will see momentarily, this formula will be recovered from the boundary conditions imposed on Chern-Simons connections at a finite radial cutoff. 

In section \ref{sec:CSBoundaryAction}, we reduced the Chern-Simons action to a boundary action depending on Gauss parameters $(\lambda, \bar{\lambda}, \Psi, \bar{\Psi}, F, \bar{F})$. Boundary conditions \eqref{zo} and \eqref{zq} can be thought of as providing solutions for $(\lambda,\bar{\lambda},\Psi,\bar{\Psi})$ in terms of $F$ and $\bar{F}$. The resulting action is then a functional of $F$ and $\bar{F}$. However, in practice, this procedure cannot be carried out analytically. Having noted this limitation, the boundary conditions equations have a beautiful interpretation: they coincide with  \eqref{eq:StressTensorDefinitionTopologicalGravity}.   

To see this, we compute the boundary stress tensor
\begin{equation}
T^{i}_a \equiv {2\pi \over e }{\delta I\over \delta e^a_{i}}\, 
\end{equation}
in terms of the Gauss parameters. The time components read
\begin{equation}
T^{t}_{+} = - {k \over  r_c e} \left(      \Psi' + \lambda^2 \Psi^2 F' +2 \Psi {\lambda'\over \lambda}       \right)
\,, \quad
T^t_-  = {k \over r_c  e} \left(  \bar{\Psi}' +\bar{\lambda}^2\bar{\Psi}^2\bar{F}' +2 \bar{\Psi} {\bar{\lambda}'\over \bar{\lambda}}   \right)
\,.
\end{equation}
Using these formulas, the differential equations imposed  by the boundary conditions \eqref{zo} can be written as
\begin{equation}\label{eq:SpaceTdef}\begin{aligned}
e^{+}_{x} - \lambda^2 F'  + { r_c\over k} \epsilon_{xi}\epsilon^{+a}T^{i}_a =& 0
\,, \\
e^{-}_{x} - \bar{\lambda}^2 \bar{F}'  + { r_c\over k } \epsilon_{xi}\epsilon^{-a}T^{i}_a =& 0
\,.
\end{aligned}
\end{equation}
These are precisely the spatial components of equation \eqref{eq:StressTensorDefinitionTopologicalGravity} upon making the identification
\begin{equation}\label{eq:FromLambdaToEx}
\lambda^2  = {\tilde{e}_x^+ \over F'}
\,, \quad
\bar{\lambda}^2 =   {\tilde{e}_x^- \over \bar{F}'} \,,
\end{equation}
and
\begin{equation}
{r_c\over k} = \pi \lambda_{T\overline{T}}\,, \quad \text{or} \quad r_c = {\pi \over 6}\lambda_{T\overline{T}} c\,.
\end{equation}
Formulas \eqref{eq:SpaceTdef} only capture the definition of the time components of the deformed stress tensor. The space components in terms of the connections \eqref{aama} read explicitly
\begin{equation}\begin{split}
T^x_- =& -{1\over \tilde{e}} {k\over r_c }  \bar{A}^+_t \vert_{r_c} =  -{1\over \tilde{e}} {k\over r_c }  f^+_t
\,, \\
T^x_+ =& {1\over \tilde{e}} {k\over r_c } A^-_t \vert_{r_c} = {1\over \tilde{e}} {k\over r_c }  f^-_t \,.
\end{split}
\end{equation}
The time components of the connections must obey the boundary condition \eqref{aamc}, which we repeat here
\begin{equation}
(E-f)^{\pm}_{t} \vert_{r_c} = \pm e^{\pm}_{t}\,.
\end{equation}
We can think of this boundary condition as fixing $E^{\pm}$ in terms of the fixed boundary vielbein $e$ and the one-forms $f^{\pm}$, which remain unfixed. Even though the time components of $f^{\pm}$ remain unfixed, they do not appear explicitly in the reduced action, given that the time components of the connection are Lagrange multipliers. However, a physical meaning can be attributed to $f_t^{\pm}$. For this, we introduce the time components of an undeformed vielbein $\tilde{e}^{\pm}_{t}$ and relabel as follows
\begin{equation}\label{eq:FromfoEt}
f^{\pm}_t =\tilde{e}^{\pm}_{t} - e^{\pm}_{t}\,.
\end{equation}
The holographic stress tensor formula can then be recast in terms of $\tilde{e}^{\pm}_{t}$ instead of $f_t^{\pm}$. The result is
\begin{equation}\label{eq:TimeTdef}\begin{split}
e^{+}_{t} - \tilde{e}^+_t  + {r_c\over k} \epsilon_{ti}\epsilon^{+a}\tilde{T}^{i}_a =& 0
\,, \\
e^{-}_{t} -  \tilde{e}^-_t +{ r_c\over k } \epsilon_{ti}\epsilon^{-a}\tilde{T}^{i}_a =& 0\,.
\end{split}
\end{equation}
These are precisely the time components of the definition of the deformed stress tensor in a $T\overline{T}$ deformed theory. In summary, we conclude that the Dirichlet boundary conditions imposed at $r=r_c$ together with the definition of the holographic stress tensor have a nice interpretation in the context of a theory deformed by a coupling to topological gravity. This is achieved by identifying our Gauss parameters $\lambda$ and $\bar{\lambda}$ with the space components of an undeformed zweibein $\tilde{e}^{\pm}$ as written in \eqref{eq:FromLambdaToEx}, as well as identifying the time components of  $f^{\pm}$ with the time components of such zweibein as written in \eqref{eq:FromfoEt}. The fixed zweibein $e$ at $r=r_c$ plays the role of the deformed zweibein.

 The connection between Chern-Simons theory with finite cutoff and $T\overline{T}$-deformed CFT is even more apparent at the level of the action. Evaluating the action in terms of $F$, $\bar{F}$, and the zweibein $\tilde{e}^{\pm}$ introduced here as a relabeling of $\lambda$, $\bar{\lambda}$, and $f_t^{\pm}$, we find
\begin{equation}\label{eq:ActionResultCutOffCurved}
I[F,\bar{F},\tilde{e}^{\pm}; e] = I_0[F,\bar{F}; \tilde{e}^{\pm}] + I_{\text{grav}}  + I_{\text{extra}} \,.
\end{equation}
The first term is the Wick rotated action \eqref{zs} we found in the main text when studying the reduced action of AdS$_3$ gravity with a curved background at $r=0$. The second term is a coupling between $e$ and $\tilde{e}$. Explicitly,
\begin{equation}
 I_{\text{grav}}  = {1\over 16\pi G r_c} \int_{\partial M_3} d^2x \, \epsilon^{ij}\epsilon_{ab}\left(e- \tilde{e}  \right)_{i}^a \left(e- \tilde{e}  \right)_{j}^b \,.
\end{equation}
This matches the topological coupling \eqref{eq:DeformedActionTopological} introduced above as a mechanism to deform the original theory by the $T\overline{T}$ operator. The last term in  \eqref{eq:ActionResultCutOffCurved} reads
\begin{equation}
\label{qqq}
I_{\text{extra}} = {1\over 16\pi G} \int_{\partial M_3} d^2 x \, \text{det}(e) \, {\left(  \tilde{\omega}_x(e) - \omega_x(\tilde{e})   \right)^2\over \tilde{e}^+_{x}\tilde{e}^-_x} \,.
\end{equation}

We now show that this term vanishes on-shell, which does not affect the value of the deformed stress tensor. 
We do so by computing the flatness equations of the Chern-Simons theory at the cutoff boundary $r=r_c$.
We relabel the parameters $\lambda$, $\bar{\lambda}$ by introducing the space components of an undeformed zweibein $\tilde{e}_x$, as explained in formulas \eqref{eq:FromLambdaToEx}. We also relabel $f^{\pm}_t$ in terms of $\tilde{e}_t$ as written in \eqref{eq:FromfoEt}. We therefore expect the on-shell conditions at $r=r_c$ to involve the zweibeins $\tilde{e}$ and $e$, the functions $F$ and $\bar{F}$, and the spin connection at the Dirichlet boundary $\omega$.

Interestingly, when using the boundary conditions \eqref{zo}, the field strength components do not depend on $F$ and $\bar{F}$ explicitly. They involve exclusively $\tilde{e}$, $e$, and  $\omega$. Explicitly, the components of the field strength are evaluated to the following:
\begin{equation}\label{eq:EOMgeometric}
\begin{split}
 \operatorname{Tr}\left[  (\mathcal{F}-\bar{\mathcal{F}})_{xt} L_{\mp 1}  \right]  &= \left( e^{\pm}\wedge \omega \pm de^{\pm} \right)_{tx}   \,, \\
- \operatorname{Tr}\left[ \bar{\mathcal{F}}_{xt} L_{1}  \right]  &= \left( \tilde{e}^{-}\wedge \omega- d\tilde{e}^{-} \right)_{tx}   \,, \\
 \operatorname{Tr}\left[  \mathcal{F}_{xt} L_{-1}  \right]  &= \left( \tilde{e}^{+}\wedge \omega + d\tilde{e}^{+} \right)_{tx}   \,, \\
2r_c  \operatorname{Tr}\left[  \mathcal{F}_{xt} L_{0}  \right]  &= \left(  2 \tilde{e}^+\wedge (e^--\tilde{e}^-)   - d\omega  \right)_{tx}   \,, \\
2r_c  \operatorname{Tr}\left[  \bar{\mathcal{F}}_{xt} L_{0}  \right]  &= \left( 2 \tilde{e}^-\wedge (e^+-\tilde{e}^+)   - d\omega \right)_{tx}   \,. \\
\end{split}
\end{equation}
An important feature of the first three lines in formulas \eqref{eq:EOMgeometric} is that on-shell, the zweibein $\tilde{e}$ obeys the relation
\begin{equation}
\omega_x (e)\equiv {1\over e} \tilde{\epsilon}^{ij} \partial_{i}e_{j,a} e^a_x= {1\over \tilde{e}} \tilde{\epsilon}^{ij} \partial_{i}\tilde{e}_{j,a}
\tilde{e}^a_x\equiv \omega_x (\tilde{e}) \,.
\end{equation}
This implies \eqref{qqq} vanishing.

To summarize, in this appendix, we showed that Chern-Simons theory with curved cutoff boundary can be understood on-shell as coupling the theory at an asymptotic boundary at $r=0$ to topological gravity. While conceptually satisfying, the action \eqref{eq:DeformedActionTopological} is not very practical for direct computation because the boundary conditions \eqref{zo} cannot be solved analytically.

\section{Gibbons-Hawking-York term in Chern-Simons}
\label{sec:GHY_CS}
Here, we will show how the Gibbons-Hawking-York (GHY) term can be written in terms of the Chern-Simons variables $A$ and $\overline A$. As an intermediate step, we will first write this term in terms of the vielbein and spin connection. Though the Chern-Simons description only applies in 3 dimensions, translating from the metric to vielbein description is not simplified in 3 dimensions, so we perform that portion of the calculation in arbitrary dimension.\footnote{Throughout this appendix we work in Euclidean signature, but to obtain the Lorentzian result it is sufficient to negate the overall sign of \eqref{C:a} which propagates to negating the overall sign of the final results, \eqref{C:e}, \eqref{C:i}, and \eqref{C:l}.}

On a $D = d + 1$ dimensional spacetime $M$, the GHY term is given by
\begin{equation}
\label{C:a}
S_{\text{bndy}} = -\frac{1}{16\pi G}\int_{\partial M_{d+1}}2\sqrt{h}d^dx K\,,
\end{equation}
where $h$ is the induced metric on the boundary, and $K$ is the trace of the extrinsic curvature. If $n$ is the outward-pointing normal to $\partial M_{d+1}$ normalized so $n^\mu n_\mu \equiv \sigma = \pm1$, then $K_{\mu\nu} = h^\lambda_{\mu}\nabla_\lambda n_{\nu}$. By writing the projection down to $\partial M_{d+1}$ as $h_\nu^\mu = \delta_\nu^\mu - \sigma n^\mu n_\nu$, we obtain the identity
\begin{equation}
    \label{C:b}
K = g^{\mu\nu}K_{\mu\nu} = \nabla^\mu n_\mu - \sigma n^\mu n^\nu\nabla_\mu n_\nu.
\end{equation}

The final term here is equal to $\frac{1}{2}\sigma n^\nu \nabla_\nu (n^\mu n_\mu)$, which is zero so long as we choose an extension of $n_\mu$ off $\partial M$ which is everywhere normalized. We will assume here that we have chosen such an extension.

Using lower-case Latin letters for flat Lorentz indices, we may write $\nabla_\mu n^\mu = \nabla_a n^a = e^\mu_a \nabla_\mu n^a$ so
\begin{equation}
\begin{aligned}
\label{C:c}
K &= e^\mu_a\left(\delta_b^a\partial_\mu + \omega_{\mu b}^a\right)n^b \\
&= \partial_a n^a + e^\mu_a\omega^a_{\mu b}n^b\,.
\end{aligned}
\end{equation}

The second term admits a nice coordinate-independent representation in terms of the vielbein and spin connection, leading us to
\begin{equation}
\begin{aligned}
    \label{C:e}
    S_{\text{bndy}} &= -\frac{1}{16\pi G}\int_{\partial M_{d+1}}2\sqrt{h}d^d x\left(e^\mu_a \omega^a_{\mu b}n^b + \partial_a n^a\right) \\
&= -\frac{1}{16\pi G}\int_{\partial M_{d+1}}\left[-\frac{1}{(d-1)!} \epsilon_{ab c_2\ldots c_d}\omega^{ab}\wedge e^{c_2}\wedge\cdots \wedge e^{c_d} + 2\sqrt{h}d^d x\partial_a n^a\right]\,.
\end{aligned}
\end{equation}

To show this final equality, it is sufficient to note that $\omega^{ab} = e^\mu_c \omega^{ab}_\mu e^c$ and the identity
\begin{equation}
    \label{C:f}
\int_{\partial M_{d+1}}e^c\wedge e^{c_2} \wedge \cdots\wedge e^{c_d} = \int_{\partial M_{d+1}}n_d\epsilon^{dcc_2\cdots c_d}\sqrt{h}d^d x
\end{equation}

The $\epsilon_{abc_2\ldots c_d}\omega^{ab}\wedge e^{c_2}\wedge\ldots\wedge e^{c_d}$ term in $S_{\text{bndy}}$ has a relatively simple form, but to obtain this in the way we have here is non-trivial. Instead, we could have motivated it by starting from the first-order vielbein formulation of gravity, for example, \cite{Freedman:2012zz}, in which we write the bulk portion of the action as
\begin{equation}
\begin{aligned}
    \label{C:g}
    S_{\text{bulk}} &= -\frac{1}{16\pi G}\int_{M_{d+1}}\frac{\epsilon_{abc_2\ldots c_d}}{(d-1)!}\left(R^{ab} - \frac{2\Lambda}{d(d + 1)}e^a\wedge e^b\right)\wedge e^{c_2}\wedge \cdots \wedge e^{c_d}\,,
\end{aligned}
\end{equation}
where $R^{ab} \equiv d\omega^{ab} + \omega^a{}_c \wedge \omega^{cb}$. This is identically equal to the usual Einstein-Hilbert action. This form makes it clear that the only derivative appears in the curvature, so upon variation, the boundary term is given by
\begin{equation}
    \label{C:h}
    \theta = -\frac{1}{16\pi G}\frac{\epsilon_{abc_2\ldots c_d}}{(d-1)!}\delta \omega^{ab}\wedge e^{c_2}\wedge \cdots \wedge e^{c_d}
\end{equation}

which is compatible with Dirichlet boundary conditions on the spin connection, not the metric or vielbein. To make the variational principle compatible with Dirichlet boundary condition on the vielbein, it would be sufficient to add a term like $\sim \epsilon_{abc_2\ldots c_d}\omega^{ab}\wedge e^{c_2}\wedge\ldots\wedge e^{c_d}$, which is precisely the coordinate-independent term we found in our calculation of the GHY boundary term. The remaining term $\sim \partial_a n^a$ is also compatible with Dirichlet boundary conditions on the vielbein because its variation can be shown to be independent of the normal derivatives of $\delta n_a$, which could, in principle, have state dependence through the flat index.

Specializing now to $D = 3$, the boundary action \eqref{C:e} becomes
\begin{equation}
    \label{C:i}
    S_{\text{bndy}} = -\frac{1}{16\pi G}\int_{\partial M_3}\left[-\epsilon_{ab c}\omega^{ab}\wedge e^{c} + 2\sqrt{h}d^2 x\partial_a n^a\right]
\end{equation}
so upon writing $\omega_a = \frac{1}{2}\epsilon_{abc}\omega^{bc}$ and converting to the Chern-Simons connections
\begin{equation}
    \label{C:j}
    A^a = \omega^a + e^a,\ \ \ \ \overline A^a = \omega^a - e^a\,,
\end{equation}
we find
\begin{equation}
\label{C:k}
-\epsilon_{abc}\omega^{ab}\wedge e^c = 2\operatorname{Tr}(A\wedge\overline A)\,.
\end{equation}

Hence, the GHY term in the Chern-Simons variables may be written
\begin{equation}
\label{C:l}
S_{\text{bndy}} = -\frac{1}{16\pi G}\int_{\partial M_3}\left[2\operatorname{Tr}(A\wedge \overline A) + 2\sqrt{h}d^2x\partial_a n^a\right]\,.
\end{equation}

It should also be noted that when transforming the bulk action into Chern-Simons variables, another factor of $ \operatorname{Tr}(A\wedge \overline A)$ appears from a total derivative in the bulk action. The boundary action presented here is only equal to the GHY part and does not include this additional contribution.

\section{Integrals}

In this appendix, we review how to perform the integrals which appear in our loop computations. Starting in section \ref{Integrals:1-loop}, we review how to perform a slight generalization of the entire class of integrals that appear in our one-loop calculations. In section \ref{Integrals:fourier}, we demonstrate how to perform a class of Fourier transform within dimensionally-regularized integrals. Section \ref{app:2Loopff} displays the details of the two-loop self-energy calculation, since this integral cannot be reduced to the integrals that appear in the one-loop calculations. Finally, in section \ref{Integrals:compare}, we perform an example calculation showing how a perturbative calculation using the propagator \eqref{ih} relates to the calculation from the covariant rule \eqref{ii}.

\subsection{Relevant one-loop integrals}\label{Integrals:1-loop}

Here, we review how to perform some of the integrals that appear in one-loop calculations, which take the generic form
\begin{equation}
\label{D:3}
I_{n,m}(r; \Delta) \equiv \int\frac{d^dk}{(2\pi)^d}\frac{(k_z)^n(k_{\overline z})^m}{[k^2 + \Delta]^r}\,.
\end{equation}

In the process, we will also review how to perform some other standard integrals in dimensional regularization.

We understand the numerator of the integrand in \eqref{D:3} as a particular tensor product of momenta components, much like
\begin{equation}
    \label{D:4}
    \int\frac{d^d k}{(2\pi)^d}\frac{k_\mu k_\nu}{[k^2 + \Delta]^r} \propto \delta_{\mu\nu}
\end{equation}

or its generalization to an arbitrary product of components in the numerator. With this in mind, we will think of the $d$-dimensional domain of integration as containing a 2-dimensional subspace on which we choose the complex coordinates $k_z$ and $k_{\overline z}$. As a result, the $d$-dimensional inner product will be given by $p\cdot q = 2(p_z q_{\overline z} + p_{\overline z}q_z) + p_\perp \cdot q_\perp$ where $p_\perp$ and $q_\perp$ are the components of $p$ and $q$ orthogonal to the two-dimensional subspace we have singled out.

This setup allows us to produce a generating function for the integrals $I_{n,m}$ by first noting the identity
\begin{equation}
\begin{aligned}
    \label{D:5}
   & \frac{(k_z)^n(k_{\overline z})^m}{[k^2 + 2p\cdot k + \Delta]^r} \\&= \frac{\Gamma(1 - r)}{\Gamma(n + m - r + 1)}\left(\frac{1}{4}\frac{\partial}{\partial p_{\overline z}}\right)^n\left(\frac{1}{4}\frac{\partial}{\partial p_z}\right)^m\frac{1}{[k^2 + 2p\cdot k + \Delta]^{r - n - m}}
\end{aligned}
\end{equation}
and writing
\begin{equation}
\begin{aligned}
    \label{D:6}
   & I_{n,m}(r;\Delta) \\&= \frac{\Gamma(1 - r)}{\Gamma(n + m - r + 1)}\left(\frac{1}{4}\frac{\partial}{\partial p_{\overline z}}\right)^n\left(\frac{1}{4}\frac{\partial}{\partial p_z}\right)^m\int\frac{d^d k}{(2\pi)^d}\frac{1}{[k^2 + 2 p \cdot k + \Delta]^{r - n - m}}\Bigg|_{p_z, p_{\overline z}=0} \cr
&= \frac{\Gamma(1 - r)}{\Gamma(n + m - r + 1)}\left(\frac{1}{4}\frac{\partial}{\partial p_{\overline z}}\right)^n\left(\frac{1}{4}\frac{\partial}{\partial p_z}\right)^m I_{0,0}(r - n - m; \Delta - p^2)\Bigg|_{p_z, p_{\overline z}=0}
\end{aligned}
\end{equation}

so we can generate all the $I_{n,m}$ in terms of $I_{0,0}$ and its derivatives.

To perform the integral $I_{0,0}$ we write 
\begin{equation}
\begin{aligned}
    \label{D:1}
    I_{0,0}(r; \Delta) &= \int\frac{d^dk}{(2\pi)^d}\frac{1}{(k^2 + \Delta)^r} \cr
&= \int\frac{d^d k}{(2\pi)^d}\frac{1}{\Gamma(r)}\int_0^\infty dx x^{r-1}e^{-x(k^2 + \Delta)} \cr
&= \frac{1}{(4\pi)^{\frac{d}{2}}}\frac{\Delta^{\frac{d}{2} - r}}{\Gamma(r)}\Gamma \left(r - \frac{d}{2}\right)\,,
\end{aligned}
\end{equation}

where in the second line, we have used the identity
\begin{equation}
\label{D:2}
\frac{1}{\alpha^z} = \frac{1}{\Gamma(z)}\int_0^\infty dx x^{z-1}e^{-\alpha x}
\end{equation}

and then finally performed the remaining Gaussian integral in $k$, identifying the remaining integral over $x$ as being the gamma function.

Putting everything together, we find
\begin{equation}
\label{D:7}
I_{n,m}(r; \Delta) = \frac{\Gamma \left(r - 2n - \frac{d}{2} \right)}{(4\pi)^{\frac{d}{2}}\Gamma(r)}\left(\frac{1}{4}\frac{\partial}{\partial p_{\overline z}}\right)^n\left(\frac{1}{4}\frac{\partial}{\partial p_z}\right)^m(\Delta - 4p_zp_{\overline z})^{\frac{d}{2} - r + 2n}\Bigg|_{p_z, p_{\overline z}=0}\,.
\end{equation}

From this generating function, we can also note that only rotationally invariant integrands will be nonzero. That is, $I_{n,m}\propto \delta_{n,m}$.

Since many of our diagrams have two propagators carrying momenta, the special case $I_{n,n}(2; \Delta)$ will be particularly important. These integrals can always be put into the form
\begin{equation}
\label{D:8}
I_{n,n}(2; \Delta) = \frac{Z_{2,n}}{(4\pi)^{\frac{d}{2}}}\Gamma \left(2 - \frac{d}{2} \right)\Delta^{\frac{d}{2} + n - 2}\,,
\end{equation}
where the coefficients $Z_{2,n}$ depend only on $d$ and $n$. The first few of these coefficients are given by
\begin{equation}
\label{D:9}
Z_{2,0} = 1,\ \ \ \ Z_{2,1} = \frac{1}{2(2 - d)},\ \ \ \ Z_{2,2} = -\frac{1}{2d(2-d)},\ \ \ \ Z_{2,3} = \frac{3}{4d(2 - d)(2 + d)}\,.
\end{equation}

The above integral allows us to perform all one-loop integrations in section \ref{corsec}.

\subsection{Fourier transform identities}\label{Integrals:fourier}

We have also found it useful to compute the Fourier transform in $d$ dimensions of functions with the form $k_z^m k_{\overline z}^n (k^2)^s$, which we find to be
\begin{equation}
\begin{aligned}
\label{D:a1}
R_{m,n}^s(x) &\equiv \int\frac{d^d k}{(2\pi)^d} e^{ik\cdot x}(k^2)^s k_z^m k_{\overline z}^n \cr
&= (-i)^{m + n}\frac{4^s}{\pi^{d/2}}\frac{\Gamma \left(s+\frac{d}{2} \right)}{\Gamma(-s)}\partial_z^m\partial_{\overline z}^n(z\overline z + x_\perp^2)^{-s-d/2}.
\end{aligned}
\end{equation}

To show this, we will take the same approach as we did for the one-loop integrals and obtain a generating function for them. To this end, we first perform the Fourier transform
\begin{equation}
\begin{aligned}
    \label{D:10}
    R^s_{0,0}(x) &= \int\frac{d^dk}{(2\pi)^d}e^{ik\cdot x}(k^2)^s \\
&= \int\frac{d^dk}{(2\pi)^d}e^{ik\cdot x}\frac{1}{\Gamma(-s)}\int_0^\infty d\alpha \alpha^{-s-1}e^{-\alpha k^2} \\
&= \frac{1}{(4\pi)^{\frac{d}{2}}\Gamma(-s)}\int_0^\infty d\alpha \alpha^{-s-1-\frac{d}{2}}e^{-\frac{x\cdot x}{4\alpha}} \\
&= \frac{4^s}{\pi^{\frac{d}{2}}}\frac{\Gamma \left(s+\frac{d}{2} \right)}{\Gamma(-s)}\frac{1}{(x\cdot x)^{s+\frac{d}{2}}}\,,
\end{aligned}
\end{equation}

where we have performed the Gaussian integral over $k$ and reidentified the result as a Gamma function after rescaling the integration variable to $\beta = \frac{x\cdot x}{4\alpha}$.

With this, we complete our calculation by writing
\begin{equation}
\begin{aligned}
\label{D:11}
R_{m,n}^s(x) &= \left(\frac{1}{i}\partial_z\right)^{m}\left(\frac{1}{i}\partial_{\overline z}\right)^n \int\frac{d^d k}{(2\pi)^d}e^{ik\cdot x}(k^2)^s \\
&= \left(\frac{1}{i}\partial_z\right)^{m}\left(\frac{1}{i}\partial_{\overline z}\right)^n R_{0,0}^s(z\overline z + x_\perp^2) \\
&= (-i)^{m + n}\frac{4^s}{\pi^{\frac{d}{2}}}\frac{\Gamma \left(s+\frac{d}{2}\right)}{\Gamma(-s)}\partial_z^m\partial_{\overline z}^n(z\overline z + x_\perp^2)^{-s-\frac{d}{2}}\,,
\end{aligned}
\end{equation}

where we have assumed $x$ to have raised index and introduced a shorthand in which the two complex coordinates of $x$ are denoted by $z$ and $\overline z$ so $x\cdot x = z \overline z + x_\perp^2$. This establishes the claimed form result of the Fourier transform.

By expanding $R_{m,n}^s$ on both sides in a power series in $s$ and matching terms, the expansion
\begin{equation}
\label{D:11v2}
(k^2)^s = \sum_{\ell=0}^\infty \frac{1}{\ell!}(\ln k^2)^\ell s^\ell
\end{equation}
allows us to generate the Fourier transform of functions with the form $k_z^m k_{\overline z}^n (\ln k^2)^\ell$ as well. Of particular note, in this thesis are the following special cases:
\begin{equation}\label{D:12}
\begin{split}
\int\frac{d^2k}{(2\pi)^2}e^{ik\cdot x}k^n_zk_{\overline z}^m\ln k^2 =& -\frac{(-1)^{\frac{n+m}{2}}}{\pi}\frac{\Gamma(m+1)\Gamma(n+1)}{z^{n+1}\overline z^{m+1}} \,,\\
\int\frac{d^2k}{(2\pi)^2}e^{ik\cdot x}k^n_zk_{\overline z}^m(\ln k^2)^2 =& \frac{2 (-1)^{\frac{n+m}{2}}}{\pi} \frac{\Gamma(n+1)\Gamma(m+1)}{z^{n+1}\overline z^{m+1}}  \left(2\gamma - H_m - H_n + \ln (z\overline z)^2\right)\,, 
\end{split}
\end{equation}
where $\gamma = \Gamma'(1)$ is the Euler-Mascheroni constant and 
\begin{equation}
    H_n = \sum^n_{k=1} \frac{1}{k}
\end{equation}
is the $n^\text{th}$ harmonic number. In particular, these two integrals \eqref{D:12} appear when writing the stress tensor correlator \eqref{jj} in position space.

\subsection{\texorpdfstring{$\langle f'(p)f'(-p) \rangle$}{<f'(p) f'(-p)>} propagator at two-loop order}\label{app:2Loopff}
In this appendix, we compute the following integral, which appears in the calculation of the propagator at two-loop order
\begin{equation}
\label{eq:090ikpk}
I = \int {d^2 k\over (2\pi)^2} \int {d^2 k'\over (2\pi)^2} {k_{\bar{z}}^2 k_{\bar{z}}^{\prime 2}  (p_z-k_z-k_z')^2   \over k^2 k^{\prime 2}(p-k-k')^2 }\,.
\end{equation}
We start by using Feynman parameters to write \eqref{eq:090ikpk} as 
\begin{equation}\begin{split}
I =& 
\int {d^2 k\over (2\pi)^2} \int {d^2 k'\over (2\pi)^2} 
{k_{\bar{z}}^2 k_{\bar{z}}^{\prime 2}  (p_z-k_z-k'_z)^2  }  \\
&\times   \int_0^1 du \int^{1-u}_0 dv   {\Gamma(3)\over \Gamma(1)^3}  {1\over (  u k^2 + v k^{\prime 2} + (1-u-v)(p-k-k')^2        )^3}\,.
\end{split}
\end{equation}
We then  change momentum variables, noting that
\begin{equation}
  u k^2 + v k^{\prime 2} + (1-u-v)(p-k-k')^2   =  \alpha q^2 + \alpha' q^{\prime 2} + \gamma p^2\,, 
\end{equation}
with
\begin{equation}
\alpha = 1-v\,, \quad \alpha' = { (u+v)(1-v) - u^2  \over 1-v}\,, \quad \gamma = {  u v (1-u-v) \over (u+v)(1-v) - u^2  }\,,
\end{equation}
and
\begin{equation}\label{E:18}
q = k + {1-u-v \over 1-v} (k'-p)\,, \quad q' = k' - {   u(1-u-v) \over (u+v)(1-v) -u^2  } p\,.
\end{equation}
The change of momenta variables from $k,k'$ to $q,q'$ has a trivial Jacobian. To continue, we convert the denominator to an exponential using a Schwinger parameter as in formula \eqref{D:2}. Explicitly,
\begin{equation}
{1\over ( \alpha q^2 + \alpha' q^{\prime 2} + \gamma p^2       )^3}  = {1\over \Gamma(3)} \int_0^{\infty} dU \, U^2 e^{-U (\alpha q^2 + \alpha' q^{\prime 2} + \gamma p^2  )}\,.
\end{equation}
We now have 
\begin{equation}\begin{split}
I
=&
\int {d^2 q\over (2\pi)^2} \int {d^2 q'\over (2\pi)^2} 
{k_{\bar{z}}^2 k_{\bar{z}}^{\prime 2}  (p-k-k')_z^2  }  \\
&\times   \int_0^1 du \int^{1-u}_0 dv  \int_0^{\infty} dU \, U^2 e^{-U (\alpha q^2 + \alpha' q^{\prime 2} + \gamma p^2  )}\,,
\end{split}
\end{equation}
where in the first line $k$ and $k'$ are understood to be functions of $q$ and $q'$ using \eqref{E:18}.
The momentum integrals  are all Gaussian of the form
\begin{equation}\label{eq:NeedInt}
\int {d^2 q\over (2\pi)^2}  q_z^n q_{\bar{z}}^m e^{-U \alpha q^2 }\,.
\end{equation}
We perform these integrals via dimensional regularization. We start with the generating function
\begin{equation}\begin{split}
\tilde{G}[p, C] =& \int {d^d K \over (2\pi)^d} e^{-C (K^2 +2K\cdot p)} 
= {1 \over (4\pi C)^{d\over 2}} e^{C p^2 }\,.
\end{split}
\end{equation}
Noting that 
\begin{equation}
K \cdot p =  2 K_z p_{\bar{z}} + 2K_{\bar{z}} p_z + \vec{K}_\perp\cdot \vec{p}_\perp\,,
\end{equation}
we conclude
\begin{equation}\begin{split}
\left[ \left(  {-1\over 4 C} \partial_{p_{z}}  \right)^n \left(  {-1\over 4 C} \partial_{p_{\bar{z}}}  \right)^m \tilde{G}[p, C]  \right]_{p=0} = \int {d^d K \over (2\pi)^d} K_z^m K_{\bar{z}}^n e^{-C K^2} \,. 
\end{split}
\end{equation}
This allows us to compute the integrals \eqref{eq:NeedInt} explicitly as a function of the dimension $d$. Note in particular that this integral vanishes unless $m=n$, so the only formula we need is
\begin{equation}\begin{split}
\left[{1\over (4 C)^n}  \partial_{p_{z}}^n  \partial_{p_{\bar{z}}}^n \tilde{G}[p, C]  \right]_{p=0} = \int {d^d K \over (2\pi)^d}(K_z K_{\bar{z}})^n e^{-C K^2}\,. 
\end{split}
\end{equation}
After performing the Gaussian integrals, we find
\begin{equation}\begin{split}
&I= \int_0^1 du \int^{1-u}_0 dv  \int_0^{\infty} dU \, e^{-U \gamma p^2}
 {p_{\bar{z}}^2 u^2 v^2 (1-u-v)^2 \over 4 (4\pi)^d  ((u+v)(1-v)-u^2)^{4+{d\over 2}}} \\
&\times  \left(
  3 U^{-d} 
- 8 (p_z p_{\bar{z}}) U^{1-d} {u v (1-u -v) \over (u+v)(1-v)-u^2}
+ 4 (p_zp_{\bar{z}})^2 U^{2-d} {v^2 u^2 (1-u-v)^2\over  ((u+v)(1-v)-u^2)^2 } 
  \right)\,.
\end{split}
\end{equation}
The integral over the Schwinger parameter $U$ can now be performed trivially using the formula
\begin{equation}
\int_0^{\infty} dU\, U^x e^{-U \gamma p^2 }  = {\Gamma(1+x) \over (\gamma p^2)^{1+x}}\,.
\end{equation}
Before performing the Feynman integrals, we expand around $d=2+\epsilon$. We obtain
\begin{equation}\begin{split}
I
=&\int_0^1 du \int^{1-u}_0 dv 
{5\over 16\pi^2 }    {  u^3 v^3 (1-u-v)^3   \over ((u+v)(1-v) -u^2)^6   } p_z p_{\bar{z}}^3 \log p^2 \, +{\text{polynomial}}\,.
\end{split}
\end{equation}
Integration over Feynman parameters in \texttt{Mathematica} yields
\begin{equation}\begin{split}
I
=&
{1 \over 2^7 3 \pi^2 }    p_z p_{\bar{z}}^3 \log p^2 \, +{\text{polynomial}}\,,
\end{split}
\end{equation}
which is \eqref{eq:2LoopInt}.

\subsection{Diagram with generalized propagator}\label{Integrals:compare}

Here, we provide further explanation regarding our choice of propagator, discussed below \eqref{ih}. Consider the following family of propagators labeled by the parameter $\eta$,
\begin{equation}
    \label{ihzz}
    \langle f'(p) f'(-p)\rangle_0 = 32\pi G\left( {p_z^2 \over p^2}+ \eta  {p_z p_{\bar{z}} \over p^2} \right) \,,\quad \langle \bar{f}'(p) \bar{f}'(-p)\rangle_0 = 32\pi G\left( { p_{\bar{z}}^2 \over p^2}+\eta  {p_z p_{\bar{z}} \over p^2}  \right)\,.
\end{equation}
Direct inversion of the quadratic terms in the action gives $\eta=1$, while in our computations, we took $\eta=0$, claiming that this amounted to a particular Lorentz invariant renormalization scheme. To further illustrate this, we consider a typical diagram computed with general $\eta$.  

In particular, consider the one-loop contribution to $\langle f'f'\bar{f}'\bar{f}'\rangle$ in diagram \eqref{iw}. Using the generalized propagator, the diagram is proportional to the following integral:
\begin{equation}
\begin{aligned}
    \label{kva}
     I_\eta  &= \int\! {d^dk \over (2\pi)^d} {k_{\bar{z}} (\eta k_z+k_{\bar{z}}  )  \over k^2} {(k_z-p_z)(k_z+\eta k_{\bar{z}} -p_z-\eta p_{\bar{z}} )  \over (k-p)^2}
\end{aligned}
\end{equation}
\begin{equation}
\label{kva}
I_\eta  = \int\! {d^dk \over (2\pi)^d} {k_{\bar{z}} (\eta k_z+k_{\bar{z}}  )  \over k^2} {(k_z-p_z)(k_z+\eta k_{\bar{z}} -p_z-\eta p_{\bar{z}} )  \over (k-p)^2} 
\end{equation}
which, using the integral \eqref{D:8}, we compute as
\begin{equation}
\label{kvb}
I_\eta = \frac{p_z p_{\overline z}}{8\pi\varepsilon} + \frac{p_zp_{\overline z}}{96\pi}\left(6\gamma - 11 + 6\ln\left(\frac{p^2}{4\pi}\right)\right) + \frac{p_z^2 + p_{\overline z}^2}{48\pi}\eta + \frac{p_zp_{\overline z}}{96\pi}\eta^2 + \mathcal{O}(\varepsilon)\,.
\end{equation}
 The relevant observation is that the divergent and log parts of the integral are independent of $\eta$. Furthermore, the $\eta$-dependent terms are purely polynomial in the momenta and all terms which do not respect Lorentz invariance vanish if we take $\eta$ to zero. So using a general value for $\eta$  corresponds to using a different (non-Lorentz invariant) renormalization scheme. That is, if we chose a nonzero value of $\eta$, we should also include additional non-Lorentz invariant counterterms to cancel off the non-Lorentz invariant polynomial terms in \eqref{kvb}. A simpler way to obtain the same final result is to set $\eta=0$ at the outset. This feature applies to all Feynman diagrams considered in this thesis.     
\chapter{Supersymmetric Quantum Mechanics}
\label{app:susyqms}

\section{Conventions}\label{sec:conventions}
We outline our notation for the superspaces used in chapter \ref{ch:SUSY-QM}. Although the main focus of our analysis is on a supersymmetric quantum mechanics theory in $(0+1)$-dimensions, we obtain some expressions for $T\overline{T}$-type deformations by dimensionally reducing previous results for $(1+1)$-dimensional supersymmetric theories. For this reason, we begin with an overview of the conventions for $\mathcal{N} = (1,1)$ supersymmetry in two spacetime dimensions following \cite{Chang:2018dge}.

We begin by discussing two-dimensional Lorentzian field theories. We assume that these theories have coordinates $(t, x)$. When we perform dimensional reduction, we assume that the spatial coordinate $x$ parameterizes a circle with some radius $R$ so that $x \sim x + R$.

It will often be convenient to change coordinates from $(t, x)$ to light-cone coordinates:
\begin{align}\label{bispinor_light_cone}
    x^{\pm \pm} = \frac{1}{\sqrt{2}} \left( t \pm x \right) \,.
\end{align}
Here, we have adopted the bi-spinor convention, where a vector index is written as a pair of spinor indices. The derivatives with respect to the coordinates (\ref{bispinor_light_cone}) are
\begin{align}
    \partial_{\pm \pm} = \frac{1}{\sqrt{2}} \left( \partial_t \pm \partial_x \right) \,, 
\end{align}
which satisfy
\begin{align}
    \partial_{\pm \pm} x^{\pm \pm} = 1 \,, \qquad \partial_{\pm \pm} x^{\mp \mp} = 0 \,.
\end{align}
Spinor indices, which are written with early Greek letters, are raised or lowered with the epsilon tensor as
\begin{align}
    \psi_\beta = \epsilon_{\beta \alpha} \psi^{\alpha} \,,
\end{align}
where we take $\epsilon_{+-} = 1$ so $\epsilon_{-+} = -1$, $\epsilon^{+-} = -1$, $\epsilon^{-+} = 1$. For instance, this implies that:
\begin{align}\label{spinor_raise_lower}
    \psi^- = \psi_+ \,, \qquad \psi^+ = - \psi_- \,. 
\end{align}
For two-dimensional theories with $\mathcal{N} = (1,1)$ supersymmetry, we write the Grassmann coordinates as $\theta^{\pm}$. The supercovariant derivatives with respect to these anticommuting coordinates are
\begin{align}
    D_{\pm} = \frac{\partial}{\partial \theta^{\pm}} + \theta^{\pm} \partial_{\pm \pm} \,.
\end{align}
These satisfy
\begin{align}
    D_\pm D_\pm = \partial_{\pm \pm} \,, \qquad \{ D_+ , D_- \} = 0 \,.
\end{align}
We will also be interested in discussing theories of supersymmetric quantum mechanics in $(0+1)$-dimensions, so next, we describe how to perform this reduction and match conventions between the two theories.

When we reduce from $(1+1)$-dimensional field theory to $(0+1)$-dimensional quantum mechanics, we assume that all quantities are independent of the spatial direction $x$. Operationally, one can achieve this by setting $\partial_x \equiv 0$ everywhere, which amounts to making the replacement $\partial_{\pm \pm} = \frac{1}{\sqrt{2}} \partial_t$. We will re-scale our time coordinate $t$ to eliminate the factor of $\frac{1}{\sqrt{2}}$ and instead write $\partial_{\pm \pm} = \partial_t$

We note that making this replacement leads to expressions that have unbalanced numbers of $+$ and $-$ indices, like $D_+ = \frac{\partial}{\partial \theta^+} + \theta^+ \partial_t$. Although such an expression would not exhibit the correct properties under Lorentz transformation in a $(1+1)$-dimensional theory, in our reduced $(0+1)$-dimensional theory, there is no notion of spin nor Lorentz symmetry. Performing the dimensional reduction in this way, therefore, yields a consistent set of conventions.

It will be convenient to write the superspace of the $\mathcal{N} = 2$ supersymmetric quantum mechanics theory in complex coordinates, which more closely matches the conventions in the literature. We first Wick-rotate our time coordinate,\footnote{We will be somewhat cavalier about real versus imaginary time. All formulas in $2d$ field theory will be Lorentzian and involve real times $t$, but upon dimensional reduction to quantum mechanics, we eventually rotate $t \to it$ to match more common conventions. However, we continue to use the symbol $t$ rather than $\tau$ in this context for simplicity.} sending $t \to i t$, so that the supercovariant derivatives are
\begin{align}
    D_{\pm} = \frac{\partial}{\partial \theta^{\pm}} - i \theta^{\pm} \partial_t \,.
\end{align}
Next, we perform the change of variables
\begin{align}\label{theta_complex_change_of_variables}
    \theta = \frac{1}{\sqrt{2}} \left( \theta^+ - i \theta^- \right) \,, \qquad \bar{\theta} = \frac{1}{\sqrt{2}} \left( \theta^+ + i \theta^- \right) \,,
\end{align}
so that
\begin{align}
    D = \frac{1}{\sqrt{2}} \left( D_+ + i D_- \right) = \frac{\partial}{\partial \theta} - i \bar{\theta} \partial_t \,, \quad \bar{D} = \frac{1}{\sqrt{2}} \left( D_+ - i D_- \right) = \frac{\partial}{\partial \bar{\theta}} - i \theta \partial_t \,.
\end{align}
The new supercovariant derivatives satisfy the canonical algebra
\begin{align}\label{ddbar_algebra}
    \{ D , \bar{D} \} = - 2 i \partial_t \,, 
\end{align}
with $D^2 = \bar{D}^2 = 0$.

The rotation from real to complex Grassmann coordinates will introduce a factor of $i$ in the measure since
\begin{align}
    d \theta \, d \bar{\theta} = i \, d \theta^+ \, d \theta^- \,,
\end{align}
but this is compensated by the factor of $i$ arising from the Wick rotation $d t \to i \, d t$. 

Finally, in section \ref{sec:n_equals_one}, we briefly discuss the $\mathcal{N} = 1$ version of our deformation.

In $\mathcal{N} = 1$ superspace we have a single anticommuting coordinate $\theta$, along with a corresponding supercovariant derivative
\begin{align}
    D = \frac{\partial}{\partial \theta} - i \theta \frac{\partial}{\partial t} \,,
\end{align}
which satisfies the algebra
\begin{align}
    \{ D, D \} = - 2 i \partial_t \,.
\end{align}

\section{Change of coordinates to complex supercharges}\label{app:change_to_complex}

In this appendix, we carry out the change of variables to express our SUSY-QM deformation $f(\mathcal{Q}_+, \mathcal{Q}_-)$ of (\ref{scsquare_reduction_final}) in complex coordinates, ultimately arriving at the expression (\ref{scsquare_reduction_final_complex}) for $f(\mathcal{Q}, \bar{\mathcal{Q}})$. This is a straightforward application of the change of variables described in equations (\ref{theta_complex_change_of_variables}) - (\ref{ddbar_algebra}) of appendix \ref{sec:conventions}, but because it involves some on-shell manipulations, we have moved the calculation to this appendix to avoid cluttering the main body.

We shift to complex supercovariant derivatives via
\begin{align}
    D = \frac{1}{\sqrt{2}} \left( D_+ + i D_- \right)  \,, \quad \bar{D} = \frac{1}{\sqrt{2}} \left( D_+ - i D_- \right) \,, 
\end{align}
and similarly rotate the supercurrents via
\begin{align}
    \mathcal{Q} = \frac{1}{\sqrt{2}} \left( \mathcal{Q}_- + i \mathcal{Q}_+ \right)   \,, \quad \bar{\mathcal{Q}} = \frac{1}{\sqrt{2}} \left( \mathcal{Q}_- - i \mathcal{Q}_+ \right) \,.
\end{align}
Note that since $\mathcal{Q}_{\pm}$ are fermionic, one has
\begin{align}\label{QQbar_conversion}
    \mathcal{Q} \bar{\mathcal{Q}} &= \frac{1}{2} \left( \mathcal{Q}_-^2 - i \mathcal{Q}_- \mathcal{Q}_+ + i \mathcal{Q}_+ \mathcal{Q}_- + \mathcal{Q}_+^2 \right) \nonumber \\
    &= i \mathcal{Q}_+ \mathcal{Q}_- \,.
\end{align}
Next, we compute the supercovariant derivatives. The combination $\bar{D} \mathcal{Q}$ is
\begin{align}
    \bar{D} \mathcal{Q} &= \frac{1}{2} \left( D_+ - i D_- \right) \left( \mathcal{Q}_- + i \mathcal{Q}_+ \right) \nonumber \\
    &= \frac{1}{2} \Big[ D_+ \mathcal{Q}_- + i D_+ \mathcal{Q}_+ - i D_- \mathcal{Q}_- + D_- \mathcal{Q}_+  \Big]  \,, 
\end{align}
or after using the conservation equation $D_+ \mathcal{Q}_- + D_- \mathcal{Q}_+ = 0$ and the on-shell condition that $D_+ \mathcal{Q}_+ = - D_- \mathcal{Q}_-$,
\begin{align}
    \bar{D} \mathcal{Q} = i D_+ \mathcal{Q}_+ \,.
\end{align}
Likewise,
\begin{align}
    D \bar{\mathcal{Q}} &= \frac{1}{2} \left( D_+ + i D_- \right) \left( \mathcal{Q}_- - i \mathcal{Q}_+ \right) \nonumber \\
    &= \frac{1}{2} \left( D_+ \mathcal{Q}_- - i D_+ \mathcal{Q}_+ + i D_- \mathcal{Q}_- + D_- \mathcal{Q}_+ \right) \,,
\end{align}
and, again, this can be written on-shell as
\begin{align}\label{DpQp_to_Dbar_Q}
    D \bar{\mathcal{Q}} = - i D_+ \mathcal{Q}_+ .
\end{align}
Thus, we see that the new complex supercurrents satisfy the conservation equation $\bar{D} \mathcal{Q} + D \bar{\mathcal{Q}} = 0$, since
\begin{align}\label{new_complex_conservation}
    \bar{D} \mathcal{Q} + D \bar{\mathcal{Q}} = i D_+ \mathcal{Q}_+ - i D_+ \mathcal{Q}_+ = 0
\end{align}
when the equations of motion are satisfied.

We now return to the expression $f(\mathcal{Q}_+, \mathcal{Q}_-)$ defining our deformation, which can now be written in terms of complex coordinates as
\begin{align}\label{first_step_to_QQbar}
    \int \, dt \, d \theta^+ \, d \theta^- \, \frac{\mathcal{Q}_+ \mathcal{Q}_-}{4 \lambda D_+ \mathcal{Q}_+ + 1} = \int \, dt \, d \theta^+ \, d \theta^- \, \frac{-i \mathcal{Q} \bar{\mathcal{Q}}}{- 4 i \lambda \bar{D} \mathcal{Q} + 1 } \,.
\end{align}
We would now like to eliminate the factors of $i$ that have appeared in (\ref{first_step_to_QQbar}). One factor arises from the change of measure via $d \theta \, d \bar{\theta} = i \, d \theta^+ \, d \theta^-$. A second factor arises because, as pointed out in the discussion below (\ref{susy_Q_def}), there is a relative factor of $i$ arising between the natural expressions appearing in the Noether procedures which define $\mathcal{Q}, \bar{\mathcal{Q}}$ as opposed to $\mathcal{Q}_+, \mathcal{Q}_-$. Therefore, to obtain an appropriate matching, we will rescale:
\begin{align}
    \mathcal{Q} \longrightarrow - i \mathcal{Q} \,, \qquad \bar{\mathcal{Q}} \longrightarrow - i \bar{\mathcal{Q}} \,.
\end{align}
After incorporating these two factors, we find
\begin{align}
    \int \, dt \, d \theta^+ \, d \theta^- f ( \mathcal{Q}_+ , \mathcal{Q}_- ) = \int \, dt \, d \theta \, d \bar{\theta} \frac{\mathcal{Q} \bar{\mathcal{Q}}}{- 4 \lambda \bar{D} \mathcal{Q} + 1} \,.
\end{align}
Finally, we scale out an overall factor of $\frac{1}{2}$ to write
\begin{align}
    f ( \mathcal{Q}_+ , \mathcal{Q}_- ) &\sim \frac{\mathcal{Q} \bar{\mathcal{Q}}}{\frac{1}{2} - 2 \lambda \bar{D} \mathcal{Q}} \, \nonumber \\
    &\equiv f ( \mathcal{Q} , \bar{\mathcal{Q}} ) \,,
\end{align}
where $\sim$ indicates proportionality on-shell (as we have used conservation equations to relate $D_+ \mathcal{Q}_+$ to $\bar{D} \mathcal{Q}$). We chose to rescale by this prefactor to make the right side more closely match (\ref{gross_flow_eqn}). This is the form quoted in (\ref{scsquare_reduction_final_complex}).

\section{Dimensional reduction without trace flow equation}\label{app:no_trace_flow}

As we have pointed out in the main body of this thesis, the deformation (\ref{gross_flow_eqn}) is very convenient for deforming quantum mechanical theories that descend from $2d$ CFTs via dimensional reduction. However, for theories with potential, the trace flow equation \eqref{trace_flow} fails, and we cannot use this expression for the reduced $T\overline{T}$ deformation. In this case, our only recourse is to directly study the $T\overline{T}$-deformed field theory in two dimensions, then compactify one spatial direction on a circle and truncate to the lowest Fourier mode.

In this appendix, we will obtain the Hamiltonian for such a theory by first solving the $2d$ flow equation and then performing the circle compactification only at the final step. Suppose we begin with an undeformed Lagrangian
\begin{align}
    \mathcal{L}_E ( \lambda = 0 , \phi ) = \frac{1}{2} \partial^\mu \phi \partial_\mu \phi + V (\phi )\,, 
\end{align}
with a positive sign on the potential because we work in Euclidean signature for now. The deformed Lagrangian at finite $\lambda$ appears in equation (2.8) of \cite{Bonelli:2018kik} (see also \cite{Cavaglia:2016oda}) as
\begin{align}
    \mathcal{L}_E ( \lambda, \phi )&=  - \frac{1}{2 \lambda} \left(\frac{1 - 2 \lambda V(\phi)}{1 - \lambda V(\phi)}\right) \nonumber \\&+ \frac{1}{2 \lambda} \sqrt{ \left(\frac{1 - 2 \lambda V(\phi)}{1 - \lambda V(\phi)}\right)^2 + 2 \lambda \left(\frac{ \partial^\mu \phi \partial_\mu \phi + 2 V(\phi) }{1 - \lambda V(\phi)} \right) }\,.
\end{align}
Again, here, the metric appearing in the $\partial^\mu \phi \partial_\mu \phi$ contraction is $\delta^\mu{}_\nu$ because we are in Euclidean signature. The prescription for rotating back to Minkowski signature is to multiply the Lagrangian by an overall minus sign, then to invert the sign on the time derivative of $\phi$, giving
\begin{align}
    \mathcal{L}_M ( \lambda, \phi ) = \frac{1}{2 \lambda} \left( \frac{1 - 2 \lambda V(\phi)}{1 - \lambda V(\phi)} \right) - \frac{1}{2 \lambda} \sqrt{ \left(\frac{1 - 2 \lambda V(\phi)}{1 - \lambda V(\phi)}\right)^2 + 2 \lambda \bigg(\frac{ \phi^{\prime 2}  - \dot{\phi}^2 + 2 V(\phi) }{1 - \lambda V(\phi)}\bigg) }\,.
    \label{deformed_minkowski}
\end{align}
Here we used $\dot{\phi} = \frac{\partial \phi}{\partial t}$ and $\phi' = \frac{\partial \phi}{\partial x}$. We can study the behavior of $\mathcal{L}_M$ in a few limits:
\begin{align}
    \mathcal{L}_M ( \lambda \to 0 , \phi ) &= \frac{1}{2} \dot{\phi}^2 - \frac{1}{2} \phi^{\prime 2} - V ( \phi ) , \nonumber \\
    \mathcal{L}_M ( \lambda, \phi ) \Big\vert_{\dot{\phi} = \phi' = 0} &=- \frac{V(\phi)}{1 - \lambda V(\phi)} , \nonumber \\
    \mathcal{L}_M ( \lambda , \phi ) \Big\vert_{V(\phi)=0} &= \frac{1}{2 \lambda} \bigg( 1 - \sqrt{1 + 2 \lambda \big( \phi^{\prime 2} - \dot{\phi}^2 \big) } \bigg) = \frac{1}{2\lambda} + \mathcal{L}_{\text{Nambu-Goto}}\,.
\end{align}
To write this as a Hamiltonian, we will resort to Legendre transform. The conjugate momentum to $\phi$ is
\begin{align}
    \Pi = \frac{\partial \mathcal{L}}{\partial \dot{\phi}} = \frac{\dot{\phi}}{\sqrt{1 - 2 \lambda \left( 1 - \lambda V(\phi) \right) \big( \dot{\phi}^2 - \phi^{\prime 2} \big) } } \,.
    \label{conjugate_momentum}
\end{align}
The relation (\ref{conjugate_momentum}) can be inverted to find
\begin{align}
    \dot{\phi} = \Pi \cdot \sqrt{ \frac{ 1 + 2 \lambda ( 1 - \lambda V(\phi) ) \phi^{\prime 2}}{1 + 2 \lambda ( 1 -  \lambda V(\phi) ) \Pi^2 } } \,.
    \label{phidot_to_pi}
\end{align}
The Hamiltonian is then defined by the Legendre transformation
\begin{align}
    \mathcal{H} = \Pi \dot{\phi} - \mathcal{L}\,, 
\end{align}
after replacing all instances of $\dot{\phi}$ with $\Pi$ using (\ref{phidot_to_pi}). This gives
\begin{align}
   \mathcal{H} = \phi^{\prime 2} \cdot &\sqrt{ \frac{1 + 2 \lambda ( 1 -  \lambda V(\phi) ) \Pi^2 }{1 + 2 \lambda ( 1 - \lambda V(\phi) ) \phi^{\prime 2} }} \nonumber + \frac{1}{2 \lambda ( 1 - \lambda V(\phi) ) } \cdot \sqrt{ \frac{1 + 2 \lambda ( 1 -  \lambda V(\phi) ) \Pi^2 }{1 + 2 \lambda ( 1 - \lambda V(\phi) ) \phi^{\prime 2} }}\\
    & + \frac{V(\phi)}{1 - \lambda V(\phi)} - \frac{1}{2 \lambda ( 1 - \lambda V(\phi) ) } \,.
    \label{2d_ham}
\end{align}
The dependence on $\Pi^2$ is masked by the terms involving square roots. Near $\lambda = 0$, (\ref{2d_ham}) is
\begin{align}
    \mathcal{H} = \frac{1}{2} \Pi^2 + \frac{1}{2} \phi^{\prime 2} + V ( \phi ) + \mathcal{O} ( \lambda )\,, 
\end{align}
which is the expected Hamiltonian for a scalar field with a potential.

Next, we would like to put the coordinate $x$ on a circle of radius $R$, Fourier-expand the $x$-dependence of $\phi(x,t)$, and integrate the Hamiltonian $H$ over the circle to obtain a quantum-mechanical Hamiltonian associated with the modes $\phi^{(n)} ( t )$. We expand $ \phi ( x, t ) $ in modes as
\begin{align}
    \phi ( x, t ) = \sum_{n=0}^{\infty} \left( \phi^{(n)}_c ( t ) \cos \left( \frac{2  \pi n}{R} \, x \right) + \phi^{(n)}_s ( t ) \sin \left( \frac{2  \pi  n}{R} \, x \right) \right)\,. 
    \label{phi_fourier}
\end{align}
Inserting (\ref{phi_fourier}) into (\ref{2d_ham}) and integrating over the circle would, in principle, leave us with a Hamiltonian for infinitely many interacting particles $\phi^{(n)}_{c} ( t )$ and $\phi^{(n)}_{s}$ in quantum mechanics. Such an analysis seems intractable in general, so for simplicity, let us restrict to the zero-momentum sector\footnote{Restricting to the zero-momentum sector also allowed us to use the implicit solution (\ref{2d_tt_burgers_implicit}) to the inviscid Burgers' equation, which was pointed out in \cite{Cavaglia:2016oda}.}
\begin{align}
    \phi ( x , t ) \equiv \phi ( t ) \,.
\end{align}
This gives us a Hamiltonian
\begin{align}
    H = \frac{ \sqrt{ 1 +2 \lambda ( 1 -  \lambda V(\phi) ) \Pi^2 }}{2 \lambda \left( 1 - \lambda V(\phi) \right)} + \frac{V(\phi)}{1 - \lambda V(\phi)} - \frac{1}{2 \lambda ( 1 - \lambda V(\phi) ) } \,.
    \label{reduced_hamiltonian}
\end{align}
For small $\lambda$, \eqref{reduced_hamiltonian} looks like
\begin{align}
    H = \frac{1}{2} \Pi^2 + V ( \phi ) + \lambda \left( V ( \phi )^2 - \frac{1}{4} \Pi^4 \right) + \frac{1}{4} \lambda^2 \left( 4 V ( \phi )^3 + \Pi^4 V(\phi) + \Pi^6 \right) + \mathcal{O} ( \lambda^3 )\,, 
\end{align}
The leading term is the usual Hamiltonian $H = \frac{p^2}{2m} + V(\phi)$ if we identify $p = \Pi$, $m = 1$. But this usual Hamiltonian receives an infinite series of corrections, which affect both the kinetic and potential terms (and mix them). The purely kinetic part of \eqref{reduced_hamiltonian} reduces to
\begin{align}
    H \Big\vert_{V(\phi) = 0} = \frac{-1 + \sqrt{1 + 2 \lambda \Pi^2} }{2 \lambda} \,, 
\end{align}
which is a $(0+1)$-dimensional analog of the Nambu-Goto action. If we alternatively set $\Pi=0$ and consider the pure potential piece, from \eqref{reduced_hamiltonian} we find
\begin{align}\label{V_over_one_minus_lambda_V}
    H \Big\vert_{\Pi = 0} = \frac{V(\phi)}{1 - \lambda V(\phi)} \,.
\end{align}
This looks identical to the result of deforming a pure-potential Hamiltonian by the function $f(H) = H^2$, rather than the more complicated operator (\ref{gross_flow_eqn}), which is equivalent to $T\overline{T}$ for theories that descend from deformations of $2d$ CFTs. To be explicit, if we consider the flow equation:
\begin{align}
    \frac{\partial H}{\partial \lambda} = H^2 \,,
\end{align}
with initial condition $H(0) = H_0$, then the solution is trivially
\begin{align}
    H ( \lambda ) = \frac{H_0}{1 - \lambda H_0} \,.
\end{align}
At low momentum, where the kinetic term can be neglected, and the undeformed Hamiltonian is approximately the pure potential $H = V(\phi)$, this is (\ref{V_over_one_minus_lambda_V}). In particular, the Hamiltonian diverges when $V(\phi) = \frac{1}{\lambda}$. This is purely a classical statement about the solution to an $f(H)$-type flow equation, which is not necessarily indicative of the structure of the quantum theory.

However, the possible presence of poles is interesting and hints at a modification of the vacuum structure of the theory. For the moment, we will allow ourselves to speculate about the physical implications of the existence of such poles if, indeed, they persist at the quantum level.

We mention a few explicit potentials by way of examples. For instance, suppose we begin with the harmonic oscillator potential $V_0 ( \phi ) = m^2 \phi^2$, where we take $m=1$ for simplicity. The potential deforms as follows:
\begin{align}
    \raisebox{-0.5\height}{\includegraphics[width=0.4\linewidth]{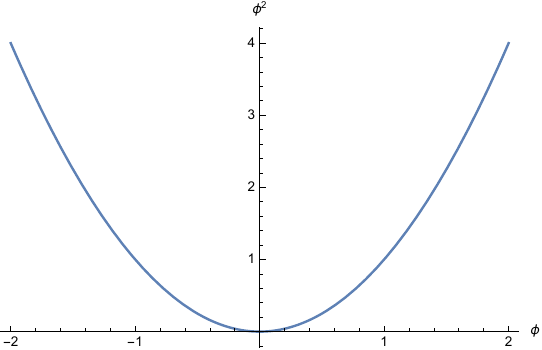}} \longrightarrow \; \raisebox{-0.5\height}{\includegraphics[width=0.4\linewidth]{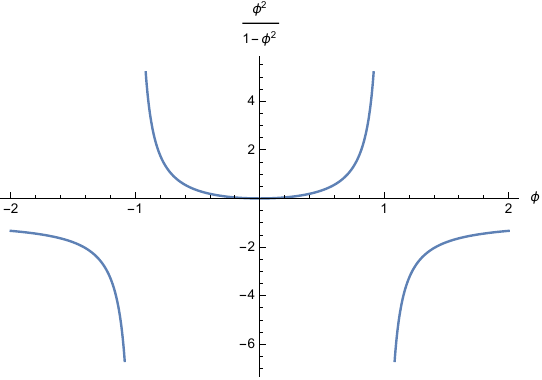}}
\end{align}
It is very natural to ask what has happened to the basis of eigenfunctions after applying this deformation. The undeformed potential is the usual harmonic oscillator, whose eigenstates are Hermite polynomials. However, the deformed potential has infinite barriers at $\phi = \pm 1$. One might believe that there is a complete set of eigenfunctions for the deformed potential, which are forced to vanish at $\phi = \pm 1$. The regions $| \phi | > 1$ seem to have been ``cut off'' from the theory by applying this deformation.

Another interesting case to consider is a linear potential $V ( \phi ) = \phi$.
\begin{align}
    \raisebox{-0.5\height}{\includegraphics[width=0.4\linewidth]{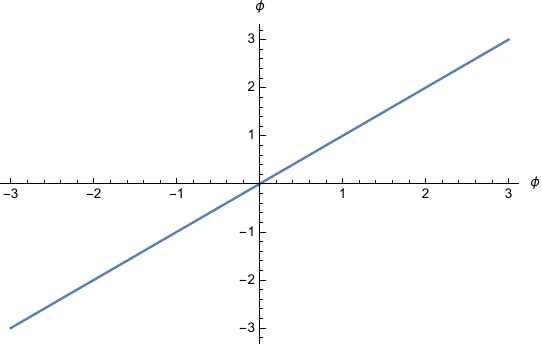}} \longrightarrow \;  \raisebox{-0.5\height}{\includegraphics[width=0.4\linewidth]{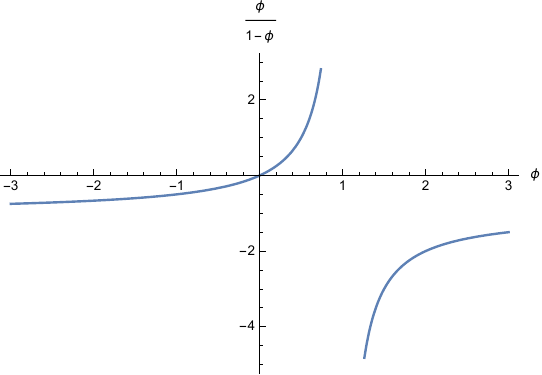}}
\end{align}
Now, the change is even more drastic: the undeformed linear potential had eigenstates, which were Airy functions, but they were non-normalizable because the potential was unbounded below. The deformation has now inserted a hard cutoff at $\phi = 1$. To the left of this cutoff, the potential is bounded below as $V(\phi) > -1$. Has the $T\overline{T}$ deformation ``cured'' the non-normalizability of the linear potential? If so, is there a relationship between the undeformed eigenstates (Airy functions) and the eigenstates of the deformed potential?

For a third example, the double-well potential $V(\phi) = \left( 1 - \phi^2 \right)^2$ deforms as
\begin{align}\label{double_well}
    \raisebox{-0.5\height}{\includegraphics[width=0.4\linewidth]{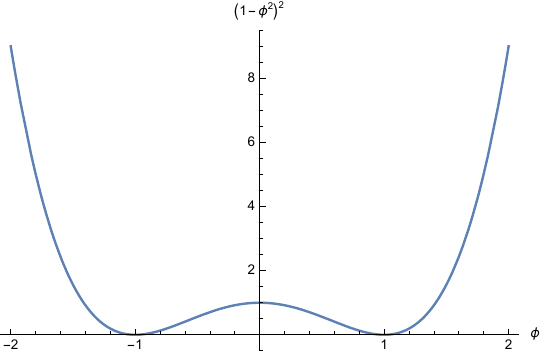}} \longrightarrow \; \raisebox{-0.5\height}{\includegraphics[width=0.4\linewidth]{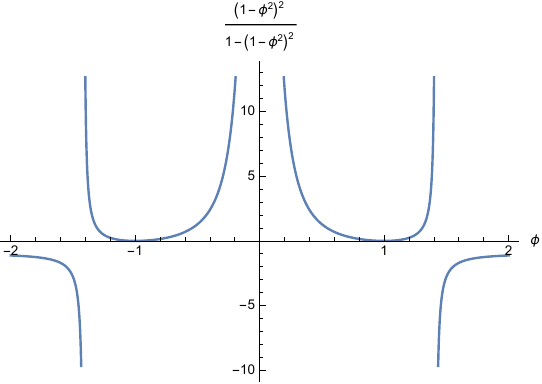}}
\end{align}
Now, there is a pole at $\phi = 0$, so the two wells have become separated by an infinite potential barrier. Again, one might wonder what has happened to the Hilbert space. Is there still a complete basis of eigenfunctions, but now localized to each of the disconnected wells?

The above examples are presented only in the context of ordinary quantum mechanics without any supersymmetry. However, one might hope that the presence of some SUSY might be useful in learning about the fate of the Hilbert space after such a deformation. For instance, the spectrum of ground states in a supersymmetric theory exhibits a great deal of structure, and one can extract data about it using index-like quantities. Is there some calculation in SUSY-QM theory that is sensitive to the fact that the two ground states in the double well (\ref{double_well}) may have been ``cut off'' from one another in the deformed theory? It would be exciting to gain a better understanding of the Hilbert space of these deformed quantum mechanics theories and to understand whether $T\overline{T}$ or $f(\mathcal{Q}, \bar{\mathcal{Q}})$ indeed has effects on the infrared structure of the kind described here.

\chapter{AdS$_3$ with $T\overline{T}$-Deformed Boundary Conditions}
\label{sec:AdS3TTApp}

In section \ref{sec:AdS3RootTT}, we used several methods that have been developed for studying $\mathrm{AdS}_3$ gravity with $T\overline{T}$-deformed boundary conditions, both in the metric formalism \cite{Guica:2019nzm} and in the Chern-Simons formalism \cite{He:2020hhm}. To make the present work self-contained, we review some aspects of these methods in this appendix, which are also useful for our analysis of root-$T\overline{T}$ deformed boundary conditions. We refer the reader to the original works for further details and to the related work \cite{Llabres:2019jtx} for additional results in the Chern-Simons formalism.

\section{Metric formalism}\label{app:TT_bc_pde_soln}

We recall that the modified metric $\gamma_{\alpha \beta}^{(\lambda)}$ and stress tensor $T_{\alpha \beta}^{(\lambda)}$ corresponding to a boundary $T\overline{T}$ deformation satisfy (\ref{TT_varied_flow}) which was re-derived in the main text. By equating the coefficients of the independent terms on both sides of this equation, one arrives at a set of partial differential equations for the deformed quantities. These differential equations were first analyzed in \cite{Guica:2008mu}, where it was shown that they can be written as
\begin{equation}
\label{eq:PDEsTT}
\frac{\partial \gamma_{\alpha \beta}}{\partial \lambda} = -2 \hat{T}_{\alpha \beta}\,, \quad \frac{\partial \hat{T}_{\alpha \beta}}{\partial \lambda} = - \hat{T}_{\alpha \gamma} \hat{T}_\beta{}^\gamma\,, \quad \frac{\partial ( \hat{T}_{\alpha \gamma} \hat{T}_\beta{}^\gamma  )}{\partial \lambda} = 0\,.
\end{equation}
Here we have omitted the $(\lambda)$ superscripts on $\gamma_{\alpha \beta}^{(\lambda)}$ and $\widehat{T}_{\alpha \beta}^{(\lambda)} = T_{\alpha \beta}^{(\lambda)} - \gamma_{\alpha \beta}^{(\lambda)} T^{(\lambda)\rho}{}_\rho$.

The solutions of \eqref{eq:PDEsTT} are \eqref{TT_deformed_gamma_T}. In terms of the Fefferman-Graham quantities, the deformed boundary metric and stress tensor are
\begin{equation}\label{appendix_deformed_metric}
\begin{aligned}
\gamma^{(\lambda)}_{\alpha \beta} &= g^{(0)}_{\alpha \beta} - \frac{2\lambda}{8 \pi G \ell} g^{(2)}_{\alpha \beta} +  \frac{\lambda^2 }{\left(8 \pi G \ell \right)^2}  g^{(2)}_{\alpha \rho} g^{(2)}_{\sigma \beta} \gamma^{(0)\rho\sigma}
\\&= g^{(0)}_{\alpha \beta} - \lambda g^{(2)}_{\alpha \beta} + \lambda^2 g^{(4)}_{\alpha \beta}\,,
\end{aligned}
\end{equation}
and
\begin{equation}
\begin{aligned}\label{appendix_deformed_T}
\widehat{T}^{(\lambda)}_{\alpha \beta} &= \widehat{T}^{(0)}_{\alpha \beta} - \lambda \widehat{T}^{(0)}_{\alpha \rho} \, \widehat{T}^{(0)}_{\sigma \beta} \gamma^{(0) \rho \sigma}
\\&=\frac{1}{8 \pi G \ell} g^{(2)}_{\alpha \beta} - \frac{\lambda}{(8 \pi G \ell)^2} g^{(2)}_{\alpha \rho} g^{(2)}_{\sigma \beta} g^{(0)\rho \sigma}
\\&= \frac{1}{2} \left( g^{(2)}_{\alpha \beta} - 2\lambda g^{(4)}_{\alpha \beta} \right)\,,
\end{aligned}
\end{equation}
where we used \eqref{eq:g4} and work in conventions such that $4\pi G \ell=  1$. For the bad sign of the deformation parameter, these modified asymptotic boundary conditions can be interpreted as Dirichlet boundary conditions at a finite radial coordinate $\rho_c = - \frac{\lambda}{4 \pi G \ell}$.\footnote{One can see by straightforward algebra that the asymptotic conditions (\ref{appendix_deformed_metric}) are equivalent to fixing the induced metric to be $g_{\alpha \beta}^{(0)}$ at this value of $\rho_c$ if $\lambda < 0$. Another way to determine the relation between the bulk cutoff $\rho_c$ and the $T\overline{T}$ coupling $\lambda$ is using the trace flow equation $T^\alpha{}_{\alpha} \propto \lambda \det T_{\alpha \beta}$ \cite{McGough:2016lol,Kraus:2018xrn,Hartman:2018tkw}.}
Although we are primarily interested in the good sign of the deformation, it is convenient to express various quantities in terms of $\rho_c$. We have $\rho_c < 0$, and in this context, $\rho_c$ cannot be interpreted as a physical value of the coordinate $\rho$. Thus
\begin{equation}\label{app_T_rhoc}
\gamma^{(\lambda)}_{\alpha \beta} = g^{(0)}_{\alpha \beta} + \rho_c g^{(2)}_{\alpha \beta} + \rho_c^2 g^{(4)}_{\alpha \beta}\,, \quad \widehat{T}^{(\lambda)}_{\alpha \beta} =\frac{1}{2} \left( g^{(2)}_{\alpha \beta} + 2\rho_c g^{(4)}_{\alpha \beta} \right)\,.
\end{equation}
Specializing to a Ba\~{n}ados geometry \eqref{eq:BanadosMetric}, the boundary metric in Fefferman-Graham quantities is
\begin{equation}
\label{eq:kre2}
\gamma^{(\lambda)}_{\alpha \beta} dx^\alpha \, dx^\beta = du \, dv + \rho_c \left( \mathcal{L} (u) du^2 + \bar{\mathcal{L}}(v) dv^2 \right) +\rho_c^2 \mathcal{L} (u) \bar{\mathcal{L}}(v) du \, dv \,.
\end{equation}
We express \eqref{eq:kre2} as 
\begin{equation}
\gamma^{(\lambda)}_{\alpha \beta} dx^\alpha \, dx^\beta = dU \, dV \,,
\end{equation}
where $(U, V)$ are the undeformed coordinates
\begin{equation}
\label{eq:TTcoords}
dU = du + \rho_c \bar{\mathcal{L}}(v) dv, \quad dV = dv + \rho_c \mathcal{L}(u) du\,.
\end{equation}
In matrix form, we can define the state-dependent coordinate transformation in \eqref{eq:TTcoords} and its inverse as 
\begin{equation}
\begin{aligned}
\label{eq:statedepcoordTT}
\left( \begin{array} {c}
dU\\dV
\end{array}
\right) &= \left( \begin{array}{cc}
    1    & \quad \rho_c \bar{\mathcal{L}}(v)  \\
   \rho_c \mathcal{L}(u)    & \quad 1 
    \end{array} \right) \left( \begin{array} {c}
du\\dv
\end{array}
\right), \\ \left( \begin{array} {c}
du\\dv
\end{array}
\right) &= \frac{1}{1-\rho_c^2 \mathcal{L}(u) \bar{\mathcal{L}}(v)}  \left( \begin{array}{cc}
    1    & \quad -\rho_c \bar{\mathcal{L}}(v)  \\
   -\rho_c \mathcal{L}(u)    & \quad 1 
    \end{array} \right)\left( \begin{array} {c}
dU\\dV
\end{array}
\right)\,.
\end{aligned}
\end{equation}
Using \eqref{eq:statedepcoordTT}, we can write the boundary metric $g^{(0)}_{\alpha \beta}$ in the $(U, V)$ coordinates
\begin{equation}
\begin{aligned}
\label{eq:g0TT}
g^{(0)}_{\alpha \beta} dx^\alpha \, dx^\beta &= du \, dv\\& = \frac{(dU - \rho_c \bar{\mathcal{L}}(v) dV  ) (dV - \rho_c \mathcal{L}(u) dU  )}{(1-\rho_c^2 \mathcal{L}(u) \bar{\mathcal{L}}(v) ) ^2} \,,
\end{aligned}
\end{equation}
as well as the other Fefferman-Graham quantities:
\begin{equation}
\begin{aligned}
\label{eq:g2TT}
g^{(2)}_{\alpha \beta} dx^\alpha \, dx^\beta &= \mathcal{L}(u) du^2 + \bar{\mathcal{L}}(v) dv^2 \\&=  \mathcal{L}(u) \left( \frac{dU - \rho_c \bar{\mathcal{L}}(v) }{1-\rho_c^2 \mathcal{L}(u) \bar{\mathcal{L}}(v)  } \right)^2 + \bar{\mathcal{L}}(v) \left( \frac{dV - \rho_c \mathcal{L}(v)}{1-\rho_c^2 \mathcal{L}(u) \bar{\mathcal{L}}(v)  } \right)^2
\\&= \frac{(1+\rho_c^2 \mathcal{L}(u) \bar{\mathcal{L}}(v) ) (\mathcal{L} (u) dU^2 + \bar{\mathcal{L}} dV^2  ) - 4\rho_c \mathcal{L}(u) \bar{\mathcal{L}}(v) dU \, dV    }{(1-\rho_c^2 \mathcal{L} \bar{\mathcal{L}}(v)  )^2} \,, 
\end{aligned}
\end{equation}
and
\begin{equation}
\begin{aligned}
\label{eq:g4TT}
g^{(4)}_{\alpha \beta} dx^\alpha \, dx^\beta &= \mathcal{L}(u) \bar{\mathcal{L}}(v) du \, dv\\& =\mathcal{L}(u) \bar{\mathcal{L}}(v) \frac{(dU - \rho_c \bar{\mathcal{L}}(v) dV  ) (dV - \rho_c \mathcal{L}(u) dU  )}{(1-\rho_c^2 \mathcal{L}(u) \bar{\mathcal{L}}(v) ) ^2}\,.
\end{aligned}
\end{equation}
Proving \eqref{eq:g4TT} is straightforward:
\begin{equation}
\begin{aligned}
 g^{(4)}_{\alpha \beta} &= \frac{1}{4} g^{(2)}_{\alpha \rho} g^{(2)}_{\sigma \beta} g^{(0)\rho \sigma} \\&=\frac{1}{4} \left( \begin{array}{cc}
    \mathcal{L}(u)  & \quad  0 \\
    0  & \quad \bar{\mathcal{L}}(v)
 \end{array} \right)   \left( \begin{array}{cc}
   0  & \quad  2 \\
   2  & \quad 0
 \end{array} \right)   \left( \begin{array}{cc}
    \mathcal{L}(u)  & \quad  0 \\
    0  & \quad \bar{\mathcal{L}}(v)
 \end{array} \right)  \\& =\mathcal{L}(u) \bar{\mathcal{L}}(v) g^{(0)}_{ \alpha \beta}\,.
 \end{aligned}
\end{equation}
Substituting the expressions for $g^{(2)}_{\alpha \beta}$ and $g^{(4)}_{\alpha \beta}$ in \eqref{eq:g2TT} and \eqref{eq:g4TT} into the result (\ref{app_T_rhoc}) for the trace-reversed deformed stress tensor $\widehat{T}^{(\lambda)}_{\alpha \beta}$, we find that
\begin{equation}
\begin{aligned}
\widehat{T}^{(\lambda)}_{\alpha \beta} dx^\alpha \, dx^\beta &= \frac{1}{2} \left( g^{(2)}_{\alpha \beta} +2 \rho_c g^{(4)}_{\alpha \beta} \right) dx^\alpha \, dx^\beta 
\\&= \frac{\mathcal{L}(u) dU^2 + \bar{\mathcal{L}}(v) dV^2 - 2\rho_c \mathcal{L}(u) \bar{\mathcal{L}}(v) dU \, dV  }{2 (1-\rho_c^2 \mathcal{L} (u) \bar{\mathcal{L}}(v)  )} \,, 
\end{aligned}
\end{equation}
and trace-reversing to obtain the deformed stress tensor in the $(U, V)$ coordinates yields
\begin{equation}
\label{eq:TcomponentsTT}
T_{\alpha \beta}^{(\lambda)} dx^\alpha \, dx^\beta = \frac{\mathcal{L}(u) dU^2 + \bar{\mathcal{L}}(v) dV^2 + 2\rho_c \mathcal{L}(u) \bar{\mathcal{L}}(v) dU \, dV  }{2 (1-\rho_c^2 \mathcal{L}(u) \bar{\mathcal{L}}(v)  )}\,.
\end{equation}
It is straightforward to show that \eqref{eq:TcomponentsTT} obeys the $T\overline{T}$ trace flow equation and is conserved:
\begin{equation}
    \partial_V T^{(\lambda)}_{UU} + \partial_U T^{(\lambda)}_{VU} =  \partial_V T^{(\lambda)}_{UV} + \partial_U T^{(\lambda)}_{VV}   = 0\,.
\end{equation}
From this Fefferman-Graham analysis, we have therefore determined the deformed black hole solutions for constant $(\mathcal{L}(u), \bar{\mathcal{L}}(v)) \equiv (\mathcal{L}_\lambda, \bar{\mathcal{L}}_\lambda)$. In terms of the temporal and angular coordinates $\phi$ and $T$, \eqref{eq:TcomponentsTT} becomes 
\begin{equation}
\begin{aligned}
&T_{\alpha \beta}^{(\lambda)} dx^\alpha \, dx^\beta \\&= \frac{(\mathcal{L}_\lambda + \bar{\mathcal{L}}_\lambda -2 \rho_c \mathcal{L}_\lambda \bar{\mathcal{L}}_\lambda )dT^2+(\mathcal{L}_\lambda + \bar{\mathcal{L}}_\lambda +2 \rho_c \mathcal{L}_\lambda \bar{\mathcal{L}}_\lambda ) d\phi^2 + 2 (\mathcal{L}_\lambda - \bar{\mathcal{L}}_\lambda) d\phi \, dT}{2 (1-\rho_c^2 \mathcal{L}_\lambda \bar{\mathcal{L}}_\lambda  )} \,, 
\end{aligned}
\end{equation}
where $(U, V) = (\phi + T, \phi - T)$. Therefore, in the $(T, \phi)$ coordinates and restoring factors of $4 \pi G \ell$, we find that the deformed energy and angular momentum are
\begin{equation}
\begin{aligned}
\label{eq:rohioe2}
E_\lambda &= \int^R_0 d\phi~ T^{(\lambda)}_{TT} = \frac{R (\mathcal{L}_\lambda+ \bar{\mathcal{L}}_\lambda - 2 \rho_c \mathcal{L}_\lambda \bar{\mathcal{L}}_\lambda )}{8 \pi G \ell (1-\rho_c^2 \mathcal{L}_\lambda \bar{\mathcal{L}}_\lambda )}\,, \\ J_\lambda &= \int^R_0 d\phi~ T^{(\lambda)}_{T\phi} =  \frac{R (\mathcal{L}_\lambda - 
 \bar{\mathcal{L}}_\lambda)}{8 \pi G \ell (1-\rho_c^2 \mathcal{L}_\lambda \bar{\mathcal{L}}_\lambda )}\,.
\end{aligned}
\end{equation}
The functions $(\mathcal{L}_\lambda, \bar{\mathcal{L}}_\lambda)$ are fixed in terms of  $(\mathcal{L}_0, \bar{\mathcal{L}}_0)$ by equating the undeformed and deformed angular momenta and event horizon areas \cite{Guica:2019nzm}. This is possible because the $T\overline{T}$ flow preserves the boundary theory's degeneracy of states, which implies that the horizon area of the black hole is unchanged by the deformation. The angular momentum is holographically dual to the momentum $P_n$ of the state in the field theory, which is quantized in units of $\frac{1}{R}$ and thus cannot flow with the deformation parameter because $\lambda$ is continuous. We expect that these two assumptions should hold for \emph{any} stress tensor deformation of the boundary field theory (including root-$T\overline{T}$) since any flow equation for the spectrum, which is driven by a function of only energies and momenta will also preserve degeneracies.

We have already determined the angular momentum, so we now consider the horizon areas. The event horizon in the Fefferman-Graham gauge is at 
\begin{equation}
\label{eq:eventhorizonU}
\rho_h = \frac{1}{\sqrt{\mathcal{L}_0 \bar{\mathcal{L}}_0  }} \,, 
\end{equation}
and \eqref{eq:BanadosMetric} evaluated at \eqref{eq:eventhorizonU} is
\begin{equation}
\begin{aligned}
&ds^2|_{\rho_h = (\mathcal{L}_0 \bar{\mathcal{L}}_0 )^{-\frac{1}{2}} }  = \frac{\ell^2 \mathcal{L}_0 \bar{\mathcal{L}}_0 }{4} d\rho^2  \\&+ \left( \sqrt{\mathcal{L}_0} - \sqrt{\bar{\mathcal{L}}_0} \right)^2 dT^2 + \left( \sqrt{\mathcal{L}_0} + \sqrt{\bar{\mathcal{L}}_0} \right)^2 d\phi^2 +2 \left( \mathcal{L}_0 - \bar{\mathcal{L}}_0 \right) dTd\phi\,.
\end{aligned}
\end{equation}
For the deformed black hole metric, substituting \eqref{eq:g0TT} - \eqref{eq:g4TT} into \eqref{eq:BanadosMetric} evaluated at the event horizon 
\begin{equation}
\rho_h= \frac{1}{\sqrt{\mathcal{L}_\lambda \bar{\mathcal{L}}_\lambda  }} \,, 
\end{equation}
we obtain
\begin{equation}
\begin{aligned}
ds^2|_{\rho_h = (\mathcal{L}_\lambda \bar{\mathcal{L}}_\lambda  )^{-\frac{1}{2}}} &= \frac{\ell^2 \mathcal{L}_\lambda \bar{\mathcal{L}}_\lambda}{4} d\rho^2 \\&+ \frac{\left( \sqrt{\mathcal{L}_\lambda} - \sqrt{\bar{\mathcal{L}}_\lambda }   \right)^2 dT^2 + \left( \sqrt{\mathcal{L}_\lambda} + \sqrt{\bar{\mathcal{L}}_\lambda } \right)^2 d\phi^2 + 2 (\mathcal{L}_\lambda - \bar{\mathcal{L}}_\lambda  ) d\phi dT }{ \left(1+\rho_c \sqrt{\mathcal{L}_\lambda \bar{\mathcal{L}}_\lambda }\right)^2} \,, 
\end{aligned}
\end{equation}
which has an event horizon area
\begin{equation}
A^{(\lambda)} = \int^R_0 d\phi \sqrt{g_{\phi \phi}}|_{\rho_h = (\mathcal{L}_\lambda \bar{\mathcal{L}}_\lambda )^{-\frac{1}{2}} }  = R \frac{\sqrt{\mathcal{L}_\lambda} + \sqrt{\bar{\mathcal{L}}_\lambda}}{1+\rho_c \sqrt{\mathcal{L}_\lambda \bar{\mathcal{L}}_\lambda }}\,.
\end{equation}
Equating the undeformed and deformed event horizon areas and angular momenta, we arrive at the constraints for $(\mathcal{L}_\lambda, \bar{\mathcal{L}}_\lambda)$,
\begin{equation}
\label{eq:AreaAngularMomentumTT}
\sqrt{\mathcal{L}_0} + \sqrt{\bar{\mathcal{L}}_0} = \frac{\sqrt{\mathcal{L}_\lambda} + \sqrt{\bar{\mathcal{L}}_\lambda}   }{1+\rho_c \sqrt{\mathcal{L}_\lambda \bar{\mathcal{L}}_\lambda }}, \quad \mathcal{L}_0 - \bar{\mathcal{L}}_0 = \frac{ \mathcal{L}_\lambda - \bar{\mathcal{L}}_\lambda     }{1 -  \rho_c^2 \mathcal{L}_\lambda \bar{\mathcal{L}}_\lambda  }\,.
\end{equation}
The solution to \eqref{eq:AreaAngularMomentumTT} is
\begin{equation}
\begin{aligned}
\label{eq:solsTT1}
\mathcal{L}_\lambda &= \resizebox{.8\hsize}{!}{$\frac{- \left(1 + \rho_c (\mathcal{L}_0   -   \bar{\mathcal{L}}_0  )  \right) \sqrt{\rho_c^2 \left( \mathcal{L}_0 - \bar{\mathcal{L}}_0 \right)^2 - 2 \rho_c \left( \mathcal{L}_0 + \bar{\mathcal{L}}_0 \right) + 1  }  + \rho_c^2 \left( \mathcal{L}_0 - \bar{\mathcal{L}}_0  \right)^2 - 2\rho_c \bar{\mathcal{L}}_0 + 1  }{2 \rho_c^2 \mathcal{L}_0},$}\\
\bar{\mathcal{L}}_\lambda &= \resizebox{.8\hsize}{!}{$ \frac{- \left(1 - \rho_c (\mathcal{L}_0   -   \bar{\mathcal{L}}_0  )  \right) \sqrt{\rho_c^2 \left( \mathcal{L}_0 - \bar{\mathcal{L}}_0 \right)^2 - 2 \rho_c \left( \mathcal{L}_0 + \bar{\mathcal{L}}_0 \right) + 1  }  + \rho_c^2 \left( \mathcal{L}_0 - \bar{\mathcal{L}}_0  \right)^2 - 2\rho_c \mathcal{L}_0 + 1  }{2 \rho_c^2 \bar{\mathcal{L}}_0}\,.$}
\end{aligned}
\end{equation}
Substituting \eqref{eq:solsTT1} into the energy equation \eqref{eq:rohioe2}, we arrive at the well-established $T\overline{T}$-deformed energy expressed in terms of the field theory energy $E_0$ and momentum $P_0$,
\begin{equation}
\begin{aligned}
E_\lambda &= \frac{R}{8 \pi G \ell \rho_c} \left(1 - \sqrt{1 - 2 \rho_c \left(\mathcal{L}_0 + \bar{\mathcal{L}}_0\right)  + \rho_c^2 \left( \mathcal{L}_0 - \bar{\mathcal{L}}_0 \right)^2  } \right)
\\&= \frac{R}{2\lambda} \left( \sqrt{1 + \frac{4 \lambda E_0}{R} + \frac{4 \lambda^2 P_0^2}{R^2}  } -1\right)\,,
\end{aligned}
\end{equation}
where the undeformed energy $E_0$, angular momentum $J_0$ (which corresponds to the momentum $P_0$ in the CFT), and deformation parameter with units restored are
\begin{equation}
    E_0 = \frac{R}{8 \pi G \ell} (\mathcal{L}_0 + \bar{\mathcal{L}}_0)\,, \quad J_0 = \frac{R}{8 \pi G \ell} \left( \mathcal{L}_0 - \bar{\mathcal{L}}_0 \right) = P_0\,, \quad \lambda =- 4 \pi G \ell \rho_c \,.
\end{equation}

\section{Chern-Simons formalism}
\label{app:A2}
To obtain the $T\overline{T}$-deformed Chern-Simons connections, we use the coordinate transformation in \eqref{eq:statedepcoordTT} to obtain
\begin{equation}
\begin{aligned}
A(\rho_c)&= -\frac{1}{2\rho} L_0 d\rho + \frac{1}{\ell} \left( -\sqrt{\rho} \mathcal{L}_\lambda L_{-1} + \frac{1}{\sqrt{\rho}} L_1 \right) \left( \frac{dU - \rho_c \bar{\mathcal{L}}_\lambda dV}{1-\rho_c^2 \mathcal{L}_\lambda \bar{\mathcal{L}}_\lambda  } \right)\,,\\
\bar{A}(\rho_c) &= \frac{1}{2\rho} L_0 d\rho + \frac{1}{\ell} \left(\frac{1}{\sqrt{\rho}} L_{-1} - \sqrt{\rho} \bar{\mathcal{L}}_\lambda L_1 \right) \left( \frac{dV - \rho_c \mathcal{L}_\lambda dU  }{1-\rho_c^2 \mathcal{L}_\lambda \bar{\mathcal{L}}_\lambda} \right)\,.
\end{aligned}
\end{equation}
We can see that the deformed gauge fields obey a mixed boundary condition:
\begin{equation}
\rho_c \bar{\mathcal{L}}_\lambda A_U(\rho_c) + A_V(\rho_c) = 0\,, \quad \bar{A}_U(\rho_c) + \rho_c \mathcal{L}_\lambda \bar{A}_V(\rho_c) = 0\,.
\end{equation}
To convert the connections from the $(U, V)$ coordinates to the $(T, \phi)$ coordinates, we recall that
\begin{equation}
    \begin{aligned}
        A &= A_\alpha dx^\alpha 
        \\&= A_U dU + A_V dV
        \\&= \left( A_U + A_V \right) d\phi+\left( A_U - A_V \right) dT \,,  
    \end{aligned}
\end{equation}
yielding 
\begin{equation}
\begin{aligned}
A_\phi = A_U + A_V, \quad A_T = A_U - A_V\,, \quad \bar{A}_\phi = \bar{A}_U + \bar{A}_V\,, \quad \bar{A}_T = \bar{A}_U - \bar{A}_V\,.
\end{aligned}
\end{equation}
Hence
\begin{equation}
A_\phi (\rho_c) = \frac{1}{\ell} \frac{1-\rho_c \bar{\mathcal{L}}_\lambda}{1- \rho_c^2 \mathcal{L}_\lambda  \bar{\mathcal{L}}_\lambda } \left( -\sqrt{\rho} \mathcal{L}_\lambda  L_{-1} + \frac{1}{\sqrt{\rho}}L_1 \right) \,, \quad A_T (\rho_c) =  \frac{1+\rho_c \bar{\mathcal{L}}_\lambda}{1-\rho_c \bar{\mathcal{L}}_\lambda} A_\phi(\rho_c) \,, 
\end{equation}
and 
\begin{equation}
\bar{A}_\phi(\rho_c) = \frac{1}{\ell} \frac{1 - \rho_c \mathcal{L}_\lambda}{1-\rho_c^2 \mathcal{L}_\lambda \bar{\mathcal{L}}_\lambda}  \left(\frac{1}{\sqrt{\rho}} L_{-1} - \sqrt{\rho} \bar{\mathcal{L}}_\lambda L_1 \right) \,, \quad   \bar{A}_T(\rho_c) = - \frac{1+\rho_c \mathcal{L}_\lambda}{1- \rho_c \mathcal{L}_\lambda}  \bar{A}_\phi (\rho_c)\,.
\end{equation}
The boundary connections obey a similar relation as the bulk connections
\begin{equation}
\begin{aligned}
\label{eq:TTdeformedboundaryA}
a_\phi (\rho_c) =  \frac{1}{\ell} \frac{1-\rho_c\bar{\mathcal{L}}_\lambda}{1- \rho_c^2 \mathcal{L}_\lambda  \bar{\mathcal{L}}_\lambda } \left( -\mathcal{L}_\lambda L_{-1} + L_1\right)\,, \quad a_T (\rho_c) =  \frac{1+\rho_c \bar{\mathcal{L}}_\lambda}{1-\rho_c \bar{\mathcal{L}}_\lambda} a_\phi(\rho_c) \,, 
\end{aligned}
\end{equation}
and
\begin{equation}
\label{eq:TTdeformedboundaryAb}
\bar{a}_\phi (\rho_c) =  \frac{1}{\ell} \frac{1 - \rho_c \mathcal{L}_\lambda}{1-\rho_c^2 \mathcal{L}_\lambda \bar{\mathcal{L}}_\lambda} \left(  L_{-1} -  \bar{\mathcal{L}}_\lambda L_1   \right)\,, \quad \bar{a}_T(\rho_c) = - \frac{1 + \rho_c \mathcal{L}_\lambda}{1-\rho_c \mathcal{L}_\lambda} \bar{a}_\phi(\rho_c)\,.
\end{equation}
We may also study the black hole entropy and horizon areas using these deformed connections in the same way as we did in the root-$T\overline{T}$ deformed case around equation \eqref{eq:Entropy}. The analogs of the matrices $(\lambda_\phi, \bar{\lambda}_\phi)$ in (\ref{root_TT_lambda_matrices}), which are simply the diagonalized versions of $(a_\phi, \bar{a}_\phi)$, for the $T\overline{T}$-deformed connections \eqref{eq:TTdeformedboundaryAb}, are
\begin{equation}
    \begin{aligned}
        \lambda_\phi &= \frac{1}{\ell} \left( \begin{array}{cc}
           \frac{\left( 1- \rho_c \bar{\mathcal{L}}_\lambda \right) \sqrt{\mathcal{L}_\lambda} }{1-\rho_c^2 \mathcal{L}_\lambda \bar{\mathcal{L}}_\lambda   }  &\quad 0  \\
          0   & \quad -  \frac{\left( 1- \rho_c \bar{\mathcal{L}}_\lambda \right) \sqrt{\mathcal{L}_\lambda} }{1-\rho_c^2 \mathcal{L}_\lambda \bar{\mathcal{L}}_\lambda   }
        \end{array} \right)   \,, \\
        \bar{\lambda}_\phi &= \frac{1}{\ell} \left( \begin{array}{cc}
         -  \frac{\left( 1- \rho_c \mathcal{L}_\lambda \right) \sqrt{\bar{\mathcal{L}}_\lambda} }{1-\rho_c^2 \mathcal{L}_\lambda \bar{\mathcal{L}}_\lambda   }  &\quad 0  \\
          0   & \quad   \frac{\left( 1- \rho_c \mathcal{L}_\lambda \right) \sqrt{\bar{\mathcal{L}}_\lambda} }{1-\rho_c^2 \mathcal{L}_\lambda \bar{\mathcal{L}}_\lambda   }
        \end{array} \right) \,.
    \end{aligned}
\end{equation}
Using the equation $S = C \operatorname{Tr} \left( (\lambda_\phi - \bar{\lambda}_\phi)L_0 \right)$ for the entropy, which we quoted in \eqref{eq:Entropy}, we find an expression for the deformed entropy $S^{(\lambda)}$:
\begin{equation}
\label{eq:TTentropies}
     S^{(\lambda)} = \frac{C}{\ell} \left(  \frac{\sqrt{\mathcal{L}_\lambda} + \sqrt{\bar{\mathcal{L}}_\lambda} }{1+\rho_c \sqrt{\mathcal{L}_\lambda \bar{\mathcal{L}}_\lambda} }  \right)\,.
\end{equation}
Setting (\ref{eq:TTentropies}) equal to the undeformed entropy
\begin{align}
    S^{(0)} = \frac{C}{\ell} \left( \sqrt{\mathcal{L}_0} + \sqrt{\bar{\mathcal{L}}_0} \right) \,, 
\end{align}
then reproduces the area equation \eqref{eq:AreaAngularMomentumTT}.

Following \cite{He:2020hhm}, we can now read off the variation of the boundary action, which is compatible with the relations (\ref{eq:TTdeformedboundaryA}) and (\ref{eq:TTdeformedboundaryAb}) for the deformed boundary connections:
\begin{align}\label{app_CS_deformed_delta_S}
    \delta S = -\frac{\ell}{8 \pi G} \int_{\partial M} dT \, d\phi \, &\Bigg( \operatorname{Tr} \left[ \left( a_T(\rho_c) - \frac{1+\rho_c \bar{\mathcal{L}}_\lambda   }{1-  \rho_c \bar{\mathcal{L}}_\lambda    } a_\phi (\rho_c) \right) \delta a_\phi (\rho_c) \right] \nonumber \\
    &\qquad - \operatorname{Tr} \left[ \left( \bar{a}_T (\rho_c) + \frac{1 + \rho_c \mathcal{L}_\lambda  }{1 - \rho_c \mathcal{L}_\lambda} \bar{a}_\phi (\rho_c)  \right) \delta \bar{a}_\phi (\rho_c)   \right] \Bigg) \,.
\end{align}
We see that, when the constraints (\ref{eq:TTdeformedboundaryA}) and (\ref{eq:TTdeformedboundaryAb}) are satisfied, the variation (\ref{app_CS_deformed_delta_S}) collapses to $\delta S_{\text{bdry}} = 0$. This guarantees a well-defined variational principle.

To determine this boundary action in terms of $\mathcal{L}_\lambda$, $\bar{\mathcal{L}}_\lambda$, and $\rho_c$, we must first evaluate the variations of the boundary connections. The variations of \eqref{eq:TTdeformedboundaryA} and \eqref{eq:TTdeformedboundaryAb}  are
\begin{equation}
\begin{aligned}
\label{eq:variationsTT}
\delta a_\phi (\rho_c) &= \frac{(1 -  \rho_c \bar{\mathcal{L}}_\lambda ) \left( L_{-1} - \rho_c^2 \bar{\mathcal{L}}_\lambda L_1 \right) \delta \mathcal{L}_\lambda  -\rho_c \left(  \mathcal{L}_\mu L_{-1} - L_1   \right) \left( 1-\rho_c \mathcal{L}_\lambda \right) \delta \bar{\mathcal{L}}_\lambda}{\ell \left(1 - \rho_c^2 \mathcal{L}_\lambda \bar{\mathcal{L}}_\lambda \right)^2}\,,\\
\delta \bar{a}_\phi (\rho_c) &= \frac{- \rho_c \left(L_{-1} - \bar{\mathcal{L}}_\mu L_1 \right) \left( 1 -\rho_c \bar{\mathcal{L}}_\lambda \right) \delta \mathcal{L}_\lambda -\left( \rho_c^2 \mathcal{L}_\mu L_{-1} - L_1 \right)  \left( 1 - \rho_c \mathcal{L}_\lambda \right)  \delta \bar{\mathcal{L}}_\lambda }{\ell \left( 1 - \rho_c^2 \mathcal{L}_\lambda \bar{\mathcal{L}}_\lambda \right)^2}\,.
\end{aligned}
\end{equation}
The variation of the boundary piece is
\begin{equation}
\begin{aligned}
\label{eq:boundarypieceTT}
\delta S_{\text{bdry}}=  \frac{\ell}{8 \pi G} \int_{\partial M} dT \, d\phi & \bigg(    \frac{1+\rho_c\bar{\mathcal{L}}_\lambda   }{1-   \rho_c \bar{\mathcal{L}}_\lambda    } \operatorname{Tr} \left(a_\phi (\rho_c)  \delta a_\phi (\rho_c)\right) \\& + \frac{1 + \rho_c  \mathcal{L}_\lambda  }{1 - \rho_c \mathcal{L}_\lambda} \operatorname{Tr} \left( \bar{a}_\phi (\rho_c)   \delta \bar{a}_\phi (\rho_c)  \right)  \bigg) \,, 
\end{aligned}
\end{equation}
 and using \eqref{eq:variationsTT}, the traces evaluate to
\begin{equation}
    \begin{aligned}
    \label{eq:TTtraces}
 \operatorname{Tr} \left(a_\phi (\rho_c)  \delta a_\phi (\rho_c)\right) &= \resizebox{.6\hsize}{!}{$ \frac{(1- \rho_c \bar{\mathcal{L}}_\lambda )^2 (1 + \rho_c^2 \mathcal{L}_\lambda \bar{\mathcal{L}}_\lambda)  \delta \mathcal{L}_\lambda - 2\rho_c \mathcal{L}_\lambda (1-\rho_c \mathcal{L}_\lambda)(1-\rho_c \bar{\mathcal{L}}_\lambda )   \delta \bar{\mathcal{L}}_\lambda }{\ell^2(1-\rho_c^2 \mathcal{L}_\lambda \bar{\mathcal{L}}_\lambda  )^3}\,,$} \\
            \operatorname{Tr} \left(\bar{a}_\phi (\rho_c)  \delta \bar{a}_\phi (\rho_c)\right) &= \resizebox{.6\hsize}{!}{$\frac{ -2\rho_c \bar{\mathcal{L}}_\lambda (1-\rho_c \mathcal{L}_\lambda)(1-\rho_c \bar{\mathcal{L}}_\lambda)   \delta \mathcal{L}_\lambda + (1- \rho_c \mathcal{L}_\lambda)^2 (1+\rho_c^2 \mathcal{L}_\lambda \bar{\mathcal{L}}_\lambda) \delta \bar{\mathcal{L}}_\lambda }{\ell^2 (1-\rho_c^2 \mathcal{L}_\lambda \bar{\mathcal{L}}_\lambda  )^3} \,.$}
    \end{aligned}
\end{equation}
 Substituting \eqref{eq:TTtraces} into \eqref{eq:boundarypieceTT}, the varied boundary action in terms of $\delta \mathcal{L}_\lambda$ and $\delta \bar{\mathcal{L}}_\lambda$ is
\begin{equation}
    \delta S_{\text{bdry}} = \frac{1}{8 \pi G \ell} \int_{\partial M} dT \, d\phi \left( \left(\frac{1- \rho_c \bar{\mathcal{L}}_\lambda  }{1-\rho_c^2 \mathcal{L}_\lambda \bar{\mathcal{L}}_\lambda  }\right)^2 \delta \mathcal{L}_\lambda   + \left(\frac{1- \rho_c\mathcal{L}_\lambda  }{1-\rho_c^2 \mathcal{L}_\lambda \bar{\mathcal{L}}_\lambda  }\right)^2 \delta \bar{\mathcal{L}}_\lambda \right) \,, 
\end{equation}
from which $S_{\text{bdry}}$ can be read off as
\begin{equation}
\label{eq:boundaryactionTT}
    S_{\text{bdry}} = \frac{1}{8 \pi G \ell} \int_{\partial M} dT \, d\phi \, \frac{\mathcal{L}_\lambda  + \bar{\mathcal{L}}_\lambda - 2 \rho_c \mathcal{L}_\lambda  \bar{\mathcal{L}}_\lambda }{1 - \rho_c^2 \mathcal{L}_\lambda \bar{\mathcal{L}}_\lambda}\,.
\end{equation}
After integration over $\phi$ in \eqref{eq:boundaryactionTT}, we find that
\begin{align}
    S_{\text{bdry}} &=  \int \, d T \, \frac{R \left( \mathcal{L}_\lambda  + \bar{\mathcal{L}}_\lambda - 2 \rho_c \mathcal{L}_\lambda  \bar{\mathcal{L}}_\lambda  \right) }{8\pi G \ell (1 - \rho_c^2 \mathcal{L}_\lambda \bar{\mathcal{L}}_\lambda)} \, \nonumber \\
    &=  \int \, d T \, E_\lambda \,, 
\end{align}
where we have used the expression for $E_\lambda$ in \eqref{eq:rohioe2}. Therefore, the boundary Lagrangian density in the Chern-Simons formalism agrees with the deformed mass (or energy) of the bulk spacetime as computed in the metric formalism.

We emphasize again that it was not clear \emph{a priori} that the boundary Chern-Simons action would necessarily reproduce the mass of the deformed spacetime. Although this is true in the undeformed theory, after adding a boundary deformation that implements mixed boundary conditions in the bulk, one needs to compute the Hamiltonian to argue that the Chern-Simons boundary action will agree with the deformed spacetime mass in general. However, in this case, we have seen by explicit computation that the two agree, at least for the class of Ba\~{n}ados-type solutions we are considering.

\chapter{Perturbative Actions and Modified Scalar Feynman Diagrams}
\section{Perturbative \texorpdfstring{$f(T^\alpha{}_\alpha, T^{\alpha \beta} T_{\alpha \beta})$}{TT}-Deformed Actions}\label{app:TTn}

Throughout this paper, we have considered various deformations which are constructed from the energy-momentum tensor. Although the most important examples within this class are the $T\overline{T}$ and root-$T\overline{T}$ flows, it appears that \emph{general} stress tensor deformations nonetheless have interesting properties -- for instance, we have shown that every parameterized family of interacting $2d$ chiral boson theories which enjoys non-manifest Lorentz invariance admits an interpretation as a stress tensor deformation. This is a $2d$ analog of similar theorems about $4d$ theories of duality-invariant electrodynamics \cite{Ferko:2023wyi} or $6d$ chiral tensor theories \cite{Ferko:2024zth}.

Motivated by these observations, one may wish to study $2d$ deformations by other functions of the energy-momentum tensor, besides the ones considered in the body of this manuscript. One way to do this is to solve the resulting flow equations perturbatively, i.e. order-by-order in the deformation parameter. In this appendix we will use $g$ for the parameter of a general stress tensor flow, which is not to be confused with the metric $g_{\alpha \beta}$ or its determinant.

Let us therefore consider the following general class of operators in $2d$ which can be expressed in terms of the two independent Lorentz scalars that can be built from the stress tensor, namely $\operatorname{Tr} ( T ) = T^\alpha{}_\alpha$ and $\operatorname{Tr} ( T^2 ) = T^{\alpha \beta} T_{\alpha \beta}$:
\begin{equation}\label{eq:TTn}
 f( T^\alpha{}_\alpha, T^{\alpha \beta} T_{\alpha \beta})\, .
\end{equation}
We note that all higher traces of the stress tensor, $\operatorname{Tr}( T^n )$ for $n > 2$, can be expressed in terms of these two lower traces. Given such an operator, we wish to study the flow equation\footnote{One can also consider flows driven by a function $f$ which has explicit dependence on the deformation parameter $g$. For instance, the so-called $T\overline{T} + \Lambda_2$ deformation is defined by performing a $T\overline{T}$ deformation and then activating a cosmological constant proportional to $\frac{1}{\lambda}$. See \cite{Gorbenko:2018oov, Lewkowycz:2019xse, Coleman:2021nor, Shyam:2021ciy, Torroba:2022jrk, Batra:2024kjl} for further details.}
\begin{equation}
\label{eq:TTn op}
    \frac{\partial S^{(g)}}{\partial g} = \int d^2x \, E \,  f( T^\alpha_\alpha, T^{\alpha \beta} T_{\alpha \beta}) \, .
\end{equation}
The solution to \eqref{eq:TTn op} can be written as a series expansion,
\begin{equation}\label{perturbative_S}
      S^{(g)} = S^{(0)} +\sum^\infty_{m=1} \frac{g^m}{m}\int d^2x \, E \,   f( T^\alpha_\alpha, T^{\alpha \beta} T_{\alpha \beta})_{m-1} \, ,
\end{equation}
where we write $ f( T^\alpha{}_\alpha, T^{\alpha \beta} T_{\alpha \beta})_{m}$ for the term of order $g^{m}$ in the expression for the $f$ operator computed from the action at order $g^{m-1}$. Because each term in this expansion only depends on the data of lower-order terms, one can build up the solution iteratively in powers of $g$.

As in section \ref{sec:classical} of the main text, we will work in the tetrad formalism with a Lorentzian tangent-space metric and with spacetime coordinates $x^\alpha = ( t, \theta )$. A general spacetime metric can therefore be expanded in terms of vielbeins as
\begin{equation}
    g_{\alpha \beta} =E^a{}_\alpha E^b{}_\beta \eta_{ab} = -\left( \begin{array}{cc}
     2 E^+{}_t E^-{}_t    & \quad   E^+{}_t  E^-{}_\theta+  E^-{}_t  E^+{}_\theta \\
       E^+{}_t  E^-{}_\theta+  E^-{}_t  E^+{}_\theta  & \quad  2 E^+{}_\theta E^-{}_\theta  
    \end{array} \right) \, .
\end{equation}
The stress tensor associated with a general action $S$, which has been coupled to gravity using the vielbeins $E^a{}_\alpha$, can be written as
\begin{equation}\label{perturbative_stress_tensor}
    T^\alpha{}_\beta= - \frac{1}{E} \frac{\partial S}{\partial E^a{}_\alpha} E^a{}_\beta = - \frac{1}{E}\left( \begin{array}{cc}
      \frac{\partial S}{\partial E^+{}_t} E^+{}_t + \frac{\partial S}{\partial E^-{}_t} E^-{}_t   & \quad \frac{\partial S}{\partial E^+{}_\theta} E^+{}_t + \frac{\partial S}{\partial E^-{}_\theta} E^-{}_t  \\ \frac{\partial S}{\partial E^+{}_t} E^+{}_\theta + \frac{\partial S}{\partial E^-{}_t} E^-{}_\theta
         & \quad     \frac{\partial S}{\partial E^+{}_\theta} E^+{}_\theta + \frac{\partial S}{\partial E^-{}_\theta} E^-{}_\theta 
    \end{array} \right)\,.
\end{equation}
We will use the general expression (\ref{perturbative_stress_tensor}) for the stress tensor, along with the expansion (\ref{perturbative_S}), to perturbatively solve the flow equation (\ref{eq:TTn op}) for various choices of the $ f$ operator. 

We begin by finding perturbative solutions for some of the flow equations considered in the main text, before generalizing to other deformations which were not considered in the body. In our examples, we compute the stress tensor (\ref{perturbative_stress_tensor}) using the vielbein formalism due to computational speed in \texttt{Mathematica}, but we note that the metric formalism gives identical results.

\emph{Root-$T\overline{T}$ Perturbative Flow for Multiple Bosons}

For our first example, we will consider the perturbative solution to the root-$T\overline{T}$ flow equation for an arbitrary number of non-chiral bosons $\varphi^i$, $i = 1, \ldots, N$. This flow equation was first solved in closed-form in \cite{Ferko:2022cix}.

We take a seed action which describes $N$ free massless bosons in Lorentzian signature,
\begin{equation}
\begin{aligned}\label{eq:seedapp}
        S^{(0)} &=  \frac{1}{2} \int d^2x \sqrt{-g} g^{\alpha \beta} \partial_\alpha \varphi^i \partial_\beta \varphi^i
    \\&= \int d^2x \frac{ E^-{}_\theta E^+{}_\theta \dot{\varphi}^i \dot{\varphi}^i + E^-{}_t E^+{}_t \varphi^{\prime i} \varphi^{\prime i} - \left( E^-{}_\theta E^+{}_t + E^-{}_t E^+{}_\theta \right) \dot{\varphi}^i \varphi^{\prime i}  }{E} \, ,
\end{aligned}
\end{equation}
which have been coupled to gravity using the tetrad formalism. We then deform using the root-$T\overline{T}$ operator, which corresponds to the general $ f( T^\alpha{}_\alpha, T^{\alpha \beta} T_{\alpha \beta})$ operator of equation (\ref{eq:TTn}) being
\begin{equation}
    f( T^\alpha{}_\alpha, T^{\alpha \beta} T_{\alpha \beta}) = \mathcal{R}^{(\gamma)} = \frac{1}{\sqrt{2}} \sqrt{ T^{\alpha \beta} T_{\alpha \beta} - \frac{1}{2} \left( T^\alpha{}_\alpha \right)^2 } \, .
\end{equation}
In this case, the perturbative solution (\ref{perturbative_S}) to the flow equation takes the form
\begin{equation}\label{eq:ansatzLorentz}
    S^{(\gamma)} = S^{(0)} +\sum^\infty_{m=1} \frac{\gamma^m}{m}\int d^2x E~ \mathcal{R}^{(\gamma)}_{m-1} \, .
\end{equation}
Following the conventions in the main text, we use the symbol $\gamma$ for the flow parameter of a root-$T\overline{T}$ deformation, rather than the variable $g$ which stood for the parameter in a general deformation above.

The first few terms in this perturbative expansion are
\begin{equation}
\begin{aligned}\label{eq:coeffsLorentz}
    \mathcal{R}^{(\gamma)}_0|_{\text{flat}} &= \frac{1}{2} \sqrt{\left( \dot{\varphi}^i - \varphi^{\prime i} \right) \left( \dot{\varphi}^i - \varphi^{\prime i} \right) \left( \dot{\varphi}^j + \varphi^{\prime j} \right) \left( \dot{\varphi}^j + \varphi^{\prime j} \right)  }\,, \\  \mathcal{R}^{(\gamma)}_1|_{\text{flat}} &= \frac{1}{2} \left( -\dot{\varphi}^i \dot{\varphi}^i +\varphi^{\prime i} \varphi^{\prime i} \right) \,, \quad 
      \mathcal{R}^{(\gamma)}_2|_{\text{flat}} = \frac{1}{2}    \mathcal{R}^{(\gamma)}_0|_{\text{flat}}\,, \quad
      \mathcal{R}^{(\gamma)}_3|_{\text{flat}} = \frac{1}{6}     \mathcal{R}^{(\gamma)}_1|_{\text{flat}}\,, \\  
          \mathcal{R}^{(\gamma)}_4|_{\text{flat}} &=  \frac{1}{24}     \mathcal{R}^{(\gamma)}_0|_{\text{flat}}\,, \quad \mathcal{R}^{(\gamma)}_5|_{\text{flat}} = \frac{1}{120} \mathcal{R}^{(\gamma)}_1|_{\text{flat}}\,,
\end{aligned}
\end{equation}
where ``flat'' means that we have set the vielbeins to their flat-space values \eqref{flat_vielbeins}.

We note that the quantities appearing in (\ref{eq:coeffsLorentz}) can be written in terms of the manifestly Lorentz-invariant combinations
\begin{equation}
\begin{aligned}
\left( \partial_\mu \varphi^{i} \partial^{\nu} \varphi^{i} \right) \left( \partial_\nu \varphi^{j} \partial^{\mu} \varphi^{j} \right)   &= \left( -\dot{\varphi}^i \dot{\varphi}^j + \varphi^{\prime i} \varphi^{\prime j}\right) \left( -\dot{\varphi}^i \dot{\varphi}^j + \varphi^{\prime i} \varphi^{\prime j}\right)
\\&= (\dot{\varphi}^i \dot{\varphi}^i)^2 + (\varphi^{\prime i} \varphi^{\prime i})^2 -2 (\dot{\varphi}^{i} \varphi^{\prime i})^2  \, , 
\end{aligned}
\end{equation}
and
\begin{equation}
\begin{aligned}
   & 2 \left( \partial_\mu \varphi^{i} \partial^{\nu} \varphi^{i} \right) \left( \partial_\nu \varphi^{j} \partial^{\mu} \varphi^{j} \right) - \left( \partial_\mu \varphi^{i} \partial^{\mu} \varphi^{i} \right)^2 \\&= \left( \dot{\varphi}^i \dot{\varphi}^i \right)^2 + \left( \varphi^{\prime i} \varphi^{\prime i} \right)^2 - 4 \left( \dot{\varphi}^i \varphi^{\prime i} \right)^2 + 2 \dot{\varphi}^i \dot{\varphi}^i \varphi^{\prime j} \varphi^{\prime j}
    \\&=\left( \dot{\varphi}^i - \varphi^{\prime i} \right) \left( \dot{\varphi}^i - \varphi^{\prime i} \right) \left( \dot{\varphi}^j + \varphi^{\prime j} \right) \left( \dot{\varphi}^j + \varphi^{\prime j} \right) \,.
\end{aligned}
\end{equation}
In terms of these quantities, one finds that the perturbative expansion to the flow equation converges to the solution \eqref{modscalar_lorentz_invariant_lagrangian},

\begin{equation}
\begin{aligned}
   &  S^{(\gamma)} = S^{(0)} + \int dt \, d\theta \, \bigg( \gamma   \mathcal{R}^{(\gamma)}_0|_{\text{flat}} + \frac{\gamma^2}{2}   \mathcal{R}^{(\gamma)}_1|_{\text{flat}} + \frac{\gamma^3}{6}   \mathcal{R}^{(\gamma)}_0|_{\text{flat}} \\& + \frac{\gamma^4}{24}   \mathcal{R}^{(\gamma)}_1|_{\text{flat}} +\frac{\gamma^5}{120}   \mathcal{R}^{(\gamma)}_0|_{\text{flat}}+ \frac{\gamma^6}{720}   \mathcal{R}^{(\gamma)}_1|_{\text{flat}}  + \cdots  \bigg)
    \\&= \frac{1}{2} \int dt \, d\theta \, \bigg [ \partial_\alpha \varphi^{i} \partial^{\alpha} \varphi^{i}  \left(1 + \frac{\gamma^2}{2}+ \frac{\gamma^4}{24} + \frac{\gamma^6}{720} + \mathcal{O}(\gamma^8) \right) \\&+ \sqrt{   2 \left( \partial_\mu \varphi^{i} \partial^{\nu} \varphi^{i} \right) \left( \partial_\nu \varphi^{j} \partial^{\mu} \varphi^{j} \right) - \left( \partial_\mu \varphi^{i} \partial^{\mu} \varphi^{i} \right)^2 } \left(  \gamma + \frac{\gamma^3}{6} + \frac{\gamma^5}{120} + \mathcal{O}(\gamma^7) \right)  \bigg] 
    \\&= \frac{1}{2} \int dt \, d\theta \, \bigg[  \cosh (\gamma) \partial_\alpha \varphi^{i} \partial^{\alpha} \varphi^i \\& + \sinh (\gamma)\sqrt{    2 \left( \partial_\mu \varphi^{i} \partial^{\nu} \varphi^{i} \right) \left( \partial_\nu \varphi^{j} \partial^{\mu} \varphi^{j} \right) - \left( \partial_\mu \varphi^{i} \partial^{\mu} \varphi^{i} \right)^2} \bigg] \, .
\end{aligned}
\end{equation}

\emph{Root-$T\overline{T}$ Perturbative Flow for Chern-Simons}

An almost identical calculation can be performed to study the perturbative root-$T\overline{T}$ deformation of the Chern-Simons boundary action given in \eqref{free_CS_bdry}. The first few terms in the expansion are
\begin{equation}
\begin{aligned}
    \mathcal{R}^{(\gamma)}_0|_{\text{flat}} &= \frac{1}{4\pi} \sqrt{ \left( k^{ij} A_{iw} A_{jw} + \bar{k}^{\overline{i} \overline{j}} \bar{A}_{\overline{i}w} \bar{A}_{\overline{j}w} \right) \left( k^{mn} A_{m\bar{w}} A_{n\bar{w}} + \bar{k}^{\overline{m} \overline{n}} \bar{A}_{\overline{m}\bar{w}} \bar{A}_{\overline{n}\bar{w}} \right) }\,, \\
      \mathcal{R}^{(\gamma)}_1|_{\text{flat}} &=  -\frac{1}{4\pi}  \left(k^{ij} A_{i w} A_{j \bar{w}}+\bar{k}^{\overline{i} \overline{j}} \bar{A}_{\overline{i} w} \bar{A}_{\overline{j} \bar{w}}\right)\,, \\ 
      \mathcal{R}^{(\gamma)}_2|_{\text{flat}} &= \frac{1}{2}    \mathcal{R}^{(\gamma)}_0|_{\text{flat}}\,, \quad    \mathcal{R}^{(\gamma)}_3|_{\text{flat}} = \frac{1}{6}    \mathcal{R}^{(\gamma)}_1|_{\text{flat}}\,,
\end{aligned}
\end{equation}
where now ``flat'' means that we have set the vielbeins equal to the values \eqref{euclidean_flat} appropriate for a flat \emph{Euclidean} tangent space metric, following the conventions of section \ref{sec:cs}. 

Therefore, the root-$T\overline{T}$-deformed Chern-Simons boundary action is
\begin{equation}
\begin{aligned}
   & I^{(\gamma)}_{\partial \mathcal{M}_3} \\&=  \frac{i}{8 \pi} \int_{\partial \mathcal{M}_3}  dw d\bar{w} \left( -\left(k^{ij} A_{i w} A_{j \bar{w}}+\bar{k}^{\overline{i} \overline{j}} \bar{A}_{\overline{i} w} \bar{A}_{\overline{j} \bar{w}} \right) \left(1 + \frac{\gamma^2}{2} + \mathcal{O}(\gamma^4) \right)\right)
    \\&+ \frac{i}{8 \pi}\int_{\partial \mathcal{M}_3}  dw d\bar{w} \sqrt{ \left( k^{ij} A_{iw} A_{jw} + \bar{k}^{\overline{i} \overline{j}} \bar{A}_{\overline{i}w} \bar{A}_{\overline{j}w} \right) \left( k^{mn} A_{m\bar{w}} A_{n\bar{w}} + \bar{k}^{\bar{m} \bar{n}} \bar{A}_{\bar{m}\bar{w}} \bar{A}_{\bar{n}\bar{w}} \right) } \\& \cdot \left( \gamma + \frac{\gamma^3}{6} + \mathcal{O}(\gamma^5) \right)
    \\&= \frac{i}{8 \pi } \int_{\partial \mathcal{M}_3}  dw d\bar{w} \bigg[- \cosh (\gamma) \left(k^{ij} A_{i w} A_{j \bar{w}}+\bar{k}^{\overline{i} \overline{j}} \bar{A}_{\overline{i} w} \bar{A}_{\overline{j} \bar{w}} \right)  \\&+ \sinh (\gamma)  \sqrt{ \left( k^{ij} A_{iw} A_{jw} + \bar{k}^{\overline{i} \overline{j}} \bar{A}_{\overline{i}w} \bar{A}_{\overline{j}w} \right) \left( k^{mn} A_{m\bar{w}} A_{n\bar{w}} + \bar{k}^{\bar{m} \bar{n}} \bar{A}_{\bar{m}\bar{w}} \bar{A}_{\bar{n}\bar{w}} \right) }   \bigg] \, .
\end{aligned}
\end{equation}

\emph{$T\overline{T}$ Perturbative Flow for a Single Boson}

For our next example, we will consider the irrelevant $T\overline{T}$ flow rather than the marginal root-$T\overline{T}$ flow. For simplicity, we will restrict to a deformation of a single bosonic field $\varphi$ whose seed action is that of a free massless field. From the general $ f( T^\alpha{}_\alpha, T^{\alpha \beta} T_{\alpha \beta})$ deformation of (\ref{eq:TTn}), we recover the usual $T\overline{T}$ deformation by taking
\begin{equation}
    f( T^\alpha{}_\alpha, T^{\alpha \beta} T_{\alpha \beta}) = -\frac{1}{2} \left(  T^\alpha{}_\beta T^\beta{}_\alpha-  (T^{\alpha}{}_\alpha)^2 \right) \, .
\end{equation}
Evaluating a few of the terms in the perturbative expansion, we find
\begin{equation}
\begin{aligned}
 T\overline{T}_{0}|_{\text{flat}} &=- \frac{1}{4} \left( - \dot{\varphi}^2 + \varphi^{\prime 2 } \right)^2\,, \quad  T\overline{T}_{1}|_{\text{flat}} = \frac{1}{2}\left( - \dot{\varphi}^2 + \varphi^{\prime 2 } \right)^3\,, \\   T\overline{T}_{2}|_{\text{flat}} &= -\frac{15}{16}\left( - \dot{\varphi}^2 + \varphi^{\prime 2 } \right)^4 \, .
\end{aligned}
\end{equation}
This series expansion then converges to the well-known $T\overline{T}$-deformed action,
\begin{equation}
\begin{aligned}
    S^{(\lambda)} &= \int dt d\theta  \bigg[ \frac{1}{2} \left( - \dot{\varphi}^2 + \varphi^{\prime 2 } \right) - \frac{\lambda}{4} \left( - \dot{\varphi}^2 + \varphi^{\prime 2 } \right)^2\\&+\frac{\lambda^2}{4}\left( - \dot{\varphi}^2 + \varphi^{\prime 2 } \right)^3 -\frac{5 \lambda^3}{16}\left( - \dot{\varphi}^2 + \varphi^{\prime 2 }\right)^4 +\cdots \bigg]
    \\&= \int dt d\theta \frac{1}{2\lambda}\bigg[\sqrt{1+2\lambda \left(  - \dot{\varphi}^2 + \varphi^{\prime 2 } \right)} -1\bigg]\,.
\end{aligned}
\end{equation}

\newpage

\emph{$T\overline{T}^{\frac{1}{3}}$ Perturbative Flow for Multiple Bosons}

Next we turn our attention to a deformation which was not considered in the body of this manuscript. Consider a deformation by the relevant $T\overline{T}^{\frac{1}{3}}$ operator, which we define by
\begin{equation}
    T\overline{T}^{\frac{1}{3}} =  \frac{1}{2} \left(\frac{1}{2} \left(T^\alpha{}_\beta T^\beta{}_\alpha-  (T^{\alpha}{}_\alpha)^2 \right)  \right)^{\frac{1}{3}}\,. 
\end{equation}

We will again consider a seed action for $N$ massless free bosons, given in equation \eqref{eq:seedapp}. The perturbative expansion for the $T\overline{T}^{\frac{1}{3}}$-deformed action takes the form
\begin{equation}
    S^{(\lambda)} = S^{(0)} +\sum^\infty_{m=1} \frac{\lambda^m}{m}\int d^2x E~ T\overline{T}^{\frac{1}{3}}_{m-1} \, ,
\end{equation}
and a few of the coefficients are
\begin{equation}
\begin{aligned}
     T\overline{T}^{\frac{1}{3}}_0|_{\text{flat}} &= \frac{1}{2^{\frac{5}{3}}} \bigg[\left( \dot{\varphi}^i - \varphi^{\prime i} \right) \left( \dot{\varphi}^i - \varphi^{\prime i} \right) \left( \dot{\varphi}^j + \varphi^{\prime j} \right) \left( \dot{\varphi}^j + \varphi^{\prime j} \right)\bigg]^{\frac{1}{3}}\,, \\    T\overline{T}^{ \frac{1}{3}}_1|_{\text{flat}} &= \frac{-\dot{\varphi}^{i} \dot{\varphi}^{i} + \varphi^{\prime i}  \varphi^{\prime i} }{9 \cdot 2^{\frac{1}{3}} \bigg[ \left( \dot{\varphi}^i - \varphi^{\prime i} \right) \left( \dot{\varphi}^i - \varphi^{\prime i} \right) \left( \dot{\varphi}^j + \varphi^{\prime j} \right) \left( \dot{\varphi}^j + \varphi^{\prime j} \right) \bigg]^{\frac{1}{3}}}\,, \\
          T\overline{T}^{\frac{1}{3}}_2|_{\text{flat}} &= - \frac{\left( \dot{\varphi}^i \dot{\varphi}^i \right)^2 + (\varphi^{\prime i} \varphi^{\prime i} )^2 + 12 (\dot{\varphi}^i \varphi^{\prime i} )^2 - 14 \dot{\varphi}^i \dot{\varphi}^i \varphi^{\prime j} \varphi^{\prime j} }{216 \bigg[  \left( \dot{\varphi}^i - \varphi^{\prime i} \right) \left( \dot{\varphi}^i - \varphi^{\prime i} \right) \left( \dot{\varphi}^j + \varphi^{\prime j} \right) \left( \dot{\varphi}^j + \varphi^{\prime j} \right)  \bigg]}\,.
\end{aligned}
\end{equation}
For this deformation, it does not seem possible to find an all-orders closed-form solution to the flow equation, but the perturbative $T\overline{T}^{\frac{1}{3}}$-deformed action to $\mathcal{O} \left( \lambda^{3} \right)$ is
\begin{equation}
\begin{aligned}
    S^{(\lambda)} &= \int dt \, d\theta \, \bigg( \frac{1}{2} \left( - \dot{\varphi}^i  \dot{\varphi}^i +  \varphi^{\prime i} \varphi^{\prime i}  \right)  \\&+ \frac{\lambda}{2^{\frac{5}{3}}} \bigg[\left( \dot{\varphi}^i - \varphi^{\prime i} \right) \left( \dot{\varphi}^i - \varphi^{\prime i} \right) \left( \dot{\varphi}^j + \varphi^{\prime j} \right) \left( \dot{\varphi}^j + \varphi^{\prime j} \right)\bigg]^{\frac{1}{3}}\\&+ \frac{\lambda^2}{9 \cdot 2^{\frac{4}{3}}} \frac{-\dot{\varphi}^{i} \dot{\varphi}^{i} + \varphi^{\prime i}  \varphi^{\prime i} }{\bigg[ \left( \dot{\varphi}^i - \varphi^{\prime i} \right) \left( \dot{\varphi}^i - \varphi^{\prime i} \right) \left( \dot{\varphi}^j + \varphi^{\prime j} \right) \left( \dot{\varphi}^j + \varphi^{\prime j} \right) \bigg]^{\frac{1}{3}}} \\& - \frac{\lambda^3}{648} \frac{\left( \dot{\varphi}^i \dot{\varphi}^i \right)^2 + (\varphi^{\prime i} \varphi^{\prime i} )^2 + 12 (\dot{\varphi}^i \varphi^{\prime i} )^2 - 14 \dot{\varphi}^i \dot{\varphi}^i \varphi^{\prime j} \varphi^{\prime j} }{  \left( \dot{\varphi}^i - \varphi^{\prime i} \right) \left( \dot{\varphi}^i - \varphi^{\prime i} \right) \left( \dot{\varphi}^j + \varphi^{\prime j} \right) \left( \dot{\varphi}^j + \varphi^{\prime j} \right)  } + \cdots \bigg)\,. 
\end{aligned}
\end{equation}

\emph{$f( T^\alpha{}_\alpha, T^{\alpha \beta} T_{\alpha \beta})$ Perturbative Flow for Multiple Bosons}

To conclude this appendix, we note that one can also study the perturbative solution to the flow driven by the $ f(T^\alpha{}_\alpha, T^{\alpha \beta} T_{\alpha \beta}) = f(z, x)$ operator of equation \eqref{eq:TTn} for an \emph{arbitrary} function $f$. We again take the initial condition for the flow to be the action \eqref{eq:seedapp} for $N$ free massless bosons. The first few terms in the perturbative expansion are
\begin{equation}
\begin{aligned}\label{eq:coeffsttgeneral}
      f(T\alpha{}_\alpha, T^{\alpha \beta} T_{\alpha \beta})_0|_{\text{flat}} &= f\left(x \right)\,, \quad 
       f( T^\alpha{}_\alpha, T^{\alpha \beta} T_{\alpha \beta})_1|_{\text{flat}} = 4x y \left( \frac{\partial f(x)}{\partial x} \right)^2 \,, 
\end{aligned}
\end{equation}
where
\begin{equation}
    \begin{aligned}
x&= \frac{1}{2} \left( \dot{\varphi}^i - \varphi^{\prime i} \right) \left( \dot{\varphi}^i - \varphi^{\prime i} \right) \left( \dot{\varphi}^j + \varphi^{\prime j} \right) \left( \dot{\varphi}^j + \varphi^{\prime j} \right) \\&= \frac{1}{2} \left(  2 \left( \partial_\mu \varphi^{i} \partial^{\nu} \varphi^{i} \right) \left( \partial_\nu \varphi^{j} \partial^{\mu} \varphi^{j} \right) - \left( \partial_\mu \varphi^{i} \partial^{\mu} \varphi^{i} \right)^2 \right)\,, \\ 
y&=   -\dot{\varphi}^i \dot{\varphi}^i +\varphi^{\prime i} \varphi^{\prime i} = \partial_\mu \varphi^i \partial^\mu \varphi^i\,.
    \end{aligned}
\end{equation}
The perturbative action at $\mathcal{O}(g^2)$ is
\begin{equation}
\begin{aligned}\label{eq:generalperturbative action}
    S^{(g)} &= \int dt d\theta \bigg[ \frac{y}{2} + g f\left(x \right) + 2g^2 xy \left( \frac{\partial f (x)}{\partial x} \right)^2  + \cdots \bigg]\,.
\end{aligned}
\end{equation}
Furthermore, to summarize in the table below, one can check equation \eqref{eq:generalperturbative action} recovers the correct coefficients at $\mathcal{O}(g^2)$ for the perturbative actions describing $N$ free massless bosons considered in this appendix.

\begin{figure*}[htbp]
	\centering
{\includegraphics[width=.95\linewidth]{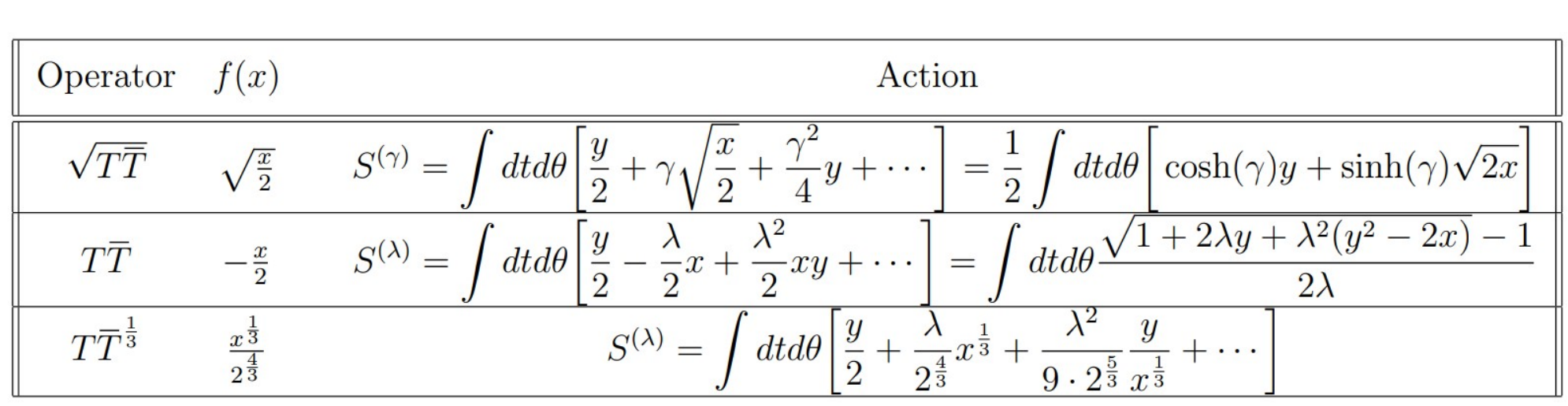}}
\end{figure*}

In principle, one could also study the perturbative quantization of these more general $ f(T^\alpha{}_\alpha, T^{\alpha \beta} T_{\alpha \beta})$-deformed scalar models. For instance, one could use the background field expansion and determine their Feynman rules as done in section \ref{sec:cian} for the Modified Scalar theory, or study canonical quantization following section \ref{sec:Quantization of first-order deformed}.

\section{Details of Feynman Diagram Calculations}\label{app:feynman}

In this appendix, we collect the technical details of certain evaluations of Feynman diagrams which occur in the analysis of section \ref{sec:cian}.

\subsection{One-loop, $2$-vertex calculation} \label{sec:scalar_field_1_loop_2_vertex_calc}

Let us first focus on the divergence structure of the diagram $\mathcal{D}_2$ of equation (\ref{D2_diagram}), which we repeat here for convenience:
\begin{align}\label{first_diagram}
    \mathcal{D}_2 \; = \; \raisebox{-0.5\height}{\includegraphics[width=0.38\linewidth]{diagrams/divergent_contribution.pdf}} \; .
\end{align}
As we mentioned around equation (\ref{simpler_integral_body}), the value of this diagram can be expressed in terms of the simpler quantity
\begin{align}
    {\mathcal{I}}^{\mu \nu \rho \tau}_2 = \int \frac{d^d\ell}{\left( 2\pi \right)^{d}} \frac{( \ell + q )^{\nu}\ell^{\mu} ( \ell + q )^{\tau} \ell^{\rho} }{\ell^2 \left( \ell + q \right)^2} \, .
\end{align}
All of the dependence on loop momenta is encoded within ${\mathcal{I}}^{\mu \nu \rho \tau}_2$, which we will also write as ${\mathcal{I}}_2$ with indices suppressed for simplicity. From the value of $\mathcal{I}_2$, the original diagram $\mathcal{D}_2$ is recovered from the expression (\ref{I2_to_D2}), which only involves additional dependence on the classical background via the tensor $P_{\mu\nu}^{ij}$ and an additional integration over the momentum $q$. Therefore, in order to study the divergences arising from the loop, it suffices to perform dimensional regularization of the quantity $\mathcal{I}_2$.

Expanding out the products and introducing a Feynman parameter $x$ in order to resolve the denominator, we find
\begin{align}
    \mathcal{I}_2 &= \int \frac{d^d\ell}{\left( 2\pi \right)^{d}} \frac{\ell^{\nu}\ell^{\mu} \ell^{\tau} \ell^{\rho} + \ell^{\nu} \ell^{\mu} q^{\tau} \ell^{\rho} + q^{\nu} \ell^{\mu} \ell^{\tau} \ell^{\rho}  + q^{\nu} \ell^{\mu} q^{\tau} \ell^{\rho}}{\ell^2 \left( \ell + q \right)^2} \nonumber \\
    &=\int \frac{d^d\ell}{\left( 2\pi \right)^{d}} \int_0^{1} d^dx \frac{\ell^{\nu}\ell^{\mu} \ell^{\tau} \ell^{\rho} + \ell^{\nu} \ell^{\mu} q^{\tau} \ell^{\rho} + q^{\nu} \ell^{\mu} \ell^{\tau} \ell^{\rho}  + q^{\nu} \ell^{\mu} q^{\tau} \ell^{\rho}}{\left[ \ell^2 \left( 1 - x \right)  +  x\left( \ell + q \right)^2 \right]^2} \nonumber \\
    &= \int \frac{d^d\ell}{\left( 2\pi \right)^{d}} \int_0^{1} d^dx\frac{\ell^{\nu}\ell^{\mu} \ell^{\tau} \ell^{\rho} + \ell^{\nu} \ell^{\mu} q^{\tau} \ell^{\rho} + q^{\nu} \ell^{\mu} \ell^{\tau} \ell^{\rho}  + q^{\nu} \ell^{\mu} q^{\tau} \ell^{\rho}}{\left[ \ell^2  + x \left( 2\ell_{\mu} q^{\mu} + q^2 \right) \right]^2} \nonumber \\
    &= \int \frac{d^d\ell}{\left( 2\pi \right)^{d}} \int_0^{1} d^dx\frac{\ell^{\nu}\ell^{\mu} \ell^{\tau} \ell^{\rho} + \ell^{\nu} \ell^{\mu} q^{\tau} \ell^{\rho} + q^{\nu} \ell^{\mu} \ell^{\tau} \ell^{\rho}  + q^{\nu} \ell^{\mu} q^{\tau} \ell^{\rho}}{\left[ \left( \ell^{\mu} + xq^{\mu}  \right) ^2  +x \left( 1 - x \right)q^2  \right]^2} \, .
\end{align}
In the final step, we have completed the square in the denominator by adding and subtracting $q^2 x^2$. We now shift the integration variable from $\ell^\mu$ to
\begin{align}
    \ell^{\prime \mu} = \ell^\mu - x q^\mu \, ,
\end{align}
which causes the denominator to become even in $\ell^\prime$, and thus terms in the numerator which are odd in $\ell^{\prime \mu}$ will vanish by symmetry. We immediately drop the primes on $\ell^{\prime \mu}$ and write the surviving terms as
\begin{align}
    {\mathcal{I}}_2 &= \int \frac{d^d \ell}{\left( 2\pi \right)^{d}} \int_0^{1} d^dx \Bigg[ \nonumber \frac{\ell^{\nu} \ell^{\mu} \ell^{\tau} \ell^{\rho}}{\left( \ell^2  + q^2x\left( 1 - x \right)  \right)^2} + \frac{x^2  \ell^{\nu} q^{\mu} \ell^{\tau} q^{\rho} }{\left( \ell^2 + q^2 x \left( 1 - x \right)  \right)^2} \\& \nonumber + \frac{\left( x^2 - 2x + 1 \right) q^{\nu} \ell^{\mu} q^{\tau} \ell^{\rho}}{\left( \ell^2  + q^2x\left( 1 - x \right)  \right)^2}  \\
    & +  \frac{\left( x^2 - x  \right) \left( q^{\nu} q^{\mu} \ell^{\tau} \ell^{\rho} + q^{\nu} \ell^{\mu} \ell^{\tau} q^{\rho} + \ell^{\nu} \ell^{\mu} q^{\tau} q^{\rho} + \ell^{\nu} q^{\mu} q^{\tau} \ell^{\rho} \right) }{\left[ \ell^2  + q^2x \left( 1 - x \right)  \right]^2}  \nonumber \\&+ \frac{\left( x^{4} -2x^3 + x^2 \right)  q^{\nu} q^{\mu} q^{\tau} q^{\rho}}{\left( \ell^2  + q^2x\left( 1 - x \right)  \right)^2} \Bigg] \, \nonumber \\
    &= \int \frac{d^d \ell}{\left( 2\pi \right)^{d}} \int_0^{1} d^dx \Bigg[ \frac{\ell^{\nu} \ell^{\mu} \ell^{\tau} \ell^{\rho}}{\left( \ell^2  + q^2x\left( 1 - x \right)  \right)^2} + \frac{ x^2  \ell^{\nu} q^{\mu} \ell^{\tau} q^{\rho} }{\left( \ell^2 + q^2 x \left( 1 - x \right)  \right)^2} \nonumber\\&+ \frac{\left( 1 - x \right)^2 q^{\nu} \ell^{\mu} q^{\tau} \ell^{\rho}}{\left[ \ell^2  + q^2x\left( 1 - x \right)  \right]^2}  \nonumber\\
    &+ \frac{ x\left( 1 -x \right) \left( q^{\nu} q^{\mu} \ell^{\tau} \ell^{\rho} + q^{\nu} \ell^{\mu} \ell^{\tau} q^{\rho} +  \ell^{\nu} \ell^{\mu} q^{\tau} q^{\rho} + \ell^{\nu} q^{\mu} q^{\tau} \ell^{\rho} \right) }{\left[ \ell^2  + q^2x \left( 1 - x \right)  \right]^2}  \nonumber \\&+ \frac{\left[ x^2 \left( 1 - x \right)^2 \right]  q^{\nu} q^{\mu} q^{\tau} q^{\rho}}{\left[ \ell^2  + q^2x\left( 1 - x \right)  \right]^2} \Bigg] \,.
\end{align}

By a symmetry argument similar to the one discussed around equations (\ref{sym_replacements_1}) and (\ref{eq:symmetrization}), within the integral we can replace products of loop momenta with symmetrized combinations of metric tensors:
\begin{align}\label{sym_replacements_app}
    \ell^{\mu} \ell^{\nu} &\to \frac{1}{d} \ell^2 g^{\mu \nu} \, , \nonumber \\
    \ell^{\mu} \ell^{\nu} \ell^{\rho} \ell^{\tau} &\to \frac{1}{d \left( d + 2 \right) } \ell^4 \left( g^{\mu \nu} g^{\rho \tau} + g^{\mu \rho} g^{\nu \tau} + g^{\mu \tau} g^{\nu \rho} \right) \, .
\end{align}
Applying the replacements (\ref{sym_replacements_app}), the integral ${\mathcal{I}}_2$ becomes
\begin{align}\label{app_A1_intermediate}
    {\mathcal{I}}_2 &= \int \frac{d^d\ell}{\left( 2\pi \right)^{d}} \int_0^{1} d^dx \Bigg( \frac{\ell^{4}}{d \left( d + 2 \right) }\frac{g^{\mu \nu} g^{\rho \tau} + g^{\mu\tau} g^{\nu \rho} + g^{\mu \rho} g^{\nu \tau}  }{\left[ \ell^2  + q^2x\left( 1 - x \right)  \right]^2} \nonumber \\& + \frac{\ell^2}{d}\frac{x^2 g^{\nu \tau}q^{\mu} q^{\rho} }{\left[ \ell^2  + q^2x \left( 1 - x \right)  \right]^2} \nonumber\\
    &+ \frac{\ell^2}{d}\frac{ \left( 1 - x \right)^2  g^{\mu \rho} q^{\nu} q^{\tau} }{\left[ \ell^2  + q^2x\left( 1 - x \right)  \right]^2} \nonumber \\&+ \frac{\ell^2}{d}\frac{x\left( 1 -x \right) \left( q^{\nu} q^{\mu} g^{\tau\rho} + g^{\mu \tau} q^{\nu} q^{\rho} + g^{\nu \mu} q^{\tau} q^{\rho} + g^{\nu \rho}q^{\mu} q^{\tau} \right) }{\left[ \ell^2  + q^2x \left( 1 - x \right)  \right]^2} \nonumber\\
    &\qquad + \frac{ x^2 \left( 1 - x \right)^2  q^{\nu} q^{\mu} q^{\tau} q^{\rho}}{\left[ \ell^2  + q^2x\left( 1 - x \right)  \right]^2} \Bigg) \, .
\end{align}
It will be convenient to make use of the standard result
\begin{align}\label{first_anselmi}
    \int \frac{d^d\ell}{\left( 2\pi \right)^{d}} \frac{\ell^{2\beta}}{\left( \ell^2 - \Delta^2 \right)^{\alpha}} = i \left( -1 \right)^{\alpha + \beta} \frac{\Gamma \left( \beta + \frac{d}{2} \right) \Gamma \left( \alpha - \beta - \frac{d}{2} \right) }{\left( 4\pi \right)^{\frac{d}{2}} \Gamma \left( \alpha \right) \Gamma \left( \frac{d}{2} \right)  } \Delta^{2 \left( \frac{d}{2} - \alpha + \beta \right) } \, ,
\end{align}
which can be found, for instance, in equation (A.4) in \cite{Anselmi:2019pdm}. Using (\ref{first_anselmi}) with
\begin{align}\label{delta_sq_defn}
    \Delta^2 = -q^2 x\left( 1 - x \right) 
\end{align}
in equation (\ref{app_A1_intermediate}), we find
\begin{align}
    {\mathcal{I}}_2 \nonumber &= \frac{i}{\left( 4\pi \right)^{\frac{d}{2}} \Gamma \left( 2 \right) \Gamma \left( \frac{d}{2} \right)  } \int_0^{1} d^dx \Bigg( \frac{\Delta^{d} \Gamma \left( 2 + \frac{d}{2} \right) \Gamma \left( -\frac{d}{2} \right) }{d \left( d + 2 \right) }\left( g^{\mu \nu} g^{\rho \tau} + g^{\mu \rho} g^{\nu \tau} + g^{\mu \tau} g^{\nu \rho} \right)  \nonumber\\
    &+\frac{\Delta^{d - 2}\Gamma \left( 1 + \frac{d}{2} \right) \Gamma \left( 1 - \frac{d}{2} \right)}{d}   x \left( 1 - x \right)  \left( q^{\nu} q^{\mu} g^{\tau \rho} + g^{\mu \tau} q^{\nu} q^{\rho} +  g^{\nu \mu} q^{\tau} q^{\rho} + g^{\nu \rho} q^{\mu} q^{\tau} \right) \nonumber\\
    & + \frac{\Delta^{d-2} \Gamma \left( 1 + \frac{d}{2} \right) \Gamma \left( 1 - \frac{d}{2} \right)}{d} x^2 g^{\nu \tau} q^{\mu} q^{\rho} \nonumber \\&+\frac{\Delta^{d-2}\Gamma \left( 1 + \frac{d}{2} \right) \Gamma \left( 1 - \frac{d}{2} \right)}{d}  \left( 1 - x \right)^2  g^{\mu \rho} q^{\nu} q^{\tau}\nonumber \\
    & +\Delta^{d-4}\Gamma \left( \frac{d}{2} \right) \Gamma \left( 2 - \frac{d}{2} \right)  \left[ x^2 \left( 1 - x \right)^2  \right] q^{\nu} q^{\mu} q^{\tau} q^{\rho} \Bigg) \, .
\end{align}
Using gamma function identities and some algebra, one can simplify this to
\begin{align}
    {\mathcal{I}}_2 &= \frac{i\Gamma \left( -\frac{d}{2} \right)}{\left( 4\pi \right)^{\frac{d}{2}}} \int_0^{1} d^dx \Bigg( \frac{\Delta^{d}}{4} \left( g^{\mu \nu} g^{\rho \tau} +   g^{\mu \rho} g^{\nu \tau} + g^{\mu \tau} g^{\nu \rho} \right) -\frac{d\Delta^{d-2}}{4} x^2 g^{\nu \tau} q^{\mu} q^{\rho}  \nonumber \\
    & -\frac{d\Delta^{d-2}}{4} \left( 1 - x \right)^2 g^{\mu \rho} q^{\nu} q^{\tau} \nonumber \\&- \frac{d\Delta^{d - 2}}{4} x \left( 1 - x \right) \left( q^{\nu} q^{\mu} g^{\tau \rho} + g^{\mu\tau} q^{\nu} q^{\rho} + g^{\nu \mu} q^{\tau} q^{\rho} + g^{\nu \rho} q^{\mu} q^{\tau} \right) \nonumber\\
    &+ \frac{d\left( d - 2 \right) \Delta^{d-4}}{4} x^2 \left( 1 - x \right)^2 q^{\nu} q^{\mu} q^{\tau} q^{\rho} \Bigg) \, .
\end{align}
After substituting in for $\Delta^2$ using the definition (\ref{delta_sq_defn}), we can now evaluate the resulting integrals using the formula
\begin{align}
    \int_0^{1} d^dx x^{\alpha - 1} \left( 1 - x \right)^{\beta - 1} &= \frac{\Gamma \left( \alpha \right) \Gamma \left( \beta \right) }{\Gamma \left( \alpha + \beta \right) } = \mathrm{B} ( \alpha, \beta ) \, , 
\end{align}
which we recognize as the definition of the beta function $\mathrm{B} ( \alpha , \beta )$. By doing this, we find
\begin{align}\label{appendix_intermediate_two}
   & {\mathcal{I}}_2 = \frac{i\Gamma \left( -\frac{d}{2} \right)}{4\left( 4\pi \right)^{\frac{d}{2}}} \int_0^{1} d^dx \Bigg( \nonumber \\& d \left( d - 2 \right) q^{d-4}\left[ -x\left( 1 - x \right)   \right]^{\frac{d}{2}} q^{\nu} q^{\mu} q^{\tau} q^{\rho} - dq^{d-2}  \left( -x \right)^{\frac{d}{2}-1}\left( 1 - x \right)^{\frac{d}{2}+1} g^{\mu \rho} q^{\nu} q^{\tau} \nonumber\\
    & + q^{d} \left[ -x\left( 1 - x \right) \right]^{\frac{d}{2}} \left( g^{\mu \nu} g^{\rho \tau} + g^{\mu \rho} g^{\nu \tau} + g^{\mu \tau} g^{\nu \rho} \right)  \nonumber \\& -dq^{d-2} \left[ \left( -x \right)^{\frac{d}{2}+1} \left( 1 - x \right)^{\frac{d}{2}-1}\right] g^{\nu \tau} q^{\mu} q^{\rho}  \nonumber \\
    &+dq^{d - 2} \left[-x \left( 1 - x \right)  \right]^{\frac{d}{2}} \left( q^{\nu} q^{\mu} g^{\tau \rho} + g^{\mu \tau} q^{\nu} q^{\rho} + g^{\nu \mu} q^{\tau} q^{\rho} + g^{\nu \rho} q^{\mu} q^{\tau} \right)   \Bigg) \nonumber \\
    &= \frac{i \left( -1 \right)^{\frac{d}{2}}\Gamma \left( -\frac{d}{2} \right) }{4\left( 4\pi \right)^{\frac{d}{2}}} \Bigg( q^{d} \frac{\Gamma \left( \frac{d}{2} + 1 \right)^2}{\Gamma \left( d + 2 \right) } \left( g^{\mu \nu} g^{\rho \tau} + g^{\mu \rho} g^{\nu \tau} + g^{\mu \tau} g^{\nu \rho} \right)   \nonumber \\
    &+dq^{d-2} \frac{\Gamma \left( \frac{d}{2} \right) \Gamma \left( \frac{d}{2} + 2 \right) }{\Gamma \left( d + 2 \right) } \left[ g^{\mu \rho} q^{\nu} q^{\tau} + g^{\nu \tau} q^{\mu} q^{\rho} \right] + d\left( d - 2 \right) q^{d - 4} \frac{\Gamma\left( \frac{d}{2} + 1 \right)^2}{\Gamma \left( d + 2 \right)} q^{\nu} q^{\mu} q^{\tau} q^{\rho} \nonumber \\
    &+dq^{d-2} \frac{\Gamma \left( \frac{d}{2} + 1 \right)^2}{\Gamma \left( d + 2 \right) } \left( q^{\nu} q^{\mu} g^{\tau \rho} + g^{\mu \tau} q^{\nu}q^{\rho} +  g^{\nu \mu} q^{\tau} q^{\rho} + g^{\nu \rho} q^{\mu} q^{\tau} \right)  \Bigg) \, .
\end{align}
Note that each term in (\ref{appendix_intermediate_two}) scales as $q^d$, as expected. Factoring out the gamma functions, we have found
\begin{align}
    {\mathcal{I}}_2 &= \frac{i \left( -1 \right)^{\frac{d}{2}}\Gamma \left( -\frac{d}{2} \right) }{4\left( 4\pi \right)^{\frac{d}{2}}} \frac{\Gamma \left( \frac{d}{2} + 1 \right)^2}{\Gamma \left( d + 2 \right) } \Bigg[ q^{d}  \left( g^{\mu \nu} g^{\rho \tau} + g^{\mu \rho} g^{\nu \tau} + g^{\mu \tau} g^{\nu \rho} \right) \nonumber \\& + d \left( d - 2 \right) q^{d - 4}q^{\nu} q^{\mu} q^{\tau} q^{\rho}+ dq^{d-2} \left( q^{\nu} q^{\mu} g^{\tau \rho} + g^{\mu \tau} q^{\nu} q^{\rho} + g^{\nu \mu} q^{\tau} q^{\rho} + g^{\nu \rho} q^{\mu} q^{\tau} \right)  \nonumber \\
    & +\left( d + 2 \right) q^{d-2} \left( g^{\mu \rho} q^{\nu} q^{\tau} + g^{\nu \tau} q^{\mu} q^{\rho} \right)  \Bigg] \, .
\end{align}
Finally, to perform dimensional regularization, we set the spacetime dimension to $d = 2 + 2 \epsilon$ and take $\epsilon \to 0$ using the limiting behavior
\begin{align}
    \Gamma \left( -1-\epsilon \right) = \frac{1}{\epsilon} - \gamma + 1 +  \mathcal{O}\left( \epsilon \right)
\end{align}
for the gamma functions. Keeping only divergent terms, we arrive at the final expression
\begin{align}\label{two_vertex}
    {\mathcal{I}}_2 &= \left( \frac{1}{\epsilon} \right) \frac{-i}{24\left( 4\pi \right)} \bigg[ q^{2}  \left( g^{\mu \nu} g^{\rho \tau}  + g^{\mu \rho} g^{\nu \tau} + g^{\mu \tau} g^{\nu \rho} \right) \nonumber \\& +2 \left( q^{\nu} q^{\mu} g^{\tau \rho} + g^{\mu \tau} q^{\nu} q^{\rho}+ g^{\nu \mu} q^{\tau} q^{\rho} + g^{\nu \rho} q^{\mu} q^{\tau} \right) \nonumber \\
    &+ 4\left( g^{\mu \rho} q^{\nu} q^{\tau} + g^{\nu \tau} q^{\mu} q^{\rho} \right) \bigg] \, .
\end{align}
This completes the evaluation of the divergent contribution from ${\mathcal{I}}_2$, which justifies the result (\ref{divergent_final_body}) which was quoted in the body of the paper.

\subsection{One-loop, $n$-vertex calculation}\label{app:one_loop_n_vertex}

The $n$-vertex diagram $\mathcal{D}_n$ can be computed via a generalization of the method used in appendix \ref{sec:scalar_field_1_loop_2_vertex_calc}. We again write $\ell$ for the loop momentum and we label the external momenta as $q_{i}$, for $i = 0, \ldots, n-1$, with momentum conservation implying that
\begin{align}
    q_{n-1} = - \sum_{i=0}^{n-2} q_{i} \, .
\end{align}
As we did with $\mathcal{D}_2$ in equation (\ref{D2_stripped}), let us strip off various factors of $P_{\mu\nu}{}^{ij}$ to write
\begin{align}\label{Dn_defn}
    \mathcal{D}_n &= \left( - 2 \tanh ( \gamma ) \right)^N \nonumber \\& \cdot \int \left(\frac{d^d q_0}{ ( 2 \pi )^d } P_{(\mu_1 \mu_2)}^{i_1 i_2} ( q_0 ) \right) \left(\frac{d^d q_0}{ ( 2 \pi )^d } P_{(\mu_3 \mu_4)}^{i_2 i_3} ( q_1 ) \right) \cdots \left(\frac{d^d q_{n-2}}{(2 \pi)^d} P_{(\mu_{2n-3} \mu_{2n-2})}^{i_{n-1} i_{n}} ( q_{n-2} ) \right) \nonumber \\
    &\cdot P_{(\mu_{2n-1} \mu_{2n})}^{i_{n} i_{1}} ( q_{n-1} ) \left( {\mathcal{I}}_n \right)^{\mu_1 \mu_2 \ldots \mu_{2n-1} \mu_{2n}} \, ,
\end{align}
where ${\mathcal{I}}_n$ is the simpler integral
\begin{align}\label{In_defn}\hspace{-10pt}
   & \left( {\mathcal{I}}_n \right)^{\mu_1 \mu_2 \ldots \mu_{2n-1} \mu_{2n} } \nonumber \\ &= \int \frac{d^d \ell}{( 2\pi )^d} \, \left( \prod_{i = 0}^{n-1} \left( \ell + \sum_{j=0}^{i} q_{j}  \right)^{-2} \right) \left( \prod_{k=1}^{n} \left( \ell + \sum_{j=0}^{k-1} q_{j} \right)^{\mu_{2k-1}} \left( \ell + \sum_{j=0}^{k-1} q_{j}  \right)^{\mu_{2k}} \right) \, .
\end{align}
We will further break up $\mathcal{I}_n$ into pieces and evaluate each piece in turn. Let us write the integrand of (\ref{In_defn}) as a product
\begin{align}\label{In_PV}
    \left( {\mathcal{I}}_n \right)^{\mu_1 \mu_2 \ldots \mu_{2n-1} \mu_{2n} } = \int \frac{d^d \ell}{( 2\pi )^d} \, \mathcal{P}_n \mathcal{V}^{\mu_1 \mu_2 \ldots \mu_{2n-1} \mu_{2n} } \, , 
\end{align}
where the symbol
\begin{align}
    \mathcal{P}_n &= \prod_{i = 0}^{n-1} \left( \ell + \sum_{j=0}^{i} q_{j}  \right)^{-2}
\end{align}
refers to the collection of all factors in $\mathcal{I}_n$ which come from propagators, and the symbol
\begin{align}\label{Vn_defn}
    \mathcal{V}^{\mu_1 \mu_2 \ldots \mu_{2n-1} \mu_{2n} } &= \prod_{k=1}^{n} \left( \ell + \sum_{j=0}^{k-1} q_{j} \right)^{\mu_{2k-1}} \left( \ell + \sum_{j=0}^{k-1} q_{j}  \right)^{\mu_{2k}} \, ,
\end{align}
which we will sometimes abbreviate as $\mathcal{V}$, refers to the pieces coming from vertex factors.

In order to highlight the divergence structure of the diagram $\mathcal{D}_n$, we will focus on performing the loop momentum integral of various terms appearing in the product $\mathcal{P} \mathcal{V}$ of equation (\ref{In_PV}), and neglect the additional structure arising from the contraction with the various tensors $P_{\mu \nu}^{ij}$ to obtain $\mathcal{D}_n$ in (\ref{Dn_defn}).

Let us begin by simplifying the product $\mathcal{P}$ of $n$ propagator factors. In general, we can write the product of $n$ propagators using a Feynman parameterization:
\begin{align}
    \prod_{i=0}^{n-1} A_{i}^{-1} &= \int_0^{1} \left( \prod_{i=0}^{n-1} d^dx_i \right) \delta \left( \sum_{i=0}^{n-1} x_{i} - 1  \right) \frac{\left( n - 1 \right)!}{\left[ \sum_{i}^{}x_{i} A_{i}  \right]^{n} } \, .
\end{align}
The product of propagators inside the loop can thus be expressed as 
\begin{align}
    \mathcal{P} = \int_0^{1} \left( \prod_{i=0}^{n-1} d^dx_{i} \right) \delta \left( \sum_{i=0}^{n-1} x_{i} - 1  \right) \left[ \sum_{i=0}^{n-1}x_{i} \left( \ell + \sum_{j=0}^{i} q_{j} \right)^2   \right]^{-n} \, .
\end{align}
As $\sum_{i}^{} x_{i} = 1$, we can expand and reduce the square bracketed term to
\begin{align}\hspace{-10pt}
  &  \left[ \sum_{i}^{}x_{i} \left( \ell + \sum_{j=0}^{i} q_{j} \right)^2   \right]^{-n} \nonumber \\ &= \left[ \ell^2  + \sum_{i}^{}x_{i} \left( 2\ell_{\mu} \sum_{j=0}^{i} q_{j}^{\mu}+ \left( \sum_{j=0}^{i} q_{j} \right)^2  \right)   \right]^{-n} \nonumber \\
    &=  \left[ \left( \ell + \sum_{i}^{} \sum_{j=0}^{i} x_{i} q_{j}  \right)^2 - \left( \sum_{i}^{} \sum_{j=0}^{i} x_{i} q_{j} \right)^2  + \sum_{i}^{}x_{i}   \left( \sum_{j=0}^{i} q_{j} \right)^2  \right]^{-n} \, .
\end{align}
We change variables in the loop momentum by shifting
\begin{align}\label{loop_momentum_shift}
    \ell^\mu \to \ell^\mu - \sum_{i=0}^{n-1} \sum_{j=0}^{i} x_{i} q^\mu_{j} \, ,
\end{align}
so that the bracketed expression becomes
\begin{align}
    \left[ \sum_{i}^{}x_{i} \left( \ell + \sum_{j=0}^{i} q_{j} \right)^2   \right]^{-n} &=  \left[ \ell^2 - \left( \sum_{i}^{} \sum_{j=0}^{i} x_{i} q_{j} \right)^2  + \sum_{i}^{}x_{i}   \left( \sum_{j=0}^{i} q_{j} \right)^2  \right]^{-n} \nonumber \\
    &=  \left[ \ell^2 - \Delta^2  \right]^{-n} \, ,
\end{align}
where we have defined
\begin{align}
    \Delta^2 &= \left( \sum_{i}^{} \sum_{j=0}^{i} x_{i} q_{j} \right)^2  - \sum_{i}^{}x_{i}   \left( \sum_{j=0}^{i} q_{j} \right)^2 \, .
\end{align}
Overall this allows us to write the propagators as
\begin{align}
    \mathcal{P} = (n - 1)! \int_0^{1} \left( \prod_{i=0}^{n-1} d^dx_{i} \right) \delta \left( \sum_{i=0}^{n-1} x_{i} - 1  \right) \left( \ell^2 - \Delta^2  \right)^{-n} \, .
\end{align}
Next let us turn to the contributions from the vertices in equation (\ref{Vn_defn}), which yield factors of momenta in the numerator of the integrand. Under the change of variables (\ref{loop_momentum_shift}) which renders the denominator of $\mathcal{P}$ quadratic in $\ell$, the vertex factor contribution becomes
\begin{align}
   & \mathcal{V}^{\mu_1 \mu_2 \ldots \mu_{2n-1} \mu_{2n} } \nonumber \\&=  \prod_{k=1}^{n} \left( \ell - \sum_{i=0}^{n-1} \sum_{j=0}^{i} x_{i} q_{j}  + \sum_{j=0}^{k-1} q_{j} \right)^{\mu_{2k-1}} \left( \ell - \sum_{i=0}^{n-1} \sum_{j=0}^{i} x_{i} q_{j}+ \sum_{j=0}^{k-1} q_{j}  \right)^{\mu_{2k}} \, .
\end{align}
We expand this product in descending powers of $\ell$ as only powers $\ell^{2n}$ and $\ell^{2n-2}$ will lead to divergent terms. We have that
\begin{align}\label{V_formula}
    \mathcal{V}^{\mu_1 \mu_2 \ldots \mu_{2n-1} \mu_{2n} } &= \prod_{k=1}^{n} \ell^{\mu_{2k-1}} \ell^{\mu_{2k}} + \sum_{a=1}^{2n} \sum_{b>a}^{2n} \left( \prod_{c \neq a,b}^{2n} \ell^{\mu_c} \right)  f^{\mu_{a}} \left( x,q,a \right) f^{\mu_{b}} \left( x,q,b  \right) + \mathcal{O}\left( \ell^{2n-4} \right) \, ,
\end{align}
where we have defined for brevity
\begin{align}\label{cursed_defn}
    f^{\mu} \left( x,q,a \right) = \sum_{i=0}^{n-1} \sum_{j=0}^{i} x_{i} q_{j}^{\mu}  + \sum_{j=0}^{\lfloor \frac{a-1}{2} \rfloor} q_{j}^{\mu} \, .
\end{align}

Next we will replace products of loop momenta using the generalized symmetrization rule of equation (\ref{eq:symmetrization}). To ease notation, let us write
\begin{align}\label{shorthand_metric}
    g^{\mu_1 \ldots \mu_n} = g^{(\mu_1 \mu_2 } \cdots g^{\mu_{n-1} \mu_{n})} \, 
\end{align}
for the symmetrized combination of derivatives appearing in this expression. When no confusion is possible, we will also write $g^{\{ \mu \}}$ for (\ref{shorthand_metric}), where $\{ \mu \}$ is understood to refer to a multi-index $\{ \mu \} = \mu_1 \ldots \mu_n$. With this notation, the replacement rule becomes
\begin{align}
    \prod_{i=1}^{n} \ell^{\mu_{2i- 1 }} \ell^{\mu_{2i}} \to \frac{\ell^{2n} \left( d -2 \right)!!}{\left( d-2 + 2n \right)!!} g^{\mu_{1}\ldots \mu_{2n}} \, .
\end{align}
This transforms the vertex factor contribution to
\begin{align}\label{vertex_intermediate}
    \mathcal{V}^{\mu_1 \mu_2 \ldots \mu_{2n-1} \mu_{2n} } &=  \frac{\ell^{2n} \left( d - 2 \right)!!}{ \left( d  - 2 + 2n \right)!!} g^{\mu_1 \cdots \mu_{2n}} \nonumber\\
    &\quad + \frac{\ell^{2n-2} \left( d - 2 \right)!!}{\left( d - 4 + 2n \right)!!}\sum_{a=1}^{2n} \sum_{b>a}^{2n} g^{\{ \mu \neq \mu_a, \mu_b \}}   f^{\mu_a} \left( x,q,a \right)  f^{\mu_{b}} \left( x,q,b \right) + \mathcal{O}\left( \ell^{2n-4} \right) \, .
\end{align}
In equation (\ref{vertex_intermediate}), we have written $g^{\{ \mu \neq \mu_a, \mu_b \}}$ to refer to a product of the form (\ref{shorthand_metric}) in which the multi-index $\{ \mu \}$ runs over all possible values except for the two indices $\mu_a$ and $\mu_b$, which are excluded.

Let us now combine the pieces and identify the divergent terms in $\mathcal{D}_n$. We can evaluate
\begin{align}\hspace{-10pt}\label{Dn_Itilde_intermediate}
    &\left( {\mathcal{I}}_{n} \right)^{\mu_1 \mu_2 \ldots \mu_{2n}} \nonumber \\
    &\quad = \int \frac{d^d \ell}{( 2\pi )^d} \mathcal{P} \mathcal{V}^{\mu_1 \mu_2 \ldots \mu_{2n}} \, \nonumber \\
    &\quad = \int \frac{d^d \ell}{( 2\pi )^d} \left( n - 1 \right)!\int_0^{1} \left( \prod_{i=0}^{n-1} d^dx_{i} \right) \delta \left( \sum_{i=0}^{n-1} x_{i} - 1  \right) \left( \ell^2 - \Delta^2  \right)^{-n}  \nonumber \\& \cdot \Bigg( \frac{\ell^{2n} \left( d - 2 \right)!!}{\left( d  - 2 + 2n \right)!!} g^{\mu_1 \ldots \mu_{2n}}
 + \frac{\ell^{2n-2} \left( d - 2 \right)!!}{\left( d - 4+ 2n  \right)!!}\sum_{a=1}^{2n} \sum_{b>a}^{2n} g^{\{ \mu \neq \mu_a, \mu_b \}}   f^{\mu_a} \left( x,q,a \right)  f^{\mu_{b}} \left( x,q,b \right) \nonumber \\&+ \mathcal{O}\left( \ell^{2n-4} \right) \Bigg) \, ,
\end{align}
in terms of the known integral
\begin{align}\label{known_integral}
    \int \frac{d^d \ell}{\left( 2\pi \right)^{d}} \frac{\ell^{2\beta}}{\left( \ell^2 - \Delta^2 \right)^{\alpha}} &= i \left( -1 \right)^{\alpha + \beta} \frac{\Gamma \left( \beta + \frac{d}{2} \right) \Gamma \left( \alpha - \beta - \frac{d}{2} \right) }{\left( 4\pi \right)^{\frac{d}{2}} \Gamma \left( \alpha \right) \Gamma \left( \frac{d}{2} \right)  } \Delta^{2 \left( \frac{d}{2} - \alpha + \beta \right) } \, .
\end{align}
First let us justify why the terms of order $\ell^{2n-4}$ and lower in equation (\ref{Dn_Itilde_intermediate}) will not give divergent contributions. A term proportional to $\ell^{2n-4}$ in the parentheses of (\ref{Dn_Itilde_intermediate}), after multiplying the propagator factor $( \ell^2 - \Delta^2 )^{-n}$, gives a term in the integrand of the form (\ref{known_integral}) with $\alpha = n$ and $\beta = n - 2$. Such a term gives a contribution
\begin{align}\label{known_integral_finite}
    \int \frac{d^d\ell}{\left( 2\pi \right)^{d}} \frac{\ell^{2 ( n - 2 ) }}{\left( \ell^2 - \Delta^2 \right)^{n}} &= i \left( -1 \right)^{n +  ( n - 2 )} \frac{\Gamma \left( n - 2 + \frac{d}{2} \right) \Gamma \left( 2 - \frac{d}{2} \right) }{\left( 4\pi \right)^{\frac{d}{2}} \Gamma \left( n \right) \Gamma \left( \frac{d}{2} \right)  } \Delta^{2 \left( \frac{d}{2} - n + ( n - 2 ) \right) } \, .
\end{align}
In the limit as $d \to 2$, the two factors of gamma functions in the numerator of (\ref{known_integral_finite}) tend to $\Gamma ( n - 1 )$ and $\Gamma ( 1 )$, which are both finite since $n > 2$. Since we are only interested in computing the divergent contributions arising from these diagrams, we ignore these terms. Similarly, any terms of lower order in $\ell$ can be evaluated in the same way but with even smaller values of $\beta$, which also lead to finite contributions from the gamma functions.

Let us therefore focus on the divergent terms. The term in the integrand proportional to $\ell^{2n}$ in equation (\ref{Dn_Itilde_intermediate}) takes the form (\ref{known_integral}) with $\alpha = \beta = n$. Similarly, the term in the integrand that scales as $\ell^{2n-2}$ is of the form (\ref{known_integral}) with $\alpha = n$ and $\beta = n - 1$. Evaluating the loop momentum integrals then gives 
\begin{align}\label{In_final}
    \left( {\mathcal{I}}_{n} \right)^{\mu_1 \mu_2 \ldots \mu_{2n}} &= \left( n - 1 \right)!\int_0^{1} \left( \prod_{i=0}^{n-1} dx_{i} \right) \delta \left( \sum_{i=0}^{n-1} x_{i} - 1  \right)  \nonumber \\
    &\quad \cdot \Bigg( \frac{i\left( d - 2 \right)!!}{ \left( d  - 2 + 2n\right)!!} g^{\mu_1 \cdots \mu_{2n}} \frac{\Gamma \left( n + \frac{d}{2} \right)  }{\left( 4\pi \right)^{\frac{d}{2}} \Gamma \left( n \right) \Gamma \left( \frac{d}{2} \right)} \Gamma \left( -\frac{d}{2} \right)\Delta^{ d }  \nonumber \\
    &\quad + \frac{i\left( d - 2 \right)!!}{\left( d  - 4+ 2n \right)!!}\sum_{a=1}^{2n} \sum_{b>a}^{2n} g^{\{ \mu \neq \mu_a, \mu_b \}}   f^{\mu_a} \left( x,q,a \right)  f^{\mu_{b}} \left( x,q,b \right) \nonumber \\
    &\quad \cdot \frac{\Gamma \left( n-1+\frac{d}{2} \right) }{\left( 4\pi \right)^{\frac{d}{2}} \Gamma \left( n \right) \Gamma \left( \frac{d}{2} \right)  } \Gamma \left( 1 - \frac{d}{2} \right) \Delta^{d - 2} \Bigg) \, .
\end{align}
To better analyze the divergence structure, it is useful to define shorthand notation for the two terms appearing in (\ref{In_final}), which we call $\mathbf{C}_{2n}^{\mu_1 \ldots \mu_{2n}}$ and $\mathbf{D}_{2n}^{\mu_1 \ldots \mu_{2n}}$:
\begin{align}
    \mathbf{C}_{2n}^{\mu_1 \ldots \mu_{2n}} &\equiv \frac{i\left( d - 2 \right)!!}{ \left( d - 2 + 2n \right)!!} g^{\mu_1 \cdots \mu_{2n}}  \frac{\Gamma \left( n + \frac{d}{2} \right)  }{\left( 4\pi \right)^{\frac{d}{2}} \Gamma \left( n \right) \Gamma \left( \frac{d}{2} \right)} \Gamma \left( -\frac{d}{2} \right)\Delta^{d} \, \, \nonumber \\
     \mathbf{D}_{2n}^{\mu_1 \ldots \mu_{2n}} &\equiv \frac{i\left( d - 2 \right)!!}{\left( d - 4 + 2n \right)!!}\sum_{a=1}^{2n} \sum_{b>a}^{2n} g^{\{ \mu \neq \mu_a, \mu_b \}}   f^{\mu_a} \left( x,q,a \right)  f^{\mu_{b}} \left( x,q,b \right)\nonumber \\
    &\qquad \cdot \frac{\Gamma \left( n-1+\frac{d}{2} \right) }{\left( 4\pi \right)^{\frac{d}{2}} \Gamma \left( n \right) \Gamma \left( \frac{d}{2} \right)  } \Gamma \left( 1 - \frac{d}{2} \right) \Delta^{d - 2} \, .
\end{align}
Both contributions $\mathbf{C}_{2n}^{\mu_1 \ldots \mu_{2n}}$ and $\mathbf{D}_{2n}^{\mu_1 \ldots \mu_{2n}}$ scale as $q^d$ and contain a polynomial in $x_{i}$ of degree $d$. In $\mathbf{C}_{2n}^{\mu_1 \ldots \mu_{2n}}$, this $q^{d}$ dependence is contained within $\Delta^{d}$, and for $\mathbf{D}_{2n}^{\mu_1 \ldots \mu_{2n}}$, the power of $q^{d-2}$ from $\Delta^{d-2}$ is compensated by two factors of $q$, one of which sits in each function $f^\mu$.

Therefore, we conclude that in the limit $d \to 2$, all such $n$ vertex diagrams have the same general structure as the $2$-vertex diagram which we saw in (\ref{two_vertex}). In particular, both $\mathbf{C}_{2n}^{\mu_1 \ldots \mu_{2n}}$ and $\mathbf{D}_{2n}^{\mu_1 \ldots \mu_{2n}}$ generate divergences of the form $\frac{1}{\epsilon}$ because they are proportional to $\Gamma \left( - \frac{d}{2} \right)$ and $\Gamma \left( 1 - \frac{d}{2} \right)$, respectively.

We conclude that
\begin{align}
    \mathbf{C}_{2n}^{\mu_1 \ldots \mu_{2n}}&\sim \frac{1}{\epsilon} q^2 g^{\mu_1 \cdots \mu_{2n}}  \, , \nonumber \\
    \mathbf{D}_{2n}^{\mu_1 \ldots \mu_{2n}} &\sim \frac{1}{\epsilon} \sum_{a=1}^{2n} \sum_{b>a}^{2n} q^{\mu_a} q^{\mu_b} g^{\{ \mu \neq \mu_a, \mu_b \}}  \, .
\end{align}

\subsection{$m$-loop, two-vertex calculation} \label{sec:scalar_field_n_loop_2_vertex_calc}

In this appendix, we will show how to evaluate the integral over one of the $m$ loop momenta $\ell_i$ which appear in the expression for the two-vertex, $m$-loop diagram of equation (\ref{two_vertex_n_loop}). It suffices to integrate over the final momentum $\ell_{m}$, since the result may then be iterated to evaluate the other $m-1$ integrals.

Therefore, let us focus on performing the integration over $\ell_m$ in the quantity $\mathfrak{L}_{m, 2}$ of equation (\ref{frak_L_defn}). Specifically, we will compute the quantity
\begin{align}\label{one_integral_Lm2_defn}
    L_{m, 2} &= \int d^d\ell_{m} \frac{\ell_{m}^{\nu_{m+1}} \ell_{m}^{\tau_{m+1}} \left( \ell_{m} - \ell_{m-1} \right)^{\nu_{m}} \left( \ell_{m} - \ell_{m-1} \right)^{\tau_{m}}}{\ell_{m}^2 \left( \ell_{m} - \ell_{m-1} \right)^2} \, .
\end{align}
This object $L_{m, 2}$ is proportional to the remaining integrand that one finds by performing the integral over $\ell_m$ in the definition of $\mathfrak{L}_{m, 2}$. As we will see, after obtaining an expression for $L_{m,2}$, this result can be used recursively to evaluate $\mathfrak{L}_{m, 2}$ itself.

We notice the integral in equation (\ref{one_integral_Lm2_defn}) is exactly of the form of the one appearing in the $1$-loop, $2$-vertex diagram which we evaluated in appendix \ref{sec:scalar_field_1_loop_2_vertex_calc}. Proceeding in the same way, we introduce a Feynman parameter $x$ to write
\begin{align}
    L_{m, 2} &= \int d^d\ell_{m} dx \frac{\ell_{m}^{\nu_{m+1}} \ell_{m}^{\tau_{m+1}} \left( \ell_{m} - \ell_{m-1} \right)^{\nu_{m}} \left( \ell_{m} - \ell_{m-1} \right)^{\tau_{m}}}{\left[ \left( 1 - x \right) \ell_{m}^2  + x \left( \ell_{m} - \ell_{m-1} \right)^2 \right]^2} \nonumber \\
    &= \int d^d\ell_{m} dx \frac{\ell_{m}^{\nu_{m+1}} \ell_{m}^{\tau_{m+1}} \left( \ell_{m} - \ell_{m-1} \right)^{\nu_{m}} \left( \ell_{m} - \ell_{m-1} \right)^{\tau_{m}}}{\left[ \ell_{m}^2  + x \left(  2\ell_{m} \cdot \ell_{m-1}- \ell_{m-1}^2 \right) \right]^2} \nonumber \\
    &= \int d^d\ell_{m} dx \frac{\ell_{m}^{\nu_{m+1}} \ell_{m}^{\tau_{m+1}} \left( \ell_{m} - \ell_{m-1} \right)^{\nu_{m}} \left( \ell_{m} - \ell_{m-1} \right)^{\tau_{m}}}{\left[ \left( \ell_{m} + x \ell_{m-1} \right)^2 - x^2 \ell_{m-1}^2  + x \ell_{m-1}^2 \right]^2} \, ,
\end{align}
or after shifting the integration variable as $\ell_{m} \to \ell_{m} - x \ell_{m-1}$,
\begin{align}
    &L_{m, 2} = \int d^d\ell_{m} dx \nonumber  \\& \frac{\left( \ell_{m} - x\ell_{m-1} \right)^{\nu_{m+1}} \left( \ell_{m} - x\ell_{m-1} \right) ^{\tau_{m+1}} \left( \ell_{m} - \left( 1 - x \right) \ell_{m-1} \right)^{\nu_{m}} \left( \ell_{m} - \left( 1 - x \right) \ell_{m-1} \right)^{\tau_{m}}}{\left[ \ell_{m}^2 + x \left( 1 - x \right)  \ell_{m-1}^2  \right]^2} \, .
\end{align}
We may keep only even powers of $\ell_{m}$ in the integrand, as odd powers vanish by symmetry:
\begin{equation}
\begin{aligned}
    L_{m, 2} &= \int d^d\ell_{m} dx \Bigg( \frac{  \ell_{m}^{\nu_{m+1}} \ell_{m}^{\tau_{m+1}} \ell_{m}^{\nu_{m}} \ell_{m}^{\tau_{m}}  + x^2 \ell_{m-1}^{\nu_{m+1}} \ell_{m-1}^{\tau_{m+1}} \ell_{m}^{\nu_{m}} \ell_{m}^{\tau_{m}} + x \left( 1 - x \right)  \ell_{m}^{\nu_{m+1}} \ell_{m-1}^{\tau_{m+1}} \ell_{m}^{\nu_{m}} \ell_{m-1}^{\tau_{m}} }{\left[ \ell_{m}^2 + x \left( 1 - x \right)  \ell_{m-1}^2  \right]^2}  \\
    &+ \frac{x \left( 1 - x \right)  \ell_{m-1}^{\nu_{m+1}} \ell_{m}^{\tau_{m+1}} \ell_{m-1}^{\nu_{m}} \ell_{m}^{\tau_{m}} + \left( 1 - x \right)^2 \ell_{m}^{\nu_{m+1}} \ell_{m}^{\tau_{m+1}} \ell_{m-1}^{\nu_{m}} \ell_{m-1}^{\tau_{m}}}{\left[ \ell^2_{m} + x\left( 1 - x \right)\ell_{m-1}^2 \right]^2} \\&+ \frac{x^2 \left( 1 - x \right)^2 \ell_{m-1}^{\nu_{m+1}} \ell_{m-1}^{\tau_{m+1}} \ell_{m-1}^{\nu_{m}} \ell_{m-1}^{\tau_{m}} }{\left[ \ell^2_{m} + x\left( 1 - x \right)\ell_{m-1}^2 \right]^2} \Bigg) \, .
\end{aligned}
\end{equation}
We now replace products of $\ell_m^\mu$ with powers of $\ell_m$ and symmetrized metric factors, following the generalized symmetrization rule (\ref{eq:symmetrization}), which yields
\begin{align}
L_{m, 2} &= \int d^d\ell_{m} dx \Bigg( \frac{ \frac{\ell_{m}^{4}}{d \left( d + 2 \right) }  g^{(\nu_{m+1} \tau_{m+1}} g^{\nu_{m} \tau_{m} )}  + x^2 \frac{\ell_{m}^2}{d} g^{\mu_{m} \nu_{m}}\ell_{m-1}^{\nu_{m+1}} \ell_{m-1}^{\tau_{m+1}}  }{\left[ \ell_{m}^2 + x \left( 1 - x \right)  \ell_{m-1}^2  \right]^2} \nonumber \\&+  \frac{ x \left( 1 - x \right)  \frac{\ell_{m}^2}{d} g^{\nu_{m+1} \nu_{m}} \ell_{m-1}^{\tau_{m+1}}  \ell_{m-1}^{\tau_{m}} }{\left[ \ell_{m}^2 + x \left( 1 - x \right)  \ell_{m-1}^2  \right]^2} \nonumber \\
    &+ \frac{x \left( 1 - x \right)  \frac{\ell_{m}^2}{d} g^{\tau_{m+1} \nu_{m}}\ell_{m-1}^{\nu_{m+1}} \ell_{m-1}^{\tau_{m}}  + \left( 1 - x \right)^2 \frac{\ell_{m}^2}{d}  g^{{\nu_{m+1}}{\tau_{m+1}}} \ell_{m-1}^{\nu_{m}} \ell_{m-1}^{\tau_{m}} }{\left[ \ell^2_{m} + x\left( 1 - x \right)\ell_{m-1}^2 \right]^2} \nonumber \\&+  \frac{ x^2 \left( 1 - x \right)^2 \ell_{m-1}^{\nu_{m+1}} \ell_{m-1}^{\tau_{m+1}} \ell_{m-1}^{\nu_{m}} \ell_{m-1}^{\tau_{m}} }{\left[ \ell^2_{m} + x\left( 1 - x \right)\ell_{m-1}^2 \right]^2} \Bigg) \, .
\end{align}
Splitting the numerator up, we can once again apply the standard formula (\ref{known_integral}) with 
\begin{align}\label{deltasq_app_defn}
    \Delta^2 = - x \left(  1 - x \right) \ell_{m-1}^2 \, ,
\end{align}
which gives
\begin{align}
    L_{m, 2} &= \frac{i}{\left( 4\pi \right)^{\frac{d}{2}} \Gamma \left( \frac{d}{2} \right) }\int_{0}^{1} dx \Bigg( \frac{1}{d \left( d + 2 \right) } g^{(\nu_{m+1} \tau_{m+1}} g^{\nu_{m} \tau_{m})}  \Gamma \left( 2 + \frac{d}{2} \right) \Gamma \left( -\frac{d}{2} \right) \Delta^{d}  \nonumber \\
    &+ \frac{x^2}{d} g^{\mu_{m} \nu_{m}} \ell_{m-1}^{\nu_{m+1}} \ell_{m-1}^{\tau_{m+1}} \Gamma \left( 1 + \frac{d}{2} \right) \Gamma \left( 1 - \frac{d}{2} \right) \Delta^{d-2}  \nonumber\\
    & + \frac{x\left( 1 - x \right) }{d} g^{\nu_{m+1} \nu_{m}} \ell_{m-1}^{\tau_{m+1}} \ell_{m-1}^{\tau_{m}} \Gamma \left( 1 + \frac{d}{2} \right) \Gamma \left( 1 - \frac{d}{2} \right) \Delta^{d-2}  \nonumber\\
    & + \frac{x\left( 1 - x \right) }{d} g^{\tau_{m+1} \nu_{m}} \ell_{m-1}^{\nu_{m+1}} \ell_{m-1}^{\tau_{m}} \Gamma \left( 1 + \frac{d}{2} \right) \Gamma \left( 1 - \frac{d}{2} \right) \Delta^{d-2}  \nonumber\\
    &+ \frac{\left( 1 - x \right)^2 }{d} g^{\nu_{m+1} \tau_{m+1}} \ell_{m-1}^{\nu_{m}} \ell_{m-1}^{\tau_{m}} \Gamma \left( 1 + \frac{d}{2} \right) \Gamma \left( 1 - \frac{d}{2} \right) \Delta^{d-2}  \nonumber\\
    &+ x^2 \left( 1 - x \right)^2  \ell_{n-2}^{\nu_{m+1}} \ell_{m-1}^{\tau_{m+1}} \ell_{m-1}^{\nu_{m}} \ell_{m-1}^{\tau_{m}} \Gamma \left( \frac{d}{2} \right) \Gamma \left( 2 - \frac{d}{2} \right) \Delta^{d-4} \Bigg) \, .
\end{align}
Replacing $\Delta$ using its definition in equation (\ref{deltasq_app_defn}), we see that each term contains $d$ overall factors of loop momenta:
\begin{align}
    L_{m, 2} &= \frac{i}{\left( 4\pi \right)^{\frac{d}{2}} \Gamma \left( \frac{d}{2} \right) }\int_{0}^{1} dx \Bigg( \frac{ \Gamma \left( 2 + \frac{d}{2} \right) \Gamma \left( -\frac{d}{2} \right)}{d \left( d + 2 \right) }  g^{(\nu_{m+1} \tau_{m+1}} g^{\nu_{m} \tau_{m})}  \left( x\left( 1 - x \right) \right)^{\frac{d}{2}} \ell_{m-1}^{d}   \nonumber \\
    &+ \frac{\Gamma \left( 1 + \frac{d}{2} \right) \Gamma \left( 1 - \frac{d}{2} \right) }{d} g^{\mu_{m} \nu_{m}} \ell_{m-1}^{\nu_{m+1}} \ell_{m-1}^{\tau_{m+1}} x^{\frac{d}{2}+1}\left( 1 - x \right)^{\frac{d}{2} - 1} \ell_{m-1}^{d-2} \nonumber\\
    &+ \frac{\Gamma \left( 1 + \frac{d}{2} \right) \Gamma \left( 1 - \frac{d}{2} \right)}{d} g^{\nu_{m+1} \nu_{m}} \ell_{m-1}^{\tau_{m+1}} \ell_{m-1}^{\tau_{m}}  \left( x\left( 1- x \right)  \right)^{\frac{d}{2}} \ell_{m-1}^{d-2} \nonumber\\
    & + \frac{\Gamma \left( 1 + \frac{d}{2} \right) \Gamma \left( 1 - \frac{d}{2} \right) }{d} g^{\tau_{m+1} \nu_{m}}  \left( x\left( 1- x \right)  \right)^{\frac{d}{2}} \ell_{m-1}^{\nu_{m+1}} \ell_{m-1}^{\tau_{m}}\ell_{m-1}^{d-2} \nonumber\\
    &+ \frac{\Gamma \left( 1 + \frac{d}{2} \right) \Gamma \left( 1 - \frac{d}{2} \right)}{d} g^{\nu_{m+1} \tau_{m+1}}   x^{\frac{d}{2}-1}\left( 1- x \right)^{\frac{d}{2}+1}  \ell_{m-1}^{\nu_{m}} \ell_{m-1}^{\tau_{m}}\ell_{m-1}^{d-2} \nonumber\\
    &+  \Gamma \left( \frac{d}{2} \right) \Gamma \left( 2 - \frac{d}{2} \right) \left( x\left( 1- x \right)  \right)^{\frac{d}{2}} \ell_{m-1}^{\nu_{m+1}} \ell_{m-1}^{\tau_{m+1}} \ell_{m-1}^{\nu_{m}} \ell_{m-1}^{\tau_{m}}\ell_{m-1}^{d-4} \Bigg) \, .
\end{align}
We now use the identity $\Gamma \left( 1 + x \right) = x \Gamma \left( x \right)$ to factor and cancel the gamma functions, and finally evaluate the Feynman integrals, giving the result
\begin{align}
    L_{m, 2} &= \frac{i\Gamma \left( -\frac{d}{2} \right)}{\left( 4\pi \right)^{\frac{d}{2}} \Gamma \left( d + 2 \right) } \Bigg( \frac{ \left( 1 + \frac{d}{2} \right) \frac{d}{2}}{d \left( d + 2 \right) }   g^{(\nu_{m+1} \tau_{m+1}} g^{\nu_{m} \tau_{m})}   \Gamma \left( \frac{d}{2}+1 \right)^2 \ell_{m-1}^{d}   \nonumber \\
    &\qquad \qquad \qquad + \frac{\left( \frac{d}{2} \right) \left( -\frac{d}{2} \right)  }{d} g^{\mu_{m} \nu_{m}} \ell_{m-1}^{\nu_{m+1}} \ell_{m-1}^{\tau_{m+1}}  \Gamma \left( \frac{d}{2} + 2 \right) \Gamma \left( \frac{d}{2} \right) \ell_{m-1}^{d-2} \nonumber\\
    &\qquad \qquad \qquad + \frac{\frac{d}{2}\left( -\frac{d}{2} \right) }{d} g^{\nu_{m+1} \nu_{m}} \Gamma \left( \frac{d}{2} + 1 \right)^2 \ell_{m-1}^{\tau_{m+1}} \ell_{m-1}^{\tau_{m}} \ell_{m-1}^{d-2} \nonumber\\
    &\qquad \qquad \qquad + \frac{\frac{d}{2}\left( -\frac{d}{2} \right) }{d} g^{\tau_{m+1} \nu_{m}}   \Gamma \left( \frac{d}{2} + 1 \right)^2 \ell_{m-1}^{\nu_{m+1}} \ell_{m-1}^{\tau_{m}}\ell_{m-1}^{d-2} \nonumber\\
    &\qquad \qquad \qquad + \frac{\frac{d}{2}\left( -\frac{d}{2} \right) }{d} g^{\nu_{m+1} \tau_{m+1}}   \Gamma \left( \frac{d}{2} + 2 \right) \Gamma \left( \frac{d}{2} \right)  \ell_{m-1}^{\nu_{m}} \ell_{m-1}^{\tau_{m}}\ell_{m-1}^{d-2} \nonumber\\
    &\qquad \qquad \qquad +  \left( 1-\frac{d}{2} \right) \left( -\frac{d}{2} \right)   \Gamma \left( \frac{d}{2} + 1 \right)^2 \ell_{m-1}^{\nu_{m+1}} \ell_{m-1}^{\tau_{m+1}} \ell_{m-1}^{\nu_{m}} \ell_{m-1}^{\tau_{m}}\ell_{m-1}^{d-4} \Bigg) \, .
\end{align}
We see that all terms scale as $\ell_{m-1}^{d}$ and the only divergent gamma function is $\Gamma \left( -\frac{d}{2} \right) $. This establishes the result used in the body of section \ref{sec:cian}, in the text above equation (\ref{final_n_loop_2_vertex_dimreg}), which can then be applied iteratively to evaluate the remaining loop integrals.

\renewcommand{\refname}{References} 
\bibliographystyle {unsrt}
\bibliography{demo2_ref}    

\end {document}